\documentclass[12pt]{article}

\makeatletter
\let\SuppMakeTitleInternal\@maketitle
\makeatother

\newlength{\SuppLayoutLeft}
\newlength{\SuppLayoutTop}
\newlength{\SuppLayoutWidth}
\newlength{\SuppLayoutHeight}
\usepackage{geometry}
\usepackage[T1]{fontenc} 
\usepackage{setspace}
\usepackage{tocloft} 
\usepackage{indentfirst} 
\usepackage{amsthm}
\usepackage[hidelinks]{hyperref}
\usepackage{graphics,graphicx,amssymb,mathtools}
\usepackage{caption}
\usepackage{rotating}
\usepackage{adjustbox}
\usepackage{enumerate}
\usepackage{booktabs}
\usepackage{mathrsfs}  
\usepackage{url} 
\usepackage{xcolor}
\usepackage{bm}
\usepackage{siunitx}
\usepackage{booktabs}
\usepackage{footnote}
\usepackage{tablefootnote}
\usepackage[flushleft]{threeparttable}
\usepackage{multirow}
\usepackage{xr} 
\onecolumn 
\usepackage{subcaption}
\usepackage{natbib}
\usepackage{footmisc}
\usepackage{dsfont}
\usepackage{dsfont}
\usepackage{amssymb}
\usepackage{algorithm}
\usepackage{algorithmic}
\usepackage{array}
\usepackage{threeparttable}

\usepackage{hyperref}
\hypersetup{colorlinks=true,linkcolor=blue, citecolor=blue, linktocpage}

\allowdisplaybreaks

\let\hat\widehat
\let\tilde\widetilde

\newcommand{\para}{\delta}
\newcommand{\Para}{\Delta}

\newtheorem{prop}{Proposition}

\newtheorem{definition}{Definition}

\newtheorem{theorem}{Theorem}

\newcommand{\bt}{{\mathbf t}}

\DeclareMathOperator*{\argmax}{arg\,max}
\DeclareMathOperator*{\argmin}{arg\,min}
\usepackage{url}

\makeatletter
\date{} 

\begin{document}
\addtocontents{toc}{\protect\setcounter{tocdepth}{-1}}

\def\spacingset#1{\renewcommand{\baselinestretch}%
{#1}\small\normalsize} \spacingset{1}
\title{A Dynamic Factor Model for Multivariate Counting Process Data}

  \author{Fangyi Chen$^1$, Hok Kan Ling$^{2}$ and Zhiliang Ying$^1$ ~\\~\\$^1$\footnotesize\textit{Department of Statistics, Columbia University}~\\$^2$\footnotesize\textit{Department of Mathematics and Statistics, Queen's University}}
  \maketitle

\bigskip
\begin{abstract}
    We propose a dynamic multiplicative factor model for process data arising from complex problem-solving items, an emerging type of data in large-scale educational assessment. The proposed model can be viewed as an extension of the classical frailty models developed in survival analysis for multivariate recurrent event times, but with two important distinctions: (i) the factor (frailty) is of primary interest; (ii) covariates are internal and embedded in the factor. It allows us to explore low-dimensional structure with meaningful interpretation. We show that the proposed model is generically identifiable and that the maximum likelihood estimators are consistent and asymptotically normal. Furthermore, to obtain a parsimonious model and to improve the interpretation of parameters, variable selection and estimation for both fixed and random effects are developed through suitable penalisation. The computation is carried out using a stochastic EM algorithm with elliptical slice sampling in the stochastic E-step and coordinate descent in the M-step. Simulation studies demonstrate that the proposed approach effectively recovers the true structure. The proposed method is applied to the analysis of the log file of an item from the Programme for the International Assessment of Adult Competencies (PIAAC), and meaningful relationships are identified.
\end{abstract}

\noindent%
{\it Keywords:} Educational measurement; Generalised linear factor model; Multivariate event time data; Process data; Proportional intensity model 
\vfill
 %
\newpage 
\spacingset{1.9} 

\section{Introduction}
This paper is motivated by the need for statistical modelling and analysis of process data, which often consist of sequences of events of the same or different types and are commonly encountered in many disciplines, including biomedical studies, marketing research, and educational assessment.
In biomedical studies, it is of interest to examine the joint occurrence of different kinds of diseases and their relationship with covariates such as treatment assignments, demographic characteristics and exposure histories, among others. In marketing research, one may be interested in customers' purchasing patterns and their relationship to baseline demographic characteristics, dynamically collected covariate processes and interventions such as advertisements and promotions. Analysing such data is complicated by the dynamic nature of both the events of interest and the covariate processes. Furthermore, the data are often heterogeneous and contain a large number of different types of events and covariate processes. Our main goal here is to propose a model for the joint analysis of such data, motivated by the emergence of large-scale computer-based assessment in educational research. 

Computer-based assessments that involve interactive environments, such as simulation-based or scenario-based assessments, have become increasingly popular.
For example, the Organisation for Economic Co-operation and Development (OECD) has been administering interactive and scenario-based questions in the Programme for International Student Assessment (PISA) and the Programme for the International Assessment of Adult Competencies (PIAAC). In the US, the National Assessment of Educational Progress (NAEP) has been using interactive computer tasks in science and in technology and engineering literacy in recent years \cite[]{nichols2012high,bergner2019process,pellegrino2021naep,jiang2021using,jiang2023using}. At the same time, technological advances now allow the action sequences generated while solving a problem, together with their timestamps, to be recorded in log files. These process data could provide new insights into individual characteristics, such as test takers’ motivation, engagement, persistence, and problem-solving strategies, whereas traditional task analysis and scoring normally focus only on the final task outcomes. For instance, \cite{lee2014using} used response times to filter for test taker motivation and \cite{halpin2017measuring} measured student engagement in collaboration using process data. 
Because of the potential benefits and the additional information that can be obtained from analysing process data, related research has recently received considerable attention in the educational measurement literature \cite[]{hao2015analyzing, he2016analyzing, zhu2016using, shu2017item, liu2018analysis,qin2019advances,fischer2020mining,he2021leveraging,wang2023subtask,zhang2023accurate}.

We propose to handle process data by viewing the data as a multivariate counting process, specified through a dynamic multiplicative factor model. There is a substantial literature in survival analysis on the modelling and analysis of multivariate event time data; see, for example, \cite{vaupel1979impact, prentice1981regression, wei1989regression, lee1992cox, liang1993modelling, yashin1995correlated, parner1998asymptotic, vaida2000proportional,yin2005class,cook2007statistical,zeng2007semiparametric,zeng2010general,sun2013statistical,brilleman2019joint,zeng2021maximum,xu2023marginal}. These approaches mostly rely on the use of marginal models or frailty (random effects) models. Marginal models are used to avoid directly modelling the dependence and to link the events of interest to covariates, whereas frailty terms are included to model unobserved heterogeneity and dependence among different event types. In both cases, the primary focus is on regression effects: marginal-model effects are interpreted as population-average effects, whereas frailty-model effects are interpreted as subject-specific effects. On the other hand, in educational and psychological measurement applications, factor analysis and interpretation of the factors are integral components \citep{reckase200618}. In fact, in measurement models, the factors are the main target of interest.

To understand individuals' problem-solving processes, it is natural and necessary to use previous actions (events) as (internal) covariates for subsequent actions and to encode factors into actions; for internal covariates, see \cite{kalbfleisch2011statistical}.  As such, marginal models are not applicable, and standard frailty models are also not suitable. Our proposed model includes internal covariates and encodes factors through these covariates, resulting in a dynamic multiplicative factor model.

As in all other factor models, establishing identifiability is a fundamental and often challenging issue. This can be especially hard when internal covariates are present. In fact, to the best of our knowledge, there are no results in the survival analysis literature on the identifiability of mixed-effects models when internal covariates are present. The main contributions of the present paper are to propose a dynamic multiplicative factor model and to establish identifiability results. 
In addition, we obtain the maximum likelihood estimator of the model parameters and establish its consistency and asymptotic normality. Furthermore, we propose a method to deal with variable selection in both the regression and factor components.

The rest of the paper is organised as follows. In Section~\ref{sect:model}, we introduce the notation and propose our model. In Section~\ref{sect:theory}, we discuss the issue of identifiability and provide sufficient conditions under which the proposed model is generically identifiable and the maximum likelihood estimator is consistent and asymptotically normal. Moreover, we develop a variable selection procedure via suitable penalisation, and establish selection consistency and the oracle property of the resulting estimator. Section~\ref{sect:algorithm} presents the computational algorithms. The proposed method is applied to the 2012 PIAAC data in Section~\ref{sect:real_data}, and simulation studies are reported in Section~\ref{sect:simulation}. Section~\ref{sect:discussion} gives some concluding remarks. Technical proofs and additional simulation results are provided in the Supplementary Materials.

\section{Notation and Model Specification}\label{sect:model}
Suppose there are $J$ possible types of events, and let $\mathcal{J} = \{1,\ldots,J \}$ denote the set of event types. Formally, process data consist of observations of the form $\{(a_1,t_1),\ldots, (a_m,t_m) \}$, where $a_k\in\mathcal{J}$ denotes the type of the $k$th event, $t_k\in\mathbb{R}_+$ denotes the corresponding timestamp, and the timestamps satisfy $t_k<t_{k+1}$. Here, $m$ denotes the number of events for a subject. The data consist of independent observations from $n$ subjects.

For the $i$th subject, let $\boldsymbol{X}_{ij}(\cdot)$ and $\boldsymbol{Z}_{ij}(\cdot)$ denote the $L_{1j}$- and $L_{2j}$-dimensional left-continuous covariate processes corresponding to the fixed effects and random effects for the $j$th event type, $i=1,\ldots,n$, $j=1,\ldots,J$. Let $N_{ij}^{\ast}(t)$ be the number of events of type $j$ that have occurred over $[0,t]$. Let $C_{ij}$ denote the right-censoring time for the $j$th event type and $\boldsymbol{N}_i(t) = (N_{i1}(t),\ldots,N_{iJ}(t))^{\top}$, where $N_{ij}(\cdot)=N^*_{ij}(\cdot \land C_{ij})$ corresponds to the observed part of the counting process of the $j$th event type. Let $\mathscr{F}_t=\sigma\{N_{ij}(s), \boldsymbol{X}_{ij}(s), \boldsymbol{Z}_{ij}(s), Y_{ij}(s), i=1,\ldots,n,j=1,\ldots,J;~0 \leq s \leq t\}$ be the filtration. Here, $Y_{ij}(\cdot)$ is the observed at-risk indicator process, which is predictable with respect to $\mathscr{F}_t$ \citep{andersen1993statistical}. We specify the intensity of the observed counting process $N_{ij}(t)$ in counting-process form, with $Y_{ij}(t)$ ensuring that the intensity is zero whenever subject $i$ is not at risk for event type $j$. The intensity function takes the form
\begin{equation}\label{eq:our_model}
\lambda_{ij}(t|\mathscr{F}_{t^-};\boldsymbol{\theta}_i) = \lambda_{j0}(t)Y_{ij}(t)e^{ \boldsymbol{\beta}^{\top}_j \boldsymbol{X}_{ij}(t) + \boldsymbol{\theta}^{\top}_i  \mathbf{A}^{\top}_j \boldsymbol{Z}_{ij}(t)},
\end{equation}
where $\boldsymbol{\beta}_j$ is an $L_{1j}$-dimensional vector of regression coefficients for the event-specific fixed effects, $\lambda_{j0}(\cdot)$ is the event-specific baseline intensity function, which is common to all subjects, $\boldsymbol{\theta}_i$ is the subject-specific $K$-dimensional random effect, and $\mathbf{A}_j$ is an event-specific $L_{2j} \times K$ factor loading matrix.

Note that model (\ref{eq:our_model}) contains many well-known models in survival analysis as special cases. 
\begin{enumerate}[(i)]
\item When $L_{2j}=0$, the model simplifies to the multivariate proportional hazards model \cite[]{andersen1982cox}:
\begin{equation*}
\lambda_{ij}(t|\mathscr{F}_{t^-}) = \lambda_{j0}(t)Y_{ij}(t)e^{\boldsymbol{\beta}^{\top}_j \boldsymbol{X}_{ij}(t) }, \quad j=1,\ldots,J.
\end{equation*}
\item When $K = L_{2j}$ and $\mathbf{A}_j$ is the identity matrix, this corresponds to a multivariate proportional hazards model with random effects:
\begin{equation*}
\lambda_{ij}(t|\mathscr{F}_{t^-}; \boldsymbol{\theta}_i) = \lambda_{j0}(t)Y_{ij}(t)e^{\boldsymbol{\beta}_j^{\top} \boldsymbol{X}_{ij}(t) + \boldsymbol{\theta}^{\top}_i \boldsymbol{Z}_{ij}(t)}, \quad j=1,\ldots,J.
\end{equation*}
In particular, when $\lambda_{j0}(t) \equiv \lambda_0(t)$, it is a model for clustered survival data \cite[]{vaida2000proportional}, where $i$ indexes the cluster and $j$ indexes the observation.
\item When $J=1$, $L_{21}=1$, $Z_{i1}(t) \equiv 1$, and $K=1$,  it reduces to the standard frailty model \cite[]{vaupel1979impact}:
\begin{equation*}
\lambda_i(t|\mathscr{F}_{t^-}; \theta_i) = \lambda_0(t)Y_{i}(t)e^{\boldsymbol{\beta}^{\top} \boldsymbol{X}_i(t) + \theta_i} = \tilde{\theta}_i \lambda_0(t) Y_{i}(t)e^{\boldsymbol{\beta}^{\top} \boldsymbol{X}_i(t)},
\end{equation*}
where $\tilde{\theta}_i := e^{\theta_i}$.
\item When $L_{2j} = 1, Z_{ij}(t) \equiv 1$, and $K=1$, it reduces to the shared frailty model \cite[]{hougaard2000analysis}:
\begin{equation*}
\lambda_{ij}(t|\mathscr{F}_{t^-};\theta_i) = \lambda_{j0}(t)Y_{ij}(t)e^{\boldsymbol{\beta}^{\top}_j \boldsymbol{X}_{ij}(t) + a_j \theta_i}, \quad j=1,\ldots,J.
\end{equation*}
\item When $L_{1j} = 0$ and $L_{2j} = 1$ with $Z_{ij}(t) \equiv 1$, it reduces to a factor model for multivariate counting processes:
\begin{equation*}
\lambda_{ij}(t|\mathscr{F}_{t^-};\boldsymbol{\theta}_i) = \lambda_{j0}(t)Y_{ij}(t)e^{\boldsymbol{a}^{\top}_j \boldsymbol{\theta}_i},\quad j=1,\ldots,J,
\end{equation*}
which further reduces to a Poisson factor model when all the baseline functions are constant; see, for example, \cite{wedel2003factor}.
\end{enumerate}

For simplicity, we consider only the case in which the baseline intensity functions are constant, that is, $\log\lambda_{j0}(t)\equiv\beta_{j0}$, and the random effects follow a multivariate normal distribution $\mathcal{N}(\boldsymbol{0},\mathbf{\Sigma})$. The model and likelihood can be adapted to accommodate a time-varying parametric baseline intensity, such as a Weibull or Gompertz intensity. 
Sensitivity analyses using event-specific piecewise-constant log-baseline
specifications in both the simulation study and the real-data application
are reported in Section~S.2 of the Supplementary Materials.
A semiparametric extension is also feasible under additional structural assumptions; see Section~S.8.2 of the Supplementary Materials.

Our model differs from standard multivariate event time models \citep{parner1998asymptotic,zeng2007semiparametric} in two important aspects. First, in order to study the subject-specific behavioural structure from process data, actions from each subject are incorporated as covariates; that is, a subject's earlier action influences his/her subsequent actions through the intensity function. Since actions are themselves modelled as the outcome of the counting process, they act as internal covariates in our model. Internal covariates are considerably more complex and subtle to handle than external covariates. In particular, existing results on model identifiability developed for external covariates do not carry over to settings with internal covariates. Second, the random-effect component (factors) is the primary object of interest in our model, whereas the fixed-effect component (regression parameters) is usually the focus in standard event time models.

Recently, \citet{chen2025dynamic} develop a semiparametric dynamic factor model for high-dimensional recurrent-event data. Their model places a low-rank factor structure on the marginal rate functions that vary continuously over time. This factor structure primarily facilitates dimension reduction and exploratory interpretation, while their theoretical analysis focuses on kernel-smoothed estimation and the associated convergence rates.
The present paper addresses a very different modelling objective. Model \eqref{eq:our_model} specifies conditional intensities that incorporate dynamics arising from internal covariates constructed from the action history prior to the events.
These internal covariates are crucial for understanding the behavioural transition mechanisms as shown in the data analysis example in Section \ref{sect:real_data}. Consequently, the factor structure has a more mechanism-oriented role and supports confirmatory analysis, while still permitting exploratory interpretation of the estimated loadings.
Broadly speaking, the framework of \cite{chen2025dynamic} is closely related to functional data analysis, whereas the present work is more naturally aligned with sequential analysis and adaptive experimentation that are especially relevant to adaptive control and learning (see, for example, \cite{tang2019reinforcement}).

\section{Main Theoretical Results}\label{sect:theory}

Let $\boldsymbol{\para}\in\mathbb{R}^d$ denote the vector that stacks all free parameters in $\boldsymbol{\beta}$, $\mathbf{A}$, and $\mathbf{\Sigma}$, where $\boldsymbol{\beta} = \{\beta_{j0},\; \boldsymbol{\beta}_j:j=1,\ldots,J\}$ and $\mathbf{A} =(\mathbf{A}_1^{\top},\ldots,\mathbf{A}_J^{\top})^{\top}$. Let $\boldsymbol{\para}_0$ denote the true value of $\boldsymbol{\para}$. To study model identifiability and the asymptotic behaviour of the maximum likelihood estimator, we impose the following conditions:
\begin{enumerate}[(a)]
\item $\boldsymbol{\para}_0$ lies in the interior of a known compact set $\mathbf{\Para}\subset \mathbb{R}^d$ such that $\boldsymbol{\Sigma}^{-1}$ has uniformly bounded entries.

\item \label{cond:D1} For $i=1,\ldots,n$, $j=1,\ldots,J$, the covariate processes $\boldsymbol{X}_{ij}(\cdot)$ and $\boldsymbol{Z}_{ij}(\cdot)$ are elementwise uniformly bounded by a constant $M>0$.

\item \label{cond:D2} After a prespecified permutation of the rows in $\mathbf A$ and a suitable permutation of its columns, the first $K$ rows of $\mathbf A$ form an identity matrix.

\item \label{cond:D3} For fixed $j,l\in\{1, \ldots, J\}$, if there exist $\nu$ and $\boldsymbol{\mu}$ such that $\nu+\boldsymbol{\mu}^{\top} \boldsymbol{X}_{ij}(t)=0$ for every $i=1,\ldots,n$ and $0\leq t\leq C_{ij}$, then $\nu=0$ and $\boldsymbol{\mu}=\mathbf{0}$; if there exists a matrix $\mathbf{U}$ such that ${\boldsymbol{Z}^{\top}_{i j}}(t)\mathbf{U} \boldsymbol{Z}_{i l}(s)=0$ for every $i=1,\ldots,n$ and $0\leq t,s\leq C_{ij}\land C_{il}$, then $\mathbf{U}=\mathbf{0}$.
\item \label{cond:D4} For $i=1,\ldots,n$, $j=1,\ldots,J$, $\boldsymbol{X}_{ij}(\cdot)$ and $\boldsymbol{Z}_{ij}(\cdot)$ are piecewise constant on $[0,C_{ij}]$. Furthermore, the distributions of $\boldsymbol{X}_{ij}(t+0)$ and $\boldsymbol{Z}_{ij}(t+0)$ given $\mathscr{F}_t$ do not depend on the model parameters for any given $i=1,\ldots,n$, $j=1,\ldots,J$ and $0\leq t\leq C_{ij}$. 
\item For $i=1,\ldots,n$, $j=1,\ldots,J$, the censoring time $C_{ij}$ is uniformly bounded by a constant $\tau>0$. Furthermore, the conditional probability of $C_{ij}>t$ does not depend on the model parameters given $\mathscr{F}_t$ and $\boldsymbol{\theta}_i$ for any $0\leq t\leq\tau$.
\end{enumerate}
Condition (a) is standard for maximum likelihood estimation. Condition (b) is also standard when dealing with time-dependent covariates. Among other things, it guarantees the existence of the information matrix. Condition (c) anchors the rotation and scaling of matrices $\mathbf{A}$ and $\mathbf{\Sigma}$, and is also standard in multidimensional item response theory; see, for example, \cite{sun2016latent} and \cite{beguin2001mcmc}, and the references therein. In practice, we may not impose the scaling restriction and only require a diagonal submatrix of $\mathbf{A}$, in which case the scaling is imposed on $\mathbf{\Sigma}$ instead. The first part of Condition (d) is specific to the parametric model adopted in this paper and rules out collinearity between the baseline component and the fixed-effect covariates. The second part of Condition (d) requires sufficient variation in the random-effect covariates to identify the loading–covariance structure. The first part of Condition (e) is necessary in the presence of internal covariates. A counterexample presented in Section~S.3.3 of the Supplementary Materials shows that the model may fail to be generically identifiable even when the covariates evolve linearly, thereby illustrating the necessity of the first part of Condition (e), i.e., the piecewise-constant assumption. The second part of Condition (e) guarantees that the covariate processes do not provide extra information about the model parameters apart from the multivariate counting process. Without the second part of Condition (e), the likelihood function constructed below becomes a partial likelihood function and the resulting inferential procedures remain valid \citep{wong1986theory}. Condition (f) pertains to the assumption of independent and noninformative censoring \citep{nielsen1992counting}. A joint model for recurrent events and censoring can be used to accommodate informative censoring.


Under these conditions, the likelihood function for the parameters $\boldsymbol{\para}=(\boldsymbol{\beta},\mathbf{A},\mathbf{\Sigma})$ in model (\ref{eq:our_model}) can be expressed as
\begin{align}\label{eq:likelihood}
	L_n(\boldsymbol{\para}&| \textbf{N}, \textbf{X}, \textbf{Z}) = \prod^n_{i=1} \int_{\mathbb{R}^K}   \exp \bigg\{\sum_{j=1}^J \int^{\infty}_0Y_{ij}(t)\big(\beta_{j0} + \boldsymbol{\beta}^{\top}_j \boldsymbol{X}_{ij}(t) + \boldsymbol{\theta}^{\top} \mathbf{A}_j^{\top} \boldsymbol{Z}_{ij}(t)\big)\mathrm{d}N_{ij}(t) \bigg\}  \notag\\
	&\times \exp \bigg\{  -\sum_{j=1}^J\int^{\infty}_0 Y_{ij}(t)\exp\left( \beta_{j0} + \boldsymbol{\beta}^{\top}_j \boldsymbol{X}_{ij}(t) + \boldsymbol{\theta}^{\top} \mathbf{A}_j^{\top} \boldsymbol{Z}_{ij}(t) \right)  \mathrm{d}t  \bigg\}   \phi_K (\boldsymbol{\theta};\boldsymbol{0},\mathbf{\Sigma}) \mathrm{d}\boldsymbol{\theta},
\end{align}
where $(\textbf{N}, \textbf{X}, \textbf{Z}) := \{N_{ij}(s), \boldsymbol{X}_{ij}(s), \boldsymbol{Z}_{ij}(s) : 0 \leq s \leq C_{ij},\, i=1,\ldots,n,\, j=1,\ldots,J\}$ and $\phi_K(\cdot;\boldsymbol{0},\mathbf{\Sigma})$ is the multivariate normal density with mean vector $\boldsymbol{0}$ and covariance matrix $\boldsymbol{\Sigma}$.

Due to the complexity caused by the internal covariates, identifiability is a challenging issue. A simple counterexample shows that the resulting model becomes non-identifiable for certain parameter configurations in the presence of internal covariates; see Section~S.3.1 in the Supplementary Materials. To exclude such singular cases, we adopt the concept of generic identifiability; see \cite{10.1214/09-AOS689}.
\begin{definition}[Generic Identifiability]\label{def:gen_iden}
    Model (\ref{eq:our_model}) is said to be generically identifiable if there exists a set $\boldsymbol{\mathcal{V}}\subset\mathbf{\Para}$ of zero Lebesgue measure such that for any $\boldsymbol{\para}\in \mathbf{\Para}\setminus\boldsymbol{\mathcal{V}}$, if there exists $\tilde{\boldsymbol{\para}}\in \mathbf{\Para}$ satisfying $L_1(\boldsymbol{\para} |\textbf{N}, \textbf{X}, \textbf{Z})=L_1(\tilde{\boldsymbol{\para}}|\textbf{N}, \textbf{X}, \textbf{Z})$ with probability one, then $\boldsymbol{\beta}=\tilde{\boldsymbol{\beta}}$ and $(\mathbf{A}, \mathbf{\Sigma}) \sim(\widetilde{\mathbf{A}}, \widetilde{\mathbf{\Sigma}})$, i.e., there exists a permutation matrix $\mathbf{Q}$ such that $\mathbf{A}\mathbf{Q}^{\top}=\widetilde{\mathbf{A}}$ and $\mathbf{Q}\mathbf{\Sigma} \mathbf{Q}^{\top}=\widetilde{\mathbf{\Sigma}}$.
\end{definition}
The following theorem establishes the generic identifiability of model (\ref{eq:our_model}).
\begin{theorem}\label{thm_identifiability} 
    Under Conditions (c)-(f), model (\ref{eq:our_model}) is generically identifiable.
\end{theorem}


Identifiability typically guarantees the consistency of parameter estimation \citep{wald1949note}. Proving Theorem \ref{thm_identifiability}  is challenging due to several factors: (i) the presence of internal covariates could significantly reduce the richness of the data space; (ii) the likelihood function (\ref{eq:likelihood}) does not have an explicit form and, as a result, a Laplace-type approximation is needed to handle the integral; (iii) the intensity functions of different event types are mixed together in the likelihood; (iv) the presence of a low-rank factor structure introduces additional complexity. Note that existing identifiability results (e.g., \cite{parner1998asymptotic} and \cite{zeng2007semiparametric}) require the covariates to be external and, therefore, are not applicable to the present setting.

To establish the asymptotic normality of the maximum likelihood estimator, the Fisher information for model (\ref{eq:our_model}) must be nonsingular, as stated in the following theorem.
\begin{theorem}\label{thm_information}
    Under Conditions (b)-(f), the Fisher information matrix
    \begin{align*}
        \mathbf{I}(\boldsymbol{\para}):=\mathbb{E}\left[\left\{\frac{\partial}{\partial \boldsymbol{\para}} \log L_1(\boldsymbol{\para} | \textbf{N}, \textbf{X}, \textbf{Z})\right\}\left\{\frac{\partial}{\partial \boldsymbol{\para}} \log L_1(\boldsymbol{\para} | \textbf{N}, \textbf{X}, \textbf{Z})\right\}^{\top}\right]
    \end{align*}
    is finite and strictly positive definite at $\boldsymbol{\para}=\boldsymbol{\para}_0\in\boldsymbol{\Para}\setminus \boldsymbol{\mathcal{V}}$, where $\boldsymbol{\mathcal{V}}$ is a set with zero Lebesgue measure as in Definition \ref{def:gen_iden}.
\end{theorem}
Let $\hat{\boldsymbol{\para}}$ be the MLE of model (\ref{eq:our_model}). Based on Theorems \ref{thm_identifiability} and \ref{thm_information}, we obtain the following result on the consistency and asymptotic normality of $\hat{\boldsymbol{\para}}$.
\begin{theorem}\label{thm_normality}
	Under Conditions (a)-(f), $\hat{\boldsymbol{\para}}$ is consistent and asymptotically normal; that is, $\hat{\boldsymbol{\para}}\rightarrow\boldsymbol{\para}_0$ in probability and $\sqrt{n}(\hat{\boldsymbol{\para}}-\boldsymbol{\para}_0)\rightarrow\mathcal{N}(\boldsymbol{0},\mathbf{I}^{-1}(\boldsymbol{\para}_0))$ in distribution.
\end{theorem}

After obtaining $\hat{\boldsymbol{\para}}$ using the EM-type algorithm described in Section~\ref{sect:algorithm}, the standard errors of the parameter estimates can be computed from the square roots of the diagonal elements of $(n \hat{\mathbf{I}}(\hat{\boldsymbol{\para}}))^{-1}$, where $n \hat{\mathbf{I}}(\hat{\boldsymbol{\para}})$ is an approximation of the observed Fisher information matrix (see Section~4.3 of \cite{mclachlan2007algorithm}), given by
\begin{align}\label{eq:empirical_observed_Info}
  n \hat{\mathbf{I}}(\hat{\boldsymbol{\para}}) :=&  
  \sum^n_{i=1} 
    \boldsymbol{S}_{\text{observed}} (\hat{\boldsymbol{\para}}|\boldsymbol{N}_i, \boldsymbol{X}_i, \boldsymbol{Z}_i) 
     \boldsymbol{S}_{\text{observed}} (\hat{\boldsymbol{\para}}|\boldsymbol{N}_i, \boldsymbol{X}_i, \boldsymbol{Z}_i)^\top \nonumber \\
   =&\sum^n_{i=1} \mathbb{E}  \left(\boldsymbol{S}(\hat{\boldsymbol{\para}}|\boldsymbol{N}_i, \boldsymbol{X}_i, \boldsymbol{Z}_i, \boldsymbol{\theta}_i) \big| \boldsymbol{N}_i, \boldsymbol{X}_i, \boldsymbol{Z}_i \right)
    \mathbb{E}  \left(\boldsymbol{S}(\hat{\boldsymbol{\para}}|\boldsymbol{N}_i, \boldsymbol{X}_i, \boldsymbol{Z}_i, \boldsymbol{\theta}_i)^\top \big| \boldsymbol{N}_i, \boldsymbol{X}_i, \boldsymbol{Z}_i \right),
\end{align}
where $\boldsymbol{X}_i = \{\boldsymbol{X}_{i1},\ldots,\boldsymbol{X}_{iJ}\}$ and $\boldsymbol{Z}_i = \{\boldsymbol{Z}_{i1},\ldots,\boldsymbol{Z}_{iJ}\}$.
Here, $\boldsymbol{S}_{\text{observed}} (\hat{\boldsymbol{\para}}|\boldsymbol{N}_i, \boldsymbol{X}_i, \boldsymbol{Z}_i)$ and $\boldsymbol{S}(\hat{\boldsymbol{\para}}|\boldsymbol{N}_i, \boldsymbol{X}_i, \boldsymbol{Z}_i, \boldsymbol{\theta}_i)$ denote the score functions of the observed data and the complete data for the $i$th subject, respectively. Specifically, 
\begin{equation*}
    \boldsymbol{S}_{\text{observed}} (\hat{\boldsymbol{\para}}|\boldsymbol{N}_i, \boldsymbol{X}_i, \boldsymbol{Z}_i) = \frac{\partial}{\partial \boldsymbol{\para}} \log L_1(\hat{\boldsymbol{\para}}| \boldsymbol{N}_i, \boldsymbol{X}_i, \boldsymbol{Z}_i),
\end{equation*}
where $L_1$ is defined in \eqref{eq:likelihood} with $n = 1$. The complete-data likelihood for the $i$th subject is 
\begin{align}
			  L^{(\text{complete})}(\boldsymbol{\para}| \boldsymbol{N}_i, \boldsymbol{X}_i, \boldsymbol{Z}_i, \boldsymbol{\theta}_i) 
             &=  \prod^J_{j=1} \bigg[ \prod^{n_{ij}}_{m=1} e^{ \beta_{j0} + \boldsymbol{\beta}^{\top}_j \boldsymbol{X}_{ij}(t_{ijm}) + \boldsymbol{\theta}^{\top}_i \mathbf{A}_j^{\top} \boldsymbol{Z}_{ij}(t_{ijm}) }  \label{eq:complete_data_loglike}  \\ 
			 &\quad \times\exp \bigg\{ -\int^\infty_0   Y_{ij}(t) e^{ \beta_{j0} + \boldsymbol{\beta}^{\top}_j \boldsymbol{X}_{ij}(t) + \boldsymbol{\theta}^{\top}_i \mathbf{A}_j^{\top} \boldsymbol{Z}_{ij}(t) } \mathrm{d}t \bigg\}  \bigg]  \phi_K(\boldsymbol{\theta}_i ; \boldsymbol{0}, \mathbf{\Sigma}), \nonumber
		\end{align}
  where  $t_{ij1},\ldots,t_{ijn_{ij}}$ are the event times for the $j$th event type of the $i$th subject.
  The corresponding score function is defined as
\begin{equation*}
    \boldsymbol{S}(\hat{\boldsymbol{\para}}|\boldsymbol{N}_i, \boldsymbol{X}_i, \boldsymbol{Z}_i, \boldsymbol{\theta}_i) = \frac{\partial}{\partial \boldsymbol{\para}} \log L^{(\text{complete})}(\hat{\boldsymbol{\para}}|\boldsymbol{N}_i, \boldsymbol{X}_i, \boldsymbol{Z}_i, \boldsymbol{\theta}_i).
\end{equation*}

The expectations in \eqref{eq:empirical_observed_Info} are taken with respect to $\boldsymbol{\theta}_i$ and can be approximated by Monte Carlo integration based on posterior samples of $\boldsymbol{\theta}_i$ generated via the elliptical slice sampling algorithm described in Section~\ref{sect:algorithm}. The approach in \eqref{eq:empirical_observed_Info} is simpler to implement than the method based on the missing-information identity \citep{louis1982finding}, as it avoids computing the Hessian of the complete-data likelihood. This leads to a more efficient and stable procedure, especially when the number of parameters is large.


Since process data are structurally complex, one may consider a large number of potential covariates in both the fixed-effects and random-effects components of the model. It is therefore important to effectively and efficiently determine a subset of significant variables. Furthermore, a sparse factor loading matrix could lead to better interpretation and understanding of the factors. Sparse estimation of factor loadings has been studied in \cite{choi2010penalized,ning2011sparse,hirose2015sparse,sun2016latent}. Let
\begin{equation*}
    \mathcal{I} \subset \{ (j,l,k):j=1,\ldots,J, l=1,\ldots,L_{2j},k=1,\ldots,K\}
\end{equation*}
denote the index set corresponding to the diagonal elements of the diagonal submatrix of the loading matrix $\mathbf A$ that are left unpenalised to ensure identifiability.
In this connection, we consider the penalised likelihood
\begin{equation}\label{eq:penalized_likelihood}
l_{n,p}^{(\boldsymbol{\gamma})}(\boldsymbol{\para}| \textbf{N}, \textbf{X}, \textbf{Z}) := \log L_n(\boldsymbol{\para}| \textbf{N}, \textbf{X}, \textbf{Z})  -  n \bigg\{ \sum^J_{j=1}\sum^{L_{1j}}_{l=1} p_{\gamma_1} (|\beta_{jl}|)  + \sum_{(j,l,k) \in \mathcal{I}^c} p_{\gamma_2}(|a_{jlk}|) \bigg\}
\end{equation}
for simultaneous variable selection and estimation, where $\beta_{jl}$ is the 
$l$-th entry of $\boldsymbol{\beta}_{j}$, $a_{jlk}$ is the $(l,k)$-entry of 
$\mathbf{A}_j$, and $l=1,\ldots,L_{2j}$ indexes the rows within the 
event-specific matrix $\mathbf{A}_j$, rather than the rows of the stacked matrix
$\mathbf{A}=(\mathbf{A}_1^\top,\ldots,\mathbf{A}_J^\top)^\top$. 
Here, $p_{\gamma_0}(\cdot)$ is a suitably chosen penalty function, and 
$\gamma_1,\gamma_2$ are tuning parameters that could depend on $n$. 
The penalised estimator is then defined as 
$\hat{\boldsymbol{\para}}_{\text{pen}}^{(\boldsymbol{\gamma})} 
:= \argmax_{\boldsymbol{\para}\in\boldsymbol{\Para}} 
l_{n,p}^{(\boldsymbol{\gamma})}
(\boldsymbol{\para}\mid \textbf{N},\textbf{X},\textbf{Z})$. Since the nonconcave penalties of \citet{fan2001variable} and \citet{zhang2010nearly} have been shown to possess desirable oracle properties, we adopt the smoothly clipped absolute deviation (SCAD) penalty \cite[]{fan2001variable}
\begin{equation*}
p'_{\gamma_0}(x) = \gamma_0 \bigg\{ I(x \leq \gamma_0) + \frac{ (a \gamma_0 - x)_+}{(a-1)\gamma_0} I(x > \gamma_0) \bigg\}
\end{equation*}
for some $a > 2$, $\gamma_0>0$, and $x >0$. 
Following \cite{fan2001variable}, we choose $a = 3.7$. Note that we do not penalise the intercept parameters $\beta_{j0}$ and the parameters in $\mathbf{\Sigma}$.

Write $\boldsymbol{\para}_0=(\boldsymbol{\para}_{10}^{\top},\boldsymbol{\para}_{20}^{\top})^{\top}$ and $\hat{\boldsymbol{\para}}_{\text{pen}}^{(\boldsymbol{\gamma})}=((\hat{\boldsymbol{\para}}_{1,\text{pen}}^{(\boldsymbol{\gamma})})^{\top},(\hat{\boldsymbol{\para}}_{2,\text{pen}}^{(\boldsymbol{\gamma})})^{\top})^{\top}$. Without loss of generality, we assume that $\boldsymbol{\para}_{20}=\boldsymbol{0}$. Under the penalised likelihood (\ref{eq:penalized_likelihood}), the following theorem establishes the consistency of variable selection and the asymptotic normality of parameter estimation.
\begin{theorem}\label{thm:oracle}
	Under Conditions (a)-(f), suppose that $\gamma_{1},\gamma_{2}\rightarrow0$ and $\sqrt{n} \gamma_{1},\sqrt{n}\gamma_{2} \rightarrow \infty$ as $n \rightarrow \infty$. Then, for $\hat{\boldsymbol{\para}}_{\text{pen}}^{(\boldsymbol{\gamma})}=((\hat{\boldsymbol{\para}}_{1,\text{pen}}^{(\boldsymbol{\gamma})})^{\top},(\hat{\boldsymbol{\para}}_{2,\text{pen}}^{(\boldsymbol{\gamma})})^{\top})^{\top}$, we have 
	\begin{enumerate}[(i)]
		\item Selection consistency: $\mathbb{P}(\hat{\boldsymbol{\para}}_{2,\text{pen}}^{(\boldsymbol{\gamma})} = \boldsymbol{0})\rightarrow 1$ as $n \rightarrow \infty$.
		\item Asymptotic normality (oracle): $\sqrt{n}( \hat{\boldsymbol{\para}}_{1,\text{pen}}^{(\boldsymbol{\gamma})} - \boldsymbol{\para}_{10}) \rightarrow  \mathcal{N}(\boldsymbol{0}, \mathbf{I}_1^{-1}(\boldsymbol{\para}_{10}))$ in distribution, where $\mathbf{I}_1(\boldsymbol{\para}_{10})$ is the Fisher information matrix with known $\boldsymbol{\para}_{20}=\boldsymbol{0}$.
	\end{enumerate}
\end{theorem}
Theorem \ref{thm:oracle} allows us to compute standard errors for the parameter estimates in the same manner as in the case without penalisation, using only the nonzero estimates. Similar results also hold when a nonconstant parametric baseline is considered, and the corresponding computational algorithm can be modified accordingly.

\section{Implementation}\label{sect:algorithm}

To maximise (\ref{eq:penalized_likelihood}) for a specific value of $\boldsymbol{\gamma} = (\gamma_1,\gamma_2)^{\top}$, we could, in principle, apply the expectation-maximisation algorithm \citep{dempster1977maximum} by treating $\boldsymbol{\theta}_i$, $i=1,\ldots,n$, as the missing data. In the E-step, we compute the expectation of the complete-data log-likelihood with respect to the conditional distribution of the missing data given the observed data. In the present case, there is no closed-form expression for this conditional expectation. Hence, numerical approximation of the E-step or stochastic versions of the expectation-maximisation algorithm may be used instead. For low-dimensional random effects, such as the univariate case, Gaussian quadrature may be applied to approximate the integrals in the E-step. Here, we use the stochastic expectation-maximisation algorithm \citep{celeux1985sem}, with elliptical slice sampling \citep{murray2010elliptical} in the stochastic E-step. Elliptical slice sampling is directly applicable because the conditional distribution of each $\boldsymbol{\theta}_i$ is proportional to its observed-data likelihood contribution multiplied by the multivariate normal density $\phi_K(\boldsymbol{\theta}_i;\boldsymbol{0},\mathbf{\Sigma})$. The algorithm exploits this Gaussian structure and does not require a proposal variance or acceptance-rate tuning. In the M-step, the complete-data penalised log-likelihood based on the updated $\boldsymbol{\theta}_i$ is maximised using the coordinate descent algorithm developed for generalised linear models with convex penalties \citep{friedman2010regularization}. The stochastic expectation-maximisation algorithm iterates between the stochastic E-step and the M-step until convergence.

We outline the estimation algorithm using the stochastic EM algorithm with elliptical slice sampling and coordinate descent. Let $(\boldsymbol{\beta}^{(t)},\mathbf{A}^{(t)},\mathbf{\Sigma}^{(t)})$ and $\boldsymbol{\theta}^{(t)}=(\boldsymbol{\theta}_1^{(t)},\ldots,\boldsymbol{\theta}_n^{(t)})$ denote the parameter estimates and simulated latent factors at iteration
$t$, respectively. At iteration $t+1$, the following steps are performed.

\begin{enumerate}[(a)]

\item \textit{Stochastic E-step via elliptical slice sampling.}
For each $i=1,\ldots,n$, define
\begin{align*}
\ell_i^{(t)}(\boldsymbol{\theta})
=
\sum_{j=1}^J
\Bigg[
&\sum_{m=1}^{n_{ij}}
\left\{
\beta_{j0}^{(t)}
+
(\boldsymbol{\beta}_j^{(t)})^\top
\boldsymbol{X}_{ij}(t_{ijm})
+
\boldsymbol{\theta}^\top
(\mathbf{A}_j^{(t)})^\top
\boldsymbol{Z}_{ij}(t_{ijm})
\right\}
\\
&-
\int_0^\infty
Y_{ij}(s)
\exp\left\{
\beta_{j0}^{(t)}
+
(\boldsymbol{\beta}_j^{(t)})^\top
\boldsymbol{X}_{ij}(s)
+
\boldsymbol{\theta}^\top
(\mathbf{A}_j^{(t)})^\top
\boldsymbol{Z}_{ij}(s)
\right\}
\,\mathrm{d}s
\Bigg].
\end{align*}
Here, $\ell_i^{(t)}(\boldsymbol{\theta})$ is the conditional
log-likelihood contribution of subject $i$ as a function of
$\boldsymbol{\theta}$, excluding the multivariate normal density of
the latent factor. Hence, conditional on the observed data and the
current parameter estimates, the posterior density of
$\boldsymbol{\theta}_i$ is proportional to
$\exp\{\ell_i^{(t)}(\boldsymbol{\theta}_i)\}
\phi_K(\boldsymbol{\theta}_i;
\boldsymbol{0},\mathbf{\Sigma}^{(t)})$.
The Gaussian component is incorporated through the elliptical proposal.
One elliptical slice sampling transition is performed as follows.

\begin{enumerate}[(i)]

\item Generate an auxiliary vector
$\boldsymbol{\nu}_i
\sim
\mathcal{N}_K(\boldsymbol{0},\mathbf{\Sigma}^{(t)})$.

\item Generate $U_i\sim\operatorname{Unif}(0,1)$ and define the
log-slice threshold
$h_i
=
\ell_i^{(t)}(\boldsymbol{\theta}_i^{(t)})
+
\log U_i$.

\item Generate an initial angle
$\omega_i\sim\operatorname{Unif}(0,2\pi)$ and initialize the angular
bracket by
$\omega_{i,\min}=\omega_i-2\pi$ and
$\omega_{i,\max}=\omega_i$.

\item Construct the candidate point
$\boldsymbol{\theta}_i^*
=
\boldsymbol{\theta}_i^{(t)}\cos(\omega_i)
+
\boldsymbol{\nu}_i\sin(\omega_i)$.

\item If
$\ell_i^{(t)}(\boldsymbol{\theta}_i^*)\geq h_i$, set
$\boldsymbol{\theta}_i^{(t+1)}=\boldsymbol{\theta}_i^*$ and terminate
the update for subject $i$. Otherwise, set
$\omega_{i,\min}=\omega_i$ if $\omega_i<0$ and
$\omega_{i,\max}=\omega_i$ if $\omega_i>0$. Keeping
$\boldsymbol{\nu}_i$ and $h_i$ fixed, generate a new angle
$\omega_i\sim
\operatorname{Unif}(\omega_{i,\min},\omega_{i,\max})$
and repeat steps (iv)--(v) until a candidate satisfying
$\ell_i^{(t)}(\boldsymbol{\theta}_i^*)\geq h_i$ is obtained.

\end{enumerate}

To compute the integral 
  \begin{equation*}
      \int^\infty_0 Y_{ij}(t) e^{ \beta_{j0} + \boldsymbol{\beta}^{\top}_j \boldsymbol{X}_{ij}(t) + \boldsymbol{\theta}^{\top}_i \mathbf{A}_j^{\top} \boldsymbol{Z}_{ij}(t) } \mathrm{d}t
  \end{equation*}
  in $\ell_i^{(t)}(\boldsymbol{\theta})$, observe that $Y_{ij}(\cdot)$ is an at-risk indicator taking values $0$ or $1$, and that both $\boldsymbol{X}_{ij}(\cdot)$ and $\boldsymbol{Z}_{ij}(\cdot)$ are piecewise constant. Hence, the integral can be decomposed into a finite sum over the intervals on which these covariates remain constant. This allows the integral to be evaluated in closed form.

  \item \textit{M-step via coordinate descent.} We maximise 
	\begin{equation}\label{eq:M-step_objective}
	\sum^n_{i=1} \log L^{(\text{complete})}(\boldsymbol{\beta}, \mathbf{A}, \mathbf{\Sigma}| \boldsymbol{N}_i, \boldsymbol{X}_i, \boldsymbol{Z}_i, \boldsymbol{\theta}^{(t+1)}_i) -  
	n \bigg\{ \sum^J_{j=1}\sum^{L_{1j}}_{l=1} p_{\gamma_1} (|\beta_{jl}|)  + \sum_{(j,l,k) \in \mathcal{I}^c} p_{\gamma_2} (|a_{jlk}|) \bigg\}.
	\end{equation}
	Denote 
	\begin{align*}
		\Psi_j(\beta_{j0}, \boldsymbol{\beta}_j, \mathbf{A}_j| \boldsymbol{\theta}^{(t+1)})
        &= \sum^n_{i=1} \bigg[ \sum^{n_{ij}}_{m=1} \bigg\{ \beta_{j0} + \boldsymbol{\beta}^{\top}_j \boldsymbol{X}_{ij}(t_{ijm}) + (\boldsymbol{\theta}^{(t+1)}_i)^{\top} \mathbf{A}_j^{\top} \boldsymbol{Z}_{ij}(t_{ijm}) \bigg\} \\
		& \quad \quad \quad - \int^\infty_0 Y_{ij}(t) e^{ \beta_{j0} + \boldsymbol{\beta}^{\top}_j \boldsymbol{X}_{ij}(t) + (\boldsymbol{\theta}^{(t+1)}_i)^{\top} \mathbf{A}_j^{\top} \boldsymbol{Z}_{ij}(t) } \mathrm{d}t \bigg].
	\end{align*}
	Since $\mathbf{\Sigma}$ is not penalised, maximising (\ref{eq:M-step_objective}) is equivalent to maximising the following terms separately:
	\begin{eqnarray}\label{eq:M-step_separate}
	\Psi_j(\beta_{j0}, \boldsymbol{\beta}_j, \mathbf{A}_j| \boldsymbol{\theta}^{(t+1)}) -  n \bigg\{ \sum^{L_{1j}}_{l=1} p_{\gamma_1} (|\beta_{jl}|)  +  \sum_{(l,k):(j,l,k) \in \mathcal{I}^c} p_{\gamma_2}(|a_{jlk}|) \bigg\}, \quad \text{for } j=1,\ldots,J, 
	\end{eqnarray}
	and
	\begin{equation*}
	\sum^n_{i=1} \log \phi_K(\boldsymbol{\theta}_i^{(t+1)}; \boldsymbol{0}, \mathbf{\Sigma}).
	\end{equation*}

 \item Iterate (a) and (b) until convergence and use the average of the last $B$ iterations as the estimates.
\end{enumerate}

To maximise (\ref{eq:M-step_separate}), we apply the coordinate descent algorithm to update each parameter. In each update, we form a quadratic approximation of $\Psi_j$ with respect to that parameter at the current value. In addition, we apply a local linear approximation \cite[]{zou2008one} to the SCAD penalty:
\begin{equation*}
p_\gamma(|x|) \approx p_\gamma (|x_0|) + p'_\gamma(|x_0|)(|x| - |x_0|) \quad \text{for } x \approx x_0.
\end{equation*} 
The resulting univariate maximisation problem has a closed-form solution. Specifically, we first update $\beta_{j0}$ (recall we do not penalise the baseline intercept) by
\begin{equation*}
\beta_{j0}^{(t+1)} \leftarrow \beta_{j0}^{(t)} - \frac{ \partial_{\beta_{j0}} \Psi_j (\beta_{j0}^{(t)}, \boldsymbol{\beta}_j^{(t)}, \mathbf{A}_j^{(t)}| \boldsymbol{\theta}^{(t+1)}) }
{ \partial_{\beta_{j0}}^2 \Psi_j (\beta_{j0}^{(t)}, \boldsymbol{\beta}_j^{(t)}, \mathbf{A}_j^{(t)}| \boldsymbol{\theta}^{(t+1)} ) },
\end{equation*}
where $\partial \Psi_j$ and $\partial^2 \Psi_j$ denote the first and second derivatives of $\Psi_j$ with respect to one of the parameters $\beta_{j0},\beta_{jl}$ or $a_{jlk}$ as indicated by the subscripts, respectively. Denote $\boldsymbol{\beta}^{(t,l)}_j = (\beta^{(t+1)}_{j1},\ldots, \beta^{(t+1)}_{j,l-1},\beta^{(t)}_{jl},\ldots, \beta^{(t)}_{j L_{1j}} )$ and $\Psi_j^{(t,l)} = \Psi_j(\beta^{(t+1)}_{j0}, \boldsymbol{\beta}^{(t,l)}_{j}, \mathbf{A}^{(t)}_j| \boldsymbol{\theta}^{(t+1)} )$. Then, 
we update $\beta_{jl}$, $l=1,\ldots,L_{1j}$ by
\begin{equation*}
\beta_{jl}^{(t+1)} \leftarrow - \frac{ \mathcal{T}\left( \partial_{\beta_{jl}} \Psi_j^{(t,l)} - \beta_{jl}^{(t)} \partial_{\beta_{jl}}^2 \Psi_j^{(t,l)},  np'_{\gamma_1}(|\beta^{(t)}_{jl}| )\right)}{ \partial_{\beta_{jl}}^2 \Psi_j^{(t,l)} },
\end{equation*} 
where $\mathcal{T}$ is the soft-thresholding operator \cite[]{donoho1994ideal} defined as $\mathcal{T}(x,\gamma) := \text{sgn}(x)(|x|- \gamma)_+$. The updating procedure of $a_{jlk}$ is similar to that for $\beta_{jl}$ and is therefore omitted.


\subsection{Choice of regularisation parameters}
For $\boldsymbol{x} \in \mathbb{R}^q$ and $\boldsymbol{C} \in \mathbb{R}^{d_1 \times d_2}$, define 
$\mathcal{S}_v(\boldsymbol{x}) :=     (I( x_1 \neq 0),\ldots, I( x_q \neq 0))^\top$ as the binary support vector of $\boldsymbol{x}$, and let $\mathcal{S}_m(\boldsymbol{C})$ denote the binary support matrix whose $(j,k)$-th entry is $I(C_{jk} \neq 0)$. To select the regularisation parameters $\boldsymbol{\gamma} = (\gamma_1, \gamma_2)^\top$, we use the Bayesian information criterion (BIC; \cite{schwarz1978estimating}). Specifically, for each candidate $\boldsymbol{\gamma}$, we obtain the penalised estimator $(\hat{\boldsymbol{\beta}}_\text{pen}^{(\boldsymbol{\gamma})}, \hat{\mathbf{A}}_\text{pen}^{(\boldsymbol{\gamma})}, \hat{\boldsymbol{\Sigma}}_\text{pen}^{(\boldsymbol{\gamma})})$ together with the corresponding support structures $\mathcal{S}_v\bigl(\hat{\boldsymbol{\beta}}_\text{pen}^{(\boldsymbol{\gamma})}\bigr)$ and $\mathcal{S}_m\bigl(\hat{\mathbf{A}}_\text{pen}^{(\boldsymbol{\gamma})}\bigr)$. The BIC at this value of $\boldsymbol{\gamma}$ is computed as 
\begin{equation}
\text{BIC}(\boldsymbol{\gamma}) =  \min_{ (\boldsymbol{\beta},\mathbf{A},\boldsymbol{\Sigma}) : \mathcal{S}_v(\boldsymbol{\beta}) = \mathcal{S}_v(\hat{\boldsymbol{\beta}}_\text{pen}^{(\boldsymbol{\gamma})}), \mathcal{S}_m(\mathbf{A}) = \mathcal{S}_m(\hat{\mathbf{A}}_\text{pen}^{(\boldsymbol{\gamma})})} \left\{ -2 \log L_n(\boldsymbol{\beta}, \mathbf{A}, \boldsymbol{\Sigma} | \textbf{N}, \textbf{X}, \textbf{Z}) + \log(n) p \right\},
\end{equation}
where $p$ is the total number of parameters and is equal to $\|\hat{\boldsymbol{\beta}}_\text{pen}^{(\boldsymbol{\gamma})}\|_0 + \|\hat{\mathbf{A}}_\text{pen}^{(\boldsymbol{\gamma})}\|_0 + K(K-1)/2$. Here, $\|\cdot\|_0$ denotes the $L_0$ norm. Specifically, $\|\hat{\boldsymbol{\beta}}_\text{pen}^{(\boldsymbol{\gamma})}\|_0 = J + \sum^J_{j=1} \sum^{L_{1j}}_{l=1} I(\hat{\beta}_{\text{pen}, jl}^{(\boldsymbol{\gamma})} \neq 0)$ and 
$\| \hat{\mathbf{A}}_\text{pen}^{(\boldsymbol{\gamma})}\|_0 = \sum^J_{j=1} \sum^{L_{2j}}_{l=1} \sum^K_{k=1} I( \hat{a}_{\text{pen}, jlk}^{(\boldsymbol{\gamma})} \neq 0)$. In practice, $\text{BIC}(\boldsymbol{\gamma})$ is evaluated over a grid $\mathcal{G}$, whose points are uniformly spaced on the log scale, and the regularisation parameter vector is selected as $\boldsymbol{\gamma}^* =\argmin_{\boldsymbol{\gamma} \in \mathcal{G}} \text{BIC}(\boldsymbol{\gamma})$.

\subsection{Choice of factor dimension}

In practice, one approach to selecting the factor dimension is to compare a prespecified collection of candidate factor structures using the BIC. For each candidate structure satisfying the identifiability conditions, we fit the model and select the regularisation parameters by minimising the BIC over a prespecified grid of $(\gamma_1,\gamma_2)$ values. We then compare the resulting minimum BIC values across the candidate models and select the factor dimension $K$ associated with the model having the smallest resulting minimum BIC. This procedure is illustrated in Section~\ref{sect:real_data}.

\section{Application to PIAAC data}\label{sect:real_data}
The Programme for the International Assessment of Adult Competencies (PIAAC) \citep{schleicher2008piaac} develops and conducts the Survey of Adult Skills. This international survey is conducted in over $40$ countries and measures adults' proficiency in information-processing skills, literacy, numeracy and problem solving in technology-rich environments (PSTRE) \citep{organisation2012literacy}. The proposed method is applied to an item in the PSTRE domain. The data used here comprise responses from $3,713$ adults in the United States, the United Kingdom, Ireland, Japan, and the Netherlands who answered all the items in the PSTRE domain in PIAAC 2012.

\begin{figure}[H]
		\centering
	\includegraphics[width=10cm]{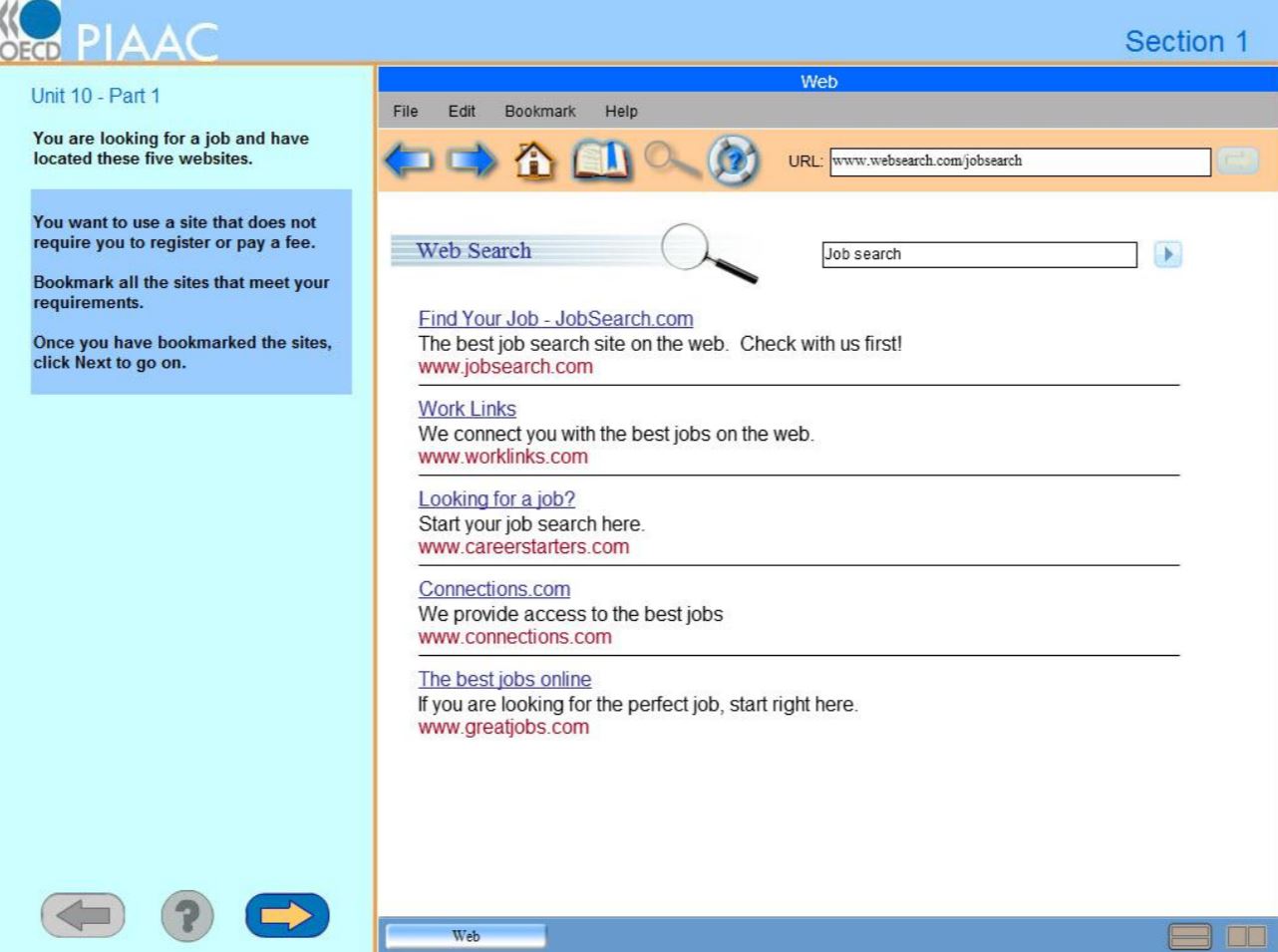}
	\caption{Screenshot of the sample item provided on the OECD website.}
	\label{jpg:sample_item}
\end{figure}

\begin{figure}[H]
		\centering
	\includegraphics[width=10cm]{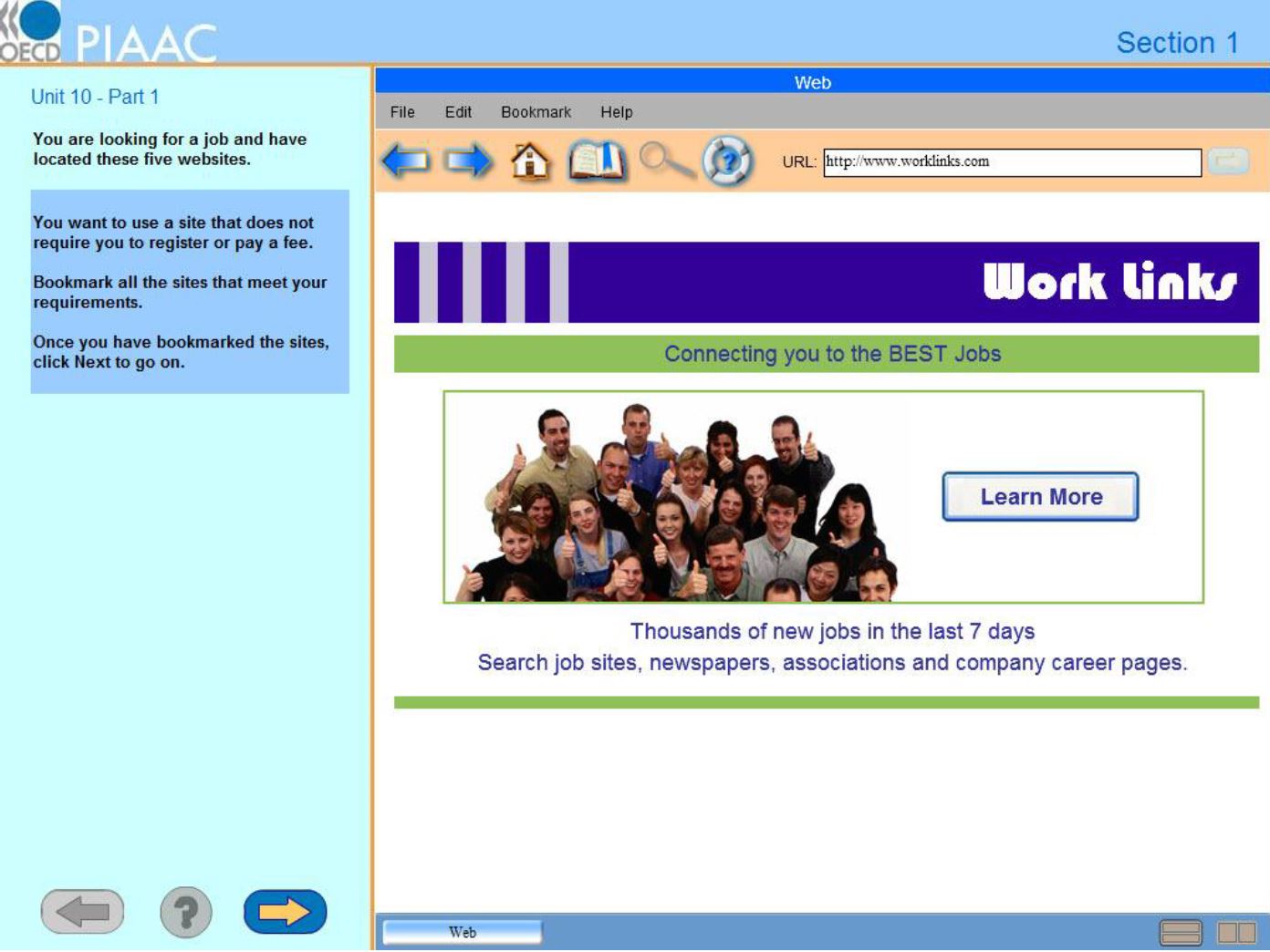}
	\caption{Screenshot of the sample item provided on the OECD website.}
	\label{jpg:sample_item_2}
\end{figure}

The actual item is confidential, but a sample item similar to the real data is available on the PIAAC website of the Organisation for Economic Co-operation and Development (OECD). In both the actual and the sample items, test takers are required to browse through websites containing various links and buttons and to evaluate the information provided therein. Two screenshots of the sample item are shown in Figures \ref{jpg:sample_item} and \ref{jpg:sample_item_2}. Figure \ref{jpg:sample_item} shows the first page that test takers see. They are required to access and evaluate information relating to job search in a simulated environment that resembles a real-world website. In particular, they can click on the links and perform actions such as going back and forth. If they click on the second link, ``Work Links'', they are directed to the page shown in Figure \ref{jpg:sample_item_2}. The test takers can then click the button ``Learn More'' to obtain further information. The task requires the test takers to choose an answer from a pull-down menu. However, some test takers may not select any option and may simply proceed to the next item.
Table~\ref{table:event_types_real} summarises the event types in the actual item and their corresponding meanings, comprising $25$ event types. In the data, there is no censoring, and all test takers completed the task; in particular, each subject’s final event is recorded as Next$_{\text{OK}}$.

Three examples of the process data for the actual item recorded in the log file are shown in Tables~\ref{table:example_real_data1}--\ref{table:example_real_data3}. Each column corresponds to a specific event (e.g., opening a website, going back to the previous page, or submitting a final answer), and each row records the time at which each event occurred. In the first example (Table~\ref{table:example_real_data1}), the test taker visited the five websites sequentially (W1--W5) without clicking any additional links. The sequence of ``Back'' events indicates that the test taker returned to the main page (which contains links to the five websites) after each visit. At the end of the sequence, the test taker opened the response panel ($\text{R}_\text{Open}$), selected the option ($\text{R}_\text{2}$), and submitted Website 2 as the final answer (Next, $\text{Next}_\text{OK}$). In the second example (Table~\ref{table:example_real_data2}), the test taker visited only Websites 2 and 4, returning to each twice. The log shows repeated patterns of visiting and leaving these two websites, followed by opening and closing the response panel ($\text{R}_\text{Open}$ and $\text{R}_\text{Close}$). Finally, the test taker selected Website 4 as the final answer ($\text{R}_\text{Open}$, $\text{R}_\text{4}$, Next, $\text{Next}_\text{OK}$). In the third example (Table~\ref{table:example_real_data3}), the test taker visited only Website 1 (W1) and then clicked an additional link within Website 1 ($\text{W1}_\text{M}$). Two back actions were performed to return to the main page: the first returned the test taker from $\text{W1}_\text{M}$ back to $\text{W1}$, and the second from $\text{W1}$ to the main page. The test taker then selected a response as in Example 1.

\begin{table}[H]
  \centering
  \tabcolsep=0.16cm
  \begin{tabular}{lcccccccccccccc}
			\hline
		Event & W1 & Back & W2 & Back & W3 & Back & W4 & Back & W5 & Back & $\text{R}_\text{Open}$ & $\text{R}_2$ & Next & $\text{Next}_\text{OK}$ \\ 
	\hline
		Time & 14 & 21 & 33 & 35 & 37 & 39 & 41 & 46 & 47 & 50 & 53 & 59 & 65 & 67 \\ \hline
  \end{tabular}
\caption{Example 1 of process data from the real dataset. Time is rounded to the nearest second.}
  	\label{table:example_real_data1}
\end{table}

\begin{table}[H]
  \centering
  \tabcolsep=0.16cm
  \begin{tabular}{lcccccccccccccc}
			\hline
		Event & W4 & Back & W2 & Back & W4 & Back & W2 & Back & $\text{R}_\text{Open}$ & $\text{R}_{\text{Close}}$ & $\text{R}_\text{Open}$ & $\text{R}_\text{4}$ & Next & $\text{Next}_\text{OK}$ \\ 
	\hline
		Time & 9 & 13 & 15 & 19 & 21 & 27 & 31 & 34 & 36 & 38 & 42 & 44 & 46 & 48  \\ \hline
  \end{tabular}
\caption{Example 2 of process data from the real dataset. Time is rounded to the nearest second.}
  	\label{table:example_real_data2}

\end{table}

\begin{table}[H]
  \centering
  \tabcolsep=0.16cm
  \begin{tabular}{lcccccccccccccc}
			\hline
		Event & W1 & $\text{W1}_\text{M}$ & Back & Back &  $\text{R}_\text{Open}$ &  $\text{R}_\text{2}$ & Next & $\text{Next}_\text{OK}$ \\ 
	\hline
		Time & 8 & 14 & 16 & 17 & 26 & 28 & 29 & 30   \\ \hline
  \end{tabular}
\caption{Example 3 of process data from the real dataset. Time is rounded to the nearest second.}
  	\label{table:example_real_data3}
    \end{table}

Due to the nature of the item, the two most recent events are expected to have a substantial influence on the next event. Therefore, we include information on the past two events in the covariate processes. These are internal, time-varying covariate processes: they are updated after each observed action and summarise the recent event history used to model the intensities of subsequent events. Specifically, the same covariate processes are used for both the fixed and random effects across all event types; that is, $X_{ijl}(\cdot) \equiv Z_{ijl}(\cdot) \equiv X_{il}(\cdot)$ for each $j = 1, \ldots, J$ and $l = 1, \ldots, L$, where $L_{1j} = L_{2j} = L$. Setting the same covariate processes for both the fixed and random effects across all events provides a relatively robust modelling strategy, as the penalisation can perform variable selection automatically, eliminating the need to prespecify different subsets of covariates for each event.
Recall that we have $25$ event types. For the $i$th subject, define $X_{il}(t) = 1$ for $l = 1, \ldots, 24$ (excluding the terminating event, which is not used in the covariate processes) if the most recent event prior to time $t$ is of type $l$; otherwise, let $X_{il}(t) = 0$. In addition, for each $l = 1, \ldots, 5$, define $X_{i,l+24}(t) = 1$ if the most recent event is Back and the second most recent event is W$l$; otherwise, let $X_{i,l+24}(t) = 0$. When $X_{i,l+24}(t) = 1$, it indicates that the test taker has just returned to the main page from one of the five websites. For concreteness, we also label these covariate processes using their event names. For example, we write $\text{W1}$ instead of $X_{i,1}$ and $\text{W$l$}, \text{Back}$ for $X_{i,l+24}$ for $l = 1,\ldots,5$. The full list of covariate processes is given in the first column of Table \ref{table:coef_matrix}. We use the notation $a \rightarrow \lambda_b$ to denote the effect of covariate process $a$ on the intensity function of event type $b$.

We explore a model with up to three factors. As discussed in Section~\ref{sect:theory}, identifiability requires constraints on the factor loading matrix: each factor must have at least one row that loads exclusively on that factor. Specifically, we impose constraints on the effects W2 $\rightarrow$ $\lambda_{\text{W2}_{\text{A}}}$, W2 $\rightarrow$ $\lambda_{\text{Back}}$, and W2, Back $\rightarrow $ $\lambda_{\text{W1}}$, where we refer to the corresponding factors as Factor 1, Factor 2 and Factor 3, respectively.
For instance, the factor loading for W2 $\rightarrow$ $\lambda_{\text{W2}_\text{A}}$ is unpenalised in the first dimension, while its loadings in the second and third dimensions are fixed at zero. The constraints defining Factors 1 and 2 capture distinct behavioural patterns that most frequently occur after event W2, and since the second website is the correct answer, the model is centred around it. The constraint on the third dimension reflects the tendency of test takers to move forward to the next webpage rather than returning to a previous one.

By incorporating random effects and applying variable selection, we can assess whether these behavioural patterns are correlated across webpages. To determine the appropriate number of latent factors, we fit eight models: one without any factors, three with a single factor, three with two factors, and one with all three factors. The models are summarised in Table~\ref{table:BIC_real_data}.
\begin{table}[t]
\centering
\begin{tabular}{lll}
\toprule
Model & Factors included & {BIC} \\
\midrule
$\mathcal{M}_0$ & None & 385848 \\
$\mathcal{M}_1$ & Factor 1 & 382416 \\
$\mathcal{M}_2$ & Factor 2 & 382771 \\
$\mathcal{M}_3$ & Factor 3 & 383021 \\
$\mathcal{M}_{1, 2}$ & Factor 1 + Factor 2 & 380485 \\
$\mathcal{M}_{1, 3}$ & Factor 1 + Factor 3 & 380903 \\
$\mathcal{M}_{2, 3}$ & Factor 2 + Factor 3 & 381302 \\
$\mathcal{M}_{1, 2, 3}$ & Factor 1 + Factor 2 + Factor 3 & \textbf{379796} \\
\bottomrule
\end{tabular}
\caption{Model comparison by BIC. The boldface value indicates the smallest BIC, corresponding to the model that includes all three factors.}
\label{table:BIC_real_data}
\end{table}

Under this setting, we apply the proposed method to the PIAAC data. For each of the eight models, we maximise the penalised likelihood over a grid of penalty parameter pairs $(\gamma_1, \gamma_2)$, where $\gamma_1$ controls the penalty for the fixed effects and $\gamma_2$ controls the penalty for the random effects. Table~\ref{table:BIC_real_data} reports, for each model, the minimum BIC achieved over the grid of $(\gamma_1, \gamma_2)$, where $\log(\gamma_1)$ is uniformly spaced over $[-8, -6]$ with 20 grid points and $\log(\gamma_2)$ is uniformly spaced over $[-7, -4.7]$ with 30 grid points.

We find that the model including all three factors ($\mathcal{M}_{1,2,3}$) yields the smallest BIC. Among the single-factor models ($\mathcal{M}_1$, $\mathcal{M}_2$, $\mathcal{M}_3$), Factor 1 appears to be the most influential, followed by Factor 2. This ordering is consistent with the comparison among the two-factor models ($\mathcal{M}_{1,2}$, $\mathcal{M}_{1,3}$, and $\mathcal{M}_{2,3}$), where combinations involving Factor 1 consistently produce lower BIC values. The model without any factors ($\mathcal{M}_0$) has the largest BIC, highlighting the importance of incorporating latent factors to capture unobserved heterogeneity and achieve a better model fit.

In the following, we focus on the results for $\mathcal{M}_{1,2,3}$. The BIC is minimised at the penalty parameter pair $(\gamma_1, \gamma_2) = (0.000961, 0.00482)$. Since the two values are of different magnitudes, this indicates the necessity of using different penalty parameters for the fixed and random components. Estimation for a given pair of penalty parameters took approximately 7 minutes on a laptop equipped with an Intel i9-12900HK CPU (2.50 GHz), while evaluations across multiple penalty pairs were performed in parallel on a computing cluster.

\begin{table}[htbp]
\centering
\resizebox{\textwidth}{!}{
\begin{threeparttable}
\begin{tabular}{lrrrrrrrr}
\toprule
 & $\lambda_{\text{W1}}$ & $\lambda_{\text{W2}}$ & $\lambda_{\text{W3}}$ & $\lambda_{\text{W4}}$ & $\lambda_{\text{W5}}$ & $\lambda_{\text{Back}}$ & $\lambda_{\text{Next}}$ & $\lambda_{\text{Web}}$ \\
\midrule
$\beta_0$ & -3.83 (0.02) & -5.75 (0.07) & -6.94 (0.13) & -6.59 (0.11) & -7.43 (0.15) & -9.60 (0.76) & -6.45 (0.08) & -8.30 (0.17) \\
W1 & 0 & 0 & 0 & 0 & 0 & 6.12 (0.76) & -2.09 (0.63) & 2.04 (0.21) \\
W1$_\text{M}$ & 0 & 0 & 0 & 0 & 0 & 7.34 (0.78) & 0.58 (0.36) & 4.31 (0.24) \\
W2 & 0 & 0 & 0 & 0 & 0 & 6.74 (0.76) & 0.25 (0.14) & 1.18 (0.27) \\
W2$_\text{A}$ & 0 & 0 & 0 & 0 & 0 & 7.45 (0.83) & 0 & 0.80 (0.63) \\
W3 & 0 & 0 & 0 & 0 & 0 & 6.90 (0.76) & -0.28 (0.21) & 0 \\
W3$_\text{A}$ & 0 & 0 & 0 & 0 & 0 & 7.74 (1.20) & 0 & 2.96 (0.65) \\
W3$_{\text{O}1}$ & 0 & 0 & 0 & 0 & 0 & 6.55 (0.82) & 0 & 0 \\
W3$_{\text{O}2}$ & 0 & 0 & 0 & 0 & 0 & 6.69 (0.94) & 0 & 0 \\
W4 & 0 & 0 & 0 & 0 & 0 & 6.79 (0.76) & -0.19 (0.18) & 0.71 (0.31) \\
W5 & 0 & 0 & 0 & 0 & 0 & 6.83 (0.76) & -0.71 (0.25) & 0.31 (0.37) \\
W5$_\text{O}$ & 0 & 0 & 0 & 0 & 0 & 6.78 (0.84) & 0 & 0 \\
Next & -4.23 (1.05) & -1.84 (0.76) & 0 & -1.04 (1.06) & 0 & 0 & 0 & 0 \\
Next$_\text{Cancel}$ & -0.10 (1.52) & 0 & 0 & 0 & 0 & 6.74 (0.85) & 2.69 (0.28) & 3.60 (0.49) \\
R$_1$ & -0.73 (0.62) & 0 & 0 & 0 & 0 & 5.03 (0.87) & 4.78 (0.17) & 0 \\
R$_2$ & -3.70 (1.02) & 0.78 (0.23) & 1.27 (0.33) & 0 & 0 & 4.90 (0.78) & 5.29 (0.09) & 0.78 (0.55) \\
R$_3$ & -0.71 (0.63) & 0 & 0 & 0 & 0 & 4.64 (1.03) & 5.23 (0.18) & 0 \\
R$_4$ & -3.02 (23.38) & 0 & 0 & 1.56 (0.33) & 0 & 4.60 (0.82) & 5.28 (0.14) & 1.24 (0.56) \\
R$_5$ & -0.67 (1.37) & 0 & 0 & 0 & 0 & 4.76 (0.95) & 4.97 (0.17) & 0 \\
R$_\text{Open}$ & -1.89 (0.12) & -0.13 (0.13) & -0.05 (0.24) & 0.16 (0.23) & -1.11 (0.53) & 4.22 (0.77) & -1.20 (0.26) & 0 \\
R$_\text{Close}$ & 1.34 (0.10) & 1.45 (0.15) & 1.38 (0.28) & 1.48 (0.23) & 1.38 (0.36) & 6.07 (0.77) & 1.76 (0.19) & 2.08 (0.36) \\
Back & -0.99 (0.16) & 3.27 (0.10) & 3.44 (0.17) & 2.95 (0.16) & 2.52 (0.19) & 6.64 (0.79) & 0.32 (0.17) & 0 \\
Forward & 0 & 0 & 0 & 0 & 0 & 7.57 (0.77) & 1.79 (0.33) & 3.24 (0.44) \\
Home & -0.49 (2.10) & 2.93 (0.50) & 3.60 (4.28) & 3.46 (0.72) & 3.91 (4.07) & 0 & 0 & 0 \\
Web & 0 & 0 & 0 & 0 & 0 & 7.41 (0.77) & 1.84 (0.27) & 5.76 (0.18) \\
W1, Back & -0.88 (0.26) & 0.89 (0.08) & -1.17 (0.24) & -3.12 (0.52) & -1.35 (0.29) & -3.18 (0.25) & -1.15 (0.54) & 0 \\
W2, Back & -0.87 (0.31) & -2.12 (0.14) & 1.62 (0.12) & -0.34 (0.16) & -0.83 (0.18) & -3.50 (0.25) & 0 & 0 \\
W3, Back & -0.71 (0.25) & -1.95 (0.21) & -1.79 (0.20) & 2.26 (0.13) & 0 & -3.74 (0.34) & -1.04 (0.54) & 0 \\
W4, Back & -0.50 (0.23) & -1.42 (0.14) & -2.16 (0.31) & -1.35 (0.22) & 3.34 (0.13) & -3.98 (0.34) & 0 & 0 \\
W5, Back & 0.48 (0.19) & -0.30 (0.09) & -1.44 (0.21) & -0.78 (0.20) & -2.13 (0.34) & -3.46 (0.25) & 0 & 0 \\
\bottomrule
\end{tabular}
\caption{Partial results for the estimated regression coefficients (standard errors in parentheses). The columns correspond to the intensity functions for different event types. The first row represents the log baseline intensity, while the other rows correspond to the covariate processes.}
\label{table:coef_matrix}
\end{threeparttable}
}
\end{table}

Table~\ref{table:coef_matrix} reports the estimated fixed-effect regression coefficients for the intensity functions of the events W1, W2, W3, W4, W5, Back, Next, and Web. These coefficients quantify how the evolving action history, represented by the internal covariate processes, affects the intensities of subsequent events. Below, we summarise several notable findings from Table~\ref{table:coef_matrix}.

The webpage-related coefficients in the intensity of the Back event, $\lambda_{\text{Back}}(\cdot)$, are large and positive. This reflects the logical navigation pattern that test takers must return to the main page before accessing other links. Moreover, the coefficients for $\text{W1}_\text{M}$, $\text{W2}_\text{A}$, and $\text{W3}_\text{A}$ are slightly larger than those for the other web links, suggesting that these pages contain less information, allowing test takers to finish reading and click Back more quickly.

The coefficients for $\text{R}_1 \rightarrow$ $\lambda_{\text{Next}}$, $\ldots$, $\text{R}_5 \rightarrow$ $\lambda_{\text{Next}}$ are all positive and relatively large, indicating that once test takers have selected a response, they are more likely to proceed immediately to submit it by clicking Next.
%

For the covariate process Web, the strongest effects are observed on the intensities of Back and Web itself. This suggests that some test takers may have initially assumed that clicking Web would return them to the previous or main page. After realising that this was not the case, they often followed the click with a Back event.

The coefficients for W$i$, Back $\rightarrow$ $\lambda_{\text{W}j}$, for $i,j = 1, \ldots, 5$, reveal a clear sequential examination pattern in browsing behaviour. Specifically, the coefficients for W$i$, Back $\rightarrow$ $\lambda_{\text{W}j}$ are positive when $i = 1, \ldots, 4$ and $j = i + 1$, and negative (with one zero) when $i = 1, \ldots, 4$ and $j \neq i + 1$. For instance, the coefficients for W1, Back $\rightarrow$ $\lambda_{\text{W}i}(\cdot)$, $i = 1, \ldots, 5$, are $-0.88$, $0.89$, $-1.17$, $-3.12$, and $-1.35$, respectively. This pattern indicates that after returning to the main page from the first website, test takers are more likely to proceed sequentially to the second website rather than selecting another link or revisiting the same page. 
Such sequential examination behaviour has also been documented in studies of search result navigation. For example, \citet{klockner2004depth} identified two typical browsing strategies: a depth-first strategy, where users examine each item in the list in order and decide sequentially whether to open it, and a breadth-first strategy, where users scan through multiple items before making a selection. Similarly, click models can be categorised according to whether they adhere to the sequential examination hypothesis \citep{wang2015incorporating}.

The internal covariate processes also enter the random-effect component, allowing the factor loadings to capture subject-specific heterogeneity in the effects of previous actions on subsequent event intensities. Partial results for the random effects are presented in Table~\ref{table:factor_loadings}. We begin by discussing the findings related to the first dimension. Recall that the effect of W2 $\rightarrow$ $\lambda_{\text{W2}_\text{A}}$ is constrained to load only on the first dimension. In this dimension, many related transitions exhibit factor loadings with the same sign as that of W2 $\rightarrow$ $\lambda_{\text{W2}_\text{A}}$. These include W1 $\rightarrow$ $\lambda_{\text{W1}_\text{M}}$, W3 $\rightarrow$ $\lambda_{\text{W3}_\text{A}}$, W3 $\rightarrow$ $\lambda_{\text{W3}_\text{O1}}$, and $\text{W3}_\text{O1}$ $\rightarrow$ $\lambda_{\text{W3}_\text{O2}}$. Moreover, the loadings for these transitions are either zero or negligible in the other two dimensions. We interpret this group of relationships as reflecting information-seeking behaviour involving actions in which test takers explore additional webpages for relevant information.
Furthermore, the loadings of $\text{R}_\text{Open}$ $\rightarrow$ $\lambda_{\text{R}_i}$, for $i = 1, \ldots, 5$, suggest that these information-seeking actions are positively associated with the selection of the correct answer. Another noteworthy finding is that the factor loading for Next $\rightarrow$ $\lambda_{\text{Next}_\text{Cancel}}$ has the opposite sign to that of W2 $\rightarrow$ $\lambda_{\text{W2}_\text{A}}$, indicating that test takers tend to exhibit greater confidence when visiting $\text{W1}_\text{M}$, $\text{W2}_\text{A}$, $\text{W3}_\text{A}$, $\text{W3}_\text{O1}$, and $\text{W3}_\text{O2}$.

The second dimension is primarily associated with the Back event. In particular, the factor loadings for W$j$ $\rightarrow$ $\lambda_{\text{Back}}$, for $j = 1, \ldots, 5$, are of comparable magnitude and share the same sign, suggesting a common latent factor underlying these transitions.
The third dimension, on the other hand, is mainly linked to the sequential examination pattern observed in the fixed effects. Specifically, the loadings for W$i$, Back $\rightarrow$ $\lambda_{\text{W}j}$ are positive when $j = i + 1$ for $i = 1, \ldots, 4$, and are zero or negative when $j \neq i + 1$. This indicates that the sequential browsing patterns across different webpages are positively correlated through the third factor.

\section{Simulation Study}\label{sect:simulation}
In this section, we conduct simulations under a setting that is similar to, but slightly simpler than, the real data example. Suppose that on the main page of the item there are three links to different websites, and each website contains an additional link leading to a secondary page that provides further information. Within the item, a test taker can click on these links and navigate back and forth using the browser. To answer the question, the test taker uses a pull-down menu to select one of the three websites as their response. The test taker completes the item by clicking the ``Next'' button, and then confirms the choice by clicking ``OK'' or ``Cancel''. In total, there are 15 event types (see Table~\ref{table:event_types_simulation}), and an example of the corresponding process data is provided in Table~\ref{table:example}.

The event times are generated according to the intensity functions of our proposed model 
\begin{equation*}
\lambda_{ij}(t|\mathscr{F}_{t^-};\boldsymbol{\theta}_i) = Y_{ij}(t)e^{ \beta_{j0} + \boldsymbol{\beta}^{\top}_j \boldsymbol{X}_{ij}(t) + \boldsymbol{\theta}^{\top}_i  \mathbf{A}^{\top}_j \boldsymbol{Z}_{ij}(t)}, \quad j=1,\ldots,J.
\end{equation*}
Specifically, the same covariate processes are used for both the fixed and random effects across all event types. That is, $X_{ijl}(\cdot) \equiv Z_{ijl}(\cdot) \equiv X_{il}(\cdot)$ for each $j = 1, \ldots, J$ and $l = 1, \ldots, L$. For the $i$th subject, define $X_{il}(t) = 1$ for $l=1, \ldots, 14$ if the most recent event prior to time $t$ is of the $l$th event type; otherwise, set $X_{il}(t) = 0$. Also, for each $l=1, \ldots, 3$, let $X_{i,l+14}(t) = 1$ if the last event is Back and the second most recent event is W$l$; otherwise, set  $X_{i,l+14}(t) = 0$. These are covariate processes that include information from the past two events. 
For instance, using the example in Table~\ref{table:example}, $X_{\text{W2}}(t) = 1$ when $t \in (15, 25]$, $X_{\text{Back}}(t) = 1$ when $t \in (25, 28] \cup (36, 42]$ and $X_{\text{W2, Back}}(t) = 1$ when $t \in (25, 28]$; here, the subscripts $i$ are omitted and the event type names are used for clarity. 

In the simulation setting, there are $23$ nonzero parameters for the fixed effects and there are $3$ dimensions in the random effects, with $13$ nonzero factor loadings. Details of the parameter setting are given in Section~S.1 in the Supplementary Materials.

To simulate event sequences from this model, we proceed as follows. For each subject, a subject-specific random effect is first generated from a multivariate normal distribution. Because the covariate processes are constant on each inter-event interval (their change points occur only at event times), the intensity for each event type is constant on that interval. Under a constant intensity, the waiting time to the next event follows an exponential distribution with rate equal to that intensity. Therefore, to determine the next event, we draw $J$ independent exponential random variables, each with a rate equal to the corresponding intensity. If the event type is not currently possible, then we set the waiting time as infinity. The next event type is chosen as the one attaining the minimum of these exponential draws. After the event occurs, we update the covariate processes and recompute the intensities based on the new state and event history. We then repeat the process to generate the next event until the absorbing state $\text{Next}_{\text{OK}}$ is reached. To mimic the real data, we do not include censoring in this simulation study.

We first assess the performance of the penalised estimator and the selection
of tuning parameters using the BIC. Let $\mathcal{P}$ denote the index set
of parameters that are subject to penalisation. Thus, the baseline
parameters, the parameters in $\boldsymbol{\Sigma}$, and the loading
parameters that are left unpenalised for identifiability are excluded from
$\mathcal{P}$. We evaluate the recovery of the true
penalised structure using the following criteria:
\begin{enumerate}
    \item $C_0 = 1$ if there exists a penalty parameter pair
    $\boldsymbol{\gamma} = (\gamma_1,\gamma_2)^\top$ such that
    \[
    \{j\in\mathcal{P}:
    \hat{\boldsymbol{\para}}_{\mathrm{pen},j}^{(\boldsymbol{\gamma})}\neq0\}
    =
    \{j\in\mathcal{P}:\boldsymbol{\para}_{0,j}\neq0\},
    \]
    and $C_0=0$ otherwise.

    \item $C_1 = 1$ if the penalty parameter pair $\boldsymbol{\gamma}$
    chosen using the BIC satisfies
    \[
    \{j\in\mathcal{P}:
    \hat{\boldsymbol{\para}}_{\mathrm{pen},j}^{(\boldsymbol{\gamma})}\neq0\}
    =
    \{j\in\mathcal{P}:\boldsymbol{\para}_{0,j}\neq0\},
    \]
    and $C_1=0$ otherwise.

    \item True positive rate (TPR):
    \begin{equation*}
    \mathrm{TPR}
    =
    \frac{
    |\{j\in\mathcal{P}:
    \hat{\boldsymbol{\para}}_{\mathrm{pen},j}^{(\boldsymbol{\gamma})}\neq0,
    \ \boldsymbol{\para}_{0,j}\neq0\}|
    }{
    |\{j\in\mathcal{P}:\boldsymbol{\para}_{0,j}\neq0\}|
    }.
    \end{equation*}

    \item False positive rate (FPR):
    \begin{equation*}
    \mathrm{FPR}
    =
    \frac{
    |\{j\in\mathcal{P}:
    \hat{\boldsymbol{\para}}_{\mathrm{pen},j}^{(\boldsymbol{\gamma})}\neq0,
    \ \boldsymbol{\para}_{0,j}=0\}|
    }{
    |\{j\in\mathcal{P}:\boldsymbol{\para}_{0,j}=0\}|
    }.
    \end{equation*}
\end{enumerate}

For computing TPR and FPR, $\hat{\boldsymbol{\para}}_{\text{pen},j}^{(\boldsymbol{\gamma})}$ is the estimator corresponding to the penalty parameter pair that minimises the BIC. Table~\ref{table:eva} reports the results for these evaluation criteria, averaged over 100 independent simulations. As the sample size increases, the probability that the BIC selects the correct model also increases. Moreover, when the true model is not selected, the nonzero parameters are consistently estimated as nonzero, and only a very small number of zero parameters are mistakenly identified as nonzero.

We also evaluate the bias of the estimates, the accuracy of the standard error formula, and the coverage probability. For each penalised parameter, these summaries are computed over the replications in which that parameter is estimated as nonzero; the unpenalised parameters are summarised over all replications. The results, presented in the Supplementary Materials, indicate that the biases are small for most parameters, as typically expected under penalised estimation with SCAD, with only a few parameters showing noticeably larger bias. The estimated standard errors closely align with the empirical standard deviations of the estimates and yield satisfactory coverage probabilities, except in cases where the biases are relatively large.

Estimation for a given pair of penalty parameters took approximately $22$, $47$, and $97$ seconds for $n = 500, 1000, 2000$, respectively, on a laptop equipped with an Intel i9\text{-}12900HK CPU (2.50\,GHz). Evaluations across multiple penalty pairs were carried out in parallel on a computing cluster.

In the Supplementary Materials, we provide additional simulation results, including a scenario with an additional censoring variable. We also demonstrate that when the true underlying mechanism involves random effects, estimation based solely on fixed effects can produce biased results and can lead to incorrect model identification.

\section{Discussion}\label{sect:discussion}

In this article, we propose a dynamic multiplicative model with random coefficients (factors) for multivariate event time data. We develop a methodology for parameter estimation and variable selection in both the regression and factor components. We establish theoretical results on model identifiability and on the nondegeneracy of the Fisher information, which are key to proving consistency and asymptotic normality. We provide sufficient conditions under which the maximum likelihood estimator is consistent and asymptotically normal. By introducing a suitable penalty, we obtain a parsimonious model and improve the interpretability of the parameters therein, where theoretical properties for the penalised estimator are also established. 

In our model, there are several ways to incorporate group structure among covariates. First, one can define a covariate process that itself represents the combined effects of multiple covariates. For example, in our analysis we included a covariate process indicating whether the two preceding actions were (i) clicking on a particular website and (ii) subsequently clicking the ``Back'' button. Second, analogous to a factor model, group structure among covariates can be explored through the signs and magnitudes of the factor loadings. For instance, if two covariate processes have positive loadings of similar magnitude, this may suggest that they belong to the same group. This approach is primarily exploratory in nature. Third, when the group structure of covariates is known a priori, one may employ a group lasso--type penalty that shrinks or removes entire groups of regression coefficients. This approach allows us to select whether certain groups of covariates should be included in the model.

The parametric assumption in model (\ref{eq:our_model}) is used for simplicity because of the relatively large number of parameters in our application. It is also reasonable for event time data when the time span is relatively short, as in the process data example. The intensity function is modelled through internal covariates because the occurrence of a certain event will likely lead to the occurrence of another event.


The proposed method is applied to the 2012 PIAAC data. Our method finds meaningful relationships among different types of events that can help in understanding both the task design and the behaviour of subjects when attempting to solve a problem. Furthermore, the proposed method can be applied to exploratory analyses, confirmatory analyses, or combinations of the two, by imposing constraints on the loading matrix.


Although the PIAAC example only contains one item, the method can be readily extended to handle multiple items. Specifically, suppose that we have $S$ items and, for each item, there are $J_s$ event types. Then, model (\ref{eq:our_model}) becomes
\begin{eqnarray*}
	\lambda_{isj}(t|\mathscr{F}_{t^-}, \boldsymbol{\theta}_i) = \lambda_{sj0}(t) Y_{isj}(t) e^{  \boldsymbol{\beta}^{\top}_{sj}\boldsymbol{X}_{isj}(t) + \boldsymbol{\theta}^{\top}_i \mathbf{A}^{\top}_{sj} \boldsymbol{Z}_{isj}(t) } \quad s=1,\ldots,S,\quad j=1, \ldots, J_s,
\end{eqnarray*}
with $\boldsymbol{\theta}_i \sim  \mathcal{N}(\boldsymbol{0}, \mathbf{\Sigma})$, where $\boldsymbol{\beta}_{sj}$ and $\mathbf{A}_{sj}$ are, respectively, the vector of coefficients for the fixed effects and the loading matrix for the random effects for the $j$th event type in the $s$th item. For the $i$th subject, $\boldsymbol{X}_{isj}$ and $\boldsymbol{Z}_{isj}$ are two vectors of covariate processes for the $j$th event type in the $s$th item, and $\boldsymbol{\theta}_i$ is the subject-specific latent variable that is common across all items and event types. The corresponding likelihood function remains the same as (\ref{eq:likelihood}), except that the integrand is replaced by a product of $S$ terms, each corresponding to a specific item.

Similar process data also arise from online personalised learning systems, which consist of assessments and interventions; see, for example, \cite{wang2018tracking} and \cite{tang2019reinforcement}. The model and method proposed here may be modified to provide an alternative to the commonly used hidden Markov models by incorporating time-dependent random effects. Regression models with time-varying coefficients have been studied in \cite{guo2022inference}. It is also of interest to extend the current model to latent space models with longitudinally observed network data; see \cite{he2025semiparametric}.


\spacingset{1.6} 
\begin{table}[H]
	\centering
 \setlength\extrarowheight{4pt}
		\def~{\hphantom{0}}
  {
	\begin{tabular}{lp{10cm}}
		Event Type & Meaning \\ 
		\hline
		W$i$ ($i=1,\ldots,5$) & Click the link to the $i$th webpage \\ 
		$\text{W1}_\text{M}$ & Click the ``More'' link in the first webpage \\ 
		$\text{W}i_\text{A}$ ($i=2,3$) & Click the ``Author'' link in the $i$th webpage \\ 
		$\text{W}3_{\text{O}i}$ ($i=1,2$) & Click the $i$th ``Order'' link in the third webpage \\ 
		$\text{W}5_\text{O}$ & Click the ``Order'' link in the fifth webpage \\ 
		Next & Click the ``Next'' button\\ 
		$\text{Next}_\text{Cancel}$ & Click the ``Cancel'' button in the pop-up window that appears after clicking the ``Next'' button \\ 
		$\text{R}_i$ ($i=1,\ldots,5$) & Choose the $i$th website as the answer \\ 
		$\text{R}_\text{Open}$ & Click on the pull-down menu for choosing an answer \\ 
		$\text{R}_\text{Close}$ & Close the pull-down menu for choosing an answer without choosing an answer \\
		Back & Click the back arrow in the toolbar \\ 
		Forward & Click the forward arrow in the toolbar\\ 
		Home & Click the home button in the toolbar\\ 
		Web & Click the Web environment icon \\ 
		$\text{Next}_\text{OK}$ & Click the ``OK'' button in the pop-up window that appears after clicking the ``Next'' button  (the terminating event) \\ 
		\hline
	\end{tabular}

 \caption{Event types and their meanings in the real-data analysis.}
 	\label{table:event_types_real}
}
\end{table}

\spacingset{1.2} 
\setlength{\tabcolsep}{14pt}
\begin{table}[H]
	\centering
    
	\def~{\hphantom{0}}

	\begin{tabular}{lllll}
		\hline
		Covariate & Event & Factor 1 & Factor 2 & Factor 3 \\ 
		\hline
		Next & Next$_\text{Cancel}$ & 0.81 (0.17) & -0.11 (0.19) & 0.08 (0.23) \\ 
		R$_\text{Open}$ & R$_1$ & 0.71 (0.17) & 0 & 0.74 (0.15) \\ 
		R$_\text{Open}$ & R$_2$ & -0.79 (0.06) & 0.39 (0.06) & 0.16 (0.06) \\ 
		R$_\text{Open}$ & R$_3$ & 0 & 0.04 (0.12) & 0 \\ 
		R$_\text{Open}$ & R$_4$ & 0.51 (0.11) & 0.57 (0.07) & 0 \\ 
		R$_\text{Open}$ & R$_5$ & 0.52 (0.23) & 0 & 0 \\ 
		W1 & Back & 0 & 0.48 (0.06) & -0.1 (0.07) \\ 
		W1 & W1$_\text{M}$ & -1.31 (0.12) & -0.03 (0.11) & 0 \\ 
		W1, Back & W2 & 0.29 (0.09) & 0.06 (0.07) & 0.3 (0.09) \\ 
		W1, Back & W3 & 0.75 (0.19) & 0.03 (0.16) & -0.82 (0.24) \\ 
		W1, Back & W4 & 0 & 0 & -1.79 (0.35) \\ 
		W2 & Back & 0 & 0.48 (0.04) & 0 \\ 
		W2 & W2$_\text{A}$ & -2.12 (0.2) & 0 & 0 \\ 
		W2, Back & W1 & 0 & 0 & -1.02 (0.22) \\ 
		W2, Back & W2 & -0.04 (0.21) & 0 & -0.12 (0.17) \\ 
		W2, Back & W3 & 0.97 (0.12) & 0 & 0.41 (0.13) \\ 
		W2, Back & W4 & 0.96 (0.17) & 0.35 (0.14) & -0.33 (0.18) \\ 
		W3 & Back & -0.03 (0.07) & 0.57 (0.05) & 0 \\ 
		W3 & W3$_\text{A}$ & -2.55 (0.37) & -0.37 (0.25) & 0 \\ 
		W3 & W3$_\text{O1}$ & -1.14 (0.31) & 0 & 0 \\ 
		W3, Back & W2 & -0.76 (0.22) & 0 & -1.31 (0.19) \\ 
		W3, Back & W4 & 0.75 (0.14) & 0.21 (0.1) & 0.45 (0.13) \\ 
		W3$_\text{O1}$ & W3$_\text{O2}$ & -0.97 (0.39) & 0 & 0 \\ 
		W4 & Back & -0.11 (0.06) & 0.44 (0.05) & 0.04 (0.07) \\ 
		W4, Back & W2 & -0.34 (0.14) & 0 & -0.64 (0.15) \\ 
		W4, Back & W3 & 0 & 0 & -1.46 (0.25) \\ 
		W4, Back & W5 & 0.27 (0.15) & -0.05 (0.12) & 1.14 (0.15) \\ 
		W5 & Back & -0.14 (0.06) & 0.45 (0.06) & 0.06 (0.06) \\ 
		W5, Back & W1 & 0 & 0 & 0.17 (0.2) \\ 
		W5, Back & W2 & -0.34 (0.09) & -0.07 (0.08) & 0 \\ 
		W5, Back & W3 & 0 & 0 & -1.09 (0.23) \\ 
		W5, Back & W4 & 0.53 (0.16) & 0.34 (0.16) & -0.99 (0.2) \\ 
		\hline
	\end{tabular}
\caption{Partial results for the estimated factor loadings for the real data analysis. The values outside the parentheses are the estimated factor loadings, and the values in parentheses are the corresponding standard errors.}

\label{table:factor_loadings}
\end{table}

\spacingset{1.7} 

\begin{table}[H]
	\centering
		\def~{\hphantom{0}}
  {
	\begin{tabular}{lp{9cm}}
		Event Type (Simulation) & Meaning \\ 
		\hline
		W$i$ ($i=1,\ldots,3$) & Click the link to the $i$th webpage \\ 
		W$i_\text{M}$ ($i=1, \ldots, 3$) & Click the ``More'' link in the $i$th webpage \\ 
		Next & Click the ``Next'' button\\ 
		Next$_\text{Cancel}$ & Click the ``Cancel'' button in the pop-up window that appears after clicking the ``Next'' button \\ 
		R$_i$ ($i=1,\ldots,3$) & Choose the $i$th website as the answer \\ 
		R$_\text{Open}$ & Click on the pull-down menu for choosing an answer \\ 
		R$_\text{Close}$ & Close the pull-down menu for choosing an answer without choosing an answer \\
		Back & Click the back arrow in the toolbar \\ 
		Forward & Click the forward arrow in the toolbar\\ 
		Next$_\text{OK}$ & Click the ``OK'' button in the pop-up window that appears after clicking the ``Next'' button  (the terminating event) \\ 
		\hline
	\end{tabular}
}
\caption{Event types and their meanings in the simulation studies.}
	\label{table:event_types_simulation}
\end{table}
\begin{table}[H]
  \centering
  \tabcolsep=0.14cm
  \begin{tabular}{lccccccccccccc}
			\hline
		Event & W$2$ & Back & W1 & W1$_\text{M}$ & Back & Back & W3 & R$_\text{Open}$ & R$_3$ & Next & Next$_\text{OK}$ \\ 
	\hline
		Time &  15 & 25 & 28 & 34 & 36 & 38 & 42 & 45 & 50 & 52 & 53\\
		\hline
  \end{tabular}

	\caption{Example of process data in the simulation studies.}
  	\label{table:example}

\end{table}

\spacingset{1.5} 
\begin{table}[H]
	\centering
 \setlength\extrarowheight{-3pt}
	\def~{\hphantom{0}}
	{
		\begin{tabular}{rrrrr}
			\hline
			& $C_0$ & $C_1$ & TPR & FPR ($\times 10^{-2}$)\\ 
			\hline
			$n = 500$ &  0.85 & 0.60 & 1.00 & 0.13 \\ 
			$n = 1000$ &  0.96 & 0.78 & 1.00 & 0.05 \\ 
			$n = 2000$ & 0.99  & 0.83 & 1.00 & 0.03 \\ 
			\hline
		\end{tabular}
	}
\caption{Evaluation criteria in the simulation studies based on 100 independent replications. The table reports the proportions $C_0$ and $C_1$, the true positive rate (TPR), and the false positive rate (FPR, multiplied by $10^{-2}$).}

	\label{table:eva}
\end{table}

\spacingset{0.9} 
\bibliographystyle{apalike}
\bibliography{reference}

\addtocontents{toc}{\protect\setcounter{tocdepth}{3}%
  \protect\setlength{\cftsecnumwidth}{3em}%
  \protect\setlength{\cftsubsecnumwidth}{\cftsecnumwidth}}
\newgeometry{left=\SuppLayoutLeft,top=\SuppLayoutTop,
  textwidth=\SuppLayoutWidth,textheight=\SuppLayoutHeight}
\setcounter{section}{0}
\renewcommand{\thesection}{S.\arabic{section}}
\renewcommand{\theequation}{S.\arabic{equation}}
\newcommand{\mat}[1]{\mathbf{#1}}
\renewcommand{\vec}[1]{\boldsymbol{#1}}
\newcommand{\e}{\mathbb{E}}
\newcommand{\p}{\mathbb{P}}
\newcommand{\R}{\mathbb{R}}
\newcommand{\D}{\mathrm{d}}
\newcommand{\bX}{\boldsymbol{X}}
\newcommand{\bZ}{\boldsymbol{Z}}

\newtheorem{thm}{Theorem}[section]
\newtheorem{lem}{Lemma}[section]
\newtheorem{defin}{Definition}[section]
\newtheorem{asum}{Assumption}[section]
\newtheorem{coro}{Corollary}[section]
\renewcommand\qedsymbol{$\square$}
\newcommand\placeqed{\nobreak\enspace$\square$}
\newtheorem{innercustomthm}{Theorem}
\newenvironment{customthm}[1]
  {\renewcommand\theinnercustomthm{#1}\innercustomthm}
  {\endinnercustomthm}
\newtheorem{innercustomprop}{Proposition}
\newenvironment{customprop}[1]
  {\renewcommand\theinnercustomprop{#1}\innercustomprop}
  {\endinnercustomprop}
  \newtheorem{innercustomcor}{Corollary}
\newenvironment{customcor}[1]
  {\renewcommand\theinnercustomcor{#1}\innercustomcor}
  {\endinnercustomcor}
\newtheorem{innercustomlem}{Lemma}
\newenvironment{customlem}[1]
  {\renewcommand\theinnercustomlem{#1}\innercustomlem}
  {\endinnercustomlem}
\def\spacingset#1{\renewcommand{\baselinestretch}%
{#1}\small\normalsize} \spacingset{1}
\spacingset{1.5}
\begingroup
\makeatletter
\def\@title{Supplementary Materials for ``A Dynamic Factor Model for Multivariate Counting Process Data''}
\def\@author{\vspace{-7ex}}
\def\@date{\vspace{-7ex}}
\SuppMakeTitleInternal
\makeatother
\endgroup
\tableofcontents
\addtolength{\SuppLayoutHeight}{-0.3in}
\addtolength{\SuppLayoutTop}{+0.6in}
\newgeometry{left=\SuppLayoutLeft,top=\SuppLayoutTop,
  textwidth=\SuppLayoutWidth,textheight=\SuppLayoutHeight}
\spacingset{1.5} 
\newenvironment{coloredtext}[1]{%
  \par\color{#1}%
}{%
  \par%
}
\setcounter{table}{0}
\renewcommand{\thetable}{S\arabic{table}}
The Supplementary Materials are organised as follows.
Section~\ref{sec:simulation} presents additional simulation studies beyond those reported in the main paper. Section~\ref{sect:sensitivity} presents sensitivity analyses under event-specific piecewise-constant baseline intensity specifications for both the simulation study and the real-data application.
Section~\ref{sec:identifiability} provides further discussion of the challenges involved in establishing model identifiability in the presence of internal covariates. 
Sections~\ref{sec:thm1}--\ref{sec:thm4} contain the detailed proofs of Theorems~\ref{thm_identifiability}--\ref{thm:oracle} in the main paper. 
Finally, Section~\ref{sec:extension} discusses possible extensions to parametric models with nonconstant baseline intensity functions, as well as a semiparametric extension.
    
\section{Simulation setting and results}\label{sec:simulation}
Tables \ref{table:true_coef_sim_fixed}--\ref{table_covariance} report the parameter settings in the simulation studies. The scaling constraints are imposed on the covariance matrix and we do not need to constrain the scaling of the loading matrices. 


Tables~\ref{table:bias_se_cp_n500}--\ref{table:bias_se_cp_n2000}
report the simulation results for $n=500$, $1000$, and $2000$,
respectively. In each table, the left block reports the baseline parameters
and fixed effects, while the right block reports the factor loadings and
covariance parameters. To make the parameter labels directly interpretable,
we write $\beta_{b;a}$ for the coefficient of covariate process $a$ in the
intensity function for event type $b$, and $a^{(k)}_{b;a}$ for the
corresponding loading on Factor $k$.

For each simulated dataset, the selected model is the structure associated
with the minimum BIC. Because multiple pairs of penalty parameters may yield
the same selected structure, we report the original penalised estimates from
the pair having the smallest sum of the two penalty parameters among all
pairs that yield the selected structure. Thus, the estimates reported below
are not the refitted estimates used to calculate the BIC.

Bias is $100$ times the difference between the mean estimate and the true value; 
SE is $100$ times the average estimated standard error; SD is $100$ times the 
sample standard deviation of the estimates; and CP is the empirical coverage 
probability of the normal-approximation $95\%$ confidence interval. For a 
penalised parameter, these summaries are computed over the replications in 
which that parameter is estimated as nonzero.

Overall, Tables~\ref{table:bias_se_cp_n500}--\ref{table:bias_se_cp_n2000}
show that the estimation performance improves as the sample size increases.
For most parameters, the absolute bias and empirical SD decrease with $n$,
the average estimated SE is broadly comparable to the empirical SD, and the
coverage probability is reasonably close to the nominal level. The baseline
parameters, most fixed effects, factor loadings, and covariance parameters
are estimated with relatively small bias, particularly when $n=1000$ or
$2000$. A few fixed-effect parameters and one Factor 1 loading nevertheless exhibit
noticeably larger finite-sample bias or variability, or lower coverage
probability when $n=500$, as discussed below.

In particular, as shown in Table~\ref{table:bias_se_cp_n500}, the fixed-effect
parameters
$\beta_{\mathrm{W2};\,(\mathrm{W2},\mathrm{Back})}$,
$\beta_{\mathrm{W3};\,(\mathrm{W1},\mathrm{Back})}$, and
$\beta_{\mathrm{W3};\,(\mathrm{W3},\mathrm{Back})}$
have relatively large biases and low coverage probabilities when $n=500$.
These parameters represent the effects
\begin{equation*}
(\mathrm{W2},\mathrm{Back})\rightarrow\lambda_{\mathrm{W2}},
\qquad
(\mathrm{W1},\mathrm{Back})\rightarrow\lambda_{\mathrm{W3}},
\qquad
(\mathrm{W3},\mathrm{Back})\rightarrow\lambda_{\mathrm{W3}},
\end{equation*}
respectively, and all have a true value of $-2$. Consequently, the simulation
setting reduces the intensities of these transitions and makes the
corresponding events relatively infrequent. For example, following the
sequence $(\mathrm{W1},\mathrm{Back})$, the fixed-effect coefficient in the
intensity of W3 is $-2$, whereas the corresponding coefficient in the
intensity of W2 is $2$. Thus, after returning from W1, the fixed-effect
component encourages a test taker to proceed to W2 and discourages a direct
transition to W3.

As an empirical diagnostic, we generated $100$ independent datasets with
$n=500$. The average numbers of occurrences of
$(\mathrm{W2},\mathrm{Back})\rightarrow\mathrm{W2}$,
$(\mathrm{W1},\mathrm{Back})\rightarrow\mathrm{W3}$, and
$(\mathrm{W3},\mathrm{Back})\rightarrow\mathrm{W3}$ were $21.04$, $24.24$,
and $19.39$, respectively. By comparison,
$(\mathrm{W1},\mathrm{Back})\rightarrow\mathrm{W1}$ occurred an average of
$64.74$ times, and the corresponding parameter
$\beta_{\mathrm{W1};\,(\mathrm{W1},\mathrm{Back})}$ does not exhibit similarly
poor estimation performance in Table~\ref{table:bias_se_cp_n500}. These
results support the explanation that the relatively small numbers of relevant
events provide limited information for estimating the three parameters. In
addition, shrinkage induced by the SCAD penalty can contribute to
finite-sample bias, particularly for coefficients associated with
infrequently observed covariate patterns and transitions, for which the
likelihood contains relatively little information. The estimates summarised
in the simulation tables are the original penalised estimates, rather than
estimates obtained by refitting the selected model without a penalty.

The Bias, SE, and SD columns in
Tables~\ref{table:bias_se_cp_n500}--\ref{table:bias_se_cp_n2000} are
multiplied by $100$. Therefore, the reported biases of $31.37$, $29.57$,
and $39.09$ for $n=500$ correspond to actual biases of $0.3137$, $0.2957$,
and $0.3909$, respectively. As shown in
Table~\ref{table:bias_se_cp_n1000}, the corresponding reported biases
decrease to $15.70$, $13.40$, and $19.34$ when $n=1000$, corresponding to
actual biases of $0.1570$, $0.1340$, and $0.1934$, respectively. They
decrease further to $3.76$, $4.44$, and $7.24$ when $n=2000$, corresponding
to actual biases of $0.0376$, $0.0444$, and $0.0724$, respectively, as shown
in Table~\ref{table:bias_se_cp_n2000}. This progressive reduction is
consistent with the general improvement in estimation performance as the
sample size and the amount of information available for these transitions
increase.

A similar finite-sample phenomenon is observed for the Factor 1 loading
$a^{(1)}_{\mathrm{R}_3;\,\mathrm{R}_{\mathrm{Open}}}$, which has a relatively
large bias and low coverage probability when $n=500$. This loading describes
the heterogeneous effect of R$_{\mathrm{Open}}$ on the intensity of R$_3$.
To examine the amount of information available for estimating this loading,
we recorded the numbers of transitions corresponding to the four Factor 1
loadings reported in Table~\ref{table:bias_se_cp_n500}. Across the $100$
simulated datasets with $n=500$, the transition
R$_{\mathrm{Open}}\rightarrow$R$_3$ occurred an average of $132.47$ times.
By comparison, W2$\rightarrow$W2$_{\mathrm M}$,
W3$\rightarrow$W3$_{\mathrm M}$, and
W1$\rightarrow$W1$_{\mathrm M}$ occurred on average $245.74$, $321.61$,
and $379.80$ times, respectively. Thus, substantially fewer relevant
transitions are observed for
$a^{(1)}_{\mathrm{R}_3;\,\mathrm{R}_{\mathrm{Open}}}$ than for the other
Factor 1 loadings, which helps explain its greater finite-sample variability
and poorer coverage. This interpretation is also consistent with the
improvement as the sample size increases. On the original parameter scale,
its bias, average estimated standard error, empirical standard deviation, and
coverage probability are $-0.110$, $0.128$, $0.206$, and $0.71$ at $n=500$,
respectively. These become $0.001$, $0.088$, $0.099$, and $0.97$ at
$n=1000$, and $-0.001$, $0.060$, $0.060$, and $0.97$ at $n=2000$.

\begin{sidewaystable}[htbp]
\centering
\footnotesize 
\setlength{\tabcolsep}{4pt} 
\renewcommand{\arraystretch}{1.0} 
\begin{tabular}{l|rrrrrrrrrrrrrrr} 
\toprule
& $\lambda_{\text{W1}}$
& $\lambda_{\text{W1}_M}$
& $\lambda_{\text{W2}}$
& $\lambda_{\text{W2}_M}$
& $\lambda_{\text{W3}}$
& $\lambda_{\text{W3}_M}$
& $\lambda_{\text{Next}}$
& $\lambda_{\text{Next}_\text{Cancel}}$
& $\lambda_{\text{R}_1}$
& $\lambda_{\text{R}_2}$
& $\lambda_{\text{R}_3}$
& $\lambda_{\text{R}_{\text{Open}}}$
& $\lambda_{\text{R}_{\text{Close}}}$
& $\lambda_{\text{Back}}$
& $\lambda_{\text{Next}_{\text{OK}}}$
 \\
\midrule
$\beta_0$ & -4 & -5 & -5 & -5 & -5 & -4 & -7 & -4 & -5 & -5 & -5 & -6 & -5 & -7 & -2 \\
\midrule
W1 & . & . & . & . & . & . & . & . & . & . & . & . & . & 3 & . \\
W1$_\text{M}$ & . & . & . & . & . & . & . & . & . & . & . & . & . & 5 & . \\
W2 & . & . & . & . & . & . & . & . & . & . & . & . & . & 3 & . \\
W2$_\text{M}$ & . & . & . & . & . & . & . & . & . & . & . & . & . & 5 & . \\
W3 & . & . & . & . & . & . & . & . & . & . & . & . & . & 3 & . \\
W3$_\text{M}$ & . & . & . & . & . & . & . & . & . & . & . & 2 & . & 5 & . \\
Next & . & . & . & . & . & . & . & . & . & . & . & . & . & . & . \\
Next$_\text{Cancel}$ & . & . & . & . & . & 4 & . & . & . & . & . & . & . & . & . \\
R$_1$ & . & . & . & . & . & . & 5 & . & . & . & . & . & . & . & . \\
R$_2$ & . & . & . & . & . & . & 5 & . & . & . & . & . & . & . & . \\
R$_3$ & . & . & . & . & . & . & 5 & . & . & . & . & . & . & . & . \\
R$_\text{Open}$ & . & . & . & . & . & . & . & . & . & . & . & . & . & . & . \\
R$_\text{Close}$ & . & . & . & . & . & . & . & . & . & . & . & . & . & . & . \\
Back & 1 & . & 1 & . & 1 & . & . & . & . & . & . & . & . & 3 & . \\
W1, Back & -2 & . & 2 & . & -2 & . & . & . & . & . & . & . & . & . & . \\
W2, Back & -1 & . & -2 & . & 2 & . & . & . & . & . & . & . & . & . & . \\
W3, Back & . & . & . & . & -2 & . & . & . & . & . & . & 2 & . & . & . \\
\bottomrule
\end{tabular}
\caption{Simulation setting for the fixed effects. The columns correspond to the intensity functions for different event types. The first row represents the constant baseline, and the remaining rows correspond to the covariate processes. The numbers are the regression coefficients, and dots (.) indicate zero entries.}
\label{table:true_coef_sim_fixed}

\end{sidewaystable}

\begin{sidewaystable}[htbp]
\centering
\footnotesize 
\setlength{\tabcolsep}{4pt} 
\renewcommand{\arraystretch}{1.0} 
\begin{tabular}{l|rrrrrrrrrrrrrrr} 
\toprule
& $\lambda_{\text{W1}}$
& $\lambda_{\text{W1}_M}$
& $\lambda_{\text{W2}}$
& $\lambda_{\text{W2}_M}$
& $\lambda_{\text{W3}}$
& $\lambda_{\text{W3}_M}$
& $\lambda_{\text{Next}}$
& $\lambda_{\text{Next}_\text{Cancel}}$
& $\lambda_{\text{R}_1}$
& $\lambda_{\text{R}_2}$
& $\lambda_{\text{R}_3}$
& $\lambda_{\text{R}_{\text{Open}}}$
& $\lambda_{\text{R}_{\text{Close}}}$
& $\lambda_{\text{Back}}$
& $\lambda_{\text{Next}_{\text{OK}}}$
 \\
\midrule
W1              & . & 2 & . & . & . & . & . & . & . & . & . & . & . & . & . \\
W1$_\text{M}$   & . & . & . & . & . & . & . & . & . & . & . & . & . & . & . \\
W2              & . & . & . & 2 & . & . & . & . & . & . & . & . & . & . & . \\
W2$_\text{M}$   & . & . & . & . & . & . & . & . & . & . & . & . & . & . & . \\
W3              & . & . & . & . & . & 2 & . & . & . & . & . & . & . & . & . \\
W3$_\text{M}$   & . & . & . & . & . & . & . & . & . & . & . & . & . & . & . \\
Next            & . & . & . & . & . & . & . & . & . & . & . & . & . & . & . \\
Next$_\text{Cancel}$ 
                & . & . & . & . & . & . & . & . & . & . & . & . & . & . & . \\
R$_1$           & . & . & . & . & . & . & . & . & . & . & . & . & . & . & . \\
R$_2$           & . & . & . & . & . & . & . & . & . & . & . & . & . & . & . \\
R$_3$           & . & . & . & . & . & . & . & . & . & . & . & . & . & . & . \\
R$_\text{Open}$ & . & . & . & . & . & . & . & . & . & . & 1 & . & . & . & . \\
R$_\text{Close}$& . & . & . & . & . & . & . & . & . & . & . & . & . & . & . \\
Back            & . & . & . & . & . & . & . & . & . & . & . & . & . & . & . \\
W1, Back        & . & . & . & . & . & . & . & . & . & . & . & . & . & . & . \\
W2, Back        & . & . & . & . & . & . & . & . & . & . & . & . & . & . & . \\
W3, Back        & . & . & . & . & . & . & . & . & . & . & . & . & . & . & . \\
\bottomrule
\end{tabular}
\caption{Simulation setting for the first factor. The columns correspond to the intensity functions for different event types. The rows correspond to the covariate processes. The numbers are the factor loadings, and dots (.) indicate zero entries.}
\label{table:true_coef_sim_factor1}

\end{sidewaystable}

\begin{sidewaystable}[htbp]
\centering
\footnotesize 
\setlength{\tabcolsep}{4pt} 
\renewcommand{\arraystretch}{1.0} 
\begin{tabular}{l|rrrrrrrrrrrrrrr} 
\toprule
& $\lambda_{\text{W1}}$
& $\lambda_{\text{W1}_M}$
& $\lambda_{\text{W2}}$
& $\lambda_{\text{W2}_M}$
& $\lambda_{\text{W3}}$
& $\lambda_{\text{W3}_M}$
& $\lambda_{\text{Next}}$
& $\lambda_{\text{Next}_\text{Cancel}}$
& $\lambda_{\text{R}_1}$
& $\lambda_{\text{R}_2}$
& $\lambda_{\text{R}_3}$
& $\lambda_{\text{R}_{\text{Open}}}$
& $\lambda_{\text{R}_{\text{Close}}}$
& $\lambda_{\text{Back}}$
& $\lambda_{\text{Next}_{\text{OK}}}$
 \\
\midrule
W1              & . & . & . & . & . & . & . & . & . & . & . & . & . & 1 & . \\
W1$_\text{M}$   & . & . & . & . & . & . & . & . & . & . & . & . & . & 1 & . \\
W2              & . & . & . & . & . & . & . & . & . & . & . & . & . & 1 & . \\
W2$_\text{M}$   & . & . & . & . & . & . & . & . & . & . & . & . & . & 1 & . \\
W3              & . & . & . & . & . & . & . & . & . & . & . & . & . & 1 & . \\
W3$_\text{M}$   & . & . & . & . & . & . & . & . & . & . & . & . & . & 1 & . \\
Next            & . & . & . & . & . & . & . & . & . & . & . & . & . & . & . \\
Next$_\text{Cancel}$ 
                & . & . & . & . & . & . & . & . & . & . & . & . & . & . & . \\
R$_1$           & . & . & . & . & . & . & . & . & . & . & . & . & . & . & . \\
R$_2$           & . & . & . & . & . & . & . & . & . & . & . & . & . & . & . \\
R$_3$           & . & . & . & . & . & . & . & . & . & . & . & . & . & . & . \\
R$_\text{Open}$ & . & . & . & . & . & . & . & . & . & . & . & . & . & . & . \\
R$_\text{Close}$& . & . & . & . & . & . & . & . & . & . & . & . & . & . & . \\
Back            & . & . & . & . & . & . & . & . & . & . & . & . & . & . & . \\
W1, Back        & . & . & . & . & . & . & . & . & . & . & . & . & . & . & . \\
W2, Back        & . & . & . & . & . & . & . & . & . & . & . & . & . & . & . \\
W3, Back        & . & . & . & . & . & . & . & . & . & . & . & . & . & . & . \\
\bottomrule
\end{tabular}
\caption{Simulation setting for the second factor. The columns correspond to the intensity functions for different event types. The rows correspond to the covariate processes. The numbers are the factor loadings, and dots (.) indicate zero entries.}
\label{table:true_coef_sim_factor2}

\end{sidewaystable}

\begin{sidewaystable}[htbp]
\centering
\footnotesize 
\setlength{\tabcolsep}{4pt} 
\renewcommand{\arraystretch}{1.0} 
\begin{tabular}{l|rrrrrrrrrrrrrrr} 
\toprule
& $\lambda_{\text{W1}}$
& $\lambda_{\text{W1}_M}$
& $\lambda_{\text{W2}}$
& $\lambda_{\text{W2}_M}$
& $\lambda_{\text{W3}}$
& $\lambda_{\text{W3}_M}$
& $\lambda_{\text{Next}}$
& $\lambda_{\text{Next}_\text{Cancel}}$
& $\lambda_{\text{R}_1}$
& $\lambda_{\text{R}_2}$
& $\lambda_{\text{R}_3}$
& $\lambda_{\text{R}_{\text{Open}}}$
& $\lambda_{\text{R}_{\text{Close}}}$
& $\lambda_{\text{Back}}$
& $\lambda_{\text{Next}_{\text{OK}}}$
 \\
\midrule
W1              & . & . & . & . & . & . & . & . & . & . & . & . & . & . & . \\
W1$_\text{M}$   & . & . & . & . & . & . & . & . & . & . & . & . & . & . & . \\
W2              & . & . & . & . & . & . & . & . & . & . & . & . & . & . & . \\
W2$_\text{M}$   & . & . & . & . & . & . & . & . & . & . & . & . & . & . & . \\
W3              & . & . & . & . & . & . & . & . & . & . & . & . & . & . & . \\
W3$_\text{M}$   & . & . & . & . & . & . & . & . & . & . & . & . & . & . & . \\
Next            & . & . & . & . & . & . & . & . & . & . & . & . & . & . & . \\
Next$_\text{Cancel}$ 
                & . & . & . & . & . & . & . & . & . & . & . & . & . & . & . \\
R$_1$           & . & . & . & . & . & . & . & . & . & . & . & . & . & . & . \\
R$_2$           & . & . & . & . & . & . & . & . & . & . & . & . & . & . & . \\
R$_3$           & . & . & . & . & . & . & . & . & . & . & . & . & . & . & . \\
R$_\text{Open}$ & . & . & . & . & . & . & . & . & . & . & . & . & . & . & . \\
R$_\text{Close}$& . & . & . & . & . & . & . & . & . & . & . & . & . & . & . \\
Back            & . & . & . & . & . & . & . & . & . & . & . & . & . & . & . \\
W1, Back        & . & . & 1 & . & . & . & . & . & . & . & . & . & . & . & . \\
W2, Back        & . & . & . & . & 1 & . & . & . & . & . & . & . & . & . & . \\
W3, Back        & 1 & . & . & . & . & . & . & . & . & . & . & . & . & . & . \\
\bottomrule
\end{tabular}
\caption{Simulation setting for the third factor. The columns correspond to the intensity functions for different event types. The rows correspond to the covariate processes. The numbers are the factor loadings, and dots (.) indicate zero entries.}
\label{table:true_coef_sim_factor3}

\end{sidewaystable}

\begin{table}[H]
	\centering
	\tabcolsep=0.11cm
 \setlength\extrarowheight{-5pt}

	\begin{tabular}{c|ccc}
\hline
 & $\theta_1$ & $\theta_2$ & $\theta_3$\\
  \hline
		$\theta_1$ & 1 & -0.3  &0.3\\
  	$\theta_2$& -0.3 & 1  & -0.3\\
   	$\theta_3$ & 0.3 & -0.3  &1\\
    \hline
	\end{tabular}
     \caption{Simulation setting for the covariance matrix of the random effects.}
\label{table_covariance}
\end{table}

\begin{table}[!htbp]
\centering
\begin{minipage}[t]{0.59\textwidth}
\vspace{0pt}
\centering
{\small\textbf{Baseline parameters and fixed effects}}

\vspace{2pt}
\footnotesize
\setlength{\tabcolsep}{1.1pt}
\renewcommand{\arraystretch}{1.2}
\begin{adjustbox}{max width=\linewidth}
\begin{tabular}{@{}lrr@{\hspace{4.5pt}}r@{\hspace{4.5pt}}r@{\hspace{4.5pt}}r@{}}
\toprule
Parameter & True & Bias & SE & SD & CP \\
\midrule
\multicolumn{6}{l}{\textit{Baseline parameters}} \\
\addlinespace[1pt]
$\beta_{\mathrm{W1},0}$ & -4 & -1.23 & 5.42 & 5.11 & 0.96 \\
$\beta_{\mathrm{W1}_{\mathrm M},0}$ & -5 & -0.39 & 6.86 & 5.59 & 0.99 \\
$\beta_{\mathrm{W2},0}$ & -5 & -0.97 & 9.03 & 7.77 & 0.98 \\
$\beta_{\mathrm{W2}_{\mathrm M},0}$ & -5 & -0.12 & 7.69 & 6.78 & 0.98 \\
$\beta_{\mathrm{W3},0}$ & -5 & -0.18 & 8.97 & 8.46 & 0.96 \\
$\beta_{\mathrm{W3}_{\mathrm M},0}$ & -4 & -0.35 & 5.52 & 5.63 & 0.96 \\
$\beta_{\mathrm{Next},0}$ & -7 & 0.64 & 7.52 & 6.80 & 0.94 \\
$\beta_{\mathrm{Next}_{\mathrm{Cancel}},0}$ & -4 & -1.74 & 13.33 & 12.85 & 0.98 \\
$\beta_{\mathrm{R}_1,0}$ & -5 & 0.86 & 10.23 & 9.47 & 0.95 \\
$\beta_{\mathrm{R}_2,0}$ & -5 & 0.29 & 10.29 & 8.14 & 0.98 \\
$\beta_{\mathrm{R}_3,0}$ & -5 & 4.38 & 11.66 & 12.75 & 0.89 \\
$\beta_{\mathrm{R}_{\mathrm{Open}},0}$ & -6 & -0.27 & 4.89 & 4.66 & 0.96 \\
$\beta_{\mathrm{R}_{\mathrm{Close}},0}$ & -5 & -2.11 & 10.58 & 8.80 & 0.99 \\
$\beta_{\mathrm{Back},0}$ & -7 & -3.04 & 21.69 & 17.00 & 0.99 \\
$\beta_{\mathrm{Next}_{\mathrm{OK}},0}$ & -2 & 3.94 & 6.15 & 9.97 & 0.93 \\
\midrule
\multicolumn{6}{l}{\textit{Fixed effects}} \\
\addlinespace[1pt]
$\beta_{\mathrm{W1};\,\mathrm{Back}}$ & 1 & 1.16 & 7.17 & 7.21 & 0.96 \\
$\beta_{\mathrm{W1};\,(\mathrm{W1},\mathrm{Back})}$ & -2 & -1.97 & 14.98 & 13.55 & 1.00 \\
$\beta_{\mathrm{W1};\,(\mathrm{W2},\mathrm{Back})}$ & -1 & -0.60 & 10.68 & 9.52 & 0.98 \\
$\beta_{\mathrm{W2};\,\mathrm{Back}}$ & 1 & -1.88 & 11.07 & 10.56 & 0.96 \\
$\beta_{\mathrm{W2};\,(\mathrm{W1},\mathrm{Back})}$ & 2 & 4.57 & 9.72 & 9.63 & 0.91 \\
$\beta_{\mathrm{W2};\,(\mathrm{W2},\mathrm{Back})}$ & -2 & 31.37 & 26.42 & 21.65 & 0.75 \\
$\beta_{\mathrm{W3};\,\mathrm{Back}}$ & 1 & -8.34 & 12.55 & 14.67 & 0.87 \\
$\beta_{\mathrm{W3};\,(\mathrm{W1},\mathrm{Back})}$ & -2 & 29.57 & 24.66 & 25.91 & 0.77 \\
$\beta_{\mathrm{W3};\,(\mathrm{W2},\mathrm{Back})}$ & 2 & 10.98 & 11.49 & 13.65 & 0.83 \\
$\beta_{\mathrm{W3};\,(\mathrm{W3},\mathrm{Back})}$ & -2 & 39.09 & 27.02 & 25.61 & 0.66 \\
$\beta_{\mathrm{Next};\,\mathrm{Next}_{\mathrm{Cancel}}}$ & 4 & 0.21 & 16.97 & 14.62 & 0.99 \\
$\beta_{\mathrm{Next};\,\mathrm{R}_1}$ & 5 & 0.14 & 13.25 & 12.86 & 0.94 \\
$\beta_{\mathrm{Next};\,\mathrm{R}_2}$ & 5 & -1.11 & 13.20 & 12.63 & 0.97 \\
$\beta_{\mathrm{Next};\,\mathrm{R}_3}$ & 5 & 0.25 & 12.60 & 11.40 & 0.98 \\
$\beta_{\mathrm{R}_{\mathrm{Open}};\,\mathrm{W3}_{\mathrm M}}$ & 2 & -0.22 & 12.82 & 12.01 & 0.97 \\
$\beta_{\mathrm{R}_{\mathrm{Open}};\,(\mathrm{W3},\mathrm{Back})}$ & 2 & -0.18 & 10.67 & 9.80 & 0.96 \\
$\beta_{\mathrm{Back};\,\mathrm{W1}}$ & 3 & 2.56 & 22.83 & 18.79 & 0.97 \\
$\beta_{\mathrm{Back};\,\mathrm{W1}_{\mathrm M}}$ & 5 & 2.16 & 23.04 & 19.14 & 0.99 \\
$\beta_{\mathrm{Back};\,\mathrm{W2}}$ & 3 & 2.93 & 22.98 & 17.50 & 0.99 \\
$\beta_{\mathrm{Back};\,\mathrm{W2}_{\mathrm M}}$ & 5 & 2.84 & 23.40 & 19.76 & 0.99 \\
$\beta_{\mathrm{Back};\,\mathrm{W3}}$ & 3 & 2.89 & 23.40 & 19.52 & 0.99 \\
$\beta_{\mathrm{Back};\,\mathrm{W3}_{\mathrm M}}$ & 5 & 2.37 & 23.01 & 19.64 & 0.98 \\
$\beta_{\mathrm{Back};\,\mathrm{Back}}$ & 3 & 2.89 & 22.06 & 18.20 & 0.98 \\
\bottomrule
\end{tabular}
\end{adjustbox}
\end{minipage}
\hfill
\begin{minipage}[t]{0.40\textwidth}
\vspace{0pt}
\centering
\begin{adjustbox}{max width=\linewidth}
{\small\textbf{Factor loadings and covariance parameters}}
\end{adjustbox}

\vspace{2pt}
\footnotesize
\setlength{\tabcolsep}{1.1pt}
\renewcommand{\arraystretch}{1.2}
\begin{adjustbox}{max width=\linewidth}
\begin{tabular}{@{}lrr@{\hspace{4.5pt}}r@{\hspace{4.5pt}}r@{\hspace{4.5pt}}r@{}}
\toprule
Parameter & True & Bias & SE & SD & CP \\
\midrule
\multicolumn{6}{l}{\textit{Factor 1 loadings}} \\
\addlinespace[1pt]
$a^{(1)}_{\mathrm{W1}_{\mathrm M};\,\mathrm{W1}}$ & 2 & -2.49 & 12.15 & 17.71 & 0.97 \\
$a^{(1)}_{\mathrm{W2}_{\mathrm M};\,\mathrm{W2}}$ & 2 & -0.83 & 14.19 & 14.47 & 0.91 \\
$a^{(1)}_{\mathrm{W3}_{\mathrm M};\,\mathrm{W3}}$ & 2 & 0.24 & 13.11 & 13.32 & 0.95 \\
$a^{(1)}_{\mathrm{R}_3;\,\mathrm{R}_{\mathrm{Open}}}$ & 1 & -11.03 & 12.82 & 20.56 & 0.71 \\
\midrule
\multicolumn{6}{l}{\textit{Factor 2 loadings}} \\
\addlinespace[1pt]
$a^{(2)}_{\mathrm{Back};\,\mathrm{W1}}$ & 1 & 0.99 & 6.77 & 5.58 & 0.99 \\
$a^{(2)}_{\mathrm{Back};\,\mathrm{W1}_{\mathrm M}}$ & 1 & 1.38 & 8.09 & 6.54 & 1.00 \\
$a^{(2)}_{\mathrm{Back};\,\mathrm{W2}}$ & 1 & 1.13 & 7.00 & 7.03 & 0.94 \\
$a^{(2)}_{\mathrm{Back};\,\mathrm{W2}_{\mathrm M}}$ & 1 & 3.14 & 9.47 & 8.42 & 0.97 \\
$a^{(2)}_{\mathrm{Back};\,\mathrm{W3}}$ & 1 & 1.17 & 7.93 & 7.39 & 0.97 \\
$a^{(2)}_{\mathrm{Back};\,\mathrm{W3}_{\mathrm M}}$ & 1 & 0.93 & 8.26 & 7.69 & 0.96 \\
\midrule
\multicolumn{6}{l}{\textit{Factor 3 loadings}} \\
\addlinespace[1pt]
$a^{(3)}_{\mathrm{W1};\,(\mathrm{W3},\mathrm{Back})}$ & 1 & 0.45 & 10.78 & 9.84 & 0.95 \\
$a^{(3)}_{\mathrm{W2};\,(\mathrm{W1},\mathrm{Back})}$ & 1 & 1.56 & 7.67 & 6.83 & 0.98 \\
$a^{(3)}_{\mathrm{W3};\,(\mathrm{W2},\mathrm{Back})}$ & 1 & 1.12 & 8.22 & 7.64 & 0.96 \\
\midrule
\multicolumn{6}{l}{\textit{Covariance parameters}} \\
\addlinespace[1pt]
$\Sigma_{12}$ & -0.30 & -1.25 & 6.27 & 8.74 & 0.96 \\
$\Sigma_{13}$ & 0.30 & 2.21 & 8.63 & 9.65 & 0.94 \\
$\Sigma_{23}$ & -0.30 & -0.14 & 7.92 & 7.82 & 0.95 \\
\bottomrule
\end{tabular}
\end{adjustbox}
\end{minipage}
\caption{Simulation results for $n=500$.}
\label{table:bias_se_cp_n500}
\end{table}

\clearpage
\begin{table}[!htbp]
\centering
\begin{minipage}[t]{0.59\textwidth}
\vspace{0pt}
\centering
{\small\textbf{Baseline parameters and fixed effects}}

\vspace{2pt}
\footnotesize
\setlength{\tabcolsep}{1.1pt}
\renewcommand{\arraystretch}{1.2}
\begin{adjustbox}{max width=\linewidth}
\begin{tabular}{@{}lrr@{\hspace{4.5pt}}r@{\hspace{4.5pt}}r@{\hspace{4.5pt}}r@{}}
\toprule
Parameter & True & Bias & SE & SD & CP \\
\midrule
\multicolumn{6}{l}{\textit{Baseline parameters}} \\
\addlinespace[1pt]
$\beta_{\mathrm{W1},0}$ & -4 & -0.46 & 3.69 & 3.52 & 0.97 \\
$\beta_{\mathrm{W1}_{\mathrm M},0}$ & -5 & -0.46 & 4.55 & 3.96 & 0.98 \\
$\beta_{\mathrm{W2},0}$ & -5 & -0.44 & 6.13 & 6.26 & 0.95 \\
$\beta_{\mathrm{W2}_{\mathrm M},0}$ & -5 & -0.55 & 5.06 & 4.65 & 0.97 \\
$\beta_{\mathrm{W3},0}$ & -5 & -0.54 & 6.12 & 5.91 & 0.98 \\
$\beta_{\mathrm{W3}_{\mathrm M},0}$ & -4 & -0.28 & 3.65 & 4.02 & 0.94 \\
$\beta_{\mathrm{Next},0}$ & -7 & -0.44 & 5.12 & 4.80 & 0.95 \\
$\beta_{\mathrm{Next}_{\mathrm{Cancel}},0}$ & -4 & -0.95 & 8.98 & 8.40 & 0.96 \\
$\beta_{\mathrm{R}_1,0}$ & -5 & -0.74 & 6.97 & 7.16 & 0.94 \\
$\beta_{\mathrm{R}_2,0}$ & -5 & -0.25 & 6.98 & 6.99 & 0.93 \\
$\beta_{\mathrm{R}_3,0}$ & -5 & -0.13 & 8.14 & 9.11 & 0.92 \\
$\beta_{\mathrm{R}_{\mathrm{Open}},0}$ & -6 & -0.39 & 3.30 & 3.31 & 0.97 \\
$\beta_{\mathrm{R}_{\mathrm{Close}},0}$ & -5 & -1.46 & 7.07 & 6.95 & 0.96 \\
$\beta_{\mathrm{Back},0}$ & -7 & -0.44 & 14.46 & 13.06 & 1.00 \\
$\beta_{\mathrm{Next}_{\mathrm{OK}},0}$ & -2 & 0.70 & 3.66 & 6.06 & 0.95 \\
\midrule
\multicolumn{6}{l}{\textit{Fixed effects}} \\
\addlinespace[1pt]
$\beta_{\mathrm{W1};\,\mathrm{Back}}$ & 1 & 0.35 & 4.80 & 4.70 & 0.97 \\
$\beta_{\mathrm{W1};\,(\mathrm{W1},\mathrm{Back})}$ & -2 & -0.88 & 9.78 & 9.66 & 0.94 \\
$\beta_{\mathrm{W1};\,(\mathrm{W2},\mathrm{Back})}$ & -1 & 0.66 & 6.94 & 6.74 & 0.97 \\
$\beta_{\mathrm{W2};\,\mathrm{Back}}$ & 1 & -0.81 & 7.42 & 7.55 & 0.92 \\
$\beta_{\mathrm{W2};\,(\mathrm{W1},\mathrm{Back})}$ & 2 & 1.83 & 6.30 & 6.68 & 0.91 \\
$\beta_{\mathrm{W2};\,(\mathrm{W2},\mathrm{Back})}$ & -2 & 15.70 & 17.24 & 19.85 & 0.86 \\
$\beta_{\mathrm{W3};\,\mathrm{Back}}$ & 1 & -4.61 & 8.43 & 12.99 & 0.88 \\
$\beta_{\mathrm{W3};\,(\mathrm{W1},\mathrm{Back})}$ & -2 & 13.40 & 16.28 & 23.52 & 0.85 \\
$\beta_{\mathrm{W3};\,(\mathrm{W2},\mathrm{Back})}$ & 2 & 6.02 & 7.54 & 13.64 & 0.82 \\
$\beta_{\mathrm{W3};\,(\mathrm{W3},\mathrm{Back})}$ & -2 & 19.34 & 17.87 & 23.61 & 0.79 \\
$\beta_{\mathrm{Next};\,\mathrm{Next}_{\mathrm{Cancel}}}$ & 4 & 1.28 & 11.33 & 10.49 & 0.97 \\
$\beta_{\mathrm{Next};\,\mathrm{R}_1}$ & 5 & 0.56 & 8.98 & 9.73 & 0.95 \\
$\beta_{\mathrm{Next};\,\mathrm{R}_2}$ & 5 & -0.03 & 8.96 & 9.53 & 0.92 \\
$\beta_{\mathrm{Next};\,\mathrm{R}_3}$ & 5 & 0.32 & 8.48 & 8.99 & 0.93 \\
$\beta_{\mathrm{R}_{\mathrm{Open}};\,\mathrm{W3}_{\mathrm M}}$ & 2 & -0.03 & 8.73 & 8.19 & 0.96 \\
$\beta_{\mathrm{R}_{\mathrm{Open}};\,(\mathrm{W3},\mathrm{Back})}$ & 2 & 0.26 & 7.00 & 6.97 & 0.95 \\
$\beta_{\mathrm{Back};\,\mathrm{W1}}$ & 3 & 0.44 & 15.25 & 14.22 & 0.96 \\
$\beta_{\mathrm{Back};\,\mathrm{W1}_{\mathrm M}}$ & 5 & 0.11 & 15.37 & 14.10 & 0.99 \\
$\beta_{\mathrm{Back};\,\mathrm{W2}}$ & 3 & 0.89 & 15.33 & 13.16 & 0.99 \\
$\beta_{\mathrm{Back};\,\mathrm{W2}_{\mathrm M}}$ & 5 & 0.66 & 15.56 & 14.14 & 0.96 \\
$\beta_{\mathrm{Back};\,\mathrm{W3}}$ & 3 & 0.42 & 15.58 & 14.18 & 0.98 \\
$\beta_{\mathrm{Back};\,\mathrm{W3}_{\mathrm M}}$ & 5 & 0.49 & 15.33 & 14.07 & 0.98 \\
$\beta_{\mathrm{Back};\,\mathrm{Back}}$ & 3 & 0.36 & 14.69 & 14.13 & 0.97 \\
\bottomrule
\end{tabular}
\end{adjustbox}
\end{minipage}
\hfill
\begin{minipage}[t]{0.40\textwidth}
\vspace{0pt}
\centering
\begin{adjustbox}{max width=\linewidth}
{\small\textbf{Factor loadings and covariance parameters}}
\end{adjustbox}

\vspace{2pt}
\footnotesize
\setlength{\tabcolsep}{1.1pt}
\renewcommand{\arraystretch}{1.2}
\begin{adjustbox}{max width=\linewidth}
\begin{tabular}{@{}lrr@{\hspace{4.5pt}}r@{\hspace{4.5pt}}r@{\hspace{4.5pt}}r@{}}
\toprule
Parameter & True & Bias & SE & SD & CP \\
\midrule
\multicolumn{6}{l}{\textit{Factor 1 loadings}} \\
\addlinespace[1pt]
$a^{(1)}_{\mathrm{W1}_{\mathrm M};\,\mathrm{W1}}$ & 2 & 0.51 & 8.23 & 8.57 & 0.94 \\
$a^{(1)}_{\mathrm{W2}_{\mathrm M};\,\mathrm{W2}}$ & 2 & 0.18 & 9.35 & 10.51 & 0.90 \\
$a^{(1)}_{\mathrm{W3}_{\mathrm M};\,\mathrm{W3}}$ & 2 & 0.96 & 8.76 & 9.43 & 0.89 \\
$a^{(1)}_{\mathrm{R}_3;\,\mathrm{R}_{\mathrm{Open}}}$ & 1 & 0.11 & 8.76 & 9.85 & 0.97 \\
\midrule
\multicolumn{6}{l}{\textit{Factor 2 loadings}} \\
\addlinespace[1pt]
$a^{(2)}_{\mathrm{Back};\,\mathrm{W1}}$ & 1 & 0.62 & 4.44 & 4.14 & 0.98 \\
$a^{(2)}_{\mathrm{Back};\,\mathrm{W1}_{\mathrm M}}$ & 1 & 1.08 & 5.34 & 4.29 & 0.97 \\
$a^{(2)}_{\mathrm{Back};\,\mathrm{W2}}$ & 1 & 0.79 & 4.60 & 4.38 & 0.96 \\
$a^{(2)}_{\mathrm{Back};\,\mathrm{W2}_{\mathrm M}}$ & 1 & 1.42 & 6.10 & 5.10 & 1.00 \\
$a^{(2)}_{\mathrm{Back};\,\mathrm{W3}}$ & 1 & 0.72 & 5.15 & 4.89 & 0.97 \\
$a^{(2)}_{\mathrm{Back};\,\mathrm{W3}_{\mathrm M}}$ & 1 & 0.45 & 5.43 & 5.15 & 0.97 \\
\midrule
\multicolumn{6}{l}{\textit{Factor 3 loadings}} \\
\addlinespace[1pt]
$a^{(3)}_{\mathrm{W1};\,(\mathrm{W3},\mathrm{Back})}$ & 1 & 0.08 & 6.93 & 7.35 & 0.91 \\
$a^{(3)}_{\mathrm{W2};\,(\mathrm{W1},\mathrm{Back})}$ & 1 & 0.36 & 4.92 & 4.89 & 0.94 \\
$a^{(3)}_{\mathrm{W3};\,(\mathrm{W2},\mathrm{Back})}$ & 1 & 0.35 & 5.32 & 7.01 & 0.94 \\
\midrule
\multicolumn{6}{l}{\textit{Covariance parameters}} \\
\addlinespace[1pt]
$\Sigma_{12}$ & -0.30 & -0.62 & 4.25 & 3.92 & 0.97 \\
$\Sigma_{13}$ & 0.30 & 1.50 & 5.80 & 8.80 & 0.96 \\
$\Sigma_{23}$ & -0.30 & -0.61 & 5.21 & 5.50 & 0.95 \\
\bottomrule
\end{tabular}
\end{adjustbox}
\end{minipage}
\caption{Simulation results for $n=1000$.}
\label{table:bias_se_cp_n1000}
\end{table}

\clearpage
\begin{table}[!htbp]
\centering
\begin{minipage}[t]{0.59\textwidth}
\vspace{0pt}
\centering
{\small\textbf{Baseline parameters and fixed effects}}

\vspace{2pt}
\footnotesize
\setlength{\tabcolsep}{1.1pt}
\renewcommand{\arraystretch}{1.2}
\begin{adjustbox}{max width=\linewidth}
\begin{tabular}{@{}lrr@{\hspace{4.5pt}}r@{\hspace{4.5pt}}r@{\hspace{4.5pt}}r@{}}
\toprule
Parameter & True & Bias & SE & SD & CP \\
\midrule
\multicolumn{6}{l}{\textit{Baseline parameters}} \\
\addlinespace[1pt]
$\beta_{\mathrm{W1},0}$ & -4 & 0.05 & 2.57 & 2.30 & 0.96 \\
$\beta_{\mathrm{W1}_{\mathrm M},0}$ & -5 & -0.38 & 3.12 & 2.95 & 0.96 \\
$\beta_{\mathrm{W2},0}$ & -5 & -0.24 & 4.25 & 4.60 & 0.96 \\
$\beta_{\mathrm{W2}_{\mathrm M},0}$ & -5 & 0.01 & 3.51 & 3.52 & 0.97 \\
$\beta_{\mathrm{W3},0}$ & -5 & -0.07 & 4.25 & 4.78 & 0.93 \\
$\beta_{\mathrm{W3}_{\mathrm M},0}$ & -4 & -0.32 & 2.52 & 2.97 & 0.88 \\
$\beta_{\mathrm{Next},0}$ & -7 & 0.05 & 3.54 & 3.19 & 0.94 \\
$\beta_{\mathrm{Next}_{\mathrm{Cancel}},0}$ & -4 & -1.00 & 6.22 & 5.93 & 0.97 \\
$\beta_{\mathrm{R}_1,0}$ & -5 & -0.38 & 4.84 & 4.48 & 0.95 \\
$\beta_{\mathrm{R}_2,0}$ & -5 & 0.14 & 4.83 & 4.06 & 0.98 \\
$\beta_{\mathrm{R}_3,0}$ & -5 & 0.27 & 5.64 & 5.75 & 0.94 \\
$\beta_{\mathrm{R}_{\mathrm{Open}},0}$ & -6 & -0.19 & 2.28 & 2.28 & 0.94 \\
$\beta_{\mathrm{R}_{\mathrm{Close}},0}$ & -5 & -1.04 & 4.88 & 5.52 & 0.92 \\
$\beta_{\mathrm{Back},0}$ & -7 & -0.48 & 10.07 & 9.51 & 0.96 \\
$\beta_{\mathrm{Next}_{\mathrm{OK}},0}$ & -2 & 0.18 & 2.38 & 3.03 & 0.97 \\
\midrule
\multicolumn{6}{l}{\textit{Fixed effects}} \\
\addlinespace[1pt]
$\beta_{\mathrm{W1};\,\mathrm{Back}}$ & 1 & -0.24 & 3.34 & 2.99 & 0.97 \\
$\beta_{\mathrm{W1};\,(\mathrm{W1},\mathrm{Back})}$ & -2 & -0.32 & 6.67 & 6.37 & 0.97 \\
$\beta_{\mathrm{W1};\,(\mathrm{W2},\mathrm{Back})}$ & -1 & 0.62 & 4.73 & 4.81 & 0.95 \\
$\beta_{\mathrm{W2};\,\mathrm{Back}}$ & 1 & -0.10 & 5.16 & 5.43 & 0.93 \\
$\beta_{\mathrm{W2};\,(\mathrm{W1},\mathrm{Back})}$ & 2 & 0.31 & 4.35 & 4.76 & 0.91 \\
$\beta_{\mathrm{W2};\,(\mathrm{W2},\mathrm{Back})}$ & -2 & 3.76 & 11.93 & 15.87 & 0.88 \\
$\beta_{\mathrm{W3};\,\mathrm{Back}}$ & 1 & -1.56 & 5.81 & 7.29 & 0.90 \\
$\beta_{\mathrm{W3};\,(\mathrm{W1},\mathrm{Back})}$ & -2 & 4.44 & 11.18 & 12.68 & 0.89 \\
$\beta_{\mathrm{W3};\,(\mathrm{W2},\mathrm{Back})}$ & 2 & 2.08 & 5.20 & 5.90 & 0.92 \\
$\beta_{\mathrm{W3};\,(\mathrm{W3},\mathrm{Back})}$ & -2 & 7.24 & 12.16 & 17.68 & 0.84 \\
$\beta_{\mathrm{Next};\,\mathrm{Next}_{\mathrm{Cancel}}}$ & 4 & 0.36 & 7.85 & 7.70 & 0.94 \\
$\beta_{\mathrm{Next};\,\mathrm{R}_1}$ & 5 & -0.25 & 6.22 & 6.04 & 0.97 \\
$\beta_{\mathrm{Next};\,\mathrm{R}_2}$ & 5 & -0.11 & 6.22 & 6.47 & 0.96 \\
$\beta_{\mathrm{Next};\,\mathrm{R}_3}$ & 5 & 0.26 & 5.87 & 6.09 & 0.95 \\
$\beta_{\mathrm{R}_{\mathrm{Open}};\,\mathrm{W3}_{\mathrm M}}$ & 2 & 0.27 & 6.07 & 6.05 & 0.97 \\
$\beta_{\mathrm{R}_{\mathrm{Open}};\,(\mathrm{W3},\mathrm{Back})}$ & 2 & 0.19 & 4.81 & 4.76 & 0.94 \\
$\beta_{\mathrm{Back};\,\mathrm{W1}}$ & 3 & 0.85 & 10.62 & 10.71 & 0.94 \\
$\beta_{\mathrm{Back};\,\mathrm{W1}_{\mathrm M}}$ & 5 & 0.87 & 10.71 & 10.34 & 0.99 \\
$\beta_{\mathrm{Back};\,\mathrm{W2}}$ & 3 & 1.17 & 10.67 & 9.84 & 0.97 \\
$\beta_{\mathrm{Back};\,\mathrm{W2}_{\mathrm M}}$ & 5 & 0.79 & 10.83 & 10.72 & 0.95 \\
$\beta_{\mathrm{Back};\,\mathrm{W3}}$ & 3 & 0.50 & 10.85 & 10.82 & 0.96 \\
$\beta_{\mathrm{Back};\,\mathrm{W3}_{\mathrm M}}$ & 5 & 0.56 & 10.68 & 10.62 & 0.97 \\
$\beta_{\mathrm{Back};\,\mathrm{Back}}$ & 3 & 0.63 & 10.24 & 10.00 & 0.96 \\
\bottomrule
\end{tabular}
\end{adjustbox}
\end{minipage}
\hfill
\begin{minipage}[t]{0.40\textwidth}
\vspace{0pt}
\centering
\begin{adjustbox}{max width=\linewidth}
{\small\textbf{Factor loadings and covariance parameters}}
\end{adjustbox}

\vspace{2pt}
\footnotesize
\setlength{\tabcolsep}{1.1pt}
\renewcommand{\arraystretch}{1.2}
\begin{adjustbox}{max width=\linewidth}
\begin{tabular}{@{}lrr@{\hspace{4.5pt}}r@{\hspace{4.5pt}}r@{\hspace{4.5pt}}r@{}}
\toprule
Parameter & True & Bias & SE & SD & CP \\
\midrule
\multicolumn{6}{l}{\textit{Factor 1 loadings}} \\
\addlinespace[1pt]
$a^{(1)}_{\mathrm{W1}_{\mathrm M};\,\mathrm{W1}}$ & 2 & -0.36 & 5.64 & 5.46 & 0.96 \\
$a^{(1)}_{\mathrm{W2}_{\mathrm M};\,\mathrm{W2}}$ & 2 & -0.43 & 6.44 & 6.36 & 0.95 \\
$a^{(1)}_{\mathrm{W3}_{\mathrm M};\,\mathrm{W3}}$ & 2 & 0.15 & 6.03 & 6.64 & 0.93 \\
$a^{(1)}_{\mathrm{R}_3;\,\mathrm{R}_{\mathrm{Open}}}$ & 1 & -0.14 & 6.03 & 5.98 & 0.97 \\
\midrule
\multicolumn{6}{l}{\textit{Factor 2 loadings}} \\
\addlinespace[1pt]
$a^{(2)}_{\mathrm{Back};\,\mathrm{W1}}$ & 1 & 0.71 & 3.03 & 2.96 & 0.95 \\
$a^{(2)}_{\mathrm{Back};\,\mathrm{W1}_{\mathrm M}}$ & 1 & 1.25 & 3.65 & 3.42 & 0.95 \\
$a^{(2)}_{\mathrm{Back};\,\mathrm{W2}}$ & 1 & 0.59 & 3.15 & 2.76 & 0.97 \\
$a^{(2)}_{\mathrm{Back};\,\mathrm{W2}_{\mathrm M}}$ & 1 & 1.33 & 4.17 & 3.60 & 0.97 \\
$a^{(2)}_{\mathrm{Back};\,\mathrm{W3}}$ & 1 & 0.89 & 3.51 & 3.73 & 0.92 \\
$a^{(2)}_{\mathrm{Back};\,\mathrm{W3}_{\mathrm M}}$ & 1 & 0.83 & 3.69 & 3.65 & 0.98 \\
\midrule
\multicolumn{6}{l}{\textit{Factor 3 loadings}} \\
\addlinespace[1pt]
$a^{(3)}_{\mathrm{W1};\,(\mathrm{W3},\mathrm{Back})}$ & 1 & 0.00 & 4.70 & 5.08 & 0.92 \\
$a^{(3)}_{\mathrm{W2};\,(\mathrm{W1},\mathrm{Back})}$ & 1 & 0.54 & 3.38 & 3.71 & 0.93 \\
$a^{(3)}_{\mathrm{W3};\,(\mathrm{W2},\mathrm{Back})}$ & 1 & 0.82 & 3.60 & 3.58 & 0.95 \\
\midrule
\multicolumn{6}{l}{\textit{Covariance parameters}} \\
\addlinespace[1pt]
$\Sigma_{12}$ & -0.30 & -0.34 & 2.95 & 3.15 & 0.93 \\
$\Sigma_{13}$ & 0.30 & 0.33 & 4.05 & 4.13 & 0.96 \\
$\Sigma_{23}$ & -0.30 & -0.68 & 3.58 & 3.99 & 0.90 \\
\bottomrule
\end{tabular}
\end{adjustbox}
\end{minipage}
\caption{Simulation results for $n=2000$.}
\label{table:bias_se_cp_n2000}
\end{table}

\clearpage

\subsection{Additional simulation results}

\subsubsection{Omission of fixed- or random-effect components}
We consider the same simulation setting as in the main manuscript with $n = 500$, and we design an additional experiment to illustrate the importance of modelling both fixed and random effects. Specifically, we examine two misspecified scenarios: (i) the model is fitted without any random effects, and (ii) the model is fitted without any fixed effects. In each case, the model is estimated over a grid of penalty parameters for the remaining components, and the final model is selected by minimising the BIC. These two misspecified models are then compared with the correctly specified model that includes both fixed and random effects.

For scenario (i), across 100 independent replications, the fitted model always includes at least one nonzero estimate for the fixed effects where the true coefficient is zero, with an average of 4.16 such falsely identified parameters. This occurs because omitting random effects forces the fixed effects to capture heterogeneity that should instead be explained by the random-effect components. The average bias of the fixed-effect estimates is 0.0217. In contrast, under the correctly specified model, the average number of falsely identified parameters is 0.69 and the average bias is 0.00758.

For scenario (ii), across 100 independent replications, the fitted model yields an average of 55.3 falsely identified parameters. This substantial inflation is due to the complete absence of fixed effects, which causes the random effects to pick up a large amount of structure that should be explained by fixed-effect components. In contrast, under the correctly specified model, the average number of falsely identified parameters is 1.92. Because the misspecified model generates such a large number of false positives, we do not report bias comparisons for this case.

Although the true positive rates are close to $1$ in all scenarios, these results demonstrate that omitting either the fixed‐effect or random‐effect component leads to inflated false positive rates and biased estimates when the data exhibit heterogeneous structure induced by the random effects.

\subsubsection{Scenarios with right censoring}
We consider the same simulation setting as in the main manuscript with $n = 500$. Here, we introduce an additional censoring variable for each subject that is independent of the event process. The censoring time is generated from an exponential distribution with rate parameter $e^{-7}$ (in the same time units as the event times). On average, about $30\%$ of the subjects have at least one event observed before censoring.

The model is estimated over a grid of penalty parameters, and the final model is selected using the smallest BIC, as in the case without censoring. Across $100$ independent replications, we find that the true positive rate is $0.964$ and the false positive rate is $0.0343$. The average absolute bias of all parameters is $0.012$. For comparison, in the case without censoring, the true positive rate is $1$ and the false positive rate is $0.0013$, and the average absolute bias is $0.0037$. These results indicate that the presence of censoring introduces some loss of efficiency in variable selection and estimation accuracy, which is expected because parts of the event processes are unobserved for some of the subjects. Nevertheless, the true positive rate remains high and the increases in false positive rate and bias are relatively moderate, suggesting that the proposed method remains robust in the presence of moderate censoring.


\section{Sensitivity analysis using piecewise-constant baseline intensities}\label{sect:sensitivity}
\subsection{Simulation under piecewise-constant baseline intensities}

We conduct an additional simulation study to examine the performance of the
proposed estimation procedure under an alternative time-varying baseline
intensity specification. We consider the same simulation setting as in the main
manuscript with $n=500$, except that the constant baseline intensity for each
event type is replaced by an event-specific piecewise-constant baseline
intensity with three intervals. All fixed-effect coefficients, factor
loadings, the covariance matrix of the random effects, the at-risk mechanism,
and the history-dependent covariate processes remain unchanged. This
experiment therefore examines whether the proposed estimation framework can
be extended to accommodate baseline intensities that vary over time.

To construct the time-varying baseline intensities, we first generate a
pilot dataset of size $3000$ under the original simulation setting and use
the event-specific empirical tertiles of the event times as the two cut
points for each event type. These cut points are then fixed and used in all
$100$ Monte Carlo replications. Relative to the original constant baseline
intensity, we consider decreasing, increasing, and hump-shaped temporal
profiles, with multiplicative factors $(1.35,1,0.65)$,
$(0.65,1,1.35)$, and $(0.75,1.50,0.75)$, respectively, over the three
intervals. Thus, the data-generating model incorporates time variation in
the baseline intensities while the remaining components of the model are
kept unchanged.

The estimation procedure is modified to accommodate the interval-specific
baseline parameters. We use the same stochastic EM algorithm with
elliptical slice sampling for the latent random effects, while the
event-specific baseline parameters are updated in the M-step. In this
simulation, the fitted model uses the same fixed event-specific cut points
as the data-generating model. This allows us to focus on the performance of
the proposed procedure under a time-varying baseline specification without
introducing the separate issue of selecting the number or locations of the
cut points.

Variable selection is performed over the same type of two-dimensional
penalty grid as in the main simulation study. For each replication, the
selected structure is determined by the minimum BIC, where an unpenalised
refit of each selected structure is used only for calculating the BIC. When
multiple penalty pairs yield the BIC-selected structure, the original
penalised estimates from the pair with the smallest sum of the two penalty
parameters are used for reporting the estimation results.

Across the $100$ replications, the average true positive rate is $0.9986$
and the average false positive rate is $0.00194$, indicating that almost all
truly nonzero coefficients are selected while the proportion of truly zero
coefficients incorrectly selected remains very small. Tables~\ref{table:piecewise_bias_se_cp_baseline_n500} and
\ref{table:piecewise_bias_se_cp_other_n500} report the estimation results.
The interval-specific log-baseline parameters generally have small biases,
the average estimated standard errors are broadly comparable to the
empirical standard deviations, and the coverage probabilities are generally
close to the nominal level. The fixed effects, factor loadings, and
covariance parameters exhibit a pattern similar to that observed in the
original simulation study. Most parameters have relatively small biases and satisfactory coverage
probabilities. The relatively poorer finite-sample performance is concentrated
in the same fixed-effect coefficients associated with infrequently observed
transitions and in the Factor 1 loading
$a^{(1)}_{\mathrm{R}_3;\,\mathrm{R}_{\mathrm{Open}}}$, which also exhibits
greater variability and lower coverage in the original simulation study.
Thus, these features appear to reflect limited finite-sample information for
particular parameters rather than the piecewise-constant baseline
specification.

Overall, the proposed estimation procedure continues to perform
satisfactorily under the event-specific piecewise-constant baseline
specification, with strong variable-selection performance and generally
satisfactory estimation and inference. Together with the real-data
sensitivity analysis below, these results provide numerical evidence that
the performance of the proposed estimation and variable-selection procedure
is not specific to the constant-baseline specification.

\begin{table}[!htbp]
\centering
\scriptsize
\setlength{\tabcolsep}{4.0pt}
\renewcommand{\arraystretch}{0.88}
\begin{threeparttable}
\begin{tabular}{lrrrrr}
\toprule
Parameter & True & Bias & SE & SD & CP \\
\midrule
\multicolumn{6}{l}{\textit{Interval 1}} \\
\addlinespace[1pt]
$\beta_{0,\mathrm{W1},1}$ & -3.70 & 0.22 & 5.87 & 5.48 & 0.96 \\
$\beta_{0,\mathrm{W1}_{\mathrm M},1}$ & -4.70 & -0.36 & 9.19 & 8.94 & 0.96 \\
$\beta_{0,\mathrm{W2},1}$ & -4.70 & -0.72 & 9.62 & 8.21 & 0.98 \\
$\beta_{0,\mathrm{W2}_{\mathrm M},1}$ & -4.70 & -0.72 & 10.82 & 9.33 & 0.97 \\
$\beta_{0,\mathrm{W3},1}$ & -4.70 & 0.20 & 9.51 & 8.81 & 0.98 \\
$\beta_{0,\mathrm{W3}_{\mathrm M},1}$ & -3.70 & -0.04 & 7.71 & 7.12 & 0.99 \\
$\beta_{0,\mathrm{Next},1}$ & -7.43 & 1.24 & 10.94 & 11.25 & 0.95 \\
$\beta_{0,\mathrm{Next}_{\mathrm{Cancel}},1}$ & -4.29 & 0.64 & 25.11 & 22.63 & 0.97 \\
$\beta_{0,\mathrm{R}_1,1}$ & -5 & 1.18 & 20.17 & 19.89 & 0.96 \\
$\beta_{0,\mathrm{R}_2,1}$ & -5 & -2.85 & 19.30 & 16.88 & 0.98 \\
$\beta_{0,\mathrm{R}_3,1}$ & -5 & 4.97 & 20.72 & 21.28 & 0.91 \\
$\beta_{0,\mathrm{R}_{\mathrm{Open}},1}$ & -6.43 & -0.25 & 8.63 & 7.74 & 0.96 \\
$\beta_{0,\mathrm{R}_{\mathrm{Close}},1}$ & -5.29 & -1.19 & 23.94 & 21.13 & 0.99 \\
$\beta_{0,\mathrm{Back},1}$ & -6.70 & -9.01 & 24.19 & 24.47 & 0.97 \\
$\beta_{0,\mathrm{Next}_{\mathrm{OK}},1}$ & -2.43 & -1.08 & 10.66 & 10.04 & 0.98 \\
\midrule
\multicolumn{6}{l}{\textit{Interval 2}} \\
\addlinespace[1pt]
$\beta_{0,\mathrm{W1},2}$ & -4 & 0.14 & 8.45 & 7.09 & 0.97 \\
$\beta_{0,\mathrm{W1}_{\mathrm M},2}$ & -5 & -0.48 & 9.38 & 9.03 & 0.95 \\
$\beta_{0,\mathrm{W2},2}$ & -5 & -0.81 & 11.57 & 9.89 & 0.98 \\
$\beta_{0,\mathrm{W2}_{\mathrm M},2}$ & -5 & 0.04 & 10.75 & 9.28 & 0.98 \\
$\beta_{0,\mathrm{W3},2}$ & -5 & 2.55 & 11.78 & 9.61 & 0.99 \\
$\beta_{0,\mathrm{W3}_{\mathrm M},2}$ & -4 & -1.60 & 8.35 & 7.63 & 0.98 \\
$\beta_{0,\mathrm{Next},2}$ & -7 & 0.23 & 9.90 & 10.39 & 0.96 \\
$\beta_{0,\mathrm{Next}_{\mathrm{Cancel}},2}$ & -3.59 & -1.79 & 18.78 & 17.92 & 0.97 \\
$\beta_{0,\mathrm{R}_1,2}$ & -5 & 1.62 & 17.27 & 16.18 & 0.95 \\
$\beta_{0,\mathrm{R}_2,2}$ & -5 & 1.50 & 16.83 & 14.01 & 0.97 \\
$\beta_{0,\mathrm{R}_3,2}$ & -5 & 0.66 & 17.66 & 16.32 & 0.97 \\
$\beta_{0,\mathrm{R}_{\mathrm{Open}},2}$ & -6 & 0.49 & 7.30 & 6.52 & 0.96 \\
$\beta_{0,\mathrm{R}_{\mathrm{Close}},2}$ & -4.59 & 1.02 & 13.99 & 13.98 & 0.96 \\
$\beta_{0,\mathrm{Back},2}$ & -7 & -8.57 & 24.11 & 24.38 & 0.97 \\
$\beta_{0,\mathrm{Next}_{\mathrm{OK}},2}$ & -2 & -0.72 & 8.80 & 8.49 & 0.98 \\
\midrule
\multicolumn{6}{l}{\textit{Interval 3}} \\
\addlinespace[1pt]
$\beta_{0,\mathrm{W1},3}$ & -4.43 & 0.52 & 10.21 & 8.59 & 0.98 \\
$\beta_{0,\mathrm{W1}_{\mathrm M},3}$ & -5.43 & -1.17 & 13.18 & 10.84 & 1.00 \\
$\beta_{0,\mathrm{W2},3}$ & -5.43 & -0.18 & 13.69 & 11.36 & 0.99 \\
$\beta_{0,\mathrm{W2}_{\mathrm M},3}$ & -5.43 & -0.91 & 15.89 & 13.58 & 0.96 \\
$\beta_{0,\mathrm{W3},3}$ & -5.43 & 1.62 & 14.34 & 11.85 & 0.99 \\
$\beta_{0,\mathrm{W3}_{\mathrm M},3}$ & -4.43 & -1.98 & 12.49 & 10.41 & 0.97 \\
$\beta_{0,\mathrm{Next},3}$ & -6.70 & -0.15 & 10.76 & 9.40 & 0.96 \\
$\beta_{0,\mathrm{Next}_{\mathrm{Cancel}},3}$ & -4.29 & -3.45 & 31.94 & 28.62 & 0.96 \\
$\beta_{0,\mathrm{R}_1,3}$ & -5 & -0.35 & 17.64 & 15.28 & 0.98 \\
$\beta_{0,\mathrm{R}_2,3}$ & -5 & -0.35 & 19.34 & 15.39 & 0.98 \\
$\beta_{0,\mathrm{R}_3,3}$ & -5 & 3.72 & 18.58 & 16.11 & 0.97 \\
$\beta_{0,\mathrm{R}_{\mathrm{Open}},3}$ & -5.70 & 0.87 & 8.47 & 6.94 & 0.97 \\
$\beta_{0,\mathrm{R}_{\mathrm{Close}},3}$ & -5.29 & -5.40 & 23.03 & 21.34 & 0.97 \\
$\beta_{0,\mathrm{Back},3}$ & -7.43 & -8.36 & 24.35 & 24.25 & 0.97 \\
$\beta_{0,\mathrm{Next}_{\mathrm{OK}},3}$ & -1.70 & -1.18 & 9.30 & 10.50 & 0.99 \\
\bottomrule
\end{tabular}
\begin{tablenotes}[flushleft]
\footnotesize
\item We write $\beta_{0j,h}$ for the interval-specific log-baseline parameter for event type $j$ in interval $h$. Bias is $100$ times the difference between the mean estimate and the true value; SE is $100$ times the average estimated standard error; SD is $100$ times the sample standard deviation; and CP is the empirical coverage probability of the normal-approximation $95\%$ confidence interval.
\end{tablenotes}
\end{threeparttable}
\caption{Simulation results for the interval-specific log-baseline parameters under the piecewise-constant log-baseline specification with $n=500$.}
\label{table:piecewise_bias_se_cp_baseline_n500}
\end{table}

\begin{table}[!htbp]
\centering
\begin{minipage}[t]{0.59\textwidth}
\vspace{0pt}
\centering
{\small\textbf{Fixed effects}}

\vspace{2pt}
\footnotesize
\setlength{\tabcolsep}{1.1pt}
\renewcommand{\arraystretch}{1.2}
\begin{adjustbox}{max width=\linewidth}
\begin{tabular}{@{}lrr@{\hspace{4.5pt}}r@{\hspace{4.5pt}}r@{\hspace{4.5pt}}r@{}}
\toprule
Parameter & True & Bias & SE & SD & CP \\
\midrule
$\beta_{\mathrm{W1};\,\mathrm{Back}}$ & 1 & -0.20 & 8.19 & 7.72 & 0.95 \\
$\beta_{\mathrm{W1};\,(\mathrm{W1},\mathrm{Back})}$ & -2 & -3.30 & 15.73 & 15.78 & 0.97 \\
$\beta_{\mathrm{W1};\,(\mathrm{W2},\mathrm{Back})}$ & -1 & 0.90 & 10.79 & 11.10 & 0.96 \\
$\beta_{\mathrm{W2};\,\mathrm{Back}}$ & 1 & -2.23 & 11.88 & 10.21 & 0.99 \\
$\beta_{\mathrm{W2};\,(\mathrm{W1},\mathrm{Back})}$ & 2 & 4.54 & 9.58 & 10.86 & 0.89 \\
$\beta_{\mathrm{W2};\,(\mathrm{W2},\mathrm{Back})}$ & -2 & 31.34 & 26.57 & 25.52 & 0.75 \\
$\beta_{\mathrm{W3};\,\mathrm{Back}}$ & 1 & -8.78 & 13.26 & 12.90 & 0.89 \\
$\beta_{\mathrm{W3};\,(\mathrm{W1},\mathrm{Back})}$ & -2 & 31.63 & 25.46 & 32.38 & 0.75 \\
$\beta_{\mathrm{W3};\,(\mathrm{W2},\mathrm{Back})}$ & 2 & 10.52 & 11.25 & 13.71 & 0.82 \\
$\beta_{\mathrm{W3};\,(\mathrm{W3},\mathrm{Back})}$ & -2 & 36.02 & 27.03 & 30.98 & 0.71 \\
$\beta_{\mathrm{Next};\,\mathrm{Next}_{\mathrm{Cancel}}}$ & 4 & 1.23 & 17.87 & 16.22 & 0.99 \\
$\beta_{\mathrm{Next};\,\mathrm{R}_1}$ & 5 & -0.06 & 13.94 & 13.40 & 0.97 \\
$\beta_{\mathrm{Next};\,\mathrm{R}_2}$ & 5 & 0.47 & 13.89 & 14.01 & 0.94 \\
$\beta_{\mathrm{Next};\,\mathrm{R}_3}$ & 5 & -0.01 & 13.28 & 12.36 & 0.98 \\
$\beta_{\mathrm{R}_{\mathrm{Open}};\,\mathrm{W3}_{\mathrm M}}$ & 2 & 0.59 & 13.67 & 12.71 & 0.97 \\
$\beta_{\mathrm{R}_{\mathrm{Open}};\,(\mathrm{W3},\mathrm{Back})}$ & 2 & 0.29 & 11.25 & 11.07 & 0.96 \\
$\beta_{\mathrm{Back};\,\mathrm{W1}}$ & 3 & 8.33 & 25.10 & 24.88 & 0.97 \\
$\beta_{\mathrm{Back};\,\mathrm{W1}_{\mathrm M}}$ & 5 & 9.66 & 25.43 & 25.49 & 0.98 \\
$\beta_{\mathrm{Back};\,\mathrm{W2}}$ & 3 & 9.24 & 25.22 & 25.24 & 0.97 \\
$\beta_{\mathrm{Back};\,\mathrm{W2}_{\mathrm M}}$ & 5 & 9.32 & 25.51 & 25.47 & 0.96 \\
$\beta_{\mathrm{Back};\,\mathrm{W3}}$ & 3 & 9.02 & 25.58 & 25.97 & 0.97 \\
$\beta_{\mathrm{Back};\,\mathrm{W3}_{\mathrm M}}$ & 5 & 9.04 & 25.11 & 25.52 & 0.97 \\
$\beta_{\mathrm{Back};\,\mathrm{Back}}$ & 3 & 8.77 & 24.33 & 24.38 & 0.96 \\
\bottomrule
\end{tabular}
\end{adjustbox}
\end{minipage}
\hfill
\begin{minipage}[t]{0.40\textwidth}
\vspace{0pt}
\centering
\begin{adjustbox}{max width=\linewidth}
{\small\textbf{Factor loadings and covariance parameters}}
\end{adjustbox}

\vspace{2pt}
\footnotesize
\setlength{\tabcolsep}{1.1pt}
\renewcommand{\arraystretch}{1.2}
\begin{adjustbox}{max width=\linewidth}
\begin{tabular}{@{}lrr@{\hspace{4.5pt}}r@{\hspace{4.5pt}}r@{\hspace{4.5pt}}r@{}}
\toprule
Parameter & True & Bias & SE & SD & CP \\
\midrule
\multicolumn{6}{l}{\textit{Factor 1 loadings}} \\
\addlinespace[1pt]
$a^{(1)}_{\mathrm{W1}_{\mathrm M};\,\mathrm{W1}}$ & 2 & 1.32 & 12.40 & 11.95 & 0.96 \\
$a^{(1)}_{\mathrm{W2}_{\mathrm M};\,\mathrm{W2}}$ & 2 & -0.04 & 13.94 & 11.78 & 0.97 \\
$a^{(1)}_{\mathrm{W3}_{\mathrm M};\,\mathrm{W3}}$ & 2 & 1.19 & 12.80 & 12.74 & 0.96 \\
$a^{(1)}_{\mathrm{R}_3;\,\mathrm{R}_{\mathrm{Open}}}$ & 1 & -9.45 & 13.34 & 22.64 & 0.75 \\
\midrule
\multicolumn{6}{l}{\textit{Factor 2 loadings}} \\
\addlinespace[1pt]
$a^{(2)}_{\mathrm{Back};\,\mathrm{W1}}$ & 1 & 1.35 & 6.68 & 5.71 & 0.96 \\
$a^{(2)}_{\mathrm{Back};\,\mathrm{W1}_{\mathrm M}}$ & 1 & 0.68 & 8.13 & 6.89 & 0.99 \\
$a^{(2)}_{\mathrm{Back};\,\mathrm{W2}}$ & 1 & -0.55 & 6.86 & 5.55 & 0.98 \\
$a^{(2)}_{\mathrm{Back};\,\mathrm{W2}_{\mathrm M}}$ & 1 & -0.64 & 9.37 & 8.05 & 0.98 \\
$a^{(2)}_{\mathrm{Back};\,\mathrm{W3}}$ & 1 & 0.47 & 7.83 & 6.06 & 0.98 \\
$a^{(2)}_{\mathrm{Back};\,\mathrm{W3}_{\mathrm M}}$ & 1 & -0.45 & 7.75 & 6.41 & 0.97 \\
\midrule
\multicolumn{6}{l}{\textit{Factor 3 loadings}} \\
\addlinespace[1pt]
$a^{(3)}_{\mathrm{W1};\,(\mathrm{W3},\mathrm{Back})}$ & 1 & 0.35 & 10.68 & 12.31 & 0.97 \\
$a^{(3)}_{\mathrm{W2};\,(\mathrm{W1},\mathrm{Back})}$ & 1 & 0.64 & 7.42 & 6.51 & 0.98 \\
$a^{(3)}_{\mathrm{W3};\,(\mathrm{W2},\mathrm{Back})}$ & 1 & -2.99 & 8.32 & 13.19 & 0.91 \\
\midrule
\multicolumn{6}{l}{\textit{Covariance parameters}} \\
\addlinespace[1pt]
$\Sigma_{12}$ & -0.30 & -0.17 & 6.30 & 5.86 & 0.97 \\
$\Sigma_{13}$ & 0.30 & -0.29 & 8.45 & 9.50 & 0.91 \\
$\Sigma_{23}$ & -0.30 & -1.39 & 7.62 & 9.56 & 0.95 \\
\bottomrule
\end{tabular}
\end{adjustbox}
\end{minipage}
\begin{minipage}{\textwidth}
\vspace{2pt}
\footnotesize
\end{minipage}
\caption{Simulation results for the fixed effects, factor loadings, and covariance parameters under the piecewise-constant log-baseline specification with $n=500$.}
\label{table:piecewise_bias_se_cp_other_n500}
\end{table}


\clearpage

\subsection{Real-data sensitivity analysis}

We further examine the sensitivity of the real-data results to the baseline
intensity specification by considering an event-specific piecewise-constant
log-baseline model. Specifically, we assess whether the main conclusions
concerning the internal covariate effects and latent-factor structure are
sensitive to the constant-baseline specification used in the main analysis.

We retain the three-factor model $\mathcal{M}_{1,2,3}$, the same internal
covariate processes, and the same identifiability constraints on the factor
loading matrices. The only modification is to replace the constant
log-baseline parameter for each event type by three interval-specific
parameters. Specifically, for event type $j$, let $\kappa_{j1}$ and
$\kappa_{j2}$ denote the empirical $1/3$ and $2/3$ quantiles of its observed
event times, and specify
\[
\log \lambda_{j0}(t)
=
\begin{cases}
\beta_{0j,1}, & 0\leq t\leq \kappa_{j1},\\
\beta_{0j,2}, & \kappa_{j1}<t\leq \kappa_{j2},\\
\beta_{0j,3}, & t>\kappa_{j2}.
\end{cases}
\]
The cut points are therefore determined separately for each event type and
are held fixed throughout the analysis. All interval-specific baseline
parameters are unpenalised. This three-interval specification is used as a
representative time-varying parametric baseline for the sensitivity analysis;
we do not perform model selection over the number or locations of the cut
points.

Estimation follows the same stochastic EM procedure as in the main analysis,
with elliptical slice sampling used for the latent factors and the
interval-specific baseline parameters updated in the M-step. The
regularisation parameters for the fixed effects and factor loadings are
selected by minimising the BIC over the same $20\times30$ penalty grid used
in the main analysis.  As in the main analysis, an
unpenalised refit of the selected structure is used only for calculating the
BIC, while the reported estimates and standard errors are based on the
penalised fit.

Tables~\ref{table:coef_matrix_piecewise} and
\ref{table:factor_loadings_piecewise} report the same subsets of fixed-effect
coefficients and factor loadings as Tables~5 and
7 of the main paper, so that the substantive
results can be compared directly. The interval-specific baseline estimates
for all event types are additionally reported in
Table~\ref{table:all_piecewise_baseline_estimates} for completeness.

The principal fixed-effect patterns identified in the main analysis are
preserved under this alternative baseline specification. In particular, the
webpage-related covariates continue to have large positive effects on the
intensity of Back. The estimated coefficients for W1,
W1$_{\mathrm M}$, W2, W2$_{\mathrm A}$, W3, and
W3$_{\mathrm A}$ on $\lambda_{\mathrm{Back}}(\cdot)$ range from
$5.42$ to $6.56$. Likewise, the coefficients for
R$_1,\ldots,$R$_5$ on the intensity of Next remain large and positive,
ranging from $4.86$ to $5.26$. These findings agree with the behavioural
relationships identified under the constant-baseline specification.

The sequential browsing pattern discussed in the main analysis is also
retained. After returning from W1, the coefficient for
W1, Back $\rightarrow\lambda_{\mathrm{W2}}$ is $0.95$, whereas the
corresponding coefficients for moving to W1, W3, W4, and W5 are negative.
Similarly, the coefficients for
W2, Back $\rightarrow\lambda_{\mathrm{W3}}$,
W3, Back $\rightarrow\lambda_{\mathrm{W4}}$, and
W4, Back $\rightarrow\lambda_{\mathrm{W5}}$ are $1.59$, $2.24$, and
$3.33$, respectively. Thus, the conclusion that the recent action history is
informative about subsequent behaviour does not depend on imposing a
constant baseline intensity.

The main features of the latent-factor interpretation are also retained.
For Factor 1, the loadings for
W1 $\rightarrow\lambda_{\mathrm{W1}_{\mathrm M}}$,
W2 $\rightarrow\lambda_{\mathrm{W2}_{\mathrm A}}$,
W3 $\rightarrow\lambda_{\mathrm{W3}_{\mathrm A}}$, and
W3 $\rightarrow\lambda_{\mathrm{W3}_{\mathrm{O}1}}$ are
$-1.68$, $-2.26$, $-2.82$, and $-1.26$, respectively, preserving the
common direction associated with information-seeking behaviour. For Factor
2, the loadings for W$j\rightarrow\lambda_{\mathrm{Back}}$,
$j=1,\ldots,5$, are $0.57$, $0.62$, $0.73$, $0.59$, and $0.59$,
respectively. The sequential browsing structure also remains evident in
Factor 3: the loadings for
W1, Back $\rightarrow\lambda_{\mathrm{W2}}$,
W2, Back $\rightarrow\lambda_{\mathrm{W3}}$,
W3, Back $\rightarrow\lambda_{\mathrm{W4}}$, and
W4, Back $\rightarrow\lambda_{\mathrm{W5}}$ are all positive.

There are differences in some individual selected coefficients and factor
loadings, as expected when the baseline specification and the resulting
penalised model selection are changed. Nevertheless, the principal
history-dependent relationships and latent behavioural patterns on which
the interpretation of the real-data analysis is based remain qualitatively
similar. This sensitivity analysis therefore supports the stability of the
main substantive conclusions under an alternative time-varying parametric
baseline specification.

\begin{table}[htbp]
\centering
\resizebox{\textwidth}{!}{
\begin{threeparttable}
\begin{tabular}{lrrrrrrrr}
\toprule
 & $\lambda_{\text{W1}}$ & $\lambda_{\text{W2}}$ & $\lambda_{\text{W3}}$ & $\lambda_{\text{W4}}$ & $\lambda_{\text{W5}}$ & $\lambda_{\text{Back}}$ & $\lambda_{\text{Next}}$ & $\lambda_{\text{Web}}$ \\
\midrule
$\beta_{0,1}$ & -4.04 (0.04) & -5.76 (0.08) & -6.87 (0.12) & -6.66 (0.11) & -7.40 (0.15) & -8.91 (0.81) & -6.46 (0.08) & -8.16 (0.18) \\
$\beta_{0,2}$ & -3.22 (0.04) & -5.68 (0.09) & -6.85 (0.13) & -6.27 (0.12) & -7.36 (0.17) & -8.57 (0.82) & -6.46 (0.10) & -8.19 (0.18) \\
$\beta_{0,3}$ & -4.23 (0.06) & -5.65 (0.09) & -7.39 (0.13) & -6.12 (0.12) & -7.70 (0.18) & -8.35 (0.82) & -6.34 (0.11) & -8.60 (0.18) \\
\midrule
W1 & 0 & 0 & 0 & 0 & 0 & 5.42 (0.82) & -0.39 (0.16) & 1.95 (0.21) \\
W1$_\text{M}$ & 0 & 0 & 0 & 0 & 0 & 6.50 (0.84) & 0.58 (0.36) & 4.30 (0.25) \\
W2 & 0 & 0 & 0 & 0 & 0 & 5.85 (0.82) & 0.24 (0.14) & 1.28 (0.27) \\
W2$_\text{A}$ & 0 & 0 & 0 & 0 & 0 & 6.26 (0.88) & 0 & 1.03 (0.63) \\
W3 & 0 & 0 & 0 & 0 & 0 & 6.01 (0.82) & -0.31 (0.21) & 0 \\
W3$_\text{A}$ & 0 & 0 & 0 & 0 & 0 & 6.56 (1.15) & 0 & 3.18 (0.63) \\
W3$_{\text{O}1}$ & 0 & 0 & 0 & 0 & 0 & 5.44 (0.92) & 0 & 0 \\
W3$_{\text{O}2}$ & 0 & 0 & 0 & 0 & 0 & 5.51 (0.99) & 0 & 0 \\
W4 & 0 & 0 & 0 & 0 & 0 & 5.78 (0.82) & -0.23 (0.18) & 0.92 (0.30) \\
W5 & 0 & 0 & 0 & 0 & 0 & 5.79 (0.82) & -0.76 (0.25) & 0.66 (0.38) \\
W5$_\text{O}$ & 0 & 0 & 0 & 0 & 0 & 5.59 (0.89) & 0 & 0 \\
Next & -3.87 (1.05) & -1.93 (0.74) & 0 & -1.41 (1.05) & 0 & 0 & 0 & 0 \\
Next$_\text{Cancel}$ & 0 & 0 & 0 & 0 & 0 & 6.08 (0.85) & 2.67 (0.28) & 3.72 (0.50) \\
R$_1$ & -0.44 (0.65) & 0 & 0 & 0 & 0 & 3.98 (0.91) & 4.86 (0.17) & 0 \\
R$_2$ & -3.31 (1.03) & 0.70 (0.23) & 1.50 (0.33) & 0 & 0 & 3.80 (0.84) & 5.26 (0.10) & 1.06 (0.57) \\
R$_3$ & -0.32 (0.61) & 0 & 0 & 0 & 0 & 3.54 (1.03) & 5.22 (0.16) & 0 \\
R$_4$ & -2.67 (22.39) & 0 & 0 & 1.26 (0.34) & 0 & 3.52 (0.86) & 5.21 (0.13) & 1.55 (0.57) \\
R$_5$ & -0.26 (1.27) & 0 & 0 & 0 & 0 & 3.28 (1.00) & 4.96 (0.18) & 0 \\
R$_\text{Open}$ & -1.61 (0.12) & -0.19 (0.14) & 0 & 0 & -1.02 (0.53) & 3.11 (0.83) & -1.23 (0.26) & 0 \\
R$_\text{Close}$ & 1.31 (0.10) & 1.44 (0.16) & 1.44 (0.27) & 1.34 (0.23) & 1.41 (0.37) & 5.20 (0.83) & 1.77 (0.18) & 2.12 (0.37) \\
Back & -0.62 (0.17) & 3.21 (0.10) & 3.64 (0.16) & 2.77 (0.15) & 2.64 (0.20) & 5.64 (0.85) & 0.28 (0.17) & 0 \\
Forward & 0 & 0 & 0 & 0 & 0 & 6.67 (0.83) & 1.76 (0.34) & 3.39 (0.45) \\
Home & 0 & 2.85 (0.53) & 3.88 (2.53) & 3.10 (0.75) & 4.04 (3.10) & 0 & 0 & 0 \\
Web & 0 & 0 & 0 & 0 & 0 & 6.70 (0.82) & 1.83 (0.27) & 5.80 (0.18) \\
W1, Back & -1.03 (0.26) & 0.95 (0.08) & -1.53 (0.26) & -3.17 (0.59) & -1.40 (0.29) & -2.97 (0.25) & -1.11 (0.54) & 0 \\
W2, Back & -0.95 (0.33) & -2.06 (0.13) & 1.59 (0.11) & -0.32 (0.16) & -0.83 (0.18) & -3.51 (0.26) & 0 & 0 \\
W3, Back & -0.65 (0.25) & -1.95 (0.21) & -1.76 (0.20) & 2.24 (0.13) & 0 & -3.73 (0.34) & -1.02 (0.54) & 0 \\
W4, Back & -0.42 (0.23) & -1.28 (0.14) & -2.14 (0.32) & -1.54 (0.22) & 3.33 (0.13) & -4.06 (0.34) & 0 & 0 \\
W5, Back & 0.54 (0.19) & -0.26 (0.09) & -1.32 (0.22) & -0.91 (0.20) & -2.10 (0.34) & -3.57 (0.26) & 0 & 0 \\
\bottomrule
\end{tabular}
\caption{Partial results of the estimated fixed-effect regression coefficients under the piecewise-constant log-baseline specification (standard errors in parentheses). The columns correspond to the intensity functions for different event types. The first three rows report the interval-specific log-baseline estimates, with event-specific cut points given by the empirical tertiles of the observed event times. The remaining rows correspond to the covariate processes.}
\label{table:coef_matrix_piecewise}
\end{threeparttable}
}
\end{table}

\spacingset{1.2}
\setlength{\tabcolsep}{14pt}
\begin{table}[H]
\centering
\begin{tabular}{lllll}
\toprule
Covariate & Event & Factor 1 & Factor 2 & Factor 3 \\
\midrule
Next & Next$_\text{Cancel}$ & 0.05 (0.23) & -0.53 (0.15) & 0 \\
R$_\text{Open}$ & R$_1$ & 0.47 (0.18) & -0.05 (0.13) & 0.67 (0.15) \\
R$_\text{Open}$ & R$_2$ & -0.77 (0.06) & 0.43 (0.06) & 0.15 (0.05) \\
R$_\text{Open}$ & R$_3$ & 0 & 0 & 0 \\
R$_\text{Open}$ & R$_4$ & 0.54 (0.11) & 0.85 (0.08) & 0 \\
R$_\text{Open}$ & R$_5$ & 0.04 (0.24) & 0 & 0 \\
W1 & Back & 0.03 (0.07) & 0.57 (0.05) & -0.12 (0.06) \\
W1 & W1$_\text{M}$ & -1.68 (0.14) & 0.10 (0.12) & 0 \\
W1, Back & W2 & 0.29 (0.08) & 0.07 (0.07) & 0.30 (0.08) \\
W1, Back & W3 & 0.32 (0.18) & 0.07 (0.14) & -1.09 (0.22) \\
W1, Back & W4 & 0 & 0 & -1.78 (0.39) \\
W2 & Back & 0 & 0.62 (0.05) & 0 \\
W2 & W2$_\text{A}$ & -2.26 (0.21) & 0 & 0 \\
W2, Back & W1 & 0 & 0 & -1.08 (0.23) \\
W2, Back & W2 & 0 & -0.03 (0.14) & -0.07 (0.16) \\
W2, Back & W3 & 0.81 (0.11) & 0 & 0.35 (0.06) \\
W2, Back & W4 & 1.09 (0.17) & 0.45 (0.12) & -0.11 (0.17) \\
W3 & Back & -0.05 (0.08) & 0.73 (0.05) & 0 \\
W3 & W3$_\text{A}$ & -2.82 (0.43) & -0.16 (0.25) & 0 \\
W3 & W3$_\text{O1}$ & -1.26 (0.31) & 0 & 0 \\
W3, Back & W2 & -0.69 (0.21) & 0 & -1.33 (0.18) \\
W3, Back & W4 & 0.79 (0.14) & 0.26 (0.10) & 0.51 (0.12) \\
W3$_\text{O1}$ & W3$_\text{O2}$ & -0.14 (0.31) & 0 & 0 \\
W4 & Back & -0.15 (0.07) & 0.59 (0.04) & 0 \\
W4, Back & W2 & -0.08 (0.13) & 0 & -0.47 (0.14) \\
W4, Back & W3 & 0 & 0 & -1.50 (0.23) \\
W4, Back & W5 & 0.19 (0.14) & -0.21 (0.11) & 1.01 (0.14) \\
W5 & Back & -0.16 (0.06) & 0.59 (0.05) & 0 \\
W5, Back & W1 & 0 & 0 & 0.11 (0.19) \\
W5, Back & W2 & -0.26 (0.09) & -0.12 (0.08) & 0 \\
W5, Back & W3 & 0 & 0 & -1.08 (0.22) \\
W5, Back & W4 & 0.65 (0.16) & 0.45 (0.14) & -0.92 (0.18) \\
\bottomrule
\end{tabular}
\caption{Partial results of the estimated factor loadings under the piecewise-constant log-baseline specification (standard errors in parentheses). The same transitions as in Table~\ref{table:factor_loadings} are displayed to facilitate comparison with the constant-baseline analysis. }
\label{table:factor_loadings_piecewise}
\end{table}
\begin{table}[htbp]
\centering
\small
\setlength{\tabcolsep}{12pt}
\begin{threeparttable}
\begin{tabular}{lccc}
\toprule
Event type & $\widehat{\beta}_{0j,1}$ & $\widehat{\beta}_{0j,2}$ & $\widehat{\beta}_{0j,3}$ \\
\midrule
W1 & -4.04 (0.04) & -3.22 (0.04) & -4.23 (0.06) \\
W1$_{\mathrm{M}}$ & -6.85 (0.27) & -5.98 (0.27) & -5.16 (0.25) \\
W2 & -5.76 (0.08) & -5.68 (0.09) & -5.65 (0.09) \\
W2$_{\mathrm{A}}$ & -6.49 (0.35) & -5.96 (0.34) & -5.45 (0.33) \\
W3 & -6.87 (0.12) & -6.85 (0.13) & -7.39 (0.13) \\
W3$_{\mathrm{A}}$ & -6.53 (0.58) & -5.79 (0.55) & -5.35 (0.62) \\
W3$_{\mathrm{O}1}$ & -5.88 (0.47) & -5.09 (0.47) & -5.44 (0.50) \\
W3$_{\mathrm{O}2}$ & -8.51 (0.57) & -7.82 (0.52) & -7.86 (0.51) \\
W4 & -6.66 (0.11) & -6.27 (0.12) & -6.12 (0.12) \\
W5 & -7.40 (0.15) & -7.36 (0.17) & -7.70 (0.18) \\
Next & -6.46 (0.08) & -6.46 (0.10) & -6.34 (0.11) \\
Next$_{\mathrm{Cancel}}$ & -4.21 (0.21) & -4.46 (0.23) & -5.36 (0.25) \\
R$_1$ & -5.06 (0.16) & -5.59 (0.18) & -5.71 (0.19) \\
R$_2$ & -3.49 (0.06) & -2.84 (0.06) & -2.62 (0.07) \\
R$_3$ & -5.47 (0.13) & -5.06 (0.14) & -5.47 (0.15) \\
R$_4$ & -4.71 (0.14) & -4.03 (0.12) & -3.53 (0.12) \\
R$_5$ & -5.87 (0.19) & -5.51 (0.22) & -6.11 (0.21) \\
R$_{\mathrm{Open}}$ & -4.59 (0.04) & -4.20 (0.05) & -3.59 (0.04) \\
R$_{\mathrm{Close}}$ & -2.87 (0.06) & -3.91 (0.06) & -4.27 (0.06) \\
Back & -8.91 (0.81) & -8.57 (0.82) & -8.35 (0.82) \\
Forward & -8.85 (0.26) & -8.28 (0.25) & -8.50 (0.23) \\
Home & -9.23 (0.40) & -8.84 (0.87) & -9.37 (0.46) \\
W5$_{\mathrm{O}}$ & -7.72 (0.45) & -7.53 (0.52) & -7.70 (0.49) \\
Web & -8.16 (0.18) & -8.19 (0.18) & -8.60 (0.18) \\
Next$_{\mathrm{OK}}$ & -1.17 (0.57) & -1.04 (0.58) & -0.94 (0.58) \\
\bottomrule
\end{tabular}
\begin{tablenotes}[flushleft]
\footnotesize
\item Entries are the estimated interval-specific log-baseline intensities, with standard errors in parentheses. For each event type, the three intervals are defined using its own empirical tertiles of the observed event times.
\end{tablenotes}
\caption{Estimated piecewise-constant log-baseline parameters for all event types.}
\label{table:all_piecewise_baseline_estimates}
\end{threeparttable}
\end{table}
\newpage
\section{Further discussion on model identifiability}\label{sec:identifiability}
In this section, we provide additional discussion of the challenges of establishing identifiability for the proposed model when internal covariates are present. In Section~\ref{S211}, we illustrate the necessity of adopting the concept of generic identifiability \citep{10.1214/09-AOS689} by presenting a simple counterexample in which the model is generically identifiable but not identifiable in the conventional sense. The next two sections further clarify how identifiability with internal covariates differs from the settings studied in the existing survival analysis literature, where only external covariates are considered. Specifically, Section~\ref{S21} establishes identifiability results when our model involves only external covariates, while Section~\ref{S22} explains why these arguments fail once internal covariates are introduced. We then construct a counterexample in Section~\ref{S22} showing that the model is not generically identifiable in the presence of internal covariates without the piecewise-constant assumption, highlighting the necessity of the first part of Condition (e).

\subsection{The necessity of adopting generic identifiability}\label{S211}
To justify the use of the concept of generic identifiability, we construct a simple counterexample in which the model satisfies all Conditions (c)-(f) but is not identifiable in the conventional sense. Consider the setting with only one event type. Let $[0,\tau]$ be the observation period and $t_1$ be the time point of the first event or $\tau$ if no event occurs. Suppose that the covariate processes $\boldsymbol{X}(t),\boldsymbol{Z}(t)\in\mathbb{R}^3$ are defined as  $\boldsymbol{X}(t)=\boldsymbol{g}(t)I(t\leq t_1)$ and $\boldsymbol{Z}(t)=\boldsymbol{g}(t)I(t\leq t_1)$, where $\boldsymbol{g}(t): \mathbb{R}_+\rightarrow \mathbb{R}^3$ is a deterministic function, defined by
    \begin{align*}
        \boldsymbol{g}(t)=\begin{cases}(1,0,0)^{\top}&t\leq \tau/3,\\
        (0,1,0)^{\top}&\tau/3<t\leq 2\tau/3,\\
        (0,1,1)^{\top}&t>2\tau/3.
        \end{cases}
    \end{align*}
    Note that $\boldsymbol{Z}(\cdot)$ is an internal covariate in the sense that the observed sample path carries information about the event time.  
    
    We consider the model in which the intensity function takes the form:
    \begin{align}\label{generic}
        \lambda(t|\mathscr{F}_{t^{-}};\boldsymbol{\theta})=Y(t)\exp(\beta_{0}+\boldsymbol{\beta}^{\top}\boldsymbol{X}(t)+\boldsymbol{\theta}^{\top}\mathbf{A}^{\top}\boldsymbol{Z}(t)),
    \end{align}
    where $\boldsymbol{\theta}\sim \mathcal{N}_3(\mathbf{0},\mathbf{\Sigma})$. Suppose $Y(t)\equiv 1$ for all $t\in[0,\tau]$. It is easy to verify that Conditions (d)-(f) are met. Let $\mathbf{\Sigma},\widetilde{\mathbf{\Sigma}}\in\mathbb{R}^{3\times 3}$ be defined as 
    \begin{align*}
        \mathbf{\Sigma}=\left(\begin{array}{ccc}
            1 & 0 & 0 \\
            0 & 1 & 0 \\
            0 & 0 & 1
            \end{array}\right),~\widetilde{\mathbf{\Sigma}}=\left(\begin{array}{ccc}
            1 & 0 & 1 \\
            0 & 1 & -1 \\
            1 & -1 & 3
        \end{array}\right).
    \end{align*}
    We make the following claim.
    \begin{thm}
        Model (\ref{generic}) with parameter tuple $(\boldsymbol{\beta},\mathbf{A},\mathbf{\Sigma})$ induces the same probability measure as model (\ref{generic})  with parameter tuple $(\widetilde{\boldsymbol{\beta}},\widetilde{\mathbf{A}},\widetilde{\mathbf{\Sigma}})$, where $\beta_0=\widetilde{\beta}_0=0$, $\boldsymbol{\beta}=\widetilde{\boldsymbol{\beta}}=\mathbf{0}$ and $\mathbf{A}=\widetilde{\mathbf{A}}=\mathbf{I}_3$. Both models satisfy Conditions (c)-(f).
    \end{thm}
    \begin{proof}
        We prove that for any given outcome, the density functions under the two models are identical. Let $\mathbf{Q},\widetilde{\mathbf{Q}}\in\mathbb{R}^{3\times 3}$ be defined as
        \begin{align*}
        \mathbf{Q}=\left(\begin{array}{ccc}
        1~ & 0~ & 0 \\
        0~ & 1~ & 0 \\
        0~ & 0~ & 1
        \end{array}\right),~\widetilde{\mathbf{Q}}=\left(\begin{array}{ccc}
        1~ & 0~ & 0 \\
        0~ & 1~ & 0 \\
        1~ & -1 & 1
        \end{array}\right).
    \end{align*}
    We can verify that $\mathbf{\Sigma}=\mathbf{Q}\mathbf{Q}^{\top}$ and $\widetilde{\mathbf{\Sigma}}=\widetilde{\mathbf{Q}}\widetilde{\mathbf{Q}}^{\top}$. Since $\beta_0=\widetilde{\beta}_0=0$, $\boldsymbol{\beta}=\widetilde{\boldsymbol{\beta}}=\mathbf{0}$ and $\mathbf{A}=\widetilde{\mathbf{A}}=\mathbf{I}_3$, using a linear transformation of the multivariate normal vector, we can write the two models as follows.
    \begin{align*}
        \text{Model A:~~~}\lambda(t|\boldsymbol{\theta})=&Y(t)\exp(\boldsymbol{\theta}^{\top}\mathbf{Q}^{\top}\boldsymbol{Z}(t)),\\
        \text{Model B:~~~}\widetilde{\lambda}(t|\boldsymbol{\theta})=&Y(t)\exp(\boldsymbol{\theta}^{\top}\widetilde{\mathbf{Q}}^{\top}\boldsymbol{Z}(t)),
    \end{align*}
    where $\boldsymbol{\theta}\sim \mathcal{N}(\boldsymbol{0},\mathbf{I}_3)$. We can verify that
    \begin{align*}
        \boldsymbol{g}_1(t):=~& \mathbf{Q}^{\top}\boldsymbol{g}(t)=\begin{cases}(1,0,0)^{\top}&t\leq \tau/3,\\
        (0,1,0)^{\top}&\tau/3<t\leq 2\tau/3,\\
        (0,1,1)^{\top}&t>2\tau/3.
        \end{cases}\\
        \boldsymbol{g}_2(t):=~&\widetilde{\mathbf{Q}}^{\top}\boldsymbol{g}(t)=\begin{cases}(1,0,0)^{\top}&t\leq \tau/3,\\
        (0,1,0)^{\top}&\tau/3<t\leq 2\tau/3,\\
        (1,0,1)^{\top}&t>2\tau/3.
        \end{cases}
    \end{align*}
    We then prove that Model A and Model B induce the same density functions for any given outcome in the following two cases:
    \begin{enumerate}[(i)]
        \item Case 1: The outcome contains no event.\\[3mm] In this case, the density function of model A given the observation on $[0,\tau]$ is calculated by:
        \begin{align*}
            &\mathbb{E}_{\boldsymbol{\theta}}\left[\exp\left(-\int_{0}^{\tau}\lambda(t|\boldsymbol{\theta})\mathrm{d}t\right)\right]\\
            =&\mathbb{E}_{\boldsymbol{\theta}}\left[\exp\left(-\int_{0}^{\tau}\exp(\boldsymbol{\theta}^{\top}\boldsymbol{g}_1(t))\mathrm{d}t\right)\right]\\
            =&\mathbb{E}_{\boldsymbol{\theta}}\left[\exp\left(-\frac{\tau}{3}\left[\exp(\theta_1)+\exp(\theta_2)+\exp(\theta_2+\theta_3)\right]\right)\right].
        \end{align*}
        On the other hand, the density function of model B is given by:
        \begin{align*}
            &\mathbb{E}_{\boldsymbol{\theta}}\left[\exp\left(-\int_{0}^{\tau}\widetilde{\lambda}(t|\boldsymbol{\theta})\mathrm{d}t\right)\right]\\
            =&\mathbb{E}_{\boldsymbol{\theta}}\left[\exp\left(-\int_{0}^{\tau}\exp(\boldsymbol{\theta}^{\top}\boldsymbol{g}_2(t))\mathrm{d}t\right)\right]\\
            =&\mathbb{E}_{\boldsymbol{\theta}}\left[\exp\left(-\frac{\tau}{3}\left[\exp(\theta_1)+\exp(\theta_2)+\exp(\theta_1+\theta_3)\right]\right)\right]\\
            =&\mathbb{E}_{\boldsymbol{\theta}}\left[\exp\left(-\frac{\tau}{3}\left[\exp(\theta_1)+\exp(\theta_2)+\exp(\theta_2+\theta_3)\right]\right)\right].
        \end{align*}
        The last equality holds due to the fact that $(\theta_1,\theta_2,\theta_3)\stackrel{d}{=}(\theta_2,\theta_1,\theta_3)\sim \mathcal{N}(\boldsymbol{0},\mathbf{I}_3)$. This indicates that both density functions match.
        \item Case 2: The outcome contains at least one event.\\[3mm]
        Suppose the recurrent event times are  $0<t_1<\ldots<t_N<\tau$. Since the covariate $\boldsymbol{Z}$ becomes zero after the first event, we have $\lambda(t|\boldsymbol{\theta})=\widetilde{\lambda}(t|\boldsymbol{\theta})=1$ for all $t_1<t\leq \tau$.
        
        If $0\leq t_1\leq 2\tau/3$, it is easy to see that $\mathbf{Q}^{\top}\boldsymbol{Z}(t)=\boldsymbol{g}_1(t)I(t\leq t_1)=\boldsymbol{g}_2(t)I(t\leq t_1)=\widetilde{\mathbf{Q}}^{\top}\boldsymbol{Z}(t)$, which further implies that $\lambda(t|\boldsymbol{\theta})=\widetilde{\lambda}(t|\boldsymbol{\theta})$ for all $0\leq t\leq \tau$. Then the two density functions match. 
        
        If $2\tau/3<t_1\leq \tau$, the density function of model A is given by:
        \begin{align*}
            &\mathbb{E}_{\boldsymbol{\theta}}\left[\lambda(t_1|\boldsymbol{\theta})\exp\left(-\int_{0}^{\tau}\lambda(t|\boldsymbol{\theta})\mathrm{d}t\right)\right]\\
            =&\mathbb{E}_{\boldsymbol{\theta}}\left[\exp\left(\boldsymbol{\theta}^{\top}\boldsymbol{g}_1(t_1)-\int_{0}^{t_1}\exp(\boldsymbol{\theta}^{\top}\boldsymbol{g}_1(t))\mathrm{d}t-(\tau-t_1)\right)\right]\\
            =&\mathbb{E}_{\boldsymbol{\theta}}\left[\exp\left(\theta_2+\theta_3-\frac{\tau}{3}\exp(\theta_1)-\frac{\tau}{3}\exp(\theta_2)-(t_1-\frac{2\tau}{3})\exp(\theta_2+\theta_3)-(\tau-t_1)\right)\right].
        \end{align*}
        On the other hand, the density function of model B is given by:
        \begin{align*}
            &\mathbb{E}_{\boldsymbol{\theta}}\left[\widetilde{\lambda}(t_1|\boldsymbol{\theta})\exp\left(-\int_{0}^{\tau}\widetilde{\lambda}(t|\boldsymbol{\theta})\mathrm{d}t\right)\right]\\
            =&\mathbb{E}_{\boldsymbol{\theta}}\left[\exp\left(\boldsymbol{\theta}^{\top}\boldsymbol{g}_2(t_1)-\int_{0}^{t_1}\exp(\boldsymbol{\theta}^{\top}\boldsymbol{g}_2(t))\mathrm{d}t-(\tau-t_1)\right)\right]\\
            =&\mathbb{E}_{\boldsymbol{\theta}}\left[\exp\left(\theta_1+\theta_3-\frac{\tau}{3}\exp(\theta_1)-\frac{\tau}{3}\exp(\theta_2)-(t_1-\frac{2\tau}{3})\exp(\theta_1+\theta_3)-(\tau-t_1)\right)\right]\\
            =&\mathbb{E}_{\boldsymbol{\theta}}\left[\exp\left(\theta_2+\theta_3-\frac{\tau}{3}\exp(\theta_1)-\frac{\tau}{3}\exp(\theta_2)-(t_1-\frac{2\tau}{3})\exp(\theta_2+\theta_3)-(\tau-t_1)\right)\right].
        \end{align*}
        The last equality holds due to the fact that $(\theta_1,\theta_2,\theta_3)\stackrel{d}{=}(\theta_2,\theta_1,\theta_3)\sim \mathcal{N}(\boldsymbol{0},\mathbf{I}_3)$. This indicates that both density functions match.
    \end{enumerate}
    Hence, we have proved that both models induce the same probability measure, which implies that the model is not identifiable in certain configurations, i.e., not identifiable in the conventional sense.
    \end{proof}
The constructed counterexample shows that conventional identifiability fails for this specific configuration, thereby justifying the need to adopt the framework of generic identifiability, which excludes a measure-zero set in the parameter space that can possibly be non-identifiable.

\subsection{Identifiability when only external covariates are present}\label{S21}
When the covariate processes $\boldsymbol{X}_{ij}(\cdot)$ and $\boldsymbol{Z}_{ij}(\cdot)$ in the proposed model (1) are external, the probability measure can be written in product form: $\mu_{\mathbf{N},\mathbf{X},\mathbf{Z}}=\mu_{\mathbf{N}|\mathbf{X},\mathbf{Z}}\times \mu_{\mathbf{X},\mathbf{Z}}$. Identifiability is then studied by fixing arbitrary $\mathbf{X},\mathbf{Z}$ and integrating the conditional probability measure $\mu_{\mathbf{N}|\mathbf{X},\mathbf{Z}}$. Such models are well studied in traditional survival analysis, where the first part of Condition (e) is not required to ensure model identifiability; see \cite{zeng2007semiparametric, zeng2010general}. We formally state the identifiability result in the following theorem.
\begin{thm}
    If $\boldsymbol{X}_{ij}(\cdot)$ and $\boldsymbol{Z}_{ij}(\cdot)$ are external covariates for $i=1,\ldots,n,j=1,\ldots,J$, model (1) is identifiable under Conditions (c) and (d).
\end{thm}
\begin{proof}
    The proof is similar to that in \cite{zeng2007semiparametric} and \cite{zeng2010general}. Let $[0,\tau]$ be the observation period. Consider two competing parametric models:
    \begin{align*}
        \lambda_{j}(t|\boldsymbol{X}_{j}, \boldsymbol{Z}_j ; \boldsymbol{\theta})& =Y_j(t) \exp(\beta_{j0}+\boldsymbol{\beta}_{j}^{\top} \boldsymbol{X}_{j}(t)+\boldsymbol{\theta}^{\top}\boldsymbol{\Sigma}^{1/2} \mathbf{A}_{j}^{\top} \boldsymbol{Z}_j(t))\text{,}\\
        \widetilde{\lambda}_{j}(t|\boldsymbol{X}_{j}, \boldsymbol{Z}_j ; \boldsymbol{\theta}) &= Y_j(t) \exp(\widetilde{\beta}_{j0}+\widetilde{\boldsymbol{\beta}}_{j}^{\top} \boldsymbol{X}_{j}(t)+\boldsymbol{\theta}^{\top}\widetilde{\mathbf{\Sigma}}^{1/2} {\widetilde{\mathbf{A}}_{j}}^{\top} \boldsymbol{Z}_j(t))\text{,}
    \end{align*}
    where $\boldsymbol{\theta}\sim \mathcal{N}(\boldsymbol{0},\mathbf{I}_K)$. Suppose the two parameter sets induce the same distribution of the observed multivariate counting process.
    Without loss of generality, we only consider the case when $Y_j(t)=1$ for all $j=1,\ldots,J$ and $t\in[0,\tau]$. Fix $k_{0j}, k_{1j} \in \mathbb{N}$ for all $j=1,\ldots,J$. Consider event times  $\{ t_{j11},\ldots, t_{j1k_{1j}}\}$ and $\{t_{j1},\ldots,t_{jk_{0j}} \}$ for event type $j$, for $j=1,\ldots,J$. Then,
	\begin{eqnarray*}
		&& \int_{\boldsymbol{\theta}} \prod^J_{j=1} \bigg\{ \prod^{k_{1j}}_{k=1} \lambda_j(t_{j1k}) \times \prod^{k_{0j}}_{k=1} \lambda_j(t_{jk}) e^{-\int^{\tau}_0 \lambda_j(t) \mathrm{d}t} \bigg\} \phi_K(\boldsymbol{\theta};\boldsymbol{0},\mathbf{I}_K) \mathrm{d}\boldsymbol{\theta}\\
		&=& \int_{\boldsymbol{\theta}} \prod^J_{j=1} \bigg\{ \prod^{k_{1j}}_{k=1} \widetilde{\lambda}_j(t_{j1k}) \times \prod^{k_{0j}}_{k=1} \widetilde{\lambda}_j(t_{jk}) e^{-\int^{\tau}_0 \widetilde{\lambda}_j(t) \mathrm{d}t} \bigg\} \phi_K(\boldsymbol{\theta};\boldsymbol{0},\mathbf{I}_K) \mathrm{d}\boldsymbol{\theta}. 
	\end{eqnarray*}
	Integrating $t_{j11},\ldots, t_{j1k_{1j}}$ from $0$ to $t_{j}$ for $j=1,\ldots,J$ and integrating $t_{j1},\ldots,t_{jk_{0j}}$ from $0$ to $\tau$ for $j=1,\ldots,J$, we have
	\begin{eqnarray*}
		&& \int_{\boldsymbol{\theta}} \prod^J_{j=1} \bigg[  \bigg\{ \int^{t_{j}}_0 \lambda_j(t) \mathrm{d}t \bigg\}^{k_{1j}} \times \bigg\{ \int^{\tau}_0 \lambda_j (t) \mathrm{d}t \bigg\}^{k_{0j}} e^{-\int^{\tau}_0 \lambda_j(t) \mathrm{d}t} \bigg] \phi_K(\boldsymbol{\theta};\boldsymbol{0},\mathbf{I}_K) \mathrm{d}\boldsymbol{\theta}\\
		&=& \int_{\boldsymbol{\theta}} \prod^J_{j=1} \bigg[ \bigg\{ \int^{t_{j}}_0 \widetilde{\lambda}_j(t) \mathrm{d}t \bigg\}^{k_{1j}} \times \bigg\{\int^{\tau}_0 \widetilde{\lambda}_j (t) \mathrm{d}t \bigg\}^{k_{0j}} e^{-\int^{\tau}_0 \widetilde{\lambda}_j(t) \mathrm{d}t} \bigg] \phi_K(\boldsymbol{\theta};\boldsymbol{0},\mathbf{I}_K) \mathrm{d}\boldsymbol{\theta}. 
	\end{eqnarray*}
	Multiply both sides by $ \prod^J_{j=1}  \frac{ ( \text{i} s_{j} )^{k_{1j}}}{k_{1j}!}  \cdot\frac{1}{k_{0j}!}$, where $\text{i}$ is the imaginary unit. Summing over $k_{1j}=0,1,2,\ldots$  and $k_{0j}=0,1,2,\ldots$ for all $j=1,\ldots,J$, we get
	\begin{eqnarray*}
		\int_{\boldsymbol{\theta}} \prod^J_{j=1} e^{ \text{i} s_{j} \int^{t_{j}}_0 \lambda_j(t) \mathrm{d}t }  \phi_K(\boldsymbol{\theta};\boldsymbol{0},\mathbf{I}_K) \mathrm{d}\boldsymbol{\theta}  = 
		\int_{\boldsymbol{\theta}} \prod^J_{j=1}  e^{ \text{i} s_{j} \int^{t_{j}}_0 \widetilde{\lambda}_j(t) \mathrm{d}t }  \phi_K(\boldsymbol{\theta};\boldsymbol{0},\mathbf{I}_K) \mathrm{d}\boldsymbol{\theta}.
	\end{eqnarray*}
	 	Since this holds for any $s_{j}\in\mathbb{R}$, this implies that the distribution of $\{ \int^{t_{j}}_0 \lambda_j(t) \mathrm{d}t \}_{j=1,\ldots,J}$ and $\{ \int^{t_{j}}_0 \widetilde{\lambda}_j(t) \mathrm{d}t \}_{j=1,\ldots,J}$ are the same for any $t_j\in [0,\tau]$, where $\boldsymbol{\theta} \sim \mathcal{N}(\boldsymbol{0},\mathbf{I}_K)$. It holds that $\{ \log \lambda_j(t_{j})\}_{j=1,\ldots,J}$ and $\{ \log \widetilde{\lambda}_j(t_{j})\}_{j=1,\ldots,J}$ have the same distribution. Since $\int \boldsymbol{\theta} \phi_K(\boldsymbol{\theta};\boldsymbol{0},\mathbf{I}_K) \mathrm{d}\boldsymbol{\theta} = \boldsymbol{0}$, by considering the mean of $\log \lambda_j(t_{j})$ and $\log \widetilde{\lambda}_j(t_{j})$, we have $\beta_{j0} + \boldsymbol{\beta}_j^{\top} \boldsymbol{X}_j(t_{j}) = \widetilde{\beta}_{j0}+ {\widetilde{\boldsymbol{\beta}}_j}^{\top} \boldsymbol{X}_j(t_{j })$. Since this holds for any $t_1,\ldots,t_J\in[0,\tau]$, by Condition (d), we have $\beta_{j0} = \widetilde{\beta}_{j0}$ and $\boldsymbol{\beta}_j = \widetilde{\boldsymbol{\beta}}_j$ for all $j=1,\ldots,J$. Then $\{ \boldsymbol{\theta}^{\top} \boldsymbol{\Sigma}^{1/2}\mathbf{A}_j^{\top} \boldsymbol{Z}_j(t_{j})\}_{j=1,\ldots,J} $ has the same joint distribution as $\{ \boldsymbol{\theta}^{\top}\widetilde{\boldsymbol{\Sigma}}^{1/2} {\widetilde{\mathbf{A}}_j}^{\top} \boldsymbol{Z}_j(t_{j })\}_{j=1,\ldots,J}$. By considering the covariance matrices of these two random vectors, we have for each $j,l =1,\ldots, J$ and any $t_1,\ldots,t_J\in[0,\tau]$,
	 \begin{equation*}
	 \boldsymbol{Z}^{\top}_j(t_{j})\mathbf{A}_j\boldsymbol{\Sigma} \mathbf{A}^{\top}_{l} \boldsymbol{Z}_l(t_{l}) = 	 \boldsymbol{Z}^{\top}_j(t_{j}) \widetilde{\mathbf{A}}_j\widetilde{\boldsymbol{\Sigma}} (\widetilde{\mathbf{A}}_{l} )^{\top}\boldsymbol{Z}_l(t_{l}).
	 \end{equation*}
	 Let $\mathbf{B} = \mathbf{A}_j \mathbf{\Sigma} \mathbf{A}^{\top}_{l} - \widetilde{\mathbf{A}}_j\widetilde{\mathbf{\Sigma}} (\widetilde{\mathbf{A}}_{l} )^{\top}$. We have $\boldsymbol{Z}^{\top}_j(t_{j}) \mathbf{B} \boldsymbol{Z}_l(t_{l}) = 0 $.  Condition (d) then implies $\mathbf{B} = 0$. Hence, we have $(\mathbf{A},\mathbf{\Sigma})\sim (\widetilde{\mathbf{A}},\widetilde{\mathbf{\Sigma}})$, which indicates that model (1) is identifiable.
\end{proof}
Note also that the above proof establishes conventional identifiability of the proposed model without relying on the concept of generic identifiability. This indicates that the need to adopt the concept of generic identifiability arises from the presence of internal covariates.

\subsection{Challenges posed by internal covariates}\label{S22}
Internal covariates are entangled with the event history and thereby restrict the joint support of the event and covariate processes. Consequently, the proof technique in the previous section, which fixes the covariate processes and integrates the conditional density of the counting processes, cannot be carried over to this setting, creating fundamental challenges for establishing identifiability. 

Furthermore, we construct a counterexample showing that identifiability fails without stronger structural assumptions. This highlights the necessity of the first part of Condition (e), which requires the covariate processes to be piecewise-constant, when internal covariates are present. The counterexample shows that the model may remain non-identifiable even when the covariates are allowed to evolve in a piecewise-linear manner.

We first illustrate the construction of the counterexample using the gamma frailty model, for which explicit density functions are available, and then extend the approach to arbitrary distribution families (e.g., log-normal frailty distribution in our proposed model).

We consider a scenario where there is only one event type. Let $[0,\tau]$ be the observation period and $t_1$ be the time point of the first event or $\tau$ if no event occurs. The univariate covariate processes $X(\cdot)$ and $Z(\cdot)$ are defined by:
\begin{align*}
    X(t)=\begin{cases}
        t  & \text{ if } t\leq t_1,\\
        0 & \text{ if } t>t_1,
    \end{cases}
    \quad\quad Z(t)=\begin{cases}
        1 & \text{ if } t\leq t_1,\\
        0 & \text{ if } t>t_1.
    \end{cases}
\end{align*}
It is straightforward to verify that Conditions (c) and (d) are satisfied. Moreover, $X(\cdot)$ and $Z(\cdot)$ are internal covariates in the sense that their observed paths carry information about the observed event time. Also note that $X(\cdot)$ and $Z(\cdot)$ satisfy piecewise-linear assumption instead of piecewise-constant assumption. We then consider the model where the intensity function takes the following form:
\begin{align}\label{model_counterexample}
    \lambda(t|\mathscr{F}_{t^{-}};\theta)=\lambda_0(t)Y(t)\exp(\beta X(t)+\theta Z(t)),
\end{align}
where $\exp(\theta)\sim \Gamma(\alpha,\alpha)$, i.e., the gamma frailty model. Suppose $Y(t)\equiv 1$ for all $t\in[0,\tau]$. 

We use the gamma distribution for illustration because its Laplace transform admits an explicit form, which is stated in the following lemma.
\begin{lem}\label{lem_MGF}
    For random variable $X$ such that $\exp(X)\sim \Gamma(\alpha,\alpha)$, the following hold:
    \begin{enumerate}[(1)]
        \item $\mathbb{E}\exp(-t\exp(X))=(1+t/\alpha)^{-\alpha}$ for any $t\geq 0$.
        \item $\mathbb{E}\exp(X-t\exp(X))=(1+t/\alpha)^{-\alpha-1}$ for any $t\geq 0$.
    \end{enumerate}
\end{lem}
\begin{proof}
    \begin{enumerate}[(1)]
        \item The result is immediate in view of the moment generating function of the gamma distribution.
        \item By the result in part (1), we have
        \begin{align*}
            \mathbb{E}\exp(X-t\exp(X))=-\frac{\mathrm{d}}{\mathrm{d}t}\mathbb{E}\exp(-t\exp(X))=-\frac{\mathrm{d}}{\mathrm{d}t}\left(1+\frac{t}{\alpha}\right)^{-\alpha}=\left(1+\frac{t}{\alpha}\right)^{-\alpha-1}.
        \end{align*}
    \end{enumerate}
\end{proof}

Using Lemma \ref{lem_MGF}, we now construct a counterexample for the gamma frailty model, as stated in the following theorem.

\begin{thm}
    Model (\ref{model_counterexample}) is not generically identifiable.
\end{thm}
\begin{proof}
    We consider the following two competing models:
    \begin{align*}
        \text{Model A:~~~}\lambda(t|\mathscr{F}_{t^{-}};\theta)=&\lambda_0(t)Y(t)\exp(\beta X(t)+\theta Z(t)),\\
        \text{Model B:~~~}\widetilde{\lambda}(t|\mathscr{F}_{t^{-}};\widetilde{\theta})=&\widetilde{\lambda}_0(t)Y(t)\exp(\widetilde{\beta} X(t)+\widetilde{\theta} Z(t)),
    \end{align*}
    where $\exp(\theta)\sim \Gamma(\alpha,\alpha)$ and $\exp(\widetilde{\theta})\sim \Gamma(\widetilde{\alpha},\widetilde{\alpha})$. For any given $\beta\neq 1$, $\alpha$, we let $\widetilde{\beta}=\beta+1$, $\widetilde{\alpha}=\alpha/2$ and $\widetilde{\lambda}_0(t)=\lambda_0(t)=\alpha\exp((1-\beta)t)$. We then prove that both models induce the same probability measure, i.e., for any given outcome, the density functions in both models are identical. Consider the following two cases:
    \begin{enumerate}[(i)]
        \item Case 1: The outcome contains no event.\\[3mm]
        By Lemma \ref{lem_MGF}, the density under model $A$ in this case is given by:
        \begin{align*}
            &\mathbb{E}_{\theta}\left[\exp\left(-\int_{0}^{\tau}\lambda(t|\mathscr{F}_{t^{-}};\theta)\mathrm{d}t\right)\right]\\
            =&\mathbb{E}_{\theta}\left[\exp\left(-\int_{0}^{\tau}\lambda_0(t)\exp\left(\beta X(t)+\theta Z(t)\right)\mathrm{d}t\right)\right]\\
            =&\mathbb{E}_{\theta}\left[\exp\left(-\exp(\theta)\int_{0}^{\tau}\alpha\exp((1-\beta)t)\cdot \exp\left(\beta t\right)\mathrm{d}t\right)\right]\\
            =&\left(1+\frac{\int_{0}^{\tau}\alpha\exp((1-\beta)t)\cdot \exp\left(\beta t\right)\mathrm{d}t}{\alpha}\right)^{-\alpha}\\
            =&\left(1+\int_{0}^{\tau}\exp(t)\mathrm{d}t\right)^{-\alpha}\\
            =&\exp(-\alpha\tau).
        \end{align*}
        By Lemma \ref{lem_MGF}, the density under model $B$ is given by:
        \begin{align*}
            &\mathbb{E}_{\widetilde{\theta}}\left[\exp\left(-\int_{0}^{\tau}\widetilde{\lambda}(t|\mathscr{F}_{t^{-}};\widetilde{\theta})\mathrm{d}t\right)\right]\\
            =&\mathbb{E}_{\widetilde{\theta}}\left[\exp\left(-\int_{0}^{\tau}\widetilde{\lambda}_0(t)\exp(\widetilde{\beta} X(t)+\widetilde{\theta} Z(t))\mathrm{d}t\right)\right]\\
            =&\mathbb{E}_{\widetilde{\theta}}\left[\exp\left(-\exp(\widetilde{\theta})\int_{0}^{\tau}\alpha\exp((1-\beta)t)\cdot \exp(\widetilde{\beta} t)\mathrm{d}t\right)\right]\\
            =&\left(1+\frac{\int_{0}^{\tau}\alpha\exp((1-\beta)t)\cdot \exp((\beta+1) t)\mathrm{d}t}{\widetilde{\alpha}}\right)^{-\widetilde{\alpha}}\\
            =&\left(1+\frac{\int_{0}^{\tau}\alpha\exp(2t)\mathrm{d}t}{\alpha/2}\right)^{-\alpha/2}\\
            =&\left(1+2\int_{0}^{\tau}\exp(2t)\mathrm{d}t\right)^{-\alpha/2}\\
            =&\exp(-\alpha\tau),
        \end{align*}
        which matches the density under model A.
        \item Case 2: The outcome contains at least one event.\\[3mm]
        Suppose the recurrent event times are  $0<t_1<\ldots<t_{N}<\tau$. By Lemma \ref{lem_MGF}, the density under model $A$ is given by:
        \begin{align*}
            &\mathbb{E}_{\theta}\left[\exp\left(-\int_{0}^{\tau}\lambda(t|\mathscr{F}_{t^{-}};\theta)\mathrm{d}t\right)\prod_{k=1}^N\lambda(t_k|\mathscr{F}_{t^{-}};\theta)\right]\\
            =&\mathbb{E}_{\theta}\left[\exp\left(-\int_{0}^{\tau}\lambda_0(t)\exp\left(\beta X(t)+\theta Z(t)\right)\mathrm{d}t\right)\prod_{k=1}^N\lambda_0(t_k)\exp\left(\beta X(t_k)+\theta Z(t_k)\right)\right]\\
            =&\mathbb{E}_{\theta}\left[\exp\left(-\exp(\theta)\int_{0}^{t_1}\lambda_0(t)\exp\left(\beta X(t)\right)\mathrm{d}t-\int_{t_1}^{\tau}\lambda_0(t)\mathrm{d}t\right)\cdot \exp\left(\beta X(t_1)+\theta\right)\cdot \prod_{k=1}^N \lambda_0(t_k)\right]\\
            =&\mathbb{E}_{\theta}\Bigg[\exp\left(-\exp(\theta)\int_{0}^{t_1}\alpha\exp((1-\beta)t)\cdot\exp\left(\beta t\right)\mathrm{d}t-\int_{t_1}^{\tau}\alpha\exp((1-\beta)t)\mathrm{d}t\right)\\
            &\quad\exp\left(\beta t_1+\theta\right)\Bigg]\cdot \prod_{k=1}^N(\alpha\exp((1-\beta)t_k))\\
            =&\alpha^N\exp\left(\beta t_1+(1-\beta)\sum_{k=1}^Nt_k-\alpha\int_{t_1}^{\tau}\exp((1-\beta)t)\mathrm{d}t\right)\mathbb{E}_{\theta}\left[\exp\left(\theta-\alpha\exp(\theta)\int_{0}^{t_1}\exp\left(t\right)\mathrm{d}t\right)\right]\\
            =&\alpha^N \exp\left(t_1+(1-\beta)\sum_{k=2}^Nt_k-\frac{\alpha}{1-\beta}[\exp((1-\beta)\tau)-\exp((1-\beta)t_1)]\right)\left(1+\frac{\alpha\int_{0}^{t_1}\exp(t)\mathrm{d}t}{\alpha}\right)^{-\alpha-1}\\
            =&\alpha^N \exp\left(t_1+(1-\beta)\sum_{k=2}^Nt_k-\frac{\alpha}{1-\beta}[\exp((1-\beta)\tau)-\exp((1-\beta)t_1)]-(\alpha+1)t_1\right)\\
            =&\alpha^N \exp\left((1-\beta)\sum_{k=2}^Nt_k-\frac{\alpha}{1-\beta}[\exp((1-\beta)\tau)-\exp((1-\beta)t_1)]-\alpha  t_1\right).
        \end{align*}
        By Lemma \ref{lem_MGF}, the density under model $B$ is given by
        \begin{align*}
            &\mathbb{E}_{\widetilde{\theta}}\left[\exp\left(-\int_{0}^{\tau}\widetilde{\lambda}(t|\mathscr{F}_{t^{-}};\widetilde{\theta})\mathrm{d}t\right)\prod_{k=1}^N\widetilde{\lambda}(t_k|\mathscr{F}_{t^{-}};\widetilde{\theta})\right]\\
            =&\mathbb{E}_{\widetilde{\theta}}\left[\exp\left(-\int_{0}^{\tau}\widetilde{\lambda}_0(t)\exp(\widetilde{\beta} X(t)+\widetilde{\theta} Z(t))\mathrm{d}t\right)\prod_{k=1}^N\widetilde{\lambda}_0(t_k)\exp(\widetilde{\beta} X(t_k)+\widetilde{\theta} Z(t_k))\right]\\
            =&\mathbb{E}_{\widetilde{\theta}}\left[\exp\left(-\exp(\widetilde{\theta})\int_{0}^{t_1}\widetilde{\lambda}_0(t)\exp(\widetilde{\beta} X(t))\mathrm{d}t-\int_{t_1}^{\tau}\widetilde{\lambda}_0(t)\mathrm{d}t\right)\cdot \exp(\widetilde{\beta} X(t_1)+\widetilde{\theta})\cdot \prod_{k=1}^N \widetilde{\lambda}_0(t_k)\right]\\
            =&\mathbb{E}_{\widetilde{\theta}}\Bigg[\exp\left(-\exp(\widetilde{\theta})\int_{0}^{t_1}\alpha\exp((1-\beta)t)\cdot\exp\left((\beta+1) t\right)\mathrm{d}t-\int_{t_1}^{\tau}\alpha\exp((1-\beta)t)\mathrm{d}t\right)\\
            &\exp((\beta+1) t_1+\widetilde{\theta})\Bigg]\cdot \prod_{k=1}^N(\alpha\exp((1-\beta)t_k))\\
            =&\alpha^N\exp\left((\beta+1) t_1+(1-\beta)\sum_{k=1}^Nt_k-\alpha\int_{t_1}^{\tau}\exp((1-\beta)t)\mathrm{d}t\right)\mathbb{E}_{\widetilde{\theta}}\left[\exp\left(\widetilde{\theta}-\alpha\exp(\widetilde{\theta})\int_{0}^{t_1}\exp\left(2t\right)\mathrm{d}t\right)\right]\\
            =&\alpha^N \exp\left(2t_1+(1-\beta)\sum_{k=2}^Nt_k-\frac{\alpha}{1-\beta}[\exp((1-\beta)\tau)-\exp((1-\beta)t_1)]\right)\left(1+\frac{\alpha\int_{0}^{t_1}\exp(2t)\mathrm{d}t}{\widetilde{\alpha}}\right)^{-\widetilde{\alpha}-1}\\
            =&\alpha^N \exp\left(2t_1+(1-\beta)\sum_{k=2}^Nt_k-\frac{\alpha}{1-\beta}[\exp((1-\beta)\tau)-\exp((1-\beta)t_1)]\right)\left(1+\frac{\alpha\int_{0}^{t_1}\exp(2t)\mathrm{d}t}{\alpha/2}\right)^{-\alpha/2-1}\\
            =&\alpha^N \exp\left(2t_1+(1-\beta)\sum_{k=2}^Nt_k-\frac{\alpha}{1-\beta}[\exp((1-\beta)\tau)-\exp((1-\beta)t_1)]-2(\alpha/2+1)t_1\right)\\
            =&\alpha^N \exp\left((1-\beta)\sum_{k=2}^Nt_k-\frac{\alpha}{1-\beta}[\exp((1-\beta)\tau)-\exp((1-\beta)t_1)]-\alpha  t_1\right),
        \end{align*}
        which matches the density under model A.
    \end{enumerate}
    Hence both models induce the same probability measure, which indicates that model (\ref{model_counterexample}) is not generically identifiable (indeed, it is not identifiable at almost any given $(\beta,\alpha)$).
\end{proof}
We then summarise the high-level idea of this counterexample and demonstrate how this construction can be generalised to any choice of frailty distribution (e.g., the log-normal frailty distribution in our proposed model). Specifically, we let $X(t)=f(t)I(t\leq t_1)$ with $f(t)$ being any given strictly increasing smooth function with $f(0)=0$ (e.g., linear function in the previous counterexample), $Z(t)=I(t\leq t_1)$ and $\lambda_0(t)=\widetilde{\lambda}_0(t)$. We consider the following two competing models.
\begin{align*}
        \text{Model A:~~~}\lambda(t|\mathscr{F}_{t^{-}};\theta)=&\lambda_0(t)Y(t)\exp(\beta X(t)+\theta Z(t)),\\
        \text{Model B:~~~}\widetilde{\lambda}(t|\mathscr{F}_{t^{-}};\widetilde{\theta})=&\widetilde{\lambda}_0(t)Y(t)\exp(\widetilde{\beta} X(t)+\widetilde{\theta} Z(t)),
    \end{align*}

First, we observe that the two models yield identical intensity functions after the first event (since $X(\cdot)$ and $Z(\cdot)$ become zero and only the baseline hazard functions remain), and thus no additional information can be obtained to distinguish the two models starting from the second event. 

We then show that the first event also provides no additional information for distinguishing the two models. That is, if the model is not identifiable in the case where no event occurs, then the first event cannot help distinguish the two models. The reason is that
\begin{align*}
    &\mathbb{E}_{\theta}\left[\lambda(t|\mathscr{F}_{t^{-}};\theta)\exp\left(-\int_{0}^t\lambda(s|\mathscr{F}_{s^{-}};\theta)\mathrm{d}s\right)\right]\\
    =&-\frac{\mathrm{d}}{\mathrm{d}t}\mathbb{E}_{\theta}\left[\exp\left(-\int_{0}^t{\lambda}(s|\mathscr{F}_{s^{-}};\theta)\mathrm{d}s\right)\right]\\
    =&-\frac{\mathrm{d}}{\mathrm{d}t}\mathbb{E}_{\widetilde{\theta}}\left[\exp\left(-\int_{0}^t\widetilde{\lambda}(s|\mathscr{F}_{s^{-}};\widetilde{\theta})\mathrm{d}s\right)\right]\quad(\text{suppose both density functions match when no event occurs})\\
    =&\mathbb{E}_{\widetilde{\theta}}\left[\widetilde{\lambda}(t|\mathscr{F}_{t^{-}};\widetilde{\theta})\exp\left(-\int_{0}^t\widetilde{\lambda}(s|\mathscr{F}_{s^{-}};\widetilde{\theta})\mathrm{d}s\right)\right].
\end{align*}
The above equation shows that the likelihood function with one event has the same mathematical form as the time derivative of the likelihood function with no event. If the two likelihood functions are identical when no event happens, then they are still identical when exactly one event happens.
In other words, the first event also does not provide any additional information to distinguish between the two competing models. 

Combining the above discussions, if the model is identifiable, the two models should already be distinguishable in the case where no event occurs.
Since $Z(\cdot)$ equals 1 before the first event, the likelihood function in such a case is exactly the Laplace transform of the frailty distribution. When the two Laplace transforms coincide, i.e.,
\begin{align}\label{eq_counter1}
    \mathbb{E}_{\theta}\left[\exp\left(-\exp(\theta)\int_{0}^{t}\lambda_0(s)\exp(\beta f(s))\mathrm{d}s\right)\right]=\mathbb{E}_{\widetilde{\theta}}\left[\exp\left(-\exp(\widetilde{\theta})\int_{0}^{t}\lambda_0(s)\exp(\widetilde{\beta} f(s))\mathrm{d}s\right)\right],
\end{align}
then the two competing models induce the same probability measure. Hence our main goal in the counterexample is to find a solution $\lambda_0(t)=\widetilde{\lambda}_0(t)$ that makes these two Laplace transforms identical.

We choose two different frailty distributions $\theta$ and $\widetilde{\theta}$ such that $\mathbb{E}(\exp(\theta))=\mathbb{E}(\exp(\widetilde{\theta}))=1$ and $\operatorname{Var}(\exp(\theta))\neq \operatorname{Var}(\exp(\widetilde{\theta}))$. Suppose $\phi(s)=\mathbb{E}_{\theta}\left[\exp\left(-s\exp(\theta)\right)\right]$ and $\widetilde{\phi}(s)=\mathbb{E}_{\widetilde{\theta}}[\exp(-s\exp(\widetilde{\theta}))]$ are two functions defined on $[0,\infty)$. Since $\phi(0)=\widetilde{\phi}(0)=1$ and both functions are strictly decreasing, there exists $\psi=\widetilde{\phi}^{-1}\circ \phi$ such that $\psi(0)=0$ and $\psi$ is strictly increasing. Hence, (\ref{eq_counter1}) is equivalent to 
\begin{align*}
    \int_{0}^{t}\lambda_0(s)\exp(\widetilde{\beta} f(s))\mathrm{d}s=\psi\left(\int_{0}^{t}\lambda_0(s)\exp(\beta f(s))\mathrm{d}s\right).
\end{align*}
Note that the two sides already match when $t=0$. Hence, we only need to match their derivatives, i.e.,
\begin{align*}
    \lambda_0(t)\exp(\widetilde{\beta} f(t))=&\frac{\mathrm{d}}{\mathrm{d}t}\int_{0}^{t}\lambda_0(s)\exp(\widetilde{\beta} f(s))\mathrm{d}s\\
    =&\frac{\mathrm{d}}{\mathrm{d}t}\psi\left(\int_{0}^{t}\lambda_0(s)\exp(\beta f(s))\mathrm{d}s\right)\\
    =&\psi^{\prime}\left(\int_{0}^{t}\lambda_0(s)\exp(\beta f(s))\mathrm{d}s\right)\lambda_0(t)\exp(\beta f(t)),
\end{align*}
which is equivalent to 
\begin{align}\label{arbitrary}
    \exp((\widetilde{\beta}-\beta)f(t))=\psi^{\prime}\left(\int_{0}^{t}\lambda_0(s)\exp(\beta f(s))\mathrm{d}s\right).
\end{align}
Given that $\mathbb{E}(\exp(\theta))=\mathbb{E}(\exp(\widetilde{\theta}))=1$ and $\operatorname{Var}(\exp(\theta))\neq \operatorname{Var}(\exp(\widetilde{\theta}))$, we can show that $\psi^{\prime}(0)=1$ and $\psi^{\prime\prime}(0)\neq 0$. Without loss of generality, we assume $\psi^{\prime\prime}(0) > 0$. Consequently, $\psi^{\prime}$ is strictly increasing in a small neighbourhood of $0$, which ensures the existence of an inverse function. For any given $\beta\in\mathbb{R}$, by letting $\widetilde{\beta}=\beta+1$ in (\ref{arbitrary}), we have
\begin{align*}
    \int_{0}^{t}\lambda_0(s)\exp(\beta f(s))\mathrm{d}s=(\psi^{\prime})^{-1}(\exp(f(t))).
\end{align*}
Since both $f$ and $\psi^{\prime}$ are strictly increasing, the right-hand side is also strictly increasing and smooth, given the smoothness of $f$ and $\psi^{\prime}$. By the Newton--Leibniz formula, this yields a positive solution for $\lambda_0(t)$ on a neighbourhood of $0$ since both sides equal $0$ at $t=0$. Hence, we construct two competing models with different parameter values that induce the same probability measure, implying that the model is not generically identifiable.

Note that this construction method is valid as long as $f(t)$ is strictly increasing (or decreasing) in a small neighbourhood of $0$, which illustrates the necessity of the first part of Condition (e), i.e., the piecewise-constant assumption.

\section{Proof of Theorem \ref{thm_identifiability}}\label{sec:thm1}
In Section~\ref{S22}, we have already discussed the challenges in establishing identifiability posed by internal covariates. Under such challenges, the identification procedure is limited to a pointwise argument since the conditioning argument is invalid. Specifically, for two competing models with parameters $\boldsymbol{\delta}$ and $\widetilde{\boldsymbol{\delta}}$, since both models induce the same probability measure, i.e., $\mu_{\boldsymbol{\delta}}=\mu_{\widetilde{\boldsymbol{\delta}}}$, there exists a set $\mathcal{A}\subset \Omega$ with $\mathbb{P}(\mathcal{A})=1$, such that the density functions under both models are identical on the set $\mathcal{A}$, i.e., $p_{\boldsymbol{\delta}}(\omega)=p_{\widetilde{\boldsymbol{\delta}}}(\omega)$ for any $\omega\in\mathcal{A}$. The main goal in the proof of Theorem \ref{thm_identifiability} is to show that $\lambda_{\boldsymbol{\delta}}(t,\omega|\boldsymbol{\theta})=\lambda_{\widetilde{\boldsymbol{\delta}}}(t,\omega|\boldsymbol{\theta})$ up to rotations and permutations by analysing the density functions on any given single outcome $\omega\in\mathcal{A}$. Then by combining this result over all $\omega\in\mathcal{A}$, we can fix the permutations, which completes the proof of Theorem \ref{thm_identifiability}.
\subsection{Sketch of the proof for Theorem \ref{thm_identifiability}}
We outline the main steps in the proof of Theorem~1. A more detailed version, along with a complete proof, is given in Section~\ref{S33}.\\[3mm]
    \noindent \textbf{Step 1: }For any $\omega\in\mathcal{A}$, matching the likelihood functions leads to
    \begin{align}\label{sketch_1.1}
        &\int \Big[\prod_{j=1}^{J} \prod_{s \leq t} \lambda_{j}(s,\omega|\boldsymbol{\theta})^{\Delta N_{j}(s,\omega)}\Big]\exp\Big(-\sum_{j=1}^J \int_0^t \lambda_j(s,\omega|\boldsymbol{\theta})\mathrm{d}s\Big) \phi_K(\boldsymbol{\theta} ; \boldsymbol{0},\mathbf{I}_K) \mathrm{d}\boldsymbol{\theta}\notag\\
        =&\int \Big[\prod_{j=1}^{J} \prod_{s \leq t} \widetilde{\lambda}_{j}(s,\omega|\boldsymbol{\theta})^{\Delta N_{j}(s,\omega)}\Big]\exp\Big(-\sum_{j=1}^J \int_0^t \widetilde{\lambda}_j(s,\omega|\boldsymbol{\theta})\mathrm{d}s\Big) \phi_K(\boldsymbol{\theta} ;\boldsymbol{0}, \mathbf{I}_K) \mathrm{d} \boldsymbol{\theta},
    \end{align}
    where the random effect $\boldsymbol{\theta}$ is $K$-variate standard normal, and
    \begin{align*}     \lambda_j(t,\omega|\boldsymbol{\theta})&=\exp(\beta_{j0}+\boldsymbol{\beta}_{j}^{\top} \boldsymbol{X}_{j}(t,\omega)+\boldsymbol{\theta}^{\top}\boldsymbol{\Sigma}^{1/2} \mathbf{A}_{j}^{\top} \boldsymbol{Z}_{j}(t,\omega)),\\       \widetilde{\lambda}_j(t,\omega|\boldsymbol{\theta})&=\exp(\widetilde{\beta}_{j0}+\widetilde{\boldsymbol{\beta}}_{j}^{\top} \boldsymbol{X}_{j}(t,\omega)+\boldsymbol{\theta}^{\top} \widetilde{\boldsymbol{\Sigma}}^{1/2}\widetilde{\mathbf{A}}_{j}^{\top} \boldsymbol{Z}_{j}(t,\omega)).
    \end{align*}
   For simplicity, we omit $\bt$ in the expression. The piecewise-constant assumption in Condition (e) enables straightforward differentiation of the likelihood function. By fixing an $\omega\in\mathcal{A}$ and performing repeated differentiations with respect to $t$, equation (\ref{sketch_1.1}) leads to
    \begin{align}\label{sketch_1.2}
        &\int \Big[\prod_{j=1}^{J} \prod_{s \leq t} \lambda_{j}(s,\omega)^{\Delta N_{j}(s,\omega)}\Big]\exp\Big(-\sum_{j=1}^J \int_0^t \lambda_j(s,\omega)\mathrm{d}s\Big)\Big(\sum_{j=1}^J\lambda_{j}(t+0,\omega)\Big)^{n} \phi_K(\boldsymbol{\theta} ;\boldsymbol{0}, \mathbf{I}_K) \mathrm{d} \boldsymbol{\theta}\notag\\
        =&\int \Big[\prod_{j=1}^{J} \prod_{s \leq t} \widetilde{\lambda}_{j}(s,\omega)^{\Delta N_{j}(s,\omega)}\Big]\exp\Big(-\sum_{j=1}^J \int_0^t \widetilde{\lambda}_j(s,\omega)\mathrm{d}s\Big)\Big(\sum_{j=1}^J\widetilde{\lambda}_{j}(t+0,\omega)\Big)^{n} \phi_K(\boldsymbol{\theta} ;\boldsymbol{0},\mathbf{I}_K) \mathrm{d} \boldsymbol{\theta}.
    \end{align}
    Note that for any $t<\tau$ where the counting processes and covariate processes have no jump and any $\Delta t>0$, we can always find $\omega_m\in\mathcal{A}$ for any $m=1,\ldots,J$ such that the counting processes and covariate processes coincide with those on $\omega$ up to time $t$, while $\omega_m$ has exactly one additional event of type $m$ occurring in time interval $t+\Delta t$. By letting $\Delta t\downarrow 0$ and performing repeated differentiations with respect to $t$, equation (\ref{sketch_1.1}) leads to 
    \begin{align}\label{sketch_1.3}
        &\int \lambda_{m}(t+0,\omega)\Big[\prod_{j=1}^{J} \prod_{s \leq t} \lambda_{j}(s,\omega)^{\Delta N_{j}(s,\omega)}\Big]\exp\Big(-\sum_{j=1}^J \int_0^t \lambda_j(s,\omega)\mathrm{d}s\Big)\notag\\
        &\times\Big(\sum_{j=1}^J\lambda_{j}(t+0,\omega)\Big)^{n} \phi_K(\boldsymbol{\theta} ;\boldsymbol{0}, \mathbf{I}_K) \mathrm{d} \boldsymbol{\theta}\notag\\
        =&\int \widetilde{\lambda}_{m}(t+0,\omega)\Big[\prod_{j=1}^{J} \prod_{s \leq t} \widetilde{\lambda}_{j}(s,\omega)^{\Delta N_{j}(s,\omega)}\Big]\exp\Big(-\sum_{j=1}^J \int_0^t \widetilde{\lambda}_j(s,\omega)\mathrm{d}s\Big)\notag\\
        &\times\Big(\sum_{j=1}^J\widetilde{\lambda}_{j}(t+0,\omega)\Big)^{n} \phi_K(\boldsymbol{\theta} ;\boldsymbol{0},\mathbf{I}_K) \mathrm{d} \boldsymbol{\theta}
    \end{align}
    for any $t$, $m=1,\ldots,J$ and $n$. Here we replace $\lambda_j(t,\omega_m)$ by $\lambda_j(t,\omega)$ ($j=1,\ldots,J$) since they are assumed to be identical before time $t$. 
    
    We need to point out that (\ref{sketch_1.3}) is used solely to align the event types under the two competing models. When no events occur, the types of events cannot be distinguished through (\ref{sketch_1.2}) alone. However, the intensity functions can already be matched up to a permutation of event types using (\ref{sketch_1.2}) alone.\\[3mm]
    \textbf{Step 2: }We then prove that for any $j,j_1,j_2=1,\ldots,J$ and any $t,s$, it holds that
    \begin{align}\label{sketch_1.4}
        \beta_{j0}+\boldsymbol{\beta}_j^{\top}\boldsymbol{X}_j(t,\omega)&=\widetilde{\beta}_{j0}+\widetilde{\boldsymbol{\beta}}_j^{\top}\boldsymbol{X}_j(t,\omega),\notag\\
        \boldsymbol{Z}_{j_1}^{\top}(t,\omega)\mathbf{A}_{j_1}\boldsymbol{\Sigma} \mathbf{A}^{\top}_{j_2}\boldsymbol{Z}_{j_2}(s,\omega)&=\boldsymbol{Z}_{j_1}^{\top}(t,\omega)\widetilde{\mathbf{A}}_{j_1}\widetilde{\boldsymbol{\Sigma}} \widetilde{\mathbf{A}}^{\top}_{j_2}\boldsymbol{Z}_{j_2}(s,\omega).
    \end{align}
    Guaranteed by the first part of Condition (e), we can partition the observation period into small intervals: $[0,t_1],(t_1,t_2],\ldots$ such that $\boldsymbol{X}(\cdot)$ and $\boldsymbol{Z}(\cdot)$ remain constant on each interval. We then prove (\ref{sketch_1.4}) through induction by analysing (\ref{sketch_1.2}) and (\ref{sketch_1.3}). Suppose that (\ref{sketch_1.4}) is already verified on $[0,t_k]$. We then let $t=t_k$ in equations (\ref{sketch_1.2}) and (\ref{sketch_1.3}). For simplicity, we ignore the rotation issues (due to the multivariate normal distribution) in random effect $\boldsymbol{\theta}$. Then the induction assumption (\ref{sketch_1.4}) is equivalent to saying that $\lambda_{j}(t,\omega)$ is matched with $\widetilde{\lambda}_{j}(t,\omega)$ on $[0,t_k]$. We then match $\lambda_{j}(t_k+0,\omega)$ with $\widetilde{\lambda}_{j}(t_k+0,\omega)$ up to a permutation among $\{1,\ldots,J\}$ by analysing the asymptotic behaviours in (\ref{sketch_1.2}) as $n$ goes to infinity and fix this permutation by using (\ref{sketch_1.3}).

    Several challenging issues are addressed in Section~\ref{S33}, which details a full proof. In particular, the integrals in (\ref{sketch_1.2}) and (\ref{sketch_1.3}) do not have explicit forms. A Laplace-type argument in Proposition~\ref{prop_laplace} provides a way to obtain the asymptotic behaviours of the integrals as $n$ goes to infinity. Furthermore, the sample path of $\{\boldsymbol{Z}_j(\cdot):j=1,\ldots,J\}$ falls into two regimes that control the asymptotic orders on both sides of (\ref{sketch_1.2}) or (\ref{sketch_1.3}). The two regimes are dealt with in Proposition~\ref{prop_canonical_projection} and Proposition~\ref{prop_characterisation_equation}, respectively. Lastly, both sides of (\ref{sketch_1.2}) or (\ref{sketch_1.3}) contain $J^n$ terms due to multiple event types. 
    Proposition~\ref{prop_canonical_projection} and Proposition~\ref{prop_characterisation_equation} characterise the concentration points in the two regimes, respectively. We show that the summation can be approximated by a partial summation within a small neighbourhood of the concentration point, while the remaining terms are negligible and do not affect the analysis of the dominant terms. Arguments similar to those used in the proof of Proposition~\ref{lem_summation} are then applied to match all terms in both summations inductively, leading to equation (\ref{sketch_1.4}) on $[0,t_{k+1}]$.\\[3mm]
    \textbf{Step 3: }Guaranteed by Conditions (c) and (d), we integrate the equations (\ref{sketch_1.4}) for all $\omega\in\mathcal{A}$ to prove that $\boldsymbol{\beta}=\widetilde{\boldsymbol{\beta}}$ and $(\mathbf{A}, \boldsymbol{\Sigma}) \sim(\widetilde{\mathbf{A}}, \widetilde{\boldsymbol{\Sigma}})$, thus verifying Theorem~1.
    
\subsection{Preliminary results}
We first state some preliminary results to be used in the proof of Theorem \ref{thm_identifiability}. The proofs of these results are deferred to subsequent sections.

The following proposition provides the foundation for the identifiability argument. A key difficulty in proving Theorem \ref{thm_identifiability} is that different event types are mixed together in the likelihood function. The proof in Section~\ref{S21} separates event types by introducing different events at specific time points. However, this approach is not feasible in our proposed model with internal covariates, since introducing events may alter the covariate values. Instead, the piecewise-constant assumption in Condition (e) enables repeated differentiations in a tractable way. The high-level idea behind Theorem \ref{thm_identifiability} is therefore to distinguish event types through the asymptotic behaviour of high-order derivatives.
\begin{prop}\label{prop_likelihood}
    Under model (1) and Conditions (a)-(f), for given $\vec{\para}=(\vec{\beta},\mat{A},\mat{\Sigma})$ and $\widetilde{\vec{\para}}=(\widetilde{\vec{\beta}},\widetilde{\mat{A}},\widetilde{\mat{\Sigma}})$, denote their corresponding intensity functions by $\lambda_j(\cdot)$ and $\widetilde{\lambda}_j(\cdot),j=1,\ldots,J$. Suppose that the model with intensity functions $\lambda_j(\cdot),j=1,\ldots,J$ and the model with intensity functions $\widetilde{\lambda}_j(\cdot),j=1,\ldots,J$ induce the same probability measure. Then the following equation holds for any $n\in \mathbb{N}_0$ and any $0<t<\tau$, with probability 1:
    \begin{align*}
        &\int_{\bt} \Big[\prod_{j=1}^{J} \prod_{s \leq t} \lambda_{j}(s)^{\Delta N_{j}(s)}\Big]\exp\Big(-\sum_{j=1}^J \int_0^t \lambda_j(s)\mathrm{d}s\Big)\Big(\sum_{j=1}^J\lambda_{j}(t+0)\Big)^{n} \phi_K(\vec{\theta};\vec{0}, \mat{\Sigma}) \mathrm{d}\vec{\theta}\notag\\
        =&\int_{\bt} \Big[\prod_{j=1}^{J} \prod_{s \leq t} \widetilde{\lambda}_{j}(s)^{\Delta N_{j}(s)}\Big]\exp\Big(-\sum_{j=1}^J \int_0^t \widetilde{\lambda}_j(s)\mathrm{d}s\Big)\Big(\sum_{j=1}^J\widetilde{\lambda}_{j}(t+0)\Big)^{n} \phi_K(\vec{\theta} ;\vec{0}, \widetilde{\mat{\Sigma}}) \mathrm{d}\vec{\theta}\text{.}
    \end{align*}
\end{prop}

The following corollary follows from arguments similar to those in Proposition~\ref{prop_likelihood}. When no event occurs, the event types appearing in Proposition~\ref{prop_likelihood} are symmetric under permutations on both sides, implying that the intensity functions can only be matched up to a permutation of event types $1,\ldots,J$. By introducing a specific event at the endpoint, which does not change the covariate values prior to that time, the following corollary allows us to resolve this permutation ambiguity.
\begin{coro}\label{cor_likelihood}
    We consider the same setting as in Proposition~\ref{prop_likelihood}. Then for any $0<t<\tau$, any $m=1,\ldots,J$ and $n\in\mathbb{N}_0$, the following equation holds with probability 1:
    \begin{align*}
        &\int_{\bt} \lambda_m(t+0)\Big[\prod_{j=1}^{J} \prod_{s \leq t} \lambda_{j}(s)^{\Delta N_{j}(s)}\Big]\exp\Big(-\sum_{j=1}^J \int_0^t \lambda_j(s)\mathrm{d}s\Big)\Big(\sum_{j=1}^J\lambda_{j}(t+0)\Big)^{n} \phi_K(\vec{\theta};\vec{0}, \mat{\Sigma}) \mathrm{d}\vec{\theta}\notag\\
        =&\int_{\bt} \widetilde{\lambda}_m(t+0)\Big[\prod_{j=1}^{J} \prod_{s \leq t} \widetilde{\lambda}_{j}(s)^{\Delta N_{j}(s)}\Big]\exp\Big(-\sum_{j=1}^J \int_0^t \widetilde{\lambda}_j(s)\mathrm{d}s\Big)\Big(\sum_{j=1}^J\widetilde{\lambda}_{j}(t+0)\Big)^{n}\phi_K(\vec{\theta} ;\vec{0}, \widetilde{\mat{\Sigma}}) \mathrm{d}\vec{\theta}\text{.}
    \end{align*}
\end{coro}

Let \(t\downarrow 0\) in Proposition~\ref{prop_likelihood}, the remaining terms on both sides reduce to
\((\sum_{j=1}^J \lambda_j(t+0))^{n}\) and
\((\sum_{j=1}^J \widetilde{\lambda}_j(t+0))^{n}\), which expand into sums of \(J^n\) log-normal moments.
The following proposition shows that, by analysing the asymptotic behaviour of these two summations as \(n \to \infty\), the intensity functions can be matched up to a permutation of the event types.
The proof also highlights the high-level intuition for extending this matching argument when the induction proceeds to larger values of \(t\).
\begin{prop}\label{lem_summation}
    Let $J$ be a given positive integer. For any $1\leq i,j\leq J$, let $x_i,\widetilde{x}_{i},y_{ij},\widetilde{y}_{ij}\in \mathbb{R}^{+}$. Suppose that for any $1\leq i,j\leq J$, it holds that
        \begin{align}\label{eq_lem_5.1}
            &y_{ij}^2=y_{ji}^2\leq y_{ii}y_{jj}\text{,}\notag\\
            &\widetilde{y}_{ij}^2=\widetilde{y}_{ji}^2\leq \widetilde{y}_{ii}\widetilde{y}_{jj}\text{.}
        \end{align}
        Furthermore, suppose that the values $\left\{y_{ij}:
        1\leq i\leq j\leq J\right\}$ are distinct. Assume that the following equation holds for every $n\in\mathbb{N}$:
        \begin{align}\label{eq_lem_5.2}
            \sum_{1\leq j_1,\ldots,j_n\leq J}\left(\prod_{k=1}^{n}x_{j_k}\prod_{1\leq k_1,k_2\leq n}y_{j_{k_1}j_{k_2}}\right)=\sum_{1\leq j_1,\ldots,j_n\leq J}\left(\prod_{k=1}^{n}\widetilde{x}_{j_k}\prod_{1\leq k_1,k_2\leq n}\widetilde{y}_{j_{k_1}j_{k_2}}\right)\text{.}
        \end{align}
        Then there exists a permutation $\pi:\{1,\ldots,J\}\rightarrow \{1,\ldots,J\}$ such that for any $j,j_1,j_2=1,\ldots,J$, $x_{j}=\widetilde{x}_{\pi(j)}$ and $y_{j_1j_2}=\widetilde{y}_{\pi(j_1)\pi(j_2)}$.
\end{prop}

Another main challenge in proving Theorem~1 is that the likelihood function takes the form of an integral without an explicit closed form. We address this issue by applying a Laplace-type approximation to the integral. The result is formally stated in the following proposition.

\begin{prop}\label{prop_laplace}
    Let $\vec{\alpha}_1,\ldots,\vec{\alpha}_K\in\mathbb{R}^d$ be $d$-vectors and $w_1,\ldots,w_K$ be positive constants. For a given $\vec{\eta}\in\mathbb{R}^d$, let $f(\vec{\theta})=-\sum_{k=1}^Kw_k\exp(\vec{\alpha}_k^{\top}\vec{\theta})+\vec{\eta}^{\top}\vec{\theta}-\frac{1}{2}\vec{\theta}^{\top}\vec{\theta}$ and denote by $\hat{\vec{\theta}}$ its unique maximiser. Denote the negative Hessian matrix of function $f$ by $\mat{I}(\vec{\theta})=\mat{I}_d+\sum_{k=1}^Kw_k\exp(\vec{\alpha}_k^{\top}\vec{\theta})\vec{\alpha}_k\vec{\alpha}_k^{\top}$. Then it holds that
    \begin{align*}
        M^{-1}\cdot\frac{\exp(f(\hat{\vec{\theta}}))}{\sqrt{\operatorname{det}(\mat{I}(\hat{\vec{\theta}}))}}\leq \int_{\bt}(2\pi)^{-d/2}\exp(f(\vec{\theta}))\mathrm{d}\vec{\theta}\leq M\cdot\frac{\exp(f(\hat{\vec{\theta}}))}{\sqrt{\operatorname{det}(\mat{I}(\hat{\vec{\theta}}))}},
    \end{align*}
    where $M>0$ is a constant that does not depend on the choice of $\vec{\eta}$.
\end{prop}

Given Proposition~\ref{prop_laplace}, our main goal is to characterise the maximum value of the function $f_n$. The piecewise-constant assumption in Condition (e) ensures that $f_n$ takes the form
\[
f_n(\vec{\theta})
=
-\sum_{k=1}^K w_k \exp(\vec{\alpha}_k^{\top}\vec{\theta})
+ \vec{\eta}_n^{\top}\vec{\theta}
- \frac{1}{2}\vec{\theta}^{\top}\vec{\theta},
\]
where the first term arises from the exponential tail, the second term corresponds to the observed events, and the final term comes from the normal frailty component.  

The following proposition characterises the dominating term in the maximum value of $f_n$, based on the concept of canonical projection. We introduce the notion of canonical projection in part (1) and prove its existence and continuity. In part (2), we characterise the asymptotic behaviour of $f_n$.

\begin{prop}[Canonical Projection]\label{prop_canonical_projection}
    Let $\vec{\alpha}_1,\ldots,\vec{\alpha}_K\in\mathbb{R}^d\setminus{\{\vec{0}\}}$, i.e.,  nonzero $d$-vectors, and $P$ be the projection operator. We have the following results:
    \begin{enumerate}
    \item[(1)] For any fixed $\vec{\eta}\in \mathbb{R}^d\setminus{\{\vec{0}\}}$, there exists a (possibly empty) subset $\left\{\vec{\alpha}_{k_1},\ldots,\vec{\alpha}_{k_m}\right\}\subseteq \left\{\vec{\alpha}_1,\ldots,\vec{\alpha}_K\right\}$ and $\mathcal{H}_{\vec{\eta}}=\operatorname{span}\left\{\vec{\alpha}_{k_1},\ldots,\vec{\alpha}_{k_m}\right\}$ such that
    \begin{enumerate}
        \item[(i)] $P_{\mathcal{H}_{\vec{\eta}}}\vec{\eta}=\sum_{j=1}^m c_{k_j}\vec{\alpha}_{k_j}$ for some $c_{k_1},\ldots,c
        _{k_m}\geq 0$.
        \item[(ii)] $\vec{\alpha}^{\top}_{k}P_{\mathcal{H}_{\vec{\eta}}^{\perp}}\vec{\eta}<0$ for any $k\in\left\{1,\ldots,K\right\}\setminus \left\{k_1,\ldots,k_m\right\}$.
        \item[(iii)] $\mathcal{H}_{\vec{\eta}}$ in (i) is uniquely defined, and so is $P_{\mathcal{H}_{\vec{\eta}}^{\perp}}\vec{\eta}$; this mapping is continuous with respect to $\vec{\eta}$. We shall call $P_{\mathcal{H}_{\vec{\eta}}^{\perp}}\vec{\eta}$ the canonical projection of $\vec{\eta}$ with respect to $\{\vec{\alpha}_1,\ldots,\vec{\alpha}_K\}$.
    \end{enumerate}
    \item[(2)] Let $w_1,\ldots,w_K$ be positive constants and $\vec{\eta}_n\in\mathbb{R}^d$ be $d$-vectors such that $\lim_{n\rightarrow \infty}\vec{\eta}_n/{n}=\vec{\eta}\in\mathbb{R}^d\setminus{\{0\}}$. Define $f_n(\vec{\theta})=-\sum_{k=1}^Kw_k\exp(\vec{\alpha}_k^{\top}\vec{\theta})+\vec{\eta}_n^{\top}\vec{\theta}-\frac{1}{2}\vec{\theta}^{\top}\vec{\theta}$ and denote by $\vec{\theta}_n$ its unique maximiser. Then we have
    \begin{align*}
        \lim_{n\rightarrow \infty}\frac{\vec{\theta}_n}{n}=P_{\mathcal{H}_{\vec{\eta}}^{\perp}}\vec{\eta}\quad\text{and}\quad\lim_{n\rightarrow \infty}\frac{f_n(\vec{\theta}_n)}{n^2}=\frac{1}{2}\|P_{\mathcal{H}^{\perp}_{\vec{\eta}}}\vec{\eta}\|^2.
    \end{align*}
\end{enumerate}
\end{prop}

The summations on both sides of Proposition~\ref{prop_likelihood} contain $J^n$ terms, where $\vec{\eta}_n$ is a linear combination of $n$ terms chosen from $\vec{\eta}_1,\ldots,\vec{\eta}_J$. Proposition~\ref{prop_likelihood} shows that, rather than focusing on discrete linear combinations, we may instead consider the probability simplex
\[
\left\{(\nu_1,\ldots,\nu_J)^{\top}: \sum_{j=1}^J \nu_j = 1,\nu_1,\ldots,\nu_J\geq 0\right\},
\]
which represents the asymptotic proportions of event types $1,\ldots,J$. By the continuity property of the canonical projection, a maximiser exists on this probability simplex. The following proposition further shows that this maximiser is unique and must lie at a corner of the simplex, that is, $\nu_j = 1$ for one $j$, with all remaining components equal to zero.
\begin{prop}\label{prop_max_projection}
    Let $\vec{\alpha}_1,\ldots,\vec{\alpha}_K\in\mathbb{R}^d\setminus{\{0\}}$ and $\vec{\eta}_1,\ldots,\vec{\eta}_J\in \mathbb{R}^d$ be $d$-vectors. Let $\mathcal{G}=\{\vec{\eta}=\sum_{j=1}^J\nu_j\vec{\eta}_j:0\leq \nu_j\leq 1,\sum_{j=1}^J\nu_j=1\}$. Suppose that $\|P_{\mathcal{H}^{\perp}_{\vec{\eta}_1}}\vec{\eta}_1\|>\max_{j=2,\ldots,J}\|P_{\mathcal{H}^{\perp}_{\vec{\eta}_j}}\vec{\eta}_j\|$, where $P_{\mathcal{H}^{\perp}_{\vec{\eta}}}\vec{\eta}$ is the canonical projection of $\vec{\eta}$ with respect to $\{\vec{\alpha}_1,\ldots,\vec{\alpha}_K\}$, uniquely defined in Proposition~\ref{prop_canonical_projection}. Then $\vec{\eta}_1$ is the unique maximiser of $\|P_{\mathcal{H}^{\perp}_{\vec{\eta}}}\vec{\eta}\|$ in $\mathcal{G}$.
\end{prop}

Proposition~\ref{prop_max_projection} guarantees the uniqueness of the maximiser on the probability simplex and shows that it lies at one of the corners. The following corollary shows that the gradient at this maximiser is strictly nonzero.

\begin{coro}\label{cor_max_projection}
    Under the setting of Proposition~\ref{prop_max_projection}, it holds that $(\vec{\eta}_1-\vec{\eta}_j)^{\top}P_{\mathcal{H}^{\perp}_{\vec{\eta}_1}}\vec{\eta}_1>0$ for any $j=2,\ldots,J$.
\end{coro}

In certain configurations, the canonical projections may all vanish on the probability simplex, which occurs when each of $\vec{\eta}_1,\ldots,\vec{\eta}_J$ can be expressed as a linear combination of $\vec{\alpha}_1,\ldots,\vec{\alpha}_K$ with nonnegative coefficients. To handle this degenerate case, we introduce the notion of canonical expansion in the following proposition, which is used to characterise the asymptotic behaviour of $f_n$ under this scenario.

\begin{prop}[Canonical Expansion]\label{lem_canonical_expansion}
    Let $\vec{\alpha}_1,\ldots,\vec{\alpha}_K\in\mathbb{R}^d$ be $d$-vectors and $c_1,\ldots,c_K$ be nonnegative constants. Let $\vec{\eta}=\sum_{k=1}^K c_k\vec{\alpha}_k$. Then there exists an expansion $\vec{\eta}=\sum_{p=1}^m \widetilde{c}_{k_p}\vec{\alpha}_{k_p}$, where $\left\{\vec{\alpha}_{k_1},\ldots,\vec{\alpha}_{k_m}\right\}\subseteq\left\{\vec{\alpha}_1,\ldots,\vec{\alpha}_K\right\}$ and $\widetilde{c}_{k_1},\ldots,\widetilde{c}_{k_m}>0$, such that there exists nonzero $\vec{\epsilon}\in\mathbb{R}^d$ satisfying
    \begin{align*}
        \vec{\alpha}_{k_1}^{\top}\vec{\epsilon}=\ldots=\vec{\alpha}_{k_m}^{\top}\vec{\epsilon}=\underset{k=1,\ldots,K}{\max}\vec{\alpha}_k^{\top}\vec{\epsilon}>0.
    \end{align*}
    We call it a canonical expansion of $\vec{\eta}$ with respect to $\{\vec{\alpha}_1,\ldots,\vec{\alpha}_K\}$. Furthermore, if there exist two canonical expansions of $\vec{\eta}$ as $\vec{\eta}=\sum_{p=1}^{m_1} c_{k_p}\vec{\alpha}_{k_p}=\sum_{p=1}^{m_2} \widetilde{c}_{l_p}\vec{\alpha}_{l_p}$, then $\sum_{p=1}^{m_1} c_{k_p}=\sum_{p=1}^{m_2} \widetilde{c}_{l_p}$.
\end{prop}

The following proposition addresses a specific degenerate setting in the proof of Theorem~1 concerning the asymptotic behaviour at the maximiser when the number of certain indices in $\{1,\ldots,J\}$ is of order $\Theta(\log n)$.

\begin{prop}\label{lem_characterisation_log}
    Let $\vec{\alpha}_1,\ldots,\vec{\alpha}_K,\vec{\eta}_1,\ldots,\vec{\eta}_J,\vec{\varphi}_1,\ldots,\vec{\varphi}_{m}\in\mathbb{R}^d\setminus\{0\}$ be $d$-vectors, $1= \nu_1>\nu_2\geq\ldots\geq\nu_J>0$ and $\hat{c}_1,\ldots,\hat{c}_m,w_1,\ldots,w_K>0$ be constants. Suppose there exists a vector $\hat{\vec{\theta}}\in\mathbb{R}^d$ and a disjoint partition of the set $\{\vec{\alpha}_1,\ldots,\vec{\alpha}_K\}=\mat{U}_1\cup\ldots\cup \mat{U}_J\cup \mat{V}_{0} \cup \mat{V}_{-}$ such that:
    \begin{enumerate}
        \item[(i)] $\mat{U}_1\cup\ldots\cup \mat{U}_J\cup \mat{V}_{0}$ is linearly independent.
        \item[(ii)] For $j=1,\ldots,J$, $\vec{\eta}_j=\sum_{\vec{\alpha}_k\in \mat{U}_j}d_k\vec{\alpha}_k$ for some positive constants $d_k$. Moreover, $\vec{\alpha}_k^{\top}\hat{\vec{\theta}}=\nu_j$ for any $\vec{\alpha}_k\in \mat{U}_j$.
        \item[(iii)] $\hat{\vec{\theta}}-\sum_{j=1}^m\hat{c}_{j}\vec{\varphi}_j\in \operatorname{span}\left(\mat{U}_1\cup\ldots\cup \mat{U}_J\cup \mat{V}_{0}\right):= \mathcal{H}$. For any $\vec{\alpha}_k\in \mat{V}_0$, it holds that $\vec{\alpha}_k^{\top}\hat{\vec{\theta}}=0$. Moreover, the coefficient of $\vec{\alpha}_k$ in the expansion of $\hat{\vec{\theta}}-\sum_{j=1}^m\hat{c}_{j}\vec{\varphi}_j$ with respect to the basis $\mat{U}_1\cup\ldots\cup \mat{U}_J\cup \mat{V}_{0}$ is negative.
        \item[(iv)] $\vec{\alpha}_k^{\top}\hat{\vec{\theta}}<0$ for any $\vec{\alpha}_k\in \mat{V}_-$.
    \end{enumerate}
    For any $\vec{c}\in \mathbb{R}^m$ in a small neighbourhood of $\hat{\vec{c}}=(\hat{c}_1,\ldots,\hat{c}_m)^{\top}$, let $\bm{\xi}^{(n)}=(\xi_1^{(n)},\ldots,\xi_{J}^{(n)})^{\top}$ be a $J$-vector sequence and ${\bm{\zeta}}^{(n,\vec{c})}=({\zeta}_1^{(n,\vec{c})},\ldots,{\zeta}_m^{(n,\vec{c})})^{\top}$ be a $m$-vector sequence such that
    \begin{align*}
        \lim_{n\rightarrow \infty}\frac{\log \vec{\xi}^{(n)}}{\log n}=(\nu_1,\ldots,\nu_J)^{\top}\quad\text{and}\quad\lim_{n\rightarrow \infty}\frac{\vec{\zeta}^{(n,\vec{c})}}{\log n}=\vec{c}.
    \end{align*}
    Define
    \begin{align*}
        f_n(\vec{\theta}|\bm{\xi}^{(n)},{\bm{\zeta}}^{(n,\vec{c})})=-\sum_{k=1}^Kw_k\exp(\vec{\alpha}_k^{\top}\vec{\theta})+\left(\sum_{j=1}^J\xi_j^{(n)}\vec{\eta}_j+\sum_{j=1}^m\zeta_j^{(n,\vec{c})}\vec{\varphi}_j\right)^{\top}\vec{\theta}-\frac{1}{2}\vec{\theta}^{\top}\vec{\theta}
    \end{align*}
    and denote by $\vec{\theta}_n(\bm{\xi}^{(n)},{\bm{\zeta}}^{(n,\vec{c})})$ its unique maximiser. Then we have
    \begin{align*}
        \vec{\theta}_n(\bm{\xi}^{(n)},{\bm{\zeta}}^{(n,\vec{c})})=&\log n\left(\widetilde{\vec{\theta}}+\sum_{j=1}^m{c}_{j}P_{\mathcal{H}^{\perp}}\vec{\varphi}_j+o(1)\right),\\
        f_n(\vec{\theta}_n(\bm{\xi}^{(n)},{\bm{\zeta}}^{(n,\vec{c})})|\bm{\xi}^{(n)},{\bm{\zeta}}^{(n,\vec{c})})=&e_{n,1}+\log^2 n\left(\vec{c}^{\top}\vec{e}_{2}+\frac{1}{2}\left\|\sum_{j=1}^m{c}_{j}P_{\mathcal{H}^{\perp}}\vec{\varphi}_j\right\|^2+o(1)\right),
    \end{align*}
    where $\widetilde{\vec{\theta}}\in\mathbb{R}^d$, $e_{n,1}\in\mathbb{R}$ and $\vec{e}_2\in\mathbb{R}^m$ do not depend on $\vec{c}$.
\end{prop}

Proposition \ref{prop_characterisation_equation} presents the main result for the degenerate case in which each of $\vec{\eta}_1,\ldots,\vec{\eta}_J$ can be expressed as a linear combination of $\vec{\alpha}_1,\ldots,\vec{\alpha}_K$ with nonnegative coefficients. In contrast to the nondegenerate case, where the concentration point lies at a corner of the probability simplex, the concentration point can be an interior point, which is determined by a so-called characterisation equation.
The characterisation equation is formally defined as follows.

\begin{definition}[Characterisation Equation]
    Let $\vec{\alpha}_1,\ldots,\vec{\alpha}_K\in\mathbb{R}^d\setminus\{0\}$ be $d$-vectors.
For $k\in\left\{1,\ldots,d\right\}$, we call a set of $k$ linear equations a characterisation equation at $\left(\nu_1,\ldots,\nu_J\right)$ where $\nu_1>\ldots>\nu_J>0$ if the $l$-th $(l=1,\ldots,k)$ equation is of one of the following two types:
            \begin{itemize}
                \item Type-1 equation: $\vec{\alpha}_{j_l}^{\top}\vec{\theta}=\xi_l$, where $\xi_l\in\left\{\nu_1,\ldots,\nu_J,0\right\}$.
                \item Type-2 equation: $\vec{\alpha}_{j_l}^{\top}\vec{\theta}=\vec{\zeta}_l^{\top}\vec{\theta}$, where $\vec{\zeta}_l\in\left\{\vec{\alpha}_1,\ldots,\vec{\alpha}_K\right\}\setminus\left\{\vec{\alpha}_{j_1},\ldots,\vec{\alpha}_{j_k}\right\}$.
            \end{itemize}
            Here $\left\{\vec{\alpha}_{j_1},\ldots,\vec{\alpha}_{j_k}\right\}\subseteq\left\{\vec{\alpha}_1,\ldots,\vec{\alpha}_K\right\}$ is a set of $k$ vectors. Furthermore, the characterisation equation is required to satisfy the following conditions:
            \begin{enumerate}
                \item[(i)] $\vec{\alpha}_{j_1},\ldots,\vec{\alpha}_{j_k}$ are linearly independent.
                \item[(ii)] The equation has unique solution $\hat{\vec{\theta}}$ under the constraint: $\vec{\theta}\in \operatorname{span}\left\{\vec{\alpha}_{j_1},\ldots,\vec{\alpha}_{j_k}\right\}$.
                \item[(iii)] For $j=1,\ldots,J$, there exists a unique $(c_{j,1},\ldots,c_{j,k})^{\top}\in\mathbb{R}^k$ such that $\vec{\eta}_j$ has expansion: $\vec{\eta}_j=\sum_{l=1}^kc_{j,l}\vec{\alpha}_{j_l}$. Suppose $\vec{\alpha}_{j_{l_1}},\ldots,\vec{\alpha}_{j_{l_m}}$ are all elements in $\left\{\vec{\alpha}_{j_1},\ldots,\vec{\alpha}_{j_k}\right\}$ such that $\vec{\alpha}_{j_{l_1}}^{\top}\hat{\vec{\theta}}=\ldots=\vec{\alpha}_{j_{l_m}}^{\top}\hat{\vec{\theta}}=\nu_j$, then it holds that $c_{j,l_1},\ldots,c_{j,l_m}>0$. For any $l\in\left\{1,\ldots,k\right\}$ such that $\vec{\alpha}_{j_l}^{\top}\hat{\vec{\theta}}<\nu_j$, it holds that $c_{j,l}=0$.
                \item[(iv)] There exists a unique $(c_{1},\ldots,c_{k})^{\top}\in\mathbb{R}^k$ such that $\hat{\vec{\theta}}$ has expansion: $\hat{\vec{\theta}}=\sum_{l=1}^kc_{l}\vec{\alpha}_{j_l}$. Suppose $l_1,\ldots,l_m$ are all indices among $\left\{j_1,\ldots,j_k\right\}$ such that $\vec{\alpha}_{l_1}^{\top}\hat{\vec{\theta}}=\ldots=\vec{\alpha}_{l_m}^{\top}\hat{\vec{\theta}}=0$, then it holds that $c_{l_1},\ldots,c_{l_m}< 0$.
                \item[(v)] For any $\vec{\alpha}\in\left\{\vec{\alpha}_{j_1},\ldots,\vec{\alpha}_{j_k}\right\}$, either $\vec{\alpha}^{\top}\hat{\vec{\theta}}\in\left\{\nu_1,\ldots,\nu_J,0\right\}$ or there exists a unique $\vec{\beta}\in\left\{\vec{\alpha}_1,\ldots,\vec{\alpha}_K\right\}\setminus\left\{\vec{\alpha}_{j_1},\ldots,\vec{\alpha}_{j_k}\right\}$ such that $\vec{\alpha}^{\top}\hat{\vec{\theta}}=\vec{\beta}^{\top}\hat{\vec{\theta}}\notin\left\{\nu_1,\ldots,\nu_J,0\right\}$ and $0<\vec{\alpha}^{\top}\hat{\vec{\theta}}<\nu_1$. In the second case, there exists a unique $\left(c_{1},\ldots,c_{k}\right)^{\top}\in\mathbb{R}^k$ such that $\vec{\beta}$ has expansion: $\vec{\beta}=\sum_{l=1}^kc_{l}\vec{\alpha}_{j_l}$. For any $l\in\left\{1,\ldots,k\right\}$ such that $\vec{\alpha}_{j_l}^{\top}\hat{\vec{\theta}}=\vec{\beta}^{\top}\hat{\vec{\theta}}$, it holds that $c_{l}<0$. For any $l\in\left\{1,\ldots,k\right\}$ such that $\vec{\alpha}_{j_l}^{\top}\hat{\vec{\theta}}<\vec{\beta}^{\top}\hat{\vec{\theta}}$, it holds that $c_{l}=0$.
                \item[(vi)] For any $\vec{\beta}\in\left\{\vec{\alpha}_1,\ldots,\vec{\alpha}_K\right\}\setminus\left\{\vec{\alpha}_{j_1},\ldots,\vec{\alpha}_{j_k}\right\}$, either $\vec{\beta}^{\top}\hat{\vec{\theta}}<\min_{l=1,\ldots,k}\vec{\alpha}_{j_l}^{\top}\hat{\vec{\theta}}$ or there exists $\vec{\alpha}\in\left\{\vec{\alpha}_{j_1},\ldots,\vec{\alpha}_{j_k}\right\}$ such that $\vec{\alpha}^{\top}\hat{\vec{\theta}}=\vec{\beta}^{\top}\hat{\vec{\theta}}\notin\left\{\nu_1,\ldots,\nu_J,0\right\}$.
\end{enumerate}
\end{definition}
This characterisation equation plays a central role in matching the intensity functions in this setting.
Note that, in the nondegenerate case, the counts of indices $1,\ldots,J$ scale as $\xi_1/n,\ldots,\xi_J/n$, whereas in the degenerate case they scale as $\log \xi_1/\log n,\ldots,\log \xi_J/\log n$.
Throughout the proof, we use the convention $\log 0=0$ for notational simplicity.
The following proposition shows that the asymptotic behaviour can be characterised through this scaling and the characterisation equation.

\begin{prop}[Characterisation Equation]\label{prop_characterisation_equation}
    Let $\vec{\alpha}_1,\ldots,\vec{\alpha}_K,\vec{\eta}_1,\ldots,\vec{\eta}_J\in\mathbb{R}^d\setminus\{0\}$ be $d$-vectors, $w_1,\ldots,w_K$ and $\widetilde{\nu}_1\geq\ldots\geq \widetilde{\nu}_J>0$ be positive constants. Suppose $\vec{\eta}_1,\ldots,\vec{\eta}_J\in\{\sum_{k=1}^K c_k\vec{\alpha}_k:c_1,\ldots,c_K\geq 0\}$. Further suppose there exists a sequence $\vec{\nu}^{(m)}\rightarrow \widetilde{\vec{\nu}}$ such that
    \begin{enumerate}
        \item[(i)] For any $m\in\mathbb{N}$, it holds that $\nu_1^{(m)}>\ldots>\nu_J^{(m)}>0$.
        \item[(ii)] For any $m\in\mathbb{N}$, there exists a characterisation equation at $\vec{\nu}^{(m)}$ such that for any $1\leq i<j\leq J$ satisfying $\widetilde{\nu}_i=\widetilde{\nu}_j$, the expansions of $\vec{\eta}_i$ and $\vec{\eta}_j$ with respect to the basis of the characterisation equation contain disjoint terms.
    \end{enumerate}
    Then we can define a continuous function $\vec{\theta}(\vec{\nu})$ in a neighbourhood $\mathcal{O}$ of $\widetilde{\vec{\nu}}$ such that for any $\vec{\nu}\in\mathcal{O}$, any $\vec{\xi}^{(n)}$ satisfying
    \begin{align*}
        \lim_{n\rightarrow \infty}\frac{\log \vec{\xi}^{(n)}}{\log n}=\vec{\nu}
    \end{align*}
    and any uniformly bounded vector sequence $\{\vec{\varphi}^{(n)}\in \mathbb{R}^d:n=1,\ldots\}$, the unique maximiser $\vec{\theta}_n$ of the following function:
    \begin{align*}
        f_n(\vec{\theta})=-\sum_{k=1}^Kw_k\exp(\vec{\alpha}_k^{\top}\vec{\theta})+\left(\sum_{j=1}^J\xi_j^{(n)}\vec{\eta}_j+\vec{\varphi}^{(n)}\right)^{\top}\vec{\theta}-\frac{1}{2}\vec{\theta}^{\top}\vec{\theta}
    \end{align*}
    satisfies the following convergence result:
    \begin{align*}
        \lim_{n\rightarrow \infty}\frac{\vec{\theta}_n}{\log n}=\vec{\theta}(\vec{\nu}).
    \end{align*}
\end{prop}

\subsection{Main proof of Theorem \ref{thm_identifiability}}\label{S33}
    Throughout the proof of Theorem~1, we fix one specific $\omega\in\mathcal{A}$. For notational simplicity, we omit the dependence of the counting processes and covariate processes on the outcome $\omega$ in the presentation. We first provide a more detailed proof sketch of Theorem~1.\\[3mm]
    \textbf{Step 1:} For simplicity, we normalise the distribution of the random effect, $\vec{\theta}\sim \mathcal{N}_K(\vec{0},\mat{I}_K)$, and assume the intensities
    \begin{align*}     \lambda_j(t|\vec{X},\vec{Z},\vec{\theta})&=\exp(\beta_{j0}+\vec{\beta}_{j}^{\top} \vec{X}_{j}(t)+\vec{\theta}^{\top}\mat{\Sigma}^{1/2} \mat{A}_{j}^{\top} \vec{Z}_{j}(t)),\\       \widetilde{\lambda}_j(t|\vec{X},\vec{Z},\vec{\theta})&=\exp(\widetilde{\beta}_{j0}+\widetilde{\vec{\beta}}_{j}^{\top} \vec{X}_{j}(t)+\vec{\theta}^{\top} \widetilde{\mat{\Sigma}}^{1/2}\widetilde{\mat{A}}_{j}^{\top} \vec{Z}_{j}(t)).
    \end{align*}
    Guaranteed by Conditions (c) and (d), which preclude rotation and scaling in the factor loadings and preclude covariate collinearity, identifying the parameters in model (1) is equivalent to proving that for any $j,j_1,j_2=1,\ldots,J$ and any $0\leq t,s\leq T$, it holds that
    \begin{align}\label{ske1}
        \beta_{j0}+\vec{\beta}_j^{\top}\vec{X}_j(t)&=\widetilde{\beta}_{j0}+\widetilde{\vec{\beta}}_j^{\top}\vec{X}_j(t),\notag\\
        \vec{Z}_{j_1}^{\top}(t)\mat{A}_{j_1}\mat{\Sigma} \mat{A}^{\top}_{j_2}\vec{Z}_{j_2}(s)&=\vec{Z}_{j_1}^{\top}(t)\widetilde{\mat{A}}_{j_1}\widetilde{\mat{\Sigma}} \widetilde{\mat{A}}^{\top}_{j_2}\vec{Z}_{j_2}(s).
    \end{align}
    By ignoring the rotation in the random effect without loss of generality, proving (\ref{ske1}) is equivalent to showing that $\lambda_{j}(t)=\widetilde{\lambda}_j(t)$ for any $j=1,\ldots,J$ and $t\in[0,T]$.\\[3mm]
    \textbf{Step 2: }Guaranteed by Condition (e), we can partition $[0,T]$ into small intervals: $[0,t_1]$, $(t_1,t_2]$, $\ldots$ , $(t_k,T]$ such that $\vec{X}$ and $\vec{Z}$ remain constant on each interval. 
    We then use an induction argument to match the intensities of the two competing models. To be specific, suppose that we have identified two intensities on $[0,t_{q}]$. We first use Proposition~\ref{prop_likelihood} to prove that $\lambda_j(t_{q+1}),j=1,\ldots,J$ and $\widetilde{\lambda}_j(t_{q+1}),j=1,\ldots,J$ match up to a permutation among the indices $\{1,\ldots,J\}$. Letting $t=t_q$ in Proposition~\ref{prop_likelihood}, we have
    \begin{align*}
        &\int_{\bt} \Big[\prod_{j=1}^{J} \prod_{s \leq t} \lambda_{j}(s)^{\Delta N_{j}(s)}\Big]\exp\Big(-\sum_{i=1}^q (t_i-t_{i-1})\sum_{j=1}^J \lambda_{j}(t_i)\Big)\Big(\sum_{j=1}^J\lambda_{j}(t_{q+1})\Big)^{n}\phi_K(\vec{\theta} ;\vec{0}, \mat{I}_K) \mathrm{d} \vec{\theta}\notag\\
        =&\int_{\bt} \Big[\prod_{j=1}^{J}\prod_{s \leq t} \widetilde{\lambda}_{j}(s)^{\Delta N_{j}(s)}\Big]\exp\Big(-\sum_{i=1}^q (t_i-t_{i-1})\sum_{j=1}^J \widetilde{\lambda}_{j}(t_i)\Big)\Big(\sum_{j=1}^J\widetilde{\lambda}_{j}(t_{q+1})\Big)^{n}\phi_K(\vec{\theta} ;\vec{0}, \mat{I}_K) \mathrm{d} \vec{\theta}\text{.}
    \end{align*}
    Proving that $\lambda_j(t_{q+1})=\widetilde{\lambda}_j(t_{q+1})$ for $j=1,\ldots,J$ up to a permutation is equivalent to proving that $\mu_j=\widetilde{\mu}_j$ and $\vec{\eta}_j=\widetilde{\vec{\eta}}_j$ for $j=1,\ldots,J$ up to a permutation in the following equation under proper variable substitutions:
    \begin{align}\label{thm_sketch_1}
        &\sum_{1\leq j_1,\ldots,j_n\leq J}\int_{\bt}(2\pi)^{-\frac{K}{2}}\exp\Big(\sum_{k=1}^n \mu_{j_k}-\sum_{k=1}^{W}w_k \exp(\vec{\alpha}_k^{\top}\vec{\theta})+\big(\vec{\varphi}+\sum_{k=1}^{n}\vec{\eta}_{j_k}\big)^{\top}\vec{\theta}-\frac{1}{2}\vec{\theta}^{\top}\vec{\theta}\Big)\mathrm{d}\vec{\theta}\notag\\
        =&\sum_{1\leq j_1,\ldots,j_n\leq J}\int_{\bt}(2\pi)^{-\frac{K}{2}}\exp\Big(\sum_{k=1}^n \widetilde{\mu}_{j_k}-\sum_{k=1}^{W}w_k \exp({\vec{\alpha}}_k^{\top}\vec{\theta})+\big({\vec{\varphi}}+\sum_{k=1}^{n}\widetilde{\vec{\eta}}_{j_k}\big)^{\top}\vec{\theta}-\frac{1}{2}\vec{\theta}^{\top}\vec{\theta}\Big)\mathrm{d}\vec{\theta}.
    \end{align}
    We then group identical terms in this summation, which contains $J^n$ terms. For any $n\in\mathbb{N}_0$, define $\mathcal{O}_n=\{(\xi_2,\ldots,\xi_J)^{\top}\in\mathbb{N}_0^{J-1}:\sum_{j=2}^J\xi_j\leq n\}$. For any $n$ and $\bm{\xi}^{(n)}=(\xi_2^{(n)},\ldots,\xi_J^{(n)})^{\top}$ (each $\xi_j^{(n)}$ counts the number of occurrences of index $j$), we introduce the following notation:
    \begin{align*}
        f_n(\vec{\theta}\big|\bm{\xi}^{(n)})&=n\mu_1-\sum_{k=1}^{W}w_k \exp(\vec{\alpha}_k^{\top}\vec{\theta})+(\vec{\varphi}+n\vec{\eta}_1)^{\top}\vec{\theta}-\frac{1}{2}\vec{\theta}^{\top}\vec{\theta}-\sum_{j=2}^J\xi_j^{(n)}\left[(\vec{\eta}_1-\vec{\eta}_j)^{\top}\vec{\theta}+(\mu_1-\mu_j)\right],\\
        \widetilde{f}_n(\vec{\theta}\big|\bm{\xi}^{(n)})&=n\widetilde{\mu}_1-\sum_{k=1}^{W}w_k \exp(\vec{\alpha}_k^{\top}\vec{\theta})+(\vec{\varphi}+n\widetilde{\vec{\eta}}_1)^{\top}\vec{\theta}-\frac{1}{2}\vec{\theta}^{\top}\vec{\theta}-\sum_{j=2}^J\xi_j^{(n)}\left[(\widetilde{\vec{\eta}}_1-\widetilde{\vec{\eta}}_j)^{\top}\vec{\theta}+(\widetilde{\mu}_1-\widetilde{\mu}_j)\right],\\
        \phi_n(\bm{\xi}^{(n)})&=\int_{\bt}(2\pi)^{-\frac{K}{2}}\exp\left(f_n(\vec{\theta}\big|\bm{\xi}^{(n)})\right)\mathrm{d}\vec{\theta},\\
        \widetilde{\phi}_n(\bm{\xi}^{(n)})&=\int_{\bt}(2\pi)^{-\frac{K}{2}}\exp\left(\widetilde{f}_n(\vec{\theta}\big|\bm{\xi}^{(n)})\right)\mathrm{d}\vec{\theta},\\
        \Delta_n(\bm{\xi}^{(n)})&=\binom{n}{n-\sum_{j=2}^J \xi_j^{(n)},\xi_2^{(n)},\ldots,\xi_J^{(n)}}=\frac{n!}{(n-\sum_{j=2}^J \xi_j^{(n)})!\prod_{j=2}^J \xi_j^{(n)}!}.
    \end{align*}
    Here $\Delta_n(\bm{\xi}^{(n)})$ denotes the combinatorial factor that counts the number of identical terms. We merge the identical terms in equation (\ref{thm_sketch_1}) to get
    \begin{align}\label{thm_sketch_2}
        \sum_{\bm{\xi}^{(n)}\in\mathcal{O}_n}\Delta_n(\bm{\xi}^{(n)})\phi_n(\bm{\xi}^{(n)})=\sum_{\bm{\xi}^{(n)}\in\mathcal{O}_n}\Delta_n(\bm{\xi}^{(n)})\widetilde{\phi}_n(\bm{\xi}^{(n)}).
    \end{align}
    By Proposition~\ref{prop_laplace}, we can use a Laplace-type approximation to approximate $\phi_n$ and $\widetilde{\phi}_n$ by
    \begin{align}\label{thm_sketch_3}
        &\phi_n(\bm{\xi}^{(n)})\asymp \frac{\exp(f_n(\vec{\theta}_n(\bm{\xi}^{(n)})\big|\bm{\xi}^{(n)}))}{\sqrt{\operatorname{det}(-\nabla^2 f_n(\vec{\theta}_n(\bm{\xi}^{(n)})\big|\bm{\xi}^{(n)}))}},\notag\\
        &\widetilde{\phi}_n(\bm{\xi}^{(n)})\asymp \frac{\exp(\widetilde{f}_n(\widetilde{\vec{\theta}}_n(\bm{\xi}^{(n)})\big|\bm{\xi}^{(n)}))}{\sqrt{\operatorname{det}(-\nabla^2 \widetilde{f}_n(\widetilde{\vec{\theta}}_n(\bm{\xi}^{(n)})\big|\bm{\xi}^{(n)}))}},
    \end{align}
    where $\vec{\theta}_n(\bm{\xi}^{(n)})$ and $\widetilde{\vec{\theta}}_n(\bm{\xi}^{(n)})$ are the unique maximisers of ${f}_n(\vec{\theta}\big|\bm{\xi}^{(n)})$ and $\widetilde{f}_n(\vec{\theta}\big|\bm{\xi}^{(n)})$, respectively. In the main proof, we show that the asymptotic behaviours of $\phi_n$ and $\widetilde{\phi}_n$ are completely determined by the numerators in the approximation (\ref{thm_sketch_3}). This indicates the necessity to study the asymptotic behaviours of $\exp(f_n(\vec{\theta}_n(\bm{\xi}^{(n)})\big|\bm{\xi}^{(n)}))$ and $\exp(\widetilde{f}_n(\widetilde{\vec{\theta}}_n(\bm{\xi}^{(n)})\big|\bm{\xi}^{(n)}))$ as $n$ goes to infinity.\\[3mm]
    \textbf{Step 3: }Proposition~\ref{prop_canonical_projection} implies that the dominant terms on both sides of (\ref{thm_sketch_2}) occur at the point where $\|P_{\mathcal{H}^{\perp}_{\vec{\eta}}}\vec{\eta}\|$ reaches its maximum among all convex combinations of $\vec{\eta}_1,\ldots,\vec{\eta}_J$. The existence of this maximum is guaranteed by the continuity property of the canonical projection established in Proposition~\ref{prop_canonical_projection}.
    
    Moreover, Proposition~\ref{prop_canonical_projection} also implies that $\|P_{\mathcal{H}^{\perp}_{\vec{\eta}}}\vec{\eta}\|=0$ if and only if there exist $c_{1},\ldots,c_{W}\geq 0$ such that $\vec{\eta}=\sum_{k=1}^W c_{k}\vec{\alpha}_{k}$. Consequently, the asymptotic result in Proposition~\ref{prop_canonical_projection} alone is insufficient to analyse the asymptotic behaviour of (\ref{thm_sketch_3}) when all of $\vec{\eta}_1,\ldots,\vec{\eta}_J$ can be expressed as linear combinations of $\vec{\alpha}_1,\ldots,\vec{\alpha}_W$ with nonnegative coefficients, since any convex combination of $\vec{\eta}_1,\ldots,\vec{\eta}_J$ is then also a linear combination of $\vec{\alpha}_1,\ldots,\vec{\alpha}_W$ with nonnegative coefficients. Hence, we divide the remainder of the proof into three cases:
    \begin{enumerate}
        \item[(i)] \textbf{Case 1: }$\max_{j=1,\ldots,J}\|P_{\mathcal{H}^{\perp}_{\vec{\eta}_j}}\vec{\eta}_j\|>0$ and $\max_{j=1,\ldots,J}\|P_{\mathcal{H}^{\perp}_{\widetilde{\vec{\eta}}_j}}\widetilde{\vec{\eta}}_j\|>0$.
        \item[(ii)] \textbf{Case 2: }$\max_{j=1,\ldots,J}\|P_{\mathcal{H}^{\perp}_{\vec{\eta}_j}}\vec{\eta}_j\|=0$ and $\max_{j=1,\ldots,J}\|P_{\mathcal{H}^{\perp}_{\widetilde{\vec{\eta}}_j}}\widetilde{\vec{\eta}}_j\|=0$.
        \item[(iii)] \textbf{Case 3: }$\max_{j=1,\ldots,J}\|P_{\mathcal{H}^{\perp}_{\vec{\eta}_j}}\vec{\eta}_j\|>0$ and $\max_{j=1,\ldots,J}\|P_{\mathcal{H}^{\perp}_{\widetilde{\vec{\eta}}_j}}\widetilde{\vec{\eta}}_j\|=0$ or $\max_{j=1,\ldots,J}\|P_{\mathcal{H}^{\perp}_{\vec{\eta}_j}}\vec{\eta}_j\|=0$ and $\max_{j=1,\ldots,J}\|P_{\mathcal{H}^{\perp}_{\widetilde{\vec{\eta}}_j}}\widetilde{\vec{\eta}}_j\|>0$.
    \end{enumerate}
    Since we are currently matching the intensities up to a permutation among $\{1,\ldots,J\}$, we assume WLOG that $\|P_{\mathcal{H}^{\perp}_{\vec{\eta}_1}}\vec{\eta}_1\|=\max_{j=1,\ldots,J}\|P_{\mathcal{H}^{\perp}_{\vec{\eta}_j}}\vec{\eta}_j\|$ and $\|P_{\mathcal{H}^{\perp}_{\widetilde{\vec{\eta}}_{1}}}\widetilde{\vec{\eta}}_{1}\|=\max_{j=1,\ldots,J}\|P_{\mathcal{H}^{\perp}_{\widetilde{\vec{\eta}}_j}}\widetilde{\vec{\eta}}_j\|$.\\[3mm]
    \textbf{Case 1: }Under the generic identifiability framework, we can assume that $\|P_{\mathcal{H}^{\perp}_{\vec{\eta}_1}}\vec{\eta}_1\|$ is the unique maximiser, which further indicates that $\|P_{\mathcal{H}^{\perp}_{\widetilde{\vec{\eta}}_{1}}}\widetilde{\vec{\eta}}_{1}\|$ is also the unique maximiser through proper matching. Proposition~\ref{prop_max_projection} indicates that $\|P_{\mathcal{H}^{\perp}_{\vec{\eta}_1}}\vec{\eta}_1\|$ attains the unique maximum of $\|P_{\mathcal{H}^{\perp}_{\vec{\eta}}}\vec{\eta}\|$ where $\vec{\eta}$ can be any convex combination of $\vec{\eta}_1,\ldots,\vec{\eta}_J$.
    Combined with the continuity property of canonical projection, the dominant terms on both sides of (\ref{thm_sketch_2}) appear when $\xi^{(n)}/n$ falls into a small neighbourhood around $(0,\ldots,0)^{\top}$ (the counts of indices $2,\ldots,J$ are close to zero). Then the remaining proof is sketched as follows:\\[3mm]
    \textbf{Step 1.1: }For each side of (\ref{thm_sketch_2}), we define the concentration point in hypercube $\mathcal{G}_0=\widetilde{\mathcal{G}}_0=[0,1]^{J-1}$ in the following way: For each $\bm{\xi}^{(n)}\in\mathcal{O}_n$, we scale it by $\bm{\xi}^{(n)}/n$, which will fall into $[0,1]^{J-1}$. For the current hypercube $\mathcal{G}_k$ (or $\widetilde{\mathcal{G}}_k$), we partition it into $2^{J-1}$ equally-sized hypercubes and divide the sum (\ref{thm_sketch_2}) within the hypercube into $2^{J-1}$ partial sums. Then we choose a hypercube such that the partial sum within the hypercube attains the maximum among all $2^{J-1}$ partial sums infinitely often. In this way, we can construct two nested hypercube sequences $\{\mathcal{G}_k:k\in\mathbb{N}\}$ and $\{\widetilde{\mathcal{G}}_k:k\in\mathbb{N}\}$ for both sides of (\ref{thm_sketch_2}). By the nested interval theorem, we can obtain two unique concentration points $(\nu_2,\ldots,\nu_J)^{\top}$ and $(\widetilde{\nu}_2,\ldots,\widetilde{\nu}_J)^{\top}$. By the construction method of the hypercube sequences, we can approximate the complete summation on both sides of (\ref{thm_sketch_2}) by the partial sums within the hypercube at layer $k$ up to a constant ratio for $n$ infinitely often. Then we have
    \begin{align}\label{thm_sketch_4}
        \sum_{\bm{\xi}^{(n)}\in\mathcal{O}_n}\Delta_n(\bm{\xi}^{(n)})\phi_n(\bm{\xi}^{(n)})\mathbf{1}\Big(\frac{1}{n}\bm{\xi}^{(n)}\in\mathcal{G}_k\Big)\asymp\sum_{\bm{\xi}^{(n)}\in\mathcal{O}_n}\Delta_n(\bm{\xi}^{(n)})\widetilde{\phi}_n(\bm{\xi}^{(n)})\mathbf{1}\Big(\frac{1}{n}\bm{\xi}^{(n)}\in\widetilde{\mathcal{G}}_k\Big)
    \end{align}
    for any layer $k$. Such partition is performed to ensure that the asymptotic behaviours of all terms within the small hypercube are similar, as guaranteed by continuity property of canonical projection.\\[3mm]
    \textbf{Step 1.2: }We prove that the two concentration points can only be $(0,\ldots,0)^{\top}$ by contradiction. If $\bm{\xi}^{(n)}/n$ converges to any given point in $[0,1]^{J-1}$ other than $(0,\ldots,0)^{\top}$, there will be a difference of order $\mathcal{O}(n^2)$ between $f_n(\vec{\theta}_n(\mathbf{0})|\mathbf{0})$ and $f_n(\vec{\theta}_n(\bm{\xi}^{(n)})|\bm{\xi}^{(n)})$ by Proposition~\ref{prop_canonical_projection} and \ref{prop_max_projection}. Note that the order of log combinatorial factor $\log \Delta_n$ is $\mathcal{O}(n\log n)$ at most, this implies that the partial summation around $(0,\ldots,0)^{\top}$ has higher order than the partial summation around that given point, which contradicts the construction of the concentration point. Hence, in the following steps, we only focus on the partial summations within the neighbourhood of $(0,\ldots,0)^{\top}$ on both sides of (\ref{thm_sketch_2}).\\[3mm]
    \textbf{Step 1.3: }We then rank all terms in the partial summations on both sides of (\ref{thm_sketch_4}) according to their asymptotic order. Take the left side of (\ref{thm_sketch_4}) as an example. Note that the asymptotic orders of two different terms are compared in the sense that $\bm{\xi}^{(n)}$ are fixed while $n$ goes to infinity, i.e., $\bm{\xi}^{(n)}\equiv \bm{\xi}$. Heuristically, if we approximate the maximiser $\vec{\theta}_n(\bm{\xi})$ by $\vec{\theta}_n(\mathbf{0})$, then the difference between $f_n(\vec{\theta}_n(\bm{\xi})\big|\bm{\xi})$ and $f_n(\vec{\theta}_n(\mathbf{0})\big|\mathbf{0})$ is as:
    \begin{align}\label{thm_sketch_5}
        f_n(\vec{\theta}_n(\mathbf{0})\big|\mathbf{0})-f_n(\vec{\theta}_n(\bm{\xi})\big|\bm{\xi})\approx&~ f_n(\vec{\theta}_n(\mathbf{0})\big|\mathbf{0})-f_n(\vec{\theta}_n(\mathbf{0})\big|\bm{\xi})\notag\\
        =&~ \sum_{j=2}^J\xi_j(\mu_1-\mu_j)+\sum_{j=2}^J\xi_j(\vec{\eta}_1-\vec{\eta}_j)^{\top}\vec{\theta}_n(\mathbf{0})\notag\\
        \approx&~n\sum_{j=2}^J\xi_j(\vec{\eta}_1-\vec{\eta}_j)^{\top}P_{\mathcal{H}_{\vec{\eta}_1}^{\perp}}\vec{\eta}_1\notag\\
        :=&~ nT(\bm{\xi})>0,
    \end{align}
    where the strict positivity of $T(\bm{\xi})$ is guaranteed by Corollary~\ref{cor_max_projection}. On the other hand, the difference between the logarithms of the combinatorial factors or determinants is of order $o(n)$. This implies that the asymptotic ranking of $\Delta_n(\bm{\xi})\phi_n(\bm{\xi})$ is equivalent to the ranking of $T(\bm{\xi})$ in increasing order. To identify the model, it suffices to rank finitely many terms in descending order and match the corresponding values of $T(\bm{\xi})$. By choosing layer index $k$ to be sufficiently large, we can ensure that the ordering induced by $T(\bm{\xi})$ represents the ordering of $\Delta_n(\bm{\xi})\phi_n(\bm{\xi})$ when approximating $\vec{\theta}_n(\bm{\xi})$ by $\vec{\theta}_n(\mathbf{0})$, due to the continuity of the canonical projection. This validates our heuristic argument.\\
    \indent We then prove that every term can dominate the summation of all terms with lower rank. Specifically, we use an induction argument to match every term of the same rank on both sides of (\ref{thm_sketch_2}) by proving that the dominant terms in the remaining summations are strictly equal. Then we eliminate the dominant terms from equation (\ref{thm_sketch_2}) and continue the induction. This inductive method enables us to match every term on both sides of (\ref{thm_sketch_2}).\\[3mm]
    \textbf{Step 1.4: }We use Corollary~\ref{cor_likelihood} to obtain equations similar to (\ref{thm_sketch_2}) and (\ref{thm_sketch_4}). The concentration points on both sides will remain the same. Since the added terms $\lambda_m$ and $\widetilde{\lambda}_m$ are of the same event type on both sides in Corollary~\ref{cor_likelihood}, this enables us to fix the permutation among event types $\{1,\ldots,J\}$.\\[3mm]
    \textbf{Case 2: }In this case, all of $\vec{\eta}_1,\ldots,\vec{\eta}_J$ can be expressed as a linear combination of $\vec{\alpha}_1,\ldots,\vec{\alpha}_W$ with nonnegative coefficients. Hence, we should apply Proposition~\ref{prop_characterisation_equation} instead of Proposition~\ref{prop_canonical_projection} to distinguish the asymptotic order in the summation of (\ref{thm_sketch_2}). 
    Proposition~\ref{prop_characterisation_equation} resembles Proposition~\ref{prop_canonical_projection} as it guarantees the continuity property of the asymptotic behaviour of the maximiser. We sketch the proof as follows:\\[3mm]
    \textbf{Step 2.1: }For each $\bm{\xi}^{(n)}\in\mathcal{O}_n$, we scale it by $\log \bm{\xi}^{(n)}/\log n$ (define $\log 0=0$), which will fall into $[0,1]^{J-1}$. We then construct two concentration points: $(\nu_2,\ldots,\nu_J)^{\top}$ and $(\widetilde{\nu}_2,\ldots,\widetilde{\nu}_J)^{\top}$ and the corresponding hypercube sequence $\{\mathcal{G}_k:k\in\mathbb{N}\}$ and $\{\widetilde{\mathcal{G}}_k:k\in\mathbb{N}\}$ by the same method as in Case 1. Similarly for any $k\in\mathbb{N}$ we have
    \begin{align}\label{thm_sketch_6}
        \sum_{\bm{\xi}^{(n)}\in\mathcal{O}_n}\Delta_n(\bm{\xi}^{(n)})\phi_n(\bm{\xi}^{(n)})\mathbf{1}\Big(\frac{1}{\log n}\log \bm{\xi}^{(n)}\in\mathcal{G}_k\Big)\asymp\sum_{\bm{\xi}^{(n)}\in\mathcal{O}_n}\Delta_n(\bm{\xi}^{(n)})\widetilde{\phi}_n(\bm{\xi}^{(n)})\mathbf{1}\Big(\frac{1}{\log n}\log \bm{\xi}^{(n)}\in\widetilde{\mathcal{G}}_k\Big)
    \end{align}
    We assume WLOG that $\nu_2\geq \ldots\geq\nu_{p}>\nu_{p+1}=\ldots=\nu_{J}=0$ and $\widetilde{\nu}_2\geq \ldots\geq \widetilde{\nu}_{\widetilde{p}}>\widetilde{\nu}_{\widetilde{p}+1}=\ldots=\widetilde{\nu}_J=0$.\\[3mm]
    \textbf{Step 2.2: }We use the concept of canonical expansion in Proposition~\ref{lem_canonical_expansion} to decide the main direction on both sides of (\ref{thm_sketch_2}):
    Suppose that the canonical expansions of $\vec{\eta}_1,\ldots,\vec{\eta}_J,\widetilde{\vec{\eta}}_1,\ldots,\widetilde{\vec{\eta}}_J$ are as $\vec{\eta}_j=\sum_{k=1}^{m_j}c_{j,k}\vec{\alpha}_{j,k}$ and $\widetilde{\vec{\eta}}_j=\sum_{k=1}^{\widetilde{m}_j}\widetilde{c}_{j,k}\widetilde{\vec{\alpha}}_{j,k}$. We assume WLOG that $\sum_{k=1}^{m_1}c_{1,k}$ and $\sum_{k=1}^{\widetilde{m}_1}\widetilde{c}_{1,k}$ are the unique maximisers among $\sum_{k=1}^{m_j}c_{j,k}$ and $\sum_{k=1}^{\widetilde{m}_j}\widetilde{c}_{j,k}$ respectively. Then we prove that $\nu_2,\ldots,\nu_J$, $\widetilde{\nu}_2,\ldots,\widetilde{\nu}_J$ are bounded away from 1 by arguments similar to those used in Step 1.2.\\[3mm]
    \textbf{Step 2.3: }In this step, we still analyse both sides of (\ref{thm_sketch_6}) separately. We take the left side as an example. For $j=2,\ldots,p$, we characterise the relationship between $\nu_j$ and $\vec{\eta}_j$ by arguments similar to those used in Step 1.2 of Case 1 through the construction method of concentration point. Since $0<\nu_j<1$ is an interior point, we can derive equation between $\nu_j$ and $\vec{\eta}_j$ by first-order equation in the asymptotic sense:
    \begin{align}\label{thm_sketch.12}
        1-\nu_j=(\vec{\eta}_1-\vec{\eta}_j)^{\top}\vec{\theta}(\nu_2,\ldots,\nu_J),
    \end{align}
    where $\vec{\theta}(\nu_2,\ldots,\nu_J)$ is defined in Proposition~\ref{prop_characterisation_equation}. Note that here we omit the dependency on $\nu_1$ since $\nu_1\equiv 1$. For $j=p+1,\ldots,J$ (which is on the boundary), we can only derive the one-sided inequality:
    \begin{align*}
        1\leq (\vec{\eta}_1-\vec{\eta}_j)^{\top}\vec{\theta}(\nu_2,\ldots,\nu_J).
    \end{align*}
    However, we can assume that the strict inequality holds under the generic identifiability framework. Then we use the same method as in Case 1 to rank $\Delta_n(\bm{\xi}^{(n)})\phi_n(\bm{\xi}^{(n)})$. We fix $(\xi_{p+1}^{(n)},\ldots,\xi_J^{(n)})^{\top}\equiv (\xi_{p+1},\ldots,\xi_J)^{\top}:= \bm{\xi}$. By approximating $\vec{\theta}_n(\bm{\xi}^{(n)})$ by $\vec{\theta}_n(\xi_2^{(n)},\ldots,\xi_p^{(n)},\mathbf{0})$, we have the following estimation similar to (\ref{thm_sketch_5}):
    \begin{align}\label{thm_sketch_7}
        &f_n(\vec{\theta}_n(\xi_2^{(n)},\ldots,\xi_p^{(n)},\mathbf{0})\big|\xi_2^{(n)},\ldots,\xi_p^{(n)},\mathbf{0})-f_n(\vec{\theta}_n(\bm{\xi})\big|\bm{\xi})\notag\\
        \approx&f_n(\vec{\theta}_n(\xi_2^{(n)},\ldots,\xi_p^{(n)},\mathbf{0})\big|\xi_2^{(n)},\ldots,\xi_p^{(n)},\mathbf{0})-f_n(\vec{\theta}_n(\xi_2^{(n)},\ldots,\xi_p^{(n)},\mathbf{0})\big|\bm{\xi})\notag\\
        \approx& \log n\sum_{j=p+1}^{J}\xi_j(\vec{\eta}_1-\vec{\eta}_j)^{\top}\vec{\theta}(\nu_2,\ldots,\nu_J)
    \end{align}
    Note that the combinatorial factor should also be taken into consideration in Case 2. We have the following estimation by the Stirling formula:
    \begin{align}\label{thm_sketch_8}
        \log\Delta_n(\xi_2^{(n)},\ldots,\xi_p^{(n)},\mathbf{0})-\log\Delta_n(\bm{\xi}^{(n)})\approx-\log n\sum_{j=p+1}^{J}\xi_j
    \end{align}
    (\ref{thm_sketch_7}) and (\ref{thm_sketch_8}) imply that the ranking of $\Delta_n(\bm{\xi}^{(n)})\phi_n(\bm{\xi}^{(n)})$ is equivalent to ranking $T(\bm{\xi}):= \sum_{j=p+1}^{J}\xi_j[(\vec{\eta}_1-\vec{\eta}_j)^{\top}\vec{\theta}(\nu_2,\ldots,\nu_J)-1]$. The continuity result in Proposition~\ref{prop_characterisation_equation} ensures that any single term within the hypercube can represent all terms in the hypercube with ``small'' error as long as we choose layer index $k$ large enough. Since we only need to identify finitely many terms in the ranking to prove identifiability, we can similarly prove that the summation in (\ref{thm_sketch_6}) can be separated in order, where every term can dominate the summation of all terms with lower rank.\\[3mm]
    \textbf{Step 2.4: }We then prove that the two concentration points are identical. For simplicity, we only discuss the case when $\nu_2> \ldots>\nu_{p}$ and $\widetilde{\nu}_2> \ldots> \widetilde{\nu}_{\widetilde{p}}$ in the sketch. By Corollary~\ref{cor_likelihood}, we have
    \begin{align}\label{thm_sketch_11}
        &\sum_{1\leq j_1,\ldots,j_n\leq J}\int_{\bt}(2\pi)^{-\frac{K}{2}}\exp\Big(\mu_{m}+\vec{\eta}_m^{\top}\vec{\theta}+\sum_{k=1}^n \mu_{j_k}-\sum_{k=1}^{W}w_k \exp(\vec{\alpha}_k^{\top}\vec{\theta})+\big(\vec{\varphi}+\sum_{k=1}^{n}\vec{\eta}_{j_k}\big)^{\top}\vec{\theta}-\frac{1}{2}\vec{\theta}^{\top}\vec{\theta}\Big)\mathrm{d}\vec{\theta}\notag\\
        =&\sum_{1\leq j_1,\ldots,j_n\leq J}\int_{\bt}(2\pi)^{-\frac{K}{2}}\exp\Big(\widetilde{\mu}_{m}+\widetilde{\vec{\eta}}_m^{\top}\vec{\theta}+\sum_{k=1}^n \widetilde{\mu}_{j_k}-\sum_{k=1}^{W}w_k \exp({\vec{\alpha}}_k^{\top}\vec{\theta})+\big({\vec{\varphi}}+\sum_{k=1}^{n}\widetilde{\vec{\eta}}_{j_k}\big)^{\top}\vec{\theta}-\frac{1}{2}\vec{\theta}^{\top}\vec{\theta}\Big)\mathrm{d}\vec{\theta}.
    \end{align}
    We can easily construct the same concentration points for equation (\ref{thm_sketch_11}). Heuristically, both sides of (\ref{thm_sketch_11}) are multiplied by $\exp(\log n(\vec{\eta}_m^{\top}\vec{\theta}(\nu_2,\ldots,\nu_J)+o(1)))$ and $\exp(\log n(\widetilde{\vec{\eta}}_m^{\top}\widetilde{\vec{\theta}}(\widetilde{\nu}_2,\ldots,\widetilde{\nu}_J)+o(1)))$ around the concentration points. Under the generic identifiability framework, we can assume that $\vec{\eta}_1^{\top}\vec{\theta}(\nu_2,\ldots,\nu_J),\ldots,\vec{\eta}_J^{\top}\vec{\theta}(\nu_2,\ldots,\nu_J)$ are distinct. Hence we can match the two concentration points by (\ref{thm_sketch.12}). Then we can match $\lambda_{1}(t_{q+1}),\ldots,\lambda_{p}(t_{q+1})$ with $\widetilde{\lambda}_{1}(t_{q+1}),\ldots,\widetilde{\lambda}_{p}(t_{q+1})$ by arguments similar to those used in Case 1.\\[3mm]
    \textbf{Step 2.5: }We use the same method as in Step 1.3 of Case 1 to match the rest terms.\\[3mm]
    \textbf{Step 2.6: }We use the same method as in Step 1.4 of Case 1 to fix the permutation.\\[3mm]
    \textbf{Case 3: }This case leads to a contradiction since the summations on both sides of (\ref{thm_sketch_2}) have different asymptotic orders according to the discussions in Case 1 and 2.\\
    
    Now we provide the formal proof of Theorem~1.
\renewcommand*{\proofname}{Proof of Theorem \ref{thm_identifiability}}
    \begin{proof}
    For simplicity, we ignore the censoring time and assume that the study period is $[0,T]$. In the following proof, we compare the likelihood function for a subject with a given sample path $\omega\in\mathcal{A}$ on the time interval $[0,T]$ under two competing parametric models. Denote the intensity functions under two competing parametric models as
    \begin{align*}
        &\lambda_{j}(t|\vec{X}_{j}, \vec{Z}_j ; \vec{\theta})=\exp(\beta_{j0}+\vec{\beta}_{j}^{\top} \vec{X}_{j}(t)+\vec{\theta}^{\top}\mat{\Sigma}^{1/2} \mat{A}_{j}^{\top} \vec{Z}_j(t))\text{,}\\
        &\widetilde{\lambda}_{j}(t|\vec{X}_{j}, \vec{Z}_j ; \vec{\theta})=\exp(\widetilde{\beta}_{j0}+\widetilde{\vec{\beta}}_{j}^{\top} \vec{X}_{j}(t)+\vec{\theta}^{\top}\widetilde{\mat{\Sigma}}^{1/2} {\widetilde{\mat{A}}_{j}}^{\top} \vec{Z}_j(t))\text{.}
    \end{align*}
    where $\vec{\theta}\sim \mathcal{N}_K(0,\mat{I}_K)$. Suppose the two parameter sets induce the same distribution of the observed multivariate counting process.
    For notational simplicity, denote $\mu_j(t)=\beta_{j0}+\vec{\beta}_{j}^{\top} \vec{X}_{j}(t)$ and $\widetilde{\mu}_j(t)=\widetilde{\beta}_{j0}+\widetilde{\vec{\beta}}_{j}^{\top} \vec{X}_{j}(t)$ for $j=1,\ldots,J$. Condition (e) guarantees that $[0,T]$ can be divided into $v$ finite intervals: $[0,t_1],(t_1,t_2],\ldots,(t_{v-1},t_v]$ such that $\vec{X}_j$ and $\vec{Z}_j$ remain constant on each interval, which implies that $\mu_j$, $\widetilde{\mu}_j$, $\lambda_j$ and $\widetilde{\lambda}_j$ are all constant within these partitioned intervals. We then use an induction argument to prove that for any $j,j_1,j_2$ and $0\leq t\leq s\leq T$, it holds that
    \begin{align}\label{eq_thm_1.1}
        {\mu}_j(t)&=\widetilde{\mu}_j(t),\notag\\
        \vec{Z}_{j_1}^{\top}(t)\mat{A}_{j_1}\mat{\Sigma} \mat{A}^{\top}_{j_2}\vec{Z}_{j_2}(s)&=\vec{Z}_{j_1}^{\top}(t)\widetilde{\mat{A}}_{j_1}\widetilde{\mat{\Sigma}} \widetilde{\mat{A}}^{\top}_{j_2}\vec{Z}_{j_2}(s).
    \end{align}
    We first prove that (\ref{eq_thm_1.1}) holds on interval $[0,t_1]$. Choose $t=0$ in Proposition~\ref{prop_likelihood}, for any $n\in\mathbb{N}_0$ we have
    \begin{align}\label{eq_thm_1.2}
        &\int_{\bt} \Big(\sum_{j=1}^J\lambda_{j}(0)\Big)^{n} \phi_K(\vec{\theta} ;\vec{0}, \mat{I}_K) \mathrm{d} \vec{\theta}=\int_{\bt} \Big(\sum_{j=1}^J\widetilde{\lambda}_{j}(0)\Big)^{n} \phi_K(\vec{\theta} ;\vec{0}, \mat{I}_K) \mathrm{d} \vec{\theta}\text{.}
    \end{align}
    By explicit integration of (\ref{eq_thm_1.2}), we have
    \begin{align}\label{eq_thm_1.3}
        &\sum_{1\leq j_1,\ldots,j_n\leq J}\exp\Bigg(\sum_{k=1}^n \mu_{j_k}(0)+\frac{1}{2}\Big[\sum_{k=1}^n \mat{A}^{\top}_{j_k}\vec{Z}_{j_k}(0)\Big]^{\top}\mat{\Sigma}\Big[\sum_{k=1}^n \mat{A}^{\top}_{j_k}\vec{Z}_{j_k}(0)\Big]\Bigg)\notag\\
        =&\sum_{1\leq j_1,\ldots,j_n\leq J}\exp\Bigg(\sum_{k=1}^n \widetilde{\mu}_{j_k}(0)+\frac{1}{2}\Big[\sum_{k=1}^n \widetilde{\mat{A}}^{\top}_{j_k}\vec{Z}_{j_k}(0)\Big]^{\top}\widetilde{\mat{\Sigma}}\Big[\sum_{k=1}^n \widetilde{\mat{A}}^{\top}_{j_k}\vec{Z}_{j_k}(0)\Big]\Bigg).
    \end{align}
    For any $j,j_1,j_2=1,\ldots,J$, we introduce the following notation: $x_j=\exp(\mu_j(0))$, $\widetilde{x}_j=\exp(\widetilde{\mu}_j(0))$, $y_{j_1j_2}=\exp(\frac{1}{2}\vec{Z}_{j_1}^{\top}(0)\mat{A}_{j_1}\mat{\Sigma} \mat{A}^{\top}_{j_2}\vec{Z}_{j_2}(0))$ and $\widetilde{y}_{j_1j_2}=\exp(\frac{1}{2}\vec{Z}_{j_1}^{\top}(0)\widetilde{\mat{A}}_{j_1}\widetilde{\mat{\Sigma}} \widetilde{\mat{A}}^{\top}_{j_2}\vec{Z}_{j_2}(0))$. If at least one of $\vec{Z}_1(0),\ldots,\vec{Z}_J(0)$ is zero, e.g., $\vec{Z}_J(0)=\vec{0}$, then by Corollary~\ref{cor_likelihood}, for any $n\in\mathbb{N}_0$ we have
    \begin{align}\label{eq_thm_1.4}
        &\int_{\bt} \exp(\mu_{J}(0))\Big(\sum_{j=1}^J\lambda_{j}(0)\Big)^{n} \phi_K(\vec{\theta} ;\vec{0}, \mat{I}_K) \mathrm{d} \vec{\theta}=\int_{\bt} \exp(\widetilde{\mu}_{J}(0))\Big(\sum_{j=1}^J\widetilde{\lambda}_{j}(0)\Big)^{n} \phi_K(\vec{\theta} ;\vec{0}, \mat{I}_K) \mathrm{d} \vec{\theta}\text{.}
    \end{align}
    By (\ref{eq_thm_1.2}) and (\ref{eq_thm_1.4}) we have $x_J=\widetilde{x}_J$ and $y_{1J}=\ldots=y_{JJ}=\widetilde{y}_{1J}=\ldots=\widetilde{y}_{JJ}=1$. Then equation (\ref{eq_thm_1.3}) is equivalent to
    \begin{align*}
        &\sum_{1\leq j_1,\ldots,j_n\leq J-1}\exp\Bigg(\sum_{k=1}^n \mu_{j_k}(0)+\frac{1}{2}\Big[\sum_{k=1}^n \mat{A}^{\top}_{j_k}\vec{Z}_{j_k}(0)\Big]^{\top}\mat{\Sigma}\Big[\sum_{k=1}^n \mat{A}^{\top}_{j_k}\vec{Z}_{j_k}(0)\Big]\Bigg)\notag\\
        =&\sum_{1\leq j_1,\ldots,j_n\leq J-1}\exp\Bigg(\sum_{k=1}^n \widetilde{\mu}_{j_k}(0)+\frac{1}{2}\Big[\sum_{k=1}^n \widetilde{\mat{A}}^{\top}_{j_k}\vec{Z}_{j_k}(0)\Big]^{\top}\widetilde{\mat{\Sigma}}\Big[\sum_{k=1}^n \widetilde{\mat{A}}^{\top}_{j_k}\vec{Z}_{j_k}(0)\Big]\Bigg).
    \end{align*}
    for any $n\in\mathbb{N}_0$. This means that we can still apply the same analysis to the remaining $J-1$ event types. So we assume WLOG that $\vec{Z}_1(0),\ldots,\vec{Z}_J(0)$ are all nonzero. Hence by excluding a set of measure zero in the parameter space, we can assume that $\{{y}_{j_1j_2}:1\leq j_1\leq j_2\leq J\}$ are distinct. Then by Proposition~\ref{lem_summation}, there exists a permutation $\pi$: $\{1,\ldots,J\}\rightarrow\{1,\ldots,J \}$ such that for any $1\leq i,j\leq J$, $x_{i}=\widetilde{x}_{\pi(i)}$ and $y_{ij}=\widetilde{y}_{\pi(i)\pi(j)}$. Hence $\{\widetilde{y}_{j_1j_2}:1\leq j_1\leq j_2\leq J\}$ are also distinct. On the other hand, for any $j=1,\ldots,J$, Corollary~\ref{cor_likelihood} indicates that
    \begin{align}\label{eq_thm_1.5}
        &\int_{\bt} \lambda_j(0)\Big(\sum_{j=1}^J\lambda_{j}(0)\Big)^{n} \phi_K(\vec{\theta} ;\vec{0}, \mat{I}_K) \mathrm{d} \vec{\theta}=\int_{\bt} \widetilde{\lambda}_j(0)\Big(\sum_{j=1}^J\widetilde{\lambda}_{j}(0)\Big)^{n} \phi_K(\vec{\theta} ;\vec{0}, \mat{I}_K) \mathrm{d} \vec{\theta}\text{.}
    \end{align}
    By explicit integration of (\ref{eq_thm_1.5}), for any $n\in\mathbb{N}_0$ we have
    \begin{align}\label{eq_thm_1.6}
        \sum_{1\leq j_1,\ldots,j_n\leq J}&\exp\Bigg(\sum_{k=1}^n \mu_{j_k}(0)+\Big[\sum_{k=1}^n \mat{A}^{\top}_{j_k}\vec{Z}_{j_k}(0)\Big]^{\top}\mat{\Sigma} \mat{A}^{\top}_{j}\vec{Z}_{j}(0)
        \notag\\ &+\frac{1}{2}\Big[\sum_{k=1}^n \mat{A}^{\top}_{j_k}\vec{Z}_{j_k}(0)\Big]^{\top}\mat{\Sigma}\Big[\sum_{k=1}^n \mat{A}^{\top}_{j_k}\vec{Z}_{j_k}(0)\Big]+\mu_{j}(0)+\frac{1}{2}\vec{Z}_{j}^{\top}(0)\mat{A}_{j}\mat{\Sigma} \mat{A}^{\top}_{j}\vec{Z}_{j}(0)\Bigg)\notag\\
        =\sum_{1\leq j_1,\ldots,j_n\leq J}&\exp\Bigg(\sum_{k=1}^n \widetilde{\mu}_{j_k}(0)+\Big[\sum_{k=1}^n \widetilde{\mat{A}}^{\top}_{j_k}\vec{Z}_{j_k}(0)\Big]^{\top}\widetilde{\mat{\Sigma}} \widetilde{\mat{A}}^{\top}_{j}\vec{Z}_{j}(0)
        \notag\\ &+\frac{1}{2}\Big[\sum_{k=1}^n \widetilde{\mat{A}}^{\top}_{j_k}\vec{Z}_{j_k}(0)\Big]^{\top}\widetilde{\mat{\Sigma}}\Big[\sum_{k=1}^n \widetilde{\mat{A}}^{\top}_{j_k}\vec{Z}_{j_k}(0)\Big]+\widetilde{\mu}_{j}(0)+\frac{1}{2}\vec{Z}_{j}^{\top}(0)\widetilde{\mat{A}}_{j}\widetilde{\mat{\Sigma}} \widetilde{\mat{A}}^{\top}_{j}\vec{Z}_{j}(0)\Bigg).
    \end{align}
    Divide (\ref{eq_thm_1.6}) by the same equation when letting $n=0$ in (\ref{eq_thm_1.6}), we obtain
    \begin{align}\label{eq_thm_1.7}
        \sum_{1\leq j_1,\ldots,j_n\leq J}\exp\Bigg(&\sum_{k=1}^n \mu_{j_k}(0)+\Big[\sum_{k=1}^n \mat{A}^{\top}_{j_k}\vec{Z}_{j_k}(0)\Big]^{\top}\mat{\Sigma} \mat{A}^{\top}_{j}\vec{Z}_{j}(0)\notag\\
        &+\frac{1}{2}\Big[\sum_{k=1}^n \mat{A}^{\top}_{j_k}\vec{Z}_{j_k}(0)\Big]^{\top}\mat{\Sigma}\Big[\sum_{k=1}^n \mat{A}^{\top}_{j_k}\vec{Z}_{j_k}(0)\Big]\Bigg)\notag\\
        =\sum_{1\leq j_1,\ldots,j_n\leq J}\exp\Bigg(&\sum_{k=1}^n \widetilde{\mu}_{j_k}(0)+\Big[\sum_{k=1}^n \widetilde{\mat{A}}^{\top}_{j_k}\vec{Z}_{j_k}(0)\Big]^{\top}\widetilde{\mat{\Sigma}} \widetilde{\mat{A}}^{\top}_{j}\vec{Z}_{j}(0)\notag\\
        &+\frac{1}{2}\Big[\sum_{k=1}^n \widetilde{\mat{A}}^{\top}_{j_k}\vec{Z}_{j_k}(0)\Big]^{\top}\widetilde{\mat{\Sigma}}\Big[\sum_{k=1}^n \widetilde{\mat{A}}^{\top}_{j_k}\vec{Z}_{j_k}(0)\Big]\Bigg).
    \end{align}
    For an arbitrary fixed $j=1,\ldots,J$ and any $m=1,\ldots,J$, denote
    \begin{align*}
        \psi_m&:=\exp(\mu_m(0)+\vec{Z}^{\top}_{m}(0)\mat{A}_{m}\mat{\Sigma} \mat{A}^{\top}_{j}\vec{Z}_{j}(0))=x_m y_{mj}^2,\\
        \widetilde{\psi}_m&:=\exp(\widetilde{\mu}_m(0)+\vec{Z}^{\top}_{m}(0)\widetilde{\mat{A}}_{m}\widetilde{\mat{\Sigma}} \widetilde{\mat{A}}^{\top}_{j}\vec{Z}_{j}(0))=\widetilde{x}_m\widetilde{y}_{mj}^2.
    \end{align*}
    By applying Proposition~\ref{lem_summation} to equation (\ref{eq_thm_1.7}), there exists a permutation $\hat{\pi}:~\left\{1,\ldots,J \right\}\rightarrow\left\{1,\ldots,J \right\}$ such that for any $1\leq i,j\leq J$, $\psi_{i}=\widetilde{\psi}_{\hat{\pi}(i)}$ and $y_{ij}=\widetilde{y}_{\hat{\pi}(i)\hat{\pi}(j)}$. Then for any $j=1,\ldots,J$, $\widetilde{y}_{\pi(j)\pi(j)}=\widetilde{y}_{\hat{\pi}(j)\hat{\pi}(j)}$. 
    Since $\widetilde{y}_{11},\ldots,\widetilde{y}_{JJ}$ are distinct, $\pi$ and $\hat{\pi}$ are identical. Then
    \begin{align*}
        \widetilde{y}_{{\pi}(j){\pi}(j)}=y_{jj}=\sqrt{\frac{\psi_j}{x_j}}=\sqrt{\frac{\widetilde{\psi}_{\pi(j)}}{\widetilde{x}_{\pi(j)}}}=\widetilde{y}_{\pi(j)j}.
    \end{align*}
    Since $\{\widetilde{y}_{j_1j_2}:1\leq j_1\leq j_2\leq J\}$ are distinct, we have $\pi(j)=j$. Since $j$ is arbitrarily chosen from $\left\{1,\ldots,J\right\}$, this implies that $\pi=id$. Hence we proved that for any $j,j_1,j_2$ and $0\leq t\leq s\leq t_1$, it holds that
    \begin{align*}
        {\mu}_j(t)&=\widetilde{\mu}_j(t),\\
        \vec{Z}_{j_1}^{\top}(t)\mat{A}_{j_1}\mat{\Sigma} \mat{A}^{\top}_{j_2}\vec{Z}_{j_2}(s)&=\vec{Z}_{j_1}^{\top}(t)\widetilde{\mat{A}}_{j_1}\widetilde{\mat{\Sigma}} \widetilde{\mat{A}}^{\top}_{j_2}\vec{Z}_{j_2}(s).
    \end{align*}
    Now suppose that (\ref{eq_thm_1.1}) is proved on time interval $[0,t_q]$. We then prove that (\ref{eq_thm_1.1}) also holds on time interval $[0,t_{q+1}]$. Denote $t_0=0$ and apply Proposition~\ref{prop_likelihood} to the case when $t=t_q$, we have
    \begin{align}\label{eq_thm_1.8}
        &\int_{\bt} \Big[\prod_{j=1}^{J} \prod_{s \leq t_q} \lambda_{j}(s)^{\Delta N_{j}(s)}\Big]\exp\Big(-\sum_{i=1}^q (t_i-t_{i-1})\sum_{j=1}^J \lambda_{j}(t_i)\Big)\Big(\sum_{j=1}^J\lambda_{j}(t_{q+1})\Big)^{n}\phi_K(\vec{\theta} ;\vec{0}, \mat{I}_K) \mathrm{d} \vec{\theta}\notag\\
        =&\int_{\bt} \Big[\prod_{j=1}^{J}\prod_{s \leq t_q} \widetilde{\lambda}_{j}(s)^{\Delta N_{j}(s)}\Big]\exp\Big(-\sum_{i=1}^q (t_i-t_{i-1})\sum_{j=1}^J \widetilde{\lambda}_{j}(t_i)\Big)\Big(\sum_{j=1}^J\widetilde{\lambda}_{j}(t_{q+1})\Big)^{n}\phi_K(\vec{\theta} ;\vec{0}, \mat{I}_K) \mathrm{d} \vec{\theta}\text{.}
    \end{align}
    To simplify the notation, for any $k=0,\ldots,q-1,~j=1,\ldots,J$, define:
    \[\begin{array}{ll}
        \vec{\varphi}=\sum_{j=1}^{J}\int_{0}^{t_q}\mat{\Sigma}^{1/2}\mat{A}^{\top}_j\vec{Z}_j(t)\mathrm{d}N_j(t),&\widetilde{\vec{\varphi}}=\sum_{j=1}^{J}\int_{0}^{t_q}\widetilde{\mat{\Sigma}}^{1/2}\widetilde{\mat{A}}^{\top}_j\vec{Z}_j(t)\mathrm{d}N_j(t),\\
        \vec{\alpha}_{kJ+j}=\mat{\Sigma}^{1/2}\mat{A}^{\top}_j\vec{Z}_j(t_{k+1}),&\widetilde{\vec{\alpha}}_{kJ+j}=\widetilde{\mat{\Sigma}}^{1/2}\widetilde{\mat{A}}^{\top}_j\vec{Z}_j(t_{k+1}),\\
        w_{kJ+j}=(t_{k+1}-t_k)\exp(\mu_j(t_{k+1})),&\widetilde{w}_{kJ+j}=(t_{k+1}-t_k)\exp(\widetilde{\mu}_j(t_{k+1})),\\
        \vec{\eta}_j=\mat{\Sigma}^{1/2}\mat{A}^{\top}_j\vec{Z}_j(t_{q+1})
        ,&\widetilde{\vec{\eta}}_j=\widetilde{\mat{\Sigma}}^{1/2}\widetilde{\mat{A}}^{\top}_j\vec{Z}_j(t_{q+1}).
    \end{array}\]
    Let $W=qJ$, then equation (\ref{eq_thm_1.8}) can be explicitly characterised as
    \begin{align*}
        &\sum_{1\leq j_1,\ldots,j_n\leq J}\int_{\bt}(2\pi)^{-\frac{K}{2}}\exp\Big(\sum_{k=1}^n \mu_{j_k}(t_{q+1})-\sum_{k=1}^{W}w_k \exp(\vec{\alpha}_k^{\top}\vec{\theta})+\big(\vec{\varphi}+\sum_{k=1}^{n}\vec{\eta}_{j_k}\big)^{\top}\vec{\theta}-\frac{1}{2}\vec{\theta}^{\top}\vec{\theta}\Big)\mathrm{d}\vec{\theta}\notag\\
        =&\sum_{1\leq j_1,\ldots,j_n\leq J}\int_{\bt}(2\pi)^{-\frac{K}{2}}\exp\Big(\sum_{k=1}^n \widetilde{\mu}_{j_k}(t_{q+1})-\sum_{k=1}^{W}\widetilde{w}_k \exp(\widetilde{\vec{\alpha}}_k^{\top}\vec{\theta})+\big(\widetilde{\vec{\varphi}}+\sum_{k=1}^{n}\widetilde{\vec{\eta}}_{j_k}\big)^{\top}\vec{\theta}-\frac{1}{2}\vec{\theta}^{\top}\vec{\theta}\Big)\mathrm{d}\vec{\theta}.
    \end{align*}
    We write ${\mu}_{j}(t_{q+1})$, $\widetilde{\mu}_{j}(t_{q+1})$ as $\mu_j$, $\widetilde{\mu}_j$ in the following proof for simplicity. Induction assumption indicates that for any $k,k_1,k_2=1,\ldots,W$, $w_{k}=\widetilde{w}_k$ and $\vec{\alpha}_{k_1}^{\top}\vec{\alpha}_{k_2}=\widetilde{\vec{\alpha}}_{k_1}^{\top}\widetilde{\vec{\alpha}}_{k_2}$. Then there exists orthogonal matrix $\mat{T}$ in $\mathbb{R}^{K\times K}$ such that $\widetilde{\vec{\alpha}}_k=\mat{T}\vec{\alpha}_k$ for any $k=1,\ldots,W$. So by changing variables we have
    \begin{align}\label{eq_thm_1.9}
        &\sum_{1\leq j_1,\ldots,j_n\leq J}\int_{\bt}(2\pi)^{-\frac{K}{2}}\exp\Big(\sum_{k=1}^n \mu_{j_k}-\sum_{k=1}^{W}w_k \exp(\vec{\alpha}_k^{\top}\vec{\theta})+\big(\vec{\varphi}+\sum_{k=1}^{n}\vec{\eta}_{j_k}\big)^{\top}\vec{\theta}-\frac{1}{2}\vec{\theta}^{\top}\vec{\theta}\Big)\mathrm{d}\vec{\theta}\notag\\
        =&\sum_{1\leq j_1,\ldots,j_n\leq J}\int_{\bt}(2\pi)^{-\frac{K}{2}}\exp\Big(\sum_{k=1}^n \widetilde{\mu}_{j_k}-\sum_{k=1}^{W}w_k \exp({\vec{\alpha}}_k^{\top}\vec{\theta})+\big({\vec{\varphi}}+\sum_{k=1}^{n}\mat{T}^{\top}\widetilde{\vec{\eta}}_{j_k}\big)^{\top}\vec{\theta}-\frac{1}{2}\vec{\theta}^{\top}\vec{\theta}\Big)\mathrm{d}\vec{\theta}.
    \end{align}
    For simplicity, we assume that $\mat{T}=\mat{I}_K$. Hence, $\mat{T}^{\top}\widetilde{\vec{\eta}}_j=\widetilde{\vec{\eta}}_j$. We assume WLOG that $\vec{\alpha}_1,\ldots,\vec{\alpha}_W$ are distinct, or we can merge the identical ones together. We also assume WLOG that $\vec{\alpha}_1,\ldots,\vec{\alpha}_W$ are nonzero, or we can eliminate the terms on both sides of (\ref{eq_thm_1.9}). For $j=1,\ldots,J$, we call that $\vec{\eta}_j$ has a degenerate expansion if $\vec{Z}_j(t_{q+1})\in \operatorname{span}\{\vec{Z}_j(t_{1}),\ldots,\vec{Z}_j(t_{q})\}$. In such case, suppose that $\vec{Z}_j(t_{q+1})=\sum_{k=1}^qc_k\vec{Z}_j(t_{k})$. Then by induction assumption, we have
    \begin{align*}
        \widetilde{\vec{\eta}}_j=\widetilde{\mat{\Sigma}}^{1/2}\widetilde{\mat{A}}^{\top}_j\left(\sum_{k=1}^qc_k\vec{Z}_j(t_{k})\right)=\sum_{k=1}^qc_k(\widetilde{\mat{\Sigma}}^{1/2}\widetilde{\mat{A}}^{\top}_j\vec{Z}_j(t_{k}))=\sum_{k=1}^qc_k(\mat{\Sigma}^{1/2}\mat{A}^{\top}_j\vec{Z}_j(t_{k}))=\vec{\eta}_j.
    \end{align*}
    Hence the degenerate expansion of $\vec{\eta}_j$ implies that $\vec{\eta}_j=\widetilde{\vec{\eta}}_j$.\\[3mm]
    Now we prove that there exists a permutation ${\pi}:~\{1,\ldots,J \}\rightarrow\{1,\ldots,J \}$ such that for any $j=1,\ldots,J$, $\mu_{j}=\widetilde{\mu}_{\pi(j)}$ and $\vec{\eta}_{j}=\widetilde{\vec{\eta}}_{\pi(j)}$. By part (1) in Proposition~\ref{prop_canonical_projection}, there exists $\mathcal{H}_{\vec{\eta}_1},\ldots,\mathcal{H}_{\vec{\eta}_J},\mathcal{H}_{\widetilde{\vec{\eta}}_1},\ldots,\\
    \mathcal{H}_{\widetilde{\vec{\eta}}_J}$ corresponding to $\vec{\eta}_1,\ldots,\vec{\eta}_J,\widetilde{\vec{\eta}}_1,\ldots,\widetilde{\vec{\eta}}_J$. We assume WLOG that
    \begin{align*}
        \|P_{\mathcal{H}^{\perp}_{\vec{\eta}_1}}\vec{\eta}_1\|=\max_{j=1,\ldots,J}\|P_{\mathcal{H}^{\perp}_{\vec{\eta}_j}}\vec{\eta}_j\|,\\
        \|P_{\mathcal{H}^{\perp}_{\widetilde{\vec{\eta}}_{1}}}\widetilde{\vec{\eta}}_{1}\|=\max_{j=1,\ldots,J}\|P_{\mathcal{H}^{\perp}_{\widetilde{\vec{\eta}}_j}}\widetilde{\vec{\eta}}_j\|.
    \end{align*}
    For any $n$ and $\bm{\xi}^{(n)}=(\xi_2^{(n)},\ldots,\xi_J^{(n)})$, define:
    \begin{align*}
        f_n(\vec{\theta}\big|\bm{\xi}^{(n)})&=n\mu_1-\sum_{k=1}^{W}w_k \exp(\vec{\alpha}_k^{\top}\vec{\theta})+(\vec{\varphi}+n\vec{\eta}_1)^{\top}\vec{\theta}-\frac{1}{2}\vec{\theta}^{\top}\vec{\theta}-\sum_{j=2}^J\xi_j^{(n)}\left[(\vec{\eta}_1-\vec{\eta}_j)^{\top}\vec{\theta}+(\mu_1-\mu_j)\right],\\
        \widetilde{f}_n(\vec{\theta}\big|\bm{\xi}^{(n)})&=n\widetilde{\mu}_1-\sum_{k=1}^{W}w_k \exp(\vec{\alpha}_k^{\top}\vec{\theta})+(\vec{\varphi}+n\widetilde{\vec{\eta}}_1)^{\top}\vec{\theta}-\frac{1}{2}\vec{\theta}^{\top}\vec{\theta}-\sum_{j=2}^J\xi_j^{(n)}\left[(\widetilde{\vec{\eta}}_1-\widetilde{\vec{\eta}}_j)^{\top}\vec{\theta}+(\widetilde{\mu}_1-\widetilde{\mu}_j)\right],\\
        \phi_n(\bm{\xi}^{(n)})&=\int_{\bt}(2\pi)^{-\frac{K}{2}}\exp\left(f_n(\vec{\theta}\big|\bm{\xi}^{(n)})\right)\mathrm{d}\vec{\theta},\\
        \widetilde{\phi}_n(\bm{\xi}^{(n)})&=\int_{\bt}(2\pi)^{-\frac{K}{2}}\exp\left(\widetilde{f}_n(\vec{\theta}\big|\bm{\xi}^{(n)})\right)\mathrm{d}\vec{\theta},\\
        \Delta_n(\bm{\xi}^{(n)})&=\binom{n}{n-\sum_{j=2}^J \xi_j^{(n)},\xi_2^{(n)},\ldots,\xi_J^{(n)}}=\frac{n!}{\left(n-\sum_{j=2}^J \xi_j^{(n)}\right)!\prod_{j=2}^J \xi_j^{(n)}!}.
    \end{align*}
    Furthermore, denote the unique maximisers of ${f}_n(\vec{\theta}\big|\bm{\xi}^{(n)})$ and $\widetilde{f}_n(\vec{\theta}\big|\bm{\xi}^{(n)})$ by $\vec{\theta}_n(\bm{\xi}^{(n)})$ and $\widetilde{\vec{\theta}}_n(\bm{\xi}^{(n)})$, respectively. For any $n\in\mathbb{N}_0$, denote $\mathcal{O}_n=\{(\xi_2,\ldots,\xi_J)^{\top}\in\mathbb{N}_0^{J-1}:\sum_{j=2}^J\xi_j\leq n\}$. Then equation (\ref{eq_thm_1.9}) turns into
    \begin{align}\label{eq_thm_1.10}
        \sum_{\bm{\xi}^{(n)}\in\mathcal{O}_n}\Delta_n(\bm{\xi}^{(n)})\phi_n(\bm{\xi}^{(n)})=\sum_{\bm{\xi}^{(n)}\in\mathcal{O}_n}\Delta_n(\bm{\xi}^{(n)})\widetilde{\phi}_n(\bm{\xi}^{(n)}).
    \end{align}
    By Proposition~\ref{prop_laplace}, for any $\bm{\xi}^{(n)}\in\mathcal{O}_n$ we have
    \begin{align}\label{eq_thm_1.11}
        &\phi_n(\bm{\xi}^{(n)})\asymp \frac{\exp(f_n(\vec{\theta}_n(\bm{\xi}^{(n)})\big|\bm{\xi}^{(n)}))}{\sqrt{\operatorname{det}(-\nabla^2 f_n(\vec{\theta}_n(\bm{\xi}^{(n)})\big|\bm{\xi}^{(n)}))}},\notag\\
        &\widetilde{\phi}_n(\bm{\xi}^{(n)})\asymp \frac{\exp(\widetilde{f}_n(\widetilde{\vec{\theta}}_n(\bm{\xi}^{(n)})\big|\bm{\xi}^{(n)}))}{\sqrt{\operatorname{det}(-\nabla^2 \widetilde{f}_n(\widetilde{\vec{\theta}}_n(\bm{\xi}^{(n)})\big|\bm{\xi}^{(n)}))}}.
    \end{align}
    Since Proposition~\ref{prop_laplace} implies that the ratio of both sides of (\ref{eq_thm_1.11}) is bounded above and away from zero, it can be ignored in the identifying procedure. The proof then falls into either of the following three cases:\\[3mm]
    \textbf{Case 1: }$\|P_{\mathcal{H}^{\perp}_{\vec{\eta}_1}}\vec{\eta}_1\|>0$ and $\|P_{\mathcal{H}^{\perp}_{\widetilde{\vec{\eta}}_{1}}}\widetilde{\vec{\eta}}_{1}\|>0$.\\[3mm]
    \textbf{Step 1: }For $\bm{\xi}^{(n)}\in\mathcal{O}_n$, we bound the denominator in (\ref{eq_thm_1.11}) by $\exp\left(\mathcal{O}(n)\right)$ uniformly.\\[3mm]
    For any $n\in\mathbb{N}$, define $\bar{\bm{\xi}}^{(n)}\in\mathcal{O}_n$ as
    \begin{align*}
        \bar{\bm{\xi}}^{(n)}=\underset{\bm{\xi}^{(n)}\in\mathcal{O}_n}{\operatorname{argmax}}~\operatorname{det}(-\nabla^2 f_n(\vec{\theta}_n(\bm{\xi}^{(n)})\big|\bm{\xi}^{(n)})).
    \end{align*}
    Since $\bar{\bm{\xi}}^{(n)}=\mathcal{O}(n)$, we can prove that $\vec{\theta}_n(\bar{\bm{\xi}}^{(n)})=\mathcal{O}(n)$ by part (2) in Proposition~\ref{prop_canonical_projection}. We can similarly define $\widetilde{\bm{\xi}}^{(n)}$ for the other side and prove that $\widetilde{\vec{\theta}}_n(\widetilde{\bm{\xi}}^{(n)})=\mathcal{O}(n)$. Hence there exists $M>0$ such that for any $\bm{\xi}^{(n)}\in\mathcal{O}_n$, it holds that
    \begin{align}\label{eq_thm_1.12}
        1\leq& \sqrt{\operatorname{det}\big(-\nabla^2 f_n(\vec{\theta}_n(\bm{\xi}^{(n)})\big|\bm{\xi}^{(n)})\big)}\leq\sqrt{\operatorname{det}\big(-\nabla^2 f_n(\vec{\theta}_n(\bar{\bm{\xi}}^{(n)})\big|\bar{\bm{\xi}}^{(n)})\big)}\leq \exp\left(Mn\right),\notag\\
        1\leq& \sqrt{\operatorname{det}\big(-\nabla^2 \widetilde{f}_n(\widetilde{\vec{\theta}}_n(\bm{\xi}^{(n)})\big|\bm{\xi}^{(n)})\big)}\leq\sqrt{\operatorname{det}\big(-\nabla^2 \widetilde{f}_n(\widetilde{\vec{\theta}}_n(\widetilde{\bm{\xi}}^{(n)})\big|\widetilde{\bm{\xi}}^{(n)})\big)}\leq \exp\left(Mn\right).
    \end{align}
    \textbf{Step 2: }Construct the concentration points on both sides of (\ref{eq_thm_1.10}).\\[3mm]
    Denote $\mathcal{G}_{0}=\mat{D}=[0,1]^{J-1}$. We define $\{\mathcal{G}_{k}:k\in\mathbb{N}_0\}$ in the following inductive method: Suppose $\mathcal{G}_{k-1}$ is constructed, we partition $\mathcal{G}_{k-1}$ into $2^{J-1}$ identical hypercubes $\mat{D}_{1}^{(k)},\ldots,\mat{D}_{2^{(J-1)}}^{(k)}$ with length $2^{-k}$ on each side. For any $n\in\mathbb{N}_0$ and $i=1,\ldots,2^{J-1}$, denote
    \begin{align*}
        S_{i,k,n}=\sum_{\bm{\xi}^{(n)}\in\mathcal{O}_n}\Delta_n(\bm{\xi}^{(n)})\phi_n(\bm{\xi}^{(n)})\mathbf{1}\Big(\frac{1}{n}\bm{\xi}^{(n)}\in \overline{\mat{D}_{i}^{(k)}}\Big).
    \end{align*}
    Then we define $\mathcal{G}_{k}=\overline{\mat{D}_{j_k}^{(k)}}$, which satisfies $S_{j_k,k,n}=\underset{i=1,\ldots,2^{J-1}}{\max}S_{i,k,n}~~i.o.$ Hence we can define a nested hypercube sequence $\{\mathcal{G}_{k}:k\in\mathbb{N}_0\}$. By nested interval theorem, there exists a unique $(\nu_2,\ldots,\nu_J)^{\top}\in [0,1]^{J-1}$ such that
    \begin{align*}
        (\nu_2,\ldots,\nu_J)\in \bigcap_{k=0}^{\infty}{\mathcal{G}_{k}}.
    \end{align*}
    We call this point the concentration point of the left side of (\ref{eq_thm_1.10}). By the definition of $(\nu_2,\ldots,\nu_J)^{\top}$, we have
    \begin{align}\label{eq_thm_1.13}
        S_{j_k,k,n}=&\sum_{\bm{\xi}^{(n)}\in\mathcal{O}_n}\Delta_n(\bm{\xi}^{(n)})\phi_n(\bm{\xi}^{(n)})\mathbf{1}\Big(\frac{1}{n}\bm{\xi}^{(n)}\in \mathcal{G}_{k}\Big)\geq 2^{-k(J-1)}\sum_{\bm{\xi}^{(n)}\in\mathcal{O}_n}\Delta_n(\bm{\xi}^{(n)})\phi_n(\bm{\xi}^{(n)})
    \end{align}
    infinitely often. We assume WLOG that (\ref{eq_thm_1.13}) holds for any $n\in\mathbb{N}$, or we can proceed with the proof along a subsequence. Similarly, we can define concentration point $(\widetilde{\nu}_2,\ldots,\widetilde{\nu}_J)^{\top}$ and $\widetilde{S}_{\widetilde{j}_k,k,n}$ for the right-hand side of (\ref{eq_thm_1.10}). Then for any $n\in\mathbb{N}$, it holds that
    \begin{align}\label{eq_thm_1.14}
        \widetilde{S}_{\widetilde{j}_k,k,n}\geq 2^{-k(J-1)}\sum_{\bm{\xi}^{(n)}\in\mathcal{O}_n}\Delta_n(\bm{\xi}^{(n)})\widetilde{\phi}_n(\bm{\xi}^{(n)}).
    \end{align}
    By (\ref{eq_thm_1.10}), (\ref{eq_thm_1.13}) and (\ref{eq_thm_1.14}), for any $k\in\mathbb{N}_0$, we have
    \begin{align}\label{eq_thm_1.15}
        \sum_{\bm{\xi}^{(n)}\in\mathcal{O}_n}\Delta_n(\bm{\xi}^{(n)})\phi_n(\bm{\xi}^{(n)})\mathbf{1}\Big(\frac{1}{n}\bm{\xi}^{(n)}\in \mathcal{G}_{k}\Big)\asymp\sum_{\bm{\xi}^{(n)}\in\mathcal{O}_n}\Delta_n(\bm{\xi}^{(n)})\widetilde{\phi}_n(\bm{\xi}^{(n)})\mathbf{1}\Big(\frac{1}{n}\bm{\xi}^{(n)}\in \widetilde{\mathcal{G}}_{k}\Big).
    \end{align}
    Hence we reduce equation (\ref{eq_thm_1.10}) to partial sums around the two concentration points.\\[3mm]
    \textbf{Step 3: }Prove that $(\nu_2,\ldots,\nu_J)^{\top}=(\widetilde{\nu}_2,\ldots,\widetilde{\nu}_J)^{\top}=(0,\ldots,0)^{\top}$.\\[3mm]
    For any $0\leq \nu_2,\ldots,\nu_J\leq 1$ and $\sum_{j=2}^J\nu_j\leq 1$, denote $\vec{\eta}(\nu_2,\ldots,\nu_J)=(1-\sum_{j=2}^J\nu_j)\vec{\eta}_1+\sum_{j=2}^J\nu_j\vec{\eta}_j$. We first prove that $(\nu_2,\ldots,\nu_J)^{\top}=(0,\ldots,0)^{\top}$. If this is not the case, i.e., $(\nu_2,\ldots,\nu_J)^{\top}\neq(0,\ldots,0)^{\top}$, then by Proposition~\ref{prop_max_projection} we have
    \begin{align*}
        \left\|P_{\mathcal{H}_{\vec{\eta}(\nu_2,\ldots,\nu_J)}^{\perp}}\vec{\eta}(\nu_2,\ldots,\nu_J)\right\|<\left\|P_{\mathcal{H}_{\vec{\eta}_1}^{\perp}}\vec{\eta}_1\right\|.
    \end{align*}
    By the continuity of canonical projection, we fix $k\in\mathbb{N}_0$ large enough such that
    \begin{align}\label{eq_thm_1.16}
        \underset{(\bar{\nu}_2,\ldots,\bar{\nu}_J)^{\top}\in\mathcal{G}_k}{\max}\left\|P_{\mathcal{H}_{\vec{\eta}(\bar{\nu}_2,\ldots,\bar{\nu}_J)}^{\perp}}\vec{\eta}(\bar{\nu}_2,\ldots,\bar{\nu}_J)\right\|^2+\delta\leq\underset{(\bar{\nu}_2,\ldots,\bar{\nu}_J)^{\top}\in\hat{\mathcal{G}}_k}{\min}\left\|P_{\mathcal{H}_{\vec{\eta}(\bar{\nu}_2,\ldots,\bar{\nu}_J)}^{\perp}}\vec{\eta}(\bar{\nu}_2,\ldots,\bar{\nu}_J)\right\|^2:= C,
    \end{align}
    where $\delta>0$ is constant and $\hat{\mathcal{G}}_k$ is the hypercube with length $2^{-k}$ on each side which contains point $(0,\ldots,0)$. Define
    \begin{align*}
        \bar{\bm{\xi}}^{(n)}=\underset{\frac{1}{n}\bm{\xi}^{(n)}\in\mathcal{G}_k}{\operatorname{argmax}}\Delta_n(\bm{\xi}^{(n)})\phi_n(\bm{\xi}^{(n)})
    \end{align*}
    and assume that
    \begin{align*}
        \lim_{n\rightarrow \infty}\frac{\bar{\bm{\xi}}^{(n)}}{n}=\left(\bar{\nu}_2,\ldots,\bar{\nu}_J\right)^{\top}\in\mathcal{G}_k.
    \end{align*}
    Then by (\ref{eq_thm_1.11}), (\ref{eq_thm_1.12}), (\ref{eq_thm_1.16}) and part (2) in Proposition~\ref{prop_canonical_projection}, for $n$ large enough, we have
    \begin{align*}
        \Delta_n(\bar{\bm{\xi}}^{(n)})\phi_n(\bar{\bm{\xi}}^{(n)})=&\exp(o(n^2))\exp\left(f_n(\vec{\theta}_n(\bar{\bm{\xi}}^{(n)})\big|\bar{\bm{\xi}}^{(n)})\right)\\
        =&\exp\left(o(n^2)+\frac{n^2}{2}\left\|P_{\mathcal{H}_{\vec{\eta}(\bar{\nu}_2,\ldots,\bar{\nu}_J)}^{\perp}}\vec{\eta}(\bar{\nu}_2,\ldots,\bar{\nu}_J)\right\|^2\right)\\
        \leq&\exp\left(o(n^2)+\frac{C-\delta}{2}n^2\right).
    \end{align*}
    Hence we have
    \begin{align}\label{eq_thm_1.17}
        \sum_{\bm{\xi}^{(n)}\in\mathcal{O}_n}\Delta_n(\bm{\xi}^{(n)})\phi_n(\bm{\xi}^{(n)})\mathbf{1}\Big(\frac{1}{n}\bm{\xi}^{(n)}\in \mathcal{G}_{k}\Big)\leq& \operatorname{card}\left(\mathcal{O}_n\right)\exp\left(o(n^2)+\frac{C-\delta}{2}n^2\right)\notag\\
        =&\exp\left(o(n^2)+\frac{C-\delta}{2}n^2\right).
    \end{align}
    Similarly we can prove that:
    \begin{align}\label{eq_thm_1.18}
        \sum_{\bm{\xi}^{(n)}\in\mathcal{O}_n}\Delta_n(\bm{\xi}^{(n)})\phi_n(\bm{\xi}^{(n)})\mathbf{1}\Big(\frac{1}{n}\bm{\xi}^{(n)}\in \hat{\mathcal{G}}_{k}\Big)\geq\exp\left(o(n^2)+\frac{C}{2}n^2\right).
    \end{align}
    However, by the definition of $\left(\nu_2,\ldots,\nu_J\right)^{\top}$, we should have
    \begin{align*}
        \sum_{\bm{\xi}^{(n)}\in\mathcal{O}_n}\Delta_n(\bm{\xi}^{(n)})\phi_n(\bm{\xi}^{(n)})\mathbf{1}\Big(\frac{1}{n}\bm{\xi}^{(n)}\in \hat{\mathcal{G}}_{k}\Big)\lesssim\sum_{\bm{\xi}^{(n)}\in\mathcal{O}_n}\Delta_n(\bm{\xi}^{(n)})\phi_n(\bm{\xi}^{(n)})\mathbf{1}\Big(\frac{1}{n}\bm{\xi}^{(n)}\in \mathcal{G}_{k}\Big),
    \end{align*}
    which contradicts (\ref{eq_thm_1.17}) and (\ref{eq_thm_1.18}). So we have $(\nu_2,\ldots,\nu_J)^{\top}=(0,\ldots,0)^{\top}$. Similarly, we can prove that $(\widetilde{\nu}_2,\ldots,\widetilde{\nu}_J)^{\top}=(0,\ldots,0)^{\top}$.\\[3mm]
    \textbf{Step 4: }Separate the order of summation on both sides of (\ref{eq_thm_1.15}).\\[3mm]
    For $0\leq\nu_2,\ldots,\nu_J\leq 1$ such that $\sum_{j=2}^J\nu_j\leq 1$, by part (2) of Proposition~\ref{prop_canonical_projection} we can define
    \begin{align*}
        \lim_{n\rightarrow \infty}\frac{\vec{\theta}_n(\lfloor n\nu_2\rfloor,\ldots,\lfloor n\nu_J\rfloor)}{n}:= \vec{\theta}(\nu_2,\ldots,\nu_J).
    \end{align*}
    Now we rank $(\vec{\eta}_1-\vec{\eta}_j)^{\top}\vec{\theta}(\mathbf{0}),j=2,\ldots,J$, in decreasing order. By excluding a set of measure zero in the parameter space, we can assume WLOG that there are no ties and $(\vec{\eta}_1-\vec{\eta}_2)^{\top}\vec{\theta}(\mathbf{0})>\ldots>(\vec{\eta}_1-\vec{\eta}_J)^{\top}\vec{\theta}(\mathbf{0})$. By part (2) in Proposition~\ref{prop_canonical_projection}, $\vec{\theta}(\mathbf{0})=P_{\mathcal{H}_{\vec{\eta}_1}^{\perp}}\vec{\eta}_1$. Then by Proposition~\ref{prop_max_projection} we have
    \begin{align*}
        (\vec{\eta}_1-\vec{\eta}_J)^{\top}\vec{\theta}(\mathbf{0})=(\vec{\eta}_1-\vec{\eta}_J)^{\top}P_{\mathcal{H}_{\vec{\eta}_1}^{\perp}}\vec{\eta}_1:= \delta>0.
    \end{align*}
    For any $\bm{\xi}=(\xi_2,\ldots,\xi_J)^{\top}\in\mathbb{N}_0^{J-1}$, denote $T(\bm{\xi})=\sum_{j=2}^J \xi_j(\vec{\eta}_1-\vec{\eta}_j)^{\top}\vec{\theta}(\mathbf{0})$. Then we rank all the components in $\{T(\bm{\xi}):\bm{\xi}\in\mathbb{N}_0^{J-1}\}$ in increasing order. For any $r\in\mathbb{N}$, let $\bm{\xi}^{(r)}$ denote the vector such that the rank of $T(\bm{\xi}^{(r)})$ is $r$. Suppose the rank of $T(0,\ldots,0,1,1)$ is $r^{\ast}$ (If $J=1$, then the result automatically holds. If $J=2$, then we assume that the rank of $T(1,1)$ is $r^{\ast}$). By excluding a set of measure zero in the parameter space, we can assume that there are no ties among $T(\bm{\xi}^{(1)}),\ldots,T(\bm{\xi}^{(r^{\ast}+1)})$. By the continuity of canonical projection proved in Proposition~\ref{prop_canonical_projection}, we fix $k$ large enough such that
    \begin{align}\label{eq_thm_1.19}
        \underset{j=2,\ldots,J}{\min}\underset{(\nu_2,\ldots,\nu_J)^{\top}\in\mathcal{G}_k}{\min}(\vec{\eta}_1-\vec{\eta}_j)^{\top}\vec{\theta}(\nu_2,\ldots,\nu_J)\geq \frac{\delta}{2}.
    \end{align}
    Now we fix $r$ such that $1\leq r\leq r^{\ast}$. By part (2) in Proposition~\ref{prop_canonical_projection} we have
    \begin{align}\label{eq_thm_1.20}
        \lim_{n\rightarrow \infty}\frac{\vec{\theta}_n(\bm{\xi}^{(r+1)})}{n}=\lim_{n\rightarrow \infty}\frac{\vec{\theta}_n(\bm{\xi}^{(r)})}{n}=P_{\mathcal{H}_{\vec{\eta}_1}^{\perp}}\vec{\eta}_1=\vec{\theta}(\mathbf{0}).
    \end{align}
    Then by (\ref{eq_thm_1.20}), for any $\widetilde{r}\in\mathbb{N}$, it holds that
    \begin{align}\label{eq_thm_1.21}
        \lim_{n\rightarrow\infty}\frac{1}{n}\sum_{j=2}^J\xi_j^{({\widetilde{r}})}(\vec{\eta}_1-\vec{\eta}_j)^{\top}\vec{\theta}_n(\bm{\xi}^{(r)})=\sum_{j=2}^J\xi_j^{({\widetilde{r}})}(\vec{\eta}_1-\vec{\eta}_j)^{\top}\vec{\theta}(\mathbf{0})=T(\bm{\xi}^{(\widetilde{r})}).
    \end{align}
    By (\ref{eq_thm_1.20}), there also holds
    \begin{align}\label{eq_thm_1.22}
        \frac{\operatorname{det}\big(-\nabla^2 f_n(\vec{\theta}_n(\bm{\xi}^{(r+1)})\big|\bm{\xi}^{(r+1)})\big)}{\operatorname{det}\big(-\nabla^2 f_n(\vec{\theta}_n(\bm{\xi}^{(r)})\big|\bm{\xi}^{(r)})\big)}=\exp\left(o(n)\right).
    \end{align}
    Moreover, by the Stirling formula we have
    \begin{align}\label{eq_thm_1.23}
        \frac{\Delta_n(\bm{\xi}^{(r+1)})}{\Delta_n(\bm{\xi}^{(r)})}\asymp& \sqrt{\frac{n-\sum_{j=2}^J\xi_j^{(r+1)}}{n-\sum_{j=2}^J\xi_j^{({r})}}}\exp\Bigg[n\log n-\left(n-\sum_{j=2}^J\xi_j^{(r+1)}\right)\log\left(n-\sum_{j=2}^J\xi_j^{(r+1)}\right)\notag\\
        &\quad\quad\quad\quad\quad\quad\quad\quad\quad\quad-n\log n+\left(n-\sum_{j=2}^J\xi_j^{({r})}\right)\log\left(n-\sum_{j=2}^J\xi_j^{({r})}\right)\Bigg]\notag\\
        \asymp&\exp\left\{\left(\sum_{j=2}^J\xi_j^{(r+1)}-\sum_{j=2}^J\xi_j^{({r})}\right)\log n\right\}\notag\\
        =& \exp(o(n)).
    \end{align}
    By (\ref{eq_thm_1.11}), (\ref{eq_thm_1.21}), (\ref{eq_thm_1.22}) and (\ref{eq_thm_1.23}), we have
    \begin{align}\label{eq_thm_1.24}
        \frac{\Delta_n(\bm{\xi}^{(r+1)})\phi_n(\bm{\xi}^{(r+1)})}{\Delta_n(\bm{\xi}^{(r)})\phi_n(\bm{\xi}^{(r)})}\lesssim&\exp(o(n))\frac{\exp\big(f_n(\vec{\theta}_n(\bm{\xi}^{(r+1)})\big|\bm{\xi}^{(r+1)})\big)}{\exp\big(f_n(\vec{\theta}_n(\bm{\xi}^{(r)})\big|\bm{\xi}^{(r)})\big)}\notag\\
        =&\exp\Big(o(n)+\sum_{j=2}^J\xi_j^{({r})}\left[(\vec{\eta}_1-\vec{\eta}_j)^{\top}\vec{\theta}_n(\bm{\xi}^{(r)})+(\mu_1-\mu_j)\right]\notag\\
        &\quad\quad\quad\quad-\sum_{j=2}^J\xi_j^{(r+1)}\left[(\vec{\eta}_1-\vec{\eta}_j)^{\top}\vec{\theta}_n(\bm{\xi}^{(r+1)})+(\mu_1-\mu_j)\right]\Big)\notag\\
        =&\exp\left(o(n)+n\left[T(\bm{\xi}^{(r)})-T(\bm{\xi}^{(r+1)})\right]\right).
    \end{align}
    Now we prove that $\Delta_n(\bm{\xi}^{(r+1)})\phi_n(\bm{\xi}^{(r+1)})$ is the largest one among all terms with rank $\geq r+1$ when $n$ is large enough. For any $n\in\mathbb{N}$, suppose
    \begin{align*}
        \bar{\bm{\xi}}^{(n)}=\underset{\bm{\xi}\in \mathcal{O}_n\setminus\{\bm{\xi}^{(l)}:l=1,\ldots,r\},\bm{\xi}/n\in\mathcal{G}_k}{\operatorname{argmax}}\Delta_n(\bm{\xi})\phi_n(\bm{\xi}).
    \end{align*}
    If $\bar{\bm{\xi}}^{(n)}$ is unbounded, we assume WLOG that $\bar{\xi}_2^{(n)}\rightarrow \infty$ and suppose that $\lim_{n\rightarrow \infty}\bar{\bm{\xi}}^{(n)}/n=\left(\bar{\nu}_2,\ldots,\bar{\nu}_J\right)^{\top}\in\mathcal{G}_k$. Then by part (2) of Proposition~\ref{prop_canonical_projection}, we have $\vec{\theta}_n(\bar{\bm{\xi}}^{(n)})/n\rightarrow \vec{\theta}\left(\bar{\nu}_2,\ldots,\bar{\nu}_J\right)$. Since it can be readily shown that $\vec{\theta}_n(\bar{\bm{\xi}}^{(n)})-\vec{\theta}_n(\bar{\xi}_2^{(n)}-1,\bar{\xi}_3^{(n)},\ldots,\bar{\xi}_J^{(n)})=\mathcal{O}(1)$, we have
    \begin{align}\label{eq_thm_1.25}
        \operatorname{det}\big(-\nabla^2 f_n(\vec{\theta}_n(\bar{\bm{\xi}}^{(n)})\big|\bar{\bm{\xi}}^{(n)})\big)\asymp\operatorname{det}\big(-\nabla^2 f_n(\vec{\theta}_n(\bar{\xi}_2^{(n)}-1,\bar{\xi}_3^{(n)},\ldots,\bar{\xi}_J^{(n)})\big|\bar{\xi}_2^{(n)}-1,\bar{\xi}_3^{(n)},\ldots,\bar{\xi}_J^{(n)})\big).
    \end{align}
    Then by (\ref{eq_thm_1.19}) and (\ref{eq_thm_1.25}), we have
    \begin{align*}
        &\frac{\Delta_n(\bar{\xi}_2^{(n)}-1,\ldots,\bar{\xi}_J^{(n)})\phi_n(\bar{\xi}_2^{(n)}-1,\ldots,\bar{\xi}_J^{(n)})}{\Delta_n(\bar{\bm{\xi}}^{(n)})\phi_n(\bar{\bm{\xi}}^{(n)})}\\
        \gtrsim&\frac{\Delta_n(\bar{\xi}_2^{(n)}-1,\ldots,\bar{\xi}_J^{(n)})}{\Delta_n(\bar{\xi}_2^{(n)},\ldots,\bar{\xi}_J^{(n)})}\frac{\exp\left(f_n(\vec{\theta}_n(\bar{\bm{\xi}}^{(n)})\big|\bar{\xi}_2^{(n)}-1,\ldots,\bar{\xi}_J^{(n)})\right)}{\exp\left(f_n(\vec{\theta}_n(\bar{\bm{\xi}}^{(n)})\big|\bar{\bm{\xi}}^{(n)})\right)}\\
        \gtrsim& \exp\left(o(n)+\left(\vec{\eta}_1-\vec{\eta}_2\right)^{\top}\vec{\theta}_n(\bar{\bm{\xi}}^{(n)})+(\mu_1-\mu_2)\right)\\
        \gtrsim&\exp\left(o(n)+n\left(\vec{\eta}_1-\vec{\eta}_2\right)^{\top}\vec{\theta}(\bar{\nu}_2,\ldots,\bar{\nu}_J)\right)\\
        \rightarrow&\infty,
    \end{align*}
    which contradicts the definition of $\bar{\bm{\xi}}^{(n)}$ since unboundedness of $\bar{\xi}_2^{(n)}$ implies that $(\bar{\xi}_2^{(n)}-1,\ldots,\bar{\xi}_J^{(n)})\notin \left\{\bm{\xi}^{(l)}:l=1,\ldots,r\right\}$ when $n$ is large. Hence $\bar{\bm{\xi}}^{(n)}$ is bounded. Then by similar argument as (\ref{eq_thm_1.24}), it can be readily shown that $\bar{\bm{\xi}}^{(n)}=\bm{\xi}^{(r+1)}$ for $n$ large enough. So for any $1\leq r\leq r^{\ast}$, by (\ref{eq_thm_1.24}) we have
    \begin{align}\label{eq_thm_1.26}
        \frac{\sum_{u\geq r+1}\Delta_n(\bm{\xi}^{(u)})\phi_n(\bm{\xi}^{(u)})\mathbf{1}\left\{\frac{1}{n}\bm{\xi}^{(u)}\in \mathcal{G}_k\right\}}{\Delta_n(\bm{\xi}^{(r)})\phi_n(\bm{\xi}^{(r)})}\lesssim&n^{J}\frac{\Delta_n(\bm{\xi}^{(r+1)})\phi_n(\bm{\xi}^{(r+1)})}{\Delta_n(\bm{\xi}^{(r)})\phi_n(\bm{\xi}^{(r)})}\notag\\
        \lesssim&n^{J}\exp\left(o(n)+n\left[T(\bm{\xi}^{(r)})-T(\bm{\xi}^{(r+1)})\right]\right)\notag\\
        \rightarrow&0
    \end{align}
    when $k$ is large enough. Similarly, we assume WLOG that $(\widetilde{\vec{\eta}}_1-\widetilde{\vec{\eta}}_2)^{\top}\widetilde{\vec{\theta}}(\mathbf{0})>\ldots>(\widetilde{\vec{\eta}}_1-\widetilde{\vec{\eta}}_J)^{\top}\widetilde{\vec{\theta}}(\mathbf{0})>0$ and define $\widetilde{T}(\bm{\xi})$. Moreover, we assume WLOG that the rank of $\widetilde{T}(0,\ldots,0,1,1)$ is no greater than $r^{\ast}$. Then we can similarly prove that for any $1\leq r \leq r^{\ast}$ and $k$ large enough,
    \begin{align}\label{eq_thm_1.27}
        \frac{\sum_{u\geq r+1}\Delta_n(\widetilde{\bm{\xi}}^{(u)})\widetilde{\phi}_n(\widetilde{\bm{\xi}}^{(u)})\mathbf{1}\{\frac{1}{n}\widetilde{\bm{\xi}}^{(u)}\in \mathcal{G}_k\}}{\Delta_n(\widetilde{\bm{\xi}}^{(r)})\widetilde{\phi}_n(\widetilde{\bm{\xi}}^{(r)})}\rightarrow 0.
    \end{align}
    \textbf{Step 5: }Prove that for any $j,j_1,j_2=1,\ldots,J$, $\mu_j=\widetilde{\mu}_j$ and $\vec{\eta}_{j_1}^{\top}\vec{\eta}_{j_2}=\widetilde{\vec{\eta}}_{j_1}^{\top}\widetilde{\vec{\eta}}_{j_2}$.\\[3mm]
    We proceed with the proof by an induction argument to show that for any $1\leq r\leq r^{\ast}$, it holds that $\bm{\xi}^{(r)}=\widetilde{\bm{\xi}}^{(r)}$, $T(\bm{\xi}^{(r)})=\widetilde{T}(\bm{\xi}^{(r)})$ and $\phi_n(\bm{\xi}^{(r)})=\widetilde{\phi}_n(\widetilde{\bm{\xi}}^{(r)})$.\\[3mm]
    For $r=1$, by (\ref{eq_thm_1.15}), (\ref{eq_thm_1.26}) and (\ref{eq_thm_1.27}) we have
    \begin{align}\label{equation2}
        \Delta_n(\bm{\xi}^{(1)})\phi_n(\bm{\xi}^{(1)})\asymp& \sum_{\bm{\xi}\in\mathcal{O}_n}\Delta_n(\bm{\xi})\phi_n(\bm{\xi})\mathbf{I}\Big(\frac{1}{n}\bm{\xi}\in \mathcal{G}_k\Big)\notag\\
        \asymp&\sum_{\bm{\xi}\in\mathcal{O}_n}\Delta_n(\bm{\xi})\widetilde{\phi}_n(\bm{\xi})\mathbf{I}\Big(\frac{1}{n}\bm{\xi}\in \mathcal{G}_k\Big)\asymp\Delta_n(\widetilde{\bm{\xi}}^{(1)})\widetilde{\phi}_n(\widetilde{\bm{\xi}}^{(1)}).
    \end{align}
    It can be readily shown that ${\bm{\xi}}^{(1)}=\widetilde{\bm{\xi}}^{(1)}=\mathbf{0}$, so we have $\phi_n(\mathbf{0})\asymp\widetilde{\phi}_n(\mathbf{0})$. We have
    \begin{align}\label{equation3}
        \phi_n(\mathbf{0})=&\int_{\bt}(2\pi)^{-\frac{K}{2}}\exp\left(n\mu_1-\sum_{k=1}^{W}w_k \exp(\vec{\alpha}_k^{\top}\vec{\theta})+(\vec{\varphi}+n\vec{\eta}_1)^{\top}\vec{\theta}-\frac{1}{2}\vec{\theta}^{\top}\vec{\theta}\right)\mathrm{d}\vec{\theta}\notag\\
        =&\exp\left(\frac{1}{2}\left\|nP_{\mathcal{H}_{\vec{\eta}_1}^{\perp}}\vec{\eta}_1+P_{\mathcal{H}_{\vec{\eta}_1}^{\perp}}\vec{\varphi}\right\|^2+n\mu_1\right)\notag\\
        &\times\int_{\bt}(2\pi)^{-\frac{K}{2}}\exp\left(-\sum_{k=1}^{W}w_k \exp(\vec{\alpha}_k^{\top}\vec{\theta}+n\vec{\alpha}_k^{\top}P_{\mathcal{H}_{\vec{\eta}_1}^{\perp}}\vec{\eta}_1+\vec{\alpha}_k^{\top}P_{\mathcal{H}_{\vec{\eta}_1}^{\perp}}\vec{\varphi})\right.\notag\\
        &\left.+\vec{\theta}^{\top}P_{\mathcal{H}_{\vec{\eta}_1}}(\vec{\varphi}+n\vec{\eta}_1)-\frac{1}{2}\vec{\theta}^{\top}\vec{\theta}\right)\mathrm{d}\vec{\theta}.
    \end{align} 
    Now define
    \begin{align*}
        f_n(\vec{\theta})=&-\sum_{k=1}^{W}w_k \exp(\vec{\alpha}_k^{\top}\vec{\theta}+n\vec{\alpha}_k^{\top}P_{\mathcal{H}_{\vec{\eta}_1}^{\perp}}\vec{\eta}_1+\vec{\alpha}_k^{\top}P_{\mathcal{H}_{\vec{\eta}_1}^{\perp}}\vec{\varphi})+\vec{\theta}^{\top}P_{\mathcal{H}_{\vec{\eta}_1}}(\vec{\varphi}+n\vec{\eta}_1)-\frac{1}{2}\vec{\theta}^{\top}\vec{\theta},\\
        \widetilde{f}_n(\vec{\theta})=&-\sum_{k=1}^{W}w_k \exp(\vec{\alpha}_k^{\top}\vec{\theta}+n\vec{\alpha}_k^{\top}P_{\mathcal{H}_{\widetilde{\vec{\eta}}_1}^{\perp}}\widetilde{\vec{\eta}}_1+\vec{\alpha}_k^{\top}P_{\mathcal{H}_{\widetilde{\vec{\eta}}_1}^{\perp}}\vec{\varphi})+\vec{\theta}^{\top}P_{\mathcal{H}_{\widetilde{\vec{\eta}}_1}}(\vec{\varphi}+n\widetilde{\vec{\eta}}_1)-\frac{1}{2}\vec{\theta}^{\top}\vec{\theta}
    \end{align*}
    and denote the unique maximiser of $f_n$ by $\hat{\vec{\theta}}_n$. By Proposition~\ref{prop_laplace}, we have
    \begin{align}\label{equation1}
        \int_{\bt}(2\pi)^{-\frac{K}{2}}\exp\left(f_n(\vec{\theta})\right)\mathrm{d}\vec{\theta}\asymp \frac{\exp(f_n(\hat{\vec{\theta}}_n))}{\sqrt{\operatorname{det}(-\nabla^2f_n(\hat{\vec{\theta}}_n))}}.
    \end{align}
    By Proposition~\ref{prop_canonical_projection}, there exists $\left\{\vec{\alpha}_{k_1},\ldots,\vec{\alpha}_{k_m}\right\}\subseteq \left\{\vec{\alpha}_1,\ldots,\vec{\alpha}_W\right\}$ such that $P_{\mathcal{H}_{\vec{\eta}_1}}\vec{\eta}_1=\sum_{j=1}^m c_{k_j}\vec{\alpha}_{k_j}$. By Lemma~\ref{lem_characterisation_equation_1}, we can assume WLOG that $\left\{\vec{\alpha}_{k_1},\ldots,\vec{\alpha}_{k_m}\right\}$ are linearly independent. Then by arguments similar to those used in the proof of Lemma~\ref{lem_characterisation_equation_3}, we can prove that $\hat{\vec{\theta}}_n/\log n\rightarrow \hat{\vec{\theta}}\in \operatorname{span}\{\vec{\alpha}_{k_1},\ldots,\vec{\alpha}_{k_m}\}$ and $\vec{\alpha}_{k_1}^{\top}\hat{\vec{\theta}}=\ldots=\vec{\alpha}_{k_m}^{\top}\hat{\vec{\theta}}=1$. Denote $\hat{\vec{\theta}}=\sum_{j=1}^m \delta_{k_j}\vec{\alpha}_{k_j}$. Then $\delta=(\delta_{k_1},\ldots,\delta_{k_m})^{\top}$ is the unique solution of linear equation
    \begin{align*}
        (\vec{\alpha}_{k_1},\ldots,\vec{\alpha}_{k_m})^{\top}(\vec{\alpha}_{k_1},\ldots,\vec{\alpha}_{k_m})\delta=\textbf{1}_m.
    \end{align*}
    The denominator in the right-hand side of (\ref{equation1}) has order $\exp(\mathcal{O}(\log n))$. Then by arguments similar to those used in the proof of Lemma~\ref{lem_characterisation_equation_3}, we expand $\log \phi_n(\mathbf{0})$ in decreasing order as
    \begin{align*}
        \log \phi_n(\mathbf{0})=&\frac{n^2}{2}\left\|P_{\mathcal{H}_{\vec{\eta}_1}^{\perp}}\vec{\eta}_1\right\|^2+n\log n\sum_{j=1}^{m}c_{k_j}+n\left[(P_{\mathcal{H}_{\vec{\eta}_1}^{\perp}}\vec{\eta}_1)^{\top}P_{\mathcal{H}_{\vec{\eta}_1}^{\perp}}\vec{\varphi}-\sum_{j=1}^{m}c_{k_j}+\sum_{j=1}^{m}c_{k_j}\log \frac{c_{k_j}}{w_{k_j}} +\mu_1\right]\\
        &-\log^2 n\left(\frac{1}{2}\sum_{j=1}^{m}\delta^2_{k_j}+o(1)\right).
    \end{align*}
    Similarly we can prove that
    \begin{align*}
        \log \widetilde{\phi}_n(\mathbf{0})=&\frac{n^2}{2}\left\|P_{\mathcal{H}_{\widetilde{\vec{\eta}}_1}^{\perp}}\widetilde{\vec{\eta}}_1\right\|^2+n\log n\sum_{j=1}^{\widetilde{m}}\widetilde{c}_{k_j}+n\left[(P_{\mathcal{H}_{\widetilde{\vec{\eta}}_1}^{\perp}}\widetilde{\vec{\eta}}_1)^{\top}P_{\mathcal{H}_{\widetilde{\vec{\eta}}_1}^{\perp}}\vec{\varphi}-\sum_{j=1}^{\widetilde{m}}\widetilde{c}_{k_j}+\sum_{j=1}^{\widetilde m}\widetilde{c}_{k_j}\log \frac{\widetilde{c}_{k_j}}{w_{k_j}}+\widetilde{\mu}_1\right]\\
        &-\log^2 n\left(\frac{1}{2}\sum_{j=1}^{\widetilde{m}}\widetilde{\delta}^2_{k_j}+o(1)\right).
    \end{align*}
    Since $\phi_n(\mathbf{0})\asymp\widetilde{\phi}_n(\mathbf{0})$ by (\ref{equation2}), we can match the coefficients of each term. In particular, we have
    \begin{align*} \vec{1}_m^{\top}\left((\vec{\alpha}_{k_1},\ldots,\vec{\alpha}_{k_m})^{\top}(\vec{\alpha}_{k_1},\ldots,\vec{\alpha}_{k_m})\right)^{-1}\vec{1}_m=&\sum_{j=1}^{m}\delta^2_{k_j}\\
        =&\sum_{j=1}^{\widetilde{m}}\widetilde{\delta}^2_{k_j}=\vec{1}_{\widetilde{m}}^{\top}\left((\widetilde{\vec{\alpha}}_{k_1},\ldots,\widetilde{\vec{\alpha}}_{k_{\widetilde{m}}})^{\top}(\widetilde{\vec{\alpha}}_{k_1},\ldots,\widetilde{\vec{\alpha}}_{k_{\widetilde{m}}})\right)^{-1}\vec{1}_{\widetilde{m}}.
    \end{align*}
    By excluding a set of measure zero in the parameter space, we can assume that among all choices (finite choices) of linearly independent subset $\left\{\vec{\alpha}_{k_1},\ldots,\vec{\alpha}_{k_m}\right\}\subseteq \left\{\vec{\alpha}_1,\ldots,\vec{\alpha}_W\right\}$, the values of $\sum_{j=1}^{m}\delta^2_{k_j}=\vec{1}_m^{\top}\left((\vec{\alpha}_{k_1},\ldots,\vec{\alpha}_{k_m})^{\top}(\vec{\alpha}_{k_1},\ldots,\vec{\alpha}_{k_m})\right)^{-1}\vec{1}_m$ are distinct. Then $\sum_{j=1}^{m}\delta^2_{k_j}=\sum_{j=1}^{\widetilde{m}}\widetilde{\delta}^2_{k_j}$ implies that $m=\widetilde{m}$, $\vec{\alpha}_{k_1}=\widetilde{\vec{\alpha}}_{k_1},\ldots,\vec{\alpha}_{k_m}=\widetilde{\vec{\alpha}}_{k_m}$ and $\delta_{k_1}=\widetilde{\delta}_{k_1},\ldots,\delta_{k_m}=\widetilde{\delta}_{k_m}$. 
    Denote
    \begin{align*}
        g_n(\vec{\theta})=&-\sum_{k=1}^mw_{j_k} \exp(\vec{\alpha}_{j_k}^{\top}\vec{\theta})+\vec{\theta}^{\top}P_{\mathcal{H}_{\vec{\eta}_1}}\vec{\varphi}+n\sum_{k=1}^mc_{j_k}\vec{\alpha}_{j_k}^{\top}\vec{\theta}-\frac{1}{2}\vec{\theta}^{\top}\vec{\theta},\\
        \widetilde{g}_n(\vec{\theta})=&-\sum_{k=1}^mw_{j_k} \exp(\vec{\alpha}_{j_k}^{\top}\vec{\theta})+\vec{\theta}^{\top}P_{\mathcal{H}_{\vec{\eta}_1}}\vec{\varphi}+n\sum_{k=1}^m\widetilde{c}_{j_k}\vec{\alpha}_{j_k}^{\top}\vec{\theta}-\frac{1}{2}\vec{\theta}^{\top}\vec{\theta}
    \end{align*}
    and denote ${\vec{\theta}}_n,\widetilde{\vec{\theta}}_n$ as the unique maximiser of $g_n$ and $\widetilde{g}_n$ respectively. We then prove that there exists constant $C>0$ such that 
    \begin{align*}
        \left|\log \int(2\pi)^{-\frac{K}{2}}\exp\left(f_n(\vec{\theta})\right)\mathrm{d}\vec{\theta}-\log \int(2\pi)^{-\frac{K}{2}}\exp\left(g_n(\vec{\theta})\right)\mathrm{d}\vec{\theta}\right|\lesssim \exp(-Cn).
    \end{align*}
    By Proposition~\ref{prop_laplace} and the fact that $\hat{\vec{\theta}}_n=\mathcal{O}(\log n)$ and $\widetilde{\vec{\theta}}_n=\mathcal{O}(\log n)$, we have
    \begin{align}\label{equation333}
        \int(2\pi)^{-\frac{K}{2}}\exp\left(f_n(\vec{\theta})\right)\mathrm{d}\vec{\theta}=&\exp(f_n(\hat{\vec{\theta}}_n)+\mathcal{O}(\log n)),\notag\\
        \int(2\pi)^{-\frac{K}{2}}\exp\left(g_n(\vec{\theta})\right)\mathrm{d}\vec{\theta}=&\exp(g_n(\widetilde{\vec{\theta}}_n)+\mathcal{O}(\log n)).
    \end{align}
    On the other hand, Since $-\nabla^2 f_n(\vec{\theta})\succeq \mat{I}_k$, by the tail probability of Gaussian variables we have
    \begin{align}\label{equation444}
        \int_{\vec{\theta}:\|\vec{\theta}-\hat{\vec{\theta}}_n\|\geq n^{3/4}}(2\pi)^{-\frac{K}{2}}\exp\left(f_n(\vec{\theta})\right)\mathrm{d}\vec{\theta}\lesssim \exp(f_n(\hat{\vec{\theta}}_n)-n^{3/2}).
    \end{align}
    For $\vec{\theta}$ such that $\|\vec{\theta}-\hat{\vec{\theta}}_n\|\geq n^{3/4}$ we have $\|\vec{\theta}-\widetilde{\vec{\theta}}_n\|\geq n^{2/3}$ when $n$ is large enough. So we also prove that
    \begin{align}\label{equation5555}
        \int_{\vec{\theta}:\|\vec{\theta}-\hat{\vec{\theta}}_n\|\geq n^{3/4}}(2\pi)^{-\frac{K}{2}}\exp\left(g_n(\vec{\theta})\right)\mathrm{d}\vec{\theta}\lesssim \exp(g_n(\widetilde{\vec{\theta}}_n)-n^{4/3}).
    \end{align}
    By (\ref{equation333}), (\ref{equation444}) and (\ref{equation5555}), there exists $C_1>0$ such that
    \begin{align}\label{equation666}
        \left|\log \int_{\vec{\theta}:\|\vec{\theta}-\hat{\vec{\theta}}_n\|\leq n^{3/4}}(2\pi)^{-\frac{K}{2}}\exp\left(f_n(\vec{\theta})\right)\mathrm{d}\vec{\theta}-\log\int(2\pi)^{-\frac{K}{2}}\exp\left(f_n(\vec{\theta})\right)\mathrm{d}\vec{\theta}\right|\lesssim \exp(-C_1n)\notag\\
        \left|\log \int_{\vec{\theta}:\|\vec{\theta}-\hat{\vec{\theta}}_n\|\leq n^{3/4}}(2\pi)^{-\frac{K}{2}}\exp\left(g_n(\vec{\theta})\right)\mathrm{d}\vec{\theta}-\log\int(2\pi)^{-\frac{K}{2}}\exp\left(g_n(\vec{\theta})\right)\mathrm{d}\vec{\theta}\right|\lesssim \exp(-C_1n)
    \end{align}
    Furthermore, since it can be readily shown that
    \begin{align*}
        f_n(\vec{\theta})=g_n(\vec{\theta})-\sum_{\vec{\alpha}_k\notin\{\vec{\alpha}_{k_1},\ldots,\vec{\alpha}_{k_m}\}}w_k \exp(\vec{\alpha}_k^{\top}\vec{\theta}+n\vec{\alpha}_k^{\top}P_{\mathcal{H}_{\vec{\eta}_1}^{\perp}}\vec{\eta}_1+\vec{\alpha}_k^{\top}P_{\mathcal{H}_{\vec{\eta}_1}^{\perp}}\vec{\varphi})
    \end{align*}
    and $\vec{\alpha}_k^{\top}P_{\mathcal{H}_{\vec{\eta}_1}^{\perp}}\vec{\eta}_1<0$ for any $\vec{\alpha}_k\notin\{\vec{\alpha}_{k_1},\ldots,\vec{\alpha}_{k_m}\}$, we can prove that there exists $C_2>0$ such that
    \begin{align}\label{equation777}
        \left|\log\int_{\vec{\theta}:\|\vec{\theta}-\hat{\vec{\theta}}_n\|\leq n^{3/4}}(2\pi)^{-\frac{K}{2}}\exp\left(f_n(\vec{\theta})\right)\mathrm{d}\vec{\theta}-\log\int_{\vec{\theta}:\|\vec{\theta}-\hat{\vec{\theta}}_n\|\leq n^{3/4}}(2\pi)^{-\frac{K}{2}}\exp\left(g_n(\vec{\theta})\right)\mathrm{d}\vec{\theta}\right|\lesssim \exp(-C_2n)
    \end{align}
    By (\ref{equation666}), (\ref{equation777}) and similar argument for $\widetilde{f}_n$ and $\widetilde{g}_n$, there exists constant $C>0$ such that
    \begin{align*}
        \left|\log \int(2\pi)^{-\frac{K}{2}}\exp\left(f_n(\vec{\theta})\right)\mathrm{d}\vec{\theta}-\log \int(2\pi)^{-\frac{K}{2}}\exp\left(g_n(\vec{\theta})\right)\mathrm{d}\vec{\theta}\right|\lesssim \exp(-Cn)\\
        \left|\log \int(2\pi)^{-\frac{K}{2}}\exp(\widetilde{f}_n(\vec{\theta}))\mathrm{d}\vec{\theta}-\log \int(2\pi)^{-\frac{K}{2}}\exp\left(\widetilde{g}_n(\vec{\theta})\right)\mathrm{d}\vec{\theta}\right|\lesssim \exp(-Cn)
    \end{align*}
    By the proof in Step 4, it can be readily shown that there exists constant $C>0$ such that
    \begin{align*}
        \frac{\sum_{\bm{\xi}^{(n)}\in\mathcal{O}_n}\Delta_n(\bm{\xi}^{(n)})\phi_n(\bm{\xi}^{(n)})-\phi_n(\mathbf{0})}{\phi_n(\mathbf{0})}\lesssim \exp(-Cn),\\
        \frac{\sum_{\bm{\xi}^{(n)}\in\mathcal{O}_n}\Delta_n(\bm{\xi}^{(n)})\widetilde{\phi}_n(\bm{\xi}^{(n)})-\widetilde{\phi}_n(\mathbf{0})}{\widetilde{\phi}_n(\mathbf{0})}\lesssim \exp(-Cn).
    \end{align*}
    Since $\sum_{\bm{\xi}^{(n)}\in\mathcal{O}_n}\Delta_n(\bm{\xi}^{(n)})\phi_n(\bm{\xi}^{(n)})=\sum_{\bm{\xi}^{(n)}\in\mathcal{O}_n}\Delta_n(\bm{\xi}^{(n)})\widetilde{\phi}_n(\bm{\xi}^{(n)})$, we have $|\log \phi_n(\mathbf{0})-\log\widetilde{\phi}_n(\mathbf{0})|\lesssim \exp(-Cn)$. 

    Now we match the terms with lower order. If we look at all terms with order no less than $\mathcal{O}(\log n)$, we have
    \begin{align*}
        \log \phi_n(\mathbf{0})=&\frac{n^2}{2}\left\|P_{\mathcal{H}_{\vec{\eta}_1}^{\perp}}\vec{\eta}_1\right\|^2+n\log n\sum_{j=1}^{m}c_{k_j}+n\left[(P_{\mathcal{H}_{\vec{\eta}_1}^{\perp}}\vec{\eta}_1)^{\top}P_{\mathcal{H}_{\vec{\eta}_1}^{\perp}}\vec{\varphi}-\sum_{j=1}^{m}c_{k_j}+\sum_{j=1}^{m}c_{k_j}\log \frac{c_{k_j}}{w_{k_j}} +\mu_1\right]\\
        &-\frac{\log^2 n}{2}\sum_{j=1}^{m}\delta^2_{k_j}
        +\log n\left(-\sum_{j=1}^{m}\delta_{k_j}\log \frac{c_{k_j}}{w_{k_j}}-\frac{m}{2}\right) +o(\log n).
    \end{align*}
    Similar expansion is also obtained for $\log \widetilde{\phi}_n(\mathbf{0})$. Then by matching coefficients, we can derive
    \begin{align*}
        \sum_{j=1}^{m}c_{k_j}=&\sum_{j=1}^{m}\widetilde{c}_{k_j},\\
        \sum_{j=1}^{m}\delta_{k_j}\log \frac{c_{k_j}}{w_{k_j}}=&\sum_{j=1}^{m}\delta_{k_j}\log \frac{\widetilde{c}_{k_j}}{w_{k_j}}.
    \end{align*}
    Following similar arguments as in \cite{shun1995laplace}, we expand $\log \phi_n(\mathbf{0})$ and $\log \widetilde{\phi}_n(\mathbf{0})$ into infinite series and match the coefficients of terms with order $n^{-l_1}\log^{l_2}n$ where $l_1,l_2\in\mathbb{N}$ and derive similar equations regarding $(c_{k_1},\ldots,c_{k_m})^{\top}$ and $(\widetilde{c}_{k_1},\ldots,\widetilde{c}_{k_m})^{\top}$. By these equations we can match each coefficient: $c_{k_1}=\widetilde{c}_{k_1},\ldots,c_{k_m}=\widetilde{c}_{k_m}$. Hence we have
    \begin{align*}
        P_{\mathcal{H}_{\vec{\eta}_1}\vec{\eta}_1}=\sum_{j=1}^{m}c_{k_j}\vec{\alpha}_{k_j}=\sum_{j=1}^{m}\widetilde{c}_{k_j}\vec{\alpha}_{k_j}=P_{\mathcal{H}_{\widetilde{\vec{\eta}}_1}}\widetilde{\vec{\eta}}_1
    \end{align*}
    and $\mu_1=\widetilde{\mu}_1$. Moreover, for $j=1,\ldots,m$ we have
    \begin{align*}
        \vec{\eta}_1^{\top}\vec{\eta}_1=&\|P_{\mathcal{H}_{\vec{\eta}_1}^{\perp}}\vec{\eta}_1\|^2+\left\|P_{\mathcal{H}_{\vec{\eta}_1}}\vec{\eta}_1\right\|^2=\|P_{\mathcal{H}_{\widetilde{\vec{\eta}}_1}^{\perp}}\widetilde{\vec{\eta}}_1\|^2+\left\|P_{\mathcal{H}_{\widetilde{\vec{\eta}}_1}}\widetilde{\vec{\eta}}_1\right\|^2=\widetilde{\vec{\eta}}_1^{\top}\widetilde{\vec{\eta}}_1,\\
        \vec{\eta}_1^{\top}\vec{\alpha}_{k_j}=&\left(P_{\mathcal{H}_{\vec{\eta}_1}}\vec{\eta}_1\right)^{\top}\vec{\alpha}_{k_j}=\left(P_{\mathcal{H}_{\widetilde{\vec{\eta}}_1}}\widetilde{\vec{\eta}}_1\right)^{\top}\vec{\alpha}_{k_j}=\widetilde{\vec{\eta}}_1^{\top}\vec{\alpha}_{k_j}.
    \end{align*}
    Now we should match the inner product between $P_{\mathcal{H}_{\vec{\eta}_1}}\vec{\eta}_1$, $P_{\mathcal{H}_{\widetilde{\vec{\eta}}_1}}\widetilde{\vec{\eta}}_1$ and vectors in $\{\vec{\alpha}_1,\ldots,\vec{\alpha}_W\}\setminus\{\vec{\alpha}_{k_1},\ldots,\vec{\alpha}_{k_m}\}$. By excluding a set of measure zero in the parameter space, we can assume that $\vec{\alpha}_k$ is the unique vector among $\{\vec{\alpha}_1,\ldots,\vec{\alpha}_W\}\setminus\{\vec{\alpha}_{k_1},\ldots,\vec{\alpha}_{k_m}\}$ such that
    \begin{align*}
      \vec{\alpha}_k=\operatorname{argmax}_{\vec{\alpha}\in \{\vec{\alpha}_1,\ldots,\vec{\alpha}_W\}\setminus\{\vec{\alpha}_{k_1},\ldots,\vec{\alpha}_{k_m}\}}\vec{\alpha}^{\top}P_{\mathcal{H}_{\vec{\eta}_1}^{\perp}}\vec{\eta}_1.
    \end{align*}
    Then by first order expansion, we have
    \begin{align*}
        \phi_n(\mathbf{0})=&\exp\left(\frac{1}{2}\left\|nP_{\mathcal{H}_{\vec{\eta}_1}^{\perp}}\vec{\eta}_1+P_{\mathcal{H}_{\vec{\eta}_1}^{\perp}}\vec{\varphi}\right\|^2+n\mu_1\right)\int_{\bt}(2\pi)^{-\frac{K}{2}}\exp\left(f_n(\vec{\theta})\right)\mathrm{d}\vec{\theta}\\
        =&\exp\left(\frac{1}{2}\left\|nP_{\mathcal{H}_{\vec{\eta}_1}^{\perp}}\vec{\eta}_1+P_{\mathcal{H}_{\vec{\eta}_1}^{\perp}}\vec{\varphi}\right\|^2+n\mu_1\right)\int_{\bt}(2\pi)^{-\frac{K}{2}}\exp\left(g_n(\vec{\theta})\right)\mathrm{d}\vec{\theta}\\
        &-\exp\left(\frac{1}{2}\left\|nP_{\mathcal{H}_{\vec{\eta}_1}^{\perp}}\vec{\eta}_1+P_{\mathcal{H}_{\vec{\eta}_1}^{\perp}}\vec{\varphi}\right\|^2+n\mu_1+n\vec{\alpha}_k^{\top}P_{\mathcal{H}_{\vec{\eta}_1}^{\perp}}\vec{\eta}_1+o(n)\right)\int_{\bt}(2\pi)^{-\frac{K}{2}}\exp\left(g_n(\vec{\theta})\right)\mathrm{d}\vec{\theta}
    \end{align*}
    Note that we can easily prove that
    \begin{align*}
        &\log \exp\left(\frac{1}{2}\left\|nP_{\mathcal{H}_{\vec{\eta}_1}^{\perp}}\vec{\eta}_1+P_{\mathcal{H}_{\vec{\eta}_1}^{\perp}}\vec{\varphi}\right\|^2+n\mu_1\right)\int_{\bt}(2\pi)^{-\frac{K}{2}}\exp\left(g_n(\vec{\theta})\right)\mathrm{d}\vec{\theta}\\
        -&\log \exp\left(\frac{1}{2}\left\|nP_{\mathcal{H}_{\vec{\eta}_1}^{\perp}}\vec{\eta}_1+P_{\mathcal{H}_{\vec{\eta}_1}^{\perp}}\vec{\varphi}\right\|^2+n\mu_1+n\vec{\alpha}_k^{\top}P_{\mathcal{H}_{\vec{\eta}_1}^{\perp}}\vec{\eta}_1+o(n)\right)\int_{\bt}(2\pi)^{-\frac{K}{2}}\exp\left(g_n(\vec{\theta})\right)\mathrm{d}\vec{\theta}\\
        =&-n\vec{\alpha}_k^{\top}\vec{\theta}(\mathbf{0})+o(n).
    \end{align*}
    This implies that we should also match those remainder terms. Moreover, if the first order remainder terms are matched on both sides, then the higher order remainder terms are also matched. Hence we insert all first order remainder terms into the ranking $\left\{T(\bm{\xi}):\bm{\xi}\in\mathbb{N}_0^{J-1}\right\}$ with value indexed by $\vec{\alpha}^{\top}\vec{\theta}(\mathbf{0})$ for all $\vec{\alpha} \in\{\vec{\alpha}_1,\ldots,\vec{\alpha}_W\}\setminus\{\vec{\alpha}_{k_1},\ldots,\vec{\alpha}_{k_m}\}$. By excluding a set of measure zero in the parameter space, we assume that there are no ties in the ranking. Then we can still match the term in the ranking in decreasing order. The new added remainder terms are matched with the remainder terms on the right-hand side in a similar fashion. For simplicity, we assume that all the first-order remainder terms have ranks higher than $T(\bm{\xi}^{(r)})$. Then by matching order in similar way, we can prove that for any $\vec{\alpha} \in \{\vec{\alpha}_1,\ldots,\vec{\alpha}_W\}\setminus\{\vec{\alpha}_{k_1},\ldots,\vec{\alpha}_{k_m}\}$, it holds that 
    $\vec{\alpha}^{\top}P_{\mathcal{H}_{\vec{\eta}_1}^{\perp}}\vec{\eta}_1=\vec{\alpha}^{\top}P_{\mathcal{H}_{\widetilde{\vec{\eta}}_1}^{\perp}}\widetilde{\vec{\eta}}_1$. Hence we have
    \begin{align*}                                              \vec{\alpha}^{\top}\vec{\eta}_1=\vec{\alpha}^{\top}P_{\mathcal{H}_{\vec{\eta}_1}^{\perp}}\vec{\eta}_1+\vec{\alpha}^{\top}P_{\mathcal{H}_{\vec{\eta}_1}}\vec{\eta}_1=\vec{\alpha}^{\top}P_{\mathcal{H}_{\widetilde{\vec{\eta}}_1}^{\perp}}\widetilde{\vec{\eta}}_1+\vec{\alpha}^{\top}P_{\mathcal{H}_{\widetilde{\vec{\eta}}_1}}\widetilde{\vec{\eta}}_1=\vec{\alpha}^{\top}\widetilde{\vec{\eta}}_1.
    \end{align*}
    Now we have proved that $\vec{\eta}_1^{\top}\vec{\alpha}_k=\widetilde{\vec{\eta}}_1^{\top}\vec{\alpha}_k$ for $k=1,\ldots,W$. Then we can easily see that $\phi_n(\mathbf{0})=\widetilde{\phi}_n(\mathbf{0})$. So the result is proved for $r=1$.\\[3mm]
    If the case is proved for $1,\ldots,r-1$, then by (\ref{eq_thm_1.10}) and induction assumption,
    \begin{align}\label{eq_thm_1.28}
        \sum_{l\geq r}\Delta_n(\bm{\xi}^{(l)})\phi_n(\bm{\xi}^{(l)})=&\sum_{l\geq 1}\Delta_n(\bm{\xi}^{(l)})\phi_n(\bm{\xi}^{(l)})-\sum_{l=1}^{r-1}\Delta_n(\bm{\xi}^{(l)})\phi_n(\bm{\xi}^{(l)})\notag\\
        =&\sum_{l\geq 1}\Delta_n(\widetilde{\bm{\xi}}^{(l)})\widetilde{\phi}_n(\widetilde{\bm{\xi}}^{(l)})-\sum_{l=1}^{r-1}\Delta_n(\widetilde{\bm{\xi}}^{(l)})\widetilde{\phi}_n(\widetilde{\bm{\xi}}^{(l)})\notag\\
        =&\sum_{l\geq r}\Delta_n(\widetilde{\bm{\xi}}^{(l)})\widetilde{\phi}_n(\widetilde{\bm{\xi}}^{(l)}).
    \end{align}
    We then use the same construction method as in Step 2 to define the concentration points for both sides of (\ref{eq_thm_1.28}) and use the same method as in Step 3 to prove that the concentration points for both sides of (\ref{eq_thm_1.28}) are also $(0,\ldots,0)^{\top}$. So for any $k\in\mathbb{N}$, we have
    \begin{align}\label{eq_thm_1.29}
        \sum_{l\geq r}\Delta_n(\bm{\xi}^{(l)})\phi_n(\bm{\xi}^{(l)})\mathbf{I}\Big(\frac{1}{n}\bm{\xi}^{(l)}\in \mathcal{G}_k\Big)\asymp&\sum_{l\geq r}\Delta_n(\bm{\xi}^{(l)})\phi_n(\bm{\xi}^{(l)})\notag\\
        =&\sum_{l\geq r}\Delta_n(\widetilde{\bm{\xi}}^{(l)})\widetilde{\phi}_n(\widetilde{\bm{\xi}}^{(l)})\notag\\
        \asymp&\sum_{l\geq r}\Delta_n(\widetilde{\bm{\xi}}^{(l)})\widetilde{\phi}_n(\widetilde{\bm{\xi}}^{(l)})\mathbf{1}\Big(\frac{1}{n}\widetilde{\bm{\xi}}^{(l)}\in \mathcal{G}_k\Big).
    \end{align}
    Then by (\ref{eq_thm_1.26}), (\ref{eq_thm_1.27}) and (\ref{eq_thm_1.29}), we have
    \begin{align}\label{eq_thm_1.30}
        \Delta_n(\bm{\xi}^{(r)})\phi_n(\bm{\xi}^{(r)})\asymp\Delta_n(\widetilde{\bm{\xi}}^{(r)})\widetilde{\phi}_n(\widetilde{\bm{\xi}}^{(r)}).
    \end{align}
    Then by arguments similar to those used in the proof of Proposition~\ref{lem_summation}, we can match $\bm{\xi}^{(r)}$ with $\widetilde{\bm{\xi}}^{(r)}$ and match $T(\bm{\xi}^{(r)})$ with $\widetilde{T}(\bm{\xi}^{(r)})$. Then by similar proof as in the case $r=1$, we can match all cross terms and prove that $\phi_n(\bm{\xi}^{(r)})=\widetilde{\phi}_n(\widetilde{\bm{\xi}}^{(r)})$. By an induction argument, we can prove that for any $1\leq r\leq r^{\ast}$, it holds that $\bm{\xi}^{(r)}=\widetilde{\bm{\xi}}^{(r)}$, $T(\bm{\xi}^{(r)})=\widetilde{T}(\bm{\xi}^{(r)})$ and $\phi_n(\bm{\xi}^{(r)})=\widetilde{\phi}_n(\widetilde{\bm{\xi}}^{(r)})$. For any $j=2,\ldots,J$, choose $\bm{\xi}=(\xi_2,\ldots,\xi_J)^{\top}$ be the vector such that
    \begin{align*}
        \xi_m=\begin{cases}1~~~&m=j\\0~~~&\text{otherwise}
        \end{cases}.
    \end{align*}
    It can be readily shown that the order of $\bm{\xi}$ is smaller than $r^{\ast}$. In the inductive proof, we matched all cross terms in $\phi_n(\bm{\xi})$, i.e., $\vec{\eta}_j^{\top}\vec{\alpha}_k=\widetilde{\vec{\eta}}_j^{\top}\vec{\alpha}_k$ for any $k=1,\ldots,W$ and $\vec{\eta}_j^{\top}\vec{\eta}_j=\widetilde{\vec{\eta}}_j^{\top}\widetilde{\vec{\eta}}_j$. Moreover we have $\mu_j=\widetilde{\mu}_j$. For any $2\leq j_1<j_2\leq J$, choose $\bm{\xi}=(\xi_2,\ldots,\xi_J)$ be the vector such that
    \begin{align*}
        \xi_m=\begin{cases}1~~~&m=j_1 \text{ or } j_2\\0~~~&\text{otherwise}
        \end{cases}.
    \end{align*}
    The rank of $\bm{\xi}$ is higher than $r^{\ast}$, by matching all cross terms, we have proved that $\vec{\eta}_{j_1}^{\top}\vec{\eta}_{j_2}=\widetilde{\vec{\eta}}_{j_1}^{\top}\widetilde{\vec{\eta}}_{j_2}$.\\[3mm]
    \textbf{Step 6: }Fix the permutation.\\[3mm]
    For notational simplicity, we permute the order of indices $\left\{1,\ldots,J\right\}$ on both sides of (\ref{eq_thm_1.10}) in the previous steps. So far, we have only proved that there exists a permutation ${\pi}:~\left\{1,\ldots,J \right\}\rightarrow\left\{1,\ldots,J \right\}$ such that for any $j,j_1,j_2=1,\ldots,J$ and $k=1,\ldots,W$, there hold $\mu_{j}=\widetilde{\mu}_{\pi(j)}$, $\vec{\eta}_j^{\top}\vec{\alpha}_k=\widetilde{\vec{\eta}}_{\pi(j)}^{\top}\vec{\alpha}_k$ and $\vec{\eta}_{j_1}^{\top}\vec{\eta}_{j_2}=\widetilde{\vec{\eta}}_{\pi(j_1)}^{\top}\widetilde{\vec{\eta}}_{\pi(j_2)}$.\\[3mm]
    We then prove that for any $j=1,\ldots,J$, $\vec{\eta}_{j}^{\top}\vec{\theta}(\mathbf{0})=\widetilde{\vec{\eta}}_{j}^{\top}\widetilde{\vec{\theta}}(\mathbf{0})$. If this is not the case, we assume WLOG that $\vec{\eta}_{j}^{\top}\vec{\theta}(\mathbf{0})>\widetilde{\vec{\eta}}_{j}^{\top}\widetilde{\vec{\theta}}(\mathbf{0})$. We redefine $\widetilde{f}_n$ and $\widetilde{\phi}_n$ as
    \begin{align*}
        \widetilde{f}_n(\vec{\theta}|\bm{\xi}^{(n)})=&n\widetilde{\mu}_{\pi(1)}-\sum_{k=1}^{W}w_k \exp(\vec{\alpha}_k^{\top}\vec{\theta})+(\vec{\varphi}+n\widetilde{\vec{\eta}}_{\pi(1)})^{\top}\vec{\theta}\\
        &-\sum_{j=2}^J\xi_j^{(n)}\left[(\widetilde{\vec{\eta}}_{\pi(1)}-\widetilde{\vec{\eta}}_{\pi(j)})^{\top}\vec{\theta}+(\widetilde{\mu}_{\pi(1)}-\widetilde{\mu}_{\pi(j)})\right]-\frac{1}{2}\vec{\theta}^{\top}\vec{\theta},\\
        \widetilde{\phi}_n(\bm{\xi}^{(n)})=&\int_{\bt}(2\pi)^{-\frac{K}{2}}\exp(\widetilde{f}_n(\vec{\theta}|\bm{\xi}^{(n)}))\mathrm{d}\vec{\theta}.
    \end{align*}
    Then $\widetilde{f}_n$ and $\widetilde{\phi}_n$ match with the notation in previous step. There exists $l\in\left\{1,\ldots,J\right\}$ such that $\pi(l)=j$. By Corollary~\ref{cor_likelihood}, we have
    \begin{align*}
        \sum_{\bm{\xi}^{(n)}\in\mathcal{O}_n}\Delta_n(\bm{\xi}^{(n)})\phi_{n+1}(\xi_2^{(n)},\ldots,\xi_j^{(n)}+1,\ldots,\xi_J^{(n)})=\sum_{\bm{\xi}^{(n)}\in\mathcal{O}_n}\Delta_n(\bm{\xi}^{(n)})\widetilde{\phi}_{n+1}(\xi_2^{(n)},\ldots,\xi_l^{(n)}+1,\ldots,\xi_J^{(n)}).
    \end{align*}
    Similarly we can prove that for any $k\in\mathbb{N}$, it holds that:
    \begin{align}\label{eq_thm_1.38}
        &\sum_{\bm{\xi}^{(n)}\in\mathcal{O}_n}\Delta_n(\bm{\xi}^{(n)})\phi_{n+1}(\xi_2^{(n)},\ldots,\xi_j^{(n)}+1,\ldots,\xi_J^{(n)})\mathbf{1}\Big(\frac{1}{n}\bm{\xi}^{(n)}\in \mathcal{G}_k\Big)\notag\\
        \asymp&\sum_{\bm{\xi}^{(n)}\in\mathcal{O}_n}\Delta_n(\bm{\xi}^{(n)})\widetilde{\phi}_{n+1}(\xi_2^{(n)},\ldots,\xi_l^{(n)}+1,\ldots,\xi_J^{(n)})\mathbf{1}\Big(\frac{1}{n}\bm{\xi}^{(n)}\in \mathcal{G}_k\Big).
    \end{align}
    Moreover, we have proved in Steps 1-5 that
    \begin{align}\label{eq_thm_1.39}
        \sum_{\bm{\xi}^{(n)}\in\mathcal{O}_n}\Delta_n(\bm{\xi}^{(n)})\phi_{n}(\bm{\xi}^{(n)})\mathbf{1}\Big(\frac{1}{n}\bm{\xi}^{(n)}\in \mathcal{G}_k\Big)\asymp\sum_{\bm{\xi}^{(n)}\in\mathcal{O}_n}\Delta_n(\bm{\xi}^{(n)})\widetilde{\phi}_{n}(\bm{\xi}^{(n)})\mathbf{1}\Big(\frac{1}{n}\bm{\xi}^{(n)}\in \mathcal{G}_k\Big).
    \end{align}
    By the continuity of $\vec{\theta}(\nu_2,\ldots,\nu_J)$, we fix $k$ large enough such that
    \begin{align}\label{eq_thm_1.40}
        \underset{(\bar{\nu}_2,\ldots,\bar{\nu}_J)^{\top}\in\mathcal{G}_k}{\min}\vec{\eta}_{j}^{\top}\vec{\theta}(\bar{\nu}_2,\ldots,\bar{\nu}_J)> \underset{(\bar{\nu}_2,\ldots,\bar{\nu}_J)^{\top}\in\mathcal{G}_k}{\max}\widetilde{\vec{\eta}}_{j}^{\top}\widetilde{\vec{\theta}}(\bar{\nu}_2,\ldots,\bar{\nu}_J).
    \end{align}
    For any $n$, we define
    \begin{align*}
        \bar{\bm{\xi}}^{(n)}=\underset{\frac{1}{n}\bm{\xi}^{(n)}\in\mathcal{G}_k}{\operatorname{argmin}}\frac{\phi_{n+1}(\xi_2^{(n)},\ldots,\xi_j^{(n)}+1,\ldots,\xi_J^{(n)})}{\phi_{n}(\bm{\xi}^{(n)})}.
    \end{align*}
    Furthermore, assume WLOG that $\bar{\bm{\xi}}^{(n)}/n\rightarrow(\bar{\nu}_2,\ldots,\bar{\nu}_J)\in\mathcal{G}_k$. So we have
    \begin{align}\label{eq_thm_1.41}
        \varliminf_{n\rightarrow \infty}\frac{1}{n}\vec{\eta}_j^{\top}\vec{\theta}_n(\bar{\bm{\xi}}^{(n)})\geq \underset{(\bar{\nu}_2,\ldots,\bar{\nu}_J)^{\top}\in\mathcal{G}_k}{\min}\vec{\eta}_{j}^{\top}\vec{\theta}(\bar{\nu}_2,\ldots,\bar{\nu}_J).
    \end{align}
    Moreover, it can be readily shown that $\vec{\theta}_n(\bar{\xi}_2^{(n)},\ldots,\bar{\xi}_j^{(n)}+1,\ldots,\bar{\xi}_J^{(n)})-\vec{\theta}_n(\bar{\bm{\xi}}^{(n)})=\mathcal{O}(1)$ and 
    \begin{align}\label{eq_thm_1.42}
        &\operatorname{det}\left(-\nabla^2f_{n+1}(\vec{\theta}_n(\bar{\xi}_2^{(n)},\ldots,\bar{\xi}_j^{(n)}+1,\ldots,\bar{\xi}_J^{(n)})\big|\bar{\xi}_2^{(n)},\ldots,\bar{\xi}_j^{(n)}+1,\ldots,\bar{\xi}_J^{(n)})\right)\notag\\
        \asymp&\operatorname{det}\left(-\nabla^2f_{n}(\vec{\theta}_n(\bar{\bm{\xi}}^{(n)})\big|\bar{\bm{\xi}}^{(n)})\right).
    \end{align}
    By (\ref{eq_thm_1.11}), (\ref{eq_thm_1.41}) and (\ref{eq_thm_1.42}), we have
    \begin{align}\label{eq_thm_1.43}
        &\frac{\sum_{\bm{\xi}^{(n)}\in\mathcal{O}_n}\Delta_n(\bm{\xi}^{(n)})\phi_{n+1}(\xi_2^{(n)},\ldots,\xi_j^{(n)}+1,\ldots,\xi_J^{(n)})\mathbf{1}\Big(\frac{1}{n}\bm{\xi}^{(n)}\in \mathcal{G}_k\Big)}{\sum_{\bm{\xi}^{(n)}\in\mathcal{O}_n}\Delta_n(\bm{\xi}^{(n)})\phi_{n}(\bm{\xi}^{(n)})\mathbf{1}\Big(\frac{1}{n}\bm{\xi}^{(n)}\in \mathcal{G}_k\Big)}\notag\\
        \geq&\frac{\phi_{n+1}(\bar{\xi}_2^{(n)},\ldots,\bar{\xi}_{j}^{(n)}+1,\ldots,\bar{\xi}_J^{(n)})}{\phi_{n}(\bar{\bm{\xi}}^{(n)})}\notag\\
        \gtrsim&\frac{\exp\left(f_{n+1}\left(\vec{\theta}_n(\bar{\bm{\xi}}^{(n)})\big|\bar{\xi}_2^{(n)},\ldots,\bar{\xi}_j^{(n)}+1,\ldots,\bar{\xi}_J^{(n)}\right)\right)}{\exp\big(f_{n+1}(\vec{\theta}_n(\bar{\bm{\xi}}^{(n)})\big|\bar{\bm{\xi}}^{(n)}\big)}\notag\\
        =&\exp\left(o(n)+n\underset{(\bar{\nu}_2,\ldots,\bar{\nu}_J)^{\top}\in\mathcal{G}_k}{\min}\vec{\eta}_{j}^{\top}\vec{\theta}(\bar{\nu}_2,\ldots,\bar{\nu}_J)\right).
    \end{align}
    Similarly, for $n$ large enough we have
    \begin{align}\label{eq_thm_1.44}
        &\frac{\sum_{\bm{\xi}^{(n)}\in\mathcal{O}_n}\Delta_n(\bm{\xi}^{(n)})\widetilde{\phi}_{n+1}(\xi_2^{(n)},\ldots,\xi_l^{(n)}+1,\ldots,\xi_J^{(n)})\mathbf{1}\Big(\frac{1}{n}\bm{\xi}^{(n)}\in \mathcal{G}_k\Big)}{\sum_{\bm{\xi}^{(n)}\in\mathcal{O}_n}\Delta_n(\bm{\xi}^{(n)})\widetilde{\phi}_{n}(\bm{\xi}^{(n)})\mathbf{1}\Big(\frac{1}{n}\bm{\xi}^{(n)}\in \mathcal{G}_k\Big)}\notag\\
        \lesssim&\exp\left(o(n)+n\underset{(\bar{\nu}_2,\ldots,\bar{\nu}_J)^{\top}\in\mathcal{G}_k}{\max}\widetilde{\vec{\eta}}_{l}^{\top}\widetilde{\vec{\theta}}(\bar{\nu}_2,\ldots,\bar{\nu}_J)\right).
    \end{align}
    Then (\ref{eq_thm_1.38}), (\ref{eq_thm_1.39}), (\ref{eq_thm_1.43}) and (\ref{eq_thm_1.44}) lead to contradiction. So for any $j=1,\ldots,J$, $\vec{\eta}_{j}^{\top}\vec{\theta}(\mathbf{0})=\widetilde{\vec{\eta}}_{\pi(j)}^{\top}\widetilde{\vec{\theta}}(\mathbf{0})$. Since $\vec{\eta}_{1}^{\top}\vec{\theta}(\mathbf{0}),\ldots,\vec{\eta}_{J}^{\top}\vec{\theta}(\mathbf{0})$ are distinct, we have $\pi=id$ and $\mu_j=\widetilde{\mu}_j$, $\vec{\eta}_j^{\top}\vec{\alpha}_k= \widetilde{\vec{\eta}}_j^{\top}\vec{\alpha}_k$ and $\vec{\eta}_{j_1}^{\top}\vec{\eta}_{j_2}=\widetilde{\vec{\eta}}_{j_1}^{\top}\widetilde{\vec{\eta}}_{j_2}$ for any $j,j_1,j_2=1,\ldots,J$ and $k=1,\ldots,W$. Hence the result is proved on $[0,t_{q+1}]$.\\[3mm]
    \textbf{Case 2: }$\|P_{\mathcal{H}^{\perp}_{\vec{\eta}_1}}\vec{\eta}_1\|=\|P_{\mathcal{H}^{\perp}_{\widetilde{\vec{\eta}}_{1}}}\widetilde{\vec{\eta}}_{1}\|=0$.\\[3mm] In this case, $\vec{\eta}_1,\ldots,\vec{\eta}_J,\widetilde{\vec{\eta}}_1,\ldots,\widetilde{\vec{\eta}}_J\in \mathcal{X}:= \{\sum_{k=1}^W c_k\vec{\alpha}_k:c_1,\ldots,c_W\geq 0\}$ by Proposition~\ref{prop_canonical_projection}. By Proposition~\ref{lem_canonical_expansion}, for any $j=1,\ldots,J$, there exists canonical expansions for $\vec{\eta}_j$ and $\widetilde{\vec{\eta}}_j$ under $\vec{\alpha}_1,\ldots,\vec{\alpha}_W$ as: $\vec{\eta}_j=\sum_{k=1}^{m_j}c_{j,k}\vec{\alpha}_{j,k}$ and $\widetilde{\vec{\eta}}_j=\sum_{k=1}^{\widetilde{m}_j}\widetilde{c}_{j,k}\widetilde{\vec{\alpha}}_{j,k}$, where the canonical expansion is unique in the sense that $\sum_{k=1}^{m_j}c_{j,k}$, $\sum_{k=1}^{\widetilde{m}_j}\widetilde{c}_{j,k}$ are uniquely determined for each $j=1,\ldots,J$.\\[3mm]
    We assume WLOG that
    \begin{align*}
        \sum_{k=1}^{m_1}c_{1,k}&=\underset{j=1,\ldots,J}{\max}\sum_{k=1}^{m_j}c_{j,k},\\
        \sum_{k=1}^{\widetilde{m}_1}\widetilde{c}_{1,k}&=\underset{j=1,\ldots,J}{\max}\sum_{k=1}^{\widetilde{m}_j}\widetilde{c}_{j,k}.
    \end{align*}
    We first discuss the case where $\sum_{k=1}^{m_1}c_{1,k}$ and $\sum_{k=1}^{\widetilde{m}_1}\widetilde{c}_{1,k}$ are the unique maximisers, respectively.\\[3mm]
    \textbf{Step 1: }For $\bm{\xi}^{(n)}\in\mathcal{O}_n$, we bound the denominator part in (\ref{eq_thm_1.11}) by $\exp\left(\mathcal{O}(\log n)\right)$ uniformly.\\[3mm]
    For any $n\in\mathbb{N}$, define
    \begin{align*}
        \bar{\bm{\xi}}^{(n)}=\underset{\bm{\xi}\in\mathcal{O}_n}{\operatorname{argmax}}\operatorname{det}\left(-\nabla^2 f_n(\vec{\theta}_n(\bm{\xi})\big|\bm{\xi})\right).
    \end{align*}
    Denote $l_n=\|\vec{\theta}_n(\bar{\bm{\xi}}^{(n)})\|$ and $\vec{\epsilon}_n=\vec{\theta}_n(\bar{\bm{\xi}}^{(n)})/l_n\rightarrow \vec{\epsilon}$. If $l_n$ is bounded, then it can be readily shown that $\operatorname{det}\left(-\nabla^2 f_n(\vec{\theta}_n(\bm{\xi}^{(n)})\big|\bm{\xi}^{(n)})\right)$ is also bounded. If $l_n$ is not bounded, we assume WLOG that $l_n\rightarrow \infty$. Since $\vec{\eta}_1,\ldots,\vec{\eta}_J\in \mathcal{X}$, we can use the same proof as in Proposition~\ref{prop_characterisation_equation} to show that $l_n=\mathcal{O}(\log n)$. So there exists $\widetilde{M}>0$ such that for any $n\in\mathbb{N}$,
    \begin{align*}
        {\max}_{k=1,\ldots,W}\vec{\alpha}_k^{\top}\vec{\theta}_n(\bar{\bm{\xi}}^{(n)})\leq \widetilde{M}\log n.
    \end{align*}
    Similarly, we define $\widetilde{\bm{\xi}}^{(n)}$ and perform the same argument. Then there exists $M>0$ such that for any $n\in\mathbb{N}$ and $\bm{\xi}^{(n)}\in\mathcal{O}_n$, it holds that
    \begin{align}\label{eq_thm_1.45}
        1\leq \sqrt{\operatorname{det}\left(-\nabla^2 f_n(\vec{\theta}_n(\bm{\xi}^{(n)})\big|\bm{\xi}^{(n)})\right)}\leq& \sqrt{\operatorname{det}\left(-\nabla^2 f_n(\vec{\theta}_n(\bar{\bm{\xi}}^{(n)})\big|\bar{\bm{\xi}}^{(n)})\right)}\leq n^{M},\notag\\
        1\leq \sqrt{\operatorname{det}\left(-\nabla^2 \widetilde{f}_n(\widetilde{\vec{\theta}}_n(\bm{\xi}^{(n)})\big|\bm{\xi}^{(n)})\right)}\leq& \sqrt{\operatorname{det}\left(-\nabla^2 \widetilde{f}_n(\widetilde{\vec{\theta}}_n(\widetilde{\bm{\xi}}^{(n)})\big|\widetilde{\bm{\xi}}^{(n)})\right)}\leq n^{M}.
    \end{align}
    \textbf{Step 2: }Construct the concentration points on both sides of (\ref{eq_thm_1.10}).\\[3mm]
    We define $\log 0=0$. For notational simplicity, for $\bm{\xi}^{(n)}=(\xi_2^{(n)},\ldots,\xi_J^{(n)})^{\top}$, denote $\log \bm{\xi}^{(n)}=(\log \xi_2^{(n)},\ldots,\log \xi_J^{(n)})^{\top}$. Let $\mathcal{G}_{0}=[0,1]^{J-1}$. We define $\left\{\mathcal{G}_{k}:k\in\mathbb{N}_0\right\}$ in the following inductive method: Suppose $\mathcal{G}_{k-1}$ is constructed, we partition $\mathcal{G}_{k-1}$ into $2^{J-1}$ identical hypercubes $\mat{D}_{1}^{(k)},\ldots,\mat{D}_{2^{(J-1)}}^{(k)}$ with length $2^{-k}$ on each side. For any $n$ and $i=1,\ldots,2^{J-1}$, denote
    \begin{align*}
        S_{i,k,n}=\sum_{\bm{\xi}^{(n)}\in\mathcal{O}_n}\Delta_n(\bm{\xi}^{(n)})\phi_n(\bm{\xi}^{(n)})\mathbf{1}\Big(\frac{1}{\log n}\log \bm{\xi}^{(n)}\in \overline{\mat{D}_{i}^{(k)}}\Big).
    \end{align*}
    We define $\mathcal{G}_{k}=\overline{\mat{D}_{j_k}^{(k)}}$, which satisfies $S_{j_k,k,n}=\underset{i=1,\ldots,2^{J-1}}{\max}S_{i,k,n}$~i.o. Then there exists a unique $(\nu_2,\ldots,\nu_J)^{\top}\in [0,1]^{J-1}$ such that
    \begin{align*}
        (\nu_2,\ldots,\nu_J)^{\top}\in \bigcap_{k=0}^{\infty}{\mathcal{G}_{k}}.
    \end{align*}
    We call this point the concentration point of the left side of (\ref{eq_thm_1.10}). Similarly we can define concentration point $(\widetilde{\nu}_2,\ldots,\widetilde{\nu}_J)^{\top}$ for the right side of (\ref{eq_thm_1.10}) and the corresponding hypercube sequence $\{\widetilde{\mathcal{G}}_k:k\in\mathbb{N}_0\}$ for the right side of (\ref{eq_thm_1.10}). For notational simplicity, for any $k\in\mathbb{N}$ and $n\in\mathbb{N}\setminus\left\{1\right\}$, define $\mathcal{E}_{k,n}$ and $\widetilde{\mathcal{E}}_{k,n}$ as
    \begin{align*}
        \mathcal{E}_{k,n}=&\big\{\bm{\xi}\in \mathcal{O}_n:\frac{1}{\log n}\log \bm{\xi}\in \mathcal{G}_k\big\},\\
        \widetilde{\mathcal{E}}_{k,n}=&\big\{\bm{\xi}\in \mathcal{O}_n:\frac{1}{\log n}\log \bm{\xi}\in \widetilde{\mathcal{G}}_k\big\}.
    \end{align*}
    Similar to Step 2 in Case 1, for any $k\in\mathbb{N}_0$, we have
    \begin{align}\label{eq_thm_1.46}
        \sum_{\bm{\xi}^{(n)}\in\mathcal{E}_{k,n}}\Delta_n(\bm{\xi}^{(n)})\phi_n(\bm{\xi}^{(n)})\asymp&\sum_{\bm{\xi}^{(n)}\in\mathcal{O}_n}\Delta_n(\bm{\xi}^{(n)})\phi_n(\bm{\xi}^{(n)})\notag\\
        =&\sum_{\bm{\xi}^{(n)}\in\mathcal{O}_n}\Delta_n(\bm{\xi}^{(n)})\widetilde{\phi}_n(\bm{\xi}^{(n)})\asymp\sum_{\bm{\xi}^{(n)}\in\widetilde{\mathcal{E}}_{k,n}}\Delta_n(\bm{\xi}^{(n)})\widetilde{\phi}_n(\bm{\xi}^{(n)}).
    \end{align}
    \textbf{Step 3: }Characterise $(\nu_2,\ldots,\nu_J)^{\top}$ and $(\widetilde{\nu}_2,\ldots,\widetilde{\nu}_J)^{\top}$.\\[3mm]
    Similar to Steps 3-5 in Case 1, we need the continuity property within the neighbourhood of the two concentration points. While the continuity in Case 1 holds without any conditions based on Proposition~\ref{prop_canonical_projection}, we need to verify two things in order to ensure the continuity in Case 2 according to Proposition~\ref{prop_characterisation_equation}. First, we need to verify the nondegeneracy condition which is needed in Proposition~\ref{prop_characterisation_equation}. However, the non-degeneracy condition is itself proved in reverse by equation (\ref{eq_thm_1.53}) that we want to obtain in Step 3, which is hard to prove without the continuity property. This urges us to shift our focus from treating the whole summation within hypercubes to treating the term at a single point. Moreover, we need to show that for $j$ such that $\nu_j=0$, $\vec{\eta}_j$ should appear only finite times in the dominant terms in the left side of (\ref{eq_thm_1.10}).\\[3mm]
    To overcome these difficulties, the sketch of Step 3 is as follows:
    \begin{enumerate}
        \item[(1)] We define the single point $\bar{\bm{\xi}}^{(k,n)}$ which achieves the largest summand $\Delta_n(\bm{\xi}^{(n)})\phi_n(\bm{\xi}^{(n)})$ in each  $\mathcal{E}_{k,n}$ and obtain the limiting point $({\nu}_2^{(k)},\ldots,{\nu}_J^{(k)})^{\top}$ in each hypercube $\mathcal{G}_k$.
        \item[(2)] For any $j$ such that $\nu_j^{(k)}>0$, we obtain the equation on $\vec{\eta}_j$ and $\nu_j^{(k)}$ based on the maximum property.
        \item[(3)] We construct a generalised characterisation equation for $\hat{\vec{\theta}}_k$.
        \item[(4)] For any $j$ such that $\nu_j^{(k)}=0$, we prove the inequality for $\vec{\eta}_j$ and $\nu_j^{(k)}$ based on the maximum property. Then we exclude the cases where the occurrence of $\vec{\eta}_j$ among $\bar{\bm{\xi}}^{(k,n)}$ is nonzero for any fixed $k$ in the generalised characterisation equation.
        \item[(5)] We verify that the generalised characterisation equation obtained in the previous step is a characterisation equation. Since $(\nu_2,\ldots,\nu_J)^{\top}\in \bigcap_{k=0}^{\infty}{\mathcal{G}_{k}}$, the limiting point $({\nu}_2^{(k)},\ldots,{\nu}_J^{(k)})^{\top}$ should also converge to $(\nu_2,\ldots,\nu_J)^{\top}$ as $k$ goes to infinity. Then the characterisation equations at $({\nu}_2^{(k)},\ldots,{\nu}_J^{(k)})^{\top}$ should converge to characterisation equation at $(\nu_2,\ldots,\nu_J)^{\top}$, which verifies the nondegeneracy condition. This implies that $\hat{\vec{\theta}}_k=\vec{\theta}(\nu_2^{(k)},\ldots,\nu_J^{(k)})$.
        \item[(6)] Finally, by the continuity property, the equality and inequality also converge to equality and inequality at $(\nu_2,\ldots,\nu_J)^{\top}$. The equality case in the inequality is eliminated after excluding a set of measure zero in the parameter space.
    \end{enumerate}
    We first characterise $(\nu_2,\ldots,\nu_J)^{\top}$. We assume WLOG that $\nu_2,\ldots,\nu_p>0$ and $\nu_{p+1}=\ldots=\nu_{J}=0$. For any $k,n\in\mathbb{N}$, denote
    \begin{align*}
        \bar{\bm{\xi}}^{(k,n)}=\underset{\bm{\xi}\in \mathcal{E}_{k,n}}{\operatorname{argmax}}\Delta_n(\bm{\xi})\phi_n(\bm{\xi}).
    \end{align*}
    For any fixed $k$, by arguments similar to those used in the proof of Proposition~\ref{prop_canonical_projection}, we can prove that $\vec{\theta}_n(\bar{\bm{\xi}}^{(k,n)})=\mathcal{O}(\log n)$. Then we denote
    \begin{align*}
        \lim_{n\rightarrow \infty}\frac{\vec{\theta}_n(\bar{\bm{\xi}}^{(k,n)})}{\log n}=\hat{\vec{\theta}}_k\quad\text{and}\quad \lim_{n\rightarrow \infty}\frac{\log \bar{\bm{\xi}}^{(k,n)}}{\log n}=(\nu_2^{(k)},\ldots,\nu_J^{(k)})^{\top}.
    \end{align*}
    Since $(\nu_2^{(k)},\ldots,\nu_J^{(k)})^{\top}\in\mathcal{G}_k$, we have $(\nu_2^{(k)},\ldots,\nu_J^{(k)})^{\top}\rightarrow (\nu_2,\ldots,\nu_J)^{\top}$ as $k$ goes to infinity by the definition of $\mathcal{G}_k$. We assume WLOG that $\nu_1^{(k)}=1$, $\nu_2^{(k)},\ldots,\nu_{p_1}^{(k)}>0$ and $\nu_{p_1+1}^{(k)}=\ldots=\nu_{J}^{(k)}=0$ for any $k\in\mathbb{N}$, where $1\leq p\leq p_1\leq J$.\\[3mm]
    \textbf{Step 3.1: }We first prove that $\sum_{j=2}^J\bar{\xi}_j^{(k,n)}\ll n$ by the method of contradiction. If this is not the case, then $\sum_{j=2}^J\bar{\xi}_j^{(k,n)}\asymp n$. We assume WLOG that $\sum_{j=2}^J\bar{\xi}_j^{(k,n)}\geq \delta n$ for $n$ large where $\delta$ is a positive constant. By the proof in Proposition~\ref{prop_canonical_projection}, we have $\max_{k=1,\ldots,W}\vec{\alpha}_k^{\top}\hat{\vec{\theta}}_k=1$. Hence we have
    \begin{align*}
        &\frac{\sum_{\bm{\xi}^{(n)}\in\mathcal{E}_{k,n}}\Delta_n(\bm{\xi}^{(n)})\phi_n(\bm{\xi}^{(n)})}{\Delta_n(\mathbf{0})\phi_n(\mathbf{0})}\\
        \leq&\exp(o(n\log n))\frac{\Delta_n(\bar{\vec{\xi}}^{(k,n)})\phi_n(\bar{\vec{\xi}}^{(k,n)})}{\Delta_n(\mathbf{0})\phi_n(\mathbf{0})}\\
        \lesssim& \frac{\exp\Big(o(n\log n)+\log n\big[\sum_{j=2}^J\bar{\xi}_j^{(k,n)}\sum_{k=1}^{m_j}c_{j,k}+\big(n-\sum_{j=2}^J\bar{\xi}_j^{(k,n)}\big)\sum_{k=1}^{m_1}c_{1,k}\big]\Big)}{\exp\Big(n\log n\sum_{k=1}^{m_1}c_{1,k}\Big)}\\
        \lesssim&\exp\left(o(n\log n)+n\log n\left[\delta\max_{j=2,\ldots,J}\sum_{k=1}^{m_j}c_{j,k}+(1-\delta)\sum_{k=1}^{m_1}c_{1,k}-\sum_{k=1}^{m_1}c_{1,k}\right]\right)\\
        =&\exp\left(o(n\log n)+\delta n\log n\left[\max_{j=2,\ldots,J}\sum_{k=1}^{m_j}c_{j,k}-\sum_{k=1}^{m_1}c_{1,k}\right]\right)\\
        \rightarrow&0,
    \end{align*}
    which contradicts (\ref{eq_thm_1.46}).\\[3mm]
    \textbf{Step 3.2: }We then prove that for any $j=1,\ldots,p_1$, it holds that
    \begin{align*}
        1-\nu_j^{(k)}=(\vec{\eta}_1-\vec{\eta}_j)^{\top}\hat{\vec{\theta}}_k.
    \end{align*}
    If this is not the case for $j$, we first discuss the case when $1-\nu_j^{(k)}<(\vec{\eta}_1-\vec{\eta}_j)^{\top}\hat{\vec{\theta}}_k$. Suppose $(\vec{\eta}_1-\vec{\eta}_j)^{\top}\hat{\vec{\theta}}_k-(1-\nu_j^{(k)})=\delta$, where $\delta>0$ is a constant. We choose $k$ large enough and fix $\widetilde{\nu}_j^{(k)}=\nu_j^{(k)}-\delta/2$. Denote $\widetilde{\bm{\xi}}^{(k,n)}=(\bar{\xi}_2^{(k,n)},\ldots,n^{\widetilde{\nu}_j^{(k)}},\ldots,\bar{\xi}_J^{(k,n)})^{\top}$. By the Stirling formula, we have
    \begin{align}\label{eq_thm_1.47}
        &\frac{\Delta_n(\widetilde{\bm{\xi}}^{(k,n)})}{\Delta_n(\bar{\bm{\xi}}^{(k,n)})}\notag\\
        \asymp&\sqrt{\frac{(n-\sum_{j=2}^J\bar{\xi}_j^{(k,n)})\prod_{j=2}^J\bar{\xi}_j^{(k,n)}}{(n-\sum_{l\neq j}\bar{\xi}_l^{(k,n)}-n^{\widetilde{\nu}_j^{(k)}})n^{\widetilde{\nu}_j^{(k)}}\prod_{l\neq j}\bar{\xi}_l^{(k,n)}}}\notag\\
        &\times\exp\Bigg(-\big(n-\underset{l\neq j}{\sum}\bar{\xi}_l^{(k,n)}-n^{\widetilde{\nu}_j^{(k)}}\big)\log \big(n-\underset{l\neq j}{\sum}\bar{\xi}_l^{(k,n)}-n^{\widetilde{\nu}_j^{(k)}}\big)\notag\\
        &-\underset{l\neq j}{\sum}\bar{\xi}_l^{(k,n)}\log \bar{\xi}_l^{(k,n)}-n^{\widetilde{\nu}_j^{(k)}}\log n^{\widetilde{\nu}_j^{(k)}}+\big(n-\sum_{j=2}^J \bar{\xi}_j^{(k,n)}\big)\log \big(n-\sum_{j=2}^J \bar{\xi}_j^{(k,n)}\big)+\sum_{j=2}^J \bar{\xi}_j^{(k,n)} \log \bar{\xi}_j^{(k,n)}\Bigg)\notag\\
        \asymp&\sqrt{\frac{\bar{\xi}_j^{(k,n)}}{n^{\widetilde{\nu}_j^{(k)}}}}\exp\left(-(\bar{\xi}_j^{(k,n)}-n^{\widetilde{\nu}_j^{(k)}})\log n+(\bar{\xi}_j^{(k,n)} -n^{\widetilde{\nu}_j^{(k)}})\log n^{\widetilde{\nu}_j^{(k)}}\right)\notag\\
        =&\exp\left(o(\log n)-(\bar{\xi}_j^{(k,n)}-n^{\widetilde{\nu}_j^{(k)}})(1-\widetilde{\nu}_j^{(k)})\log n\right).
    \end{align}
    Then by (\ref{eq_thm_1.11}), (\ref{eq_thm_1.45}) and (\ref{eq_thm_1.47}) we have
    \begin{align}\label{eq_thm_1.48}
        &\frac{\Delta_n(\widetilde{\bm{\xi}}^{(k,n)})\phi_n(\widetilde{\bm{\xi}}^{(k,n)})}{\Delta_n(\bar{\bm{\xi}}^{(k,n)})\phi_n(\bar{\bm{\xi}}^{(k,n)})}\notag\\
        \gtrsim &n^{-M}\frac{\exp\big(f_n(\vec{\theta}_n(\bar{\bm{\xi}}^{(k,n)})\big|\widetilde{\bm{\xi}}^{(k,n)})\big)}{\exp\big(f_n(\vec{\theta}_n(\bar{\bm{\xi}}^{(k,n)})\big|\bar{\bm{\xi}}^{(k,n)})\big)}\exp\big(o(\log n)-(\bar{\xi}_j^{(k,n)}-n^{\widetilde{\nu}_j^{(k)}})(1-\widetilde{\nu}_j^{(k)})\log n\big)\notag\\
        \geq &\exp\Big(\mathcal{O}(\log n)+(\bar{\xi}_j^{(k,n)}-n^{\widetilde{\nu}_j^{(k)}})\left[(\vec{\eta}_1-\vec{\eta}_{j})^{\top}\vec{\theta}_n(\bar{\bm{\xi}}^{(k,n)})+(\mu_1-\mu_j)\right]-(\bar{\xi}_j^{(k,n)}-n^{\widetilde{\nu}_j^{(k)}})(1-\widetilde{\nu}_j^{(k)})\log n\Big)\notag\\
        =&\exp\Big(\mathcal{O}(\log n)+(\bar{\xi}_j^{(k,n)}-n^{\widetilde{\nu}_j^{(k)}})\log n\big[(\vec{\eta}_1-\vec{\eta}_{j})^{\top}\hat{\vec{\theta}}_k-(1-\widetilde{\nu}_j^{(k)})\big]\Big)\notag\\
        \geq&\exp\left(\frac{\delta}{2}(\bar{\xi}_j^{(k,n)}-n^{\widetilde{\nu}_j^{(k)}})\log n\right).
    \end{align}
    Since $\log \bar{\xi}_j^{(k,n)}/\log n\rightarrow {\nu}_j^{(k)}$, it holds that $n^{\widetilde{\nu}_j^{(k)}}\ll \bar{\xi}_j^{(k,n)}$ for $n$ large enough. Suppose $({\nu}_2^{(k)},\ldots,\widetilde{\nu}_j^{(k)},\ldots,{\nu}_J^{(k)})^{\top}$ belongs to hypercube $\hat{\mathcal{G}}_k$ with length $2^{-k}$ on each side. Suppose that $k$ is chosen large enough, then we have $\hat{\mathcal{G}}_k\neq {\mathcal{G}}_k$. So by (\ref{eq_thm_1.11}), (\ref{eq_thm_1.45}) and the definition of $\bar{\bm{\xi}}^{(k,n)}$, we have
    \begin{align*}
        \frac{\sum_{\bm{\xi}^{(n)}\in\mathcal{O}_n}\Delta_n(\bm{\xi}^{(n)})\phi_n(\bm{\xi}^{(n)})\mathbf{1}\Big(\frac{1}{\log n}\log \bm{\xi}^{(n)}\in \hat{\mathcal{G}}_k\Big)}{\sum_{\bm{\xi}^{(n)}\in\mathcal{O}_n}\Delta_n(\bm{\xi}^{(n)})\phi_n(\bm{\xi}^{(n)})\mathbf{1}\Big(\frac{1}{\log n}\log \bm{\xi}^{(n)}\in {\mathcal{G}}_k\Big)}\geq n^{-J}\frac{\Delta_n(\widetilde{\bm{\xi}}^{(k,n)})\phi_n(\widetilde{\bm{\xi}}^{(k,n)})}{\Delta_n(\bar{\bm{\xi}}^{(k,n)})\phi_n(\bar{\bm{\xi}}^{(k,n)})}\gg 1,
    \end{align*}
    which contradicts the construction method of $(\nu_2,\ldots,\nu_J)^{\top}$. If $\nu_j^{(k)}<1$, then we can use the same method to prove the other side. If $\nu_j^{(k)}=1$, we have $(\bar{\xi}_2^{(k,n)},\ldots,\bar{\xi}_j^{(k,n)}+1,\ldots,\bar{\xi}_J^{(k,n)})^{\top}\in\mathcal{E}_{k,n}$ since $\sum_{j=2}^J\bar{\xi}_j^{(k,n)}\ll n$. Then we can similarly prove that
    \begin{align*}
        \frac{\Delta_n(\bar{\xi}_2^{(k,n)},\ldots,\bar{\xi}_j^{(k,n)}+1,\ldots,\bar{\xi}_J^{(k,n)})\phi_n(\bar{\xi}_2^{(k,n)},\ldots,\bar{\xi}_j^{(k,n)}+1,\ldots,\bar{\xi}_J^{(k,n)})}{\Delta_n(\bar{\bm{\xi}}^{(k,n)})\phi_n(\bar{\bm{\xi}}^{(k,n)})}\gg 1
    \end{align*}
    if $1-\nu_j^{(k)}<(\vec{\eta}_1-\vec{\eta}_j)^{\top}\hat{\vec{\theta}}_k$. This contradicts the definition of $\bar{\bm{\xi}}^{(k,n)}$. Hence we proved that for any $j=1,\ldots,p_1$ and any $k\in\mathbb{N}$, we have
    \begin{align}\label{eq_thm_1.49}
        1-\nu_j^{(k)}=(\vec{\eta}_1-\vec{\eta}_j)^{\top}\hat{\vec{\theta}}_k.
    \end{align}
    \textbf{Step 3.3: }Similar to Step 3.2, we can prove that for any $j=p_1+1,\ldots,J$, $(\vec{\eta}_1-\vec{\eta}_j)^{\top}\hat{\vec{\theta}}_k\geq 1$. We assume WLOG that for any $j=p_1+1,\ldots,p_4$, it holds that $(\vec{\eta}_1-\vec{\eta}_j)^{\top}\hat{\vec{\theta}}_k= 1$, for any $j=p_4+1,\ldots,J$, it holds that $(\vec{\eta}_1-\vec{\eta}_j)^{\top}\hat{\vec{\theta}}_k>1$ for any $k\in\mathbb{N}$. Moreover, for $j=p_1+1,\ldots,p_4$, we assume WLOG that for any $k\in\mathbb{N}$, it holds that:
        \begin{align*}
            \lim_{n\rightarrow \infty}\frac{\bar{\xi}_j^{(k,n)}}{\log n}=\begin{cases}+\infty&\text{ for }j=p_1+1,\ldots,p_2\\
                \hat{c}_{j,k}\in(0,\infty)&\text{ for }j=p_2+1,\ldots,p_3\\
                0&\text{ for }j=p_3+1,\ldots,p_4
            \end{cases}.
        \end{align*}
        For $j=p_4+1,\ldots,J$, we can easily prove that for any $k\in\mathbb{N}$, it holds that $\bar{\xi}_j^{(k,n)}=0$ for $n$ large enough.\\[3mm]
        For any fixed $k\in\mathbb{N}$, we use similar method as in the proof of Lemma~\ref{lem_characterisation_equation_3} to create a linear equation for $\hat{\vec{\theta}}_k$. The same notation as in Lemma~\ref{lem_characterisation_equation_3} is used. If $\nu_2^{(k)}>\ldots>\nu_{p_1}^{(k)}$, then we can use exactly the same method to set up equations for $\vec{\eta}_1,\ldots,\vec{\eta}_{p_1}$ and $(\nu_1^{(k)},\ldots,\nu_{p_1}^{(k)})^{\top}$. If there exists tie among $(\nu_2^{(k)},\ldots,\nu_{p_1}^{(k)})^{\top}$, for example $\nu_2^{(k)}=\nu_3^{(k)}>\nu_4^{(k)}$, then the procedure falls into the following two cases:\\[3mm]
        \textbf{Case 1: }If $\bar{\xi}_2^{(k,n)}\gg \bar{\xi}_3^{(k,n)}$ or $\bar{\xi}_3^{(k,n)}\gg \bar{\xi}_2^{(k,n)}$, we assume WLOG that the first case holds. Then we project the first-order equation on $\mathcal{H}_1$ and divide both sides by $\bar{\xi}_2^{(k,n)}$:
        \begin{align*}
            -\sum_{k:\vec{\alpha}_k\in \mathcal{E}_{2}\setminus \mathcal{E}_{1}}\frac{\exp(\vec{\alpha}_k^{\top}\hat{\vec{\theta}}_k)}{\bar{\xi}_2^{(k,n)}}P_{\mathcal{H}_{1}^{\perp}}\vec{\alpha}_k+P_{\mathcal{H}_{1}^{\perp}}\vec{\eta}_{2}=o(1).
        \end{align*}
        So the same method can be performed to obtain the expansion of $\vec{\eta}_2$ and $\vec{\eta}_2\in\mathcal{H}_2$. Then we project the first-order equation on $\mathcal{H}_2$ and divide both sides by $\bar{\xi}_3^{(k,n)}$ to obtain the expansion of $\vec{\eta}_3$ and $\vec{\eta}_3\in\mathcal{H}_3$.\\[3mm]
        \textbf{Case 2: }If $\lim_{n\rightarrow\infty}\bar{\xi}_3^{(k,n)}/\bar{\xi}_2^{(k,n)}=c\in\left(0,\infty\right)$, then we project the first-order equation on $\mathcal{H}_1$ and divide both sides by $\bar{\xi}_2^{(k,n)}$:
        \begin{align*}
            -\sum_{k:\vec{\alpha}_k\in \mathcal{E}_{2}\setminus \mathcal{E}_{1}}\frac{\exp(\vec{\alpha}_k^{\top}\hat{\vec{\theta}}_k)}{\bar{\xi}_2^{(k,n)}}P_{\mathcal{H}_{1}^{\perp}}\vec{\alpha}_k+P_{\mathcal{H}_{1}^{\perp}}(\vec{\eta}_{2}+c\vec{\eta}_3)=o(1).
        \end{align*}
        By the same method we obtain $\mathcal{H}_2$ and $\vec{\eta}_2+c\vec{\eta}_3\in \mathcal{H}_2$. If the expansion of $\vec{\eta}_2+c\vec{\eta}_3$ is the sum of the degenerate expansions of $\vec{\eta}_2$ and $\vec{\eta}_3$, then the nondegeneracy condition in Proposition~\ref{prop_characterisation_equation} is satisfied. If this is not the case, then for any $\hat{c}$ in a small neighbourhood of $c$, we consider $\bar{\bm{\xi}}^{(n,\hat{c})}=(\xi_2^{(n,\hat{c})},\ldots,\xi_J^{(n,\hat{c})})$ such that
        \begin{align*}
            \xi_2^{(n,\hat{c})}=&\frac{1}{1+\hat{c}}\left(\bar{\xi}_2^{(k,n)}+\bar{\xi}_3^{(k,n)}\right),\\
            \xi_3^{(n,\hat{c})}=&\frac{\hat{c}}{1+\hat{c}}\left(\bar{\xi}_2^{(k,n)}+\bar{\xi}_3^{(k,n)}\right),\\
            \xi_j^{(n,\hat{c})}=&\bar{\xi}_j^{(k,n)},j=4,\ldots,J.
        \end{align*}
        We then construct characterisation equation at $\bar{\bm{\xi}}^{(n,\hat{c})}$ in a similar method. There are finitely many choices of $\hat{c}$ in the neighbourhood of $c$ such that $\vec{\eta}_2+\hat{c}\vec{\eta}_3$ can be spanned by less than $K-1$ linearly independent vectors in $\{\vec{\alpha}_1,\ldots,\vec{\alpha}_W\}$. Hence we assume WLOG that for any $\hat{c}>c$ in the neighbourhood of $c$, $\vec{\eta}_2+\hat{c}\vec{\eta}_3$ is spanned by the same basis $\mat{A}$. For any $\hat{c}<c$ in the neighbourhood of $c$, $\vec{\eta}_2+\hat{c}\vec{\eta}_3$ is spanned by the same basis $\widetilde{\mat{A}}$.\\[3mm]
        \textbf{Case 2.1: }We first consider the case where the coefficients in the  expansion of $\vec{\eta}_2+c\vec{\eta}_3$ under $\mat{A}$ and $\widetilde{\mat{A}}$ both contain zero component. Note that $\mat{A}\neq \widetilde{\mat{A}}$, otherwise the linear dependency between expansion coefficients and $\hat{c}$ will imply that the coefficient has negative components on one side, which contradicts the construction method of characterisation equation. Then this implies that the coefficients in the expansion of $\vec{\eta}_2+c\vec{\eta}_3$ under basis $\mat{A}$ and the coefficients in the expansion of $\widetilde{\mat{A}}^{-1}(\vec{\eta}_2+c\vec{\eta}_3)$ under basis $\widetilde{\mat{A}}$ both contain at least one zero entry. By excluding a set of measure zero in the parameter space, this cannot happen.\\[3mm]
        \textbf{Case 2.2: }If the coefficients in the expansion of $\vec{\eta}_2+c\vec{\eta}_3$ under either $\mat{A}$ or $\widetilde{\mat{A}}$ are all nonzero. Then by the continuity of coefficients in the expansion with respect to $\hat{c}$, we have $\operatorname{span}(\mat{A})=\operatorname{span}(\widetilde{\mat{A}})$ in the small neighbourhood of $c$. By arguments similar to those used in the proof of Lemma~\ref{lem_characterisation_equation_3}, we can expand $\log \phi_n(\bar{\bm{\xi}}^{(n,\hat{c})})$ in decreasing order, it can be readily shown that the coefficient of term $n^{\nu_2}\log n$ depends linearly on $c$. On the other hand, we can easily see from the Stirling formula that the terms in $\log \Delta_n(\bar{\bm{\xi}}^{(n,\hat{c})})$ that depends on $\hat{c}$ has smaller order than $n^{\nu_2}\log n$ by the definition of $\bar{\bm{\xi}}^{(n,\hat{c})}$. Hence by the maximum property of $\bar{\bm{\xi}}^{(n,\vec{c})}=\bar{\bm{\xi}}^{(k,n)}$, the linear coefficient of $n^{\nu_2}\log n$ should be equal to zero, which leads to a contradiction when a set of measure zero in the parameter space is excluded.\\[3mm]
        By the above discussion, the expansion of $\vec{\eta}_2+c\vec{\eta}_3$ is the sum of the degenerate expansions of $\vec{\eta}_2$ and $\vec{\eta}_3$, then the nondegeneracy condition in Proposition~\ref{prop_characterisation_equation} is satisfied. So we can construct the linear equation in the same way as in Case 1.\\[3mm]
        Now we have constructed $\vec{\eta}_1,\ldots,\vec{\eta}_{p_1}\in\mathcal{H}_{p_1}$. For $j=p_1+1,\ldots,p_2$, since $\bar{\xi}_j^{(k,n)}\gg \log n$, the linear equation is created in the same way as in $j=1,\ldots,p_1$. So we can obtain $\vec{\eta}_1,\ldots,\vec{\eta}_{p_2}\in\mathcal{H}_{p_2}$.\\[3mm]
        For $j=p_3+1,\ldots,J$, since $\bar{\xi}_j^{(k,n)}\ll \log n$, the term containing $\vec{\eta}_{j}$ vanishes when dividing the first-order equation by $\log n$.\\[3mm]
        Now we project the first-order equation on $\mathcal{H}_{p_2}$ and divide both sides by $\log n$, we will get:
        \begin{align*}
            -\sum_{k:\vec{\alpha}_k\in \mathcal{E}\setminus \mathcal{E}_{p_2}}\frac{\exp(\vec{\alpha}_k^{\top}\hat{\vec{\theta}}_k)}{\log n}P_{\mathcal{H}_{p_2}^{\perp}}\vec{\alpha}_k+P_{\mathcal{H}_{p_2}^{\perp}}\left(\hat{\vec{\theta}}_k-\sum_{j=p_2+1}^{p_3}\hat{c}_{j,k}\vec{\eta}_j\right)+o(1)=\vec{0}.
        \end{align*}
        So we require $\hat{\vec{\theta}}_k-\sum_{j=p_2+1}^{p_3}\hat{c}_{j,k}\vec{\eta}_j$ to be spanned by the basis in the linear equation, which has a unique solution $\hat{\vec{\theta}}_k$. We call it a generalised characterisation equation for $\hat{\vec{\theta}}_k$ at $(\nu_2^{(k)},\ldots,\nu_J^{(k)})^{\top}$.\\[3mm]
        \textbf{Step 3.4: }Now we prove that the constructed generalised characterisation equation is a valid characterisation equation. Moreover, it holds that $p=p_4$, i.e., all four parts in $p+1,\ldots,p_4$ are excluded.\\[3mm]
        \textbf{Part 1: }For $j=p+1,\ldots,p_1$, we assume WLOG that for any $k$, $\vec{\eta}_{j}$ is expanded by the same vectors in $\left\{\vec{\alpha}_1,\ldots,\vec{\alpha}_W\right\}$, then we have $\lim_{k\rightarrow \infty}(\vec{\eta}_1-\vec{\eta}_j)^{\top}\hat{\vec{\theta}}_k=\lim_{k\rightarrow \infty}1-\nu_j^{(k)}=1$ since $(\nu_2^{(k)},\ldots,\nu_J^{(k)})^{\top}\rightarrow (\nu_2,\ldots,\nu_J)^{\top}$. By the construction method of generalised characterisation equation, this cannot happen outside a set of measure zero in the parameter space. So we have $p=p_1$.\\[3mm]
        \textbf{Part 2: }For $j=p_1+1,\ldots,p_2$, since $(\vec{\eta}_1-\vec{\eta}_{p_1+1})^{\top}\hat{\vec{\theta}}_k=\ldots=(\vec{\eta}_1-\vec{\eta}_{p_2})^{\top}\hat{\vec{\theta}}_k=1$ implies that $\vec{\eta}_{p_1+1}^{\top}\hat{\vec{\theta}}_k=\ldots=\vec{\eta}_{p_2}^{\top}\hat{\vec{\theta}}_k$, by excluding a set of measure zero in the parameter space, the expansion of $\vec{\eta}_{p_1+1},\ldots,\vec{\eta}_{p_2}$ with respect to the basis in the generalised characterisation equation should all be degenerate, which indicates that $\vec{\eta}_{p_1+1}^{\top}\hat{\vec{\theta}}_k=\ldots=\vec{\eta}_{p_2}^{\top}\hat{\vec{\theta}}_k=0$ since $\nu_{p_1+1}=\ldots=\nu_{p_2}=0$. Then $\vec{\eta}_1^{\top}\hat{\vec{\theta}}_k=1$. By excluding a set of measure zero in the parameter space, this cannot happen. So we have $p_1=p_2$.\\[3mm]
        \textbf{Part 3: }For $j=p_3+1,\ldots,p_4$, by the construction method, $\vec{\eta}_{j}$ is not involved in the generalised characterisation equation. By excluding a set of measure zero in the parameter space, $(\vec{\eta}_1-\vec{\eta}_j)^{\top}\hat{\vec{\theta}}_k=1$ cannot happen. So we have $p_3=p_4$.\\[3mm]
        \textbf{Part 4: }For arbitrary fixed $k\in\mathbb{N}$, we have already proved that for $j=p+1,\ldots,p_4$, $\lim_{n\rightarrow \infty}{\bar{\xi}_j^{(k,n)}}/{\log n}=\hat{c}_{j,k}\in(0,\infty)$. Denote $\mathcal{H}=\operatorname{span}\{\vec{\alpha}_{w}:w=1,\ldots,W, \vec{\alpha}_w^{\top}\hat{\vec{\theta}}_k\geq 0\}$.\\[3mm]
        \textbf{Case 1: }If the generalised characterisation equation contains at least one type-2 equation, then at least one of $\vec{\eta}_1,\ldots,\vec{\eta}_p$ has nondegenerate expansion in the characterisation equation. This implies that $\vec{\eta}_{p+1}^{\top}\hat{\vec{\theta}}_k,\ldots,\vec{\eta}_{p_4}^{\top}\hat{\vec{\theta}}_k$ does not depend on the value of $\hat{c}_{p+1,k},\ldots,\hat{c}_{p_4,k}$. Then by excluding a set of measure zero in the parameter space, $(\vec{\eta}_1-\vec{\eta}_{p+1})^{\top}\hat{\vec{\theta}}_k=\ldots=(\vec{\eta}_1-\vec{\eta}_{p_4})^{\top}\hat{\vec{\theta}}_k=1$ cannot happen.\\[3mm]
    \textbf{Case 2: }If the characterisation equation contains type-1 equations only. Then the conditions in Proposition~\ref{lem_characterisation_log} are satisfied. Denote $\hat{\bm{\xi}}^{(n,\vec{c})}=(\bar{\xi}_{2}^{(k,n)},\ldots,\bar{\xi}_{p}^{(k,n)},\hat{\xi}_{p+1}^{(n,{c}_{p+1})},\ldots,\hat{\xi}_{p_4}^{(n,{c}_{p_4})},0,\ldots,0)^{\top}$ where
    \begin{align*}
        \hat{\xi}_{j}^{(n,{c}_{j})}=\frac{c_j}{\hat{c}_{j,k}}\bar{\xi}_{j}^{(k,n)},j=p+1,\ldots,p_4.
    \end{align*}
    By Proposition~\ref{lem_characterisation_log}, there exist $d_{n,1},\vec{d}_2$ which does not depend on $\vec{c}$ such that for any $\vec{c}=(c_{p+1},\ldots,c_{p_4})^{\top}$ in a small neighbourhood of $\vec{\hat{c}}=(\hat{c}_{p+1,k},\ldots,\hat{c}_{p_4,k})^{\top}$, we have
    \begin{align}\label{eq_thm_1.50}
        f_n(\vec{\theta}_n(\hat{\bm{\xi}}^{(n,\vec{c})})\big|\hat{\bm{\xi}}^{(n,\vec{c})})=o(\log^2 n)+d_{n,1}+\log^2 n\left(\vec{c}^{\top}\vec{d}_{2}+\frac{1}{2}\left\|\sum_{j=p+1}^{p_4}{c}_{j}P_{\mathcal{H}^{\perp}}\vec{\eta}_j\right\|^2\right).
    \end{align}
    It can be readily shown that
    \begin{align}\label{eq_thm_1.51}
        \frac{\operatorname{det}\left(-\nabla^2 f_n(\vec{\theta}_n(\hat{\bm{\xi}}^{(n,\vec{c})})\big|\hat{\bm{\xi}}^{(n,\vec{c})})\right)}{\operatorname{det}\left(-\nabla^2 f_n(\vec{\theta}_n(\hat{\bm{\xi}}^{(n,\hat{\vec{c}})})\big|\hat{\bm{\xi}}^{(n,\hat{\vec{c}})})\right)}=\exp\left(o(\log^2 n)\right).
    \end{align}
    By the Stirling formula, we can similarly prove that
    \begin{align}\label{eq_thm_1.52}
        \frac{\Delta_n(\hat{\bm{\xi}}^{(n,\vec{c})})}{\Delta_n(\hat{\bm{\xi}}^{(n,\hat{c})})}=&\exp\left(o(\log^2 n)+\log^2 n\sum_{j=p+1}^{p_4}\left(c_{j}-\hat{c}_{j,k}\right)\right).
    \end{align}
    Then by (\ref{eq_thm_1.11}), (\ref{eq_thm_1.50}), (\ref{eq_thm_1.51}) and (\ref{eq_thm_1.52}) we have
    \begin{align*}
        \log\frac{\Delta_n(\hat{\bm{\xi}}^{(n,\vec{c})})\phi_n(\hat{\bm{\xi}}^{(n,\vec{c})})}{\Delta_n(\hat{\bm{\xi}}^{(n,\hat{c})})\phi_n(\hat{\bm{\xi}}^{(n,\hat{c})})}=o(\log^2 n)+\widetilde{d}_{n,1}+\log^2 n\left(\vec{c}^{\top}\vec{\widetilde{d}}_2+\frac{1}{2}\left\|\sum_{j=p+1}^{p_4}{c}_{j}P_{\mathcal{H}^{\perp}}\vec{\eta}_j\right\|^2\right).
    \end{align*}
    By definition of $\bar{\bm{\xi}}^{(k,n)}=\hat{\bm{\xi}}^{(n,\hat{c})}$, $\vec{c}^{\top}\vec{\widetilde{d}}_2+\frac{1}{2}\left\|\sum_{j=p+1}^{p_4}{c}_{j}P_{\mathcal{H}^{\perp}}\vec{\eta}_j\right\|^2$ should attain its maximum value at $\vec{c}=\vec{\hat{c}}$. Since the hessian matrix of this function is calculated by
    \begin{align*}
        \nabla^2\left(\vec{c}^{\top}\vec{\widetilde{d}}_2+\frac{1}{2}\left\|\sum_{j=p+1}^{p_4}{c}_{j}P_{\mathcal{H}^{\perp}}\vec{\eta}_j\right\|^2\right)=\left(P_{\mathcal{H}^{\perp}}\vec{\eta}_{p+1},\ldots,P_{\mathcal{H}^{\perp}}\vec{\eta}_{p_4}\right)^{\top}\left(P_{\mathcal{H}^{\perp}}\vec{\eta}_{p+1},\ldots,P_{\mathcal{H}^{\perp}}\vec{\eta}_{p_4}\right)\succeq \mat{0},
    \end{align*}
    this implies that the hessian matrix can only be zero matrix at $\vec{c}=\vec{\hat{c}}$, or $\vec{\hat{c}}$ cannot be its maximum. Hence $P_{\mathcal{H}^{\perp}}\vec{\eta}_{p+1}=\ldots=P_{\mathcal{H}^{\perp}}\vec{\eta}_{p_4}=0$, i.e., $\vec{\eta}_{p+1},\ldots,\vec{\eta}_{p_4}\in \mathcal{H}$. Hence $\hat{\vec{\theta}}_k=(\hat{\vec{\theta}}_k-\sum_{j=p+1}^{p_4}\hat{c}_{j,k}\vec{\eta}_j)+\sum_{j=p+1}^{p_4}\hat{c}_{j,k}\vec{\eta}_j\in\mathcal{H}$. This implies that $\hat{\vec{\theta}}_k$ is the unique solution of a characterisation equation at $(\nu_2,\ldots,\nu_J)^{\top}$. Hence, by excluding a set of measure zero in the parameter space, $(\vec{\eta}_1-\vec{\eta}_{p+1})^{\top}\hat{\vec{\theta}}_k=\ldots=(\vec{\eta}_1-\vec{\eta}_{p_4})^{\top}\hat{\vec{\theta}}_k=1$ cannot happen. So $p=p_4$. Hence for any $j=p+1,\ldots,J$ and any $k\in\mathbb{N}$, it holds that $\bar{\xi}_j^{(k,n)}=0$ for $n$ large enough.\\[3mm]
    \textbf{Step 3.5: }By the construction method of the (generalised) characterisation equation, we have already verified the nondegeneracy condition in Proposition~\ref{prop_characterisation_equation}. Then by the uniqueness result proved in Proposition~\ref{prop_characterisation_equation}, we can define $\hat{\vec{\theta}}_k$ as $\vec{\theta}(\nu_2^{(k)},\ldots,\nu_p^{(k)},0,\ldots,0)$. Since $(\nu_2^{(k)},\ldots,\nu_p^{(k)},0,\ldots,0)\rightarrow (\nu_2,\ldots,\nu_p,0,\ldots,0)$, we have $\hat{\vec{\theta}}_k=\vec{\theta}(\nu_2^{(k)},\ldots,\nu_p^{(k)},0,\ldots,0)\rightarrow \vec{\theta}(\nu_2,\ldots,\nu_p,0,\ldots,0)=\vec{\theta}(\nu_2,\ldots,\nu_J)$ by Proposition~\ref{prop_characterisation_equation}.\\[3mm]
    \textbf{Step 3.6: }Finally, by (\ref{eq_thm_1.49}) we have
    \begin{align}\label{eq_thm_1.53}
        1-\nu_j=(\vec{\eta}_1-\vec{\eta}_j)^{\top}\vec{\theta}(\nu_2,\ldots,\nu_J)
    \end{align}
    for $j=1,\ldots,p$. For $j=p+1,\ldots,J$, we have $(\vec{\eta}_1-\vec{\eta}_j)^{\top}\vec{\theta}(\nu_2,\ldots,\nu_J)\geq 1$. Since by excluding a set of measure zero in the parameter space, $(\vec{\eta}_1-\vec{\eta}_j)^{\top}\vec{\theta}(\nu_2,\ldots,\nu_J)= 1$ cannot happen, it follows that
    \begin{align}\label{eq_thm_1.54}
        (\vec{\eta}_1-\vec{\eta}_j)^{\top}\vec{\theta}(\nu_2,\ldots,\nu_J)> 1.
    \end{align}
    Similarly we assume that $\widetilde{\nu}_2,\ldots,\widetilde{\nu}_{\widetilde{p}}>0$ and $\widetilde{\nu}_{\widetilde{p}+1}=\ldots=\widetilde{\nu}_{J}=0$. Then for any $j=2,\ldots,\widetilde{p}$, we have
    \begin{align}\label{eq_thm_1.55}
        1-\widetilde{\nu}_j=(\widetilde{\vec{\eta}}_1-\widetilde{\vec{\eta}}_j)^{\top}\widetilde{\vec{\theta}}(\widetilde{\nu}_2,\ldots,\widetilde{\nu}_J).
    \end{align}
    For any $j=\widetilde{p}+1,\ldots,J$, we have
    \begin{align}\label{eq_thm_1.56}
        (\widetilde{\vec{\eta}}_1-\widetilde{\vec{\eta}}_j)^{\top}\widetilde{\vec{\theta}}(\widetilde{\nu}_2,\ldots,\widetilde{\nu}_J)> 1.
    \end{align}
    Moreover, $\vec{\theta}(\nu_2,\ldots,\nu_J)$ is continuous at ${(\nu_2,\ldots,\nu_J)^{\top}}$ with respect to $(\nu_2,\ldots,\nu_p)$ and $\widetilde{\vec{\theta}}(\widetilde{\nu}_2,\ldots,\widetilde{\nu}_J)$ is continuous at $(\widetilde{\nu}_2,\ldots,\widetilde{\nu}_J)^{\top}$ with respect to $(\widetilde{\nu}_2,\ldots,\widetilde{\nu}_{\widetilde{p}})^{\top}$.\\[3mm]
    Note that by the proof in Step 3, we can show that $\nu_2,\ldots,\nu_J,\widetilde{\nu}_2,\ldots,\widetilde{\nu}_J<1$. If this is not the case, for example $\nu_2=1$, then we have $\vec{\eta}_1^{\top}\hat{\vec{\theta}}_k=\vec{\eta}_2^{\top}\hat{\vec{\theta}}_k$ by (\ref{eq_thm_1.53}) when $k$ is large. Since $\sum_{j=2}^J\bar{\xi}_j^{(k,n)}\ll n$, we can easily see that $\vec{\eta}_1^{\top}\hat{\vec{\theta}}_k=\sum_{k=1}^{m_1}c_{1,k}$ while $\vec{\eta}_2^{\top}\hat{\vec{\theta}}_k\leq \sum_{k=2}^{m_2}c_{2,k}<\sum_{k=1}^{m_1}c_{1,k}$ by the construction method of characterisation equation in Lemma~\ref{lem_characterisation_equation_3}. This leads to a contradiction. So we have $\nu_2,\ldots,\nu_J,\widetilde{\nu}_2,\ldots,\widetilde{\nu}_J<1$.\\[3mm]
    \textbf{Step 4: }Separate the order of summation on both sides of (\ref{eq_thm_1.46}).\\[3mm] We first separate the order on the left-hand side of (\ref{eq_thm_1.46}). Denote $\bm{\nu}=(\nu_2,\ldots,\nu_J)^{\top}$ and rank $(\vec{\eta}_1-\vec{\eta}_j)^{\top}\vec{\theta}(\bm{\nu}),j=p+1,\ldots,J$ in decreasing order. By excluding a set of measure zero in the parameter space, there are no ties among $(\vec{\eta}_1-\vec{\eta}_{p+1})^{\top}\vec{\theta}(\bm{\nu}),\ldots,$ $(\vec{\eta}_1-\vec{\eta}_J)^{\top}\vec{\theta}(\bm{\nu})$. Then by (\ref{eq_thm_1.54}) we can assume WLOG that $(\vec{\eta}_1-\vec{\eta}_{p+1})^{\top}\vec{\theta}(\bm{\nu})>\ldots>(\vec{\eta}_1-\vec{\eta}_J)^{\top}\vec{\theta}(\bm{\nu})\geq 1+\delta$, where $\delta>0$ is a positive constant.\\[3mm]
    For any given $\bm{\xi}=(\xi_{p+1},\ldots,\xi_{J})^{\top}\in\mathbb{N}_0^{J-p}$ and $\bm{\bar{\nu}}=(\bar{\nu}_2,\ldots,\bar{\nu}_p)^{\top}\in (0,1)^{p-1}$, denote $T(\bm{\xi}|\bm{\bar{\nu}})=\sum_{j=p+1}^J \xi_j[(\vec{\eta}_1-\vec{\eta}_j)^{\top}\vec{\theta}(\bar{\nu}_2,\ldots,\bar{\nu}_p,0,\ldots,0)-1]$. Then we can rank all terms in $\{T(\bm{\xi}|\bm{\nu}):\bm{\xi}\in\mathbb{N}_0^{J-p}\}$ in increasing order. Let $\bm{\xi}^{(r)}$ denote the vector such that the rank of $T(\bm{\xi}^{(r)}|\bm{\nu})$ is $r$ for any $r\in\mathbb{N}$. Suppose the rank of $T(0,\ldots,0,1,1|\bm{\nu})$ is $r^{\ast}$. By excluding a set of measure zero in the parameter space, no tie exists among $T(\bm{\xi}^{(1)}|\bm{\nu}),\ldots,T(\bm{\xi}^{(r^{\ast}+1)}|\bm{\nu})$.\\[3mm]
    Since
    \begin{align*}
        T(\bm{\xi}|\bm{\nu})=\sum_{j=p+1}^J \xi_j\left((\vec{\eta}_1-\vec{\eta}_j)^{\top}\vec{\theta}(\bm{\nu})-1\right)\geq \delta\left(\sum_{j=p+1}^J \xi_j\right),
    \end{align*}
    by the continuity property in Proposition~\ref{prop_characterisation_equation}, for any $1\leq r\leq r^{\ast}$ there exists $r_{\text{max}}\in\mathbb{N}$ and $k$ large enough such that
    \begin{align}\label{eq_thm_1.57}
        \underset{\bar{\bm{\nu}}=(\bar{\nu}_2,\ldots,\bar{\nu}_p,0,\ldots,0)^{\top}\in\mathcal{G}_k}{\max}T(\bm{\xi}^{(r)}|\bar{\bm{\nu}})<\underset{\bar{\bm{\nu}}=(\bar{\nu}_2,\ldots,\bar{\nu}_p,0,\ldots,0)^{\top}\in\mathcal{G}_k}{\min}T(\bm{\xi}^{(r+1)}|\bar{\bm{\nu}})
    \end{align}
    and
    \begin{align}\label{eq_thm_1.58}
        \underset{\bar{\bm{\nu}}=(\bar{\nu}_2,\ldots,\bar{\nu}_p,0,\ldots,0)^{\top}\in\mathcal{G}_k}{\max}T(\bm{\xi}^{(r)}|\bar{\bm{\nu}})+(J+1)\leq\underset{\bar{\bm{\nu}}=(\bar{\nu}_2,\ldots,\bar{\nu}_p,0,\ldots,0)^{\top}\in\mathcal{G}_k}{\min}T(\bm{\xi}^{(r_{\text{max}}+1)}|\bar{\bm{\nu}}).
    \end{align}
    Now we fix $r$ such that $1\leq r\leq r^{\ast}$. We assume WLOG that $r^{\ast}<r_{\text{max}}$ and fix $k$ large enough. For any $r<\widetilde{r}\leq r_{\text{max}}$ and any $(\xi_2^{(n)},\ldots,\xi_{p}^{(n)},0,\ldots,0)^{\top}\in\mathcal{E}_{k,n}$, denote ${\bm{\xi}}^{(r,n)}=(\xi_2^{(n)},\ldots,\xi_{p}^{(n)},\xi_{p+1}^{(r)},\ldots,\xi_{J}^{(r)})^{\top}$ for any $n\in\mathbb{N}$, then we have
    \begin{align*}
        \lim_{n\rightarrow \infty}\frac{\log {\bm{\xi}}^{(r,n)}}{\log n}=\lim_{n\rightarrow \infty}\frac{\log {\bm{\xi}}^{(\widetilde{r},n)}}{\log n}:=&(\bar{\nu}_2,\ldots,\bar{\nu}_p,0,\ldots,0)^{\top}.
    \end{align*}
    By Proposition~\ref{prop_characterisation_equation}, we have
    \begin{align*}
        \lim_{n\rightarrow \infty}\frac{\vec{\theta}_n({\bm{\xi}}^{(r,n)})}{\log n}=\lim_{n\rightarrow \infty}\frac{\vec{\theta}_n({\bm{\xi}}^{(\widetilde{r},n)})}{\log n}=\vec{\theta}(\bar{\nu}_2,\ldots,\bar{\nu}_p,0,\ldots,0).
    \end{align*}
    Hence it holds that
    \begin{align}\label{eq_thm_1.59}
        \frac{\operatorname{det}(-\nabla^2f_n(\vec{\theta}_n({\bm{\xi}}^{(\widetilde{r},n)})\big|{\bm{\xi}}^{(\widetilde{r},n)}))}{\operatorname{det}(-\nabla^2f_n(\vec{\theta}_n({\bm{\xi}}^{(r,n)})\big|{\bm{\xi}}^{(r,n)}))}=\exp(o(\log n)).
    \end{align}
    Moreover, we have
    \begin{align}\label{eq_thm_1.60}
        \lim_{n\rightarrow \infty}\frac{\sum_{j=p+1}^J\xi_{j}^{(\widetilde{r})}(\vec{\eta}_1-\vec{\eta}_j)^{\top}\vec{\theta}_n({\bm{\xi}}^{(\widetilde{r},n)})}{\log n}=&\sum_{j=p+1}^J\xi_{j}^{(\widetilde{r})}(\vec{\eta}_1-\vec{\eta}_j)^{\top}\vec{\theta}(\bar{\nu}_2,\ldots,\bar{\nu}_p,0,\ldots,0)\notag\\
        \lim_{n\rightarrow \infty}\frac{\sum_{j=p+1}^J\xi_{j}^{(r)}(\vec{\eta}_1-\vec{\eta}_j)^{\top}\vec{\theta}_n({\bm{\xi}}^{(\widetilde{r},n)})}{\log n}=&\sum_{j=p+1}^J\xi_{j}^{(r)}(\vec{\eta}_1-\vec{\eta}_j)^{\top}\vec{\theta}(\bar{\nu}_2,\ldots,\bar{\nu}_p,0,\ldots,0).
    \end{align}
    Furthermore, by the Stirling formula, we can prove that
    \begin{align}\label{eq_thm_1.61}
        \frac{\Delta_n({\bm{\xi}}^{(\widetilde{r},n)})}{\Delta_n({\bm{\xi}}^{(r,n)})}=\exp\left\{o(\log n)+\log n\left(\sum_{j=p+1}^J\xi_j^{(\widetilde{r})}-\sum_{j=p+1}^J\xi_j^{({r})}\right)\right\}.
    \end{align}
    Then for any $r< \widetilde{r}\leq r_{\text{max}}$, by (\ref{eq_thm_1.11}), (\ref{eq_thm_1.57}), (\ref{eq_thm_1.59}), (\ref{eq_thm_1.60}) and (\ref{eq_thm_1.61}), we have
    \begin{align}\label{eq_thm_1.62}
        &\frac{\Delta_n({\bm{\xi}}^{(\widetilde{r},n)})\phi_n({\bm{\xi}}^{(\widetilde{r},n)})}{\Delta_n({\bm{\xi}}^{(r,n)})\phi_n({\bm{\xi}}^{(r,n)})}\notag\\
        \leq& \exp\left\{o(\log n)+\log n\left(\sum_{j=p+1}^J\xi_j^{(\widetilde{r})}-\sum_{j=p+1}^J\xi_j^{({r})}\right)\right\}\times\frac{\exp\big(f_n(\vec{\theta}_n({\bm{\xi}}^{(\widetilde{r},n)})\big|{\bm{\xi}}^{(\widetilde{r},n)}))\big)}{\exp\big(f_n(\vec{\theta}_n({\bm{\xi}}^{(\widetilde{r},n)})\big|{\bm{\xi}}^{(r,n)}))\big)}\notag\\
        =&\exp\Bigg\{o(\log n)+\sum_{j=p+1}^J\xi_{j}^{(\widetilde{r})}\left(\log n-(\vec{\eta}_1-\vec{\eta}_j)^{\top}\vec{\theta}_n({\bm{\xi}}^{(\widetilde{r},n)})-(\mu_1-\mu_j)\right)\notag\\
        &\quad\quad-\sum_{j=p+1}^J\xi_{j}^{(r)}\left(\log n-(\vec{\eta}_1-\vec{\eta}_j)^{\top}\vec{\theta}_n({\bm{\xi}}^{(\widetilde{r},n)})-(\mu_1-\mu_j)\right)\Bigg\}\notag\\
        \leq& \exp\left\{\log n\left(\underset{\bar{\bm{\nu}}=(\bar{\nu}_2,\ldots,\bar{\nu}_p,0,\ldots,0)^{\top}\in\mathcal{G}_k}{\max}T(\bm{\xi}^{(r)}|\bar{\bm{\nu}})-\underset{\bar{\bm{\nu}}=(\bar{\nu}_2,\ldots,\bar{\nu}_p,0,\ldots,0)^{\top}\in\mathcal{G}_k}{\min}T(\bm{\xi}^{(\widetilde{r})}|\bar{\bm{\nu}})+o(1)\right)\right\}\notag\\
        \rightarrow&0.
    \end{align}
    Now we control the terms with rank higher than $r_{\text{max}}$. For any $n,k\in\mathbb{N}$, denote
    \begin{align*}
        \hat{\bm{\xi}}^{(k,n)}=\underset{\bm{\xi}\in \mathcal{E}_{k,n},(\xi_{p+1},\ldots,\xi_{J})\notin\{(\xi_{p+1}^{(l)},\ldots,\xi_{J}^{(l)}):l=1,\ldots,r_{\text{max}}\}}{\operatorname{argmax}}\Delta_n(\bm{\xi})\phi_n(\bm{\xi}).
    \end{align*}
    By the same method as in Step 4 of Case 1, we can prove that, for $k$ large enough and any $j=p+1,\ldots,J$, it holds that
    \begin{align*}
        (\hat{\xi}_{p+1}^{(k,n)},\ldots,\hat{\xi}_{j}^{(k,n)}-1,\ldots,\hat{\xi}_J^{(k,n)})^{\top}\in\left\{(\xi_{p+1}^{(l)},\ldots,\xi_{J}^{(l)})^{\top}:l=1,\ldots,r_{\text{max}}\right\}.
    \end{align*}
    This implies that by fixing $k$ large enough, it holds that $(\hat{\xi}_{p+1}^{(k,n)},\ldots,\hat{\xi}_J^{(k,n)})^{\top}=(\xi_{p+1}^{(r_{\text{max}}+1)},\ldots,\xi_{J}^{(r_{\text{max}}+1)})^{\top}$ when $n$ is large.
    Then by (\ref{eq_thm_1.11}), (\ref{eq_thm_1.45}), (\ref{eq_thm_1.58}), (\ref{eq_thm_1.59}), (\ref{eq_thm_1.60}) and (\ref{eq_thm_1.61}), we have
    \begin{align}\label{eq_thm_1.63}
        &\frac{\Delta_n(\hat{\bm{\xi}}^{(k,n)})\phi_n(\hat{\bm{\xi}}^{(k,n)})}{\Delta_n({\bm{\xi}}^{(r,n)})\phi_n({\bm{\xi}}^{(r,n)})}\notag\\
        \leq& \exp\left\{\log n\left[\underset{\bar{\bm{\nu}}=(\bar{\nu}_2,\ldots,\bar{\nu}_p,0,\ldots,0)^{\top}\in\mathcal{G}_k}{\max}T(\bm{\xi}^{(r)}|\bar{\bm{\nu}})-\underset{\bar{\bm{\nu}}=(\bar{\nu}_2,\ldots,\bar{\nu}_p,0,\ldots,0)^{\top}\in\mathcal{G}_k}{\min}T(\bm{\xi}^{(r_{\text{max}}+1)}|\bar{\bm{\nu}})+o(1)\right]\right\}\notag\\
        \leq&\exp\big(-\log n(J+1+o(1))\big).
    \end{align}
    So for any $1\leq r\leq r^{\ast}$ and $k$ large enough, by (\ref{eq_thm_1.62}) and (\ref{eq_thm_1.63}), we have
    \begin{align}\label{eq_thm_1.64}
        &\frac{{\sum}_{l\geq r+1}{\sum}_{{\bm{\xi}}^{(l,n)}\in \mathcal{E}_{k,n}}\Delta_n({\bm{\xi}}^{(l,n)})\phi_n({\bm{\xi}}^{(l,n)})}{{\sum}_{{\bm{\xi}}^{(r,n)}\in \mathcal{E}_{k,n}}\Delta_n({\bm{\xi}}^{(r,n)})\phi_n({\bm{\xi}}^{(r,n)})}\notag\\
        \leq&\frac{{\sum}_{l=r+1,\ldots,r_{\text{max}}}{\sum}_{{\bm{\xi}}^{(l,n)}\in \mathcal{E}_{k,n}}\Delta_n({\bm{\xi}}^{(l,n)})\phi_n({\bm{\xi}}^{(l,n)})}{{\sum}_{{\bm{\xi}}^{(r,n)}\in \mathcal{E}_{k,n}}\Delta_n({\bm{\xi}}^{(r,n)})\phi_n({\bm{\xi}}^{(r,n)})}+\frac{{\sum}_{l\geq r_{\text{max}}+1}{\sum}_{{\bm{\xi}}^{(l,n)}\in \mathcal{E}_{k,n}}\Delta_n({\bm{\xi}}^{(l,n)})\phi_n({\bm{\xi}}^{(l,n)})}{{\sum}_{{\bm{\xi}}^{(r,n)}\in \mathcal{E}_{k,n}}\Delta_n({\bm{\xi}}^{(r,n)})\phi_n({\bm{\xi}}^{(r,n)})}\notag\\
        \leq&o(1)+n^{J}\cdot\frac{\Delta_n(\hat{\bm{\xi}}^{(k,n)})\phi_n(\hat{\bm{\xi}}^{(k,n)})}{\Delta_n({\bm{\xi}}^{(r,n)})\phi_n({\bm{\xi}}^{(r,n)})}\notag\\
        \leq&o(1)+n^{J}\exp(-\log n(J+1+o(1)))\notag\\
        \rightarrow&0.
    \end{align}
    For the other side of (\ref{eq_thm_1.10}), by (\ref{eq_thm_1.56}) we can assume WLOG that $(\widetilde{\vec{\eta}}_1-\widetilde{\vec{\eta}}_{\widetilde{p}+1})^{\top}\widetilde{\vec{\theta}}(\widetilde{\bm{\nu}})>\ldots>(\widetilde{\vec{\eta}}_1-\widetilde{\vec{\eta}}_J)^{\top}\widetilde{\vec{\theta}}(\widetilde{\bm{\nu}})>1$. We can similarly denote $\{\widetilde{\bm{\xi}}^{(r)}:r\in\mathbb{N}\}$ and $\widetilde{T}(\bm{\xi}|\bar{\bm{\nu}})$. We assume WLOG that the rank of $\widetilde{T}(0,\ldots,0,1,1|\widetilde{\bm{\nu}})$ is higher than $r^{\ast}$. Similarly, we can prove that for any $1\leq r\leq r^{\ast}$ and $k$ large enough we have
    \begin{align}\label{eq_thm_1.65}
        \frac{{\sum}_{l\geq r+1}{\sum}_{\widetilde{\bm{\xi}}^{(l,n)}\in \widetilde{\mathcal{E}}_{k,n}}\Delta_n(\widetilde{\bm{\xi}}^{(l,n)})\widetilde{\phi}_n(\widetilde{\bm{\xi}}^{(l,n)})}{{\sum}_{\widetilde{\bm{\xi}}^{(r,n)}\in \widetilde{\mathcal{E}}_{k,n}}\Delta_n(\widetilde{\bm{\xi}}^{(r,n)})\widetilde{\phi}_n(\widetilde{\bm{\xi}}^{(r,n)})}\rightarrow 0.
    \end{align}
    \textbf{Step 5: }Prove that $p=\widetilde{p}$ and $\nu_j=\widetilde{\nu}_j$ for $j=1,\ldots,p$.\\[3mm]
    Since $(\xi_{p+1}^{(1)},\ldots,\xi_{J}^{(1)})^{\top}=\mathbf{0}$ and $(\widetilde{\xi}_{\widetilde{p}+1}^{(1)},\ldots,\widetilde{\xi}_{J}^{(1)})^{\top}=\mathbf{0}$, by (\ref{eq_thm_1.46}), (\ref{eq_thm_1.64}) and (\ref{eq_thm_1.65}), for $k$ large enough we have
    \begin{align}\label{eq_thm_1.66}
        &\sum_{(\xi_2^{(n)},\ldots, \xi_J^{(n)})^{\top}\in \mathcal{E}_{k,n}}\Delta_n(\xi_2^{(n)},\ldots,\xi_p^{(n)},0,\ldots,0)\phi_n(\xi_2^{(n)},\ldots,\xi_p^{(n)},0,\ldots,0)\notag\\
        \asymp& \sum_{(\widetilde{\xi}_2^{(n)},\ldots, \widetilde{\xi}_J^{(n)})^{\top}\in \widetilde{\mathcal{E}}_{k,n}}\Delta_n(\widetilde{\xi}_2^{(n)},\ldots,\widetilde{\xi}_{\widetilde{p}}^{(n)},0,\ldots,0)\widetilde{\phi}_n(\widetilde{\xi}_2^{(n)},\ldots,\widetilde{\xi}_{\widetilde{p}}^{(n)},0,\ldots,0).
    \end{align}
    For notational simplicity, we only show the first $p$ and $\widetilde{p}$ subscripts on both sides. By arguments similar to those used in the proof in Case 1, we can show that
    \begin{align*}
        \sum_{(\xi_2^{(n)},\ldots, \xi_J^{(n)})^{\top}\in \mathcal{E}_{k,n}}\Delta_n(\xi_2^{(n)},\ldots,\xi_p^{(n)})\phi_n(\xi_2^{(n)},\ldots,\xi_p^{(n)})=&\exp\left(n\log n\sum_{k=1}^{m_1}c_{1,k}+n\left(\mu_1-\sum_{k=1}^{m_1}c_{1,k}+o(1)\right)\right),\\
        \sum_{(\widetilde{\xi}_2^{(n)},\ldots, \widetilde{\xi}_J^{(n)})^{\top}\in \widetilde{\mathcal{E}}_{k,n}}\Delta_n(\widetilde{\xi}_2^{(n)},\ldots,\widetilde{\xi}_{\widetilde{p}}^{(n)})\widetilde{\phi}_n(\widetilde{\xi}_2^{(n)},\ldots,\widetilde{\xi}_{\widetilde{p}}^{(n)})=&\exp\left(n\log n\sum_{k=1}^{\widetilde{m}_1}\widetilde{c}_{1,k}+n\left(\widetilde{\mu}_1-\sum_{k=1}^{\widetilde{m}_1}\widetilde{c}_{1,k}+o(1)\right)\right).
    \end{align*}
    Then by (\ref{eq_thm_1.66}) we have
    \begin{align*}
        \vec{\eta}_1^{\top}\vec{\theta}(\nu_2,\ldots,\nu_J)=\sum_{k=1}^{m_1}c_{1,k}=&\sum_{k=1}^{\widetilde{m}_1}\widetilde{c}_{1,k}=\widetilde{\vec{\eta}}_1^{\top}\widetilde{\vec{\theta}}(\widetilde{\nu}_2,\ldots,\widetilde{\nu}_J),\\
        \mu_1=&\widetilde{\mu}_1.
    \end{align*}
    So far, the result is only proved up to a permutation of the indices $\{1,\ldots,J\}$. Now we specify the permutation and suppose there exists a permutation $\pi:\{1,\ldots,J\}\rightarrow\{1,\ldots,J\}$ such that $\vec{\eta}_1^{\top}\vec{\theta}(\nu_2,\ldots,\nu_J)=\widetilde{\vec{\eta}}_{\pi(1)}^{\top}\widetilde{\vec{\theta}}(\widetilde{\nu}_2,\ldots,\widetilde{\nu}_J)$.\\[3mm]
    For $m=2,\ldots,J$, by Corollary~\ref{cor_likelihood} we have
    \begin{align*}          
        \sum_{\bm{\xi}^{(n)}\in\mathcal{O}_n}\Delta_n(\bm{\xi}^{(n)})\phi_{n+1}(\xi_{2}^{(n)},\ldots,\xi_{m}^{(n)}+1,\ldots,\xi_{J}^{(n)})=\sum_{\bm{\xi}^{(n)}\in\mathcal{O}_n}\Delta_n(\bm{\xi}^{(n)})\widetilde{\phi}_{n+1}(\xi_{2}^{(n)},\ldots,\xi_{m}^{(n)}+1,\ldots,\xi_{J}^{(n)}).
    \end{align*}
    We can show that $(\nu_2,\ldots,\nu_J)$ and $(\widetilde{\nu}_2,\ldots,\widetilde{\nu}_J)$ remain the concentration points for both sides. By arguments similar to those used in the proof of Step 6 in Case 1, for any $m=1,\ldots,J$ we can derive that $\vec{\eta}_m^{\top}\vec{\theta}(\nu_2,\ldots,\nu_J)=\widetilde{\vec{\eta}}_{\pi(m)}^{\top}\widetilde{\vec{\theta}}(\widetilde{\nu}_2,\ldots,\widetilde{\nu}_J)$. By the construction method of characterisation equation, we can see that
    \begin{align*}      
        \vec{\eta}_1^{\top}\vec{\theta}(\nu_2,\ldots,\nu_J)>&\max_{j\neq 1}\vec{\eta}_j^{\top}\vec{\theta}(\nu_2,\ldots,\nu_J),\\
        \widetilde{\vec{\eta}}_{\pi(1)}^{\top}\widetilde{\vec{\theta}}(\widetilde{\nu}_2,\ldots,\widetilde{\nu}_J)>&\max_{j\neq \pi(1)}\widetilde{\vec{\eta}}_j^{\top}\widetilde{\vec{\theta}}(\widetilde{\nu}_2,\ldots,\widetilde{\nu}_J).
    \end{align*}
    Hence we prove that $\pi(1)=1$. 
    By (\ref{eq_thm_1.53}) and (\ref{eq_thm_1.55}), for $j=2,\ldots,p$, we have
    \begin{align*}
        1-\nu_j=\vec{\eta}_1^{\top}\vec{\theta}(\nu_2,\ldots,\nu_J)-\vec{\eta}_j^{\top}\vec{\theta}(\nu_2,\ldots,\nu_J)=\widetilde{\vec{\eta}}_{\pi(1)}^{\top}\widetilde{\vec{\theta}}(\widetilde{\nu}_2,\ldots,\widetilde{\nu}_J)-\widetilde{\vec{\eta}}_{\pi(j)}^{\top}\widetilde{\vec{\theta}}(\widetilde{\nu}_2,\ldots,\widetilde{\nu}_J)=1-\widetilde{\nu}_{\pi(j)}.
    \end{align*}
    The last equality holds since if $\widetilde{\nu}_{\pi(j)}=0$, then $\widetilde{\vec{\eta}}_1^{\top}\widetilde{\vec{\theta}}(\widetilde{\nu}_2,\ldots,\widetilde{\nu}_J)-\widetilde{\vec{\eta}}_j^{\top}\widetilde{\vec{\theta}}(\widetilde{\nu}_2,\ldots,\widetilde{\nu}_J)$ should be strictly larger than 1 by (\ref{eq_thm_1.56}).
    If $\nu_2,\ldots,\nu_p$ are distinct, then we can easily see that $\pi(j)=j$ for $j=1,\ldots,p$ and $p=\widetilde{p}$ since we assumed that $1>\nu_2>\ldots>\nu_p>0$ and $1>\widetilde{\nu}_2>\ldots>\widetilde{\nu}_{\widetilde{p}}>0$. Then the result is proved.\\[3mm]
    If $\nu_2,\ldots,\nu_p$ are not distinct, for example $\nu_2=\nu_3>\ldots>\nu_p$, then we can prove that $\pi(2)=3,\pi(3)=2$ or $\pi(2)=2,\pi(3)=3$. Hence we can still show that $\nu_2=\widetilde{\nu}_2=\nu_3=\widetilde{\nu}_3$.\\[3mm]
    \textbf{Step 6: }Prove that $\mu_j=\widetilde{\mu}_j$, $\vec{\eta}_j^{\top}\vec{\alpha}_k= \widetilde{\vec{\eta}}_j^{\top}\vec{\alpha}_k$ and $\vec{\eta}_{j_1}^{\top}\vec{\eta}_{j_2}=\widetilde{\vec{\eta}}_{j_1}^{\top}\widetilde{\vec{\eta}}_{j_2}$ for any $j,j_1,j_2=1,\ldots,p$ through the dominant term in the summation.\\[3mm]
    By definition of $(\nu_2,\ldots,\nu_J)^{\top}$, we can easily show that $\mathcal{G}_k=\widetilde{\mathcal{G}}_k$, where $\mathcal{G}_k$ and $\widetilde{\mathcal{G}}_k$ are the hypercubes in layer $k$ where concentration points $(\nu_2,\ldots,\nu_J)^{\top}=(\widetilde{\nu}_2,\ldots,\widetilde{\nu}_J)^{\top}$ belong to, respectively. Hence $\mathcal{E}_{k,n}=\widetilde{\mathcal{E}}_{k,n}$ for any $n,k$.\\[3mm]
    \textbf{Case 1: }$(\nu_2,\ldots,\nu_p)^{\top}$ are distinct and $(\widetilde{\nu}_2,\ldots,\widetilde{\nu}_{\widetilde{p}})^{\top}$ are distinct. Moreover, the two characterisation equations at $(\nu_2,\ldots,\nu_p)^{\top}$ and $(\widetilde{\nu}_2,\ldots,\widetilde{\nu}_{\widetilde{p}})^{\top}$ contain only type-1 equations.\\[3mm]
    Similar to Step 5 in Case 1, we can match the terms on both sides in decreasing order. Then we can prove that $\mu_j=\widetilde{\mu}_j$, $\vec{\eta}_j^{\top}\vec{\alpha}_k= \widetilde{\vec{\eta}}_j^{\top}\vec{\alpha}_k$ and $\vec{\eta}_{j_1}^{\top}\vec{\eta}_{j_2}=\widetilde{\vec{\eta}}_{j_1}^{\top}\widetilde{\vec{\eta}}_{j_2}$ for any $j,j_1,j_2=1,\ldots,p$.\\[3mm]
    \textbf{Case 2: }$(\nu_2,\ldots,\nu_p)^{\top}$ are distinct and $(\widetilde{\nu}_2,\ldots,\widetilde{\nu}_{\widetilde{p}})^{\top}$ are distinct. At least one of the two characterisation equations at $(\nu_2,\ldots,\nu_p)^{\top}$ contains type-2 equations.\\[3mm]
    We assume WLOG that the characterisation equation at $(\nu_2,\ldots,\nu_p)^{\top}$ contains type-2 equations. Furthermore, we suppose that the term in the type-2 equation is of order $\nu$. Similar to the proof in Case 1, we can match the terms on both sides in decreasing order. Then we can see that the characterisation equation at $(\widetilde{\nu}_2,\ldots,\widetilde{\nu}_p)^{\top}$ should also contain a type-2 equation with order $\nu$. By excluding a set of measure zero in the parameter space, this indicates that the two characterisation equations should be exactly the same, which implies that $\mu_j=\widetilde{\mu}_j$, $\vec{\eta}_j^{\top}\vec{\alpha}_k= \widetilde{\vec{\eta}}_j^{\top}\vec{\alpha}_k$ and $\vec{\eta}_{j_1}^{\top}\vec{\eta}_{j_2}=\widetilde{\vec{\eta}}_{j_1}^{\top}\widetilde{\vec{\eta}}_{j_2}$ for any $j,j_1,j_2=1,\ldots,p$ and $k=1,\ldots,W$.\\[3mm]
    \textbf{Case 3: }There exists a tie among $(\nu_2,\ldots,\nu_p)^{\top}$. Moreover, the two characterisation equations at $(\nu_2,\ldots,\nu_p)^{\top}$ contain only type-1 equations.\\[3mm]
    For simplicity, consider the case where $p=3$ and $1>\nu_2=\nu_3>0$. If the expansion of $\vec{\eta}_1$ and $\widetilde{\vec{\eta}}_1$ in both characterisation equations are nondegenerate, then the two characterisation equations are determined independently of $\nu_2=\nu_3$. By excluding a set of measure zero in the parameter space, $\vec{\eta}_2^{\top}\vec{\theta}(\nu_2,\ldots,\nu_J)=\vec{\eta}_3^{\top}\vec{\theta}(\nu_2,\ldots,\nu_J)$ cannot happen. This implies that $\vec{\eta}_1$ has degenerate expansion in the characterisation equation, which indicates that $\vec{\eta}_1=\widetilde{\vec{\eta}}_1$. Since $\nu_2=\nu_3$, by (\ref{eq_thm_1.53}) and the construction method of characterisation equation, $\vec{\eta}_2$ and $\vec{\eta}_3$ should also have degenerate expansions in the characterisation equation, which indicates that $\vec{\eta}_2=\widetilde{\vec{\eta}}_2$ and $\vec{\eta}_3=\widetilde{\vec{\eta}}_3$. Now we should prove that $\mu_2=\widetilde{\mu}_2$ and $\mu_3=\widetilde{\mu}_3$. For any $n\in\mathbb{N}$, suppose
    \begin{align*}
        \hat{\bm{\xi}}^{(n)}&=\underset{\bm{\xi}=(\xi_2,\ldots,\xi_J)^{\top}\in \mathcal{E}_{k,n}:\xi_4=\ldots=\xi_J=0}{\operatorname{argmax}}~\Delta_n(\bm{\xi}){\phi}_n(\bm{\xi}),\\
        \widetilde{\bm{\xi}}^{(n)}&=\underset{\bm{\xi}=(\xi_2,\ldots,\xi_J)^{\top}\in \mathcal{E}_{k,n}:\xi_4=\ldots=\xi_J=0}{\operatorname{argmax}}~\Delta_n(\bm{\xi})\widetilde{\phi}_n(\bm{\xi}).
    \end{align*}
    By the definition of $\hat{\bm{\xi}}^{(n)}$ and $\widetilde{\bm{\xi}}^{(n)}$ we have
    \begin{align}\label{equation4}
        \Delta_n(\hat{\bm{\xi}}^{(n)})\phi_n(\hat{\bm{\xi}}^{(n)})\leq\sum_{{\bm{\xi}}^{(n)}\in \mathcal{E}_{k,n}}\Delta_n({\bm{\xi}}^{(n)})\phi_n({\bm{\xi}}^{(n)})\leq& n^p\Delta_n(\hat{\bm{\xi}}^{(n)})\phi_n(\hat{\bm{\xi}}^{(n)}),\notag\\
        \Delta_n(\widetilde{\bm{\xi}}^{(n)})\widetilde{\phi}_n(\widetilde{\bm{\xi}}^{(n)})\leq\sum_{{\bm{\xi}}^{(n)}\in \mathcal{E}_{k,n}}\Delta_n({\bm{\xi}}^{(n)})\widetilde{\phi}_n({\bm{\xi}}^{(n)})\leq& n^p\Delta_n(\widetilde{\bm{\xi}}^{(n)})\widetilde{\phi}_n(\widetilde{\bm{\xi}}^{(n)}).
    \end{align}
    Then by (\ref{eq_thm_1.66}) and (\ref{equation4}), we have
    \begin{align}\label{equation5}
        \left|\log \left[\Delta_n(\hat{\bm{\xi}}^{(n)})\phi_n(\hat{\bm{\xi}}^{(n)})\right]-\log \left[\Delta_n(\widetilde{\bm{\xi}}^{(n)})\widetilde{\phi}_n(\widetilde{\bm{\xi}}^{(n)})\right]\right|\lesssim \log n.
    \end{align}
    Similar to Step 5 in Case 1, we approximate $\log \left[\Delta_n(\hat{\bm{\xi}}^{(n)})\phi_n(\hat{\bm{\xi}}^{(n)})\right]$ and $\log \left[\Delta_n(\widetilde{\bm{\xi}}^{(n)})\widetilde{\phi}_n(\widetilde{\bm{\xi}}^{(n)})\right]$ by the Stirling formula and Proposition~\ref{prop_laplace} and expand them in an infinite series in decreasing order. Denote the unique maximisers of $f_n(\vec{\theta}|\hat{\bm{\xi}}^{(n)})$ and $\widetilde{f}_n(\vec{\theta}|\widetilde{\bm{\xi}}^{(n)})$ by $\hat{\vec{\theta}}_n$ and $\widetilde{\vec{\theta}}_n$. Suppose the expansions of $\vec{\eta}_1$, $\vec{\eta}_2$ and $\vec{\eta}_3$ in the characterisation equation are given by $\vec{\eta}_1=\sum_{k}^{m_1}c_{1,k}\vec{\alpha}_{1,k}$, $\vec{\eta}_2=\sum_{k}^{m_2}c_{2,k}\vec{\alpha}_{2,k}$ and $\vec{\eta}_3=\sum_{k}^{m_3}c_{3,k}\vec{\alpha}_{3,k}$. Define
    \[\begin{array}{ll}
        c_1=\sum_{k}^{m_1}c_{1,k}&d_1=\sum_{k}^{m_1}c_{1,k}\log\frac{c_{1,k}}{w_{1,k}},\\
        c_2=\sum_{k}^{m_2}c_{2,k}&d_2=\sum_{k}^{m_2}c_{2,k}\log\frac{c_{2,k}}{w_{2,k}},\\
        c_3=\sum_{k}^{m_3}c_{3,k}&d_3=\sum_{k}^{m_3}c_{3,k}\log\frac{c_{3,k}}{w_{3,k}}.
    \end{array}\]
    Similar to Step 5 in Case 1, we have the following approximation:
    \begin{align*}
        &\log \left[\Delta_n({\bm{\xi}}^{(n)})\phi_n({\bm{\xi}}^{(n)})\right]\\
        =&c_1n\log n+(c_1-1)n+{\xi}_2^{(n)}\left[-(c_1-1)\log n+(c_2-1)\log {\xi}_2^{(n)}-(d_1-d_2)-(c_2-1)\right]\\
        &+{\xi}_3^{(n)}\left[-(c_1-1)\log n+(c_3-1)\log {\xi}_3^{(n)}-(d_1-d_3)-(c_3-1)\right]+o(n^{\delta}),
    \end{align*}
    where $\delta>0$ is an arbitrarily small constant. Moreover, we can easily show that
    \begin{align*}
        &\log \left[\Delta_n(\hat{\xi}_2^{(n)}+1,\hat{\xi}_3^{(n)},\ldots,\hat{\xi}_J^{(n)})\phi_n(\hat{\xi}_2^{(n)}+1,\hat{\xi}_3^{(n)},\ldots,\hat{\xi}_J^{(n)})\right]-\log \left[\Delta_n(\hat{\bm{\xi}}^{(n)})\phi_n(\hat{\bm{\xi}}^{(n)})\right]\\[2mm]
        =&(1-c_1)\log n-(1-c_2)\log \hat{\xi}_2^{(n)}-(d_1-d_2)-(\mu_1-\mu_2)+o(n^{-\delta}),\\[2mm]
        &\log \left[\Delta_n(\hat{\xi}_2^{(n)}-1,\hat{\xi}_3^{(n)},\ldots,\hat{\xi}_J^{(n)})\phi_n(\hat{\xi}_2^{(n)}-1,\hat{\xi}_3^{(n)},\ldots,\hat{\xi}_J^{(n)})\right]-\log \left[\Delta_n(\hat{\bm{\xi}}^{(n)})\phi_n(\hat{\bm{\xi}}^{(n)})\right]\\[2mm]
        =&-(1-c_1)\log n+(1-c_2)\log \hat{\xi}_2^{(n)}+(d_1-d_2)+(\mu_1-\mu_2)+o(n^{-\delta}),\\[2mm]
        &\log \left[\Delta_n(\hat{\xi}_2^{(n)},\hat{\xi}_3^{(n)}+1,\ldots,\hat{\xi}_J^{(n)})\phi_n(\hat{\xi}_2^{(n)},\hat{\xi}_3^{(n)}+1,\ldots,\hat{\xi}_J^{(n)})\right]-\log \left[\Delta_n(\hat{\bm{\xi}}^{(n)})\phi_n(\hat{\bm{\xi}}^{(n)})\right]\\[2mm]
        =&(1-c_1)\log n-(1-c_3)\log \hat{\xi}_3^{(n)}-(d_1-d_3)-(\mu_1-\mu_3)+o(n^{-\delta}),\\[2mm]
        &\log \left[\Delta_n(\hat{\xi}_2^{(n)},\hat{\xi}_3^{(n)}-1,\ldots,\hat{\xi}_J^{(n)})\phi_n(\hat{\xi}_2^{(n)},\hat{\xi}_3^{(n)}-1,\ldots,\hat{\xi}_J^{(n)})\right]-\log \left[\Delta_n(\hat{\bm{\xi}}^{(n)})\phi_n(\hat{\bm{\xi}}^{(n)})\right]\\[2mm]
        =&-(1-c_1)\log n+(1-c_3)\log \hat{\xi}_3^{(n)}+(d_1-d_3)+(\mu_1-\mu_3)+o(n^{-\delta}).
    \end{align*}
    By the definition of $\hat{\bm{\xi}}^{(n)}$, we can derive a first-order-type argument as
    \begin{align*}
        \log \hat{\xi}_2^{(n)}=&\frac{(1-c_1)\log n-(d_1-d_2)-(\mu_1-\mu_2)}{1-c_2}+o(n^{-\delta}),\\
        \log \hat{\xi}_3^{(n)}=&\frac{(1-c_1)\log n-(d_1-d_3)-(\mu_1-\mu_3)}{1-c_3}+o(n^{-\delta}).
    \end{align*}
    Similarly, we can prove that
    \begin{align*}
        \log \widetilde{\xi}_2^{(n)}=&\frac{(1-c_1)\log n-(d_1-d_2)-(\mu_1-\widetilde{\mu}_2)}{1-c_2}+o(n^{-\delta}),\\
        \log \widetilde{\xi}_3^{(n)}=&\frac{(1-c_1)\log n-(d_1-d_3)-(\mu_1-\widetilde{\mu}_3)}{1-c_3}+o(n^{-\delta}).
    \end{align*}
    By the construction of characterisation equation, we have
    \begin{align*}
        1-\nu_2=&\vec{\eta}_1^{\top}\vec{\theta}(\nu_2,\ldots,\nu_J)-\vec{\eta}_2^{\top}\vec{\theta}(\nu_2,\ldots,\nu_J)=c_1-c_2\nu_2,\\
        1-\nu_3=&\vec{\eta}_1^{\top}\vec{\theta}(\nu_2,\ldots,\nu_J)-\vec{\eta}_3^{\top}\vec{\theta}(\nu_2,\ldots,\nu_J)=c_1-c_3\nu_3.
    \end{align*}
    Hence $\nu_2=(1-c_1)/(1-c_2)$ and $\nu_3=(1-c_1)/(1-c_3)$. This implies that $1>c_1>c_2=c_3$ since $\nu_2=\nu_3<1$ and $c_1>c_2\vee c_3$. For any $|k_2|\vee|k_3|\geq n^{(\nu_2+\delta)/2}$,
    \begin{align*}
        &\log \Delta_n(\hat{\xi}_2^{(n)}+k_2,\hat{\xi}_3^{(n)}+k_3,\ldots,\hat{\xi}_J^{(n)})\phi_n(\hat{\xi}_2^{(n)}+k_2,\hat{\xi}_3^{(n)}+k_3,\ldots,\hat{\xi}_J^{(n)})-\log \Delta_n(\hat{\bm{\xi}}^{(n)})\phi_n(\hat{\bm{\xi}}^{(n)})\\[2mm]
        =&o(n^{\delta})+(\hat{\xi}_2^{(n)}+k_2)\left[-(c_1-1)\log n+(c_2-1)\log (\hat{\xi}_2^{(n)}+k_2)-(d_1-d_2)-(c_2-1)\right]\\
        &-\hat{\xi}_2^{(n)}\left[-(c_1-1)\log n+(c_2-1)\log \hat{\xi}_2^{(n)}-(d_1-d_2)-(c_2-1)\right]\\
        &+(\hat{\xi}_3^{(n)}+k_3)\left[-(c_1-1)\log n+(c_2-1)\log (\hat{\xi}_3^{(n)}+k_3)-(d_1-d_3)-(c_2-1)\right]\\
        &-\hat{\xi}_3^{(n)}\left[-(c_1-1)\log n+(c_2-1)\log \hat{\xi}_3^{(n)}-(d_1-d_3)-(c_2-1)\right]\\
        =&o(n^{\delta})+k_2\left[-(c_1-1)\log n+(c_2-1)\log \hat{\xi}_2^{(n)}-(d_1-d_2)-(c_2-1)\right]\\
        &+(\hat{\xi}_2^{(n)}+k_2)\left[(c_2-1)\log \frac{\hat{\xi}_2^{(n)}+k_2}{\hat{\xi}_2^{(n)}}\right]\\
        &+k_3\left[-(c_1-1)\log n+(c_2-1)\log \hat{\xi}_3^{(n)}-(d_1-d_3)-(c_2-1)\right]\\
        &+(\hat{\xi}_3^{(n)}+k_3)\left[(c_2-1)\log \frac{\hat{\xi}_3^{(n)}+k_3}{\hat{\xi}_3^{(n)}}\right]\\
        =&o(n^{\delta})-(1-c_2)\left((\hat{\xi}_2^{(n)}+k_2)\log \frac{\hat{\xi}_2^{(n)}+k_2}{\hat{\xi}_2^{(n)}}-k_2\right)-(1-c_2)\left((\hat{\xi}_3^{(n)}+k_3)\log \frac{\hat{\xi}_3^{(n)}+k_3}{\hat{\xi}_3^{(n)}}-k_3\right).
    \end{align*}
    It can be readily shown that $(\hat{\xi}_2^{(n)}+k_2)(\log (\hat{\xi}_2^{(n)}+k_2)-\log\hat{\xi}_2^{(n)})-k_2$ and $(\hat{\xi}_3^{(n)}+k_3)(\log (\hat{\xi}_3^{(n)}+k_3)-\log\hat{\xi}_3^{(n)})-k_3$ are monotone increasing in $k_2$ and $k_3$ when $k_2\geq 0$ and $k_3\geq 0$, respectively, and are monotone decreasing in $k_2$ and $k_3$ when $k_2\leq 0$ and $k_3\leq 0$, respectively. So when $|k_2|\vee|k_3|\geq n^{(\nu_2+\delta)/2}$ (assume WLOG that $|k_2|\geq n^{(\nu_2+\delta)/2}$), we have
    \begin{align}\label{equation555}
        &\log \Delta_n(\hat{\xi}_2^{(n)}+k_2,\hat{\xi}_3^{(n)}+k_3,\ldots,\hat{\xi}_J^{(n)})\phi_n(\hat{\xi}_2^{(n)}+k_2,\hat{\xi}_3^{(n)}+k_3,\ldots,\hat{\xi}_J^{(n)})-\log \Delta_n(\hat{\bm{\xi}}^{(n)})\phi_n(\hat{\bm{\xi}}^{(n)})\notag\\
        \leq& o(n^{\delta})-(1-c_2)\left((\hat{\xi}_2^{(n)}+n^{(\nu_2+\delta)/2})\log \frac{\hat{\xi}_2^{(n)}+n^{(\nu_2+\delta)/2}}{\hat{\xi}_2^{(n)}}-n^{(\nu_2+\delta)/2}\right)\notag\\
        =&o(n^{\delta})-(1-c_2)\left((\hat{\xi}_2^{(n)}+n^{(\nu_2+\delta)/2}) \left[\frac{n^{(\nu_2+\delta)/2}}{\hat{\xi}_2^{(n)}}+\mathcal{O}\left(\frac{n^{(\nu_2+\delta)/2}}{\hat{\xi}_2^{(n)}}\right)^2\right]-n^{(\nu_2+\delta)/2}\right)\notag\\
        =&o(n^{\delta})-\frac{(1-c_2+o(1))n^{\nu_2+\delta}}{\hat{\xi}_2^{(n)}}\notag\\
        \leq&-cn^{\delta},
    \end{align}
    where $c>0$ is a constant. Similar to Step 3.2, we can also show that
    \begin{align}\label{equation6}
        \frac{\sum_{\bm{\xi}^{(n)}\in\mathcal{O}_n\setminus\mathcal{E}_{k,n}}\Delta_n(\bm{\xi}^{(n)})\phi_n(\bm{\xi}^{(n)})}{\sum_{\bm{\xi}^{(n)}\in\mathcal{O}_{n}}\Delta_n(\bm{\xi}^{(n)})\phi_n(\bm{\xi}^{(n)})}\lesssim \exp(-cn^{\delta}).
    \end{align}
    Define the set $\mathcal{A}_n$ as
    \begin{align*}
        \mathcal{A}_n=\left\{\bm{\xi}=(\xi_2,\xi_3,0,\ldots,0)^{\top}\in \mathcal{E}_{k,n}: |\xi_2-\hat{\xi}_2^{(n)}|\vee |\xi_3-\hat{\xi}_3^{(n)}|\leq n^{(\nu_2+\delta)/2} \right\}.
    \end{align*}
    Then by (\ref{eq_thm_1.64}), (\ref{equation555}) and (\ref{equation6}) we have
    \begin{align}\label{equation7}
        &\frac{\sum_{\bm{\xi}^{(n)}\in\mathcal{O}_n\setminus\mathcal{A}_{n}}\Delta_n(\bm{\xi}^{(n)})\phi_n(\bm{\xi}^{(n)})}{\sum_{\bm{\xi}^{(n)}\in\mathcal{O}_{n}}\Delta_n(\bm{\xi}^{(n)})\phi_n(\bm{\xi}^{(n)})}\notag\\
        \leq& \frac{\sum_{\bm{\xi}^{(n)}\in\mathcal{O}_n\setminus\mathcal{E}_{k,n}}\Delta_n(\bm{\xi}^{(n)})\phi_n(\bm{\xi}^{(n)})}{\sum_{\bm{\xi}^{(n)}\in\mathcal{O}_{n}}\Delta_n(\bm{\xi}^{(n)})\phi_n(\bm{\xi}^{(n)})}+\frac{\sum_{\bm{\xi}^{(n)}\in\mathcal{E}_{k,n}\setminus\mathcal{A}_{n}}\Delta_n(\bm{\xi}^{(n)})\phi_n(\bm{\xi}^{(n)})}{\sum_{\bm{\xi}^{(n)}\in\mathcal{E}_{k,n}}\Delta_n(\bm{\xi}^{(n)})\phi_n(\bm{\xi}^{(n)})}\notag\\
        \leq&\frac{\sum_{\bm{\xi}^{(n)}\in\mathcal{O}_n\setminus\mathcal{E}_{k,n}}\Delta_n(\bm{\xi}^{(n)})\phi_n(\bm{\xi}^{(n)})}{\sum_{\bm{\xi}^{(n)}\in\mathcal{O}_{n}}\Delta_n(\bm{\xi}^{(n)})\phi_n(\bm{\xi}^{(n)})}+\frac{{\sum}_{l\geq 2}{\sum}_{{\bm{\xi}}^{(l,n)}\in \mathcal{E}_{k,n}}\Delta_n({\bm{\xi}}^{(l,n)})\phi_n({\bm{\xi}}^{(l,n)})}{\sum_{\bm{\xi}^{(n)}\in\mathcal{E}_{k,n}}\Delta_n(\bm{\xi}^{(n)})\phi_n(\bm{\xi}^{(n)})}\notag\\
        &+\frac{\sum_{\bm{\xi}^{(n)}\in\mathcal{E}_{k,n}\setminus\mathcal{A}_{n}:\xi_4=\ldots=\xi_J=0}\Delta_n(\bm{\xi}^{(n)})\phi_n(\bm{\xi}^{(n)})}{\Delta_n(\hat{\bm{\xi}}^{(n)})\phi_n(\hat{\bm{\xi}}^{(n)})}\notag\\
        \lesssim& \exp(-cn^{\delta})+\exp(-c^{\prime}\log n)+\exp(-cn^{\delta})n^J\notag\\
        \lesssim& \exp(-c^{\prime}\log n).
    \end{align}
    Similarly, we can define $\widetilde{\mathcal{A}}_n$ for the right-hand side and prove that
    \begin{align}\label{equation8}
        \frac{\sum_{\bm{\xi}^{(n)}\in\mathcal{O}_n\setminus\widetilde{\mathcal{A}}_{n}}\Delta_n(\bm{\xi}^{(n)})\widetilde{\phi}_n(\bm{\xi}^{(n)})}{\sum_{\bm{\xi}^{(n)}\in\mathcal{O}_{n}}\Delta_n(\bm{\xi}^{(n)})\widetilde{\phi}_n(\bm{\xi}^{(n)})}\lesssim& \exp(-c^{\prime}\log n).
    \end{align}
    Then by (\ref{eq_thm_1.66}), (\ref{equation7}) and (\ref{equation8}) we have
    \begin{align}\label{equation12}
        \left|\log\sum_{\bm{\xi}^{(n)}\in\mathcal{A}_n}\Delta_n(\bm{\xi}^{(n)})\phi_n(\bm{\xi}^{(n)})-\log\sum_{\bm{\xi}^{(n)}\in\widetilde{\mathcal{A}}_n}\Delta_n(\bm{\xi}^{(n)})\widetilde{\phi}_n(\bm{\xi}^{(n)})\right|\lesssim \exp(-c^{\prime}\log n).
    \end{align}
    For $m=2$, by Corollary~\ref{cor_likelihood} we have
    \begin{align}\label{equation9}      
        \sum_{\bm{\xi}^{(n)}\in\mathcal{O}_n}\Delta_n(\bm{\xi}^{(n)})\phi_{n+1}(\xi_{2}^{(n)}+1,\ldots,\xi_{J}^{(n)})=\sum_{\bm{\xi}^{(n)}\in\mathcal{O}_n}\Delta_n(\bm{\xi}^{(n)})\widetilde{\phi}_{n+1}(\xi_{2}^{(n)}+1,\ldots,\xi_{J}^{(n)}).
    \end{align}
    We can also show that $(\nu_2,\ldots,\nu_J)^{\top}$ and $(\widetilde{\nu}_2,\ldots,\widetilde{\nu}_J)^{\top}$ are the concentration points for both sides. By similar arguments, we can show that
    \begin{align}\label{equation10}
        &\frac{\sum_{\bm{\xi}^{(n)}\in\mathcal{O}_n\setminus\mathcal{A}_{n}}\Delta_n(\bm{\xi}^{(n)})\phi_{n+1}(\xi_{2}^{(n)}+1,\ldots,\xi_{J}^{(n)})}{\sum_{\bm{\xi}^{(n)}\in\mathcal{O}_{n}}\Delta_n(\bm{\xi}^{(n)})\phi_{n+1}(\xi_{2}^{(n)}+1,\ldots,\xi_{J}^{(n)})}\notag\\
        \leq& \frac{\sum_{\bm{\xi}^{(n)}\in\mathcal{O}_n\setminus\mathcal{E}_{k,n}}\Delta_n(\bm{\xi}^{(n)})\phi_{n+1}(\xi_{2}^{(n)}+1,\ldots,\xi_{J}^{(n)})}{\sum_{\bm{\xi}^{(n)}\in\mathcal{O}_{n}}\Delta_n(\bm{\xi}^{(n)})\phi_{n+1}(\xi_{2}^{(n)}+1,\ldots,\xi_{J}^{(n)})}+\frac{\sum_{\bm{\xi}^{(n)}\in\mathcal{E}_{k,n}\setminus\mathcal{A}_{n}}\Delta_n(\bm{\xi}^{(n)})\phi_{n+1}(\xi_{2}^{(n)}+1,\ldots,\xi_{J}^{(n)})}{\sum_{\bm{\xi}^{(n)}\in\mathcal{E}_{k,n}}\Delta_n(\bm{\xi}^{(n)})\phi_{n+1}(\xi_{2}^{(n)}+1,\ldots,\xi_{J}^{(n)})}\notag\\
        \leq&\frac{\sum_{\bm{\xi}^{(n)}\in\mathcal{O}_n\setminus\mathcal{E}_{k,n}}\Delta_n(\bm{\xi}^{(n)})\phi_{n+1}(\xi_{2}^{(n)}+1,\ldots,\xi_{J}^{(n)})}{\sum_{\bm{\xi}^{(n)}\in\mathcal{O}_{n}}\Delta_n(\bm{\xi}^{(n)})\phi_{n+1}(\xi_{2}^{(n)}+1,\ldots,\xi_{J}^{(n)})}+\frac{{\sum}_{l\geq 2}{\sum}_{{\bm{\xi}}^{(l,n)}\in \mathcal{E}_{k,n}}\Delta_n({\bm{\xi}}^{(l,n)})\phi_{n+1}(\xi_{2}^{(n)}+1,\ldots,\xi_{J}^{(n)})}{\sum_{\bm{\xi}^{(n)}\in\mathcal{E}_{k,n}}\Delta_n(\bm{\xi}^{(n)})\phi_{n+1}(\xi_{2}^{(n)}+1,\ldots,\xi_{J}^{(n)})}\notag\\
        &+\frac{\sum_{\bm{\xi}^{(n)}\in\mathcal{E}_{k,n}\setminus\mathcal{A}_{n}:\xi_4=\ldots=\xi_J=0}\Delta_n(\bm{\xi}^{(n)})\phi_{n+1}(\xi_{2}^{(n)}+1,\ldots,\xi_{J}^{(n)})}{\Delta_n(\hat{\bm{\xi}}^{(n)})\phi_{n+1}(\hat{\xi}_{2}^{(n)}+1,\ldots,\hat{\xi}_{J}^{(n)})}\notag\\
        \lesssim& \exp(-cn^{\delta})+\exp(-c^{\prime}\log n)+\exp(-cn^{\delta}+c^{\prime\prime}\log n)n^J\notag\\
        \lesssim& \exp(-c^{\prime}\log n).
    \end{align}
    Similarly we have
    \begin{align}\label{equation11}
        \frac{\sum_{\bm{\xi}^{(n)}\in\mathcal{O}_n\setminus\widetilde{\mathcal{A}}_{n}}\Delta_n(\bm{\xi}^{(n)})\widetilde{\phi}_{n+1}(\xi_{2}^{(n)}+1,\ldots,\xi_{J}^{(n)})}{\sum_{\bm{\xi}^{(n)}\in\mathcal{O}_{n}}\Delta_n(\bm{\xi}^{(n)})\widetilde{\phi}_{n+1}(\xi_{2}^{(n)}+1,\ldots,\xi_{J}^{(n)})}\lesssim& \exp(-c^{\prime}\log n).
    \end{align}
    By (\ref{equation9}), (\ref{equation10}) and (\ref{equation11}) we have
    \begin{align}\label{equation13}
        &\left|\log\sum_{\bm{\xi}^{(n)}\in\mathcal{A}_{n}}\Delta_n(\bm{\xi}^{(n)})\phi_{n+1}(\xi_{2}^{(n)}+1,\ldots,\xi_{J}^{(n)})-\log\sum_{\bm{\xi}^{(n)}\in\widetilde{\mathcal{A}}_{n}}\Delta_n(\bm{\xi}^{(n)})\widetilde{\phi}_{n+1}(\xi_{2}^{(n)}+1,\ldots,\xi_{J}^{(n)})\right|\notag\\
        \lesssim& \exp(-c^{\prime}\log n).
    \end{align}
    For any $\bm{\xi}^{(n)}\in\mathcal{A}_{n}$, by the construction method of characterisation equation,
    \begin{align*}
        \vec{\eta}_2^{\top}\vec{\theta}_n(\bm{\xi}^{(n)})=c_2\log \xi_2^{(n)}+d_2+o(n^{-\delta})=c_2\log \hat{\xi}_2^{(n)}+d_2+o(n^{-\delta}).
     \end{align*}
    So we can prove that
    \begin{align}\label{equation14}
        &\log\sum_{\bm{\xi}^{(n)}\in\mathcal{A}_{n}}\Delta_n(\bm{\xi}^{(n)})\phi_{n+1}(\xi_{2}^{(n)}+1,\ldots,\xi_{J}^{(n)})-\log\sum_{\bm{\xi}^{(n)}\in\mathcal{A}_n}\Delta_n(\bm{\xi}^{(n)})\phi_{n}(\bm{\xi}^{(n)})\notag\\
        =&c_2\log \hat{\xi}_2^{(n)}+d_2+\mu_2+o(n^{-\delta})\notag\\
        =&c_2\nu_2\log n+c_2\frac{-(d_1-d_2)-(\mu_1-\mu_2)}{1-c_2}+\mu_2+o(n^{-\delta}).
    \end{align}
    Similarly, we have
    \begin{align}\label{equation15}
        &\log\sum_{\bm{\xi}^{(n)}\in\widetilde{\mathcal{A}}_{n}}\Delta_n(\bm{\xi}^{(n)})\widetilde{\phi}_{n+1}(\xi_{2}^{(n)}+1,\ldots,\xi_{J}^{(n)})-\log\sum_{\bm{\xi}^{(n)}\in\widetilde{\mathcal{A}}_n}\Delta_n(\bm{\xi}^{(n)})\widetilde{\phi}_{n}(\bm{\xi}^{(n)})\notag\\
        =&c_2\log \widetilde{\xi}_2^{(n)}+d_2+\widetilde{\mu}_2+o(n^{-\delta})\notag\\
        =&c_2\nu_2\log n+c_2\frac{-(d_1-d_2)-(\mu_1-\widetilde{\mu}_2)}{1-c_2}+\widetilde{\mu}_2+o(n^{-\delta}).
    \end{align}
    Then by (\ref{equation12}), (\ref{equation13}), (\ref{equation14}) and (\ref{equation15}), we have
    \begin{align*}
        c_2\frac{-(d_1-d_2)-(\mu_1-\mu_2)}{1-c_2}+\mu_2-c_2\frac{-(d_1-d_2)-(\mu_1-\widetilde{\mu}_2)}{1-c_2}-\widetilde{\mu}_2=\frac{\mu_2-\widetilde{\mu}_2}{1-c_2}=0.
    \end{align*}
    Since $0<c_2<1$, this implies $\mu_2=\widetilde{\mu}_2$. Similarly we can prove that $\mu_3=\widetilde{\mu}_3$. Then we can similarly prove that $\mu_j=\widetilde{\mu}_j$, $\vec{\eta}_j^{\top}\vec{\alpha}_k= \widetilde{\vec{\eta}}_j^{\top}\vec{\alpha}_k$ and $\vec{\eta}_{j_1}^{\top}\vec{\eta}_{j_2}=\widetilde{\vec{\eta}}_{j_1}^{\top}\widetilde{\vec{\eta}}_{j_2}$ for any $j,j_1,j_2=1,\ldots,p$ and $k=1,\ldots,W$.\\[3mm]
    \textbf{Case 4: }There exists a tie among $(\nu_2,\ldots,\nu_p)^{\top}$ or $(\widetilde{\nu}_2,\ldots,\widetilde{\nu}_{\widetilde{p}})^{\top}$. Moreover, at least one of the characterisation equations at $(\nu_2,\ldots,\nu_p)^{\top}$ and $(\widetilde{\nu}_2,\ldots,\widetilde{\nu}_{\widetilde{p}})^{\top}$ contains a type-2 equation.\\[3mm]
    By arguments similar to those used in Cases 1 and 3, we can match all terms from type-1 equations in decreasing order. Then by arguments similar to those used in Case 2, we can match the whole characterisation equation. Hence we can prove that $\mu_j=\widetilde{\mu}_j$, $\vec{\eta}_j^{\top}\vec{\alpha}_k= \widetilde{\vec{\eta}}_j^{\top}\vec{\alpha}_k$ and $\vec{\eta}_{j_1}^{\top}\vec{\eta}_{j_2}=\widetilde{\vec{\eta}}_{j_1}^{\top}\widetilde{\vec{\eta}}_{j_2}$ for any $j,j_1,j_2=1,\ldots,p$ and $k=1,\ldots,W$.\\[3mm]
    \textbf{Step 7: }Prove that $\mu_j=\widetilde{\mu}_j$, $\vec{\eta}_j^{\top}\vec{\alpha}_k= \widetilde{\vec{\eta}}_j^{\top}\vec{\alpha}_k$ and $\vec{\eta}_{j_1}^{\top}\vec{\eta}_{j_2}=\widetilde{\vec{\eta}}_{j_1}^{\top}\widetilde{\vec{\eta}}_{j_2}$ for any $j,j_1,j_2=1,\ldots,J$, $k=1,\ldots,W$ and fix the permutation.\\[3mm]
    Since the continuity of $\vec{\theta}(\nu_2,\ldots,\nu_J)$ and $\widetilde{\vec{\theta}}(\nu_2,\ldots,\nu_J)$ with respect to $\nu_2,\ldots,\nu_p$ is guaranteed by Proposition~\ref{prop_characterisation_equation}, we can use a similar induction argument as in Step 5 of Case 1 to prove that for any $1\leq r\leq r^{\ast}$,
    \begin{align*}
        &\Delta_n(\xi_2^{(n)},\ldots,\xi_p^{(n)},\xi_{p+1}^{(r)},\ldots,\xi_{J}^{(r)})\phi_n(\xi_2^{(n)},\ldots,\xi_p^{(n)},\xi_{p+1}^{(r)},\ldots,\xi_{J}^{(r)})\\
        =&\Delta_n(\widetilde{\xi}_2^{(n)},\ldots,\widetilde{\xi}_{p}^{(n)},\widetilde{\xi}_{p+1}^{(r)},\ldots,\widetilde{\xi}_{J}^{(r)})\widetilde{\phi}_n(\widetilde{\xi}_2^{(n)},\ldots,\widetilde{\xi}_{p}^{(n)},\widetilde{\xi}_{p+1}^{(r)},\ldots,\widetilde{\xi}_{J}^{(r)})
    \end{align*}
    for any $(\xi_2^{(n)},\ldots,\xi_p^{(n)})^{\top}$. Then by similar arguments, we can prove that there exists a permutation $\pi:\{1,\ldots,J\}\rightarrow\{1,\ldots,J\}$ such that $\mu_j=\widetilde{\mu}_{\pi(j)}$, $\vec{\eta}_j^{\top}\vec{\alpha}_k= \widetilde{\vec{\eta}}_{\pi(j)}^{\top}\vec{\alpha}_k$ and $\vec{\eta}_{j_1}^{\top}\vec{\eta}_{j_2}=\widetilde{\vec{\eta}}_{\pi(j_1)}^{\top}\widetilde{\vec{\eta}}_{\pi(j_2)}$ for any $j,j_1,j_2=1,\ldots,J$ and $k=1,\ldots,W$.\\[3mm]
    Finally, we use similar arguments as in Step 6 of Case 1 to prove that $\vec{\eta}_j^{\top}\vec{\theta}(\nu_2,\ldots,\nu_J)=\widetilde{\vec{\eta}}_j^{\top}\widetilde{\vec{\theta}}(\nu_2,\ldots,\nu_J)$. By excluding a set of measure zero, we can assume that $\vec{\eta}_1^{\top}\vec{\theta}(\nu_2,\ldots,\nu_J),\ldots,\vec{\eta}_J^{\top}\vec{\theta}(\nu_2,\ldots,\nu_J)$ are distinct. So we can similarly show that $\pi=id$, $\mu_j=\widetilde{\mu}_j$, $\vec{\eta}_j^{\top}\vec{\alpha}_k= \widetilde{\vec{\eta}}_j^{\top}\vec{\alpha}_k$ and $\vec{\eta}_{j_1}^{\top}\vec{\eta}_{j_2}=\widetilde{\vec{\eta}}_{j_1}^{\top}\widetilde{\vec{\eta}}_{j_2}$. So for any $j,j_1,j_2=1,\ldots,J$ and any $0\leq t\leq s\leq t_{q+1}$, we proved that
    \begin{align*}
        \mu_j(t_{q+1})=&\widetilde{\mu}_j(t_{q+1}),\\
        \vec{Z}_{j_1}^{\top}(t)\mat{A}_{j_1}\mat{\Sigma} \mat{A}^{\top}_{j_2}\vec{Z}_{j_2}(s)&=\vec{Z}_{j_1}^{\top}(t)\widetilde{\mat{A}}_{j_1}\widetilde{\mat{\Sigma}} \widetilde{\mat{A}}^{\top}_{j_2}\vec{Z}_{j_2}(s).
    \end{align*}
    If there exist multiple maximisers among $\sum_{k=1}^{m_1}c_{1,k},\ldots,\sum_{k=1}^{m_J}c_{J,k}$ or $\sum_{k=1}^{\widetilde{m}_1}\widetilde{c}_{1,k},\ldots,\sum_{k=1}^{\widetilde{m}_J}\widetilde{c}_{J,k}$, we assume WLOG that $\sum_{k=1}^{m_1}c_{1,k}=\sum_{k=1}^{m_2}c_{2,k}$ are all the maximisers among $\sum_{k=1}^{m_1}c_{1,k},\ldots,\sum_{k=1}^{m_J}c_{J,k}$. This indicates that $\vec{\eta}_1$, $\vec{\eta}_2$ have degenerate expansions. So we have $\vec{\eta}_1=\widetilde{\vec{\eta}}_1$ and $\vec{\eta}_2=\widetilde{\vec{\eta}}_2$.\\[3mm]
    We use similar method to prove the result by the following two steps:\\[3mm]
    \textbf{Step 1: }Similarly to Step 2 in Case 1, we first partition over $[0,1]^{J}$ to find the concentration point under scaling $\bm{\xi}^{(n)}/n$. Then by arguments similar to those used in Step 2 of Case 2, we can prove that this concentration point should have zero components in the $3$rd through $J$th components.\\[3mm]
    \textbf{Step 2: }Similar to Step 2 in Case 2, we then partition over $[0,1]^{J-2}$ under scaling $\log \bm{\xi}^{(n)}/\log n$ to find the concentration point. Similar arguments can be performed to characterise the concentration point. We can still construct a characterisation equation at the concentration point. Since in the characterisation equation, the expansions of $\vec{\eta}_1$ and $\vec{\eta}_2$ contain disjoint terms, we can still prove a continuity result which is similar to Proposition~\ref{prop_characterisation_equation} around the concentration point. Then similar arguments as in Steps 3-6 can be performed to prove the result.\\[3mm]
    \textbf{Case 3:} $\|P_{\mathcal{H}^{\perp}_{\vec{\eta}_1}}\vec{\eta}_1\|>0$, $\|P_{\mathcal{H}^{\perp}_{\widetilde{\vec{\eta}}_{1}}}\widetilde{\vec{\eta}}_{1}\|=0$ or $\|P_{\mathcal{H}^{\perp}_{\vec{\eta}_1}}\vec{\eta}_1\|=0$, $\|P_{\mathcal{H}^{\perp}_{\widetilde{\vec{\eta}}_{1}}}\widetilde{\vec{\eta}}_{1}\|>0$.\\[3mm]
    We only discuss the first scenario. For any $n\in\mathbb{N}$, define $\bar{\bm{\xi}}^{(n)}={\operatorname{argmax}}_{\bm{\xi}\in\mathcal{O}_n}\widetilde{\phi}_n(\bm{\xi})$ and suppose that
    \begin{align*}
        \lim_{n\rightarrow\infty}\frac{(n-\sum_{j=2}^J\bar{\xi}_j^{(n)})\widetilde{\vec{\eta}}_1+\sum_{j=2}^J\bar{\xi}_j^{(n)}\widetilde{\vec{\eta}}_j}{n}=\sum_{j=1}^{J}\nu_j\widetilde{\vec{\eta}}_j,
    \end{align*}
    where $0\leq\nu_1,\ldots,\nu_J\leq 1$ and $\sum_{j=1}^J\nu_j=1$. Since $\widetilde{\vec{\eta}}_1,\ldots,\widetilde{\vec{\eta}}_J\in \mathcal{X}$, we have $\sum_{j=1}^{J}\nu_j\widetilde{\vec{\eta}}_j\in \mathcal{X}$. So it can be readily shown that
    \begin{align*}
        P_{\mathcal{H}_{\sum_{j=1}^{J}\nu_j\widetilde{\vec{\eta}}_j}^{\perp}}\sum_{j=1}^{J}\nu_j\widetilde{\vec{\eta}}_j=0.
    \end{align*}
    Then by part (2) in Proposition~\ref{prop_canonical_projection}, we have $\widetilde{\vec{\theta}}_n(\bar{\bm{\xi}}^{(n)})/n\rightarrow 0$. Then we have
    \begin{align}\label{eq_thm_1.80}
        \sum_{\bm{\xi}^{(n)}\in\mathcal{O}_n}\Delta_n(\bm{\xi}^{(n)})\widetilde{\phi}_n(\bm{\xi}^{(n)})\leq& J^n\widetilde{\phi}_n(\bar{\bm{\xi}}^{(n)})=J^n\exp(o(n^2))=\exp(o(n^2)).
    \end{align}
    On the other hand, from the proof in Case 1, we have
    \begin{align}\label{eq_thm_1.81}
        {\phi}_n(\mathbf{0})=\exp\left(o(n^2)+\frac{n^2}{2}\|P_{\mathcal{H}_{\vec{\eta}_1}^{\perp}}\vec{\eta}_1\|^2\right).
    \end{align}
    Then (\ref{eq_thm_1.10}), (\ref{eq_thm_1.80}) and (\ref{eq_thm_1.81}) lead to a contradiction.\\[3mm]
    Hence, by induction, we prove that for any $j,j_1,j_2$ and $0\leq t\leq s\leq T$, with probability 1 it holds that
    \begin{align*}
        \beta_{j0}+\vec{\beta}_j^{\top}\vec{X}_j(t)&=\widetilde{\beta}_{j0}+\widetilde{\vec{\beta}}_j^{\top}\vec{X}_j(t),\\
        \vec{Z}_{j_1}^{\top}(t)\mat{A}_{j_1}\mat{\Sigma} \mat{A}^{\top}_{j_2}\vec{Z}_{j_2}(s)&=\vec{Z}_{j_1}^{\top}(t)\widetilde{\mat{A}}_{j_1}\widetilde{\mat{\Sigma}} \widetilde{\mat{A}}^{\top}_{j_2}\vec{Z}_{j_2}(s).
    \end{align*}
    By Condition (d), this implies that for any $j,j_1,j_2=1,\ldots,J$, $\beta_{j0}=\widetilde{\beta}_{j0}$, $\vec{\beta}_j=\widetilde{\vec{\beta}}_j$ and $\mat{A}_{j_1}\mat{\Sigma} \mat{A}^{\top}_{j_2}=\widetilde{\mat{A}}_{j_1}\widetilde{\mat{\Sigma}} \widetilde{\mat{A}}^{\top}_{j_2}$. So we have $\mat{A}\mat{\Sigma} \mat{A}^{\top}=\widetilde{\mat{A}}\widetilde{\mat{\Sigma}} \widetilde{\mat{A}}^{\top}$. By Condition (c), there exist permutation matrices $\mat{D}\in\mathbb{R}^{L\times L}$ (here $L:=\sum_{j=1}^JL_{2j}$), $\mat{C}\in\mathbb{R}^{K\times K}$ and $\widetilde{\mat{C}}\in\mathbb{R}^{K\times K}$ such that $\mat{D}\mat{A}\mat{C}=(\mat{I}_{K},\mat{R}^{\top})^{\top}$ and $\mat{D}\widetilde{\mat{A}}\widetilde{\mat{C}}=(\mat{I}_{K},\widetilde{\mat{R}}^{\top})^{\top}$. Then it can be readily shown that 
    \begin{align*}
        \mat{D}\mat{A}\mat{\Sigma} \mat{A}^{\top}\mat{D}^{\top}=(\mat{D}\mat{A}\mat{C})\mat{C}^{\top}\mat{\Sigma}\mat{C}(\mat{C}^{\top}\mat{A}^{\top}\mat{D}^{\top})=\begin{pmatrix}
            \mat{I}_K\\ \mat{R}
        \end{pmatrix}\mat{C}^{\top}\mat{\Sigma}\mat{C}(\mat{I}_{K},\mat{R}^{\top})=\begin{pmatrix}
            \mat{C}^{\top}\mat{\Sigma}\mat{C}&\mat{C}^{\top}\mat{\Sigma}\mat{C}\mat{R}^{\top}\\ \mat{R}\mat{C}^{\top}\mat{\Sigma}\mat{C}&\mat{R}\mat{C}^{\top}\mat{\Sigma}\mat{C}\mat{R}^{\top}
        \end{pmatrix}\\
        \mat{D}\widetilde{\mat{A}}\widetilde{\mat{\Sigma}} \widetilde{\mat{A}}^{\top}\mat{D}^{\top}=(\mat{D}\widetilde{\mat{A}}\widetilde{\mat{C}})\widetilde{\mat{C}}^{\top}\widetilde{\mat{\Sigma}}\widetilde{\mat{C}}(\widetilde{\mat{C}}^{\top}\widetilde{\mat{A}}^{\top}\mat{D}^{\top})=\begin{pmatrix}
            \mat{I}_K\\ \widetilde{\mat{R}}
        \end{pmatrix}\widetilde{\mat{C}}^{\top}\widetilde{\mat{\Sigma}}\widetilde{\mat{C}}(\mat{I}_{K},\widetilde{\mat{R}}^{\top})=\begin{pmatrix}
            \widetilde{\mat{C}}^{\top}\widetilde{\mat{\Sigma}}\widetilde{\mat{C}}&\widetilde{\mat{C}}^{\top}\widetilde{\mat{\Sigma}}\widetilde{\mat{C}}\widetilde{\mat{R}}^{\top}\\ \widetilde{\mat{R}}\widetilde{\mat{C}}^{\top}\widetilde{\mat{\Sigma}}\widetilde{\mat{C}}&\widetilde{\mat{R}}\widetilde{\mat{C}}^{\top}\widetilde{\mat{\Sigma}}\widetilde{\mat{C}}\widetilde{\mat{R}}^{\top}
        \end{pmatrix}.
    \end{align*}
    Denote $\mat{B}=\widetilde{\mat{C}}\mat{C}^{\top}$. Since $\mat{C}$ and $\widetilde{\mat{C}}$ are permutation matrices, $\mat{B}$ is also a permutation matrix.
    Note that $\mat{A}\mat{\Sigma} \mat{A}^{\top}=\widetilde{\mat{A}}\widetilde{\mat{\Sigma}} \widetilde{\mat{A}}^{\top}$ implies that $\mat{C}^{\top}\mat{\Sigma}\mat{C}=\widetilde{\mat{C}}^{\top}\widetilde{\mat{\Sigma}}\widetilde{\mat{C}}$, which further implies that $\mat{B}\mat{\Sigma}\mat{B}^{\top}=\widetilde{\mat{\Sigma}}$. 
    Moreover, $\mat{R}\mat{C}^{\top}\mat{\Sigma}\mat{C}=\widetilde{\mat{R}}\widetilde{\mat{C}}^{\top}\widetilde{\mat{\Sigma}}\widetilde{\mat{C}}$ implies that $\mat{R}=\widetilde{\mat{R}}$. Hence $\mat{D}\mat{A}\mat{C}=\mat{D}\widetilde{\mat{A}}\widetilde{\mat{C}}$, which implies that $\mat{A}\mat{B}^{\top}=\widetilde{\mat{A}}$, i.e., $(\mat{A},\mat{\Sigma})\sim(\widetilde{\mat{A}},\widetilde{\mat{\Sigma}})$. Hence the identifiability result is proved.
    \end{proof}
    
    \subsection{Proof of Proposition \ref{prop_likelihood} and Corollary \ref{cor_likelihood}}
    Since the two competing models induce the same probability measure, i.e., $\mu_{\boldsymbol{\delta}}=\mu_{\widetilde{\boldsymbol{\delta}}}$, there exists a set $\mathcal{A}\subset \Omega$ with $\mathbb{P}(\mathcal{A})=1$, such that the density functions under both models are identical on the set $\mathcal{A}$, i.e., $p_{\boldsymbol{\delta}}(\omega)=p_{\widetilde{\boldsymbol{\delta}}}(\omega)$ for any $\omega\in\mathcal{A}$.
    \renewcommand*{\proofname}{Proof of Proposition \ref{prop_likelihood}}
    
    \begin{proof}
    For any $t\in[0,T]$, there exists $t_0>0$ such that no events occur on the interval $(t,t+t_0]$ and the two intensity functions remain constant on $(t,t+t_0]$. Then for any $0<\Delta t<t_0$, by matching the likelihood function in the two competing models on $[0,t+\Delta t]$, we have
        \begin{align}\label{eq_prop_1.1}
            &\int_{\bt} \prod_{j=1}^{J} \Big[\prod_{s \leq t} (\lambda_{j}(s)^{\Delta N_{j}(s)} )e^{-\int_{0}^{t} \lambda_{j}(s) \mathrm{d} s} \Big]\exp\Big(-\Delta t\sum_{j=1}^J \lambda_j(t+0)\Big)\phi_K(\vec{\theta};\vec{0}, \mat{\Sigma}) \mathrm{d} \vec{\theta}\notag\\
            =&\int_{\bt} \prod_{j=1}^{J} \Big[\prod_{s \leq t} (\widetilde{\lambda}_{j}(s)^{\Delta N_{j}(s)}) e^{-\int_{0}^{t} \widetilde{\lambda}_{j}(s) \mathrm{d} s} \Big]\exp\Big(-\Delta t\sum_{j=1}^J \widetilde{\lambda}_j(t+0)\Big)\phi_K(\vec{\theta} ;\vec{0}, \widetilde{\mat{\Sigma}}) \mathrm{d} \vec{\theta}\text{.}
        \end{align}
        For any $n$, we take the $n$-th derivative of both sides of equation (\ref{eq_prop_1.1}) with respect to $\Delta t$ and let $\Delta t\downarrow 0$, then we have
        \begin{align*}
            &\int_{\bt} \Big[\prod_{j=1}^{J} \prod_{s \leq t} \lambda_{j}(s)^{\Delta N_{j}(s)}\Big]\exp\Big(-\sum_{j=1}^J \int_0^t \lambda_j(s)\mathrm{d}s\Big)\Big(\sum_{j=1}^J\lambda_{j}(t+0)\Big)^{n}\phi_K(\vec{\theta};\vec{0}, \mat{\Sigma}) \mathrm{d} \vec{\theta}\notag\\
            =&\int_{\bt} \Big[\prod_{j=1}^{J} \prod_{s \leq t} \widetilde{\lambda}_{j}(s)^{\Delta N_{j}(s)}\Big]\exp\Big(-\sum_{j=1}^J \int_0^t \widetilde{\lambda}_j(s)\mathrm{d}s\Big)\Big(\sum_{j=1}^J\widetilde{\lambda}_{j}(t+0)\Big)^{n} \phi_K(\vec{\theta} ;\vec{0}, \widetilde{\mat{\Sigma}}) \mathrm{d} \vec{\theta}\text{.}
        \end{align*} 
        Thus the proposition is proved.
        \end{proof}
    \renewcommand*{\proofname}{Proof of Corollary~\ref{cor_likelihood}}
    \begin{proof}
        Fix \(t<\tau\). By the piecewise-constant assumption in Condition (e), for each sample path \(\omega\) there exists \(h_0(\omega)>0\) such that no counting or covariate process jumps on \((t,t+h_0(\omega)]\). Hence, for \(s\in (t,t+h_0(\omega)]\),
        \[
    \lambda_j(s,\omega\mid \boldsymbol{\theta})
    =
    \lambda_j(t+0,\omega\mid \boldsymbol{\theta}),
    \qquad
    \widetilde{\lambda}_j(s,\omega\mid \boldsymbol{\theta})
    =
    \widetilde{\lambda}_j(t+0,\omega\mid \boldsymbol{\theta}).
    \]
    Let
    \[
        L_t(\omega,\boldsymbol{\theta})
        =\Big[\prod_{j=1}^{J}\prod_{s\le t}\lambda_j(s,\omega\mid\boldsymbol{\theta})^{\Delta N_j(s,\omega)}\Big]\exp\Big(-\sum_{j=1}^{J}\int_0^t \lambda_j(s,\omega\mid\boldsymbol{\theta})\,ds\Big).
    \]
    For \(m=1,\ldots,J\) and \(0<h<h_0(\omega)\), consider the path-density contribution of histories that agree with \(\omega\) on \([0,t+h)\) and have one additional event of type \(m\) at the endpoint \(t+h\). Under the first model this contribution equals
    \[
    \Psi_m(\omega,h)=\int_{\bt} L_t(\omega,\boldsymbol{\theta})\,\lambda_m(t+0,\omega\mid\boldsymbol{\theta})\,\exp\Big(-h\sum_{j=1}^{J}\lambda_j(t+0,\omega\mid\boldsymbol{\theta})\Big)\phi_K(\boldsymbol{\theta};\boldsymbol{0},\mat{\Sigma})\,d\boldsymbol{\theta},
    \]
    and analogously define \(\widetilde{\Psi}_m(\omega,h)\) for the competing model.
    Since the two models induce the same law, \(\Psi_m(\omega,h)=\widetilde{\Psi}_m(\omega,h)\) for almost every \((\omega,h)\); continuity in \(h\) then implies equality for all sufficiently small \(h\). Differentiating both sides \(n\) times with respect to \(h\) at \(h=0\) and cancelling the common factor \((-1)^n\) leads to
    \begin{align*}
            &\int_{\bt} \lambda_m(t+0)\Big[\prod_{j=1}^{J} \prod_{s \leq t} \lambda_{j}(s)^{\Delta N_{j}(s)}\Big]\exp\Big(-\sum_{j=1}^J \int_0^t \lambda_j(s)\mathrm{d}s\Big)\Big(\sum_{j=1}^J\lambda_{j}(t+0)\Big)^{n} \phi_K(\vec{\theta};\vec{0}, \mat{\Sigma}) \mathrm{d} \vec{\theta}\notag\\
            =&\int_{\bt} \widetilde{\lambda}_m(t+0)\Big[\prod_{j=1}^{J}\prod_{s \leq t} \widetilde{\lambda}_{j}(s)^{\Delta N_{j}(s)}\Big]\exp\Big(-\sum_{j=1}^J \int_0^t \widetilde{\lambda}_j(s)\mathrm{d}s\Big)\Big(\sum_{j=1}^J\widetilde{\lambda}_{j}(t+0)\Big)^{n}\phi_K(\vec{\theta} ;\vec{0}, \widetilde{\mat{\Sigma}}) \mathrm{d} \vec{\theta}\text{.}
        \end{align*}
        Hence, the result is proved.
    \end{proof}
        \subsection{Proof of Proposition \ref{lem_summation}}
        \renewcommand*{\proofname}{Proof of Proposition \ref{lem_summation}}
    \begin{proof}
    We only prove the case when the values among $\left\{\widetilde{y}_{ij}: 1\leq i\leq j\leq J\right\}$ are also distinct. We assume WLOG that $y_{11}$ is the unique largest term among $y_{11},\ldots,y_{JJ}$ since they are distinct. Furthermore, by (\ref{eq_lem_5.1}) we assume WLOG that $y_{11}>y_{12}>\ldots>y_{1J}$. Suppose that $\widetilde{y}_{j_1j_1}$ is the unique largest term among $\widetilde{y}_{11},\ldots,\widetilde{y}_{JJ}$ and suppose $\widetilde{y}_{j_1j_1}>\widetilde{y}_{j_1j_2}>\ldots>\widetilde{y}_{j_1j_J}$, where $\left\{j_1,\ldots,j_J\right\}$ is a permutation of $\left\{1,\ldots,J\right\}$. Let $\pi(1)=j_1,\ldots,\pi(J)=j_J$. In the following part, we prove that for any $j,j_1,j_2=1,\ldots,J$, $x_{j}=\widetilde{x}_{\pi(j)}$ and $y_{j_1j_2}=\widetilde{y}_{\pi(j_1)\pi(j_2)}$. For notational simplicity, we assume WLOG that $\pi(1)=1,\ldots,\pi(J)=J$.\\[3mm]
        The following proof consists of two steps. In the first step, we prove that the summations on both sides of (\ref{eq_lem_5.2}) can be separated in order, where each term dominates the summation of all terms with lower rank. In the second step, we prove that the dominant terms on both sides can match exactly. Then by an induction argument, we can match every term on both sides.\\[3mm]
        \textbf{Step 1: }For any $(\xi_2,\ldots,\xi_J)^{\top}\in \mathbb{N}_0^{J-1}$ and any $n\in\mathbb{N}$, denote
        \begin{align*}
            T(\xi_2,\ldots,\xi_J)&=\prod_{j=2}^J \left(\frac{y_{1j}}{y_{11}}\right)^{\xi_j}\text{,}\\
            S_n(\xi_2,\ldots,\xi_J)&=x_1^{n-\sum_{j=2}^J\xi_j}y_{11}^{\left(n-\sum_{j=2}^J\xi_j\right)^2}\prod_{j=2}^J\left(x_j^{\xi_j}y_{1j}^{2\xi_j\left(n-\sum_{j=2}^J\xi_j\right)}\right)\prod_{2\leq j_1,j_2\leq J}y_{j_1j_2}^{\xi_{j_1}\xi_{j_2}}.
        \end{align*}
        We rank all the components in $\left\{T(\xi_2,\ldots,\xi_J):~\xi_2,\ldots,\xi_J\in\mathbb{N}_0\right\}$ in decreasing order. For any $r\in\mathbb{N}$, let $(\xi_2^{(r)},\ldots,\xi_J^{(r)})^{\top}$ denote the vector such that $T(\xi_2^{(r)},\ldots,\xi_J^{(r)})$ has rank $r$ and denote $K_{r}=\sum_{j=2}^{J}\xi_j^{(r)}$. We assume that there are no ties in the rank (If there are ties, then similar proof can be performed by putting the tie terms together). Define
        \begin{align*}
            \Delta_{r,n}=\binom{n}{n-\sum_{j=2}^J\xi_j^{(r)},\xi_2^{(r)},\ldots,\xi_J^{(r)}}.
        \end{align*}
        Then we can simplify the summation on the left-hand side of (\ref{eq_lem_5.2}) as
        \begin{align*}
            \sum_{1\leq j_1,\ldots,j_n\leq J}\left(\prod_{k=1}^{n}x_{j_k}\prod_{1\leq k_1,k_2\leq n}y_{j_{k_1}j_{k_2}}\right)=\sum_{r\geq 1}\Delta_{r,n}S_n(\xi_2^{(r)},\ldots,\xi_J^{(r)}).
        \end{align*}
        Since $y_{11}>y_{12}>\ldots>y_{1J}$, we have
        \begin{align*}
            \frac{T(\xi_2,\ldots,\xi_J)}{T(0,\ldots,0)}\leq \left(\frac{y_{12}}{y_{11}}\right)^{\sum_{j=2}^J \xi_j}\text{.}
        \end{align*}
        Hence for any fixed $r$, there exists $r_{\text{max}}$ such that
        \begin{align}\label{eq_lem_5.3}
            \underset{u>r_{\text{max}}}{\max}T(\xi_2^{(u)},\ldots,\xi_J^{(u)})\leq\frac{1}{J}T(\xi_2^{(r)},\ldots,\xi_J^{(r)})\text{.}
        \end{align}
        It can be readily shown that
        \begin{align}\label{eq_lem_5.4}
            \sum_{r=1}^{r_{\text{max}}}\Delta_{r,n}\leq n^{r_{\text{max}}}J^{r_{\text{max}}}\text{.}
        \end{align}
        We assume WLOG that for any $u\geq r_{\text{max}}$, it holds that $K_u>K_r$. Then for any $\widetilde{r}$ such that $\widetilde{r}>r$, we discuss the following two cases:\\[3mm]
        \textbf{Case 1: }If $K_{\widetilde{r}}\geq K_{r}$, then
        \begin{align}\label{eq_lem_5.5}
            &~~~~\frac{S_n(\xi_2^{(\widetilde{r})},\ldots,\xi_J^{(\widetilde{r})})}{S_n(\xi_2^{(r)},\ldots,\xi_J^{(r)})}\notag\\
            &=\frac{x_1^{n-K_{\widetilde{r}}}y_{11}^{(n-K_{\widetilde{r}})^2}\prod_{j=2}^J x_j^{\xi_j^{(\widetilde{r})}}y_{1j}^{2\xi_j^{(\widetilde{r})}(n-K_{\widetilde{r}})}\prod_{2\leq j_1,j_2\leq J}y_{j_1j_2}^{\xi_{j_1}^{(\widetilde{r})}\xi_{j_2}^{(\widetilde{r})}}}{x_1^{n-K_{r}}y_{11}^{(n-K_{r})^2}\prod_{j=2}^J x_j^{\xi_j^{(r)}}y_{1j}^{2\xi_j^{(r)}(n-K_{r})}\prod_{2\leq j_1,j_2\leq J}y_{j_1j_2}^{\xi_{j_1}^{(r)}\xi_{j_2}^{(r)}}}\notag\\
            &\leq \left[\frac{\underset{j=1,\ldots,J}{\max}x_j}{\underset{j=1,\ldots,J}{\min}x_j}\right]^{K_{\widetilde{r}}}\left[\frac{y_{11}}{\underset{2\leq j_1,j_2\leq J}{\max}y_{j_1j_2}}\right]^{-(K_{r}-K_{\widetilde{r}})^2}\left[\frac{\underset{2\leq j_1,j_2\leq J}{\max}y_{j_1j_2}}{\underset{2\leq j_1,j_2\leq J}{\min}y_{j_1j_2}}\right]^{K_{\widetilde{r}}^2-(K_{\widetilde{r}}-K_{r})^2}\left[\frac{y_{11}^{-K_{\widetilde{r}}}\prod_{j=2}^J y_{1j}^{\xi_j^{(\widetilde{r})}}}{y_{11}^{-K_{r}}\prod_{j=2}^J y_{1j}^{\xi_j^{(r)}}}\right]^{2(n-K_{\widetilde{r}})}\notag\\
            &= C_1^{K_{\widetilde{r}}}C_2^{-(K_{r}-K_{\widetilde{r}})^2}C_3^{K_{\widetilde{r}}^2-(K_{\widetilde{r}}-K_{r})^2}\left[\frac{T(\xi_2^{(r)},\ldots,\xi_J^{(r)})}{T(\xi_2^{(\widetilde{r})},\ldots,\xi_J^{(\widetilde{r})})}\right]^{-2(n-K_{\widetilde{r}})}\text{,}
        \end{align}
        where $C_1=\underset{j=1,\ldots,J}{\max}x_j/\underset{j=1,\ldots,J}{\min}x_j$, $C_2=y_{11}/\underset{2\leq j_1,j_2\leq J}{\max}y_{j_1j_2}>1$ and $C_3=\underset{2\leq j_1,j_2\leq J}{\max}y_{j_1j_2}/\underset{2\leq j_1,j_2\leq J}{\min}y_{j_1j_2}$ are constants that do not depend on the choice of $n$, $r$ or $\widetilde{r}$.\\[3mm]
        Furthermore, for $u>r_{\text{max}}$, by (\ref{eq_lem_5.3}) and (\ref{eq_lem_5.5}), we have
        \begin{align}\label{eq_lem_5.6}
            \frac{S_n(\xi_2^{(u)},\ldots,\xi_J^{(u)})}{S_n(\xi_2^{(r)},\ldots,\xi_J^{(r)})}\leq &C_1^{K_{u}}C_2^{-(K_{r}-K_{u})^2}C_3^{K_{u}^2-(K_{u}-K_{r})^2}\left[\frac{T(\xi_2^{(r)},\ldots,\xi_J^{(r)})}{T(\xi_2^{(u)},\ldots,\xi_J^{(u)})}\right]^{-2(n-K_{u})}\notag\\
            \leq& C_1^{K_{u}}C_2^{-(K_{r}-K_{u})^2}C_3^{K_{u}^2-(K_{u}-K_{r})^2}J^{-2(n-K_{u})}\notag\\
            \leq& J^{-2n}\max_{K>K_r}\left\{C_1^{K}C_2^{-(K_{r}-K)^2}C_3^{K^2-(K-K_{r})^2}J^{2K}\right\}\notag\\
            \lesssim& J^{-2n}
        \end{align}
        since $C_2>1$.\\[3mm]
        \textbf{Case 2: }If $K_{\widetilde{r}}<K_{r}$, then
        \begin{align}\label{eq_lem_5.7}
            &~~~~\frac{S_n(\xi_2^{(\widetilde{r})},\ldots,\xi_J^{(\widetilde{r})})}{S_n(\xi_2^{(r)},\ldots,\xi_J^{(r)})}\notag\\
            &=\frac{x_1^{n-K_{\widetilde{r}}}y_{11}^{(n-K_{\widetilde{r}})^2}\prod_{j=2}^J x_j^{\xi_j^{(\widetilde{r})}}y_{1j}^{2\xi_j^{(\widetilde{r})}(n-K_{\widetilde{r}})}\prod_{2\leq j_1,j_2\leq J}y_{j_1j_2}^{\xi_{j_1}^{(\widetilde{r})}\xi_{j_2}^{(\widetilde{r})}}}{x_1^{n-K_{r}}y_{11}^{(n-K_{r})^2}\prod_{j=2}^J x_j^{\xi_j^{(r)}}y_{1j}^{2\xi_j^{(r)}(n-K_{r})}\prod_{2\leq j_1,j_2\leq J}y_{j_1j_2}^{\xi_{j_1}^{(r)}\xi_{j_2}^{(r)}}}\notag\\
            &\leq \left[\frac{\underset{j=1,\ldots,J}{\max}x_j}{\underset{j=1,\ldots,J}{\min}x_j}\right]^{K_{r}}\left[\frac{\underset{2\leq j_1,j_2\leq J}{\max}y_{j_1j_2}}{\underset{2\leq j_1,j_2\leq J}{\min}y_{j_1j_2}}\right]^{K_{r}^2}\left[\frac{y_{11}^{-K_{\widetilde{r}}}\prod_{j=2}^J y_{1j}^{\xi_j^{(\widetilde{r})}}}{y_{11}^{-K_{r}}\prod_{j=2}^J y_{1j}^{\xi_j^{(r)}}}\right]^{2(n-K_{r})}\notag\\
            &= C_1^{K_{r}}C_3^{K_{r}^2}\left[\frac{T(\xi_2^{(r)},\ldots,\xi_J^{(r)})}{T(\xi_2^{(\widetilde{r})},\ldots,\xi_J^{(\widetilde{r})})}\right]^{-2(n-K_{r})}\text{.}
        \end{align}
        By (\ref{eq_lem_5.4}), (\ref{eq_lem_5.6}) and  (\ref{eq_lem_5.7}), we have
        \begin{align}\label{eq_lem_5.8}
            &\frac{\sum_{u\geq r+1}\Delta_{u,n}S_n(\xi^{(u)}_2,\ldots,\xi^{(u)}_J)}{\Delta_{r,n}S_n(\xi^{(r)}_2,\ldots,\xi^{(r)}_J)}\notag\\
            \leq&\frac{\sum_{u=r+1,\ldots,r_{\text{max}}}\Delta_{u,n}S_n(\xi^{(u)}_2,\ldots,\xi^{(u)}_J)}{S_n(\xi^{(r)}_2,\ldots,\xi^{(r)}_J)}+\frac{\sum_{u\geq r_{\text{max}}+1}\Delta_{u,n}S_n(\xi^{(u)}_2,\ldots,\xi^{(u)}_J)}{S_n(\xi^{(r)}_2,\ldots,\xi^{(r)}_J)}\notag\\
            \lesssim& n^{r_{\text{max}}}J^{r_{\text{max}}}\left(\frac{T(\xi_2^{(r)},\ldots,\xi_J^{(r)})}{T(\xi_2^{(r+1)},\ldots,\xi_J^{(r+1)})}\right)^{-2n}+J^nJ^{-2n}\notag\\
            \rightarrow&0\text{.}
        \end{align}
    Similarly, we define $\widetilde{T}(\xi_2,\ldots,\xi_J)$, $\widetilde{S}_n(\xi_2,\ldots,\xi_J)$, $(\widetilde{\xi}_2^{(r)},\ldots,\widetilde{\xi}_J^{(r)})$ and $\widetilde{\Delta}_{r,n}$ for the right-hand side of (\ref{eq_lem_5.2}).
    Then we can prove that
    \begin{align}\label{eq_lem_5.9}
        \lim_{n\rightarrow \infty}\frac{\sum_{u> r}\widetilde{\Delta}_{u,n}\widetilde{S}_n(\widetilde{\xi}^{(u)}_2,\ldots,\widetilde{\xi}^{(u)}_J)}{\widetilde{\Delta}_{r,n}\widetilde{S}_n(\widetilde{\xi}^{(r)}_2,\ldots,\widetilde{\xi}^{(r)}_J)}= 0.
    \end{align}
    This finishes Step 1.\\[3mm]
    \textbf{Step 2: }Under the introduced notation, equation (\ref{eq_lem_5.2}) turns into
    \begin{align}\label{eq_lem_5.10}
        \sum_{u\geq 1}\Delta_{u,n}S_n(\xi^{(u)}_2,\ldots,\xi^{(u)}_J)=\sum_{u\geq 1}\widetilde{\Delta}_{u,n}\widetilde{S}_n(\widetilde{\xi}^{(u)}_2,\ldots,\widetilde{\xi}^{(u)}_J)
    \end{align}
    for any $n\in\mathbb{N}$. We then use an induction argument to prove that for any $r,n\in\mathbb{N}$:
    \begin{align*}
        (\xi^{(r)}_2,\ldots,\xi^{(r)}_J)^{\top}&=(\widetilde{\xi}^{(r)}_2,\ldots,\widetilde{\xi}^{(r)}_J)^{\top}\notag\text{,}\\
        T(\xi_2^{(r)},\ldots,\xi_J^{(r)})&=\widetilde{T}(\widetilde{\xi}_2^{(r)},\ldots,\widetilde{\xi}_J^{(r)})\text{,}\notag\\
        S_n(\xi_2^{(r)},\ldots,\xi_J^{(r)})&=\widetilde{S}_n(\widetilde{\xi}^{(r)}_2,\ldots,\widetilde{\xi}^{(r)}_J)\text{.}
    \end{align*}
    When $r=1$, by assumption it can be readily shown that $(\xi^{(1)}_2,\ldots,\xi^{(1)}_J)^{\top}=(\widetilde{\xi}^{(1)}_2,\ldots,\widetilde{\xi}^{(1)}_J)^{\top}=(0,\ldots,0)^{\top}$. By (\ref{eq_lem_5.8}) and (\ref{eq_lem_5.9}) we have
    \begin{align}\label{eq_lem_5.11}
        \lim_{n\rightarrow \infty}\frac{\sum_{u\geq 1}\Delta_{u,n}S_n(\xi^{(u)}_2,\ldots,\xi^{(u)}_J)}{\Delta_{1,n}S_n(0,\ldots,0)}=1
    \end{align}
    and
    \begin{align}\label{eq_lem_5.12}
        \lim_{n\rightarrow \infty}\frac{\sum_{u\geq 1}\widetilde{\Delta}_{u,n}\widetilde{S}_n(\widetilde{\xi}^{(u)}_2,\ldots,\widetilde{\xi}^{(u)}_J)}{\widetilde{\Delta}_{1,n}\widetilde{S}_n(0,\ldots,0)}=1\text{.}
    \end{align}
    By (\ref{eq_lem_5.10}), (\ref{eq_lem_5.11}) and (\ref{eq_lem_5.12}), we have
    \begin{align}\label{eq_lem_5.13}
        \lim_{n\rightarrow \infty}\frac{S_n(0,\ldots,0)}{\widetilde{S}_n(0,\ldots,0)}=&\lim_{n\rightarrow \infty}\frac{\Delta_{1,n}S_n(0,\ldots,0)}{\widetilde{\Delta}_{1,n}\widetilde{S}_n(0,\ldots,0)}\notag\\
        =&\lim_{n\rightarrow \infty}\frac{\Delta_{1,n}S_n(0,\ldots,0)}{\sum_{u\geq 1}\Delta_{u,n}S_n(\xi^{(u)}_2,\ldots,\xi^{(u)}_J)}\frac{\sum_{u\geq 1}\widetilde{\Delta}_{u,n}\widetilde{S}_n(\widetilde{\xi}^{(u)}_2,\ldots,\widetilde{\xi}^{(u)}_J)}{\widetilde{\Delta}_{1,n}\widetilde{S}_n(0,\ldots,0)}\notag\\
        =&1\text{.}
    \end{align}
    By the definition of $S_n$, we have 
    \begin{align*}
        \frac{S_n(0,\ldots,0)}{\widetilde{S}_n(0,\ldots,0)}=\frac{x_1^{n}y_{11}^{n^2}}{\widetilde{x}_1^{n}\widetilde{y}_{11}^{n^2}}.
    \end{align*}
    Then (\ref{eq_lem_5.13}) implies that $x_1=\widetilde{x}_1$, $y_{11}=\widetilde{y}_{11}$, $S_n(0,\ldots,0)=\widetilde{S}_n(0,\ldots,0)$ and $T(0,\ldots,0)=\widetilde{T}(0,\ldots,0)$ for any $n\in\mathbb{N}$. Hence the result is proved for $r=1$.\\[3mm]
    Suppose that the result is proved for $1,\ldots,r-1$. By induction assumption and (\ref{eq_lem_5.10}), we have
    \begin{align}\label{eq_lem_5.14}
        \sum_{u\geq r}\Delta_{u,n}S_n(\xi^{(u)}_2,\ldots,\xi^{(u)}_J)&=\sum_{u\geq 1}\Delta_{u,n}S_n(\xi^{(u)}_2,\ldots,\xi^{(u)}_J)-\sum_{1\leq u<r}\Delta_{u,n}S_n(\xi^{(u)}_2,\ldots,\xi^{(u)}_J)\notag\\
        &=\sum_{u\geq 1}\widetilde{\Delta}_{u,n}\widetilde{S}_n(\widetilde{\xi}^{(u)}_2,\ldots,\widetilde{\xi}^{(u)}_J)-\sum_{1\leq u< r}\widetilde{\Delta}_{u,n}\widetilde{S}_n(\widetilde{\xi}^{(u)}_2,\ldots,\widetilde{\xi}^{(u)}_J)\notag\\[3mm]
        &=\sum_{u\geq r}\widetilde{\Delta}_{u,n}\widetilde{S}_n(\widetilde{\xi}^{(u)}_2,\ldots,\widetilde{\xi}^{(u)}_J)\text{.}
    \end{align}
    For $j=2,\ldots,J$, define $\tau_j=\min\{r\in\mathbb{N}:\xi_{j}^{(r)}>0\}$ and $\widetilde{\tau}_j=\min\{r\in\mathbb{N}:\widetilde{\xi}_{j}^{(r)}>0\}$. Since $y_{11}>\ldots>y_{1J}$, it can be readily shown that $1=\tau_{1}<\tau_2<\ldots<\tau_{J}$. Suppose that $\tau_{l}<r\leq \tau_{l+1}$ (define $\tau_0=0$ and $\tau_{J+1}=\infty$). By induction assumption, we also have $\widetilde{\tau}_{l}<r\leq \widetilde{\tau}_{l+1}$. There exists $\widetilde{r}\in\mathbb{N}$ such that $(\xi_2^{(r)},\ldots,\xi_J^{(r)})^{\top}=(\widetilde{\xi}_2^{(\widetilde{r})},\ldots,\widetilde{\xi}_J^{(\widetilde{r})})^{\top}$. We then prove that $r=\widetilde{r}$. The proof falls into five cases:\\[3mm]
    \textbf{Case 1: }$r<\tau_{l+1}$ and $\widetilde{r}<r$. By induction assumption, we have $(\widetilde{\xi}_2^{(\widetilde{r})},\ldots,\widetilde{\xi}_J^{(\widetilde{r})})^{\top}=(\xi_2^{(\widetilde{r})},\ldots,\xi_J^{(\widetilde{r})})^{\top}$, which implies that $(\xi_2^{(\widetilde{r})},\ldots,\xi_J^{(\widetilde{r})})^{\top}=(\xi_2^{(r)},\ldots,\xi_J^{(r)})^{\top}$. This leads to a contradiction.\\[3mm]
    \textbf{Case 2: }$r<\tau_{l+1}$ and $\widetilde{r}>r$. For any $j=2,\ldots,l$, let $\xi_j=1$ and $\xi_2=\ldots=\xi_{j-1}=\xi_{j+1}=\ldots=\xi_{J}=0$. It can be readily shown that the rank of $(\xi_2,\ldots,\xi_J)^{\top}$ is exactly $\tau_j$, which is smaller than $r$. So by induction assumption, we have
    \begin{align*}
        1=\frac{T(\xi_2,\ldots,\xi_J)}{\widetilde{T}(\xi_2,\ldots,\xi_J)}=\frac{y_{1j}}{\widetilde{y}_{1j}},
    \end{align*}
    which implies that $y_{1j}=\widetilde{y}_{1j}$ for $j=1,\ldots,l$. Then by (\ref{eq_lem_5.5}), (\ref{eq_lem_5.7}) and induction assumption, it holds that
    \begin{align}\label{eq_lem_5.15}
        \frac{S_n(\xi^{(r)}_2,\ldots,\xi^{(r)}_J)}{\widetilde{S}_n(\widetilde{\xi}^{(\widetilde{r})}_2,\ldots,\widetilde{\xi}^{(\widetilde{r})}_J)}=&\frac{S_n(\xi^{(r)}_2,\ldots,\xi^{(r)}_J)}{S_n(0,\ldots,0)}\frac{\widetilde{S}_n(0,\ldots,0)}{\widetilde{S}_n(\widetilde{\xi}^{(\widetilde{r})}_2,\ldots,\widetilde{\xi}^{(\widetilde{r})}_J)}\asymp\left[\frac{T(\xi^{(r)}_2,\ldots,\xi^{(r)}_J)}{\widetilde{T}(\widetilde{\xi}^{(\widetilde{r})}_2,\ldots,\widetilde{\xi}^{(\widetilde{r})}_J)}\right]^{2n}=1
    \end{align}
    since $\xi_{l+1}^{(r)}=\ldots=\xi_{J}^{(r)}=\widetilde{\xi}_{l+1}^{(\widetilde{r})}=\ldots=\widetilde{\xi}_{J}^{(\widetilde{r})}=0$. However, by (\ref{eq_lem_5.8}), (\ref{eq_lem_5.9}) and (\ref{eq_lem_5.14}) we have
    \begin{align*}
        &\frac{\widetilde{S}_n(\widetilde{\xi}^{(\widetilde{r})}_2,\ldots,\widetilde{\xi}^{(\widetilde{r})}_J)}{S_n(\xi^{(r)}_2,\ldots,\xi^{(r)}_J)}\\
        =&\frac{\widetilde{\Delta}_{\widetilde{r},n}\widetilde{S}_n(\widetilde{\xi}^{(\widetilde{r})}_2,\ldots,\widetilde{\xi}^{(\widetilde{r})}_J)}{\Delta_{r,n}S_n(\xi^{(r)}_2,\ldots,\xi^{(r)}_J)}\\
        =&\frac{\widetilde{\Delta}_{\widetilde{r},n}\widetilde{S}_n(\widetilde{\xi}^{(\widetilde{r})}_2,\ldots,\widetilde{\xi}^{(\widetilde{r})}_J)}{\widetilde{\Delta}_{r,n}\widetilde{S}_n(\widetilde{\xi}^{(r)}_2,\ldots,\widetilde{\xi}^{(r)}_J)}\frac{\widetilde{\Delta}_{r,n}\widetilde{S}_n(\widetilde{\xi}^{(r)}_2,\ldots,\widetilde{\xi}^{(r)}_J)}{\sum_{u\geq r}\widetilde{\Delta}_{u,n}\widetilde{S}_n(\widetilde{\xi}^{(u)}_2,\ldots,\widetilde{\xi}^{(u)}_J)}\frac{\sum_{u\geq r}\Delta_{u,n}S_n(\xi^{(u)}_2,\ldots,\xi^{(u)}_J)}{\Delta_{r,n}S_n(\xi^{(r)}_2,\ldots,\xi^{(r)}_J)}\\
        \leq&\frac{\sum_{u\geq r+1}\widetilde{\Delta}_{u,n}\widetilde{S}_n(\widetilde{\xi}^{(u)}_2,\ldots,\widetilde{\xi}^{(u)}_J)}{\widetilde{\Delta}_{r,n}\widetilde{S}_n(\widetilde{\xi}^{(r)}_2,\ldots,\widetilde{\xi}^{(r)}_J)}\frac{\widetilde{\Delta}_{r,n}\widetilde{S}_n(\widetilde{\xi}^{(r)}_2,\ldots,\widetilde{\xi}^{(r)}_J)}{\sum_{u\geq r}\widetilde{\Delta}_{u,n}\widetilde{S}_n(\widetilde{\xi}^{(u)}_2,\ldots,\widetilde{\xi}^{(u)}_J)}\frac{\sum_{u\geq r}\Delta_{u,n}S_n(\xi^{(u)}_2,\ldots,\xi^{(u)}_J)}{\Delta_{r,n}S_n(\xi^{(r)}_2,\ldots,\xi^{(r)}_J)}\\
        \rightarrow&0,
    \end{align*}
    which contradicts (\ref{eq_lem_5.15}).\\[3mm]
    \textbf{Case 3: }$r<\tau_{l+1}$ and $\widetilde{r}=r$. The result is proved.\\[3mm]
    \textbf{Case 4: }$r=\tau_{l+1}$ and $r<\widetilde{\tau}_{l+1}$. By arguments similar to those used in Case 1 and Case 2, this leads to a contradiction.\\[3mm]
    \textbf{Case 5: }$r=\tau_{l+1}$ and $r=\widetilde{\tau}_{l+1}$. Then it can be readily shown that $(\xi_2^{(r)},\ldots,\xi_J^{(r)})^{\top}=(\widetilde{\xi}_2^{(r)},\ldots,\widetilde{\xi}_J^{(r)})^{\top}=(0,\ldots,0,1,0,\ldots,0)^{\top}$ where all components are 0 except that the $l$-th component is 1. This implies that $r=\widetilde{r}$.\\[3mm]
    Now we have proved that $r=\widetilde{r}$, i.e., $(\xi_2^{(r)},\ldots,\xi_J^{(r)})^{\top}=(\widetilde{\xi}_2^{(r)},\ldots,\widetilde{\xi}_J^{(r)})^{\top}$. This indicates that $\Delta_{r,n}=\widetilde{\Delta}_{r,n}$. Then by (\ref{eq_lem_5.8}), (\ref{eq_lem_5.9}) and (\ref{eq_lem_5.14}), we have
    \begin{align}\label{eq_lem_5.16}
        \lim_{n\rightarrow \infty}\frac{S_n(\xi_2^{(r)},\ldots,\xi_J^{(r)})}{\widetilde{S}_n(\widetilde{\xi}^{(r)}_2,\ldots,\widetilde{\xi}^{(r)}_J)}=&\lim_{n\rightarrow \infty}\frac{\Delta_{r,n}S_n(\xi_2^{(r)},\ldots,\xi_J^{(r)})}{\widetilde{\Delta}_{r,n}\widetilde{S}_n(\widetilde{\xi}^{(r)}_2,\ldots,\widetilde{\xi}^{(r)}_J)}\notag\\
        =&\lim_{n\rightarrow \infty}\frac{\Delta_{r,n}S_n(\xi_2^{(r)},\ldots,\xi_J^{(r)})}{\sum_{u\geq r}\Delta_{u,n}S_n(\xi_2^{(u)},\ldots,\xi_J^{(u)})}\frac{\sum_{u\geq r}\widetilde{\Delta}_{u,n}\widetilde{S}_n(\widetilde{\xi}^{(u)}_2,\ldots,\widetilde{\xi}^{(u)}_J)}{\widetilde{\Delta}_{r,n}\widetilde{S}_n(\widetilde{\xi}^{(r)}_2,\ldots,\widetilde{\xi}^{(r)}_J)}\notag\\
        =&1.
    \end{align}
    By the definition of $S_n$, there exist constants $D_1,D_2,D_3>0$ such that
    \begin{align*}
        \frac{S_n(\xi_2^{(r)},\ldots,\xi_J^{(r)})}{\widetilde{S}_n(\widetilde{\xi}^{(r)}_2,\ldots,\widetilde{\xi}^{(r)}_J)}=D_1D_2^{n}D_3^{n^2}.
    \end{align*}
    Then (\ref{eq_lem_5.16}) indicates that $D_1=D_2=D_3=1$, i.e., $S_n(\xi_2^{(r)},\ldots,\xi_J^{(r)})=\widetilde{S}_n(\widetilde{\xi}^{(r)}_2,\ldots,\widetilde{\xi}^{(r)}_J)$ for any $n$. Similarly to (\ref{eq_lem_5.15}), we have
    \begin{align*}
        \frac{S_n(\xi^{(r)}_2,\ldots,\xi^{(r)}_J)}{\widetilde{S}_n(\widetilde{\xi}^{(r)}_2,\ldots,\widetilde{\xi}^{(r)}_J)}
        \asymp\left[\frac{T(\xi^{(r)}_2,\ldots,\xi^{(r)}_J)}{\widetilde{T}(\widetilde{\xi}^{(r)}_2,\ldots,\widetilde{\xi}^{(r)}_J)}\right]^{2n}.
    \end{align*}
    Then we have $T(\xi_2^{(r)},\ldots,\xi_J^{(r)})=\widetilde{T}(\widetilde{\xi}_2^{(r)},\ldots,\widetilde{\xi}_J^{(r)})$, which finishes the proof for the case $r$.\\[3mm]
    Hence, by induction, we have proved that for any $r,n\in\mathbb{N}$:
    \begin{align*}
        (\xi^{(r)}_2,\ldots,\xi^{(r)}_J)^{\top}&=(\widetilde{\xi}^{(r)}_2,\ldots,\widetilde{\xi}^{(r)}_J)^{\top}\text{,}\\
        T(\xi_2^{(r)},\ldots,\xi_J^{(r)})&=\widetilde{T}(\widetilde{\xi}_2^{(r)},\ldots,\widetilde{\xi}_J^{(r)})\text{,}\\
        S_n(\xi_2^{(r)},\ldots,\xi_J^{(r)})&=\widetilde{S}_n(\widetilde{\xi}^{(r)}_2,\ldots,\widetilde{\xi}^{(r)}_J)
        \text{.}
    \end{align*}
    For any $j=2,\ldots,J$, let $(\xi_2,\ldots,\xi_J)^{\top}$ be the vector such that
    \begin{align*}
        \xi_m=\begin{cases}1~~~&m=j\\0~~~&\text{otherwise}
        \end{cases}.
    \end{align*}
    Then for any $n\in\mathbb{N}$ we have
    \begin{align*}
        1&=\frac{S_n(\xi_2,\ldots,\xi_J)}{\widetilde{S}_n(\xi_2,\ldots,\xi_J)}=\frac{x_jy_{1j}^{2(n-1)}y_{jj}}{\widetilde{x}_j\widetilde{y}_{1j}^{2(n-1)}\widetilde{y}_{jj}}\text{,}\\
        1&=\frac{S_n(2\xi_2,\ldots,2\xi_J)}{\widetilde{S}_n(2\xi_2,\ldots,2\xi_J)}=\frac{x_j^2y_{1j}^{4(n-2)}y^4_{jj}}{\widetilde{x}_j^2\widetilde{y}_{1j}^{4(n-2)}\widetilde{y}^4_{jj}}\text{.}
    \end{align*}
    This implies that $x_j=\widetilde{x}_j$, $y_{1j}=\widetilde{y}_{1j}$ and $y_{jj}=\widetilde{y}_{jj}$. For any $2\leq j_1<j_2\leq J$ and any $n\in\mathbb{N}$, let $(\xi_2,\ldots,\xi_J)^{\top}$ be
    \begin{align*}
        \xi_m=\begin{cases}1~~~&m=j_1 \text{ or } m=j_2\\0~~~&\text{otherwise}
        \end{cases}.
    \end{align*}
    Then we have
    \begin{align*}
        1&=\frac{S_n(\xi_2,\ldots,\xi_J)}{\widetilde{S}_n(\xi_2,\ldots,\xi_J)}=\frac{y_{j_1j_2}^2}{\widetilde{y}_{j_1j_2}^2}\text{.}
    \end{align*}
    This implies that $y_{j_1j_2}=\widetilde{y}_{j_1j_2}$. Hence the proposition is proved.
    \end{proof}
    \subsection{Proof of Proposition \ref{prop_laplace}}
    To prove Proposition \ref{prop_laplace}, we first verify the following lemma:
    \begin{lem}\label{lem_laplace}
        Let $f(\vec{x})$ be a strictly concave function on $\mathbb{R}^d$ with $\vec{0}$ as its unique maximiser. Assume that $-\nabla^2 f(\vec{x})\succeq \mat{I}_d$ holds at any point $\vec{x}\in\mathbb{R}^d$. Then for any $\delta>0$, it holds that
        \begin{align*}
            \frac{\int_{\vec{x}:\left\|\vec{x}\right\|\geq C}\exp(f(\vec{x}))\mathrm{d}\vec{x}}{\int_{\vec{x}}\exp(f(\vec{x}))\mathrm{d}\vec{x}}\leq \delta,
        \end{align*}
        where $C=C(\delta)>0$ is a constant that is independent of $f$.
    \end{lem}
    \begin{proof}[Proof of Lemma~\ref{lem_laplace}]
        We change variables to $d$-dimensional polar coordinates:
        \begin{align}\label{eq_lem_1.1}
            &\int_{\vec{x}:\left\|\vec{x}\right\|\geq C}\exp(f(\vec{x}))\mathrm{d}\vec{x}\notag\\
            =&\int_{\theta_1,\ldots,\theta_{d-1}}\left(\prod_{k=2}^{d-1}\sin^{k-1}\theta_k\right)\mathrm{d}\theta_1\ldots\mathrm{d}\theta_{d-1}\int_{r\geq C}r^{d-1}\exp(f(r\vec{\alpha}(\theta_1,\ldots,\theta_{d-1})))\mathrm{d}r,
            \notag\\
            &\int_{\vec{x}:\left\|\vec{x}\right\|\leq C}\exp(f(\vec{x}))\mathrm{d}\vec{x}\notag\\
            =&\int_{\theta_1,\ldots\theta_{d-1}}\left(\prod_{k=2}^{d-1}\sin^{k-1}\theta_k\right)\mathrm{d}\theta_1\ldots,\mathrm{d}\theta_{d-1}\int_{r\leq C}r^{d-1}\exp(f(r\vec{\alpha}(\theta_1,\ldots,\theta_{d-1})))\mathrm{d}r,
        \end{align}
        where $\left\|\vec{\alpha}(\theta_1,\ldots,\theta_{d-1})\right\|=1$. For fixed $\theta_1,\ldots,\theta_{d-1}$, since $-\nabla^2 f(\vec{x}_r^{\ast})\succeq \mat{I}_d$, we have
        \begin{align}\label{eq_lem_1.3}
            -\frac{d^2}{\mathrm{d}r^2}f(r\vec{\alpha})=&-\vec{\alpha}^{\top}\nabla^2 f(r\vec{\alpha})\vec{\alpha}\geq \left\|\vec{\alpha}\right\|^2=1.
        \end{align}
        For any $r\geq C$, it holds that $Cr+r^2/2\geq C^2+r^2/2$.
        Then by (\ref{eq_lem_1.3}) we have
        \begin{align}\label{eq_lem_1.4}
            \int_{r\geq C}r^{d-1}\exp(f(r\vec{\alpha}(\theta_1,\ldots,\theta_{d-1})))\mathrm{d}r\leq \exp(f(C\vec{\alpha}))\underbrace{\int_{r\geq C}r^{d-1}\exp\left(-\frac{1}{2}r^2\right)\mathrm{d}r}_{:=I(C)}.
        \end{align}
        On the other hand, since $ \exp(f(r\vec{\alpha}))$ is monotonically decreasing for $r\geq 0$, we have
        \begin{align}\label{eq_lem_1.5}
            \int_{0\leq r\leq C}r^{d-1}\exp(f(r\vec{\alpha}(\theta_1,\ldots,\theta_{d-1})))\mathrm{d}r\geq \exp(f(C\vec{\alpha}))\int_{0\leq r\leq C}r^{d-1}\mathrm{d}r=\frac{C^d\exp(f(C\vec{\alpha}))}{d}.
        \end{align}
        By (\ref{eq_lem_1.1}), (\ref{eq_lem_1.4}) and (\ref{eq_lem_1.5}), we have
        \begin{align*}
            \frac{\int_{\vec{x}:\left\|\vec{x}\right\|\geq C}\exp(f(\vec{x}))\mathrm{d}\vec{x}}{\int_{\vec{x}:0\leq \left\|\vec{x}\right\|\leq C}\exp(f(\vec{x}))\mathrm{d}\vec{x}}\leq \frac{d}{C^{d}}I(C).
        \end{align*}
        Note that $I(C)$ is uniformly bounded. For any $\delta>0$, we can find $C$ depending only on $\delta$ such that
        \begin{align*}
            \frac{\int_{\vec{x}:\left\|\vec{x}\right\|\geq C}\exp(f(\vec{x}))\mathrm{d}\vec{x}}{\int_{\vec{x}}\exp(f(\vec{x}))\mathrm{d}\vec{x}}\leq \delta,
        \end{align*}
        which completes the proof.
    \end{proof}

    Based on Lemma \ref{lem_laplace}, we can prove Proposition~\ref{prop_laplace}.
    \renewcommand*{\proofname}{Proof of Proposition \ref{prop_laplace}}
    \begin{proof}
    We apply Lemma~\ref{lem_laplace} to the case when $\delta=\frac{1}{2}$ and obtain the corresponding constant $C>0$. It can be readily shown that for any $\vec{\theta}\in\mathbb{R}^d$ we have
    \begin{align}\label{eq_prop_2.1}
        \exp\left(-\max_{k=1,\ldots,K}\|\vec{\alpha}_k\|\|\hat{\vec{\theta}}-\vec{\theta}\|\right)(\mat{I}(\hat{\vec{\theta}})-\mat{I}_d)\leq (\mat{I}(\vec{\theta})-\mat{I}_d)\leq \exp\left(\max_{k=1,\ldots,K}\|\vec{\alpha}_k\|\|\hat{\vec{\theta}}-\vec{\theta}\|\right) (\mat{I}(\hat{\vec{\theta}})-\mat{I}_d).
    \end{align}
    Now let
    \begin{align*}
        g_1(\vec{\theta})&=-\frac{1}{2}(\vec{\theta}-\hat{\vec{\theta}})^{\top}\left[\mat{I}_d+\exp\left(-C\max_{k=1,\ldots,K}\|\vec{\alpha}_k\|\right)(\mat{I}(\hat{\vec{\theta}})-\mat{I}_d)\right](\vec{\theta}-\hat{\vec{\theta}})+f(\hat{\vec{\theta}}),\\
        g_2(\vec{\theta})&=-\frac{1}{2}(\vec{\theta}-\hat{\vec{\theta}})^{\top}\left[\mat{I}_d+\exp\left(C\max_{k=1,\ldots,K}\|\vec{\alpha}_k\|\right)(\mat{I}(\hat{\vec{\theta}})-\mat{I}_d)\right](\vec{\theta}-\hat{\vec{\theta}})+f(\hat{\vec{\theta}})
    \end{align*}
    be strictly concave functions with maximiser $\hat{\vec{\theta}}$ and maximum value $f(\hat{\vec{\theta}})$. Then for any $\vec{\theta}\in\mathbb{R}^d$ such that $\|\vec{\theta}-\hat{\vec{\theta}}\|\leq C$, by (\ref{eq_prop_2.1}) we have
    \begin{align}\label{eq_prop_2.2}
            -\nabla^2 g_1(\vec{\theta})\preceq \mat{I}(\vec{\theta})\preceq-\nabla^2 g_2(\vec{\theta}).
        \end{align}
        Since the maximisers and maximum values are matched for $f,g_1,g_2$, by (\ref{eq_prop_2.2}) we have
        \begin{align}\label{eq_prop_2.3}
            \int_{\vec{\theta}:\|\vec{\theta}-\hat{\vec{\theta}}\|\leq C}\exp(g_2(\vec{\theta}))\mathrm{d}\vec{\theta}\leq \int_{\vec{\theta}:\|\vec{\theta}-\hat{\vec{\theta}}\|\leq C}\exp(f(\vec{\theta}))\mathrm{d}\vec{\theta}\leq \int_{\vec{\theta}:\|\vec{\theta}-\hat{\vec{\theta}}\|\leq C}\exp(g_1(\vec{\theta}))\mathrm{d}\vec{\theta}.
        \end{align}
        By the definition of $g_1$ and $g_2$, it can be readily shown that $-\nabla^2 g_1(\vec{\theta})\succeq \mat{I}_d$ and $-\nabla^2 g_2(\vec{\theta})\succeq \mat{I}_d$ for any $\vec{\theta}\in\mathbb{R}^d$. Then by the choice of $C$ and (\ref{eq_prop_2.3}), we have
        \begin{align}\label{eq_prop_2.4}
            &\frac{\int_{\bt}\exp(f(\vec{\theta}))\mathrm{d}\vec{\theta}}{\int_{\bt}\exp(g_2(\vec{\theta}))\mathrm{d}\vec{\theta}}\geq \frac{\int_{\vec{\theta}:\|\vec{\theta}-\hat{\vec{\theta}}\|\leq C}\exp(f(\vec{\theta}))\mathrm{d}\vec{\theta}}{2\int_{\vec{\theta}:\|\vec{\theta}-\hat{\vec{\theta}}\|\leq C}\exp(g_2(\vec{\theta}))\mathrm{d}\vec{\theta}}\geq \frac{1}{2},\notag\\
            &\frac{\int_{\bt}\exp(f(\vec{\theta}))\mathrm{d}\vec{\theta}}{\int_{\bt}\exp(g_1(\vec{\theta}))\mathrm{d}\vec{\theta}}\leq \frac{2\int_{\vec{\theta}:\|\vec{\theta}-\hat{\vec{\theta}}\|\leq C}\exp(f(\vec{\theta}))\mathrm{d}\vec{\theta}}{\int_{\vec{\theta}:\|\vec{\theta}-\hat{\vec{\theta}}\|\leq C}\exp(g_1(\vec{\theta}))\mathrm{d}\vec{\theta}}\leq 2.
        \end{align}
        Moreover, by the definition of $g_1$ and $g_2$ we have
        \begin{align}\label{eq_prop_2.5}
            \int_{\bt}(2\pi)^{-d/2}\exp(g_1(\vec{\theta}))\mathrm{d}\vec{\theta}&=\exp(f(\hat{\vec{\theta}}))\left[\operatorname{det}\left(\mat{I}_d+\exp\left(-C\max_{k=1,\ldots,K}\|\vec{\alpha}_k\|\right)(\mat{I}(\hat{\vec{\theta}})-\mat{I}_d)\right)\right]^{-1/2}\notag\\
            &\leq \exp\left(f(\hat{\vec{\theta}})+\frac{Cd}{2}\max_{k=1,\ldots,K}\|\vec{\alpha}_k\|\right)[\operatorname{det}(\mat{I}(\hat{\vec{\theta}}))]^{-1/2},\notag\\
            \int_{\bt}(2\pi)^{-d/2}\exp(g_2(\vec{\theta}))\mathrm{d}\vec{\theta}&=\exp(f(\hat{\vec{\theta}}))\left[\operatorname{det}\left(\mat{I}_d+\exp\left(C\max_{k=1,\ldots,K}\|\vec{\alpha}_k\|\right)(\mat{I}(\hat{\vec{\theta}})-\mat{I}_d)\right)\right]^{-1/2}\notag\\
            &\geq \exp\left(f(\hat{\vec{\theta}})-\frac{Cd}{2}\max_{k=1,\ldots,K}\|\vec{\alpha}_k\|\right)[\operatorname{det}(\mat{I}(\hat{\vec{\theta}}))]^{-1/2}.
        \end{align}
        Then by (\ref{eq_prop_2.4}) and (\ref{eq_prop_2.5}), we have
        \begin{align*}
            \frac{1}{2}\exp\left(-\frac{Cd}{2}\max_{k=1,\ldots,K}\|\vec{\alpha}_k\|\right)\leq\frac{\int_{\bt}(2\pi)^{-d/2}\exp(f(\vec{\theta}))\mathrm{d}\vec{\theta}}{\exp(f(\hat{\vec{\theta}}))/\sqrt{\operatorname{det}(\mat{I}(\hat{\vec{\theta}}))}}\leq 2\exp\left(\frac{Cd}{2}\max_{k=1,\ldots,K}\|\vec{\alpha}_k\|\right).
        \end{align*}
        Since constant $C$ does not depend on the choice of $\vec{\eta}$, the result is proved.
    \end{proof}
    
    \subsection{Proof of Proposition \ref{prop_canonical_projection}}
    \renewcommand*{\proofname}{Proof of Proposition \ref{prop_canonical_projection}}
    \begin{proof}
    We prove part (1) and (2) of Proposition~\ref{prop_canonical_projection} simultaneously and prove the uniqueness and continuity of canonical projection at the end. We first consider the case when $\left\{\vec{\eta}_n\right\}\subseteq \operatorname{span}\left\{\vec{\alpha}_1,\ldots,\vec{\alpha}_K\right\}:=\mathcal{H}_0$. For simplicity, we assume that $\mathcal{H}_0=\mathbb{R}^d$. The first-order equation corresponding to $\vec{\theta}_{n}$ is given by
    \begin{align}\label{eq_prop_3.1}
        -\sum_{k=1}^K w_k\exp(\vec{\alpha}^{\top}_k\vec{\theta}_{n})\vec{\alpha}_k+\vec{\eta}_n=\vec{\theta}_{n}.
    \end{align}
    If $\vec{\theta}_{n}=0,i.o.$, then (\ref{eq_prop_3.1}) implies that $\vec{\eta}=0$, which contradicts our assumption. So we assume WLOG that $\vec{\theta}_{n}\neq 0$ and denote $l_{n}=\left\|\vec{\theta}_{n}\right\|$, $\vec{\epsilon}_{n}=\vec{\theta}_{n}/l_{n}$. Then equation (\ref{eq_prop_3.1}) turns into
    \begin{align}\label{eq_prop_3.2}
        -\sum_{k=1}^K w_k\exp(l_{n}\vec{\alpha}^{\top}_k\vec{\epsilon}_{n})\vec{\alpha}_k+\vec{\eta}_n=l_{n}\vec{\epsilon}_{n}.
    \end{align}
    Since $\left\|\vec{\epsilon}_{n}\right\|=1$ for any $n$, we assume WLOG that $\vec{\epsilon}_{n}\rightarrow \vec{\epsilon}$ where $\vec{\epsilon}\in\mathbb{R}^d$ has norm 1, otherwise we can make arguments on a subsequence.\\[3mm]
    \textbf{Step 1: }We prove that 
    $l_{n}\rightarrow \infty$.\\[3mm]
    If this is not the case, we assume WLOG that $l_{n}\rightarrow l<\infty$. Then we have
    \begin{align*}
        \vec{\eta}_n=l_{n}\vec{\epsilon}_{n}+\sum_{k=1}^K w_k\exp(l_{n}\vec{\alpha}^{\top}_k\vec{\epsilon}_{n})\vec{\alpha}_k\rightarrow l\vec{\epsilon}+\sum_{k=1}^K w_k\exp(l\vec{\alpha}^{\top}_k\vec{\epsilon})\vec{\alpha}_k,
    \end{align*}
    where the right-hand side is finite. This implies that $\vec{\eta}=0$, which contradicts our assumption.\\[3mm]
    \textbf{Step 2: }We divide the problem into three cases regarding the sign of $\underset{k=1,\ldots,K}{\max}\vec{\alpha}_k^{\top}\vec{\epsilon}$.\\[3mm]
    \textbf{Case 1: }$\underset{k=1,\ldots,K}{\max}\vec{\alpha}_k^{\top}\vec{\epsilon}<0$.\\[3mm]
    Since $l_{n}$ goes to infinity and $\vec{\alpha}^{\top}_k\vec{\epsilon}_{n}\rightarrow \vec{\alpha}^{\top}_k\vec{\epsilon}<0$ for any $k=1,\ldots,K$, by (\ref{eq_prop_3.2}) we have
    \begin{align*}
        \left\|\vec{\eta}_n-l_{n}\vec{\epsilon}_n\right\|=\left\|\sum_{k=1}^K w_k\exp(l_{n}\vec{\alpha}^{\top}_k\vec{\epsilon}_{n})\vec{\alpha}_k\right\|\rightarrow 0.
    \end{align*}
    This indicates that $\vec{\epsilon}=\vec{\eta}/\left\|\vec{\eta}\right\|$. So we have $\underset{k=1,\ldots,K}{\max}\vec{\alpha}_k^{\top}\vec{\eta}<0$, then we choose an empty set to satisfy the conditions in part (1), i.e., $\mathcal{H}_{\vec{\eta}}=\{\vec{0}\}$ and $\mathcal{H}_{\vec{\eta}}^{\perp}=\mathbb{R}^d$.\\[3mm]
    Moreover, since $\lim_{n\rightarrow \infty}\vec{\eta}_n/{n}=\vec{\eta}$, we have
    \begin{align*}
        \lim_{n\rightarrow \infty}\frac{\vec{\theta}_n}{n}=\vec{\eta}=P_{\mathcal{H}_{\vec{\eta}}^{\perp}}\vec{\eta}
    \end{align*}
    and
    \begin{align*}
        f_n(\vec{\theta}_n)=&-\sum_{k=1}^K w_k\exp(l_{n}\vec{\alpha}^{\top}_k\vec{\epsilon}_{n})+\vec{\eta}_n^{\top}\vec{\theta}_n-\frac{1}{2}\vec{\theta}_n^{\top}\vec{\theta}_n\\
        =&o(1)+n^2\left(\vec{\eta}+o(1)\right)^{\top}\left(\vec{\eta}+o(1)\right)-\frac{n^2}{2}\left(\vec{\eta}+o(1)\right)^{\top}\left(\vec{\eta}+o(1)\right)\\
        =&\left(\frac{1}{2}\left\|P_{\mathcal{H}_{\vec{\eta}}^{\perp}}\vec{\eta}\right\|^2+o(1)\right)n^2.
    \end{align*}
    \textbf{Case 2: }$\underset{k=1,\ldots,K}{\max}\vec{\alpha}_k^{\top}\vec{\epsilon}>0$.\\[2mm]
    Assume that $\left\{\vec{\alpha}_{k_1},\ldots,\vec{\alpha}_{k_m}\right\}=\left\{\vec{\alpha}_k:\vec{\alpha}_k^{\top}\vec{\epsilon}=\underset{m=1,\ldots,K}{\max}\vec{\alpha}_m^{\top}\vec{\epsilon}\right\}$. By multiplying equation (\ref{eq_prop_3.2}) by $\vec{\epsilon}$, we have
    \begin{align}\label{eq_prop_3.3}
        -\sum_{k=1}^K w_k\exp(l_{n}\vec{\alpha}^{\top}_k\vec{\epsilon}_{n})\vec{\alpha}_k^{\top}\vec{\epsilon}+\vec{\eta}_n^{\top}\vec{\epsilon}=l_{n}\vec{\epsilon}_{n}^{\top}\vec{\epsilon}.
    \end{align}
    Since $l_n\rightarrow \infty$, we have
    \begin{align}\label{eq_prop_3.4}
        \lim_{n\rightarrow \infty}\frac{\sum_{p=1}^m w_{k_p}\exp(l_{n}\vec{\alpha}^{\top}_{k_p}\vec{\epsilon}_{n})\vec{\alpha}_{k_p}^{\top}\vec{\epsilon}}{\sum_{k=1}^K w_k\exp(l_{n}\vec{\alpha}^{\top}_k\vec{\epsilon}_{n})\vec{\alpha}_k^{\top}\vec{\epsilon}}=1.
    \end{align}
    For any $p=1,\ldots,m$, since $l_n\rightarrow \infty$ and $\vec{\alpha}^{\top}_{k_p}\vec{\epsilon}_{n}\rightarrow \vec{\alpha}^{\top}_{k_p}\vec{\epsilon}>0$, we have $l_n\ll \exp(l_{n}\vec{\alpha}^{\top}_{k_p}\vec{\epsilon}_{n})$. So by (\ref{eq_prop_3.4}), we have
    \begin{align}\label{eq_prop_3.5}
        &\lim_{n\rightarrow \infty}\frac{l_{n}\vec{\epsilon}_{n}^{\top}\vec{\epsilon}}{\sum_{k=1}^K w_k\exp(l_{n}\vec{\alpha}^{\top}_k\vec{\epsilon}_{n})\vec{\alpha}_k^{\top}\vec{\epsilon}}\notag\\
        =&\lim_{n\rightarrow \infty}\frac{\sum_{p=1}^m w_{k_p}\exp(l_{n}\vec{\alpha}^{\top}_{k_p}\vec{\epsilon}_{n})\vec{\alpha}_{k_p}^{\top}\vec{\epsilon}}{\sum_{k=1}^K w_k\exp(l_{n}\vec{\alpha}^{\top}_k\vec{\epsilon}_{n})\vec{\alpha}_k^{\top}\vec{\epsilon}}\times\frac{l_{n}\vec{\epsilon}_{n}^{\top}\vec{\epsilon}}{l_{n}}\times \frac{l_{n}}{\sum_{p=1}^m w_{k_p}\exp(l_{n}\vec{\alpha}^{\top}_{k_p}\vec{\epsilon}_{n})\vec{\alpha}_{k_p}^{\top}\vec{\epsilon}}\notag\\
        =&0.
    \end{align}
    So by (\ref{eq_prop_3.3}), (\ref{eq_prop_3.4}) and (\ref{eq_prop_3.5}), we have
    \begin{align*}
        &\lim_{n\rightarrow \infty}\frac{\sum_{p=1}^m w_{k_p}\exp(l_{n}\vec{\alpha}^{\top}_{k_p}\vec{\epsilon}_{n})\vec{\alpha}_{k_p}^{\top}\vec{\epsilon}}{\vec{\eta}_n^{\top}\vec{\epsilon}}\\
        =&\lim_{n\rightarrow \infty}\frac{\sum_{p=1}^m w_{k_p}\exp(l_{n}\vec{\alpha}^{\top}_{k_p}\vec{\epsilon}_{n})\vec{\alpha}_{k_p}^{\top}\vec{\epsilon}}{\sum_{k=1}^K w_k\exp(l_{n}\vec{\alpha}^{\top}_k\vec{\epsilon}_{n})\vec{\alpha}_k^{\top}\vec{\epsilon}}\frac{\sum_{k=1}^K w_k\exp(l_{n}\vec{\alpha}^{\top}_k\vec{\epsilon}_{n})\vec{\alpha}_k^{\top}\vec{\epsilon}}{\vec{\eta}_n^{\top}\vec{\epsilon}}\\
        =&\lim_{n\rightarrow \infty}\frac{\sum_{p=1}^m w_{k_p}\exp(l_{n}\vec{\alpha}^{\top}_{k_p}\vec{\epsilon}_{n})\vec{\alpha}_{k_p}^{\top}\vec{\epsilon}}{\sum_{k=1}^K w_k\exp(l_{n}\vec{\alpha}^{\top}_k\vec{\epsilon}_{n})\vec{\alpha}_k^{\top}\vec{\epsilon}}\frac{\sum_{k=1}^K w_k\exp(l_{n}\vec{\alpha}^{\top}_k\vec{\epsilon}_{n})\vec{\alpha}_k^{\top}\vec{\epsilon}}{\sum_{k=1}^K w_k\exp(l_{n}\vec{\alpha}^{\top}_k\vec{\epsilon}_{n})\vec{\alpha}_k^{\top}\vec{\epsilon}-l_{n}\vec{\epsilon}_{n}^{\top}\vec{\epsilon}}\\
        =&1,
    \end{align*}
    which indicates that $l_{n}=o(n)$. Divide equation (\ref{eq_prop_3.3}) by $n$, and combine with (\ref{eq_prop_3.4}), we have
    \begin{align*}
        \sum_{p=1}^m \frac{w_{k_p}\exp(l_{n}\vec{\alpha}^{\top}_{k_p}\vec{\epsilon}_{n})}{n}\vec{\alpha}_{k_p}^{\top}\vec{\epsilon}=\vec{\eta}^{\top}\vec{\epsilon}+o(1).
    \end{align*}
    Since $\vec{\alpha}^{\top}_{k_p}\vec{\epsilon}>0$, it can be readily shown that $w_{k_p}\exp(l_{n}\vec{\alpha}^{\top}_{k_p}\vec{\epsilon}_{n})/{n}$ is bounded for any $p=1,\ldots,m$. Then we assume WLOG that for any $p=1,\ldots,m$,
    \begin{align}\label{eq_prop_3.6}
        \lim_{n\rightarrow \infty}\frac{w_{k_p}\exp(l_{n}\vec{\alpha}^{\top}_{k_p}\vec{\epsilon}_{n})}{n}=c_{k_p},
    \end{align}
    where $c_{k_1},\ldots,c_{k_m}$ are nonnegative constants. Then for any $\vec{\alpha}_l\notin\left\{\vec{\alpha}_{k_1},\ldots,\vec{\alpha}_{k_m}\right\}$, it holds that $\vec{\alpha}_l^{\top}\vec{\epsilon}_{n}\leq \underset{k=1,\ldots,K}{\max}\vec{\alpha}_k^{\top}\vec{\epsilon}-\delta$ for $n$ large, where $\delta>0$ is a positive constant. So we have
    \begin{align*}
        \lim_{n\rightarrow \infty}\frac{w_{l}\exp(l_{n}\vec{\alpha}^{\top}_{l}\vec{\epsilon}_{n})}{n}\leq\lim_{n\rightarrow \infty}\frac{w_{l}\exp\left(l_{n}\left[\underset{k=1,\ldots,K}{\max}\vec{\alpha}_k^{\top}\vec{\epsilon}-\delta\right]\right)}{n}=0.
    \end{align*}
    Then by dividing first-order equation (\ref{eq_prop_3.2}) by $n$, we deduce that $\vec{\eta}=\sum_{p=1}^m c_{k_p}\vec{\alpha}_{k_p}$. Since $\vec{\eta}\neq 0$, at least one of $c_{k_1},\ldots,c_{k_m}$ is strictly positive, which indicates that
    \begin{align}\label{eq_prop_3.7}
        \lim_{n\rightarrow \infty}\frac{l_{n}}{\log n}=\frac{1}{\underset{k=1,\ldots,K}{\max}\vec{\alpha}_k^{\top}\vec{\epsilon}}.
    \end{align}
    So we choose all vectors in $\left\{\vec{\alpha}_k:k=1,\ldots,K\right\}$ to satisfy the condition in part (1). Here $\mathcal{H}_{\vec{\eta}}=\mathbb{R}^d$ and $\mathcal{H}^{\perp}_{\vec{\eta}}=\{\vec{0}\}$. Moreover, we have
    \begin{align*}
        \lim_{n\rightarrow \infty}\frac{\vec{\theta}_n}{n}=\vec{0}=P_{\mathcal{H}^{\perp}_{\vec{\eta}}}\vec{\eta}.
    \end{align*}
    Since by (\ref{eq_prop_3.7}), for any $\delta>0$, it holds that $\sum_{k=1}^K w_k\exp(l_{n}\vec{\alpha}^{\top}_k\vec{\epsilon}_{n})=o(\exp((1+\delta)\log n))=o(n^{1+\delta})$, we have
    \begin{align*}
        f_n(\vec{\theta}_n)=o(n^2)=\left(\frac{1}{2}\left\|P_{\mathcal{H}^{\perp}_{\vec{\eta}}}\vec{\eta}\right\|^2+o(1)\right)n^2.
    \end{align*}
    \textbf{Case 3: }$\underset{k=1,\ldots,K}{\max}\vec{\alpha}_k^{\top}\vec{\epsilon}=0$.\\[3mm]
    Define ${f}^{(0)}_{n}=f_n$, ${\vec{\theta}}_{n}^{(0)}=\vec{\theta}_n$ and define
    \begin{align*}
        \widetilde{f}^{(0)}_{n}(\vec{\theta})=-\sum_{k:\vec{\alpha}_k^{\top}\vec{\epsilon}=0} w_k\exp(\vec{\alpha}^{\top}_k\vec{\theta})+\vec{\eta}_n^{\top}\vec{\theta}-\frac{1}{2}\vec{\theta}^{\top}\vec{\theta}=f_{n}^{(0)}(\vec{\theta})+\sum_{k:\vec{\alpha}_k^{\top}\vec{\epsilon}<0} w_k\exp(\vec{\alpha}^{\top}_k\vec{\theta}).
    \end{align*}
    Denote the unique maximiser of $\widetilde{f}_{n}^{(0)}$ by $\widetilde{\vec{\theta}}_{n}^{(0)}$. Then the first-order equations for $\vec{\theta}_n^{(0)}$ and $\widetilde{\vec{\theta}}_n^{(0)}$ are
    \begin{align*}
        \nabla \widetilde{f}^{(0)}_{n}({\vec{\theta}}_{n}^{(0)})+\sum_{k:\vec{\alpha}_k^{\top}\vec{\epsilon}<0} w_k\exp(\vec{\alpha}^{\top}_k{\vec{\theta}}_{n}^{(0)})\vec{\alpha}_k&=\nabla f^{(0)}_{n}(\vec{\theta}_{n}^{(0)})=\vec{0},\\
        \nabla \widetilde{f}^{(0)}_{n}(\widetilde{\vec{\theta}}_{n}^{(0)})&=\vec{0}.
    \end{align*}
    Since $\nabla^2\widetilde{f}_n^{(0)}(\vec{\theta})\preceq-\mat{I}_d$ for any $\vec{\theta}$, by Taylor expansion, we have
    \begin{align*}
        \left\|\vec{\theta}_{n}^{(0)}-\widetilde{\vec{\theta}}_{n}^{(0)}\right\|\leq \left\|\nabla^2 \widetilde{f}^{(0)}_{n}(\vec{\theta}_n^{\ast})(\vec{\theta}_{n}^{(0)}-\widetilde{\vec{\theta}}_{n}^{(0)})\right\|=&\left\|\nabla\widetilde{f}^{(0)}_{n}({\vec{\theta}}_{n}^{(0)})-\nabla \widetilde{f}^{(0)}_{n}(\widetilde{\vec{\theta}}_{n}^{(0)})\right\|\\
        =&\left\|\sum_{k:\vec{\alpha}_k^{\top}\vec{\epsilon}<0} w_k\exp(l_n\vec{\alpha}^{\top}_k\vec{\epsilon}_{n})\vec{\alpha}_k\right\|\rightarrow 0
    \end{align*}
    since $l_n$ goes to infinity. Since $\left\|\vec{\theta}_{n}^{(0)}-\widetilde{\vec{\theta}}_{n}^{(0)}\right\|\rightarrow 0$, it can be readily shown that $\sum_{k:\vec{\alpha}_k^{\top}\vec{\epsilon}<0} w_k\exp(\vec{\alpha}^{\top}_k\widetilde{\vec{\theta}}_{n}^{(0)})\rightarrow 0$. Then we have
    \begin{align*}
        \widetilde{f}_n^{(0)}(\widetilde{\vec{\theta}}_n^{(0)})-f_n^{(0)}(\vec{\theta}_n^{(0)})\leq& \widetilde{f}_n^{(0)}(\widetilde{\vec{\theta}}_n^{(0)})-f_n^{(0)}(\widetilde{\vec{\theta}}_n^{(0)})=\sum_{k:\vec{\alpha}_k^{\top}\vec{\epsilon}<0} w_k\exp(\vec{\alpha}^{\top}_k\widetilde{\vec{\theta}}_{n}^{(0)})\rightarrow 0,\\
        \widetilde{f}_n^{(0)}(\widetilde{\vec{\theta}}_n^{(0)})-f_n^{(0)}(\vec{\theta}_n^{(0)})\geq& \widetilde{f}_n^{(0)}(\vec{\theta}_n^{(0)})-f_n^{(0)}(\vec{\theta}_n^{(0)})=\sum_{k:\vec{\alpha}_k^{\top}\vec{\epsilon}<0} w_k\exp(\vec{\alpha}^{\top}_k\vec{\theta}_n^{(0)})\rightarrow 0,
    \end{align*}
    which implies that $\widetilde{f}_n^{(0)}(\widetilde{\vec{\theta}}_n^{(0)})-f_n^{(0)}(\vec{\theta}_n^{(0)})\rightarrow 0$.
    Denote $\mathcal{H}_1=\operatorname{span}\left\{\vec{\alpha}_k:k=1,\ldots,K, \vec{\alpha}_k^{\top}\vec{\epsilon}=0\right\}$ and denote
    \begin{align*}
        {f}_{n}^{(1)}(\vec{\theta})=-\sum_{k:\vec{\alpha}_k^{\top}\vec{\epsilon}=0} w_k\exp(\vec{\alpha}^{\top}_k\vec{\theta})+\left(P_{\mathcal{H}_1}\vec{\eta}_n\right)^{\top}\vec{\theta}-\frac{1}{2}\vec{\theta}^{\top}\vec{\theta}.
    \end{align*}
    We then plug ${\vec{\theta}}_{n}^{(1)}:=\widetilde{\vec{\theta}}^{(0)}_{n}-P_{\mathcal{H}^{\perp}_1}\vec{\eta}_n$ into the gradient of ${f}_{n}^{(1)}$:
    \begin{align*}
        \nabla {f}_{n}^{(1)}({\vec{\theta}}_{n}^{(1)})=&-\sum_{k:\vec{\alpha}_k^{\top}\vec{\epsilon}=0} w_k\exp(\vec{\alpha}^{\top}_k(\widetilde{\vec{\theta}}_{n}^{(0)}-P_{\mathcal{H}^{\perp}_1}\vec{\eta}_n))\vec{\alpha}_k+P_{\mathcal{H}_1}\vec{\eta}_n-(\widetilde{\vec{\theta}}_{n}^{(0)}-P_{\mathcal{H}^{\perp}_1}\vec{\eta}_n)\\
        &=-\sum_{k:\vec{\alpha}_k^{\top}\vec{\epsilon}=0} w_k\exp(\vec{\alpha}^{\top}_k\widetilde{\vec{\theta}}_{n}^{(0)})\vec{\alpha}_k+\vec{\eta}_n-\widetilde{\vec{\theta}}_{n}^{(0)}\\
        &=\vec{0},
    \end{align*}
    where the last step is due to the first-order equation for $\widetilde{f}^{(0)}_{n}(\widetilde{\vec{\theta}}_{n}^{(0)})$. This implies that ${\vec{\theta}}_{n}^{(1)}$ is the maximiser for ${f}_{n}^{(1)}$. Moreover, we have
    \begin{align*}
        {f}_{n}^{(1)}({\vec{\theta}}_{n}^{(1)})=&-\sum_{k:\vec{\alpha}_k^{\top}\vec{\epsilon}=0} w_k\exp\left(\vec{\alpha}^{\top}_k\left(\widetilde{\vec{\theta}}^{(0)}_{n}-P_{\mathcal{H}^{\perp}_1}\vec{\eta}_n\right)\right)+\left(\vec{\eta}_n-P_{\mathcal{H}_1^{\perp}}\vec{\eta}_n\right)^{\top}\left(\widetilde{\vec{\theta}}^{(0)}_{n}-P_{\mathcal{H}^{\perp}_1}\vec{\eta}_n\right)\\
        &-\frac{1}{2}\left(\widetilde{\vec{\theta}}^{(0)}_{n}-P_{\mathcal{H}^{\perp}_1}\vec{\eta}_n\right)^{\top}\left(\widetilde{\vec{\theta}}^{(0)}_{n}-P_{\mathcal{H}^{\perp}_1}\vec{\eta}_n\right)\\
        =&-\sum_{k:\vec{\alpha}_k^{\top}\vec{\epsilon}=0} w_k\exp\left(\vec{\alpha}^{\top}_k\widetilde{\vec{\theta}}^{(0)}_{n}\right)+\vec{\eta}_n^{\top}\widetilde{\vec{\theta}}^{(0)}_{n}-\frac{1}{2}\left(\widetilde{\vec{\theta}}^{(0)}_{n}\right)^{\top}\widetilde{\vec{\theta}}^{(0)}_{n}+\frac{1}{2}\left\|P_{\mathcal{H}^{\perp}_1}\vec{\eta}_n\right\|^2\\
        =&\widetilde{f}_n^{(0)}(\widetilde{\vec{\theta}}_n^{(0)})+\frac{1}{2}\left\|P_{\mathcal{H}^{\perp}_1}\vec{\eta}_n\right\|^2.
    \end{align*}
    We then prove that at least one vector among $\left\{\vec{\alpha}_1,\ldots,\vec{\alpha}_K\right\}$ is eliminated in the procedure from $f_{n}^{(0)}$ to $f_{n}^{(1)}$, which is equivalent to proving that $\vec{\alpha}_1^{\top}\vec{\epsilon}=\ldots=\vec{\alpha}_K^{\top}\vec{\epsilon}=0$ cannot happen. If it is the case, we multiply (\ref{eq_prop_3.2}) by $\vec{\epsilon}$ to get
    \begin{align*}
        l_n\vec{\epsilon}_n^{\top}\vec{\epsilon}=\vec{\eta}_n^{\top}\vec{\epsilon}-\sum_{k=1}^K w_k\exp(l_{n}\vec{\alpha}^{\top}_k\vec{\epsilon}_{n})\vec{\alpha}_k^{\top}\vec{\epsilon}=\vec{\eta}_n^{\top}\vec{\epsilon}.
    \end{align*}
    Since $\vec{\eta}_n\in \operatorname{span}\left\{\vec{\alpha}_1,\ldots,\vec{\alpha}_K\right\}$, we have $\vec{\eta}_n^{\top}\vec{\epsilon}=0$. Since $l_n\rightarrow \infty$ and $\vec{\epsilon}_n^{\top}\vec{\epsilon}\rightarrow\vec{\epsilon}^{\top}\vec{\epsilon}=1$, we have $l_n\vec{\epsilon}_n^{\top}\vec{\epsilon}\rightarrow \infty$. This leads to a contradiction. So in this procedure, at least one vector is eliminated. Then we apply the same procedure on $f_{n}^{(1)}$ to discuss which of the three cases it falls into. This procedure will stop in finitely many steps, i.e., falls into Case 1 or 2 in finitely many steps. In this process, we get a sequence of $\vec{\theta}_{n}^{(0)},\ldots,\vec{\theta}_{n}^{(r)}$, $\widetilde{\vec{\theta}}_{n}^{(0)},\ldots,\widetilde{\vec{\theta}}_{n}^{(r-1)}$ and $\mathcal{H}_1,\ldots,\mathcal{H}_{r}$ such that for $p=0,\ldots,r-1$,
    \begin{align}\label{eq_prop_3.8}
        \|\vec{\theta}_{n}^{(p)}-\widetilde{\vec{\theta}}_{n}^{(p)}\|&=o(1),\notag\\
        \vec{\theta}_{n}^{(p+1)}&=\widetilde{\vec{\theta}}_{n}^{(p)}-P_{\mathcal{H}^{\perp}_{p+1}}\left(P_{\mathcal{H}_{p}}\vec{\eta}_n\right),\notag\\
        f_{n}^{(p+1)}(\bt)&=-\sum_{k:\vec{\alpha}_k\in \mathcal{H}_{p+1}} w_k\exp(\vec{\alpha}^{\top}_k\vec{\theta})+\left(P_{\mathcal{H}_{p+1}}\vec{\eta}_n\right)^{\top}\vec{\theta}-\frac{1}{2}\vec{\theta}^{\top}\vec{\theta}.
    \end{align}
    Denote $\vec{\theta}_n^{(p)}/\|\vec{\theta}_n^{(p)}\|=\vec{\epsilon}_{n}^{(p)}\rightarrow\vec{\epsilon}^{(p)}$. The procedure will fall into one of the two cases in the last step:\\[3mm]
    \textbf{Case 3.1: }$\underset{k:\vec{\alpha}_k\in \mathcal{H}_{r}}{\max}\vec{\alpha}_k^{\top}\vec{\epsilon}^{(r)}<0$.\\[2mm]
    Then by the proof in Case 1, $\vec{\theta}_{n}^{(r)}=P_{\mathcal{H}_{r}}\vec{\eta}_n+o(1)$. Combine this with (\ref{eq_prop_3.8}), we have
    \begin{align*}
        \vec{\theta}^{(0)}_{n}=o(1)+\sum_{p=0}^{r-1} P_{\mathcal{H}^{\perp}_{p+1}}\left(P_{\mathcal{H}_{p}}\vec{\eta}_n\right)+P_{\mathcal{H}_{r}}\vec{\eta}_n=\vec{\eta}_n+o(1).
    \end{align*}
    This implies that $\vec{\epsilon}^{(0)}:= \vec{\epsilon}\propto \vec{\eta}$ and $\vec{\epsilon}^{(p)}\propto P_{\mathcal{H}_{p}}\vec{\eta},~p=1,\ldots,r$. For any $k=1,\ldots,K$, there exists $p=0,\ldots,r$ such that $\vec{\alpha}_k\in \mathcal{H}_{p}\setminus \mathcal{H}_{p+1}$ (define $\mathcal{H}_{r+1}=\{\vec{0}\}$), then
    \begin{align*}
        0>\vec{\alpha}^{\top}_k\vec{\epsilon}^{(p)}\propto \vec{\alpha}^{\top}_k P_{\mathcal{H}_{p}}\vec{\eta}=\vec{\alpha}^{\top}_k\left(\vec{\eta}-P_{\mathcal{H}^{\perp}_{p}}\vec{\eta}\right)=\vec{\alpha}^{\top}_k \vec{\eta}.
    \end{align*}
    So for any $k=1,\ldots,K$, $\vec{\alpha}^{\top}_k \vec{\eta}<0$, which indicates that the problem should fall into Case 1.\\[3mm]
    \textbf{Case 3.2: }$\underset{k:\vec{\alpha}_k\in \mathcal{H}_{r}}{\max}\vec{\alpha}_k^{\top}\vec{\epsilon}^{(r)}>0$.\\[2mm]
    By the proof in Case 2, $\vec{\theta}_{n}^{(r)}=\mathcal{O}(\log(n))$. Moreover,
    \begin{align*}
        P_{\mathcal{H}_{r}}\vec{\eta}&=\sum_{j=1}^m c_{k_j}\vec{\alpha}_{k_j},~~~~c_{k_1},\ldots,c_{k_m}\geq 0,
    \end{align*}
    where $\vec{\alpha}_{k_1},\ldots,\vec{\alpha}_{k_m}$ are the remaining vectors after $r$ steps and $\mathcal{H}_{r}=\operatorname{span}\left\{ \vec{\alpha}_{k_1},\ldots,\vec{\alpha}_{k_m}\right\}$. Combine this with (\ref{eq_prop_3.8}), we have
    \begin{align*}
        \vec{\theta}_{n}=\mathcal{O}(\log n)+\sum_{p=0}^{r-1} P_{\mathcal{H}^{\perp}_{p+1}}P_{\mathcal{H}_{p}}\vec{\eta}_n=P_{\mathcal{H}^{\perp}_{r}}\vec{\eta}_n+o(n).
    \end{align*}
    Then $\vec{\epsilon}^{(0)}:= \vec{\epsilon}\propto P_{\mathcal{H}^{\perp}_{r}}\vec{\eta}$ and $\vec{\epsilon}^{(p)}\propto P_{\mathcal{H}^{\perp}_{r}}\vec{\eta}-P_{\mathcal{H}^{\perp}_{p}}\vec{\eta}$. So for any $k\in \left\{1,\ldots,K\right\}\setminus\left\{k_1,\ldots,k_m\right\}$, there exists $p=0,\ldots,r-1$ such that $\vec{\alpha}_k\in \mathcal{H}_{p}\setminus \mathcal{H}_{p+1}$. Then we have
    \begin{align*}
        0>\vec{\alpha}^{\top}_k\vec{\epsilon}^{(p)}\propto\vec{\alpha}^{\top}_k\left(P_{\mathcal{H}^{\perp}_{r}}\vec{\eta}-P_{\mathcal{H}^{\perp}_{p}}\vec{\eta}\right)=\vec{\alpha}^{\top}_k P_{\mathcal{H}^{\perp}_{r}}\vec{\eta},
    \end{align*}
    which implies that condition in part (1) is satisfied by choosing $\mathcal{H}_{\vec{\eta}}$ as $\mathcal{H}_{r}$.\\[3mm]
    Furthermore, we have
    \begin{align*}
        f_n^{(0)}(\vec{\theta}_n^{(0)})=&\sum_{p=0}^{r-1} \left[\left(f_n^{(p)}({\vec{\theta}}_n^{(p)})-\widetilde{f}_n^{(p)}(\widetilde{\vec{\theta}}_n^{(p)})\right)+\left(\widetilde{f}_n^{(p)}(\widetilde{\vec{\theta}}_n^{(p)})-f_n^{(p+1)}({\vec{\theta}}_n^{(p+1)})\right)\right]+f_n^{(r)}(\vec{\theta}_n^{(r)})\\
        =&\sum_{p=0}^{r-1} \left[o(1)+\frac{1}{2}\left\|P_{\mathcal{H}^{\perp}_{p+1}}\left(P_{\mathcal{H}_{p}}\vec{\eta}_n\right)\right\|^2\right]+o(n^2)\\
        =&\frac{1}{2}\left\|P_{\mathcal{H}^{\perp}_{r}}\vec{\eta}_n\right\|^2+o(n^2)\\[1mm]
        =&\left(\frac{1}{2}\left\|P_{\mathcal{H}^{\perp}_{\vec{\eta}}}\vec{\eta}\right\|^2+o(1)\right)n^2.
    \end{align*}
    If not all vectors of $\left\{\vec{\eta}_n\right\}$ are in $\mathcal{H}_0$, then we define $\widetilde{\vec{\eta}}_n=P_{\mathcal{H}_0}\vec{\eta}_n$ and $\lim_{n\rightarrow \infty}\widetilde{\vec{\eta}}_n/n=\widetilde{\vec{\eta}}$. Then by the previous proof, there exists $\left\{\vec{\alpha}_{k_1},\ldots,\vec{\alpha}_{k_m}\right\}\subseteq \left\{\vec{\alpha}_k:~k=1,\ldots,K\right\}$ such that the conditions in (1) are satisfied. Denote $\mathcal{H}=\operatorname{span}\left\{\vec{\alpha}_{k_1},\ldots,\vec{\alpha}_{k_m}\right\}$, then $P_{\mathcal{H}}\vec{\eta}_n=P_{\mathcal{H}}\left(P_{\mathcal{H}_0}\vec{\eta}_n\right)=P_{\mathcal{H}}\widetilde{\vec{\eta}}_n$, which implies that $P_{\mathcal{H}}\vec{\eta}=P_{\mathcal{H}}\widetilde{\vec{\eta}}$. Furthermore, for $k\notin\left\{k_1,\ldots,k_m\right\}$,
    \begin{align*}
        \vec{\alpha}_k^{\top}P_{\mathcal{H}^{\perp}}\vec{\eta}=&\vec{\alpha}_k^{\top}\left(P_{\mathcal{H}_0^{\perp}}\vec{\eta}+P_{\mathcal{H}^{\perp}}\left(P_{\mathcal{H}_0}\vec{\eta}\right)\right)=\vec{\alpha}_k^{\top}P_{\mathcal{H}^{\perp}}\widetilde{\vec{\eta}}<0.
    \end{align*}
    So $\left\{\vec{\alpha}_{k_1},\ldots,\vec{\alpha}_{k_m}\right\}$ also satisfies the conditions for $\vec{\eta}$.\\[3mm]
    Moreover, define
    \begin{align*}
        \widetilde{f}_n(\vec{\theta})=-\sum_{k=1}^K w_k\exp(\vec{\alpha}^{\top}_k\vec{\theta})+\left(P_{\mathcal{H}_0}\vec{\eta}_n\right)^{\top}\vec{\theta}-\frac{1}{2}\vec{\theta}^{\top}\vec{\theta}
    \end{align*}
    and its unique maximiser as $\widetilde{\vec{\theta}}_n$. We use similar method to prove that $\widetilde{\vec{\theta}}_n=\vec{\theta}_n-P_{\mathcal{H}_0^{\perp}}\vec{\eta}_n$ and $\widetilde{f}_n(\widetilde{\vec{\theta}}_n)=f_n(\vec{\theta}_n)-\frac{1}{2}\left\|P_{\mathcal{H}_0^{\perp}}\vec{\eta}_n\right\|^2$.
    Since $P_{\mathcal{H}_0}\vec{\eta}_n\in\mathcal{H}_0$ for any $n$, by previous proof we have $\lim_{n\rightarrow \infty}\frac{\widetilde{\vec{\theta}}_n}{n}=P_{\mathcal{H}^{\perp}}\widetilde{\vec{\eta}}$ and $\widetilde{f}_n(\widetilde{\vec{\theta}}_n)=\left(\frac{1}{2}\left\|P_{\mathcal{H}^{\perp}}\widetilde{\vec{\eta}}\right\|^2+o(1)\right)n^2$. This implies that
    \begin{align*}
        \lim_{n\rightarrow \infty}\frac{{\vec{\theta}}_n}{n}=\lim_{n\rightarrow \infty}\frac{\widetilde{\vec{\theta}}_n+P_{\mathcal{H}_0^{\perp}}\vec{\eta}_n}{n}=P_{\mathcal{H}^{\perp}}\widetilde{\vec{\eta}}+P_{\mathcal{H}_0^{\perp}}\vec{\eta}=P_{\mathcal{H}_{\vec{\eta}}^{\perp}}\left(P_{\mathcal{H}_0}\vec{\eta}\right)+P_{\mathcal{H}_0^{\perp}}\vec{\eta}=\vec{\eta}-P_{\mathcal{H}_{\vec{\eta}}}\left(P_{\mathcal{H}_0}\vec{\eta}\right)=P_{\mathcal{H}_{\vec{\eta}}^{\perp}}\vec{\eta}
    \end{align*}
    and
    \begin{align*}
        f_n(\vec{\theta}_n)=&\left(\frac{1}{2}\left\|P_{\mathcal{H}^{\perp}}\widetilde{\vec{\eta}}\right\|^2+o(1)\right)n^2+\frac{1}{2}\left\|P_{\mathcal{H}_0^{\perp}}\vec{\eta}_n\right\|^2\\
        =&\left(\frac{1}{2}\left\|P_{\mathcal{H}_{\vec{\eta}}^{\perp}}\left(P_{\mathcal{H}_0}\vec{\eta}\right)\right\|^2+\frac{1}{2}\left\|P_{\mathcal{H}_0^{\perp}}\vec{\eta}\right\|^2+o(1)\right)n^2\\
        =&\frac{1}{2}\left(\left\|P_{\mathcal{H}_0}\vec{\eta}\right\|^2-\left\|P_{\mathcal{H}_{\vec{\eta}}}\left(P_{\mathcal{H}_0}\vec{\eta}\right)\right\|^2+\left\|P_{\mathcal{H}_0^{\perp}}\vec{\eta}\right\|^2+o(1)\right)n^2\\
        =&\frac{1}{2}\left(\left\|\vec{\eta}\right\|^2-\left\|P_{\mathcal{H}_{\vec{\eta}}}\vec{\eta}\right\|^2+o(1)\right)n^2\\
        =&\left(\frac{1}{2}\left\|P_{\mathcal{H}_{\vec{\eta}}^{\perp}}\vec{\eta}\right\|^2+o(1)\right)n^2.
    \end{align*}
    Finally, we prove the uniqueness and continuity of canonical projection.\\[3mm]
    \textbf{Uniqueness: }Assume all weights $w_k\equiv 1$ and denote function
    \begin{align*}
        g_n(\vec{\theta})=&-\sum_{k=1}^K\exp(\vec{\alpha}_k^{\top}\vec{\theta})+n\vec{\eta}^{\top}\vec{\theta}-\frac{1}{2}\vec{\theta}^{\top}\vec{\theta}.
    \end{align*}
    and denote the unique maximiser of $g_n$ by ${\vec{\theta}}_n$. If there exists $\left\{\vec{\alpha}_{k_1},\ldots,\vec{\alpha}_{k_m}\right\}\subseteq \left\{\vec{\alpha}_1,\ldots,\vec{\alpha}_K\right\}$ such that
    \begin{align*}
        P_{\mathcal{H}_{\vec{\eta}}}\vec{\eta}&=\sum_{j=1}^m c_{k_j}\vec{\alpha}_{k_j},~~~~c_{k_1},\ldots,c_{k_m}\geq 0\\
        \vec{\alpha}^{\top}_{k}P_{\mathcal{H}_{\vec{\eta}}^{\perp}}\vec{\eta}&<0,~~~~~~~~~~~~~~~~\forall k\in\left\{1,\ldots,K\right\}\setminus\left\{k_1,\ldots,k_m\right\}
    \end{align*}
    where $\mathcal{H}_{\vec{\eta}}=\operatorname{span}\left\{\vec{\alpha}_{k_1},\ldots,\vec{\alpha}_{k_m}\right\}$, then we define
    \begin{align*}
        \hat{g}_n(\vec{\theta})=&-\sum_{k:\vec{\alpha}_k\in\mathcal{H}_{\vec{\eta}}}\exp(\vec{\alpha}_k^{\top}\vec{\theta})+n\vec{\eta}^{\top}\vec{\theta}-\frac{1}{2}\vec{\theta}^{\top}\vec{\theta},\\
        \widetilde{g}_n(\vec{\theta})=&-\sum_{k:\vec{\alpha}_k\in\mathcal{H}_{\vec{\eta}}}\exp(\vec{\alpha}_k^{\top}\vec{\theta})+n\left(P_{\mathcal{H}_{\vec{\eta}}}\vec{\eta}\right)^{\top}\vec{\theta}-\frac{1}{2}\vec{\theta}^{\top}\vec{\theta},
    \end{align*}
    and denote the maximisers of $\hat{g}_n$ and $\widetilde{g}_n$ by $\hat{\vec{\theta}}_n$, $\widetilde{\vec{\theta}}_n$, respectively. Following a similar argument as in part (2), we have $\vec{\theta}_n-\hat{\vec{\theta}}_n=o(1)$ and $\widetilde{\vec{\theta}}_n=\hat{\vec{\theta}}_n-nP_{\mathcal{H}_{\vec{\eta}}^{\perp}}\vec{\eta}$. Denote $\|\widetilde{\vec{\theta}}_n\|={l}_n$, $\widetilde{\vec{\theta}}_n/{l}_n=\vec{\epsilon}_n\rightarrow \vec{\epsilon}$. We then prove that $\max_{k:\vec{\alpha}_k\in\mathcal{H}_{\vec{\eta}}}\vec{\alpha}_k^{\top}\vec{\epsilon}>0$. If this is not the case, then $\max_{k:\vec{\alpha}_k\in\mathcal{H}_{\vec{\eta}}}\vec{\alpha}_k^{\top}\vec{\epsilon}\leq 0$, we multiply the first-order equation for $\widetilde{\vec{\theta}}_n$ by $\vec{\epsilon}$ and plug in the expansion of $P_{\mathcal{H}_{\vec{\eta}}}\vec{\eta}$ to get
    \begin{align*}
        -\sum_{k:\vec{\alpha}_k\in\mathcal{H}_{\vec{\eta}},\vec{\alpha}_k^{\top}\vec{\epsilon}<0}\exp(l_n\vec{\alpha}_k^{\top}\vec{\epsilon}_n)\vec{\alpha}_k^{\top}\vec{\epsilon}+n\sum_{j=1}^m c_{k_j}(\vec{\alpha}_{k_j}^{\top}\vec{\epsilon})=l_n\vec{\epsilon}_n^{\top}\vec{\epsilon}.
    \end{align*}
    Since $-\sum_{k:\vec{\alpha}_k\in\mathcal{H}_{\vec{\eta}},\vec{\alpha}_k^{\top}\vec{\epsilon}<0}\exp(l_n\vec{\alpha}_k^{\top}\vec{\epsilon}_n)\vec{\alpha}_k^{\top}\vec{\epsilon}\rightarrow 0$, $n\sum_{j=1}^m c_{k_j}(\vec{\alpha}_{k_j}^{\top}\vec{\epsilon})\leq 0$ and $l_n\vec{\epsilon}_n^{\top}\vec{\epsilon}\rightarrow \infty$, this leads to a contradiction. So we have $\max_{k:\vec{\alpha}_k\in\mathcal{H}_{\vec{\eta}}}\vec{\alpha}_k^{\top}\vec{\epsilon}>0$, then follow a similar proof as in part (1), we have $\widetilde{\vec{\theta}}_n=\mathcal{O}(\log n)$. So we have $\vec{\theta}_n=o(n)+nP_{\mathcal{H}_{\vec{\eta}}^{\perp}}\vec{\eta}$, i.e.,
    \begin{align*}
        \lim_{n\rightarrow \infty}\frac{\vec{\theta}_n}{n}=P_{\mathcal{H}_{\vec{\eta}}^{\perp}}\vec{\eta}.
    \end{align*}
    Since $\vec{\theta}_n$ is unique, this implies that $P_{\mathcal{H}_{\vec{\eta}}^{\perp}}\vec{\eta}$ is uniquely determined.\\[3mm]
    \textbf{Continuity: }For any sequence $\left\{\vec{\eta}_n\right\}$ converging to $\vec{\eta}$, i.e., $\vec{\eta}_n\rightarrow \vec{\eta}$, the problem falls into two cases:\\[3mm]
    \textbf{Case 1: }If the choice of $\mathcal{H}_{\vec{\eta}}$ is proper (satisfies the condition in part (1)) in a neighbourhood of $\vec{\eta}$, then the continuity of $P_{\mathcal{H}_{\vec{\eta}}^{\perp}}\vec{\eta}$ follows by the continuity of regular projection.\\[3mm]
    \textbf{Case 2: }If the choice of $\mathcal{H}_{\vec{\eta}}$ is not proper in any neighbourhood of $\vec{\eta}$, this implies that there exists $k_p\in\left\{k_1,\ldots,k_m\right\}$ such that $c_{k_p}=0$ in the expansion of $P_{\mathcal{H}_{\vec{\eta}}}\vec{\eta}$ due to the continuity of projection. We assume WLOG that $c_{k_j}>0$ for $j=1,\ldots,p-1,p+1,\ldots,m$. Denote $\mathcal{H}=\operatorname{span}\left\{\vec{\alpha}_{k_1},\ldots,\vec{\alpha}_{k_{p-1}},\vec{\alpha}_{k_{p+1}},\ldots,\vec{\alpha}_{k_{m}}\right\}$. Since $P_{\mathcal{H}_{\vec{\eta}}}\vec{\eta}=\sum_{j=1}^m c_{k_j}\vec{\alpha}_{k_j}$, there exists $c_{j,n}\rightarrow c_{k_j}$ for $j=1,\ldots,m$ such that
    \begin{align*}
        P_{\mathcal{H}_{\vec{\eta}}}\vec{\eta}_n&=\sum_{j=1}^m c_{j,n}\vec{\alpha}_{k_j}.
    \end{align*}
    Since $c_{p,n}\rightarrow c_{k_p}=0$ and $\sum_{j\neq p} c_{j,n}\vec{\alpha}_{k_j}\in\mathcal{H}$, we have
    \begin{align*}
        \left\|P_{\mathcal{H}_{\vec{\eta}}}\vec{\eta}_n-P_{\mathcal{H}}\vec{\eta}_n\right\|=\left\|P_{\mathcal{H}_{\vec{\eta}}}\vec{\eta}_n-P_{\mathcal{H}}\left(P_{\mathcal{H}_{\vec{\eta}}}\vec{\eta}_n\right)\right\|\leq\left\|P_{\mathcal{H}_{\vec{\eta}}}\vec{\eta}_n-\sum_{j\neq p} c_{j,n}\vec{\alpha}_{k_j}\right\|=\left\|c_{p,n}\vec{\alpha}_{k_p}\right\|\rightarrow 0.
    \end{align*}
    By continuity of projection, this implies that either $\mathcal{H}_{\vec{\eta}}$ or $\mathcal{H}$ is proper for $\left\{\vec{\eta}_n\right\}$. Since by continuity of projection, we have $\left\|P_{\mathcal{H}_{\vec{\eta}}}\vec{\eta}_n-P_{\mathcal{H}_{\vec{\eta}}}\vec{\eta}\right\|\rightarrow 0$. So we have $\left\|P_{\mathcal{H}}\vec{\eta}_n-P_{\mathcal{H}_{\vec{\eta}}}\vec{\eta}\right\|\rightarrow 0$ and $\left\|P_{\mathcal{H}_{\vec{\eta}}}\vec{\eta}_n-P_{\mathcal{H}_{\vec{\eta}}}\vec{\eta}\right\|\rightarrow 0$, which implies that $P_{\mathcal{H}_{\vec{\eta}_n}}\vec{\eta}_n$ converges to $P_{\mathcal{H}_{\vec{\eta}}}\vec{\eta}$. So $P_{\mathcal{H}_{\vec{\eta}}^{\perp}}\vec{\eta}$ is continuous with respect to $\vec{\eta}$.
    \end{proof}
    
    \subsection{Proof of Proposition \ref{prop_max_projection} and Corollary \ref{cor_max_projection}}
    \renewcommand*{\proofname}{Proof of Proposition \ref{prop_max_projection}}
    \begin{proof}
    Define $\Omega=\{(\nu_1,\ldots,\nu_J)^{\top}: \|P_{\mathcal{H}^{\perp}_{\vec{\eta}(\nu_1,\ldots,\nu_J)}}\vec{\eta}(\nu_1,\ldots,\nu_J)\|=\max_{\vec{\eta}\in\mathcal{G}}\|P_{\mathcal{H}^{\perp}_{\vec{\eta}}}\vec{\eta}\|\}$. We then prove that $\Omega$ has only one element $(1,0,\ldots,0)^{\top}$. By the continuity of canonical projection and compactness of $\mathcal{E}:=\{(\nu_1,\ldots,\nu_J)^{\top}:0\leq\nu_j\leq 1,\sum_{j=1}^J\nu_j=1\}$, $\Omega$ is non-empty. Then for any $j=2,\ldots,J$, denote $M_j=\sup\left\{\nu_j:\exists(\nu_1,\ldots,\nu_J)^{\top}\in\Omega\right\}$. Since $\mathcal{E}$ is a compact set, we can find $(\widetilde{\nu}_1,\ldots,\widetilde{\nu}_J)^{\top}\in \Omega$ such that $\widetilde{\nu}_j=M_j$. Thus $M_j<1$ since $\|P_{\mathcal{H}^{\perp}_{\vec{\eta}_j}}\vec{\eta}_j\|<\|P_{\mathcal{H}^{\perp}_{\vec{\eta}_1}}\vec{\eta}_1\|$ by assumption.\\[3mm]
        If $0<M_j<1$, then there exists $i\neq j$ such that $\widetilde{\nu}_i>0$. We denote $\vec{\eta}(\delta)=\vec{\eta}(\widetilde{\nu}_1,\ldots,\widetilde{\nu}_J)+\delta(\vec{\eta}_j-\vec{\eta}_i)$, where $\vec{\eta}(\delta)$ lies in the feasible domain for $|\delta|$ small enough. By the definition of $\Omega$, $\|P_{\mathcal{H}^{\perp}_{\vec{\eta}({\delta})}}\vec{\eta}({\delta})\|\leq \|P_{\mathcal{H}^{\perp}_{\vec{\eta}(0)}}\vec{\eta}(0)\|$. By the result in (1), there exists $\left\{\vec{\alpha}_{k_1},\ldots,\vec{\alpha}_{k_m}\right\}\subseteq \left\{\vec{\alpha}_1,\ldots,\vec{\alpha}_K\right\}$ such that $\mathcal{H}_{\vec{\eta}(0)}=\operatorname{span}\left\{\vec{\alpha}_{k_1},\ldots,\vec{\alpha}_{k_m}\right\}$ and $P_{\mathcal{H}_{\vec{\eta}(0)}}\vec{\eta}(0)=\sum_{j=1}^mc_{k_j}\vec{\alpha}_{k_j}$, where $c_{k_1},\ldots,c_{k_m}$ are nonnegative constants. Then the problem falls into either of the two cases:\\[3mm]
        \textbf{Case 1}: The choice of $\mathcal{H}_{\vec{\eta}({0})}$ is proper for $\vec{\eta}({\delta})$ when $\delta$ is in a neighbourhood of 0, then
        \begin{align}\label{eq_prop_4.1}
            \left\|P_{\mathcal{H}^{\perp}_{\vec{\eta}({\delta})}}\vec{\eta}({\delta})\right\|^2&=\left\|P_{\mathcal{H}^{\perp}_{\vec{\eta}({0})}}\vec{\eta}({\delta})\right\|^2\notag\\
            &=\left\|P_{\mathcal{H}^{\perp}_{\vec{\eta}(0)}}\vec{\eta}(0)\right\|^2+2\delta\left(P_{\mathcal{H}^{\perp}_{\vec{\eta}(0)}}\vec{\eta}(0)\right)^{\top}\left(P_{\mathcal{H}^{\perp}_{\vec{\eta}(0)}}(\vec{\eta}_j-\vec{\eta}_i)\right)+\delta^2 \left\|P_{\mathcal{H}^{\perp}_{\vec{\eta}(0)}}(\vec{\eta}_i-\vec{\eta}_j)\right\|^2.
        \end{align}
        Since $\|P_{\mathcal{H}^{\perp}_{\vec{\eta}({\delta})}}\vec{\eta}({\delta})\|$ attains maximum value at $\delta=0$, (\ref{eq_prop_4.1}) implies that
        \begin{align*}
            P_{\mathcal{H}^{\perp}_{\vec{\eta}(0)}}(\vec{\eta}_j-\vec{\eta}_i)=0.
        \end{align*}
        Then (\ref{eq_prop_4.1}) indicates that $\|P_{\mathcal{H}^{\perp}_{\vec{\eta}({\delta})}}\vec{\eta}({\delta})\|=\|P_{\mathcal{H}^{\perp}_{\vec{\eta}(0)}}\vec{\eta}(0)\|$ when $\delta$ is in a small neighbourhood of 0.\\[3mm]
        \textbf{Case 2}: The choice of $\mathcal{H}_{\vec{\eta}({0})}$ is not proper for $\vec{\eta}({\delta})$ in any neighbourhood of 0. Then by similar proof as in Proposition~\ref{prop_canonical_projection}, we assume WLOG that there exists $p\in\{1,\ldots,m\}$ such that $c_{k_p}=0$ and $c_{k_j}>0$ for $j=1,\ldots,p-1,p+1,\ldots,m$. Denote
        \begin{align*}
            \widetilde{\mathcal{H}}:= \operatorname{span}\left\{\vec{\alpha}_{k_1},\ldots,\vec{\alpha}_{k_{p-1}},\vec{\alpha}_{k_{p+1}},\ldots,\vec{\alpha}_{k_{m}}\right\}\subsetneq \mathcal{H}_{\vec{\eta}(0)}:= \mathcal{H}.
        \end{align*}
        By Lemma~\ref{lem_characterisation_equation_1}, we assume WLOG that $\left\{\vec{\alpha}_{k_j}:j=1,\ldots,m,j\neq p\right\}$ are linearly independent. Similar to the proof in Proposition~\ref{prop_canonical_projection}, either $\widetilde{\mathcal{H}}$ or $\mathcal{H}$ is proper in a small neighbourhood of 0. In the following proof, we write $\vec{\alpha}_{k_p}$ simply as $\vec{\alpha}$. Denote matrix $\widetilde{\mat{Q}}=(\vec{\alpha}_{k_1},\ldots,\vec{\alpha}_{k_{p-1}},\vec{\alpha}_{k_{p+1}},\ldots,\vec{\alpha}_{k_m})$, $\mat{Q}=(\vec{\alpha}_{k_1},\ldots,\vec{\alpha}_{k_{p-1}},\vec{\alpha}_{k_{p+1}},\ldots,\vec{\alpha}_{k_m},\vec{\alpha})$, $\widetilde{\mat{H}}=\widetilde{\mat{Q}}\left(\widetilde{\mat{Q}}^{\top}\widetilde{\mat{Q}}\right)^{-1}\widetilde{\mat{Q}}^{\top}$ and $\mat{H}=\mat{Q}\left(\mat{Q}^{\top}\mat{Q}\right)^{-1}\mat{Q}^{\top}$. Since $c_{k_p}=0$, we have $\vec{\alpha}^{\top}P_{\widetilde{\mathcal{H}}^{\perp}}\vec{\eta}(0)=0$. Then we have $\vec{\alpha}^{\top}(\mat{I}-\widetilde{\mat{H}})\vec{\eta}(0)=0$.\\[3mm]
        If $\vec{\alpha}^{\top}P_{\widetilde{\mathcal{H}}^{\perp}}(\vec{\eta}_j-\vec{\eta}_i)=0$, then
        \begin{align*}
            \vec{\alpha}^{\top}P_{\widetilde{\mathcal{H}}^{\perp}}\vec{\eta}(\delta)=\vec{\alpha}^{\top}P_{\widetilde{\mathcal{H}}^{\perp}}\vec{\eta}(0)+\delta\vec{\alpha}^{\top}P_{\widetilde{\mathcal{H}}^{\perp}}(\vec{\eta}_j-\vec{\eta}_i)=0,
        \end{align*}
        which implies that the choice of $\mathcal{H}$ is proper for $\vec{\eta}({\delta})$ when $\delta$ is in a neighbourhood of 0, which indicates that the problem should fall into Case 1.\\[3mm]
        If $\vec{\alpha}^{\top}P_{\widetilde{\mathcal{H}}^{\perp}}(\vec{\eta}_j-\vec{\eta}_i)\neq 0$, we assume WLOG that $\vec{\alpha}^{\top}P_{\widetilde{\mathcal{H}}^{\perp}}(\vec{\eta}_j-\vec{\eta}_i)> 0$. Then for $\delta\geq 0$, $\vec{\alpha}^{\top}P_{\widetilde{\mathcal{H}}^{\perp}}\vec{\eta}(\delta)=\vec{\alpha}^{\top}P_{\widetilde{\mathcal{H}}^{\perp}}\vec{\eta}(0)+\delta\vec{\alpha}^{\top}P_{\widetilde{\mathcal{H}}^{\perp}}(\vec{\eta}_j-\vec{\eta}_i)\geq 0$. This implies that for small enough $\delta\geq 0$, $\mathcal{H}_{\vec{\eta}(\delta)}$ can be chosen as $\mathcal{H}$. Then for $\delta\geq 0$ small enough,
        \begin{align}\label{eq_prop_4.2}
            \|P_{\mathcal{H}^{\perp}_{\vec{\eta}({\delta})}}\vec{\eta}({\delta})\|^2&=\left\|P_{\mathcal{H}^{\perp}}\vec{\eta}({\delta})\right\|^2\notag\\
            &=\left\|P_{\mathcal{H}^{\perp}}\vec{\eta}({0})\right\|^2+2\delta \left(P_{\mathcal{H}^{\perp}}\vec{\eta}(0)\right)^{\top}\left(P_{\mathcal{H}^{\perp}}(\vec{\eta}_j-\vec{\eta}_i)\right)+\delta^2\left\|P_{\mathcal{H}^{\perp}}(\vec{\eta}_j-\vec{\eta}_i)\right\|^2.
        \end{align}
        Since $\|P_{\mathcal{H}^{\perp}_{\vec{\eta}({\delta})}}\vec{\eta}({\delta})\|\leq \|P_{\mathcal{H}^{\perp}_{\vec{\eta}({0})}}\vec{\eta}({0})\|=\left\|P_{\mathcal{H}^{\perp}}\vec{\eta}({0})\right\|$, (\ref{eq_prop_4.2}) implies that
        \begin{align}\label{eq_prop_4.3}
            \left(P_{\mathcal{H}^{\perp}}\vec{\eta}(0)\right)^{\top}\left(P_{\mathcal{H}^{\perp}}(\vec{\eta}_j-\vec{\eta}_i)\right)\leq 0.
        \end{align}
        On the other hand, for $\delta<0$, $\vec{\alpha}^{\top}P_{\widetilde{\mathcal{H}}^{\perp}}\vec{\eta}(\delta)=\vec{\alpha}^{\top}P_{\widetilde{\mathcal{H}}^{\perp}}\vec{\eta}(0)+\delta\vec{\alpha}^{\top}P_{\widetilde{\mathcal{H}}^{\perp}}(\vec{\eta}_j-\vec{\eta}_i)< 0$. This implies that for $\delta<0$ sufficiently close to 0, $\mathcal{H}_{\vec{\eta}(\delta)}$ can be chosen as $\widetilde{\mathcal{H}}$. Since $\vec{\alpha}^{\top}(\mat{I}-\widetilde{\mat{H}})\vec{\eta}(0)=0$, we have
        \begin{align}\label{eq_prop_4.4}
            &~~~~\|P_{\mathcal{H}^{\perp}_{\vec{\eta}({\delta})}}\vec{\eta}({\delta})\|^2\notag\\
            &=\left\|P_{\widetilde{\mathcal{H}}^{\perp}}\vec{\eta}({\delta})\right\|^2\notag\\
            &=\left[\vec{\eta}(0)+\delta(\vec{\eta}_j-\vec{\eta}_i)\right]^{\top}(\mat{I}-\widetilde{\mat{H}})\left[\vec{\eta}(0)+\delta(\vec{\eta}_j-\vec{\eta}_i)\right]\notag\\
            &\geq \vec{\eta}(0)^{\top}(\mat{I}-\widetilde{\mat{H}})\vec{\eta}(0)+2\delta ((\mat{I}-\widetilde{\mat{H}})\vec{\eta}(0))^{\top}(\vec{\eta}_j-\vec{\eta}_i)\notag\\
            &=\vec{\eta}(0)^{\top}\left(\mat{I}-\mat{H}+\frac{(\mat{I}-\widetilde{\mat{H}})\vec{\alpha}\vec{\alpha}^{\top}(\mat{I}-\widetilde{\mat{H}})}{\vec{\alpha}^{\top}(\mat{I}-\widetilde{\mat{H}})\vec{\alpha}}\right)\vec{\eta}(0)+2\delta \vec{\eta}(0)^{\top}\left(\mat{I}-\mat{H}+\frac{(\mat{I}-\widetilde{\mat{H}})\vec{\alpha}\vec{\alpha}^{\top}(\mat{I}-\widetilde{\mat{H}})}{\vec{\alpha}^{\top}(\mat{I}-\widetilde{\mat{H}})\vec{\alpha}}\right)(\vec{\eta}_j-\vec{\eta}_i)\notag\\
            &=\vec{\eta}(0)^{\top}\left(\mat{I}-\mat{H}\right)\vec{\eta}(0)+2\delta \vec{\eta}(0)^{\top}\left(\mat{I}-\mat{H}\right)(\vec{\eta}_j-\vec{\eta}_i)\notag\\
            &=\left\|P_{\mathcal{H}^{\perp}}\vec{\eta}(0)\right\|^2+2\delta\left(P_{\mathcal{H}^{\perp}}\vec{\eta}(0)\right)^{\top}(P_{\mathcal{H}^{\perp}}(\vec{\eta}_j-\vec{\eta}_i)).
        \end{align}
        The third-to-last step of (\ref{eq_prop_4.4}) is due to the Sherman-Morrison formula:
        \begin{align*}
            \mat{H}=\widetilde{\mat{H}}+\frac{(\mat{I}-\widetilde{\mat{H}})\vec{\alpha}\vec{\alpha}^{\top}(\mat{I}-\widetilde{\mat{H}})}{\vec{\alpha}^{\top}(\mat{I}-\widetilde{\mat{H}})\vec{\alpha}}.
        \end{align*}
        Since $\|P_{\mathcal{H}^{\perp}_{\vec{\eta}({\delta})}}\vec{\eta}({\delta})\|\leq \|P_{\mathcal{H}^{\perp}_{\vec{\eta}({0})}}\vec{\eta}({0})\|=\left\|P_{\mathcal{H}^{\perp}}\vec{\eta}({0})\right\|$, (\ref{eq_prop_4.4}) implies that
        \begin{align}\label{eq_prop_4.5}
            \left(P_{\mathcal{H}^{\perp}}\vec{\eta}(0)\right)^{\top}P_{\mathcal{H}^{\perp}}(\vec{\eta}_j-\vec{\eta}_i)\geq 0.
        \end{align}
        Combining (\ref{eq_prop_4.3}) and (\ref{eq_prop_4.5}), we have
        \begin{align*}
            (P_{\mathcal{H}^{\perp}_{\vec{\eta}(0)}}\vec{\eta}(0))^{\top}P_{\mathcal{H}^{\perp}_{\vec{\eta}(0)}}(\vec{\eta}_j-\vec{\eta}_i)=(P_{\mathcal{H}^{\perp}}\vec{\eta}(0))^{\top}P_{\mathcal{H}^{\perp}}(\vec{\eta}_j-\vec{\eta}_i)=0.
        \end{align*} 
        Then by (\ref{eq_prop_4.2}) and (\ref{eq_prop_4.4}), it holds that $\|P_{\mathcal{H}^{\perp}_{\vec{\eta}({\delta})}}\vec{\eta}({\delta})\|= \|P_{\mathcal{H}^{\perp}_{\vec{\eta}({0})}}\vec{\eta}({0})\|$ for any $\delta$ in a small neighbourhood of 0. This implies that for $\delta>0$ small enough, we have $(\widetilde{\nu}_1,\ldots,\widetilde{\nu}_i-\delta,\ldots,\widetilde{\nu}_j+\delta,\ldots,\widetilde{\nu}_J)^{\top}\in \Omega$, which contradicts the definition of $M_j$. So $M_j=0$ for $j=2,\ldots,J$, which indicates that the element in $\Omega$ can only be $(1,0,\ldots,0)^{\top}$ since $\Omega$ is nonempty. Hence $\vec{\eta}_1$ is the unique maximiser in $\mathcal{G}$.
    \end{proof}
    
    From the proof of Proposition~\ref{prop_max_projection}, we can also prove Corollary~\ref{cor_max_projection}.
        \renewcommand*{\proofname}{Proof of Corollary \ref{cor_max_projection}}
    \begin{proof}
        Following the proof in Proposition~\ref{prop_max_projection}, since $(1,0,\ldots,0)^{\top}$ is the maximiser, by the arguments in (\ref{eq_prop_4.2}) and (\ref{eq_prop_4.4}), we have $(P_{\mathcal{H}^{\perp}_{\vec{\eta}_1}}\vec{\eta}_1-P_{\mathcal{H}^{\perp}_{\vec{\eta}_1}}\vec{\eta}_j)^{\top}P_{\mathcal{H}^{\perp}_{\vec{\eta}_1}}\vec{\eta}_1>0$ for any $j=2,\ldots,J$, or $(1,0,\ldots,0)^{\top}$ will not be the only element in $\Omega$. Then we have:
            \begin{align*}
                (\vec{\eta}_1-\vec{\eta}_j)^{\top}P_{\mathcal{H}^{\perp}_{\vec{\eta}_1}}\vec{\eta}_1=(P_{\mathcal{H}^{\perp}_{\vec{\eta}_1}}\vec{\eta}_1-P_{\mathcal{H}^{\perp}_{\vec{\eta}_1}}\vec{\eta}_j)^{\top}P_{\mathcal{H}^{\perp}_{\vec{\eta}_1}}\vec{\eta}_1>0,
            \end{align*}
            which completes the proof.
    \end{proof}
    
    \subsection{Proof of Proposition \ref{lem_canonical_expansion}}
            \renewcommand*{\proofname}{Proof of Proposition \ref{lem_canonical_expansion}}
    \begin{proof}
    Define
    \begin{align*}
        f_n(\vec{\theta})=-\sum_{k=1}^K\exp\left(\vec{\alpha}_k^{\top}\vec{\theta}\right)+n\vec{\eta}^{\top}\vec{\theta}-\frac{1}{2}\vec{\theta}^{\top}\vec{\theta}
    \end{align*}
    and denote the unique maximiser of $f_n$ by $\vec{\theta}_n$. Furthermore, suppose $l_n=\left\|\vec{\theta}_n\right\|$, $\vec{\epsilon}_n=\vec{\theta}_n/l_n\rightarrow \vec{\epsilon}$. Following the proof in Proposition~\ref{prop_canonical_projection}, we have $\underset{k=1,\ldots,K}{\max}\vec{\alpha}_k^{\top}\vec{\epsilon}>0$. Then by the proof in Proposition~\ref{prop_characterisation_equation}, there exists $\left\{\vec{\alpha}_{k_1},\ldots,\vec{\alpha}_{k_m}\right\}\subseteq\left\{\vec{\alpha}_1,\ldots,\vec{\alpha}_K\right\}$ and positive constants $c_{k_1},\ldots,c_{k_m}>0$ such that $\vec{\eta}=\sum_{p=1}^m c_{k_p}\vec{\alpha}_{k_p}$. Furthermore, it holds that $\vec{\alpha}_{k_1}^{\top}\vec{\epsilon}=\ldots=\vec{\alpha}_{k_m}^{\top}\vec{\epsilon}=\underset{k=1,\ldots,K}{\max}\vec{\alpha}_k^{\top}\vec{\epsilon}$. So the existence of canonical expansion is proved.\\[3mm]
    If there exist two canonical expansions:
    \begin{align*}
        \vec{\eta}=\sum_{p=1}^{m_1} c_{k_p}\vec{\alpha}_{k_p}=\sum_{p=1}^{m_2} \widetilde{c}_{l_p}\vec{\alpha}_{l_p},
    \end{align*}
    where $c_{k_1},\ldots,c_{k_{m_1}},\widetilde{c}_{l_1},\ldots,\widetilde{c}_{l_{m_2}}>0$ with $\vec{\epsilon}$ and $\widetilde{\vec{\epsilon}}$ satisfying the condition. Then we have
    \begin{align*}
        \vec{\eta}^{\top}\vec{\epsilon}&=\sum_{p=1}^{m_1} c_{k_p}\vec{\alpha}_{k_p}^{\top}\vec{\epsilon}=\left(\sum_{p=1}^{m_1} c_{k_p}\right)\underset{k=1,\ldots,K}{\max}\vec{\alpha}_k^{\top}\vec{\epsilon},\notag\\
        \vec{\eta}^{\top}\vec{\epsilon}&=\sum_{p=1}^{m_2} \widetilde{c}_{l_p}\vec{\alpha}_{l_p}^{\top}\vec{\epsilon}\leq\left(\sum_{p=1}^{m_2} 
        \widetilde{c}_{l_p}\right)\underset{k=1,\ldots,K}{\max}\vec{\alpha}_k^{\top}\vec{\epsilon},
    \end{align*}
    which implies that $\sum_{p=1}^{m_1} c_{k_p}\leq \sum_{p=1}^{m_2} \widetilde{c}_{l_p}$. Similarly we have
    \begin{align*}
        \vec{\eta}^{\top}\widetilde{\vec{\epsilon}}&=\sum_{p=1}^{m_1} c_{k_p}\vec{\alpha}_{k_p}^{\top}\widetilde{\vec{\epsilon}}\leq\left(\sum_{p=1}^{m_1} c_{k_p}\right)\underset{k=1,\ldots,K}{\max}\vec{\alpha}_k^{\top}\widetilde{\vec{\epsilon}},\notag\\
        \vec{\eta}^{\top}\widetilde{\vec{\epsilon}}&=\sum_{p=1}^{m_2} \widetilde{c}_{l_p}\vec{\alpha}_{l_p}^{\top}\widetilde{\vec{\epsilon}}=\left(\sum_{p=1}^{m_2} \widetilde{c}_{l_p}\right)\underset{k=1,\ldots,K}{\max}\vec{\alpha}_k^{\top}\widetilde{\vec{\epsilon}},
    \end{align*}
    which implies that $\sum_{p=1}^{m_1} c_{k_p}\geq \sum_{p=1}^{m_2} \widetilde{c}_{l_p}$. So we have $\sum_{p=1}^{m_1} c_{k_p}=\sum_{p=1}^{m_2} \widetilde{c}_{l_p}$.
    \end{proof}
    
    \subsection{Proof of Proposition \ref{lem_characterisation_log}}
    \renewcommand*{\proofname}{Proof of Proposition \ref{lem_characterisation_log}}
    \begin{proof}
    By the given conditions, $\hat{\vec{\theta}}\in \mathbb{R}^d$ is the unique solution of the linear equation:
    \begin{align*}
        \vec{\alpha}_k^{\top}\vec{\theta}=\begin{cases}
            \nu_j&\text{ if }\vec{\alpha}_k\in \mat{U}_j\\
            0&\text{ if }\vec{\alpha}_k\in \mat{V}_0
        \end{cases}
    \end{align*}
    with the constraint: $\vec{\theta}-\sum_{j=1}^m\hat{c}_{j}\vec{\varphi}_j\in \mathcal{H}$. By continuity, for any $\vec{c}$ in a small neighbourhood of $\hat{\vec{c}}$, the unique solution to the above linear equation with constraint: $\vec{\theta}-\sum_{j=1}^m{c}_{j}\vec{\varphi}_j\in \mathcal{H}$ still satisfies Conditions (i)-(iv) in Proposition~\ref{lem_characterisation_log}. We denote this unique solution by $\vec{\theta}_{\vec{c}}$. Then for any ${\bm{\zeta}}^{(n,\vec{c})}=(\zeta_1^{(n,\vec{c})},\ldots,\zeta_m^{(n,\vec{c})})^{\top}$ such that $\lim_{n\rightarrow \infty}{\bm{\zeta}}^{(n,\vec{c})}/\log n=\vec{c}$, let $\widetilde{\vec{\theta}}_{n,\vec{c}}$ be the unique solution of the following linear equation satisfying:
    \begin{enumerate}
        \item[(i)] $\widetilde{\vec{\theta}}_{n,\vec{c}}-\log n\sum_{j=1}^m{c}_{j}\vec{\varphi}_j\in\mathcal{H}$.
        \item[(ii)] For $j=1,\ldots,J$ and any $\vec{\alpha}_k\in \mat{U}_j$, it holds that $\vec{\alpha}_k^{\top}\widetilde{\vec{\theta}}_{n,\vec{c}}=\log (d_k\xi_j^{(n)}-b_k\log n)-\log w_k$, where $b_{k}$ is the coefficient of $\vec{\alpha}_k$ in the expansion of $\vec{\theta}_{\vec{c}}-\sum_{j=1}^m{c}_{j}\vec{\varphi}_j$ under basis $\mat{U}_1\cup\ldots\cup \mat{U}_J\cup \mat{V}_{0}$.
        \item[(iii)] For any $\vec{\alpha}_k\in \mat{V}_0$, it holds that $\vec{\alpha}_k^{\top}\widetilde{\vec{\theta}}_{n,\vec{c}}=\log \left(-b_k\log n\right)-\log w_k$, where $b_k<0$ is the coefficient of $\vec{\alpha}_k$ in the expansion of $\vec{\theta}_{\vec{c}}-\sum_{j=1}^m{c}_{j}\vec{\varphi}_j$ under basis $\mat{U}_1\cup\ldots\cup \mat{U}_J\cup \mat{V}_{0}$.
    \end{enumerate}
    We can easily prove that $\widetilde{\vec{\theta}}_{n,\vec{c}}/\log n\rightarrow \vec{\theta}_{\vec{c}}$. Then we plug $\widetilde{\vec{\theta}}_{n,\vec{c}}$ into the gradient of $f_n(\cdot|\bm{\xi}^{(n)},{\bm{\zeta}}^{(n,\vec{c})})$:
    \begin{align}\label{eq_lem_7.1}
        \notag&\nabla f_n(\widetilde{\vec{\theta}}_{n,\vec{c}}|\bm{\xi}^{(n)},{\bm{\zeta}}^{(n,\vec{c})})\notag\\
        =&-\sum_{\vec{\alpha}_k\in \mat{V}_{-}}w_k\exp(\vec{\alpha}_k^{\top}\widetilde{\vec{\theta}}_{n,\vec{c}})\vec{\alpha}_k-\sum_{\vec{\alpha}_k\in \mat{V}_{0}}w_k\exp(\vec{\alpha}_k^{\top}\widetilde{\vec{\theta}}_{n,\vec{c}})\vec{\alpha}_k-\sum_{j=1}^J \sum_{\vec{\alpha}_k\in \mat{U}_j}w_k\exp(\vec{\alpha}_k^{\top}\widetilde{\vec{\theta}}_{n,\vec{c}})\vec{\alpha}_k\notag\\
        &+\sum_{j=1}^J\sum_{\vec{\alpha}_k\in \mat{U}_j}d_k\xi_j^{(n)}\vec{\alpha}_k+\sum_{j=1}^m{\zeta}_j^{(n,\vec{c})}\vec{\varphi}_j-\widetilde{\vec{\theta}}_{n,\vec{c}}\notag\\
        =&o(\log n)+\log n\left(\sum_{\vec{\alpha}_k\in \mat{V}_0}b_k\vec{\alpha}_k+\sum_{j=1}^J\sum_{\vec{\alpha}_k\in \mat{U}_j}b_k\vec{\alpha}_k+\sum_{j=1}^mc_{j}\vec{\varphi}_j\right)-\widetilde{\vec{\theta}}_{n,\vec{c}}\notag\\
        =&o(\log n)+\log n \cdot\vec{\theta}_{\vec{c}}-\widetilde{\vec{\theta}}_{n,\vec{c}}\notag\\
        =&o(\log n).
    \end{align}
    Similar to the proof in Lemma~\ref{lem_characterisation_equation_2}, we have $\|\vec{\theta}_n(\bm{\xi}^{(n)},{\bm{\zeta}}^{(n,\vec{c})})-\widetilde{\vec{\theta}}_{n,\vec{c}}\|=o(\log n)$. Then by (\ref{eq_lem_7.1}) we have
    \begin{align}\label{eq_lem_7.2}
        0\leq& f_n({\vec{\theta}}_n(\bm{\xi}^{(n)},{\bm{\zeta}}^{(n,\vec{c})})|\bm{\xi}^{(n)},{\bm{\zeta}}^{(n,\vec{c})})-f_n(\widetilde{\vec{\theta}}_{n,\vec{c}}|\bm{\xi}^{(n)},{\bm{\zeta}}^{(n,\vec{c})})\notag\\
        \leq& (\nabla f_n(\widetilde{\vec{\theta}}_{n,\vec{c}}|\bm{\xi}^{(n)},{\bm{\zeta}}^{(n,\vec{c})}))^{\top}({\vec{\theta}}_n(\bm{\xi}^{(n)},{\bm{\zeta}}^{(n,\vec{c})})-\widetilde{\vec{\theta}}_{n,\vec{c}})\notag\\
        =&o(\log^2 n).
    \end{align}
    We assume WLOG that $\mat{U}_1=\{\vec{\alpha}_1,\ldots,\vec{\alpha}_{p_1}\},\ldots,\mat{U}_J=\{\vec{\alpha}_{p_{J-1}+1},\ldots,\vec{\alpha}_{p_J}\}$ and $\mat{V}_0=\{\vec{\alpha}_{p_J+1},\ldots,\vec{\alpha}_p\}$. Denote $\mat{X}=(\vec{\alpha}_1,\ldots,\vec{\alpha}_p)$ and denote $\vec{\beta}=\big(\nu_1\cdot\mathbf{1}^{\top}_{p_1},\nu_2\cdot\mathbf{1}^{\top}_{p_2-p_1},\ldots,\nu_J\cdot\mathbf{1}^{\top}_{p_J-p_{J-1}},0\cdot\mathbf{1}^{\top}_{p-p_J}\big)^{\top}$. Since $\vec{\theta}_{\vec{c}}-\sum_{j=1}^m{c}_{j}\vec{\varphi}_j\in \mathcal{H}$, suppose that $\vec{\theta}_{\vec{c}}-\sum_{j=1}^m{c}_{j}\vec{\varphi}_j=\mat{X}\vec{\alpha}$  for $\vec{\alpha}\in\mathbb{R}^p$, then the following linear equation holds:
    \begin{align*}
        \vec{\beta}=\mat{X}^{\top}\vec{\theta}_{\vec{c}}=\mat{X}^{\top}\left(\mat{X}\vec{\alpha}+\sum_{j=1}^m{c}_{j}\vec{\varphi}_j\right),
    \end{align*}
    which implies that $\vec{\alpha}=(\mat{X}^{\top}\mat{X})^{-1}\left(\vec{\beta}-\sum_{j=1}^m{c}_{j}\mat{X}^{\top}\vec{\varphi}_j\right)$. So we have 
    \begin{align*}
        \vec{\theta}_{\vec{c}}=\sum_{j=1}^m{c}_{j}\vec{\varphi}_j+\mat{X}(\mat{X}^{\top}\mat{X})^{-1}\left(\vec{\beta}-\sum_{j=1}^m{c}_{j}\mat{X}^{\top}\vec{\varphi}_j\right):= \widetilde{\vec{\theta}}+\sum_{j=1}^m{c}_{j}P_{\mathcal{H}^{\perp}}\vec{\varphi}_j.
    \end{align*}
    Here $\widetilde{\vec{\theta}}=\mat{X}(\mat{X}^{\top}\mat{X})^{-1}\vec{\beta}$ does not depend on $\vec{c}$. Then we have
    \begin{align}\label{eq_lem_7.3}
        &\left(\sum_{j=1}^m{c}_{j}\vec{\varphi}_j\right)^{\top}\vec{\theta}_{\vec{c}}-\frac{1}{2}\vec{\theta}_{\vec{c}}^{\top}\vec{\theta}_{\vec{c}}\notag\\
        =&-\frac{1}{2}\left(\sum_{j=1}^m{c}_{j}\vec{\varphi}_j+\mat{X}(\mat{X}^{\top}\mat{X})^{-1}\left(\vec{\beta}-\sum_{j=1}^m{c}_{j}\mat{X}^{\top}\vec{\varphi}_j\right)\right)^{\top}\left(-\sum_{j=1}^m{c}_{j}\vec{\varphi}_j+\mat{X}(\mat{X}^{\top}\mat{X})^{-1}\left(\vec{\beta}-\sum_{j=1}^m{c}_{j}\mat{X}^{\top}\vec{\varphi}_j\right)\right)\notag\\
        =&\frac{1}{2}\left\|\sum_{j=1}^m{c}_{j}P_{\mathcal{H}^{\perp}}\vec{\varphi}_j\right\|^2-\frac{1}{2}\vec{\beta}^{\top}(\mat{X}^{\top}\mat{X})^{-1}\vec{\beta}+\vec{\beta}^{\top}(\mat{X}^{\top}\mat{X})^{-1}\sum_{j=1}^m{c}_{j}\mat{X}^{\top}\vec{\varphi}_j.
    \end{align}
    Hence by the definition of $\widetilde{\vec{\theta}}_{n,\vec{c}}$, (\ref{eq_lem_7.2}) and (\ref{eq_lem_7.3}), we have 
    \begin{align*}
        &f_n(\vec{\theta}_n(\bm{\xi}^{(n)},{\bm{\zeta}}^{(n,\vec{c})})|\bm{\xi}^{(n)},{\bm{\zeta}}^{(n,\vec{c})})\\
        =&o(\log^2 n)-\sum_{\vec{\alpha}_k\in \mat{V}_{-}}w_k\exp(\vec{\alpha}_k^{\top}\widetilde{\vec{\theta}}_{n,\vec{c}})-\sum_{\vec{\alpha}_k\in \mat{V}_{0}}w_k\exp(\vec{\alpha}_k^{\top}\widetilde{\vec{\theta}}_{n,\vec{c}})-\sum_{j=1}^J \sum_{\vec{\alpha}_k\in \mat{U}_j}w_k\exp(\vec{\alpha}_k^{\top}\widetilde{\vec{\theta}}_{n,\vec{c}})\\
       &+\sum_{j=1}^J\xi_{j}^{(n)}\sum_{\vec{\alpha}_k\in \mat{U}_j}d_k\big[\log (d_k\xi_{j}^{(n)}-b_k\log n)-\log w_k\big]+\sum_{j=1}^m\zeta_j^{(n,\vec{c})}\vec{\varphi}_j^{\top}\widetilde{\vec{\theta}}_{n,\vec{c}}-\frac{1}{2}\widetilde{\vec{\theta}}_{n,\vec{c}}^{\top}\widetilde{\vec{\theta}}_{n,\vec{c}}\\
        =&o(\log^2 n)+\sum_{j=1}^J\xi_j^{(n)}\sum_{\vec{\alpha}_k\in \mat{U}_j}\left(-d_k+d_k\log (d_k \xi_j^{(n)})-d_k\log w_k\right)\\
        &+\log^2 n\left(\left(\sum_{j=1}^m{c}_{j}\vec{\varphi}_j\right)^{\top}\vec{\theta}_{\vec{c}}-\frac{1}{2}\vec{\theta}_{\vec{c}}^{\top}\vec{\theta}_{\vec{c}}\right)\\
        :=& e_{n,1}+\log^2 n\left(\vec{c}^{\top}\vec{e}_{2}+\frac{1}{2}\left\|\sum_{j=1}^m{c}_{j}P_{\mathcal{H}^{\perp}}\vec{\varphi}_j\right\|^2+o(1)\right),
    \end{align*}
    where $e_{n,1},\vec{e}_2$ do not depend on $\vec{c}$.
    \end{proof}
        \subsection{Proof of Proposition \ref{prop_characterisation_equation}}
        \noindent To verify Proposition~\ref{prop_characterisation_equation}, we first prove the following three lemmas.
        \begin{lem}\label{lem_characterisation_equation_1}
            Let $\vec{\alpha}_1,\ldots,\vec{\alpha}_K\in\mathbb{R}^d$ be $d$-vectors and $c_1,\ldots,c_K$ be nonnegative constants. Let $\vec{\xi}=\sum_{k=1}^K c_k\vec{\alpha}_k$. Then there exists $\{\vec{\alpha}_{k_1},\ldots,\vec{\alpha}_{k_m}\}\subseteq \{\vec{\alpha}_1,\ldots,\vec{\alpha}_K\}$ such that $\vec{\xi}=\sum_{p=1}^m \widetilde{c}_{k_p}\vec{\alpha}_{k_p}$, where $\widetilde{c}_{k_1},\ldots,\widetilde{c}_{k_m}$ are positive constants and $\vec{\alpha}_{k_1},\ldots,\vec{\alpha}_{k_m}$ are linearly independent.
        \end{lem}
        \begin{lem}\label{lem_characterisation_equation_2}
            Let $\vec{\alpha}_1,\ldots,\vec{\alpha}_K\in\mathbb{R}^d\setminus\{\vec{0}\}$ be distinct $d$-vectors and let $c_1,\ldots,c_M> 0$ be positive constants. Then the vector $\hat{\vec{\theta}}\in\mathbb{R}^d$ that satisfies the following condition is unique if it exists:
            \begin{enumerate}
                \item[(i)] For $k=1,\ldots,M$, it holds that $\vec{\alpha}_k^{\top}\hat{\vec{\theta}}=c_k$.
                \item[(ii)] There exists $\{\vec{\alpha}_{j_1},\ldots,\vec{\alpha}_{j_p}\}\subseteq\{\vec{\alpha}_{M+1},\ldots,\vec{\alpha}_{K}\}$ such that
                \begin{enumerate}
                    \item[(a)] $\{\vec{\alpha}_1,\ldots,\vec{\alpha}_M,\vec{\alpha}_{j_1},\ldots,\vec{\alpha}_{j_p}\}$ are linearly independent.
                    \item[(b)] $\hat{\vec{\theta}}\in \operatorname{span}\{\vec{\alpha}_1,\ldots,\vec{\alpha}_M,\vec{\alpha}_{j_1},\ldots,\vec{\alpha}_{j_p}\}$. For any $m=1,\ldots,p$, the coefficient of $\vec{\alpha}_{j_m}$ in the expansion of $\hat{\vec{\theta}}$ under basis $\{\vec{\alpha}_1,\ldots,\vec{\alpha}_M,\vec{\alpha}_{j_1},\ldots,\vec{\alpha}_{j_p}\}$ is negative.
                    \item[(c)] For $\vec{\alpha}\in \{\vec{\alpha}_{j_1},\ldots,\vec{\alpha}_{j_p}\}$, it holds that $\vec{\alpha}^{\top}\hat{\vec{\theta}}=0$.
                    \item[(d)] For $\vec{\alpha}\in \{\vec{\alpha}_{M+1},\ldots,\vec{\alpha}_{K}\}\setminus\{\vec{\alpha}_{j_1},\ldots,\vec{\alpha}_{j_p}\}$, it holds that $\vec{\alpha}^{\top}\hat{\vec{\theta}}<0$.
                \end{enumerate}
            \end{enumerate}
        \end{lem}
        \begin{lem}\label{lem_characterisation_equation_3}
            Let $\vec{\alpha}_1,\ldots,\vec{\alpha}_K,\vec{\eta}_1,\ldots,\vec{\eta}_J\in\mathbb{R}^d\setminus\{\vec{0}\}$ be $d$-vectors, $w_1,\ldots,w_K$ and $1\geq \widetilde{\nu}_1>\ldots> \widetilde{\nu}_J>0$ be positive constants. Suppose $\vec{\eta}_1,\ldots,\vec{\eta}_J\in \mathcal{X}:=\{\sum_{k=1}^K c_k\vec{\alpha}_k:c_1,\ldots,c_K\geq 0\}$. Then we can define continuous $\vec{\theta}(\nu_1,\ldots,\nu_J)$ in a neighbourhood $\mathcal{O}$ of $(\widetilde{\nu}_1,\ldots,\widetilde{\nu}_J)^{\top}$ such that for any $(\nu_1,\ldots,\nu_J)^{\top}\in\mathcal{O}$, any $(\xi_1^{(n)},\ldots,\xi_J^{(n)})^{\top}$ satisfying
    \begin{align*}
        \lim_{n\rightarrow \infty}\frac{(\log \xi_1^{(n)},\ldots,\log \xi_J^{(n)})}{\log n}=(\nu_1,\ldots,\nu_J),
    \end{align*}
    the unique maximiser $\vec{\theta}_n$ of the following function:
    \begin{align*}
        f_n(\vec{\theta})=-\sum_{k=1}^Kw_k\exp(\vec{\alpha}_k^{\top}\vec{\theta})+\left(\sum_{j=1}^J\xi_j^{(n)}\vec{\eta}_j\right)^{\top}\vec{\theta}-\frac{1}{2}\vec{\theta}^{\top}\vec{\theta}
    \end{align*}
    satisfies the following convergence result:
    \begin{align*}
        \lim_{n\rightarrow \infty}\frac{\vec{\theta}_n}{\log n}=\vec{\theta}(\nu_1,\ldots,\nu_J).
    \end{align*}
    \end{lem}
    
        \renewcommand*{\proofname}{Proof of Lemma \ref{lem_characterisation_equation_1}}
    \begin{proof}
        We assume WLOG that $c_1,\ldots,c_K>0$. If $\vec{\alpha}_1,\ldots,\vec{\alpha}_K$ are linearly independent, then the result is proved. If not, then there exists $1\leq k_1<\ldots<k_m\leq K$ and nonzero constants $b_{k_1},\ldots,b_{k_m}$ such that
            \begin{align*}
                \sum_{p=1}^m b_{k_p}\vec{\alpha}_{k_p}=0.
            \end{align*}
            We assume WLOG that $c_{k_1}/{b_{k_1}}\leq\ldots\leq c_{k_m}/{b_{k_m}}$ and divide the problem into two cases:\\[3mm]
            \textbf{Case 1: }If $0<c_{k_1}/{b_{k_1}}\leq\ldots\leq c_{k_m}/{b_{k_m}}$, then $b_{k_1},\ldots,b_{k_m}>0$. We expand $\vec{\alpha}_{k_1}$ in terms of $\vec{\alpha}_{k_2},\ldots,\vec{\alpha}_{k_m}$ and obtain
            \begin{align*}
                \sum_{p=1}^m {c}_{k_p}\vec{\alpha}_{k_p}&=\sum_{p=2}^m {c}_{k_p}\vec{\alpha}_{k_p}-\frac{c_{k_1}}{b_{k_1}}\left(\sum_{p=2}^mb_{k_p}\vec{\alpha}_{k_p}\right)\\
                &=\sum_{p=2}^m b_{k_p}\left(\frac{c_{k_p}}{b_{k_p}}-\frac{c_{k_1}}{b_{k_1}}\right)\vec{\alpha}_{k_p},
            \end{align*}
            where $b_{k_p}(\frac{c_{k_p}}{b_{k_p}}-\frac{c_{k_1}}{b_{k_1}})\geq 0$ for $p=2,\ldots,m$.\\[3mm]
            \textbf{Case 2: }If there exists $1\leq q\leq m$ such that $c_{k_q}/b_{k_q}<0<c_{k_{q+1}}/b_{k_{q+1}}$ (If all terms are negative, then $q=m$). We expand $\vec{\alpha}_{k_q}$ in terms of $\vec{\alpha}_{k_1},\ldots,\vec{\alpha}_{k_{q-1}},\vec{\alpha}_{k_{q+1}},\ldots,\vec{\alpha}_{k_m}$ and obtain
            \begin{align*}
                \sum_{p=1}^m {c}_{k_p}\vec{\alpha}_{k_p}&=\sum_{p\neq q} {c}_{k_p}\vec{\alpha}_{k_p}-\frac{c_{k_q}}{b_{k_q}}\left(\sum_{p\neq q}b_{k_p}\vec{\alpha}_{k_p}\right)\\
                &=\sum_{p\neq q} c_{k_p}\frac{\left(\frac{c_{k_p}}{b_{k_p}}\right)-\left(\frac{c_{k_q}}{b_{k_q}}\right)}{\left(\frac{c_{k_p}}{b_{k_p}}\right)}\vec{\alpha}_{k_p}.
            \end{align*}
            It can be readily shown that for any $p\neq q$, it holds that $[(\frac{c_{k_p}}{b_{k_p}})-(\frac{c_{k_q}}{b_{k_q}})]/(\frac{c_{k_p}}{b_{k_p}})\geq 0$. So in either case, we can expand $\vec{\xi}$ by at most $K-1$ vectors chosen from $\{\vec{\alpha}_1,\ldots,\vec{\alpha}_K\}$ with positive coefficients. This implies that we can continue the procedure and it will end over finite steps. Then the final remaining vectors and coefficients satisfy the condition.
    \end{proof}
    
    \renewcommand*{\proofname}{Proof of Lemma \ref{lem_characterisation_equation_2}}
    \begin{proof}
        Define
            \begin{align*}
                f_n(\vec{\theta})=-\sum_{k=1}^K\exp(\vec{\alpha}_k^{\top}\vec{\theta})+\left(\sum_{k=1}^M n^{c_k}\vec{\alpha}_k\right)^{\top}\vec{\theta}-\frac{1}{2}\vec{\theta}^{\top}\vec{\theta}
            \end{align*}
            and denote the unique maximiser of $f_n$ by $\vec{\theta}_n$. By arguments similar to those used in Proposition~\ref{prop_canonical_projection}, we can prove that $\vec{\theta}_n=\mathcal{O}(\log n)$.\\[3mm]
            If $\hat{\vec{\theta}}\in\mathbb{R}^d$ satisfies all the conditions, then we let $\widetilde{\vec{\theta}}_n$ be the unique solution of the following system:
            \begin{enumerate}
                \item[(i)] $\widetilde{\vec{\theta}}_n\in \operatorname{span}\{\vec{\alpha}_1,\ldots,\vec{\alpha}_M,\vec{\alpha}_{j_1},\ldots,\vec{\alpha}_{j_p}\}$.
                \item[(ii)] For $k=1,\ldots,M$, let $\vec{\alpha}_k^{\top}\widetilde{\vec{\theta}}_n=\log (n^{c_k}-b_k\log n)=\log (n^{\vec{\alpha}_k^{\top}\hat{\vec{\theta}}}-b_k\log n)$, where $b_k$ is the coefficient of $\vec{\alpha}_{k}$ in the expansion of $\hat{\vec{\theta}}$ under basis $\{\vec{\alpha}_1,\ldots,\vec{\alpha}_M,\vec{\alpha}_{j_1},\ldots,\vec{\alpha}_{j_p}\}$.
                \item[(iii)] For $m=1,\ldots,p$, let $\vec{\alpha}_{j_m}^{\top}\widetilde{\vec{\theta}}_n=\log (-\zeta_m\log n)$, where $\zeta_{m}$ is the coefficient of $\vec{\alpha}_{j_m}$ in the expansion of $\hat{\vec{\theta}}$ under basis $\{\vec{\alpha}_1,\ldots,\vec{\alpha}_M,\vec{\alpha}_{j_1},\ldots,\vec{\alpha}_{j_p}\}$.
            \end{enumerate}
            Then it can be readily shown that $\widetilde{\vec{\theta}}_n/\log n\rightarrow \hat{\vec{\theta}}$. Now we plug $\widetilde{\vec{\theta}}_n$ into the gradient of $f_n$:
            \begin{align}\label{eq_lem_3.1}
                \nabla f_n(\widetilde{\vec{\theta}}_n)=&-\sum_{k=1}^M(n^{c_k}-b_k\log n)\vec{\alpha}_k+\sum_{m=1}^p\zeta_m\log n~\vec{\alpha}_{j_m}+\sum_{k=1}^Mn^{c_k}\vec{\alpha}_k\notag\\
                &~~~~~~-\log n\left[\sum_{k=1}^M
                b_k\vec{\alpha}_k+\sum_{m=1}^p\zeta_m\vec{\alpha}_{j_m}\right]-(\widetilde{\vec{\theta}}_n-\log n~\hat{\vec{\theta}})\notag\\
                =&-(\widetilde{\vec{\theta}}_n-\log n~\hat{\vec{\theta}})\notag\\
                =&o(\log n).
            \end{align}
            By Taylor expansion, we have
            \begin{align}\label{eq_lem_3.2}
                \vec{0}=\nabla f_n({\vec{\theta}}_n)=\nabla f_n(\widetilde{\vec{\theta}}_n)+\nabla^2f_n(\vec{\theta}_n^{\ast})({\vec{\theta}}_n-\widetilde{\vec{\theta}}_n).
            \end{align}
            Since $-\nabla^2f_n(\vec{\theta}_n^{\ast})\succeq \mat{I}_d$, (\ref{eq_lem_3.1}) and (\ref{eq_lem_3.2}) indicate that
            \begin{align}\label{eq_lem_3.3}
                \|{\vec{\theta}}_n-\widetilde{\vec{\theta}}_n\|=\|(\nabla^2f_n(\vec{\theta}_n^{\ast}))^{-1}\nabla f_n(\widetilde{\vec{\theta}}_n)\|\leq \|\nabla f_n(\widetilde{\vec{\theta}}_n)\|=o(\log n).
            \end{align}
            Since we have $\widetilde{\vec{\theta}}_n/\log n\rightarrow \hat{\vec{\theta}}$, (\ref{eq_lem_3.3}) implies that $\vec{\theta}_n/\log n\rightarrow \hat{\vec{\theta}}$. Since the maximiser $\vec{\theta}_n$ is unique, $\hat{\vec{\theta}}$ that satisfies the conditions is also unique if it exists.
    \end{proof}
    
    \renewcommand*{\proofname}{Proof of Lemma \ref{lem_characterisation_equation_3}}
    \begin{proof}
            We divide the proof into four steps. The sketch is as follows:
            \begin{enumerate}
                \item[(1)] In Step 1, we prove that $\vec{\theta}_n=\mathcal{O}(\log n)$ and define $\lim_{n\rightarrow \infty}\vec{\theta}_n/\log n:=\vec{\theta}(\nu_1,\ldots,\nu_J)$.
                \item[(2)] In Step 2, we introduce the concept of ``characterisation equation'' and prove that we can construct a characterisation equation at $(\nu_1,\ldots,\nu_J)^{\top}$ which has a unique solution $\vec{\theta}(\nu_1,\ldots,\nu_J)$.
                \item[(3)] In Step 3, we prove that the characterisation equation at $(\nu_1,\ldots,\nu_J)^{\top}$ is unique.
                \item[(4)] In Step 4, we prove the lemma by the uniqueness of characterisation equation.
            \end{enumerate}
            \textbf{Step 1:} Denote $l_n=\|\vec{\theta}_n\|$ and $\vec{\epsilon}_n=\vec{\theta}_n/l_n\rightarrow \vec{\epsilon}$. Following the proof in Proposition~\ref{prop_canonical_projection}, it holds that $l_n\rightarrow \infty$. We first prove that ${\max}_{k=1,\ldots,K}\vec{\alpha}_k^{\top}\vec{\epsilon}>0$.\\[3mm]
            If this is not the case, we multiply both sides of the first-order equation for $\vec{\theta}_n$ by $\vec{\epsilon}$:
            \begin{align}\label{eq_lem_4.1}
                -\sum_{k:\vec{\alpha}_k^{\top}\vec{\epsilon}<0}w_k\exp(l_n\vec{\alpha}_k^{\top}\vec{\epsilon}_n)\vec{\alpha}_k^{\top}\vec{\epsilon}+\left(\sum_{j=1}^Jn^{\nu_j}\vec{\eta}_j\right)^{\top}\vec{\epsilon}=l_n\vec{\epsilon}_n^{\top}\vec{\epsilon}.
            \end{align}
            Since $\vec{\eta}_1,\ldots,\vec{\eta}_J\in \mathcal{X}$ and ${\max}_{k=1,\ldots,K}\vec{\alpha}_k^{\top}\vec{\epsilon}\leq 0$, we have $\vec{\eta}_1^{\top}\vec{\epsilon},\ldots,\vec{\eta}_J^{\top}\vec{\epsilon}\leq 0$. Since $0
            <\nu_J<\ldots<\nu_1$, we have $\left[\sum_{j=1}^Jn^{\nu_j}\vec{\eta}_j\right]^{\top}\vec{\epsilon}\leq 0$ for $n$ large enough. On the other hand, we have
            \begin{align*}
                -\sum_{k:\vec{\alpha}_k^{\top}\vec{\epsilon}<0}w_k\exp(l_n\vec{\alpha}_k^{\top}\vec{\epsilon}_n)\vec{\alpha}_k^{\top}\vec{\epsilon}\rightarrow 0\quad\text{and}\quad l_n\vec{\epsilon}_n^{\top}\vec{\epsilon}\rightarrow\infty,
            \end{align*}
            which contradicts (\ref{eq_lem_4.1}). So $\underset{k=1,\ldots,K}{\max}\vec{\alpha}_k^{\top}\vec{\epsilon}>0$. Then following the proof in Proposition~\ref{prop_canonical_projection}, we have
            \begin{align*}
                \lim_{n\rightarrow \infty}\frac{l_n}{\log n}=\frac{\nu_1}{{\max}_{k=1,\ldots,K}\vec{\alpha}_k^{\top}\vec{\epsilon}}.
            \end{align*}
            So we have
            \begin{align*}
                \lim_{n\rightarrow \infty}\frac{\vec{\theta}_n}{\log n}=\lim_{n\rightarrow \infty}\frac{l_n}{\log n}\vec{\epsilon}_n=\frac{\nu_1}{{\max}_{k=1,\ldots,K}\vec{\alpha}_k^{\top}\vec{\epsilon}}\vec{\epsilon}:= \vec{\theta}(\nu_1,\ldots,\nu_J).
            \end{align*}
            \textbf{Step 2: }Recall the definition of characterisation equation: For $k\in\left\{1,\ldots,d\right\}$, we call a set of $k$ linear equations a characterisation equation at $\left(\nu_1,\ldots,\nu_J\right)$ where $\nu_1>\ldots>\nu_J>0$ if the $l$-th $(l=1,\ldots,k)$ equation is of one of the following two types:
            \begin{itemize}
                \item Type-1 equation: $\vec{\alpha}_{j_l}^{\top}\vec{\theta}=\xi_l$, where $\xi_l\in\left\{\nu_1,\ldots,\nu_J,0\right\}$.
                \item Type-2 equation: $\vec{\alpha}_{j_l}^{\top}\vec{\theta}=\vec{\zeta}_l^{\top}\vec{\theta}$, where $\vec{\zeta}_l\in\left\{\vec{\alpha}_1,\ldots,\vec{\alpha}_K\right\}\setminus\left\{\vec{\alpha}_{j_1},\ldots,\vec{\alpha}_{j_k}\right\}$.
            \end{itemize}
            Here $\left\{\vec{\alpha}_{j_1},\ldots,\vec{\alpha}_{j_k}\right\}\subseteq\left\{\vec{\alpha}_1,\ldots,\vec{\alpha}_K\right\}$ is a set of $k$ vectors. Furthermore, the characterisation equation is required to satisfy the following conditions:
            \begin{enumerate}
                \item[(i)] $\vec{\alpha}_{j_1},\ldots,\vec{\alpha}_{j_k}$ are linearly independent.
                \item[(ii)] The equation has a unique solution $\hat{\vec{\theta}}$ under constraint: $\vec{\theta}\in \operatorname{span}\left\{\vec{\alpha}_{j_1},\ldots,\vec{\alpha}_{j_k}\right\}$.
                \item[(iii)] For $j=1,\ldots,J$, there exists a unique $(c_{j,1},\ldots,c_{j,k})^{\top}\in\mathbb{R}^k$ such that $\vec{\eta}_j$ has expansion: $\vec{\eta}_j=\sum_{l=1}^kc_{j,l}\vec{\alpha}_{j_l}$. Suppose $\vec{\alpha}_{j_{l_1}},\ldots,\vec{\alpha}_{j_{l_m}}$ are all elements in $\left\{\vec{\alpha}_{j_1},\ldots,\vec{\alpha}_{j_k}\right\}$ such that $\vec{\alpha}_{j_{l_1}}^{\top}\hat{\vec{\theta}}=\ldots=\vec{\alpha}_{j_{l_m}}^{\top}\hat{\vec{\theta}}=\nu_j$, then it holds that $c_{j,l_1},\ldots,c_{j,l_m}>0$. For any $l\in\left\{1,\ldots,k\right\}$ such that $\vec{\alpha}_{j_l}^{\top}\hat{\vec{\theta}}<\nu_j$, it holds that $c_{j,l}=0$.
                \item[(iv)] There exists a unique $(c_{1},\ldots,c_{k})^{\top}\in\mathbb{R}^k$ such that $\hat{\vec{\theta}}$ has expansion: $\hat{\vec{\theta}}=\sum_{l=1}^kc_{l}\vec{\alpha}_{j_l}$. Suppose $l_1,\ldots,l_m$ are all indices among $\left\{j_1,\ldots,j_k\right\}$ such that $\vec{\alpha}_{l_1}^{\top}\hat{\vec{\theta}}=\ldots=\vec{\alpha}_{l_m}^{\top}\hat{\vec{\theta}}=0$, then it holds that $c_{l_1},\ldots,c_{l_m}< 0$.
                \item[(v)] For any $\vec{\alpha}\in\left\{\vec{\alpha}_{j_1},\ldots,\vec{\alpha}_{j_k}\right\}$, either $\vec{\alpha}^{\top}\hat{\vec{\theta}}\in\left\{\nu_1,\ldots,\nu_J,0\right\}$ or there exists a unique $\vec{\beta}\in\left\{\vec{\alpha}_1,\ldots,\vec{\alpha}_K\right\}\setminus\left\{\vec{\alpha}_{j_1},\ldots,\vec{\alpha}_{j_k}\right\}$ such that $\vec{\alpha}^{\top}\hat{\vec{\theta}}=\vec{\beta}^{\top}\hat{\vec{\theta}}\notin\left\{\nu_1,\ldots,\nu_J,0\right\}$ and $0<\vec{\alpha}^{\top}\hat{\vec{\theta}}<\nu_1$. In the second case, there exists a unique $\left(c_{1},\ldots,c_{k}\right)^{\top}\in\mathbb{R}^k$ such that $\vec{\beta}$ has expansion: $\vec{\beta}=\sum_{l=1}^kc_{l}\vec{\alpha}_{j_l}$. For any $l\in\left\{1,\ldots,k\right\}$ such that $\vec{\alpha}_{j_l}^{\top}\hat{\vec{\theta}}=\vec{\beta}^{\top}\hat{\vec{\theta}}$, it holds that $c_{l}<0$. For any $l\in\left\{1,\ldots,k\right\}$ such that $\vec{\alpha}_{j_l}^{\top}\hat{\vec{\theta}}<\vec{\beta}^{\top}\hat{\vec{\theta}}$, it holds that $c_{l}=0$.
                \item[(vi)] For any $\vec{\beta}\in\left\{\vec{\alpha}_1,\ldots,\vec{\alpha}_K\right\}\setminus\left\{\vec{\alpha}_{j_1},\ldots,\vec{\alpha}_{j_k}\right\}$, either $\vec{\beta}^{\top}\hat{\vec{\theta}}<\min_{l=1,\ldots,k}\vec{\alpha}_{j_l}^{\top}\hat{\vec{\theta}}$ or there exists $\vec{\alpha}\in\left\{\vec{\alpha}_{j_1},\ldots,\vec{\alpha}_{j_k}\right\}$ such that $\vec{\alpha}^{\top}\hat{\vec{\theta}}=\vec{\beta}^{\top}\hat{\vec{\theta}}\notin\left\{\nu_1,\ldots,\nu_J,0\right\}$.
            \end{enumerate}
            Then we construct a characterisation equation at $(\nu_1,\ldots,\nu_J)^{\top}$ with unique solution $\vec{\theta}(\nu_1,\ldots,\nu_J)$. For notational simplicity, we simplify $\vec{\theta}(\nu_1,\ldots,\nu_J)$ as $\hat{\vec{\theta}}$. We also assume that $(\xi_1^{(n)},\ldots,\xi_J^{(n)})^{\top}=(n^{\nu_1},\ldots,n^{\nu_J})^{\top}$ in this step.\\[3mm]
            The first-order equation for $\vec{\theta}_n$ is given by
            \begin{align}\label{eq_lem_4.2}
                -\sum_{k=1}^Kw_k\exp(\vec{\alpha}_k^{\top}\vec{\theta}_n)\vec{\alpha}_k+\sum_{j=1}^Jn^{\nu_j}\vec{\eta}_j=\vec{\theta}_n.
            \end{align}
            For $j=1,\ldots,J$, denote $\mathcal{E}_j=\{\vec{\alpha}_k\in\left\{\vec{\alpha}_1,\ldots,\vec{\alpha}_K\right\}:\vec{\alpha}_k^{\top}\hat{\vec{\theta}}\geq \nu_j\}$. We construct set $\mathcal{G}$ and $\mathcal{G}_j,j=1,\ldots,J$ in the following inductive way:\\[3mm]
            \textbf{Step 2.1: }For $j=1$, following the proof in Proposition~\ref{prop_canonical_projection} we have $\max_{k=1,\ldots,K}\vec{\alpha}_k^{\top}\hat{\vec{\theta}}=\nu_1$ and there exists $\left\{\vec{\alpha}_{k_1},\ldots,\vec{\alpha}_{k_m}\right\}\subseteq\mathcal{E}_1$ and positive constants $c_{k_1},\ldots,c_{k_m}>0$ such that $\vec{\eta}_1=\sum_{p=1}^m c_{k_p}\vec{\alpha}_{k_p}$.\\[3mm]
            We choose a maximal linearly independent subset of $\mathcal{E}_1$ which contains $\left\{\vec{\alpha}_{k_1},\ldots,\vec{\alpha}_{k_m}\right\}$ to enter set $\mathcal{G}$. Then we have constructed $\mathcal{G}_1$ with linearly independent vectors and $\operatorname{span}\left(\mathcal{G}_1\right)=\operatorname{span}\left(\mathcal{E}_1\right):= \mathcal{H}_1$. Furthermore, $\vec{\eta}_1\in \mathcal{H}_1$.\\[3mm]
            \textbf{Step 2.2: }If $\mathcal{E}_i,\mathcal{G}_i,\mathcal{H}_i$ are constructed for $i=1,\ldots,j-1$ and $\vec{\eta}_1,\ldots,\vec{\eta}_{j-1}\in\mathcal{H}_{j-1}$, we project first-order equation (\ref{eq_lem_4.2}) onto $\mathcal{H}_{j-1}^{\perp}$ and divide both sides by $n^{\nu_j}$ to get
            \begin{align}\label{eq_lem_4.3}
                -\sum_{k:\vec{\alpha}_k\in \mathcal{E}_{j}\setminus \mathcal{E}_{j-1}}w_k\frac{\exp(\vec{\alpha}_k^{\top}\vec{\theta}_n)}{n^{\nu_j}}P_{\mathcal{H}_{j-1}^{\perp}}\vec{\alpha}_k+P_{\mathcal{H}_{j-1}^{\perp}}\vec{\eta}_{j}=o(1).
            \end{align}
            By Lemma~\ref{lem_characterisation_equation_1}, for any $n$, we can choose a linearly independent subset $\{\vec{\beta}_1^{(n)},\ldots,\vec{\beta}_m^{(n)}\}$ from $\{P_{\mathcal{H}_{j-1}^{\perp}}\vec{\alpha}_k:\vec{\alpha}_k\in \mathcal{E}_{j}\setminus \mathcal{E}_{j-1}\}$ such that there exists $c_{1}^{(n)},\ldots,c_{m}^{(n)}>0$ satisfying
            \begin{align*}
                -\sum_{k:\vec{\alpha}_k\in \mathcal{E}_{j}\setminus \mathcal{E}_{j-1}}w_k\frac{\exp(\vec{\alpha}_k^{\top}\vec{\theta}_n)}{n^{\nu_j}}P_{\mathcal{H}_{j-1}^{\perp}}\vec{\alpha}_k=-\sum_{k=1}^mc_{k}^{(n)}\vec{\beta}_{k}^{(n)}.
            \end{align*}
            Since the choice of $\{\vec{\beta}_1^{(n)},\ldots,\vec{\beta}_m^{(n)}\}$ has only finite possibilities, we assume WLOG that the same set is chosen for any $n$, i.e., $\{\vec{\beta}_1^{(n)},\ldots,\vec{\beta}_m^{(n)}\}:=\{\vec{\beta}_1,\ldots,\vec{\beta}_m\}$ for any $n$ (or we should analyse along a subsequence). Then we have $\sum_{k=1}^mc_{k}^{(n)}\vec{\beta}_{k}=P_{\mathcal{H}_{j-1}^{\perp}}\vec{\eta}_j+o(1)$. Since $\vec{\beta}_1,\ldots,\vec{\beta}_m$ are linearly independent, $(c_{1}^{(n)},\ldots,c_{m}^{(n)})^{\top}$ is bounded. We assume WLOG that $(c_{1}^{(n)},\ldots,c_{m}^{(n)})^{\top}\rightarrow (c_{1},\ldots,c_{m})^{\top}$ (or we should analyse along a subsequence). This implies that $P_{\mathcal{H}_{j-1}^{\perp}}\vec{\eta}_j=\sum_{k=1}^mc_{k}\vec{\beta}_{k}=\sum_{k=1}^pc_{l_k}\vec{\beta}_{l_k}$ where $c_{l_1},\ldots,c_{l_p}$ are strictly positive. We first choose $\vec{\beta}_{l_1},\ldots,\vec{\beta}_{l_p}$ to enter set $\mathcal{G}$. 
            
            For the remaining vectors in $\mathcal{E}_{j}\setminus \mathcal{E}_{j-1}$, we rank their inner product with $\hat{\vec{\theta}}$ in decreasing order and perform the following procedure: For each vector, if the vector is linearly independent with the current vectors in $\mathcal{G}$, then we let it enter set $\mathcal{G}$, otherwise we discard it. Eventually, we obtain $\mathcal{G}_{j}$ and $\mathcal{H}_j=\operatorname{span}\left(\mathcal{G}_{j}\right)$ satisfying $\mathcal{E}_j\subseteq\mathcal{H}_j$ and $\vec{\eta}_j\in \mathcal{H}_j$. Furthermore, we know that if vector $\vec{\alpha}\in \mathcal{E}_{j}\setminus \mathcal{E}_{j-1}$ is involved in the expansion of $\vec{\eta}_j$ under $\mathcal{G}_{j}$, the coefficient of $\vec{\alpha}$ in the expansion is strictly positive since $c_{l_1},\ldots,c_{l_p}$ are strictly positive. Then by this inductive method, we obtain $\mathcal{G}_1\subseteq \ldots\subseteq\mathcal{G}_J:=\mathcal{G}$. By the construction method of $\mathcal{G}_J$ and $\mathcal{H}_J$, we know that $\mathcal{G}=\mathcal{G}_J\subseteq\mathcal{E}_J$ and $\vec{\eta}_j\in \mathcal{H}_j,j=1,\ldots,J$.\\[3mm]
            \textbf{Case 1: }If $\operatorname{card}\left(\mathcal{G}\right)=d$.\\[3mm]
            For any $\vec{\alpha}$ in $\mathcal{G}$ such that $\vec{\alpha}^{\top}\hat{\vec{\theta}}\notin\left\{\nu_1,\ldots,\nu_J\right\}$, there exists $j\in\left\{2,\ldots,J\right\}$ such that $\nu_{j-1}>\vec{\alpha}^{\top}\hat{\vec{\theta}}>\nu_{j}$. (\ref{eq_lem_4.3}) indicates that
            \begin{align}\label{eq_lem_4.4}
                \sum_{k:\vec{\alpha}_k\in \mathcal{E}_{j}\setminus \mathcal{E}_{j-1}}w_k\frac{\exp(\vec{\alpha}_k^{\top}\vec{\theta}_n)}{n^{\nu_j}}P_{\mathcal{H}_{j-1}^{\perp}}\vec{\alpha}_k=\mathcal{O}(1).
            \end{align}
            Since $\exp(\vec{\alpha}^{\top}\vec{\theta}_n)/n^{\nu_j}\gg 1$, this indicates that there exists $\vec{\beta}_1,\ldots,\vec{\beta}_m\in \mathcal{E}_{j}\setminus \mathcal{E}_{j-1}$ such that
            \begin{enumerate}
                \item[(i)] $\vec{\beta}_{1}^{\top}\hat{\vec{\theta}},\ldots,\vec{\beta}_{m}^{\top}\hat{\vec{\theta}}\geq \vec{\alpha}^{\top}\hat{\vec{\theta}}$.
                \item[(ii)] $P_{\mathcal{H}_{j-1}^{\perp}}\vec{\alpha},P_{\mathcal{H}_{j-1}^{\perp}}\vec{\beta}_1,\ldots,P_{\mathcal{H}_{j-1}^{\perp}}\vec{\beta}_m$ are linearly dependent, i.e., $\vec{\alpha}\in \operatorname{span}\left(\left\{\vec{\beta}_1,\ldots,\vec{\beta}_m\right\}\cup\mathcal{G}_{j-1} \right)$.
            \end{enumerate}
            Otherwise, term
            \begin{align*}
                w\frac{\exp(\vec{\alpha}^{\top}\vec{\theta}_n)}{n^{\nu_j}}P_{\mathcal{H}_{j-1}^{\perp}}\vec{\alpha}
            \end{align*}
            cannot cancel with other terms in \eqref{eq_lem_4.4}.
            By the construction, before $\vec{\alpha}$ enters set $\mathcal{G}$, it holds that $\operatorname{span}\{\vec{\alpha}_k:\vec{\alpha}_k^{\top}\hat{\vec{\theta}}>\vec{\alpha}^{\top}\hat{\vec{\theta}}\}\subseteq \operatorname{span}\left(\mathcal{G}\right)$. This implies that for any $\vec{\beta}_{k}^{\top}\hat{\vec{\theta}}>\vec{\alpha}^{\top}\hat{\vec{\theta}},k=1,\ldots,m$, $\vec{\beta}_k$ is already contained in $\mathcal{G}$. So there exists at least one $\vec{\beta}_k$ among $\vec{\beta}_1,\ldots,\vec{\beta}_m$ that did not enter set $\mathcal{G}$ and satisfies $\vec{\beta}_k^{\top}\hat{\vec{\theta}}=\vec{\alpha}^{\top}\hat{\vec{\theta}}$. This is because if $\vec{\beta}_k^{\top}\hat{\vec{\theta}}>\vec{\alpha}^{\top}\hat{\vec{\theta}}$, by construction $\vec{\beta}_k$ should enter $\mathcal{G}$ before $\vec{\alpha}$. Then $\vec{\alpha}$ should not enter $\mathcal{G}$ due to the linear dependence among $\vec{\beta}_1,\ldots,\vec{\beta}_m,\vec{\alpha}$.\\[3mm]
            For any $\vec{\alpha}\in\mathcal{G}$ such that $\vec{\alpha}^{\top}\hat{\vec{\theta}}=\nu_j\in\left\{\nu_1,\ldots,\nu_J\right\}$ but has zero coefficient in the expansion of $\vec{\eta}_j$ under $\mathcal{G}$, we project the first-order equation (\ref{eq_lem_4.2}) on $\operatorname{span}\{\mathcal{G}\setminus \vec{\alpha}\}$ and divide the equation by $n^{\nu_j-\delta}$, where $\delta>0$ is a constant such that $\nu_j-\delta>\nu_{j+1}$, then we have
            \begin{align}\label{eq_lem_4.5}
                -\sum_{k:\vec{\alpha}_k\in \mathcal{E}_{j}\setminus \operatorname{span}\{\mathcal{G}\setminus \vec{\alpha}\}}w_k\frac{\exp(\vec{\alpha}_k^{\top}\vec{\theta}_n)}{n^{\nu_j-\delta}}P_{\operatorname{span}\{\mathcal{G}\setminus \vec{\alpha}\}^{\perp}}\vec{\alpha}_k=o(1).
            \end{align}
            Since $\vec{\alpha}\in \mathcal{E}_{j}\setminus \operatorname{span}\{\mathcal{G}\setminus \{\vec{\alpha}\}\}$, we have $\exp(\vec{\alpha}^{\top}\vec{\theta}_n)/n^{\nu_j-\delta}\gg 1$. Similarly, we can prove that there exists $\vec{\beta}\notin\mathcal{G}$ such that $\vec{\alpha}^{\top}\hat{\vec{\theta}}=\vec{\beta}^{\top}\hat{\vec{\theta}}$.\\[3mm]
            Furthermore, for the above two scenarios, if we expand $\vec{\beta}$ in terms of basis $\mathcal{G}$, since in (\ref{eq_lem_4.4}) and (\ref{eq_lem_4.5}), the $\vec{\alpha}$ and $\vec{\beta}$ terms can cancel out with each other, the coefficient of $\vec{\alpha}$ in the expansion of $\vec{\beta}$ should be strictly negative.\\[3mm]
            From the construction method of $\mathcal{G}$, for any $\vec{\alpha}\in\mathcal{G}$ such that $\vec{\alpha}^{\top}\hat{\vec{\theta}}=\nu_j\in\left\{\nu_1,\ldots,\nu_J\right\}$ which is involved in the expansion of $\vec{\eta}_j$, we call $\vec{\alpha}$ type-1 element. Otherwise we call $\vec{\alpha}$ type-2 element. Then we have
            \begin{enumerate}
                \item[(i)] For any type-1 $\vec{\alpha}\in\mathcal{G}$, we have $\vec{\alpha}^{\top}\hat{\vec{\theta}}=\nu_j\in\left\{\nu_1,\ldots,\nu_J\right\}$ and the coefficient of $\vec{\alpha}$ in the expansion of $\vec{\eta}_j$ under $\mathcal{G}$ is strictly positive. Moreover, $\vec{\eta}_j\in \operatorname{span}\left(\mathcal{G}\cap\mathcal{E}_j\right)$.
                \item[(ii)] For any type-2 $\vec{\alpha}\in\mathcal{G}$, there exists $\vec{\beta}\notin \mathcal{G}$ such that $\vec{\alpha}^{\top}\hat{\vec{\theta}}=\vec{\beta}^{\top}\hat{\vec{\theta}}$. Furthermore, the coefficient of $\vec{\alpha}$ in the expansion of $\vec{\beta}$ under $\mathcal{G}$ is strictly negative.
            \end{enumerate}
            This induces the characterisation equation with solution $\hat{\vec{\theta}}$. Since the dimension of $\hat{\vec{\theta}}$ matches the number of linear equations, $\hat{\vec{\theta}}$ is the unique solution. Moreover, for any type-2 $\vec{\alpha}\in\mathcal{G}$ or $\vec{\alpha}\notin\mathcal{G}$, it holds that $\vec{\alpha}^{\top}\hat{\vec{\theta}}\notin \left\{\nu_1,\ldots,\nu_J\right\}$. For $\vec{\beta}\notin \mathcal{G}$, either $\vec{\beta}^{\top}\hat{\vec{\theta}}<\min_{\vec{\alpha}\in\mathcal{G}}\vec{\alpha}^{\top}\hat{\vec{\theta}}$ or there exists $\vec{\alpha}\in\mathcal{G}$ such that $\vec{\alpha}^{\top}\hat{\vec{\theta}}=\vec{\beta}^{\top}\hat{\vec{\theta}}\notin \left\{\nu_1,\ldots,\nu_J\right\}$. Hence all assumptions on characterisation equation are verified, which proves the existence of characterisation equation at $(\nu_1,\ldots,\nu_J)^{\top}$ with unique solution $\hat{\vec{\theta}}=\vec{\theta}(\nu_1,\ldots,\nu_J)$ when $k=d$.\\[3mm]
            \textbf{Case 2: }If $\operatorname{card}\left(\mathcal{G}\right)<d$.\\[3mm] Denote $\mathcal{E}=\{\vec{\alpha}_k:\vec{\alpha}_k^{\top}\hat{\vec{\theta}}\geq 0\}$. Then we project the first-order equation (\ref{eq_lem_4.2}) onto $\mathcal{H}_J^{\perp}$ and divide by $\log n$ to get:
            \begin{align*}
                -\sum_{k:\vec{\alpha}_k\in \mathcal{E}\setminus \mathcal{E}_J}w_k\frac{\exp(\vec{\alpha}_k^{\top}\vec{\theta}_n)}{\log n}P_{\mathcal{H}_{J}^{\perp}}\vec{\alpha}_k=P_{\mathcal{H}_{J}^{\perp}}\hat{\vec{\theta}}+o(1).
            \end{align*}
            Similarly, there exists $\vec{\beta}_1,\ldots,\vec{\beta}_m\in\mathcal{E}\setminus \mathcal{E}_J$ and negative constants $c_1,\ldots,c_m$ such that:
            \begin{enumerate}
                \item[(i)] $P_{\mathcal{H}_{J}^{\perp}}\vec{\beta}_1,\ldots,P_{\mathcal{H}_{J}^{\perp}}\vec{\beta}_m$ are linearly independent.
                \item[(ii)] $P_{\mathcal{H}_{J}^{\perp}}\hat{\vec{\theta}}=\sum_{k=1}^mc_{k}\vec{\beta}_k$.
            \end{enumerate}
            Similarly, we first let $\vec{\beta}_1,\ldots,\vec{\beta}_m$ enter set $\mathcal{G}$. Then we rank the vectors in $\mathcal{E}\setminus \mathcal{E}_J$ in decreasing order by their inner product with $\hat{\vec{\theta}}$ and decide whether each vector enters set $\mathcal{G}$ or not. Then we can construct $\mathcal{G}$ such that $\hat{\vec{\theta}}\in \operatorname{span}(\mathcal{G})$. Similarly we can prove that for any $\vec{\alpha}\in\mathcal{G}$ such that $\vec{\alpha}^{\top}\hat{\vec{\theta}}\notin \left\{\nu_1,\ldots,\nu_J,0\right\}$ or $\vec{\alpha}^{\top}\hat{\vec{\theta}}=0$ and has zero coefficient in the expansion of $\hat{\vec{\theta}}$ under basis $\mathcal{G}$, there exists $\vec{\beta}\notin \mathcal{G}$ such that $\vec{\alpha}^{\top}\hat{\vec{\theta}}=\vec{\beta}^{\top}\hat{\vec{\theta}}$.
            Note that there exists $\vec{\beta}_1,\ldots,\vec{\beta}_m$ ($\vec{\beta}\in\{\vec{\beta}_1,\ldots,\vec{\beta}_m\}$) such that
            \begin{enumerate}
                \item[(i)] $\vec{\beta}_{1}^{\top}\hat{\vec{\theta}},\ldots,\vec{\beta}_{m}^{\top}\hat{\vec{\theta}}\geq \vec{\alpha}^{\top}\hat{\vec{\theta}}$.
                \item[(ii)] $\vec{\alpha},\vec{\beta}_1,\ldots,\vec{\beta}_m$ are linearly dependent.
            \end{enumerate}
            Hence, for those $l\in\left\{1,\ldots,k\right\}$ such that $\vec{\alpha}_{j_l}^{\top}\hat{\vec{\theta}}<\vec{\beta}^{\top}\hat{\vec{\theta}}$, the coefficient of $\vec{\alpha}_{j_l}$ in the expansion of $\vec{\beta}$ is zero.
            \\[3mm]
            If $\operatorname{card}\left(\mathcal{G}\right)=k<d$, we have $k$ equations in the characterisation equation. Since we require $\hat{\vec{\theta}}\in \operatorname{span}(\mathcal{G})$, there still exists unique solution for $(\nu_1,\ldots,\nu_J)^{\top}$. Similar to Case 1, we can verify other conditions required for the characterisation equation.\\[3mm]
            \textbf{Step 3: }Now we prove the uniqueness of characterisation equation at $(\nu_1,\ldots,\nu_J)^{\top}$. We first suppose that $\operatorname{rank}\left\{\vec{\alpha}_1,\ldots,\vec{\alpha}_K\right\}=d$. Assume $w_k\equiv1$ and define
            \begin{align*}
                \widetilde{f}_n(\vec{\theta})=-\sum_{k=1}^K\exp(\vec{\alpha}_k^{\top}\vec{\theta})+\left(\sum_{j=1}^Jn^{\nu_j}\vec{\eta}_j\right)^{\top}\vec{\theta}.
            \end{align*}
            \textbf{Step 3.1: }We first prove that for $n$ large enough, $\widetilde{f}_n$ has a unique maximiser.\\[3mm]
            For any $\vec{\epsilon}\in\mathbb{R}^d$ satisfying $\left\|\vec{\epsilon}\right\|=1$, we discuss the two cases:\\[3mm]
            \textbf{Case 1: }$\max_{k=1,\ldots,K}\vec{\alpha}_k^{\top}\vec{\epsilon}>0$. Then it can be readily shown that
            \begin{align*}
                \lim_{l\rightarrow \infty}-\sum_{k=1}^K\exp(l\vec{\alpha}_k^{\top}\vec{\epsilon})+l\cdot\left(\sum_{j=1}^Jn^{\nu_j}\vec{\eta}_j^{\top}\vec{\epsilon}\right)\rightarrow -\infty.
            \end{align*}
            Furthermore, we can choose $\vec{\epsilon}$ such that $\vec{\eta}_1^{\top}\vec{\epsilon}>0$, then for $n$ large enough, it can be readily shown that
            \begin{align*}
                \sup_{l\geq 0}-\sum_{k=1}^K\exp(l\vec{\alpha}_k^{\top}\vec{\epsilon})+l\cdot\left(\sum_{j=1}^Jn^{\nu_j}\vec{\eta}_j^{\top}\vec{\epsilon}\right)>0.
            \end{align*}
            \textbf{Case 2: }$\max_{k=1,\ldots,K}\vec{\alpha}_k^{\top}\vec{\epsilon}\leq 0$. Since $\vec{\eta}_1,\ldots,\vec{\eta}_J\in \mathcal{X}$, we have $\vec{\eta}_1^{\top}\vec{\epsilon},\ldots,\vec{\eta}_J^{\top}\vec{\epsilon}\leq 0$. So we have
            \begin{align*}
                \sup_{l\geq 0}-\sum_{k=1}^K\exp(l\vec{\alpha}_k^{\top}\vec{\epsilon})+l\cdot\left(\sum_{j=1}^Jn^{\nu_j}\vec{\eta}_j^{\top}\vec{\epsilon}\right)\leq 0.
            \end{align*}
            This implies that for $n$ large enough, there exists a maximiser for $\widetilde{f}_n$. Since $\nabla^2\widetilde{f}_n$ is nonsingular, the maximiser is also unique, denoted by $\widetilde{\vec{\theta}}_n$. It can be readily shown that $\widetilde{\vec{\theta}}_n\neq 0$ for large $n$. Then we denote $l_n=\left\|\widetilde{\vec{\theta}}_n\right\|$ and $\vec{\epsilon}_n=\widetilde{\vec{\theta}}_n/l_n\rightarrow \vec{\epsilon}$. Note that the previous proof also implies that $\max_{k=1,\ldots,K}\vec{\alpha}_k^{\top}\vec{\epsilon}>0$, then we can use similar method as in Step 1 to prove that $l_n=\mathcal{O}(\log n)$. Assume $\lim_{n\rightarrow \infty}\widetilde{\vec{\theta}}_n/\log n:= \widetilde{\vec{\theta}}$ without loss of generality (or we analyse along a subsequence) and assume
            \begin{align}\label{eq_lem_4.6}
                \min_{k=1,\ldots,K}\vec{\alpha}_k^{\top}\widetilde{\vec{\theta}}=-\widetilde{M}.
            \end{align}
            \textbf{Step 3.2: }We expand the first-order equation for $\widetilde{\vec{\theta}}_n$ in terms of the basis $\mathcal{G}$ defined in characterisation equation at $(\nu_1,\ldots,\nu_J)^{\top}$.\\[3mm]
            By Step 2, there exists characterisation equation at $(\nu_1,\ldots,\nu_J)^{\top}$ with unique solution $\hat{\vec{\theta}}$. We first consider the case when $k=d$. The first-order equation for $\widetilde{\vec{\theta}}_n$ is
            \begin{align}\label{eq_lem_4.7}
                -\sum_{k=1}^K\exp(\vec{\alpha}_k^{\top}\widetilde{\vec{\theta}}_n)\vec{\alpha}_k+\sum_{j=1}^Jn^{\nu_j}\vec{\eta}_j=0.
            \end{align}
            For notational simplicity, we assume that $\{\vec{\alpha}_{j_1},\ldots,\vec{\alpha}_{j_d}\}=\{\vec{\alpha}_1,\ldots,\vec{\alpha}_d\}$ in the characterisation equation. Furthermore, assume that the first $l$ equations are of type 1 and the other $d-l$ equations are of type 2. Since by condition (\romannumeral 1), we have $\operatorname{rank}\left\{\vec{\alpha}_1,\ldots,\vec{\alpha}_d\right\}=d$, so we expand the first-order equation (\ref{eq_lem_4.7}) in terms of basis $\left\{\vec{\alpha}_1,\ldots,\vec{\alpha}_d\right\}$. Then we discuss the coefficient for every term $\vec{\alpha}_k,~k=1,\ldots,d$ in the expansion.\\[3mm]
            For $k=1,\ldots,l$, we assume that $\vec{\alpha}_k^{\top}\hat{\vec{\theta}}=\nu_{j_k}$, where $\nu_{j_k}\in\left\{\nu_1,\ldots,\nu_J,0\right\}$. By condition (\romannumeral 3), the coefficients of $\vec{\alpha}_k$ in the expansions of $\vec{\eta}_1,\ldots,\vec{\eta}_{j_k-1}$ are all zero. Moreover, the coefficient of $\vec{\alpha}_k$ in the expansion of $\vec{\eta}_{j_k}$ is positive. By condition (\romannumeral 6), for any $\vec{\alpha}\in\left\{\vec{\alpha}_{d+1},\ldots,\vec{\alpha}_{K}\right\}$ such that $\vec{\alpha}^{\top}\hat{\vec{\theta}}>\nu_{j_k}$, the coefficient of $\vec{\alpha}_k$ in the expansion of $\vec{\alpha}$ is zero. So the coefficient equation of $\vec{\alpha}_k,k=1,\ldots,l$ is
            \begin{align}\label{eq_lem_4.8}
                \sum_{j=j_k}^J c_{k,j}n^{\nu_j}-\exp(\vec{\alpha}_k^{\top}\widetilde{\vec{\theta}}_n)-\sum_{p=d+1,\ldots,K:\vec{\alpha}_p^{\top}\hat{\vec{\theta}}<\nu_{j_k}} \xi_{k,p}\exp(\vec{\alpha}_p^{\top}\widetilde{\vec{\theta}}_n)=0,
            \end{align}
            where $c_{k,j_k}>0$ and $\xi_{k,p}$ is the coefficient of $\vec{\alpha}_k$ in the expansion of $\vec{\alpha}_p$ for $p=d+1,\ldots,K$.\\[3mm]
            For $k=l+1,\ldots,d$, we assume that $\vec{\alpha}_k^{\top}\hat{\vec{\theta}}\in(\nu_{j_k},\nu_{j_k-1})$ (define $\nu_{J+1}=0$). By condition (\romannumeral 3), the coefficient of $\vec{\alpha}_k$ in the expansion of $\vec{\eta}_1,\ldots,\vec{\eta}_{j_k-1}$ is zero. By conditions (\romannumeral 5) and (\romannumeral 6), there exists a unique $\vec{\alpha}_{p_k}\in\left\{\vec{\alpha}_{d+1},\ldots,\vec{\alpha}_{K}\right\}$ such that $\vec{\alpha}_k^{\top}\hat{\vec{\theta}}=\vec{\alpha}_{p_k}^{\top}\hat{\vec{\theta}}$. Moreover, the coefficient of $\vec{\alpha}_k$ in the expansion of $\vec{\alpha}_{p_k}$ is negative. For any $\vec{\beta}\in\left\{\vec{\alpha}_{d+1},\ldots,\vec{\alpha}_{K}\right\}$ such that $\vec{\beta}^{\top}\hat{\vec{\theta}}> \vec{\alpha}_k^{\top}\hat{\vec{\theta}}$, the coefficient of $\vec{\alpha}_k$ in the expansion of $\vec{\beta}$ is zero. So the coefficient equation of $\vec{\alpha}_k,k=l+1,\ldots,d$ is
            \begin{align}\label{eq_lem_4.9}
                \sum_{j=j_k}^J c_{k,j}n^{\nu_j}-\exp(\vec{\alpha}_k^{\top}\widetilde{\vec{\theta}}_n)-\xi_k\exp(\vec{\alpha}_{p_k}^{\top}\widetilde{\vec{\theta}}_n)-\sum_{p=d+1,\ldots,K:\vec{\alpha}_p^{\top}\hat{\vec{\theta}}<\vec{\alpha}_k^{\top}\hat{\vec{\theta}}} \xi_{k,p}\exp(\vec{\alpha}_p^{\top}\widetilde{\vec{\theta}}_n)=0,
            \end{align}
            where $\xi_k<0$ is the coefficient of $\vec{\alpha}_k$ in the expansion of $\vec{\alpha}_{p_k}$ and $\xi_{k,p}$ is the coefficient of $\vec{\alpha}_k$ in the expansion of $\vec{\alpha}_p$ for $p=d+1,\ldots,K$.\\[3mm]
            \textbf{Step 3.3: }We expand $\widetilde{f}_n(\widetilde{\vec{\theta}}_n)$ into infinite series.\\[3mm]
            We first consider the case when $k=d$. We match the term of highest order each time. We first find the solution $\vec{\theta}_n^{(1)}$ to the equations matching the terms with highest order in (\ref{eq_lem_4.8}) and (\ref{eq_lem_4.9}), which are
            \begin{enumerate}
                \item[(i)] For $k=1,\ldots,l$, it holds that $c_{k,j_k}n^{\nu_{j_k}}-\exp(\vec{\alpha}_k^{\top}\vec{\theta}_n^{(1)})=0$.
                \item[(ii)] For $k=l+1,\ldots,d$, it holds that $-\exp(\vec{\alpha}_k^{\top}\vec{\theta}_n^{(1)})-\xi_k\exp(\vec{\alpha}_{p_k}^{\top}\vec{\theta}_n^{(1)})=0$.
            \end{enumerate}
            Since $c_{k,j_k}>0$ for $k=1,\ldots,l$ and $\xi_k<0$ for $k=l+1,\ldots,d$, there exists a unique solution and it can be readily shown that 
            \begin{align*}
                \lim_{n\rightarrow \infty}\frac{\vec{\theta}_n^{(1)}}{\log n}=\hat{\vec{\theta}}.
            \end{align*}
            So for every $k=1,\ldots,K$, it holds that $\exp(\vec{\alpha}_k^{\top}\vec{\theta}_n^{(1)})=c_{k}n^{\vec{\alpha}_k^{\top}\hat{\vec{\theta}}}$ where $c_{1},\ldots,c_{K}$ are positive constants.\\[3mm]
            Now we calculate $\nabla \widetilde{f}_n(\vec{\theta}_n^{(1)})$ to get the residual terms:
            \begin{enumerate}
                \item[(i)] For $k=1,\ldots,l$, the residual terms are:
                \begin{align*}
                    \sum_{j=j_k+1}^J c_{k,j}n^{\nu_j}-\sum_{p=d+1,\ldots,K:\vec{\alpha}_p^{\top}\hat{\vec{\theta}}<\nu_{j_k}} \xi_{k,p}\exp(\vec{\alpha}_p^{\top}\vec{\theta}_n^{(1)}).
                \end{align*}
                \item[(ii)] For $k=l+1,\ldots,d$, the residual terms are:
                \begin{align*}
                    \sum_{j=j_k}^J c_{k,j}n^{\nu_j}-\sum_{p=d+1,\ldots,K:\vec{\alpha}_p^{\top}\hat{\vec{\theta}}<\vec{\alpha}_k^{\top}\hat{\vec{\theta}}} \xi_{k,p}\exp(\vec{\alpha}_p^{\top}\vec{\theta}_n^{(1)}).
                \end{align*}
            \end{enumerate}
            Note that all terms in the residual are of the form $cn^{\xi}$. Hence, in the following proof, we define the order of terms with form $cn^{\xi}$ or $cn^{\xi}\log n$ as $\xi$. Then we can define the order gap for each residual (the maximum of empty set is defined as $-\infty$):
            \begin{enumerate}
                \item[(i)] For $k=1,\ldots,l$, the order gap $\delta_{k}^{(1)}$ is defined as the difference between the highest order in the residual with $\vec{\alpha}_k^{\top}\hat{\vec{\theta}}$, i.e.,
                \begin{align*}
                    \delta_{k}^{(1)}=\nu_{j_k}-\left(\max_{j=j_k+1,\ldots,J}\nu_j\right)\vee \left(\max_{p=d+1,\ldots,K:\vec{\alpha}_p^{\top}\hat{\vec{\theta}}<\nu_{j_k}}\vec{\alpha}_p^{\top}\hat{\vec{\theta}}\right).
                \end{align*}
                \item[(ii)] For $k=l+1,\ldots,d$, the order gap $\delta_{k}^{(1)}$ is defined as the difference between the highest order in the residual with $\vec{\alpha}_k^{\top}\hat{\vec{\theta}}$, i.e.,
                \begin{align*}
                    \delta_{k}^{(1)}=\vec{\alpha}_k^{\top}\hat{\vec{\theta}}-\left(\max_{j=j_k,\ldots,J}\nu_j\right)\vee \left(\max_{p=d+1,\ldots,K:\vec{\alpha}_p^{\top}\hat{\vec{\theta}}<\vec{\alpha}_k^{\top}\hat{\vec{\theta}}}\vec{\alpha}_p^{\top}\hat{\vec{\theta}}\right).
                \end{align*}
            \end{enumerate}
            We assume WLOG that $\delta_{1}^{(1)},\ldots,\delta_{d}^{(1)}$ all exist and are finite. Suppose $\delta_{m}^{(1)}=\min_{l=1,\ldots,d}\delta_{l}^{(1)}$ and suppose $q$ is the smallest integer such that $q\delta_{m}^{(1)}>\max_{l=1,\ldots,d}\delta_{l}^{(1)}$. For $k=1,\ldots,d$, let
            \begin{align}\label{eq_lem_4.10}
                \vec{\alpha}_k^{\top}\vec{\theta}_n^{(2)}=\vec{\alpha}_k^{\top}\vec{\theta}_n^{(1)}+\sum_{p=1}^qc_{k,p}n^{-p\delta_{m}^{(1)}}.
            \end{align}
            Then we determine $(c_{1,1},\ldots,c_{d,1})^{\top},\ldots,(c_{1,q},\ldots,c_{d,q})^{\top}$ in (\ref{eq_lem_4.10}) by the following inductive method:\\[3mm]
            For $p=1$, we expand $\exp(\vec{\alpha}_k^{\top}\vec{\theta}_n^{(1)}+c_{k,1}n^{-\delta_{m}^{(1)}})$ to the first order and find $(c_{1,1},\ldots,c_{d,1})^{\top}$ to cancel all terms in every residual with order gap $\delta_{m}^{(1)}$. So $(c_{1,1},\ldots,c_{d,1})^{\top}$ can be obtained by solving the following equation:
            \begin{enumerate}
                \item[(i)] For $k=1,\ldots,l$, $c_{k,1}=b_k$, where $b_k$ is a constant depending on the constants in (\ref{eq_lem_4.8}) and (\ref{eq_lem_4.9}).
                \item[(ii)] For $k=l+1,\ldots,d$, $c_{k,1}-(\sum_{l=1}^d\xi_{l,k}c_{l,1})=b_k$, where $b_k$ is a constant depending on the constants in (\ref{eq_lem_4.8}) and (\ref{eq_lem_4.9}) and $\vec{\alpha}_{p_k}=\sum_{l=1}^d\xi_{l,k}\vec{\alpha}_l$.
            \end{enumerate}
            We can write the above linear equations in matrix form: $\mat{\Xi} \vec{c}=\vec{\beta}$, where $\mat{\Xi}\in\mathbb{R}^{d\times d}$ and $\vec{c},\vec{\beta}\in\mathbb{R}^d$. Noticing that the characterisation equation can be written as:
            \begin{align*}
            \vec{\nu}=\left(\vec{\alpha}_1^{\top}\hat{\vec{\theta}},\ldots,\vec{\alpha}_l^{\top}\hat{\vec{\theta}},\vec{\alpha}_{l+1}^{\top}\hat{\vec{\theta}}-\vec{\alpha}_{p_{l+1}}^{\top}\hat{\vec{\theta}},\ldots,\vec{\alpha}_{d}^{\top}\hat{\vec{\theta}}-\vec{\alpha}_{p_{d}}^{\top}\hat{\vec{\theta}}\right)^{\top}:=\mat{\Xi}\left(\vec{\alpha}_1,\ldots,\vec{\alpha}_d\right)^{\top}\hat{\vec{\theta}}.
            \end{align*}
            Since $\left(\vec{\alpha}_1,\ldots,\vec{\alpha}_d\right)^{\top}$ is invertible and the characterisation equation has unique solution by condition (\romannumeral 2), $\mat{\Xi}$ is invertible. So equation $\mat{\Xi} \vec{c}=\vec{\beta}$ has unique solution. So we match the terms in each residual with order gap $\delta_{m}^{(1)}$.\\[3mm]
            If we have matched the terms in each residual with order gap $\delta_{m}^{(1)},\ldots,(p-1)\delta_{m}^{(1)}$, then we expand $\exp(\vec{\alpha}_k^{\top}\vec{\theta}_n^{(1)}+\sum_{l=1}^{p}c_{k,l}n^{-l\delta_{m}^{(1)}})$ to cancel out the terms with order gap $p\delta_{m}^{(1)}$. Similarly, we can prove that there exists a unique solution for $(c_{1,p},\ldots,c_{d,p})^{\top}$. So by this inductive method, we obtain solution $\vec{\theta}_n^{(2)}$ such that all terms with order gap $\delta_{m}^{(1)},\ldots,q\delta_{m}^{(1)}$ are cancelled out in each residual. By the construction method of $\vec{\theta}_n^{(2)}$, we have $\vec{\alpha}_k^{\top}\vec{\theta}_n^{(1)}-\vec{\alpha}_k^{\top}\vec{\theta}_n^{(2)}=o(1)$ for any $k=1,\ldots,K$, so we have $\vec{\theta}_n^{(1)}-\vec{\theta}_n^{(2)}=o(1)$, which indicates that
            \begin{align*}
                \lim_{n\rightarrow \infty}\frac{\bt_n^{(2)}}{\log n}=\hat{\vec{\theta}}.
            \end{align*}
            This implies that the order of each term in the residual will not change within finite procedures. Then by the assumption on integer $q$, the order gap in the $m$-th residual has decreased strictly, while the order gaps in other residuals have remained the same. Since the order gap of all terms in residuals can only be the linear combination of the order gaps in the residuals obtained in the first approximation with nonnegative integer coefficients, we can reduce the highest order in all residuals to any given level in finite steps.\\[3mm]
            Now assume that
            \begin{align}\label{eq_lem_4.11}
                \min_{k=1,\ldots,K}\vec{\alpha}_k^{\top}\hat{\vec{\theta}}=-M,
            \end{align}
            then for any given constant $C>0$, suppose that we obtain $\vec{\theta}_n^{(L)}$ over $L$ procedures satisfying
            \begin{align}\label{eq_lem_4.12}
                \left\|\nabla \widetilde{f}_n(\vec{\theta}_n^{(L)})\right\|\lesssim n^{-2d-d\left(M\vee \widetilde{M}\right)-C/2}.
            \end{align}
            By similar proof in Proposition~\ref{prop_canonical_projection}, we have
            \begin{align}\label{eq_lem_4.13}
                \max_{k=1,\ldots,K}\vec{\alpha}_k^{\top}\hat{\vec{\theta}}=\max_{k=1,\ldots,K}\vec{\alpha}_k^{\top}\widetilde{\vec{\theta}}=\nu_1.
            \end{align}
            Then by (\ref{eq_lem_4.6}), (\ref{eq_lem_4.11}) and (\ref{eq_lem_4.13}), for $n$ large enough we have
            \begin{align*}
                n^{-\widetilde{M}-\nu_1}\lesssim \min_{k=1,\ldots,K}\exp(\vec{\alpha}_k^{\top}\widetilde{\vec{\theta}}_n)&\leq\max_{k=1,\ldots,K}\exp(\vec{\alpha}_k^{\top}\widetilde{\vec{\theta}}_n)\lesssim n^{\nu_1+1},\\
                n^{-M-\nu_1}\lesssim \min_{k=1,\ldots,K}\exp(\vec{\alpha}_k^{\top}\vec{\theta}_n^{(L)})&\leq\max_{k=1,\ldots,K}\exp(\vec{\alpha}_k^{\top}\vec{\theta}_n^{(L)})\lesssim n^{\nu_1+1}.
            \end{align*}
            Then for $n$ large enough, and any $\vec{\theta}^{\ast}_n$ between $\widetilde{\vec{\theta}}_n$ and $\vec{\theta}_n^{(L)}$, we have
            \begin{align}\label{eq_lem_4.14}
                n^{-d\left(\widetilde{M}\vee M+\nu_1\right)}
                \lesssim&\big\|-\nabla^2\widetilde{f}_n(\vec{\theta}^{\ast}_n)\big\|_2\lesssim n^{d(\nu_1+1)}.
            \end{align}
            By Taylor expansion, we have
            \begin{align}\label{eq_lem_4.15}
                \vec{0}=\nabla \widetilde{f}_n(\widetilde{\vec{\theta}}_n)=\nabla \widetilde{f}_n(\vec{\theta}_n^{(L)})+\nabla^2 \widetilde{f}_n\left(\vec{\theta}_n^{\ast}\right)(\widetilde{\vec{\theta}}_n-\vec{\theta}_n^{(L)}),
            \end{align}
            where $\vec{\theta}^{\ast}_n$ is a point between $\widetilde{\vec{\theta}}_n$ and $\vec{\theta}_n^{(L)}$. By (\ref{eq_lem_4.12}), (\ref{eq_lem_4.14}) and (\ref{eq_lem_4.15}), we have
            \begin{align*}
                n^{-d\nu_1-d(M\vee \widetilde{M})}\big\|\widetilde{\vec{\theta}}_n-\vec{\theta}_n^{(L)}\big\|\lesssim \big\|-\nabla^2 \widetilde{f}_n(\vec{\theta}_n^{\ast})(\widetilde{\vec{\theta}}_n-\vec{\theta}_n^{(L)})\big\|=\big\|\nabla \widetilde{f}_n(\vec{\theta}_n^{(L)})\big\|\lesssim n^{-d(\nu_1+1)-d(M\vee \widetilde{M})-C/2}.
            \end{align*}
            This implies that $\|\widetilde{\vec{\theta}}_n-\vec{\theta}_n^{(L)}\|\lesssim n^{-d-C/2}$. Then by (\ref{eq_lem_4.14}) and $\nabla \widetilde{f}_n(\widetilde{\vec{\theta}}_n)=\vec{0}$, Taylor expansion leads to
            \begin{align*}
                \big|\widetilde{f}_n(\vec{\theta}_n^{(L)})-\widetilde{f}_n(\widetilde{\vec{\theta}}_n)\big|=\left|\frac{1}{2}(\vec{\theta}_n^{(L)}-\widetilde{\vec{\theta}}_n)^{\top}\left(-\nabla^2 \widetilde{f}_n(\vec{\theta}_n^{\ast})\right)(\vec{\theta}_n^{(L)}-\widetilde{\vec{\theta}}_n)\right|\lesssim n^{d(\nu_1+1)-2d-C}\leq n^{-C}.
            \end{align*}
            Furthermore, by the construction method of $\vec{\theta}_n^{(L)}$, all terms in the Taylor series of $\widetilde{f}_n(\vec{\theta}_n^{(L)})$ are of the form $cn^{\xi}$ or $cn^{\xi}\log n$, where the coefficients are functions depending on $\vec{\alpha}_1,\ldots,\vec{\alpha}_K$ and $\nu_1,\ldots,\nu_J$ and the power is the linear combination of $\vec{\alpha}_1^{\top}\hat{\vec{\theta}},\ldots,\vec{\alpha}_K^{\top}\hat{\vec{\theta}}$ with integer coefficients.\\[3mm]
            If $k<d$. Define
            \begin{align*}
                \bar{f}_n(\vec{\theta})=-\sum_{k=1}^K\exp(\vec{\alpha}_k^{\top}\vec{\theta})+\sum_{j=1}^Jn^{\nu_j}\vec{\eta}_j^{\top}\vec{\theta}+\sum_{p\in\{1,\ldots,k\}:\vec{\alpha}_p^{\top}\hat{\vec{\theta}}=0}\vec{\alpha}_p^{\top}\vec{\theta}
            \end{align*}
            and denote its maximiser by $\bar{\vec{\theta}}_n$. The additional term $\sum_{p\in\{1,\ldots,k\}:\vec{\alpha}_p^{\top}\hat{\vec{\theta}}=0}\vec{\alpha}_p^{\top}\vec{\theta}$ ensures that the equation matching the terms with highest order has a solution. Similarly we can prove that $\bar{\vec{\theta}}_n=\mathcal{O}(\log n)$, then we have
            \begin{align*}
                \widetilde{f}_n(\widetilde{\vec{\theta}}_n)-\bar{f}_n(\bar{\vec{\theta}}_n)\leq& \widetilde{f}_n(\widetilde{\vec{\theta}}_n)-\bar{f}_n(\widetilde{\vec{\theta}}_n)=-\sum_{p\in\{1,\ldots,k\}:\vec{\alpha}_p^{\top}\hat{\vec{\theta}}=0}\vec{\alpha}_p^{\top}\widetilde{\vec{\theta}}_n\lesssim \log n,\\
                \bar{f}_n(\bar{\vec{\theta}}_n)-\widetilde{f}_n(\widetilde{\vec{\theta}}_n)\leq& \bar{f}_n(\bar{\vec{\theta}}_n)-\widetilde{f}_n(\bar{\vec{\theta}}_n)=\sum_{p\in\{1,\ldots,k\}:\vec{\alpha}_p^{\top}\hat{\vec{\theta}}=0}\vec{\alpha}_p^{\top}\bar{\vec{\theta}}_n\lesssim \log n.
            \end{align*}
            This indicates that $|\widetilde{f}_n(\widetilde{\vec{\theta}}_n)-\bar{f}_n(\bar{\vec{\theta}}_n)|\lesssim \log n$. Hence substituting $\widetilde{f}_n$ by $\bar{f}_n$ will not lead to error of positive order for the maximum value. We then use similar method to approximate $\widetilde{f}_n(\widetilde{\vec{\theta}}_n)$ by the solution of characterisation equation. The only difference is that we require solution $\vec{\theta}_n^{(L)}\in \operatorname{span}\left\{\vec{\alpha}_1,\ldots,\vec{\alpha}_k\right\}$ for any $L\in \mathbb{N}$. Similarly, we assume that after $\widetilde{L}$ procedures, we have $|\bar{f}_n(\bar{\vec{\theta}}_n^{(\widetilde{L})})-\bar{f}_n(\bar{\vec{\theta}}_n)|\lesssim n^{-C}$.\\[3mm]
            \textbf{Step 3.4: }We prove the uniqueness of characterisation equation at $(\nu_1,\ldots,\nu_J)^{\top}$.\\[3mm]
            If there exists two characterisation equations at $(\nu_1,\ldots,\nu_J)^{\top}$ with solutions $\vec{\theta}$ and $\widetilde{\vec{\theta}}$ respectively, we assume WLOG that $k=d$ in both cases for simplicity since we only need to match the terms with positive order. Then the same function $\widetilde{f}_n$ is denoted in both cases. By procedure in Step 3.3, there exists finite $L$ such that
            \begin{align*}
                \big|\widetilde{f}_n(\vec{\theta}_n^{(L)})-\widetilde{f}_n(\widetilde{\vec{\theta}}_n^{(L)})\big|\leq \big|\widetilde{f}_n(\vec{\theta}_n^{(L)})-\widetilde{f}_n(\widetilde{\vec{\theta}}_n)\big|+\big|\widetilde{f}_n(\widetilde{\vec{\theta}}_n)-\widetilde{f}_n(\widetilde{\vec{\theta}}_n^{(L)})\big|\lesssim \log n.
            \end{align*}
            Since the Taylor series of $\widetilde{f}_n(\vec{\theta}_n^{(L)})$ and $\widetilde{f}_n(\widetilde{\vec{\theta}}_n^{(L)})$ consist of terms of the form $cn^{\xi}$ or $cn^{\xi}\log n$, where the coefficients are functions depending on $\vec{\alpha}_1,\ldots,\vec{\alpha}_K$ and $\nu_1,\ldots,\nu_J$ and the power is the linear combination of $\vec{\alpha}_1^{\top}{\vec{\theta}},\ldots,\vec{\alpha}_K^{\top}{\vec{\theta}}$ with integer coefficients, for any term of positive order, both its coefficient and its exponent must match exactly.
            This indicates that all positive terms among $\vec{\alpha}_1^{\top}\vec{\theta},\ldots,\vec{\alpha}_K^{\top}\vec{\theta}$ should match with all positive terms among $\vec{\alpha}_1^{\top}\widetilde{\vec{\theta}},\ldots,\vec{\alpha}_K^{\top}\widetilde{\vec{\theta}}$ exactly. Then the problem falls into two cases:\\[3mm]
            \textbf{Case 1: }If at least one of the two characterisation equations contains type-2 equations, we assume WLOG that the characterisation equation for $\vec{\theta}$ contains type-2 equation: $\vec{\alpha}_{j_l}^{\top}\vec{\theta}=\vec{\zeta}_l^{\top}\vec{\theta}$. By assumption (\romannumeral 5), $\vec{\alpha}_{j_l}^{\top}\vec{\theta}>0$. Then the two characterisation equations should match exactly, or the term with order $\vec{\alpha}_{j_l}^{\top}\vec{\theta}$ cannot be matched. So $\vec{\theta}=\widetilde{\vec{\theta}}$.\\[3mm]
            \textbf{Case 2: }If both characterisation equations contain only type-1 equations, then by Lemma~\ref{lem_characterisation_equation_2}, the two characterisation equations should match exactly and $\vec{\theta}=\widetilde{\vec{\theta}}$.\\[3mm]
            If $\operatorname{rank}\left\{\vec{\alpha}_1,\ldots,\vec{\alpha}_K\right\}<d$, we change variables to reduce dimension to $\operatorname{rank}\left\{\vec{\alpha}_1,\ldots,\vec{\alpha}_K\right\}$. Then the same proof is performed. Hence the uniqueness of characterisation equation is proved.\\[3mm]
            \textbf{Step 4: }Finally, we prove Lemma~\ref{lem_characterisation_equation_3} by the uniqueness of characterisation equation. Since for any $\{(\xi_1^{(n)},\ldots,\xi_J^{(n)})\}$ such that
            \begin{align*}
                \lim_{n\rightarrow \infty}\frac{(\log \xi_1^{(n)},\ldots,\log \xi_J^{(n)})^{\top}}{\log n}=(\nu_1,\ldots,\nu_J)^{\top},
            \end{align*}
            we have $\xi_1^{(n)}\gg \ldots\gg \xi_J^{(n)}\gg 1$. Then we can use the same method as in part (2) to construct characterisation equation at $(\nu_1,\ldots,\nu_J)^{\top}$. So by uniqueness of characterisation equation at $(\nu_1,\ldots,\nu_J)^{\top}$, $\vec{\theta}_n/\log n$ should converge to the same limit $\vec{\theta}(\nu_1,\ldots,\nu_J)$. Since $\nu_1>\ldots>\nu_J>0$, by varying $(\nu_1,\ldots,\nu_J)^{\top}$ in a small neighbourhood, the equations still satisfy all the conditions for the characterisation equation. By the continuity of linear equations, the solution should also be continuous when varying $(\nu_1,\ldots,\nu_J)^{\top}$ in a small neighbourhood. This implies that $\vec{\theta}(\nu_1,\ldots,\nu_J)$ is continuous at $(\nu_1,\ldots,\nu_J)^{\top}$. Hence the lemma is proved.
            \end{proof}
            
            Based on Lemma~\ref{lem_characterisation_equation_1}, Lemma~\ref{lem_characterisation_equation_2}, and Lemma~\ref{lem_characterisation_equation_3}, now we can verify Proposition~\ref{prop_characterisation_equation}.
            \renewcommand*{\proofname}{Proof of Proposition \ref{prop_characterisation_equation}}
    \begin{proof}
  Similar to Lemma~\ref{lem_characterisation_equation_3}, we can prove that $\vec{\theta}_n=\mathcal{O}(\log n)$ and denote $\vec{\theta}_n/\log n\rightarrow \vec{\theta}(\nu_1,\ldots,\nu_J)$. It can be readily shown that the existence of $\vec{\varphi}^{(n)}$ will introduce an $o(1)$ perturbation in the maximiser, which will not affect the limit of $\vec{\theta}_n/\log n$. Since the number of characterisation equations is finite if we omit the particular value of $\nu_1,\ldots,\nu_J$, we can assume that the characterisation equations have the same structure at $(\nu_1^{(n)},\ldots,\nu_J^{(n)})^{\top}$ except that the values of $(\nu_1^{(n)},\ldots,\nu_J^{(n)})^{\top}$ are different (or we analyse on a subsequence). Since $(\nu_1^{(n)},\ldots,\nu_J^{(n)})^{\top}\rightarrow (\widetilde{\nu}_1,\ldots,\widetilde{\nu}_J)^{\top}$, we can derive a limit linear equation by letting $n$ go to infinity. Since by assumption, the expansions of $\vec{\eta}_i$ and $\vec{\eta}_j$ with respect to the basis of characterisation contain disjoint terms, this implies that the limit linear equation is a valid characterisation equation in a small neighbourhood of $(\widetilde{\nu}_1,\ldots,\widetilde{\nu}_J)^{\top}$ if we allow ties among $(\nu_1,\ldots,\nu_J)^{\top}$. By the same uniqueness argument as in the proof of Lemma~\ref{lem_characterisation_equation_3}, we can show that the equation is the unique characterisation equation in the neighbourhood $\mathcal{O}$ which corresponds to the maximiser. Then by the same argument as in Lemma~\ref{lem_characterisation_equation_3}, the result is proved.
  \end{proof}

    \section{Proof of Theorem \ref{thm_information}}\label{sec:thm2}
    For notational simplicity, we assume WLOG that $\mat{\Sigma}_0=\mat{I}_K$ and let $\mat{A}$ absorb the transformation on $\mat{\Sigma}_0$. The proof of Theorem~2 follows a similar strategy to that of Theorem~1. By repeated differentiation, we identify different event types through their asymptotic behaviours. We therefore omit some routine details throughout the proof.
    \subsection{Preliminary Results}
    \noindent  We first state some preliminary results to be used in the proof of Theorem~2. The proofs of
these results are given in subsequent sections.

    The following proposition is analogous to Proposition~\ref{prop_likelihood} and provides the higher-order derivatives of the equation corresponding to the case in which the Fisher information is degenerate.
    \begin{prop}\label{prop_score}
        If the Fisher information matrix is singular at $\vec{\delta}_0=(\vec{\beta}_0,\mat{A}_0,\mat{\Sigma}_0)$, then there exists nonzero $\vec{w}=\{u_{j0}\in \mathbb{R},\vec{u}_j\in \mathbb{R}^{L_{1j}},\mat{V}_j\in \mathbb{R}^{L_{2j}\times K}:j=1,\ldots,J\}$ such that for any $t\in[0,T]$ and $n\geq 0$,
        \begin{align*}
        0=&\int_{\bt}\left[\sum_{j=1}^J\int_{0}^{t}(u_{j0}+\vec{u}_j^{\top}\vec{X}_j(s)+\vec{\theta}^{\top}\mat{V}_j^{\top}\vec{Z}_j(s))(\mathrm{d}N_j(s)-\lambda_j(s)\mathrm{d}s)\right] \notag\\
        &\times\prod_{j=1}^{J}\left[\prod_{s \leq t} (\lambda_{j}(s)^{\Delta N_{j}(s)} )e^{-\int_{0}^{t} \lambda_{j}(s) \mathrm{d} s} \right]\left(\sum_{j=1}^J\lambda_j(t+0)\right)^n\phi_K(\vec{\theta} ;\vec{0}, \mat{I}_K) \mathrm{d} \vec{\theta}\notag\\
        -&\int_{\bt}\left[\sum_{j=1}^J(u_{j0}+\vec{u}_j^{\top}\vec{X}_j(t+0)+\vec{\theta}^{\top}\mat{V}_j^{\top}\vec{Z}_j(t+0))\lambda_{j}(t+0)\right]\notag\\
        &\times\prod_{j=1}^{J}\left[\prod_{s \leq t} (\lambda_{j}(s)^{\Delta N_{j}(s)} )e^{-\int_{0}^{t} \lambda_{j}(s) \mathrm{d} s} \right]\left(\sum_{j=1}^J\lambda_j(t+0)\right)^{n-1}\phi_K(\vec{\theta} ;\vec{0}, \mat{I}_K) \mathrm{d} \vec{\theta}
    \end{align*}
    and for each $m \in\{1,\ldots,J\}$,
    \begin{align*}
        0=&\int_{\bt}\Bigg[\sum_{j=1}^J\int_{0}^{t}(u_{j0}+\vec{u}_j^{\top}\vec{X}_j(s)+\vec{\theta}^{\top}\mat{V}_j^{\top}\vec{Z}_j(s))(\mathrm{d}N_j(s)-\lambda_j(s)\mathrm{d}s)+u_{m0}+\vec{u}_m^{\top}\vec{X}_m(t+0)\notag\\
        &+\vec{\theta}^{\top}\mat{V}_m^{\top}\vec{Z}_m(t+0)\Bigg]\lambda_m(t+0)\prod_{j=1}^{J}\left[\prod_{s \leq t} (\lambda_{j}(s)^{\Delta N_{j}(s)} )e^{-\int_{0}^{t} \lambda_{j}(s) \mathrm{d} s} \right]\left(\sum_{j=1}^J\lambda_j(t+0)\right)^n\phi_K(\vec{\theta} ;\vec{0}, \mat{I}_K) \mathrm{d} \vec{\theta}\notag\\
        -&\int_{\bt}\left[\sum_{j=1}^J(u_{j0}+\vec{u}_j^{\top}\vec{X}_j(t+0)+\vec{\theta}^{\top}\mat{V}_j^{\top}\vec{Z}_j(t+0))\lambda_{j}(t+0)\right]\notag\\
        &\times\lambda_m(t+0)\prod_{j=1}^{J}\left[\prod_{s \leq t} (\lambda_{j}(s)^{\Delta N_{j}(s)} )e^{-\int_{0}^{t} \lambda_{j}(s) \mathrm{d} s} \right]\left(\sum_{j=1}^J\lambda_j(t+0)\right)^{n-1}\phi_K(\vec{\theta} ;\vec{0}, \mat{I}_K) \mathrm{d} \vec{\theta}~~\text{a.s.}
    \end{align*}
    \end{prop}

   The following proposition is analogous to Proposition~\ref{prop_laplace} and provides a Laplace-type approximation for the integral appearing in the proof of Theorem~2.
    \begin{prop}\label{prop_approximation}
        Let $\vec{\alpha}_1,\ldots,\vec{\alpha}_K,\{\vec{\xi}_n\},\vec{\gamma}\in\mathbb{R}^d$ be $d$-vectors and $w_1,\ldots,w_K$ be positive constants. Define $f_n(\vec{\theta})=-\sum_{k=1}^Kw_k\exp(\vec{\alpha}_k^{\top}\vec{\theta})+\vec{\xi}_n^{\top}\vec{\theta}-\frac{1}{2}\vec{\theta}^{\top}\vec{\theta}$ and denote its unique maximiser by $\hat{\vec{\theta}}_n$. Suppose $\vec{\gamma}^{\top}\hat{\vec{\theta}}_n\rightarrow \infty$. Denote the negative Hessian matrix of function $f_n$ at $\vec{\theta}$ by $\mat{I}(\vec{\theta})=\mat{I}_d+\sum_{k=1}^Kw_k\exp(\vec{\alpha}_k^{\top}\vec{\theta})\vec{\alpha}_k\vec{\alpha}_k^{\top}$. Then it holds that
        \begin{align*}
            M^{-1}\cdot\frac{(\vec{\gamma}^{\top}\hat{\vec{\theta}}_n)\exp(f_n(\hat{\vec{\theta}}_n))}{\sqrt{\operatorname{det}(\mat{I}(\hat{\vec{\theta}}_n))}}\leq \int_{\bt}(2\pi)^{-d/2}(\vec{\gamma}^{\top}\vec{\theta})\exp(f_n(\vec{\theta}))\mathrm{d}\vec{\theta}\leq M\cdot\frac{(\vec{\gamma}^{\top}\hat{\vec{\theta}}_n)\exp(f_n(\hat{\vec{\theta}}_n))}{\sqrt{\operatorname{det}(\mat{I}(\hat{\vec{\theta}}_n))}},
        \end{align*}
        where $M>0$ is a constant that does not depend on $n$.
    \end{prop}
    \subsection{Main Proof of Theorem \ref{thm_information}}
    \renewcommand*{\proofname}{Proof of Theorem \ref{thm_information}}
    \begin{proof}
    We first show that $\mat{I}(\vec{\delta}_0)$ is finite. Suppose $n=1$. Then the complete log-likelihood is 
        \begin{align*}
            \log \mathcal{L}_1(\vec{\delta}| \textbf{N}, \textbf{X}, \textbf{Z},\vec{\theta}) =&\sum_{j=1}^J \int^{\infty}_0\big(\beta_{j0} + \vec{\beta}_{j}^{\top} \vec{X}_{j}(t) + \vec{\theta}^{\top} \mat{A}_j^{\top} \vec{Z}_{j}(t)\big)\mathrm{d}N_{j}(t)\\
            &-\sum_{j=1}^J\int^{\infty}_0 Y_j(t)\exp\left( \beta_{j0} + \vec{\beta}_{j}^{\top} \vec{X}_{j}(t) + \vec{\theta}^{\top} \mat{A}_j^{\top} \vec{Z}_{j}(t) \right)  \mathrm{d}t.
        \end{align*}
        For any nonzero $\vec{w}=\{u_{j0}\in \mathbb{R},\vec{u}_j\in \mathbb{R}^{L_{1j}},\mat{V}_j\in \mathbb{R}^{L_{2j}\times K}:j=1,\ldots,J\}$, the score function in direction $\vec{w}$ is given by
        \begin{align*}
            \ell_{\vec{w}}=\sum_{j=1}^J \int_0^{\infty}(u_{j0}+\vec{u}_j^{\top}\vec{X}_j(t)+\vec{\theta}^{\top}\mat{V}_j^{\top}\vec{Z}_j(t))(\mathrm{d}N_j(t)-Y_j(t)\lambda_j(t)\mathrm{d}t).
        \end{align*}
        By the law of total variance, we have
        \begin{align*}
            &\operatorname{var}\left(\left\{\frac{\partial}{\partial \vec{\delta}} \log \mathcal{L}_1(\vec{\delta} | \textbf{N}, \textbf{X}, \textbf{Z})\right\}^{\top}\vec{w}\right)\\
            =&\operatorname{var}_{\textbf{N},\textbf{X}, \textbf{Z}}\mathbb{E}_{\vec{\theta}}\left(\ell_{\vec{w}}|\textbf{N}, \textbf{X}, \textbf{Z}\right)\\
            \leq&\operatorname{var}(\ell_{\vec{w}})\\
            \lesssim&\sum_{j=1}^J\mathbb{E}_{\bt}\left[\int_0^{\tau}(u_{j0}+\vec{u}_j^{\top}\vec{X}_j(t)+\vec{\theta}^{\top}\mat{V}_j^{\top}\vec{Z}_j(t))^2\exp\left( \beta_{j0} + \vec{\beta}_{j}^{\top} \vec{X}_{j}(t) + \vec{\theta}^{\top} \mat{A}_j^{\top} \vec{Z}_{j}(t) \right)\mathrm{d}t\right]\\
            \leq& C,
        \end{align*}
        where $\tau>0$ is the duration of the study. Here $C>0$ is a constant since $\vec{X},\vec{Z}$ are uniformly bounded by Condition (b). Since the choice of $\vec{w}$ is arbitrary, $\mat{I}(\vec{\delta}_0)$ is finite.\\[3mm]
        Now we use the method of contradiction to prove Theorem~2. For notational simplicity, we denote $\mu_j(t)=\beta_{j0}+\vec{\beta}_{j}^{\top} \vec{X}_{j}(t)$. Now we fix an arbitrary trajectory with positive density. Then by Condition (e), $[0,T]$ can be divided into $v$ finite intervals: $[0,t_1],(t_1,t_2],\ldots,(t_{v-1},t_v]$ such that the values of $\vec{X}$ and $\vec{Z}$ are constant on each interval. We then use an induction argument to prove that for any $j,j_1,j_2$ and $0\leq t,s\leq T$, it holds that
        \begin{align}\label{eqq4}
            u_{j0}+\vec{u}_j^{\top}\vec{X}_j(t)&=0,\notag\\
            (\mat{V}_{j_1}^{\top}\vec{Z}_{j_1}(t))^{\top}(\mat{A}_{j_2}^{\top}\vec{Z}_{j_2}(s))&=0.
        \end{align}
        We first prove that (\ref{eqq4}) holds on interval $[0,t_1]$. We choose $t=0$ in Proposition~\ref{prop_score} to get
        \begin{align}\label{eqq5}
        0=&\int_{\bt}\left[\sum_{j=1}^J(u_{j0}+\vec{u}_j^{\top}\vec{X}_j(0)+\vec{\theta}^{\top}\mat{V}_j^{\top}\vec{Z}_j(0))\lambda_{j}(0)\right]\left(\sum_{j=1}^J\lambda_j(0)\right)^{n-1}\phi_K(\vec{\theta} ;\vec{0}, \mat{I}_K) \mathrm{d} \vec{\theta}.
    \end{align}
            Letting $n$ increase by 1, by explicit integration of (\ref{eqq5}) we have
            \begin{align}\label{eqq6}
                0=&\sum_{j=1}^J\sum_{1\leq j_1,\ldots,j_n\leq J}\exp\Bigg(\mu_{j}(0)+\sum_{k=1}^n \mu_{j_k}(0)+\frac{1}{2}\Big[\sum_{k=1}^n \mat{A}^{\top}_{j_k}\vec{Z}_{j_k}(0)+\mat{A}^{\top}_{j}\vec{Z}_{j}(0)\Big]^{\top}\notag\\
                &\Big[\sum_{k=1}^n \mat{A}^{\top}_{j_k}\vec{Z}_{j_k}(0)+\mat{A}^{\top}_{j}\vec{Z}_{j}(0)\Big]\Bigg)\left(u_{j0}+\vec{u}_j^{\top}\vec{X}_j(0)+\Big[\sum_{k=1}^n \mat{A}^{\top}_{j_k}\vec{Z}_{j_k}(0)+\mat{A}^{\top}_{j}\vec{Z}_{j}(0)\Big]^{\top}\mat{V}_j^{\top}\vec{Z}_j(0)\right).
            \end{align}
            We assume WLOG that $\vec{Z}_1(0),\ldots,\vec{Z}_J(0)$ are all nonzero. By excluding a set of measure zero in the parameter space, we assume WLOG that $\{(\mat{A}^{\top}_{j_1}\vec{Z}_{j_1}(0))^{\top}\mat{A}^{\top}_{j_2}\vec{Z}_{j_2}(0):1\leq j_1\leq j_2\leq J\}$ are distinct and assume that $(\mat{A}^{\top}_{1}\vec{Z}_{1}(0))^{\top}\mat{A}^{\top}_{1}\vec{Z}_{1}(0)>\max_{j=2,\ldots,J}(\mat{A}^{\top}_{j}\vec{Z}_{j}(0))^{\top}\mat{A}^{\top}_{j}\vec{Z}_{j}(0)$. Furthermore, we assume WLOG that $(\mat{A}^{\top}_{1}\vec{Z}_{1}(0))^{\top}\mat{A}^{\top}_{1}\vec{Z}_{1}(0)>\ldots>(\mat{A}^{\top}_{1}\vec{Z}_{1}(0))^{\top}\mat{A}^{\top}_{J}\vec{Z}_{J}(0)$.\\[3mm]
            Similar to the proof in Lemma~\ref{lem_summation}, we can rank all terms in the right-hand side of (\ref{eqq6}) and prove that each term dominates the summation of all terms with lower order. For example, if $(\mat{A}^{\top}_{1}\vec{Z}_{1}(0))^{\top}\mat{V}_1^{\top}\vec{Z}_1(0)\neq 0$, we can show that
            \begin{align*}
                (n+1)\exp\left((n+1)\mu_1(0)+\frac{(n+1)^2}{2}(\mat{A}^{\top}_{1}\vec{Z}_{1}(0))^{\top}\mat{A}^{\top}_{1}\vec{Z}_{1}(0)\right)(\mat{A}^{\top}_{1}\vec{Z}_{1}(0))^{\top}\mat{V}_1^{\top}\vec{Z}_1(0)
            \end{align*}
            dominates the right-hand side of (\ref{eqq6}), which leads to a contradiction. Hence $(\mat{A}^{\top}_{1}\vec{Z}_{1}(0))^{\top}\mat{V}_1^{\top}\vec{Z}_1(0)=0$. Then we can prove that
            \begin{align*}
                \exp\left((n+1)\mu_1(0)+\frac{(n+1)^2}{2}(\mat{A}^{\top}_{1}\vec{Z}_{1}(0))^{\top}\mat{A}^{\top}_{1}\vec{Z}_{1}(0)\right)(u_{10}+\vec{u}_1^{\top}\vec{X}_1(0))
            \end{align*}
            dominates the right-hand side of (\ref{eqq6}) if $u_{10}+\vec{u}_1^{\top}\vec{X}_1(0)\neq 0$. Hence $u_{10}+\vec{u}_1^{\top}\vec{X}_1(0)= 0$. By this inductive method, we can show that for any $j,j_1,j_2=1,\ldots,J$, we have
            $u_{j0}+\vec{u}_j^{\top}\vec{X}_j(0)= 0$ and $(\mat{A}^{\top}_{j_1}\vec{Z}_{j_1}(0))^{\top}\mat{V}_{j_2}^{\top}\vec{Z}_{j_2}(0)=0$, which finishes the proof on $[0,t_1]$.\\[3mm]
            Now suppose that (\ref{eqq4}) is proved on interval $[0,t_q]$, we then prove that (\ref{eqq4}) also holds on interval $[0,t_{q+1}]$. By applying $t=t_q$ in Proposition~\ref{prop_score}, we have
    \begin{align}\label{eqq7}
        0=&\int_{\bt}\left[\sum_{j=1}^J\int_{0}^{t_q}(u_{j0}+\vec{u}_j^{\top}\vec{X}_j(s)+\vec{\theta}^{\top}\mat{V}_j^{\top}\vec{Z}_j(s))(\mathrm{d}N_j(s)-\lambda_j(s)\mathrm{d}s)\right] \notag\\
        &\times\prod_{j=1}^{J}\Big[\prod_{s \leq t_q} (\lambda_{j}(s)^{\Delta N_{j}(s)} )e^{-\int_{0}^{t_q} \lambda_{j}(s) \mathrm{d} s} \Big]\left(\sum_{j=1}^J\lambda_j(t_{q+1})\right)^n\phi_K(\vec{\theta} ;\vec{0}, \mat{I}_K) \mathrm{d} \vec{\theta}\notag\\
        -&\int_{\bt}\left[\sum_{j=1}^J(u_{j0}+\vec{u}_j^{\top}\vec{X}_j(t_{q+1})+\vec{\theta}^{\top}\mat{V}_j^{\top}\vec{Z}_j(t_{q+1}))\lambda_{j}(t_{q+1})\right]\notag\\
        &\times\prod_{j=1}^{J}\Big[\prod_{s \leq t_{q}} (\lambda_{j}(s)^{\Delta N_{j}(s)} )e^{-\int_{0}^{t_{q}} \lambda_{j}(s) \mathrm{d} s} \Big]\left(\sum_{j=1}^J\lambda_j(t_{q+1})\right)^{n-1}\phi_K(\vec{\theta} ;\vec{0}, \mat{I}_K) \mathrm{d} \vec{\theta}.
    \end{align}        
            Denote $t_0=0$. To simplify the notation, for any $k=0,\ldots,q-1,~j=1,\ldots,J$, we introduce the following notation:
        \[\begin{array}{ll}
            \vec{\varphi}=\sum_{j=1}^{J}\int_{0}^{t_q}\mat{A}^{\top}_j\vec{Z}_j(t)\mathrm{d}N_j(t),&\widetilde{\vec{\varphi}}=\sum_{j=1}^{J}\int_{0}^{t_q}\mat{V}^{\top}_j\vec{Z}_j(t)\mathrm{d}N_j(t)\\
            \vec{\alpha}_{kJ+j}=\mat{A}^{\top}_j\vec{Z}_j(t_{k+1}),& \widetilde{\vec{\alpha}}_{kJ+j}=\mat{V}^{\top}_j\vec{Z}_j(t_{k+1})\\
            w_{kJ+j}=\int_{t_k}^{t_{k+1}}\exp(\mu_j(s))\mathrm{d}s,&\widetilde{\mu}_j=u_{j0}+\vec{u}_j^{\top}\vec{X}_j(t_{q+1})\\
            \vec{\eta}_j=\mat{A}^{\top}_j\vec{Z}_j(t_{q+1}),&\widetilde{\vec{\eta}}_j=\mat{V}^{\top}_j\vec{Z}_j(t_{q+1}).
        \end{array}\]
        Denote $W=qJ$. For any $n$ and $\bm{\xi}^{(n)}=(\xi_2^{(n)},\ldots,\xi_J^{(n)})^{\top}$, we introduce the following notation:
        \begin{align*}
            f_n(\vec{\theta}\big|\bm{\xi}^{(n)})&=n\mu_1-\sum_{k=1}^{W}w_k \exp(\vec{\alpha}_k^{\top}\vec{\theta})+(\vec{\varphi}+n\vec{\eta}_1)^{\top}\vec{\theta}-\frac{1}{2}\vec{\theta}^{\top}\vec{\theta}-\sum_{j=2}^J\xi_j^{(n)}\left[(\vec{\eta}_1-\vec{\eta}_j)^{\top}\vec{\theta}+(\mu_1-\mu_j)\right],\\
            \phi_n(\bm{\xi}^{(n)})&=\int_{\bt}(2\pi)^{-\frac{K}{2}}\exp\left(f_n(\vec{\theta}\big|\bm{\xi}^{(n)})\right)\mathrm{d}\vec{\theta},\\
            \Delta_n(\bm{\xi}^{(n)})&=\binom{n}{n-\sum_{j=2}^J \xi_j^{(n)},\xi_2^{(n)},\ldots,\xi_J^{(n)}}=\frac{n!}{\left(n-\sum_{j=2}^J \xi_j^{(n)}\right)!\prod_{j=2}^J \xi_j^{(n)}!}.
        \end{align*}
        Furthermore, denote the unique maximiser of ${f}_n(\vec{\theta}\big|\bm{\xi}^{(n)})$ by $\vec{\theta}_n(\bm{\xi}^{(n)})$. For any $n\in\mathbb{N}_0$, define $\mathcal{O}_n=\{(\xi_2,\ldots,\xi_J)^{\top}\in\mathbb{N}_0^{J-1}:\sum_{j=2}^J\xi_j\leq n\}$. By induction assumption, we have $\vec{\alpha}_{k_1}^{\top}\widetilde{\vec{\alpha}}_{k_2}=\vec{\alpha}_{k}^{\top}\widetilde{\vec{\varphi}}=\widetilde{\vec{\alpha}}_k^{\top}{\vec{\varphi}}=\vec{\varphi}^{\top}\widetilde{\vec{\varphi}}=u_{j0}+\vec{u}_j^{\top}\vec{X}_j(t)=0$ for any $k,k_1,k_2=1,\ldots,W$, $j=1,\ldots,J$ and $0\leq t\leq t_q$. Then equation (\ref{eqq7}) can be explicitly characterised as
        \begin{align}\label{eqq8}
            0=&\sum_{\bm{\xi}^{(n)}\in\mathcal{O}_n}\widetilde{\vec{\varphi}}^{\top}\left(n\vec{\eta}_1-\sum_{j=2}^J\xi_j^{(n)}(\vec{\eta}_1-\vec{\eta}_j)\right)\Delta_n(\bm{\xi}^{(n)})\phi_n(\bm{\xi}^{(n)})\notag\\
            &-\sum_{\bm{\xi}^{(n)}\in\mathcal{O}_n}\sum_{k=1}^Ww_k\widetilde{\vec{\alpha}}_k^{\top}\left(n\vec{\eta}_1-\sum_{j=2}^J\xi_j^{(n)}(\vec{\eta}_1-\vec{\eta}_j)\right)\Delta_n(\bm{\xi}^{(n)})\int_{\bt}(2\pi)^{-\frac{K}{2}}\exp\left(f_{n}(\vec{\theta}\big|\bm{\xi}^{(n)})+\vec{\alpha}_k^{\top}\vec{\theta}\right)\mathrm{d}\vec{\theta}\notag\\
            &-\sum_{j=1}^J\widetilde{\mu}_j\exp(\mu_j)\sum_{\bm{\xi}^{(n-1)}\in\mathcal{O}_{n-1}}\Delta_{n-1}(\bm{\xi}^{(n-1)})\int_{\bt}(2\pi)^{-\frac{K}{2}}\exp\left(f_{n-1}(\vec{\theta}\big|\bm{\xi}^{(n-1)})+\vec{\eta}_j^{\top}\vec{\theta}\right)\mathrm{d}\vec{\theta}\notag\\
            &+\sum_{j=1}^J\sum_{k=1}^Ww_k\exp(\mu_j)\widetilde{\vec{\eta}}_j^{\top}\vec{\alpha}_k\sum_{\bm{\xi}^{(n-1)}\in\mathcal{O}_{n-1}}\Delta_{n-1}(\bm{\xi}^{(n-1)})\int_{\bt}(2\pi)^{-\frac{K}{2}}\exp\left(f_{n-1}(\vec{\theta}\big|\bm{\xi}^{(n-1)})+(\vec{\eta}_j
            +\vec{\alpha}_k)^{\top}\vec{\theta}\right)\mathrm{d}\vec{\theta}\notag\\
            &-\sum_{\bm{\xi}^{(n-1)}\in\mathcal{O}_{n-1}}\sum_{j=1}^J\exp(\mu_j)\left(\vec{\varphi}+(n-1)\vec{\eta}_1-\sum_{j=2}^J\xi_j^{(n-1)}(\vec{\eta}_1-\vec{\eta}_j)\right)^{\top}\widetilde{\vec{\eta}}_j\notag\\
            &\times\Delta_{n-1}(\bm{\xi}^{(n-1)})\int_{\bt}(2\pi)^{-\frac{K}{2}}\exp\left(f_{n-1}(\vec{\theta}\big|\bm{\xi}^{(n-1)})+\vec{\eta}_j^{\top}\vec{\theta}\right)\mathrm{d}\vec{\theta}.
        \end{align}
        By part (1) in Proposition~\ref{prop_canonical_projection}, there exists $\mathcal{H}_{\vec{\eta}_1},\ldots,\mathcal{H}_{\vec{\eta}_J}$ corresponding to $\vec{\eta}_1,\ldots,\vec{\eta}_J$. By excluding a set of measure zero in the parameter space, we assume WLOG that $\|P_{\mathcal{H}^{\perp}_{\vec{\eta}_1}}\vec{\eta}_1\|$ achieves the unique maximum among $\|P_{\mathcal{H}^{\perp}_{\vec{\eta}_1}}\vec{\eta}_1\|,\ldots,\|P_{\mathcal{H}^{\perp}_{\vec{\eta}_J}}\vec{\eta}_J\|$.  Similar to the proof in Theorem~1, we divide the problem into two cases:\\[3mm]
        \textbf{Case 1: }$\|P_{\mathcal{H}^{\perp}_{\vec{\eta}_1}}\vec{\eta}_1\|>0$.\\[3mm]
        Then by arguments similar to those used in the proof of Theorem~1, there exists linearly independent $\vec{\alpha}_{k_1},\ldots,\vec{\alpha}_{k_m}$ such that $\mathcal{H}_{\vec{\eta}_1}=\operatorname{span}\{\vec{\alpha}_{k_1},\ldots,\vec{\alpha}_{k_m}\}$ and $P_{\mathcal{H}_{\vec{\eta}_1}}\vec{\eta}_1=\sum_{j=1}^m\gamma_{k_j}\vec{\alpha}_{k_j}$. For notational simplicity, we denote the right-hand side of (\ref{eqq8}) as
        \begin{align*}
            &n\widetilde{\vec{\varphi}}^{\top}\vec{\eta}_1\phi_n(\mathbf{0})-n\sum_{j=1}^mw_{k_j}\widetilde{\vec{\alpha}}_{k_j}^{\top}\vec{\eta}_1\int_{\bt}(2\pi)^{-\frac{K}{2}}\exp\left(f_{n}(\vec{\theta}\big|\mathbf{0})+\vec{\alpha}_{k_j}^{\top}\vec{\theta}\right)\mathrm{d}\vec{\theta}\\
            &-\widetilde{\mu}_1\phi_n(\mathbf{0})+\sum_{j=1}^mw_{k_j}\widetilde{\vec{\eta}}_1^{\top}\vec{\alpha}_{k_j}\int_{\bt}(2\pi)^{-\frac{K}{2}}\exp\left(f_{n}(\vec{\theta}\big|\mathbf{0})+\vec{\alpha}_{k_j}^{\top}\vec{\theta}\right)\mathrm{d}\vec{\theta}-\left(\vec{\varphi}+n\vec{\eta}_1\right)^{\top}\widetilde{\vec{\eta}}_1\phi_n(\mathbf{0})+\mathcal{E}_n.
        \end{align*}
        We can show that there exists a constant $c>0$ such that
        \begin{align*}
            &\left|n\widetilde{\vec{\varphi}}^{\top}\vec{\eta}_1\phi_n(\mathbf{0})-n\sum_{j=1}^mw_{k_j}\widetilde{\vec{\alpha}}_{k_j}^{\top}\vec{\eta}_1\int_{\bt}(2\pi)^{-\frac{K}{2}}\exp\left(f_{n}(\vec{\theta}\big|\mathbf{0})+\vec{\alpha}_{k_j}^{\top}\vec{\theta}\right)\mathrm{d}\vec{\theta}\right.\\
            &\left.-\widetilde{\mu}_1\phi_n(\mathbf{0})+\sum_{j=1}^mw_{k_j}\widetilde{\vec{\eta}}_1^{\top}\vec{\alpha}_{k_j}\int_{\bt}(2\pi)^{-\frac{K}{2}}\exp\left(f_{n}(\vec{\theta}\big|\mathbf{0})+\vec{\alpha}_{k_j}^{\top}\vec{\theta}\right)\mathrm{d}\vec{\theta}-\left(\vec{\varphi}+n\vec{\eta}_1\right)^{\top}\widetilde{\vec{\eta}}_1\phi_n(\mathbf{0})\right|=\left|\mathcal{E}_n\right|\\
            \leq&\exp(-cn)\min_{j=1,\ldots,m}\left\{\int_{\bt}(2\pi)^{-\frac{K}{2}}\exp\left(f_{n}(\vec{\theta}\big|\mathbf{0})+\vec{\alpha}_{k_j}^{\top}\vec{\theta}\right)\mathrm{d}\vec{\theta}\right\}\land \phi_n(\mathbf{0}).
        \end{align*}
        Then by similar proof as in Theorem~1, we expand
        \begin{align*}
            &n\widetilde{\vec{\varphi}}^{\top}\vec{\eta}_1\phi_n(\mathbf{0})-n\sum_{j=1}^mw_{k_j}\widetilde{\vec{\alpha}}_{k_j}^{\top}\vec{\eta}_1\int_{\bt}(2\pi)^{-\frac{K}{2}}\exp\left(f_{n}(\vec{\theta}\big|\mathbf{0})+\vec{\alpha}_{k_j}^{\top}\vec{\theta}\right)\mathrm{d}\vec{\theta}\\
            &-\widetilde{\mu}_1\phi_n(\mathbf{0})+\sum_{j=1}^mw_{k_j}\widetilde{\vec{\eta}}_1^{\top}\vec{\alpha}_{k_j}\int_{\bt}(2\pi)^{-\frac{K}{2}}\exp\left(f_{n}(\vec{\theta}\big|\mathbf{0})+\vec{\alpha}_{k_j}^{\top}\vec{\theta}\right)\mathrm{d}\vec{\theta}-\left(\vec{\varphi}+n\vec{\eta}_1\right)^{\top}\widetilde{\vec{\eta}}_1\phi_n(\mathbf{0})
        \end{align*}
        in infinite series. By matching finite terms in decreasing order whose order differences with the leading term are smaller than $\exp(-cn)$, we can show if any of the following quantities: $\widetilde{\vec{\varphi}}^{\top}\vec{\eta}_1-\widetilde{\vec{\eta}}_1^{\top}\vec{\eta}_1$, $\widetilde{\mu}_1+\vec{\varphi}^{\top}\widetilde{\vec{\eta}}_1$, $\widetilde{\vec{\alpha}}_{k_j}^{\top}\vec{\eta}_1,j=1,\ldots,m$ and ${\vec{\alpha}}_{k_j}^{\top}\widetilde{\vec{\eta}}_1,j=1,\ldots,m$ are nonzero, by arguments similar to those used in the proof in Theorem~1, there exists $l\in\mathbb{N}$ such that
        \begin{align*}
            &\left|n\widetilde{\vec{\varphi}}^{\top}\vec{\eta}_1\phi_n(\mathbf{0})-n\sum_{j=1}^mw_{k_j}\widetilde{\vec{\alpha}}_{k_j}^{\top}\vec{\eta}_1\int_{\bt}(2\pi)^{-\frac{K}{2}}\exp\left(f_{n}(\vec{\theta}\big|\mathbf{0})+\vec{\alpha}_{k_j}^{\top}\vec{\theta}\right)\mathrm{d}\vec{\theta}\right.\\
            &\left.-\widetilde{\mu}_1\phi_n(\mathbf{0})+\sum_{j=1}^mw_{k_j}\widetilde{\vec{\eta}}_1^{\top}\vec{\alpha}_{k_j}\int_{\bt}(2\pi)^{-\frac{K}{2}}\exp\left(f_{n}(\vec{\theta}\big|\mathbf{0})+\vec{\alpha}_{k_j}^{\top}\vec{\theta}\right)\mathrm{d}\vec{\theta}-\left(\vec{\varphi}+n\vec{\eta}_1\right)^{\top}\widetilde{\vec{\eta}}_1\phi_n(\mathbf{0})\right|\\
            \geq& n^{-l}\min_{j=1,\ldots,m}\left\{\int_{\bt}(2\pi)^{-\frac{K}{2}}\exp\left(f_{n}(\vec{\theta}\big|\mathbf{0})+\vec{\alpha}_{k_j}^{\top}\vec{\theta}\right)\mathrm{d}\vec{\theta}\right\}\land \phi_n(\mathbf{0}),
        \end{align*}
        which leads to a contradiction. Hence for any $j=1,\ldots,m$ we have
        \begin{align*}
            \widetilde{\vec{\varphi}}^{\top}\vec{\eta}_1-\widetilde{\vec{\eta}}_1^{\top}\vec{\eta}_1=\widetilde{\mu}_1+\vec{\varphi}^{\top}\widetilde{\vec{\eta}}_1=\widetilde{\vec{\alpha}}_{k_j}^{\top}\vec{\eta}_1={\vec{\alpha}}_{k_j}^{\top}\widetilde{\vec{\eta}}_1=0.
        \end{align*}
        Then we use similar method as in the proof of Proposition~\ref{lem_summation} to rank all terms in the right-hand side of (\ref{eqq8}) in decreasing order. By excluding a set of measure zero in the parameter space, we can assume that there are no ties in the ranking. Then we can use similar method to show that each term dominates the summation of all terms with lower rank. Hence we can prove inductively that each term should be strictly equal to 0. By this method, we can prove that for any $j,j_1,j_2=1,\ldots,J$ and $k=1,\ldots,W$, we have
        \begin{align*}
            \widetilde{\vec{\varphi}}^{\top}\vec{\eta}_j=\widetilde{\vec{\alpha}}_k^{\top}\vec{\eta}_j={\vec{\alpha}}_k^{\top}\widetilde{\vec{\eta}}_j=\widetilde{\vec{\eta}}_{j_1}^{\top}\vec{\eta}_{j_2}=\widetilde{\mu}_j+\vec{\varphi}^{\top}\widetilde{\vec{\eta}}_j=0.
        \end{align*}
        Since ${\vec{\alpha}}_k^{\top}\widetilde{\vec{\eta}}_j=0$ for any $j=1,\ldots,J$ and $k=1,\ldots,W$, $\vec{\varphi}^{\top}\widetilde{\vec{\eta}}_j$ is also equal to zero. So we have $\widetilde{\mu}_j=0$ for any $j=1,\ldots,J$. Hence we finish the proof on $[0,t_{q+1}]$.\\[3mm]
        \textbf{Case 2: }$\|P_{\mathcal{H}^{\perp}_{\vec{\eta}_1}}\vec{\eta}_1\|=0$.\\[3mm]
        In this case, $\vec{\eta}_1,\ldots,\vec{\eta}_J\in \mathcal{X}:= \{\sum_{k=1}^Wc_k\vec{\alpha}_k:c_1,\ldots,c_W\geq 0\}$ by Proposition~\ref{prop_canonical_projection}. By Proposition~\ref{lem_canonical_expansion}, for any $j=1,\ldots,J$, there exists a canonical expansion for $\vec{\eta}_j$ under $\vec{\alpha}_1,\ldots,\vec{\alpha}_W$ as: $\vec{\eta}_j=\sum_{k=1}^{m_j}c_{j,k}\vec{\alpha}_{j,k}$, where the canonical expansion is unique in the sense that $\sum_{k=1}^{m_j}c_{j,k}$ is uniquely determined for each $j=1,\ldots,J$.\\[3mm]
        We assume WLOG that
        \begin{align*}
            \sum_{k=1}^{m_1}c_{1,k}&=\underset{j=1,\ldots,J}{\max}\sum_{k=1}^{m_j}c_{j,k}.
        \end{align*}
        We only consider the case where $\sum_{k=1}^{m_1}c_{1,k}$ is the unique largest term among $\sum_{k=1}^{m_1}c_{1,k},\ldots,\sum_{k=1}^{m_J}c_{J,k}$. By the proof in Theorem~1, we can find concentration point $(\nu_2,\ldots,\nu_J)^{\top}$ for the summation $\sum_{\bm{\xi}^{(n)}\in\mathcal{O}_n}\Delta_n(\bm{\xi}^{(n)})\phi_n(\bm{\xi}^{(n)})$. We assume WLOG that $1>\nu_2\geq \ldots\geq \nu_p>\nu_{p+1}=\ldots=\nu_{J}=0$. We only consider the case where $\nu_2,\ldots,\nu_p$ are distinct. By the proof in Theorem~1, we can construct unique characterisation equation in the neighbourhood of $(\nu_2,\ldots,\nu_J)^{\top}$, where the solution of characterisation equation has continuity property by Proposition~\ref{prop_characterisation_equation}. Denote the solution of characterisation equation by $\vec{\theta}(\nu_2,\ldots,\nu_J)$ at $(\nu_2,\ldots,\nu_J)^{\top}$ and denote the basis of characterisation equation at $(\nu_2,\ldots,\nu_J)^{\top}$ by $\{\vec{\alpha}_{j_1},\ldots,\vec{\alpha}_{j_k}\}$. Since $\vec{\eta}_1,\ldots,\vec{\eta}_J\in \operatorname{span}\{\vec{\alpha}_1,\ldots,\vec{\alpha}_W\}$, by induction assumption we can easily see that $\vec{\eta}_j^{\top}\widetilde{\vec{\alpha}}_k=\vec{\eta}_j^{\top}\widetilde{\vec{\varphi}}=0$ for any $j=1,\ldots,J$ and $k=1,\ldots,W$.\\[3mm]
        By the construction of characterisation equation, after excluding a set of measure zero in the parameter space, if there exists $1\leq\widetilde{p}\leq p$ such that the expansion of $\vec{\eta}_{\widetilde{p}}$ in the equation is nondegenerate, then the expansions of  $\vec{\eta}_{\widetilde{p}+1},\ldots,\vec{\eta}_{p}$ are all nondegenerate. Moreover, the characterisation equation contains type-1 equations only. For simplicity of proof, we consider the case where $p=2$. Moreover, we assume that the expansion of $\vec{\eta}_1$ is degenerate and the expansion of $\vec{\eta}_2$ is nondegenerate. Since $\vec{\eta}_1$ has degenerate expansion, by induction assumption we can easily show that $\widetilde{\vec{\eta}}_1^{\top}\vec{\alpha}_k=\widetilde{\vec{\eta}}_1^{\top}\vec{\varphi}=0$ for any $k=1,\ldots,W$. Then by applying $t=t_q$ in Proposition~\ref{prop_score}, we have
        \begin{align}\label{eqq9}
            &\sum_{\bm{\xi}^{(n)}\in\mathcal{O}_{n}}\Delta_{n}(\bm{\xi}^{(n)})\sum_{j=1}^J\widetilde{\mu}_j\exp(\mu_j)\int_{\bt}(2\pi)^{-\frac{K}{2}}\exp\left(f_{n}(\vec{\theta}\big|\bm{\xi}^{(n)})+\vec{\eta}_j^{\top}\vec{\theta}\right)\mathrm{d}\vec{\theta}\notag\\
            +&\sum_{\bm{\xi}^{(n)}\in\mathcal{O}_{n}}\Delta_{n}(\bm{\xi}^{(n)})\sum_{j=1}^J\exp(\mu_j)\int_{\bt}(2\pi)^{-\frac{K}{2}}\widetilde{\vec{\eta}}_j^{\top}\vec{\theta}\exp\left(f_{n}(\vec{\theta}\big|\bm{\xi}^{(n)})+\vec{\eta}_j^{\top}\vec{\theta}\right)\mathrm{d}\vec{\theta}=0.
        \end{align}
        Since the expansion of $\vec{\eta}_1$ is degenerate, i.e., $\vec{Z}_1(t_{q+1})\in \operatorname{span}\{\vec{Z}_1(t_1),\ldots,\vec{Z}_1(t_q)\}$, then by induction assumption we can easily show that $\widetilde{\vec{\eta}}_1^{\top}\vec{\alpha}_k=\widetilde{\vec{\eta}}_1^{\top}\vec{\varphi}=0$ for any $k=1,\ldots,W$. Hence
        \begin{align*}
            \sum_{\bm{\xi}^{(n)}\in\mathcal{O}_{n}}\Delta_{n}(\bm{\xi}^{(n)})\exp(\mu_1)\int_{\bt}(2\pi)^{-\frac{K}{2}}\widetilde{\vec{\eta}}_1^{\top}\vec{\theta}\exp\left(f_{n}(\vec{\theta}\big|\bm{\xi}^{(n)})+\vec{\eta}_1^{\top}\vec{\theta}\right)\mathrm{d}\vec{\theta}=0.
        \end{align*}
        Then we can prove that
        \begin{align*}
            -\sum_{\bm{\xi}^{(n)}\in\mathcal{O}_{n}}\Delta_{n}(\bm{\xi}^{(n)})\widetilde{\mu}_1\exp(\mu_1)\int_{\bt}(2\pi)^{-\frac{K}{2}}\exp\left(f_{n}(\vec{\theta}\big|\bm{\xi}^{(n)})+\vec{\eta}_1^{\top}\vec{\theta}\right)\mathrm{d}\vec{\theta}
        \end{align*}
        dominates the left-hand side of (\ref{eqq9}) if $\widetilde{\mu}_1\neq 0$, which leads to a contradiction. Hence $\widetilde{\mu}_1=0$. Then (\ref{eqq9}) turns into
        \begin{align}\label{eqq10}
            &\sum_{\bm{\xi}^{(n)}\in\mathcal{O}_{n}}\Delta_{n}(\bm{\xi}^{(n)})\sum_{j=2}^J\widetilde{\mu}_j\exp(\mu_j)\int_{\bt}(2\pi)^{-\frac{K}{2}}\exp\left(f_{n}(\vec{\theta}\big|\bm{\xi}^{(n)})+\vec{\eta}_j^{\top}\vec{\theta}\right)\mathrm{d}\vec{\theta}\notag\\
            +&\sum_{\bm{\xi}^{(n)}\in\mathcal{O}_{n}}\Delta_{n}(\bm{\xi}^{(n)})\sum_{j=2}^J\exp(\mu_j)\int_{\bt}(2\pi)^{-\frac{K}{2}}\widetilde{\vec{\eta}}_j^{\top}\vec{\theta}\exp\left(f_{n}(\vec{\theta}\big|\bm{\xi}^{(n)})+\vec{\eta}_j^{\top}\vec{\theta}\right)\mathrm{d}\vec{\theta}=0
        \end{align}
        Following similar arguments as in the proof of Corollary~\ref{cor_likelihood}, by adding a $m$-th event type at the right end point we can similarly show that for any $m=2,\ldots,J$, it holds that
        \begin{align}\label{eqq11}
            &-\sum_{\bm{\xi}^{(n+1)}\in\mathcal{O}_{n+1}}\Delta_{n+1}(\bm{\xi}^{(n+1)})\widetilde{\mu}_m\int_{\bt}(2\pi)^{-\frac{K}{2}}\exp\left(f_{n+1}(\vec{\theta}\big|\bm{\xi}^{(n+1)})\right)\mathrm{d}\vec{\theta}\notag\\
            &-\sum_{\bm{\xi}^{(n+1)}\in\mathcal{O}_{n+1}}\Delta_{n+1}(\bm{\xi}^{(n+1)})\int_{\bt}(2\pi)^{-\frac{K}{2}}\widetilde{\vec{\eta}}_m^{\top}\vec{\theta}\exp\left(f_{n+1}(\vec{\theta}\big|\bm{\xi}^{(n+1)})\right)\mathrm{d}\vec{\theta}\notag\\
            &+\sum_{\bm{\xi}^{(n)}\in\mathcal{O}_{n}}\Delta_{n}(\bm{\xi}^{(n)})\sum_{j=2}^J\widetilde{\mu}_j\exp(\mu_j)\int_{\bt}(2\pi)^{-\frac{K}{2}}\exp\left(f_{n}(\vec{\theta}\big|\bm{\xi}^{(n)})+(\vec{\eta}_j+\vec{\eta}_m)^{\top}\vec{\theta}\right)\mathrm{d}\vec{\theta}\notag\\
            &+\sum_{\bm{\xi}^{(n)}\in\mathcal{O}_{n}}\Delta_{n}(\bm{\xi}^{(n)})\sum_{j=2}^J\exp(\mu_j)\int_{\bt}(2\pi)^{-\frac{K}{2}}\widetilde{\vec{\eta}}_j^{\top}\vec{\theta}\exp\left(f_{n}(\vec{\theta}\big|\bm{\xi}^{(n)})+(\vec{\eta}_j+\vec{\eta}_m)^{\top}\vec{\theta}\right)\mathrm{d}\vec{\theta}=0.
        \end{align}
        If $\widetilde{\vec{\eta}}_{m}^{\top}\vec{\theta}(\nu_2,\ldots,\nu_J)\neq0$, by Proposition~\ref{prop_approximation} we can show that
        \begin{align}
            -\sum_{\bm{\xi}^{(n+1)}\in\mathcal{O}_{n+1}}\Delta_{n+1}(\bm{\xi}^{(n+1)})\int_{\bt}(2\pi)^{-\frac{K}{2}}\widetilde{\vec{\eta}}_m^{\top}\vec{\theta}\exp\left(f_{n+1}(\vec{\theta}\big|\bm{\xi}^{(n+1)})\right)\mathrm{d}\vec{\theta}\notag
        \end{align}
        dominates the summation on the left-hand side of (\ref{eqq11}). Then by arguments similar to those used in the proof of Theorem~1, this leads to a contradiction. Hence $\widetilde{\vec{\eta}}_{m}^{\top}\vec{\theta}(\nu_2,\ldots,\nu_J)=0$. Then we can show that
        \begin{align*}
            -\sum_{\bm{\xi}^{(n+1)}\in\mathcal{O}_{n+1}}\Delta_{n+1}(\bm{\xi}^{(n+1)})\widetilde{\mu}_m\int_{\bt}(2\pi)^{-\frac{K}{2}}\exp\left(f_{n+1}(\vec{\theta}\big|\bm{\xi}^{(n+1)})\right)\mathrm{d}\vec{\theta}
        \end{align*}
        dominates the left-hand side of (\ref{eqq11}) if $\widetilde{\mu}_m\neq 0$, which leads to a contradiction. Hence $\widetilde{\mu}_m= 0$ for any $m=1,\ldots,J$. Then equation (\ref{eqq11}) turns into
        \begin{align}\label{eqq12}
            \sum_{\bm{\xi}^{(n)}\in\mathcal{O}_{n}}\Delta_{n}(\bm{\xi}^{(n)})\sum_{j=2}^J\exp(\mu_j)\int_{\bt}(2\pi)^{-\frac{K}{2}}\widetilde{\vec{\eta}}_j^{\top}\vec{\theta}\exp\left(f_{n}(\vec{\theta}\big|\bm{\xi}^{(n)})+\vec{\eta}_j^{\top}\vec{\theta}\right)\mathrm{d}\vec{\theta}=0.
        \end{align}
        By arguments similar to those used in the proof of Theorem~1, we expand the left-hand side of (\ref{eqq12}) in decreasing order. Since we need to prove that $\widetilde{\vec{\eta}}_j^{\top}\vec{\alpha}_{k}=0$ for any $j=2,\ldots,J$ and $k=1,\ldots,W$, we only need finite equations regarding all $\widetilde{\vec{\eta}}_j^{\top}\vec{\alpha}_{k}$ after excluding a set of measure zero in the parameter space. Hence there exists $l\in\mathbb{N}$ such that we only need to match the coefficients of the terms whose order differences with the leading term which are less than $\exp(-l\log n)$. For $r\in\mathbb{N}$, denote $\hat{\bm{\xi}}^{(n)}$ and $\mathcal{A}_{r,n}$ as
        \begin{align*}
            \hat{\bm{\xi}}^{(n)}&=\underset{\bm{\xi}=(\xi_2,\ldots,\xi_J):\sum_{j=2}^J\xi_j\leq n}{\operatorname{argmax}}~\Delta_n(\bm{\xi}){\phi}_n(\bm{\xi}),\\
            \mathcal{A}_{r,n}&=\big\{\bm{\xi}^{(n)}=(\xi_2^{(n)},\ldots,\xi_J^{(n)}):|\xi_2^{(n)}-\hat{\xi}_2^{(n)}|\leq n^{(\nu_2+\delta)/2},\sum_{j=3}^J\xi_j^{(n)}\leq r \big\},
        \end{align*}
        where $\delta>0$ is a constant small enough. By the proof in Theorem~1, there exists $r^{\ast}\in\mathbb{N}$ such that
        \begin{align*}
            &\left|\sum_{\bm{\xi}^{(n)}\in\mathcal{O}_n\setminus \mathcal{A}_{r^{\ast},n}}\Delta_{n}(\bm{\xi}^{(n)})\sum_{j=2}^J\exp(\mu_j)\int_{\bt}(2\pi)^{-\frac{K}{2}}\widetilde{\vec{\eta}}_j^{\top}\vec{\theta}\exp\left(f_{n}(\vec{\theta}\big|\bm{\xi}^{(n)})+\vec{\eta}_j^{\top}\vec{\theta}\right)\mathrm{d}\vec{\theta}\right|\\
            \leq& \exp(-l\log n)\Delta_n(\hat{\bm{\xi}}^{(n)}){\phi}_n(\hat{\bm{\xi}}^{(n)}).
        \end{align*}
        Hence we just need to expand all terms in 
        \begin{align*}
            \sum_{\bm{\xi}^{(n)}\in \mathcal{A}_{r^{\ast},n}}\Delta_{n}(\bm{\xi}^{(n)})\sum_{j=2}^J\exp(\mu_j)\int_{\bt}(2\pi)^{-\frac{K}{2}}\widetilde{\vec{\eta}}_j^{\top}\vec{\theta}\exp\left(f_{n}(\vec{\theta}\big|\bm{\xi}^{(n)})+\vec{\eta}_j^{\top}\vec{\theta}\right)\mathrm{d}\vec{\theta}
        \end{align*}
        in decreasing order and match the coefficients of all terms which have order difference with $\Delta_n(\hat{\bm{\xi}}^{(n)}){\phi}_n(\hat{\bm{\xi}}^{(n)})$ smaller than $\exp(-l\log n)$. Following the expansion method in \cite{shun1995laplace} and similar method as in the proof of Theorem~1, we can show that $\widetilde{\vec{\eta}}_j^{\top}\vec{\alpha}_{k}=0$ for any $j=2,\ldots,J$ and $k=1,\ldots,W$ after excluding a set of measure zero in the parameter space. Hence for any $j,j_1,j_2=1,\ldots,J$ and $k=1,\ldots,W$ we have $\widetilde{\vec{\eta}}_j^{\top}\vec{\alpha}_{k}=\widetilde{\vec{\eta}}_{j_1}^{\top}\vec{\eta}_{j_2}=0$ since $\vec{\eta}_1,\ldots,\vec{\eta}_J\in \operatorname{span}\{\vec{\alpha}_1,\ldots,\vec{\alpha}_W\}$. This finishes the proof on $[0,t_{q+1}]$.\\[3mm]
        Hence, by induction, we prove that for any $j,j_1,j_2$ and $0<t, s<T$, it holds that
        \begin{align*}
            u_{j0}+\vec{u}_j^{\top}\vec{X}_j(t)&=0,\\
            (\mat{V}_{j_1}^{\top}\vec{Z}_{j_1}(t))^{\top}(\mat{A}_{j_2}^{\top}\vec{Z}_{j_2}(s))&=0.
        \end{align*}
        This indicates that $u_{j0}=0$, $\vec{u}_j=\vec{0}$ and $\mat{V}_{j_1}\mat{A}_{j_2}^{\top}=0$ by Condition (d). Since there exists $K$ rows among $\mat{A} =(\mat{A}_1^{\top},\ldots,\mat{A}_J^{\top})^{\top}$ which have full rank by Condition (c), we have $\mat{V}_{j}=\mat{0}$ for any $j=1,\ldots,J$, which contradicts the fact that $\vec{w}$ is nonzero. So we proved that $\mat{I}(\vec{\delta}_0)$ is finite and positive definite.
    \end{proof}
    
    \subsection{Proof of Proposition \ref{prop_score}}
    \renewcommand*{\proofname}{Proof of Proposition \ref{prop_score}}
    \begin{proof}
    If $\mat{I}(\vec{\delta}_0)$ is singular, then there exists nonzero $\vec{w}=\{u_{j0}\in \mathbb{R},\vec{u}_j\in \mathbb{R}^{L_{1j}},\mat{V}_j\in \mathbb{R}^{L_{2j}\times K}:j=1,\ldots,J\}$ such that $\left(\frac{\partial}{\partial \vec{\delta}} \log \mathcal{L}(\vec{\delta}_0|\textbf{N}, \textbf{X}, \textbf{Z})\right)^{\top}\vec{w}=0$ almost surely. Then it can be readily shown that $\left(\frac{\partial}{\partial \vec{\delta}} \mathcal{L}(\vec{\delta}_0|\textbf{N}, \textbf{X}, \textbf{Z})\right)^{\top}\vec{w}=0$ almost surely. For any $t\in[0,T]$, by integrating the above equation on $[t,T]$, we can see that $\left(\frac{\partial}{\partial \vec{\delta}} \mathcal{L}(\vec{\delta}_0|\textbf{N}, \textbf{X}, \textbf{Z})\right)^{\top}\vec{w}=0$ still holds if $\mathcal{L}(\vec{\delta}_0|\textbf{N}, \textbf{X}, \textbf{Z})$ represents the likelihood function derived on interval $[0,t]$. By explicit calculation, we have
    \begin{align}\label{eqq1_proof}
        0=&\int_{\bt}\left[\sum_{j=1}^J\int_{0}^{t}(u_{j0}+\vec{u}_j^{\top}\vec{X}_j(s)+\vec{\theta}^{\top}\mat{V}_j^{\top}\vec{Z}_j(s))\mathrm{d}N_j(s)\right]\notag\\
        &\times\prod_{j=1}^{J} \Big[\prod_{s \leq t} (\lambda_{j}(s)^{\Delta N_{j}(s)} )e^{-\int_{0}^{t} \lambda_{j}(s) \mathrm{d} s} \Big]\phi_K(\vec{\theta} ;\vec{0}, \mat{I}_K) \mathrm{d} \vec{\theta}\notag\\
        -&\int_{\bt}\left[\sum_{j=1}^J\int_{0}^{t}(u_{j0}+\vec{u}_j^{\top}\vec{X}_j(s)+\vec{\theta}^{\top}\mat{V}_j^{\top}\vec{Z}_j(s))\lambda_{j}(s)\mathrm{d}s\right]\notag\\
        &\times\prod_{j=1}^{J} \Big[\prod_{s \leq t} (\lambda_{j}(s)^{\Delta N_{j}(s)} )e^{-\int_{0}^{t} \lambda_{j}(s) \mathrm{d} s} \Big]\phi_K(\vec{\theta} ;\vec{0}, \mat{I}_K) \mathrm{d} \vec{\theta}~\text{a.s.}
    \end{align}
    For any fixed trajectory with positive density, there exists $t_0>0$ small enough such that $\vec{X}_j$ and $\vec{Z}_j$ are constant on $(t,t+t_0)$ and there are no events on $(t,t+t_0)$. For any $0<\Delta t<t_0$, we can derive
    \begin{align}\label{eqq2}
        0=&\int_{\bt}\left[\sum_{j=1}^J\int_{0}^{t}(u_{j0}+\vec{u}_j^{\top}\vec{X}_j(s)+\vec{\theta}^{\top}\mat{V}_j^{\top}\vec{Z}_j(s))\mathrm{d}N_j(s)\right]\notag\\
        &\times\prod_{j=1}^{J} \Big[\prod_{s \leq t} (\lambda_{j}(s)^{\Delta N_{j}(s)} )e^{-\int_{0}^{t+\Delta t} \lambda_{j}(s) \mathrm{d} s} \Big]\phi_K(\vec{\theta} ;\vec{0}, \mat{I}_K) \mathrm{d} \vec{\theta}\notag\\
        -&\int_{\bt}\left[\sum_{j=1}^J\int_{0}^{t+\Delta t}(u_{j0}+\vec{u}_j^{\top}\vec{X}_j(s)+\vec{\theta}^{\top}\mat{V}_j^{\top}\vec{Z}_j(s))\lambda_{j}(s)\mathrm{d}s\right]\notag\\
        &\times\prod_{j=1}^{J} \Big[\prod_{s \leq t} (\lambda_{j}(s)^{\Delta N_{j}(s)} )e^{-\int_{0}^{t+\Delta t} \lambda_{j}(s) \mathrm{d} s} \Big]\phi_K(\vec{\theta} ;\vec{0}, \mat{I}_K) \mathrm{d} \vec{\theta}.
    \end{align}
    By taking the $n$-th derivative of (\ref{eqq2}) with respect to $\Delta t$ and let $\Delta t$ go down to 0, we have
    \begin{align*}
        0=&\int_{\bt}\left[\sum_{j=1}^J\int_{0}^{t}(u_{j0}+\vec{u}_j^{\top}\vec{X}_j(s)+\vec{\theta}^{\top}\mat{V}_j^{\top}\vec{Z}_j(s))(\mathrm{d}N_j(s)-\lambda_j(s)\mathrm{d}s)\right] \notag\\
        &\times\prod_{j=1}^{J}\Big[\prod_{s \leq t} (\lambda_{j}(s)^{\Delta N_{j}(s)} )e^{-\int_{0}^{t} \lambda_{j}(s) \mathrm{d} s} \Big]\left(\sum_{j=1}^J\lambda_j(t+0)\right)^n\phi_K(\vec{\theta} ;\vec{0}, \mat{I}_K) \mathrm{d} \vec{\theta}\notag\\
        -&\int_{\bt}\left[\sum_{j=1}^J(u_{j0}+\vec{u}_j^{\top}\vec{X}_j(t+0)+\vec{\theta}^{\top}\mat{V}_j^{\top}\vec{Z}_j(t+0))\lambda_{j}(t+0)\right]\notag\\
        &\times\prod_{j=1}^{J}\Big[\prod_{s \leq t} (\lambda_{j}(s)^{\Delta N_{j}(s)} )e^{-\int_{0}^{t} \lambda_{j}(s) \mathrm{d} s} \Big]\left(\sum_{j=1}^J\lambda_j(t+0)\right)^{n-1}\phi_K(\vec{\theta} ;\vec{0}, \mat{I}_K) \mathrm{d} \vec{\theta}.
    \end{align*}
    For each $m\in\{1,\ldots,J\}$, by arguments similar to those used in the proof of Corollary~\ref{cor_likelihood}, we consider a hypothesized sample path on interval $[0,t+\Delta t]$ which has the same observed sample path on $[0,t+\Delta t)$ but has the $m$-th event happening at time $t+\Delta t$. Then by differentiation, we have
    \begin{align*}
        0=&\int_{\bt}\Bigg[\sum_{j=1}^J\int_{0}^{t}(u_{j0}+\vec{u}_j^{\top}\vec{X}_j(s)+\vec{\theta}^{\top}\mat{V}_j^{\top}\vec{Z}_j(s))(\mathrm{d}N_j(s)-\lambda_j(s)\mathrm{d}s)+u_{m0}+\vec{u}_m^{\top}\vec{X}_m(t+0)\notag\\
        &+\vec{\theta}^{\top}\mat{V}_m^{\top}\vec{Z}_m(t+0)\Bigg]\lambda_m(t+0)\prod_{j=1}^{J}\Big[\prod_{s \leq t} (\lambda_{j}(s)^{\Delta N_{j}(s)} )e^{-\int_{0}^{t} \lambda_{j}(s) \mathrm{d} s} \Big]\left(\sum_{j=1}^J\lambda_j(t+0)\right)^n\phi_K(\vec{\theta} ;\vec{0}, \mat{I}_K) \mathrm{d} \vec{\theta}\notag\\
        -&\int_{\bt}\left[\sum_{j=1}^J(u_{j0}+\vec{u}_j^{\top}\vec{X}_j(t+0)+\vec{\theta}^{\top}\mat{V}_j^{\top}\vec{Z}_j(t+0))\lambda_{j}(t+0)\right]\notag\\
        &\times\lambda_m(t+0)\prod_{j=1}^{J}\Big[\prod_{s \leq t} (\lambda_{j}(s)^{\Delta N_{j}(s)} )e^{-\int_{0}^{t} \lambda_{j}(s) \mathrm{d} s} \Big]\left(\sum_{j=1}^J\lambda_j(t+0)\right)^{n-1}\phi_K(\vec{\theta} ;\vec{0}, \mat{I}_K) \mathrm{d} \vec{\theta}~~\text{a.s.}
    \end{align*}
    Hence, the result is proved.
    \end{proof}
    \subsection{Proof of Proposition \ref{prop_approximation}}
    To prove Proposition~\ref{prop_approximation}, we first prove the following lemma:
    \begin{lem}\label{lem_approximation}
        Let $\{f_n(\vec{x})\}$ be a sequence of strictly concave functions on $\mathbb{R}^d$ with $\vec{0}$ as their unique maximisers. Assume that $-\nabla^2 f_n(\vec{x})\succeq \mat{I}_d$ holds at any point $\vec{x}\in\mathbb{R}^d$ for any $n$. Let $\vec{\gamma},\{\vec{\beta}_n\}\in\mathbb{R}^d$ be $d$-vectors such that $\vec{\gamma}^{\top}\vec{\beta}_n\rightarrow \infty$. Then for any $\delta>0$, for $n$ large enough we have
        \begin{align*}
            0<\frac{\left|\int_{\vec{x}:\left\|\vec{x}\right\|\geq C}\vec{\gamma}^{\top}(\vec{x}+\vec{\beta}_n)\exp(f_n(\vec{x}))\mathrm{d}\vec{x}\right|}{\int_{\vec{x}}\vec{\gamma}^{\top}(\vec{x}+\vec{\beta}_n)\exp(f_n(\vec{x}))\mathrm{d}\vec{x}}\leq \delta,
        \end{align*}
        where $C>0$ is a constant which only depends on $\delta$.
    \end{lem}
        \renewcommand*{\proofname}{Proof of Lemma \ref{lem_approximation}}
    \begin{proof}
        We change variables to $d$-dimensional polar coordinates:
        \begin{align}\label{eq_lem_8.1}
            &\left|\int_{\vec{x}:\left\|\vec{x}\right\|\geq C}\vec{\gamma}^{\top}(\vec{x}+\vec{\beta}_n)\exp(f_n(\vec{x}))\mathrm{d}\vec{x}\right|\notag\\
            \leq&\int_{\theta_1,\ldots,\theta_{d-1}}\left(\prod_{k=2}^{d-1}\sin^{k-1}\theta_k\right)\mathrm{d}\theta_1\ldots\mathrm{d}\theta_{d-1}\int_{r\geq C}r^{d-1}\left|\vec{\gamma}^{\top}(r\vec{\alpha}(\theta_1,\ldots,\theta_{d-1})+\vec{\beta}_n)\right|\exp(f_n(r\vec{\alpha}(\theta_1,\ldots,\theta_{d-1})))\mathrm{d}r\notag\\
            \leq&\int_{\theta_1,\ldots,\theta_{d-1}}\left(\prod_{k=2}^{d-1}\sin^{k-1}\theta_k\right)\mathrm{d}\theta_1\ldots\mathrm{d}\theta_{d-1}\int_{r\geq C}r^{d-1}\left|r\|\vec{\gamma}\|+|\vec{\gamma}^{\top}
            \vec{\beta}_n|\right|\exp(f_n(r\vec{\alpha}(\theta_1,\ldots,\theta_{d-1})))\mathrm{d}r,\notag\\
            &\int_{\vec{x}:\left\|\vec{x}\right\|\leq C}\vec{\gamma}^{\top}(\vec{x}+\vec{\beta}_n)\exp(f_n(\vec{x}))\mathrm{d}\vec{x}\notag\\
            \geq&\int_{\theta_1,\ldots,\theta_{d-1}}\left(\prod_{k=2}^{d-1}\sin^{k-1}\theta_k\right)\mathrm{d}\theta_1\ldots\mathrm{d}\theta_{d-1}\int_{r\leq C}r^{d-1}\left(\vec{\gamma}^{\top}(r\vec{\alpha}(\theta_1,\ldots,\theta_{d-1})+\vec{\beta}_n)\right)\exp(f_n(r\vec{\alpha}(\theta_1,\ldots,\theta_{d-1})))\mathrm{d}r\notag\\
            \geq&\int_{\theta_1,\ldots,\theta_{d-1}}\left(\prod_{k=2}^{d-1}\sin^{k-1}\theta_k\right)\mathrm{d}\theta_1\ldots\mathrm{d}\theta_{d-1}\int_{r\leq C}r^{d-1}\left(|\vec{\gamma}^{\top}
            \vec{\beta}_n|-C\|\vec{\gamma}\|\right)\exp(f_n(r\vec{\alpha}(\theta_1,\ldots,\theta_{d-1})))\mathrm{d}r,
        \end{align}
        where $\left\|\vec{\alpha}(\theta_1,\ldots,\theta_{d-1})\right\|=1$. For fixed $\theta_1,\ldots,\theta_{d-1}$, since $-\nabla^2 f_n(\vec{x}_r^{\ast})\succeq \mat{I}_d$, we have
        \begin{align}\label{eq_lem_8.3}
            -\frac{d^2}{\mathrm{d}r^2}f_n(r\vec{\alpha})=&-\vec{\alpha}^{\top}\nabla^2 f_n(r\vec{\alpha})\vec{\alpha}\geq \left\|\vec{\alpha}\right\|^2=1.
        \end{align}
        Hence, by (\ref{eq_lem_8.3}), we have
        \begin{align}\label{eq_lem_8.4}
            &\int_{r\geq C}r^{d-1}\left(r\|\vec{\gamma}\|+|\vec{\gamma}^{\top}
            \vec{\beta}_n|\right)\exp(f_n(r\vec{\alpha}(\theta_1,\ldots,\theta_{d-1})))\mathrm{d}r\notag\\
            \leq&\exp(f_n(C\vec{\alpha}))\int_{r\geq C}r^{d-1}\left(r\|\vec{\gamma}\|+|\vec{\gamma}^{\top}
            \vec{\beta}_n|\right)\exp\left(-\frac{1}{2}r^2\right)\mathrm{d}r\notag\\
            \leq&\exp(f_n(C\vec{\alpha}))(\|\vec{\gamma}\|+|\vec{\gamma}^{\top}
            \vec{\beta}_n|)\underbrace{\max\left\{\int_{r\geq C}r^{d-1}\exp\left(-\frac{1}{2}r^2\right)\mathrm{d}r,\int_{r\geq C}r^{d}\exp\left(-\frac{1}{2}r^2\right)\mathrm{d}r\right\}}_{:=I(C)}.
        \end{align}
        On the other hand, we have
        \begin{align}\label{eq_lem_8.5}
            &\int_{r\leq C}r^{d-1}\left(|\vec{\gamma}^{\top}
            \vec{\beta}_n|-C\|\vec{\gamma}\|\right)\exp(f_n(r\vec{\alpha}(\theta_1,\ldots,\theta_{d-1})))\mathrm{d}r\notag\\
            \geq&\left(|\vec{\gamma}^{\top}
            \vec{\beta}_n|-C\|\vec{\gamma}\|\right)\exp(f_n(C\vec{\alpha}))\int_{r\leq C}r^{d-1}\mathrm{d}r\notag\\
            =&\frac{C^d\left(|\vec{\gamma}^{\top}
            \vec{\beta}_n|-C\|\vec{\gamma}\|\right)\exp(f_n(C\vec{\alpha}))}{d}.
        \end{align}
        By (\ref{eq_lem_8.1}), (\ref{eq_lem_8.4}) and (\ref{eq_lem_8.5}), for $n$ large enough we have
        \begin{align*}
            0<\frac{\left|\int_{\vec{x}:\left\|\vec{x}\right\|\geq C}\vec{\gamma}^{\top}(\vec{x}+\vec{\beta}_n)\exp(f_n(\vec{x}))\mathrm{d}\vec{x}\right|}{\int_{\vec{x}:\left\|\vec{x}\right\|\leq C}\vec{\gamma}^{\top}(\vec{x}+\vec{\beta}_n)\exp(f_n(\vec{x}))\mathrm{d}\vec{x}}\leq \frac{d(\|\vec{\gamma}\|+|\vec{\gamma}^{\top}
            \vec{\beta}_n|)I(C)}{C^d\left(|\vec{\gamma}^{\top}
            \vec{\beta}_n|-C\|\vec{\gamma}\|\right)}=\frac{d(1+\|\vec{\gamma}\|/|\vec{\gamma}^{\top}
            \vec{\beta}_n|)I(C)}{C^d\left(1-C\|\vec{\gamma}\|/|\vec{\gamma}^{\top}
            \vec{\beta}_n|\right)}.
        \end{align*}
        Since $\vec{\gamma}^{\top}
        \vec{\beta}_n\rightarrow \infty$ and $I(C)$ is uniformly bounded, for any $\delta>0$, we can find $C$ which only depends on $\delta$ such that
        \begin{align*}
            0<\frac{\left|\int_{\vec{x}:\left\|\vec{x}\right\|\geq C}\vec{\gamma}^{\top}(\vec{x}+\vec{\beta}_n)\exp(f_n(\vec{x}))\mathrm{d}\vec{x}\right|}{\int_{\vec{x}}\vec{\gamma}^{\top}(\vec{x}+\vec{\beta}_n)\exp(f_n(\vec{x}))\mathrm{d}\vec{x}}\leq \delta
        \end{align*}
        for any $n$ large enough.
    \end{proof}

    Based on Lemma \ref{lem_approximation}, we can verify Proposition~\ref{prop_approximation}.
    
    \renewcommand*{\proofname}{Proof of Proposition \ref{prop_approximation}}
    \begin{proof}
        We prove the claim for all sufficiently large $n$.
Set
\[
Z_n:=(2\pi)^{-d/2}\int_{\bt}\exp(f_n(\vec{\theta}))\,\mathrm{d}\vec{\theta},
\qquad
J_n:=(2\pi)^{-d/2}\int_{\bt}(\vec{\gamma}^{\top}\vec{\theta})\exp(f_n(\vec{\theta}))\,\mathrm{d}\vec{\theta}.
\]
By Proposition~\ref{prop_laplace}, there exists a constant $M_0>0$, independent of $n$, such that
\[
M_0^{-1}\cdot
\frac{\exp(f_n(\hat{\vec{\theta}}_n))}
{\sqrt{\operatorname{det}(\mat{I}(\hat{\vec{\theta}}_n))}}
\leq
Z_n
\leq
M_0\cdot
\frac{\exp(f_n(\hat{\vec{\theta}}_n))}
{\sqrt{\operatorname{det}(\mat{I}(\hat{\vec{\theta}}_n))}}.
\]
Therefore, it remains to show that $J_n\asymp (\vec{\gamma}^{\top}\hat{\vec{\theta}}_n)\,Z_n$ with comparison constants independent of $n$.
Define the shifted function $g_n(\vec{x})
:=
f_n(\hat{\vec{\theta}}_n+\vec{x})-f_n(\hat{\vec{\theta}}_n)$ for $\vec{x}\in\mathbb{R}^d$.
Then $g_n$ is strictly concave, $g_n(\vec{0})=0$, $\nabla g_n(\vec{0})=\vec{0}$, and
\[
-\nabla^2 g_n(\vec{x})
=
-\nabla^2 f_n(\hat{\vec{\theta}}_n+\vec{x})
=
\mat{I}(\hat{\vec{\theta}}_n+\vec{x})
\succeq \mat{I}_d
\qquad\text{for all }\vec{x}\in\mathbb{R}^d.
\]
We next claim that there exists a constant $L>0$, depending only on $d$, such that
\begin{equation}\label{eq:prop_approx_moment}
\int_{\vec{x}}\|\vec{x}\|\,\exp(g_n(\vec{x}))\,\mathrm{d}\vec{x}
\leq
L\int_{\vec{x}}\exp(g_n(\vec{x}))\,\mathrm{d}\vec{x}
\qquad\text{for all }n.
\end{equation}
To prove this, fix $n$ and write the integral in polar coordinates. For each unit vector
$\vec{\alpha}$, let $h(r):=g_n(r\vec{\alpha})$ for $r\geq 0$.
Then $h'(0)=0$ and
\[
h''(r)=\vec{\alpha}^{\top}\nabla^2 g_n(r\vec{\alpha})\vec{\alpha}\leq -1
\qquad\text{for all }r\geq 0.
\]
Hence $h'(r)\leq -r$ for all $r\geq 0$, so $h$ is decreasing on $[0,\infty)$. In particular, for $r\geq 1$,
\[
h(r)-h(1)=\int_1^r h'(s)\,\mathrm{d}s
\leq
-\int_1^r s\,\mathrm{d}s
=
-\frac{r^2-1}{2}.
\]
Therefore,
\[
\int_1^\infty r^d e^{h(r)}\,\mathrm{d}r
\leq
e^{h(1)+1/2}\int_1^\infty r^d e^{-r^2/2}\,\mathrm{d}r.
\]
On the other hand, since $h$ is decreasing,
\[
\int_0^1 r^{d-1}e^{h(r)}\,\mathrm{d}r
\geq
e^{h(1)}\int_0^1 r^{d-1}\,\mathrm{d}r
=
\frac{e^{h(1)}}{d}.
\]
Combining the last two displays yields
\[
\int_1^\infty r^d e^{h(r)}\,\mathrm{d}r
\leq
d e^{1/2}\left(\int_1^\infty r^d e^{-r^2/2}\,\mathrm{d}r\right)
\int_0^1 r^{d-1}e^{h(r)}\,\mathrm{d}r.
\]
Integrating this inequality over the angular variables, we obtain
\[
\int_{\vec{x}:\|\vec{x}\|\geq 1}\|\vec{x}\|\,e^{g_n(\vec{x})}\,\mathrm{d}\vec{x}
\leq
L_1\int_{\vec{x}:\|\vec{x}\|\leq 1}e^{g_n(\vec{x})}\,\mathrm{d}\vec{x},
\]
where
\[
L_1:=d e^{1/2}\int_1^\infty r^d e^{-r^2/2}\,\mathrm{d}r<\infty.
\]
Also, it holds that
\[
\int_{\vec{x}:\|\vec{x}\|\leq 1}\|\vec{x}\|\,e^{g_n(\vec{x})}\,\mathrm{d}\vec{x}
\leq
\int_{\vec{x}:\|\vec{x}\|\leq 1}e^{g_n(\vec{x})}\,\mathrm{d}\vec{x}.
\]
Hence, we have that
\[
\int_{\vec{x}}\|\vec{x}\|\,e^{g_n(\vec{x})}\,\mathrm{d}\vec{x}
\leq
(1+L_1)\int_{\vec{x}}e^{g_n(\vec{x})}\,\mathrm{d}\vec{x},
\]
which proves \eqref{eq:prop_approx_moment} with $L:=1+L_1$.
Now we let $\vec{\theta}=\hat{\vec{\theta}}_n+\vec{x}$. Then
\[
Z_n
=
(2\pi)^{-d/2}e^{f_n(\hat{\vec{\theta}}_n)}
\int_{\vec{x}}e^{g_n(\vec{x})}\,\mathrm{d}\vec{x},
\]
and
\begin{align*}
J_n=
(2\pi)^{-d/2}e^{f_n(\hat{\vec{\theta}}_n)}
\int_{\vec{x}}
(\vec{\gamma}^{\top}\hat{\vec{\theta}}_n+\vec{\gamma}^{\top}\vec{x})
e^{g_n(\vec{x})}\,\mathrm{d}\vec{x}=(\vec{\gamma}^{\top}\hat{\vec{\theta}}_n)Z_n+R_n,
\end{align*}
where
\[
R_n
:=
(2\pi)^{-d/2}e^{f_n(\hat{\vec{\theta}}_n)}
\int_{\vec{x}}(\vec{\gamma}^{\top}\vec{x})e^{g_n(\vec{x})}\,\mathrm{d}\vec{x}.
\]
By \eqref{eq:prop_approx_moment}, we have that
\begin{align*}
|R_n|
&\leq
\|\vec{\gamma}\|
(2\pi)^{-d/2}e^{f_n(\hat{\vec{\theta}}_n)}
\int_{\vec{x}}\|\vec{x}\|e^{g_n(\vec{x})}\,\mathrm{d}\vec{x}\\
&\leq
L\|\vec{\gamma}\|
(2\pi)^{-d/2}e^{f_n(\hat{\vec{\theta}}_n)}
\int_{\vec{x}}e^{g_n(\vec{x})}\,\mathrm{d}\vec{x}\\
&=
L\|\vec{\gamma}\|\,Z_n.
\end{align*}
Therefore,
\[
\bigl(\vec{\gamma}^{\top}\hat{\vec{\theta}}_n-L\|\vec{\gamma}\|\bigr)Z_n
\leq
J_n
\leq
\bigl(\vec{\gamma}^{\top}\hat{\vec{\theta}}_n+L\|\vec{\gamma}\|\bigr)Z_n.
\]
Since $\vec{\gamma}^{\top}\hat{\vec{\theta}}_n\to\infty$, for all sufficiently large $n$,
\[
\frac{1}{2}(\vec{\gamma}^{\top}\hat{\vec{\theta}}_n)Z_n
\leq
J_n
\leq
\frac{3}{2}(\vec{\gamma}^{\top}\hat{\vec{\theta}}_n)Z_n.
\]
Combining this with Proposition~\ref{prop_laplace}, we obtain
\[
\frac{1}{2M_0}\cdot
\frac{(\vec{\gamma}^{\top}\hat{\vec{\theta}}_n)\exp(f_n(\hat{\vec{\theta}}_n))}
{\sqrt{\operatorname{det}(\mat{I}(\hat{\vec{\theta}}_n))}}
\leq
J_n
\leq
\frac{3M_0}{2}\cdot
\frac{(\vec{\gamma}^{\top}\hat{\vec{\theta}}_n)\exp(f_n(\hat{\vec{\theta}}_n))}
{\sqrt{\operatorname{det}(\mat{I}(\hat{\vec{\theta}}_n))}}.
\]
After absorbing the numerical factors into a single constant $M>0$, the proof is complete.
\end{proof}

\section{Proof of Theorem \ref{thm_normality}}\label{sec:thm3}
    Since the baseline intensity function is restricted to a parametric form in this paper, uniform boundedness of the covariate processes is sufficient to establish the consistency and asymptotic normality of the MLE.
    \renewcommand*{\proofname}{Proof of Theorem \ref{thm_normality}}
    \begin{proof}
    Condition (a) and the continuity of the likelihood function guarantee the existence of the MLE $\hat{\vec{\delta}}$. We then prove that the class $\{\frac{1}{n}\log \mathcal{L}_n(\vec{\delta}): \vec{\delta} \in \vec{\Delta}\}$ admits an integrable envelope function.
    Note that Conditions (a), (b), and (f) guarantee that there exists a constant $C>0$ such that for any $i=1,\ldots,n$, $j=1,\ldots,J$ and $0\leq t\leq \tau$, it holds that
    \begin{align*}
        \lambda_{ij}(t|\mathscr{F}_{t^-},\vec{\theta}_i)\leq C\exp(C\|\vec{\theta}_i\|).
    \end{align*}
    Then we can show that the likelihood function can be bounded as
    \begin{align*}
        \mathcal{L}_{n}(\vec{\delta})\leq& \prod_{i=1}^n\int_{\mathbb{R}^K}\prod_{j=1}^JC^{N_{ij}(\tau)}\exp(CN_{ij}(\tau)\|\vec{\theta}_i\|)\phi_K (\boldsymbol{\theta};\boldsymbol{0},\mathbf{\Sigma}) \mathrm{d}\boldsymbol{\theta}\\
        \leq&\prod_{i=1}^nC^{\sum_{j=1}^JN_{ij}(\tau)}\exp\left(\widetilde{C}\left[\sum_{j=1}^JN_{ij}(\tau)\right]^2\right)\quad(\text{the MGF of Gaussian distribution}).
    \end{align*}
    Therefore,
    \begin{align*}
        \frac{1}{n}\log \mathcal{L}_{n}(\vec{\delta})\lesssim& \frac{1}{n}\sum_{i=1}^n\sum_{j=1}^JN_{ij}(\tau)+\frac{1}{n}\sum_{i=1}^n\left[\sum_{j=1}^JN_{ij}(\tau)\right]^2\\
        \leq& \frac{1}{n}\sum_{i=1}^n\sum_{j=1}^JN_{ij}(\tau)+\frac{J}{n}\sum_{i=1}^n\sum_{j=1}^JN_{ij}(\tau)^2.
    \end{align*}
    Since the observation period $[0,\tau]$ is finite and the intensity function of $N_{ij}(t)$ is bounded by $C\exp(C\|\vec{\theta}_i\|)$, which has uniform bounded first and second moments, we have proved that the class $\{\frac{1}{n}\log \mathcal{L}_n(\vec{\delta}): \vec{\delta} \in \vec{\Delta}\}$ admits an integrable uniform upper bound.
    Using similar arguments, we can obtain an integrable uniform lower bound.
    Together, these bounds provide an integrable envelope, which yields a uniform law of large numbers.
    Together with the identifiability result in Theorem~1, standard arguments then imply consistency of the MLE $\hat{\vec{\delta}}$.
    
    Similarly, we can show that all third-order derivatives of $\{\frac{1}{n}\log \mathcal{L}(\vec{\delta}): \vec{\delta}\in\vec{\Delta}\}$ can be dominated by an integrable envelope function. The asymptotic normality of $\hat{\vec{\delta}}$ then follows from Theorem~2.
    \end{proof}
\section{Proof of Theorem \ref{thm:oracle}}\label{sec:thm4}
    \renewcommand*{\proofname}{Proof of Theorem \ref{thm:oracle}}
    \begin{proof}
    Under Conditions (a)--(f), the conditions (A) and (C) in \cite{fan2001variable} are verified. By the result in Theorem~2, condition (B) is also verified. Hence by arguments similar to those used in \cite{fan2001variable}, Theorem~4 is proved.
    \end{proof}

\section{Further discussion on parametric and semiparametric extensions}\label{sec:extension}
As noted in the main paper and Section~\ref{sec:identifiability} of the Supplementary Materials, a key challenge in this work is to establish model identifiability in the presence of internal covariates. The main paper establishes a general identifiability result under an additional piecewise-constant assumption. This condition arises naturally when action-based internal covariates are incorporated and is straightforward to verify, although it restricts the scope of the resulting theory.

In this section, we discuss possible extensions to models with alternative parametric baseline intensities or a nonparametric baseline intensity. Section~\ref{sec:parametric-extension} considers parametric extensions, whereas Section~\ref{sec:semiparametric-extension} considers a semiparametric extension.
    
\subsection{Parametric extension}\label{sec:parametric-extension}

The paper considers a constant baseline intensity function. Nevertheless, the proposed model can be readily extended to accommodate a nonconstant parametric baseline intensity, such as a Weibull or Gompertz intensity.

Although Condition~(d) is formulated specifically for the proposed model with a constant baseline intensity, it need not be modified when combined with the piecewise-constant assumption in Condition~(e). For example, suppose that a Weibull baseline intensity $\lambda_{j0}(t)=c_jt^{\kappa_j-1}$ is adopted. If the deterministic components of two competing intensity specifications agree, so that
\begin{align*}
    c_jt^{\kappa_j-1}\exp(\vec{\beta}_{j}^{\top} \vec{X}_{j}(t))=\widetilde{c}_jt^{\widetilde{\kappa}_j-1}\exp(\widetilde{\vec{\beta}}_{j}^{\top} \vec{X}_{j}(t)),
\end{align*}
or equivalently,
\begin{align}\label{weibull}
    \log(c_j/\widetilde c_j)+(\kappa_j-\widetilde\kappa_j)\log t+(\vec{\beta}_j-\widetilde{\vec{\beta}}_j)^\top\vec X_j(t)=0
\end{align}
for every $j=1,\ldots,J$ and $t\in(0,C_j]$, then the piecewise-constant structure of $\vec{X}_j(t)$ implies that, on each subinterval with nonempty interior on which $\vec{X}_j(t)$ is constant,
\begin{align*}
    \log(c_j/\widetilde c_j)+
(\kappa_j-\widetilde\kappa_j)\log t
+
(\vec{\beta}_j-\widetilde{\vec{\beta}}_j)^\top\vec X_j(t)
=0.
\end{align*}
Because the first and third terms in (\ref{weibull}) are constant on such a subinterval, whereas $\log t$ varies with $t$, it follows that $\kappa_j=\widetilde{\kappa}_j$ for every $j$.
Condition~(d) can then be used to identify $c_j=\widetilde{c}_j$ and $\vec{\beta}_{j}=\widetilde{\vec{\beta}}_{j}$.

Model identifiability can subsequently be established using differentiation arguments similar to those employed in the proof of Theorem~1. A nonconstant baseline intensity introduces additional higher-order derivative terms, which may be represented through Bell polynomials. However, for the parametric baseline families considered here, the $n$th-order derivatives contribute terms of order at most $\exp(\mathcal{O}(n\log n))$ and the corresponding ratios between competing derivative terms yield order gaps of at most $\exp(\mathcal{O}(\log n))$.
These contributions are asymptotically dominated by the $\exp(\mathcal{O}(n))$ order gaps used in the proof of Theorem~1 to identify the random-effect component. Consequently, the same differentiation-based identification strategy continues to apply.

The proof of Theorem~2 can be extended analogously for the same reason. The proofs of Theorems~3 and~4 require only minor modifications, because replacing the constant baseline intensity with a finite-dimensional parametric baseline does not alter the parametric nature of the model.

\subsection{Semiparametric extension}\label{sec:semiparametric-extension}

    The proposed framework can be extended to semiparametric models, although establishing the corresponding theoretical results becomes substantially more challenging. In particular, additional structural assumptions are required to ensure model identifiability. Below, we discuss one possible extension.
    The intensity function associated with subject $i$ and event type $j$ is assumed to be
    \begin{equation}\label{semimodel}
    \lambda_{ij}(t|\mathscr{F}_{t^-};\boldsymbol{\theta}_i) = \lambda_{j0}(t)Y_{ij}(t)e^{ \boldsymbol{\beta}^{\top}_j \boldsymbol{X}_{ij}(t) + \boldsymbol{\theta}^{\top}_i  \mathbf{A}^{\top}_j \boldsymbol{Z}_{ij}(t)}.\tag{1'}
    \end{equation}
    The model is identifiable under the following two additional structural conditions. The first condition is stated below.

    \begin{itemize}
        \item[(g)] For each $i=1,\ldots,n$ and $j=1,\ldots,J$, the values of $\boldsymbol{X}_{ij}(t)$ and $\boldsymbol{Z}_{ij}(t)$ are determined by the ordered sequence of event types occurring before time $t$ for any $t\in[0,\tau]$. In particular, they do not depend on the occurrence times of the preceding events. Moreover, $\boldsymbol{X}_{ij}(0)=\boldsymbol{0}$ and $\boldsymbol{Z}_{ij}(0)=\boldsymbol{0}$.
    \end{itemize}
    Condition~(g) arises naturally when the internal covariates are constructed from the observed action or event sequence. For example, in the data application presented in the main paper, each covariate process depends only on the type of the most recent event or on the types of the two most recent events. This construction therefore satisfies Condition~(g).

    Under Condition~(g), the covariate processes $\boldsymbol{X}_{ij}(\cdot)$ and $\boldsymbol{Z}_{ij}(\cdot)$ can equivalently be expressed as functions of the ordered sequence of event types:
    \begin{align*}
        \boldsymbol{X}_{ij}(t)=\boldsymbol{x}_{ij}(H_{t^-})\quad\text{and}\quad \boldsymbol{Z}_{ij}(t)=\boldsymbol{z}_{ij}(H_{t^-}).
    \end{align*}
    Here $H_{t^-}=(a_1,\ldots,a_m)$ denotes the ordered sequence of event types in $\{1,\ldots,J\}$ occurring before time $t$.
    For each $j=1,\ldots,J$, we embed $\boldsymbol{z}_{ij}\in\mathbb{R}^{L_{2j}}$ into $\breve{\boldsymbol{z}}_{ij}\in\mathbb{R}^{L_{2}}$ $(L_2=\sum_{j=1}^JL_{2j})$ by placing $\boldsymbol{z}_{ij}$ in the $j$-th block and setting all remaining blocks equal to zero.
    
    For any feasible event-type sequence $h=(a_1,\ldots,a_m)$, define its subsequences by $h_r=(a_1,\ldots,a_r)$ for $r=1,\ldots,m$, with $h_0$ denoting the empty sequence. The second condition is then stated as follows.
    \begin{itemize}
        \item[(d')] For any vectors $\boldsymbol{\mu}_j\in\mathbb{R}^{L_{1j}}$, $j=1,\ldots,J$, and symmetric matrix $\mathbf{U}\in\mathbb{R}^{L_{2}\times L_{2}}$, if the following identity holds
        \begin{align*}     \sum_{r=1}^m\boldsymbol{\mu}_{a_r}^{\top}\boldsymbol{x}_{i,a_r}(h_{r-1})+\left(\sum_{r=1}^m\breve{\boldsymbol{z}}_{i,a_r}(h_{r-1})\right)^{\top}\mathbf{U}\left(\sum_{r=1}^m\breve{\boldsymbol{z}}_{i,a_r}(h_{r-1})\right)=0
        \end{align*}
        for any $i=1,\ldots,n$ and every feasible event-type sequence $h$, then $\boldsymbol{\mu}_j=\boldsymbol{0}$ for all $j=1,\ldots,J$, and $\mathbf{U}=\boldsymbol{0}$.
    \end{itemize}
    Condition (d') is a strengthened version of Condition (d). It imposes a sufficiently rich covariate structure to ensure model identifiability under the semiparametric extension. Note that the condition is formulated under the assumption that transitions between different event types are unrestricted. If some transitions are structurally prohibited, Condition (d') must be reformulated over the set of feasible event sequences. Parameters associated with transitions that can never occur must either be removed from the model or fixed as zero in advance.
    
    We next show that the covariate construction used in our data application satisfies this condition. For simplicity, we consider the setting in which transitions between event types are unrestricted and the covariate processes $\boldsymbol{X}_{j}(\cdot)$ and $\boldsymbol{Z}_{j}(\cdot)$ depend only on the type of the most recent event. Specifically, for each $j=1,\ldots,J$, we define
    \begin{align*}
        \boldsymbol{X}_{j}(t)=\boldsymbol{Z}_{j}(t)=\begin{cases}
            \boldsymbol{0}&\text{if no event has occurred before time $t$}\\
            \boldsymbol{e}_k&\text{if the last event before time $t$ is of type $k$}.
        \end{cases}
    \end{align*}
    For any $j_1,j_2=1,\ldots,J$, consider the five sequences $h=(j_1,j_2)$, $(j_1,j_2,j_1)$, $(j_1,j_2,j_1,j_2)$, $(j_1,j_2,j_1,j_2,j_1)$, $(j_1,j_2,j_1,j_2,j_1,j_2)$. Then it holds that
    \begin{align*}
        \boldsymbol{x}_{j_2}(h_1)=&\,\boldsymbol{x}_{j_2}(h_3)=\boldsymbol{x}_{j_2}(h_5)=\boldsymbol{z}_{j_2}(h_1)=\boldsymbol{z}_{j_2}(h_3)=\boldsymbol{z}_{j_2}(h_5)=\boldsymbol{e}_{j_1},\\
        \boldsymbol{x}_{j_1}(h_2)=&\, \boldsymbol{x}_{j_1}(h_4)=\boldsymbol{z}_{j_1}(h_2)=\boldsymbol{z}_{j_1}(h_4)=\boldsymbol{e}_{j_2}.
    \end{align*}
    Then Condition (d') leads to the following five equations:
    \begin{align*}
        0=&\,\mu_{j_2}^{(j_1)}+\mathbf{U}_{(j_2-1)J+j_1,(j_2-1)J+j_1},\\
        0=&\,\mu_{j_2}^{(j_1)}+\mu_{j_1}^{(j_2)}+\mathbf{U}_{(j_2-1)J+j_1,(j_2-1)J+j_1}+\mathbf{U}_{(j_1-1)J+j_2,(j_1-1)J+j_2}\\
        &+\mathbf{U}_{(j_2-1)J+j_1,(j_1-1)J+j_2}+\mathbf{U}_{(j_1-1)J+j_2,(j_2-1)J+j_1},\\
        0=&\,2\mu_{j_2}^{(j_1)}+\mu_{j_1}^{(j_2)}+4\mathbf{U}_{(j_2-1)J+j_1,(j_2-1)J+j_1}+\mathbf{U}_{(j_1-1)J+j_2,(j_1-1)J+j_2}\\
        &+2\mathbf{U}_{(j_2-1)J+j_1,(j_1-1)J+j_2}+2\mathbf{U}_{(j_1-1)J+j_2,(j_2-1)J+j_1},\\
        0=&\,2\mu_{j_2}^{(j_1)}+2\mu_{j_1}^{(j_2)}+4\mathbf{U}_{(j_2-1)J+j_1,(j_2-1)J+j_1}+4\mathbf{U}_{(j_1-1)J+j_2,(j_1-1)J+j_2}\\
        &+4\mathbf{U}_{(j_2-1)J+j_1,(j_1-1)J+j_2}+4\mathbf{U}_{(j_1-1)J+j_2,(j_2-1)J+j_1},\\
        0=&\,3\mu_{j_2}^{(j_1)}+2\mu_{j_1}^{(j_2)}+9\mathbf{U}_{(j_2-1)J+j_1,(j_2-1)J+j_1}+4\mathbf{U}_{(j_1-1)J+j_2,(j_1-1)J+j_2}\\
        &+6\mathbf{U}_{(j_2-1)J+j_1,(j_1-1)J+j_2}+6\mathbf{U}_{(j_1-1)J+j_2,(j_2-1)J+j_1}.
    \end{align*}
    Here $\mu_{j_2}^{(j_1)}$ denotes the $j_1$-th entry of $\boldsymbol{\mu}_{j_2}$.
    Note that $\mathbf{U}_{(j_1-1)J+j_2,(j_2-1)J+j_1}=\mathbf{U}_{(j_2-1)J+j_1,(j_1-1)J+j_2}$ by symmetry.
    These equations imply that for all $j_1,j_2=1,\ldots,J$, we have
    \begin{align*}
        \mu_{j_2}^{(j_1)}=&\,\mu_{j_1}^{(j_2)}=\mathbf{U}_{(j_2-1)J+j_1,(j_2-1)J+j_1}=\mathbf{U}_{(j_1-1)J+j_2,(j_1-1)J+j_2}\\
        =&\,\mathbf{U}_{(j_2-1)J+j_1,(j_1-1)J+j_2}=\mathbf{U}_{(j_1-1)J+j_2,(j_2-1)J+j_1}=0.
    \end{align*}
    Hence, we have proved that $\boldsymbol{\mu}_{j}=\boldsymbol{0}$ for every $j=1,\ldots,J$ and $\mathbf{U}=\boldsymbol{0}$.
    Condition (d') is therefore verified.

    Under these two additional conditions, the following theorem establishes model identifiability.
    \begin{customthm}{1'}[Generic identifiability]\label{thm:iden}
        Under Conditions (c), (d'), (e), (f), (g), model (\ref{semimodel}) is generically identifiable.
    \end{customthm}
    \renewcommand*{\proofname}{Proof of Theorem \ref{thm:iden}}
    \begin{proof}
        Consider the following intensity functions under two competing semiparametric models:
    \begin{align*}
        &\lambda_{j}(t|\vec{X}_{j}, \vec{Z}_j ; \vec{\theta})=\lambda_{j0}(t)\exp(\vec{\beta}_{j}^{\top} \vec{X}_{j}(t)+\vec{\theta}^{\top}\mat{A}_{j}^{\top} \vec{Z}_j(t))\text{,}\\
        &\widetilde{\lambda}_{j}(t|\vec{X}_{j}, \vec{Z}_j ; \vec{\theta})=\widetilde{\lambda}_{j0}(t)\exp(\widetilde{\vec{\beta}}_{j}^{\top} \vec{X}_{j}(t)+\vec{\theta}^{\top}{\widetilde{\mat{A}}_{j}}^{\top} \vec{Z}_j(t))\text{.}
    \end{align*}
        Suppose the two parameter sets induce the same distribution of the observed multivariate counting process.
        Assume $Y_{j}(t)\equiv 1$ for all $j=1,\ldots,J$ and $t\in[0,\tau]$.
        For any fixed $t\in[0,\tau]$, we consider two sample paths: (i) no event occurs on $[0,t]$; (ii) exactly one event of type $k$ occurs at time $t$.
        Since $\boldsymbol{X}_j(s)=\boldsymbol{Z}_j(s)=\boldsymbol{0}$ for any $s\in[0,t]$ and $j=1,\ldots,J$ by Condition (g) in both sample paths, equality of the two likelihoods on $[0,t]$ gives
        \begin{align*}
            \prod_{j=1}^J\exp\left(-\int_{0}^t\lambda_{j0}(s)\mathrm{d}s\right)=&\,\prod_{j=1}^J\exp\left(-\int_{0}^t\widetilde{\lambda}_{j0}(s)\mathrm{d}s\right),\\
            \lambda_{k0}(t)\prod_{j=1}^J\exp\left(-\int_{0}^t\lambda_{j0}(s)\mathrm{d}s\right)=&\,\widetilde{\lambda}_{k0}(t)\prod_{j=1}^J\exp\left(-\int_{0}^t\widetilde{\lambda}_{j0}(s)\mathrm{d}s\right).
        \end{align*}
        By comparing the two equations, we can deduce that $\lambda_{j0}(t)=\widetilde{\lambda}_{j0}(t)$ for any $j=1,\ldots,J$ and almost every $t\in(0,\tau]$. When the true baseline intensity function is assumed to be left-continuous with right limits, the equality holds for any $t\in(0,\tau]$.

        We next identify the finite-dimensional parameters.
        For any ordered event-type sequence $h=(a_1,\ldots,a_m)$, we let these $m$ events occur at time $0<t_1<\ldots<t_m\leq\tau$.
        Let $t_0=0$.
        Matching the likelihood functions on $[0,t_m]$ leads to
        \begin{align}\label{semi1}
            &\int_{\mathbb{R}^K}\prod_{r=1}^m\lambda_{a_r}(t_r|\vec{X}, \vec{Z}; \vec{\theta})\exp\left(-\sum_{j=1}^J\sum_{r=1}^m\int_{t_{r-1}}^{t_r}\lambda_{j}(s|\vec{X}, \vec{Z}; \vec{\theta})\mathrm{d}s\right)\phi_K(\vec{\theta};\vec{0}, \mat{\Sigma}) \mathrm{d}\vec{\theta}\notag\\
            =&\,\int_{\mathbb{R}^K}\prod_{r=1}^m\widetilde{\lambda}_{a_r}(t_r|\vec{X}, \vec{Z}; \vec{\theta})\exp\left(-\sum_{j=1}^J\sum_{r=1}^m\int_{t_{r-1}}^{t_r}\widetilde{\lambda}_{j}(s|\vec{X}, \vec{Z}; \vec{\theta})\mathrm{d}s\right)\phi_K(\vec{\theta};\vec{0}, \widetilde{\mat{\Sigma}}) \mathrm{d}\vec{\theta}.
        \end{align}
        By Condition (g), substituting the explicit forms of the intensity functions in (\ref{semi1}) gives
        \begin{align}\label{semi2}
            &\int_{\mathbb{R}^K}\prod_{r=1}^m\lambda_{a_r0}(t_r)\exp(\vec{\beta}_{a_r}^{\top} \vec{x}_{a_r}(h_{r-1})+\vec{\theta}^{\top}\mat{A}_{a_r}^{\top} \vec{z}_{a_r}(h_{r-1}))\notag\\
            &\times\exp\left(-\sum_{j=1}^J\sum_{r=1}^m\exp(\vec{\beta}_{j}^{\top} \vec{x}_{j}(h_{r-1})+\vec{\theta}^{\top}\mat{A}_{j}^{\top} \vec{z}_{j}(h_{r-1}))\int_{t_{r-1}}^{t_r}\lambda_{j0}(s)\mathrm{d}s\right)\phi_K(\vec{\theta};\vec{0}, \mat{\Sigma}) \mathrm{d}\vec{\theta}\notag\\
            =&\,\int_{\mathbb{R}^K}\prod_{r=1}^m\widetilde{\lambda}_{a_r0}(t_r)\exp(\widetilde{\vec{\beta}}_{a_r}^{\top} \vec{x}_{a_r}(h_{r-1})+\vec{\theta}^{\top}\widetilde{\mat{A}}_{a_r}^{\top} \vec{z}_{a_r}(h_{r-1}))\notag\\
            &\times\exp\left(-\sum_{j=1}^J\sum_{r=1}^m\exp(\widetilde{\vec{\beta}}_{j}^{\top} \vec{x}_{j}(h_{r-1})+\vec{\theta}^{\top}\widetilde{\mat{A}}_{j}^{\top} \vec{z}_{j}(h_{r-1}))\int_{t_{r-1}}^{t_r}\widetilde{\lambda}_{j0}(s)\mathrm{d}s\right)\phi_K(\vec{\theta};\vec{0}, \widetilde{\mat{\Sigma}}) \mathrm{d}\vec{\theta}.
        \end{align}
        Note that we have already established that $\lambda_{a_r0}(t_r)=\widetilde{\lambda}_{a_r0}(t_r)$. Cancelling these terms and letting $t_1,\ldots,t_m$ tend to zero simultaneously while preserving their ordering, equation~\eqref{semi2} reduces to
        \begin{align}\label{semi3}
            &\int_{\mathbb{R}^K}\prod_{r=1}^m\exp(\vec{\beta}_{a_r}^{\top} \vec{x}_{a_r}(h_{r-1})+\vec{\theta}^{\top}\mat{A}_{a_r}^{\top} \vec{z}_{a_r}(h_{r-1}))\phi_K(\vec{\theta};\vec{0}, \mat{\Sigma}) \mathrm{d}\vec{\theta}\notag\\
            =&\,\int_{\mathbb{R}^K}\prod_{r=1}^m\exp(\widetilde{\vec{\beta}}_{a_r}^{\top} \vec{x}_{a_r}(h_{r-1})+\vec{\theta}^{\top}\widetilde{\mat{A}}_{a_r}^{\top} \vec{z}_{a_r}(h_{r-1}))\phi_K(\vec{\theta};\vec{0}, \widetilde{\mat{\Sigma}}) \mathrm{d}\vec{\theta}.
        \end{align}
        Using the explicit form of the Gaussian moment-generating function, equation~\eqref{semi3} can be further simplified to
        \begin{align*}
            &\exp\left(\sum_{r=1}^m\boldsymbol{\beta}_{a_r}^{\top}\boldsymbol{x}_{i,a_r}(h_{r-1})+\frac{1}{2}\left(\sum_{r=1}^m\breve{\boldsymbol{z}}_{i,a_r}(h_{r-1})\right)^{\top}\mathbf{U}\left(\sum_{r=1}^m\breve{\boldsymbol{z}}_{i,a_r}(h_{r-1})\right)\right)\\
            =&\,\exp\left(\sum_{r=1}^m\widetilde{\boldsymbol{\beta}}_{a_r}^{\top}\boldsymbol{x}_{i,a_r}(h_{r-1})+\frac{1}{2}\left(\sum_{r=1}^m\breve{\boldsymbol{z}}_{i,a_r}(h_{r-1})\right)^{\top}\widetilde{\mathbf{U}}\left(\sum_{r=1}^m\breve{\boldsymbol{z}}_{i,a_r}(h_{r-1})\right)\right).
        \end{align*}
        Here $\mathbf{U}=\mathbf{A}\mathbf{\Sigma}\mathbf{A}^{\top}\in\mathbb{R}^{L_2\times L_2}$ and $\widetilde{\mathbf{U}}=\widetilde{\mathbf{A}}\widetilde{\mathbf{\Sigma}}\widetilde{\mathbf{A}}^{\top}\in\mathbb{R}^{L_2\times L_2}$.
        Therefore, Condition (d') implies that $\vec{\beta}_j=\widetilde{\vec{\beta}}_j$ for $j=1,\ldots,J$ and $\mathbf{U}=\widetilde{\mathbf{U}}$, which in turn implies that $(\mathbf{A},\mathbf{\Sigma})\sim (\widetilde{\mathbf{A}},\widetilde{\mathbf{\Sigma}})$ by Condition (c).
    \end{proof}

    It is worth mentioning that this section discusses a model that contains only internal covariates for simplicity.
    If both internal and external covariates are included, the corresponding regularity conditions must be imposed separately. For the internal covariates, Conditions~(b), (d'), (e), and (g) are required. For the external (possibly time-dependent) covariates, we require the richness condition at time $t=0$, which corresponds to Condition~2(g) in \cite{parner1998asymptotic}.

    For the semiparametric extension, a natural approach would be to adopt the profile-likelihood approach and estimate the model parameters using the nonparametric maximum likelihood estimator (NPMLE); see \citep{zeng2007semiparametric,zeng2010general}. The invertibility of the information operator can be established by adapting the arguments in \cite{zeng2007semiparametric} and \cite{zeng2010general}.
    To derive the asymptotic properties of the semiparametric estimator, it is natural to require the covariate processes to have uniformly bounded total variation. Under Condition~(g), the piecewise-constant internal covariates can jump only when an event occurs. Together with the assumed uniform boundedness of the covariates, it is therefore sufficient to impose the following condition to ensure uniformly bounded total variation:
    \begin{itemize}
        \item [(h)] There exists a uniform constant $D<\infty$ such that $N_{ij}(\tau)\leq D$ almost surely for all subjects $i=1,\ldots,n$ and event types $j=1,\ldots,J$.
    \end{itemize}
    This condition together with Conditions (b) and (g) provides the bounded-variation control needed in the asymptotic analysis.
    Therefore, asymptotic normality can be established by adapting the arguments in \cite{zeng2007semiparametric} and \cite{zeng2010general}.
    Finally, the model-selection consistency can be similarly established.
\end{document}